\documentclass[aps,pre,reprint,superscriptaddress,longbibliography]{revtex4-1}

\usepackage{color}
\usepackage{amsmath}
\usepackage{bm}
\usepackage{amssymb}
\usepackage{wrapfig}
\usepackage{braket}
\usepackage{amsthm}
\usepackage{natbib}
\usepackage{dsfont}
\usepackage{graphicx}
\usepackage[colorlinks = true, citecolor = blue, linkcolor = blue, urlcolor=blue]{hyperref}

\definecolor{green}{rgb}{0.0, 0.5, 0.0}
\definecolor{red}{rgb}{1.0, 0.0, 0.0}
\definecolor{blue}{rgb}{0.0, 0.0, 0.5}

\DeclareFontFamily{OMX}{MnSymbolE}{}
\DeclareFontShape{OMX}{MnSymbolE}{m}{n}{
	<-6>  MnSymbolE5
	<6-7>  MnSymbolE6
	<7-8>  MnSymbolE7
	<8-9>  MnSymbolE8
	<9-10> MnSymbolE9
	<10-12> MnSymbolE10
	<12->   MnSymbolE12}{}
\DeclareSymbolFont{mnlargesymbols}{OMX}{MnSymbolE}{m}{n}
\SetSymbolFont{mnlargesymbols}{bold}{OMX}{MnSymbolE}{b}{n}
\DeclareMathDelimiter{\LL}{\mathopen}{mnlargesymbols}{'164}{mnlargesymbols}{'164}
\DeclareMathDelimiter{\rr}{\mathclose}{mnlargesymbols}{'171}{mnlargesymbols}{'171}

\usepackage{verbatim}
\usepackage[T1]{fontenc}
\usepackage[english]{babel}
\usepackage[utf8]{inputenc}

\newcommand{\Tr}{{\mathrm{Tr}}}
\newcommand{\Cmm}{\mathcal{C}_\text{MM}}
\newcommand{\Cp}{\mathcal{C}_{+}}
\newcommand{\Ctp}{\tilde{\mathcal{C}}_{+}}
\newcommand{\Ccl}{\mathcal{C}_{\text{cl}}}
\newcommand{\Cbcl}{{\overline{\mathcal{C}}_{\text{cl}}}}
\newcommand{\Ctcl}{{\tilde{\mathcal{C}}_{\text{cl}}}}

\renewcommand{\P}{{\mathcal{P}}}
\renewcommand{\L}{{\mathcal{L}}}
\renewcommand{\S}{{\mathcal{S}}}
\newcommand{\J}{{\mathcal{J}}}
\newcommand{\I}{{\mathcal{I}}}
\newcommand{\X}{{\mathcal{X}}}
\newcommand{\LMM}{{\mathcal{L}_\text{MM}}}
\newcommand{\rhoss}{{\rho_\text{ss}}}

\newcommand{\trho}{{\tilde{\rho}}}
\newcommand{\Q}{{\mathcal{Q}}}
\newcommand{\M}{{\mathcal{M}}}

\newcommand{\U}{{\mathcal{U}}}
\newcommand{\V}{{\mathcal{V}}}
\newcommand{\Ubf}{{\mathbf{U}}}
\newcommand{\G}{{\mathcal{G}}}
\newcommand{\Gbf}{{\mathbf{G}}}
\newcommand{\Ibf}{{\mathbf{I}}}
\newcommand{\Pibf}{{\mathbf{\Pi}}}

\newcommand{\C}{{\mathbf{C}}}
\newcommand{\p}{{\mathbf{p}}}
\newcommand{\pss}{\mathbf{p}_\text{ss}}
\newcommand{\ptss}{{\mathbf{\tilde{p}}_\text{ss}}}
\newcommand{\pt}{{\mathbf{\tilde{p}}}}
\newcommand{\W}{{\mathbf{W}}}
\newcommand{\Wt}{{\mathbf{\tilde{W}}}}
\newcommand{\Pbf}{{\mathbf{P}}}
\newcommand{\Pss}{{\mathbf{P}_\text{\!\!ss}}}
\newcommand{\Ptss}{{\mathbf{\tilde{P}}_\text{\!\!ss}}}
\newcommand{\Pt}{{\mathbf{\tilde{P}}}}
\newcommand{\R}{{\mathbf{R}}}
\newcommand{\Rt}{{\mathbf{\tilde{R}}}}
\newcommand{\Jt}{{\mathbf{\tilde{J}}}}
\newcommand{\Jbf}{{\mathbf{J}}}
\newcommand{\mut}{{\tilde{\mu}}}
\newcommand{\Mt}{{\mathbf{\tilde{M}}}}
\newcommand{\mt}{{\mathbf{\tilde{m}}}}
\newcommand{\m}{{\mathbf{m}}}
\newcommand{\Xt}{{\mathbf{\tilde{X}}}}
\newcommand{\xt}{{\mathbf{\tilde{x}}}}
\newcommand{\T}{{\mathbf{T}}}
\newcommand{\Tt}{{\mathbf{\tilde{T}}}}

\makeatletter
\def\@bibdataout@aps{%
	\immediate\write\@bibdataout{%
		@CONTROL{%
			apsrev41Control%
			\longbibliography@sw{%
				,author="08",editor="1",pages="1",title="0",year="1"%
			}{%
				,author="08",editor="1",pages="1",title="",year="1"%
			}%
		}%
	}%
	\if@filesw \immediate \write \@auxout {\string \citation {apsrev41Control}}\fi 
}
\makeatother

\begin{document}
\title{Theory of classical metastability in open quantum systems}

\author{Katarzyna Macieszczak}
\affiliation{TCM Group, Cavendish  Laboratory,  University  of  Cambridge,	J.  J.  Thomson  Avenue,  Cambridge  CB3  0HE,  United Kingdom}

\author{Dominic C. Rose}
\affiliation{School of Physics and Astronomy, University of Nottingham, University Park,  Nottingham NG7 2RD, United Kingdom}
\affiliation{Centre for the Mathematics and Theoretical Physics of Quantum Non-Equilibrium Systems, University of Nottingham, University Park,  Nottingham NG7 2RD, United Kingdom}

\author{Igor Lesanovsky}
\affiliation{Institut f\"ur Theoretische Physik, Universit\"at T\"ubingen, Auf der Morgenstelle 14, 72076 T\"ubingen, Germany}
\affiliation{School of Physics and Astronomy, University of Nottingham, University Park,  Nottingham NG7 2RD, United Kingdom}
\affiliation{Centre for the Mathematics and Theoretical Physics of Quantum Non-Equilibrium Systems, University of Nottingham, University Park,  Nottingham NG7 2RD, United Kingdom}

\author{Juan P. Garrahan}
\affiliation{School of Physics and Astronomy, University of Nottingham, University Park,  Nottingham NG7 2RD, United Kingdom}
\affiliation{Centre for the Mathematics and Theoretical Physics of Quantum Non-Equilibrium Systems, University of Nottingham, University Park,  Nottingham NG7 2RD, United Kingdom}

\begin{abstract}
	We present a general theory of classical metastability in open quantum systems. Metastability is a~consequence of a large separation in timescales in the dynamics, leading to the existence of a regime when states of the system appear stationary, before eventual relaxation toward a true stationary state at much larger times. In this work, we focus on the emergence of classical metastability, i.e., when metastable states of an open quantum system with separation of timescales can be approximated as probabilistic mixtures of a finite number of states. We find that a number of classical features follow from this approximation, for the manifold of metastable states, long-time dynamics between them, and symmetries of the dynamics.
Namely, those states are approximately disjoint and thus play the role of metastable phases, the relaxation toward the stationary state is approximated by a classical stochastic dynamics between them, and weak symmetries correspond to their permutations. Importantly, the classical dynamics is observed not only on average, but also at the level of  individual quantum trajectories: We show that time coarse-grained continuous measurement records can be viewed as noisy classical trajectories, while their statistics can be approximated by that of the classical dynamics. Among others, this explains how first-order dynamical phase transitions arise from metastability. 
Finally, to verify the presence of classical metastability in a given open quantum system, we develop an efficient numerical approach that delivers the set of metastable phases together with the effective classical dynamics. 
Since the proximity to a first-order dissipative phase transition manifests as metastability, the theory and tools introduced in this work can be used to investigate such transitions---which occur in the large size limit---through the metastable behavior of many-body systems of moderate sizes accessible to numerics. 
\end{abstract}


\maketitle

\tableofcontents

\section{Introduction}
With continuing advances in the control of experimental platforms used as quantum simulators, such as ultracold atomic gases, Rydberg atoms and circuit quantum-electrodynamics~\cite{Pritchard2010,Blatt2012,Britton2012,Dudin2012,Peyronel2012,Guenter2013,Schmidt2013}, a broad range of nonequilibrium phenomena of open many-body quantum systems has been observed recently. Theoretical studies have progressed via the combination of methods from atomic physics, quantum optics and condensed matter, giving rise to a range of techniques including quantum jump Monte Carlo (QJMC)~\cite{Molmer92,Dum1992,Molmer93,Plenio1998,Daley2014} simulations via tensor network~\cite{Gangat2017}, and field theoretical approaches~\cite{Torre2013,Sieberer2016,Maghrebi2016}.

Often the focus of studies on nonequilibrium open many-body quantum systems is a phase diagram of the stationary state, and the related question of the structure of dissipative phase transitions occurring in the thermodynamic limit of infinite system size. This includes whether such systems can exhibit bistability (or multistability) of the stationary state, a topic covered both theoretically~\cite{Mendoza-Arenas2016,Foss-Feig2017,Casteels2017,Letscher2017,Jin2018} and experimentally~\cite{Melo2016,Rodriguez2017}, and which order parameters are relevant for distinguishing the coexisting phases. Mean-field results often suggest multiple stationary states in the thermodynamic limit~\cite{Ates2012,Maghrebi2016,Landa2020}, however, more sophisticated (albeit still approximate) techniques such as variational approaches~\cite{Weimer2015,Weimer2015a,Overbeck2017}, perturbative expansions in lattice connectivity~\cite{Biondi2017,Landa2020}, infinite tensor network simulations~\cite{Gangat2017} or a field-theoretical analysis~\cite{Maghrebi2016} can still indicate a unique stationary state.

While it is unusual to see phase transitions at finite system sizes~\cite{Spohn1977,Evans1977,Schirmer2010,Nigro2019}, first-order phase transitions in stationary states manifest at large enough finite system sizes~\cite{Chaikin2000} through the occurrence of \emph{metastability}, i.e., distinct timescales in the evolution of the system statistics: classically, in the probability distribution over configuration space~\cite{Gaveau1987,Gaveau1998,Bovier2002,Gaveau2006,Kurchan2016}; quantum mechanically, in the density matrix~\cite{Macieszczak2016a,Rose2016}.
The statistics of such systems at long times can be understood in terms of metastable phases which generally correspond to the phases on either side of the transition being distinct from the unique stationary state for a given set of parameters. 
Therefore, already at a finite system size the structure of a possible first-order dissipative phase transition can be fully determined by investigating metastable states of the system~\cite{Rose2016,Minganti2018}, which is of particular importance for many-body open quantum systems, where exact methods are often limited to numerical simulations of finite systems of modest size. 

Metastability can also emerge in complex relaxation toward a unique stationary state, even without a phase transition present in the thermodynamic limit. This is the case in classical kinetically constrained models~\cite{Jaeckle1991,Sollich1999,Garrahan2002,Sollich2003,Binder2011,Biroli2013} and spin glasses~\cite{Binder1986} and recent open quantum generalizations of these models~\cite{Olmos2012,Lesanovsky2013,Olmos2014} and~\cite{Cugliandolo1999}. Here, the study of metastability can unfold the long-time dynamics responsible for the complex relaxation to the stationary state~\cite{Rose2020}, with metastable phases corresponding to dynamical rather than static phases.

\begin{figure}[t!]
	\includegraphics[width=0.9\columnwidth]{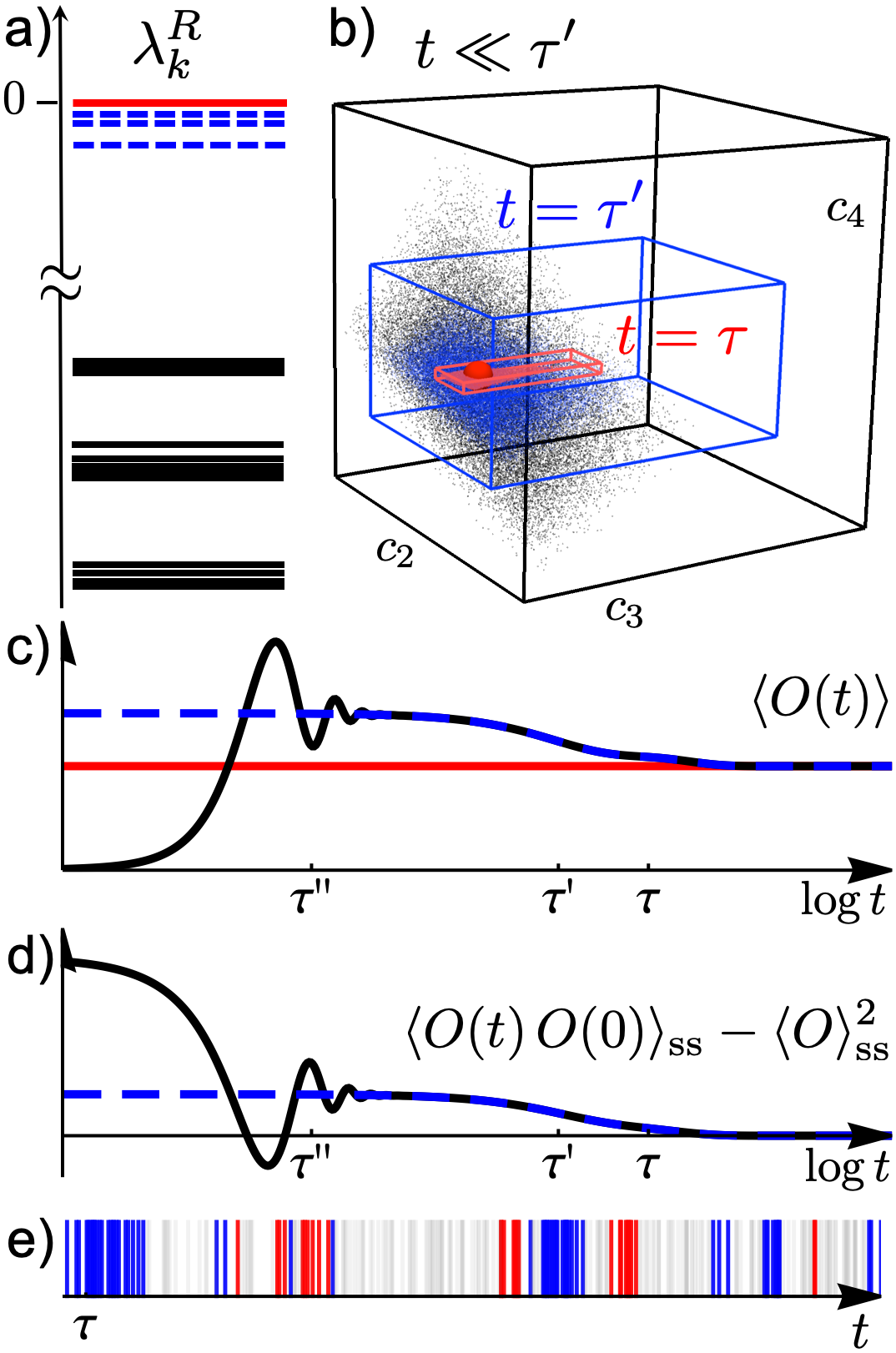}
		\vspace*{-2mm}
	\caption{\label{fig:Spectrum}\textbf{Metastability in Markovian open quantum systems}: 
		 \textbf{(a)} Metastability corresponds to  a separation in the real part of master operator spectrum, between $m\!-\!1$ slow (blue dashed) and fast modes (black solid), while the stationary state corresponds to $0$ eigenvalue (red solid); here $m=4$.
	\textbf{(b)} The manifold of metastable states is described by coefficients $c_k$ [Eq.~\eqref{Expansion}], $k=2,...,m$, of decomposition between slow modes (dots for random initial pure states). The long-time dynamics takes place within that manifold, with the exponential decay of the coefficients toward the stationary state (red sphere) [Eqs.~\eqref{Expansion1} and~\eqref{Expansion2}].
	 Metastability can be observed experimentally as a plateau in the dynamics of observable averages \textbf{(c)} or two-point correlations \textbf{(d)} appearing during the metastable regime [Eqs.~\eqref{ExpansionOB1} and~\eqref{ApproxCor}].  
	Black (solid) lines show observable dynamics, blue (dashed) lines the approximation by slow modes holding after the initial relaxation, and red (solid) lines the stationary value achieved after the final relaxation. 
		\textbf{(e)} Long timescales can also be observed in continuous measurement records, e.g., as intermittence in detection of quanta emitted due to jumps occurring in the system (two types shown in blue and red; gray---without associated quanta), with regimes of jump activity having a length comparable to the long-time relaxation timescale.  See Sec.~\ref{app:model} in the~SM for details on the model.
		\vspace*{-5mm}}
\end{figure}

For classical systems with Markovian dynamics~\cite{Gaveau1987,Gaveau1998,Bovier2002,Gaveau2006,Kurchan2016} and open quantum systems~\cite{Macieszczak2016a,Thingna2016} described by the master equation formalism~\cite{Lindblad1976,Gorini1976}, metastability necessarily requires a large separation in the spectrum of the master operator governing the system evolution.  This separation leads to metastable states residing in a space of a reduced dimension given by the slow eigenmodes of the master operator, and long-time dynamics taking place within that space. Since the slow modes themselves do not represent system states, a general structure of the metastable manifold (MM) is not known, but conjectured to feature disjoint phases, decoherence free subspaces and noiseless subsystems, while the long-time dynamics is expected to be analogous to perturbative dynamics on such states~\cite{Macieszczak2016a}.

In this work, we comprehensively prove this conjecture for \emph{classical metastability} in open quantum systems.
 We define classical metastability as the case, where metastable states can be approximated as probability distributions over a set of $m$ states, where $m-1$ is the number of slow eigenmodes in the dynamics. We then show that this definition is equivalent to a simple geometric criterion, which can be verified using  the exact diagonalization of the master operator. 
Crucially, the corresponding corrections play the role of a \emph{figure of merit} in emergent classical properties of the manifold of metastable states and its long-time dynamics. Namely, we show that, for classical metastability, $m$ states can be considered as distinct \emph{metastable phases}, as they are approximately disjoint and orthogonal to one another, while their basins of attraction form the set of $m$ \emph{order parameters} to distinguish them. Furthermore, we find that the long-time dynamics of the system can be approximated as a \emph{classical stochastic dynamics} between the metastable phases. This holds in the average system dynamics as well as in individual quantum trajectories~\cite{Gardiner2004,Daley2014}, as obtained via individual runs of an experimental system or from QJMC simulations, where classical trajectories arise via coarse graining of these in time. The classical dynamics between the metastable phases is then responsible for the occurrence of intermittence~\cite{Ates2012,Lesanovsky2013,Rose2016} or dynamical heterogeneity~\cite{Olmos2012,Olmos2014} in quantum trajectories, leading to multimodal statistics of continuous measurements and system proximity to  a first-order dynamical phase transition~\cite{Garrahan2010}. Therefore, classical metastability is a phenomenon occurring not only on average, but in dynamics of individual quantum trajectories.  All these results are also discussed in the presence of further hierarchy of relaxation timescales. Finally, while our approach does not rely on the presence of any symmetries of the dynamics~\cite{Baumgartner2008,Buvca2012,Albert2014}  (cf.~Ref.~\cite{Minganti2018}), we also show that the set of metastable phases is approximately invariant under any present symmetries. Thus, weak symmetries lead to approximate cycles of metastable phases and permutation symmetries of the classical long-time dynamics, and, as such, we find that any nontrivial continuous symmetries of slow eigenmodes of the dynamics preclude classical metastability.

To verify the classicality of metastability present in a general open quantum system and uncover the set of metastable phases together with the effective structure of long-time dynamics, we develop an efficient numerical technique, which can be further simplified  when a weak symmetry is present.
Our approach relies on the ability to diagonalize the system master operator, which is usually possible only for moderate system sizes, while metastability may become prominent only for large system sizes. To mitigate this potential issue, we show that for classical metastability accompanied by intermittence or dynamical heterogeneity in quantum trajectories, metastable phases can be extracted from quantum trajectories through the use of large-deviation methods, such as the ``thermodynamics of trajectories''~\cite{Garrahan2010,Garrahan2011,Ates2012,Lesanovsky2013,Carollo2018}. Therefore, there is potential to study classical metastability using QJMC simulations, which are generally feasible at quadratically larger system sizes than exact diagonalization of the generator.

This paper is organized as follows. 
In Sec.~\ref{sec:MM}, we review the results of Ref.~\cite{Macieszczak2016a}. 
In Sec.~\ref{sec:cMM}, we introduce the general approach to classical metastability in open quantum systems. We then discuss the resulting classical structure of the MM
in Sec.~\ref{sec:disjoint}. The effectively classical system dynamics emerging at large times is discussed in Sec.~\ref{sec:Leff}. We refine these general results considering symmetries of the system dynamics
in Sec.~\ref{sec:symmetry}. Finally, we introduce numerical approaches to unfold the structure of classical metastability in Sec.~\ref{sec:numerics}. Details and proofs of our results are presented in the Supplemental Material (SM). Our results are illustrated in Figs.~\ref{fig:Spectrum}--\ref{fig:Symmetry} with a system in proximity to an effectively classical dissipative phase transition occurring at finite size. We discuss classical metastability arising for such systems in Sec.~\ref{app:DPT} of the SM.
The application of the general methods introduced in this paper to a many-body system beyond this class is given in the accompanying paper~\cite{Rose2020} which studies in detail the metastability of the open quantum East glass model~\cite{Olmos2012}.

\section{Metastability in open quantum systems} \label{sec:MM}

We begin by reviewing the spectral theory of metastability of Ref.~\cite{Macieszczak2016a}. We then introduce a quantitative description of those results by considering the corrections to the stationarity during the metastable regime. In the next section, we build on this to understand when metastability in open quantum systems becomes classical, which is the main focus of this work.

\subsection{Dynamics of open quantum systems}\label{sec:master}

We consider a finitely dimensional open quantum system with ts average state at time $t$ described by a density matrix $\rho(t)$ evolving according to a master equation as $d\rho(t)/dt=\L[\rho(t)]$, where the master operator~\cite{Lindblad1976,Gorini1976}
\begin{equation}\label{eq:L}
\L(\rho) = -i[H,\rho]+\sum_{j}\left({J}_{j}\rho{J}_{j}^{\dagger}-\frac{1}{2}\left\{{J}_{j}^{\dagger}{J}_{j},\rho\right\}\right).
\end{equation}
Here, $H$ is the system Hamiltonian, while the jump operators ${J}_{j}$ provide coupling of the system to the surrounding environment (in this work we explicitly denote any time dependence; in particular, the master equation is time-independent).  
If the interactions between the system and the environment are associated to emissions of an energy quanta, then the action of jump operators can be detected through continuous measurements~\cite{Gardiner2004}, e.g., counting of photons emitted by atoms coupled to the vacuum electromagnetic field~\cite{Ates2012,Olmos2012,Lesanovsky2013,Olmos2014}.  Equation~\eqref{eq:L} is a general dynamics of a time-homogeneous Markovian open quantum system~\cite{Lindblad1976,Gorini1976}, which arises for systems interacting weakly with an effectively memoryless environment~\cite{Gardiner2004}.

Since the master operator $\L$ acts linearly on $\rho(t)$, the evolution can be understood in terms of its eigenmatrices ${R}_{k}$  and their corresponding eigenvalues ${\lambda}_{k}={\lambda}_{k}^{R}+i\,{\lambda}_{k}^{I}$~\footnote{It is possible that the master operator may not be completely diagonalizable, in which this case its simplest form is the Jordan normal form. Under exponentiation, this gives rise to a polynomial dependence on the time, however this is accompanied by the usual exponential evolution. Our results in Secs.~\ref{sec:cMM},~\ref{sec:disjoint},~\ref{sec:EffDynAv},~\ref{sec:symmetry}, and~\ref{sec:numerics} carry directly to that case by assuming that after the initial relaxation, any polynomial evolution of fast modes is dominated by the decaying exponential and can be neglected [cf.~Eq.~\eqref{Expansion1}], while during the metastable regime dynamics of the $m$ slow modes can be replaced by no evolution [cf.~Eq.~\eqref{Expansion}]. In Sec.~\ref{sec:QTraj}, the corrections to Eqs.~\eqref{eq:Kvar_t_approx_CL} and~\eqref{eq:kvar_ms} need to be adjusted, together with corresponding Secs.~\ref{app:var_t}, \ref{app:var_ms}, and~\ref{app:distribution} in the~SM.}.
The real parts of these eigenvalues are not greater than $0$, ${\lambda}_{k}^{R}\leq 0$, as the dynamics in Eq.~\eqref{eq:L} is (completely) positive and trace preserving; we order the eigenvalues by decreasing real part $\lambda_k^R$. In particular, zero eigenvalues correspond to stationary states~\cite{Baumgartner2008,Albert2016}. In this work, we assume a generic case of a unique stationary state $R_1={\rho}_{\rm ss}$~\cite{Spohn1977,Evans1977,Schirmer2010,Nigro2019}.
The system state at time $t$ can be then decomposed as
\begin{equation}\label{Expansion0}
\rho(t)={e}^{t\L}[\rho(0)]={\rho}_{\rm ss}+\sum_{k\geq 2}{c}_{k}{e}^{t{\lambda}_{k}}{R}_{k},
\end{equation}
where the coefficients ${c}_{k}\equiv\Tr[{L}_{k}\rho(0)]$ are bounded by the eigenvalues of $L_k$, with $L_k$ being eigenmatrices of $\L^\dagger$ normalized such that $\Tr({L}_{k}{R}_{l})={\delta}_{kl}$ (there is a freedom of choice to normalize by scaling either $R_k$ or $L_k$). The values of these coefficients for a given physical state are closely tied, such that the corresponding linear combination of $R_k$ results in a positive matrix. We refer to $L_k$ and $R_k$ as left and right eigenmatrices (eigenmodes), respectively.  Note that the trace preservation of the dynamics implies that $L_1=\mathds{1}$, and thus beyond $\rhoss$ other right eigenmatrices do not correspond to quantum states, $\Tr(R_k)=\Tr(L_1 R_k)= 0$ for $k\geq 2$.  The timescale ${\tau}$ of the final relaxation to ${\rho}_{\rm ss}$ from Eq.~\eqref{Expansion0} can be seen to depend on the gap in the spectrum, ${\tau}\geq-1/{\lambda}_{2}^{R}$.

\subsection{Spectral theory of metastability}\label{sec:spectral}

Metastability corresponds to a large separation in the real part of the spectrum~\cite{Macieszczak2016a}, ${\lambda}_{m}^{R}/{\lambda}_{m+1}^{R}\ll 1$, which denotes the ratio of eigenvalues being of a lower order than $1$; see Fig.~\ref{fig:Spectrum}\textcolor{blue}{(a)}. 
Time $t$ after the initial relaxation 
 correspond to the terms beyond the $m$-th in the sum in Eq.~\eqref{Expansion0} being negligible, ${e}^{t{\lambda}_{k}}\approx 0$ for $k\geq m+1$, and the reduced expansion 
\begin{equation}\label{Expansion1}
\rho(t)={\rho}_{\rm ss}+\sum_{k=2}^m {c}_{k} e^{t\lambda_k}{R}_{k}+...,
\end{equation}
where $...$ stands for negligible corrections [cf.~Eq.~\eqref{Expansion0}].

When the separation in the spectrum is big enough, it is possible to further consider times when decay of the remaining terms can be neglected, ${e}^{t{\lambda}_{k}^R}\approx 1$ for $k\leq m$; cf.~Ref.~\cite{Bellomo2017}. This is the \emph{metastable regime}, during which the system state is approximately stationary, i.e., metastable, as captured by
\begin{equation}\label{Expansion}
\rho(t)={\rho}_{\rm ss}+\sum_{k=2}^m {c}_{k}{R}_{k}+...\equiv\P[\rho(0)]+...,
\end{equation}
where we defined $\P$ as the projection onto the low-lying eigenmodes of the master operator, which is trace and Hermiticity preserving~\footnote{Since $\L$ is Hermiticity preserving, if $R_k$ ($L_k$) is a right (left) eigenmode of the dynamics with an eigenvalue $\lambda_k$, so is $R_k^\dagger$ ($L_k^\dagger$) with the eigenvalue $\lambda_k^*$. Thus the set of low-lying modes is invariant under the Hermitian conjugation.}. From Eq.~\eqref{Expansion} the manifold of metastable states is fully characterized by the bounded coefficients $(c_2,...,c_m)$ and thus it is $(m-1)$-dimensional. The MM is also convex, as a linear transformation of the convex set of initial states  [see Fig.~\ref{fig:Spectrum}\textcolor{blue}{(b)}]. 

At later times, only the slow modes contribute to the evolution [cf.\ Eq.~\eqref{Expansion1}]. Therefore, the dynamics toward the stationary state takes place essentially inside the MM [see Fig.~\ref{fig:Spectrum}\textcolor{blue}{(b)}],
\begin{equation}\label{Expansion2}
\rho(t)= e^{t \L_{\text{MM}}} \P[\rho(0)]+...,
\end{equation}
and is generated by [cf.~Figs.~\ref{fig:Spectrum}\textcolor{blue}{(c)} and~\ref{fig:Spectrum}\textcolor{blue}{(d)}]
\begin{equation}\label{eq:Leff}
\L_{\text{MM}}\equiv\P\L\P.
\end{equation}

Denoting by $\tau''$ the timescale of the initial relaxation, from Eq.~\eqref{Expansion1} we have  $\tau''\geq-1/{\lambda}_{m+1}^{R}$. Similarly, for $\tau'$  being the smallest timescale of the long-time dynamics we have $\tau'\leq-1/{\lambda}_{m}^{R}$ from Eq.~\eqref{Expansion2}. Then for times within the metastable regime we have $\tau''\leq t\ll \tau'$ from Eq.~\eqref{Expansion} [cf. Fig.~\ref{fig:Spectrum}\textcolor{blue}{(b)}].

Metastability can be observed in the behavior of statistical quantities such as expectation values or autocorrelations of system observables~\cite{Macieszczak2016a,Sciolla2015,Rose2016}. 
For a system observable, e.g., spin magnetization, we have
\begin{eqnarray}\nonumber
\left\langle{O}(t)\right\rangle&\equiv&\Tr\left[ O\, \rho(t)\right]=\Tr\left\{ O\, {e}^{t\L} [\rho(0)] \right\}\\\label{ExpansionOB0}
&=& \langle O\rangle_\text{ss}+\sum_{k} {b}_k\,{c}_{k}  \,{e}^{t{\lambda}_{k}},
\end{eqnarray}
where we introduced decomposition of the observable into the left eigenmodes with the coefficients $b_k\equiv \Tr( O\, R_k)$ [cf.~Eq.~\eqref{Expansion0}], and $b_1=\langle O\rangle_\text{ss}=\Tr( O\,\rhoss)$ is the static average. After the initial relaxation, the contribution from fast modes can be neglected [cf.~Eq.~\eqref{Expansion1}], 
\begin{eqnarray}\label{ExpansionOB1}
\left\langle{O}(t)\right\rangle&=& \langle O\rangle_\text{ss}+\sum_{k=2}^m {b}_k\,{c}_{k} \,{e}^{t{\lambda}_{k}} +...\\\nonumber
&=& \Tr\left\{ O\, {e}^{t\LMM} \P[\rho(0)] \right\}+...,
\end{eqnarray}
and the observable dynamics in Eq.~\eqref{ExpansionOB0} is accurately captured by the effective long-time dynamics in Eq.~\eqref{eq:Leff}. Importantly, during the metastable regime, the observable average is approximately stationary [cf.~Eq.~\eqref{Expansion1}], before the final relaxation  to $ \langle O\rangle_\text{ss}$  [see Fig.~\ref{fig:Spectrum}\textcolor{blue}{(c)}], allowing for a direct observation of the metastability. This, however, requires preparation of an initial system state different from the stationary state, $\rho(0)\neq \rhoss$, something often difficult to achieve in experimental settings. Nevertheless, for the system in the stationary state,  metastability can be observed as double-step decay in the time-autocorrelation of a system observable. This is a consequence of  the first measurement perturbing the stationary state, thus causing its subsequent evolution, which for times after the initial relaxation follows the effective dynamics [cf.~Eq.~\eqref{ExpansionOB1}],
\begin{eqnarray}\label{ApproxCor}
\langle O(t)O(0)\rangle_\text{ss}-\langle O\rangle_\text{ss}^2&=&\Tr[\mathcal{O}{e}^{t\L}\mathcal{O}(\rhoss)]- \langle O\rangle_\text{ss}^2\\\nonumber &=&\Tr[\mathcal{O}{e}^{t\L_{\text{MM}}}\P\mathcal{O}(\rhoss)]- \langle O\rangle_\text{ss}^2+..., 
\end{eqnarray}
where $\mathcal{O}$ denotes the superoperator representing the measurement of the observable $O$ on a system state~\cite{Macieszczak2016a,Rose2016}. The autocorrelation initially decays from the observable variance in the stationary state, $\langle O^2\rangle_\text{ss}-\langle O\rangle_\text{ss}^2$,  to the plateau at $\Tr[\mathcal{O}\P\mathcal{O}(\rhoss)]-\langle O\rangle_\text{ss}^2$ in the metastable regime, and afterwards to $0$ during the final relaxation [see Fig.~\ref{fig:Spectrum}\textcolor{blue}{(d)}].

\subsection{Quantitative approach}\label{sec:MM_Cmm}

In this work, we introduce a quantitative description of metastability. We later use this approach to prove our main results: emerging classical features of metastable manifold, long-time dynamics, and weak symmetries in the case of classical metastability. 

	We consider errors of the approximation of the system dynamics by the projection on the low-lying modes of the spectrum in Eq.~\eqref{Expansion} during a time regime $t''\leq t\leq t'$,
		\begin{eqnarray}\label{eq:Cmm}
			\Cmm(t'',t')&\equiv&\sup_{\rho(0)} \sup_{t''\leq t\leq t'} \lVert\rho(t)-\P[\rho(0)] \rVert\\\nonumber
			&=&   \sup_{t''\leq t\leq t'}\lVert e^{t\L} -\P\rVert,
		\end{eqnarray}
		which we refer to as the \emph{corrections to the stationarity}. Here, $\lVert X\rVert=\Tr(\sqrt{X^\dagger X})$ denotes the trace norm for an operator $X$, while for a superoperator it denotes the norm induced by the trace norm~\footnote{We restrict the discussion to the real space of Hermitian operators. In this case, the induced norm of a superoperator can be shown to be achieved for a pure state (i.e., a rank one operator). Let $X$ be a Hermitian operator with eigenvalues $x_n$ and projections on the corresponding eigenstates denoted as $\rho_n$. For a superoperator $\mathcal{Y}$, we have $\lVert\mathcal{Y}[X]\rVert/\lVert X\rVert\leq (\sum_n |x_n| \lVert\mathcal{Y}[\rho_n]\rVert) /(\sum_n |x_n|)\leq \max_n \lVert\mathcal{Y}[\rho_n]\rVert$.}. 
		For the time regime such that 
	\begin{eqnarray}\label{eq:Cmm2}
		\Cmm(t'',t')&\ll& 1,
	\end{eqnarray}
	the corrections in Eq.~\eqref{Expansion}  are negligible (note that density matrices are normalized in the trace norm), which requires $t''>-1/\lambda_{m+1}^R$ and $t'\ll -1/\lambda_m^R$ (i.e.,  $-\lambda_m^R t'\ll 1$); see Sec.~\ref{app:defMM}  in the~SM.  We refer to such a time regime as a \emph{metastable time regime}.

   We now argue that the corrections to the stationarity can be considered as the central figure of merit in the theory of metastability.
	Indeed, the \emph{corrections to the positivity} of metastable states projected on the low-lying modes are defined by the distance to the set of density matrices,
\begin{eqnarray}\label{eq:Cp}
	\Cp&\equiv&\sup_{\rho(0)} \inf_{\rho} \,\lVert\P[\rho(0)]-\rho \rVert\\\nonumber
	&=&\sup_{\rho(0)}\, \lVert \P[\rho(0)]\rVert-1 \equiv \lVert \P\rVert-1,
\end{eqnarray}
with $\rho$ and $\rho(0)$ being density matrices~\cite{Note3} (see Sec.~\ref{app:MM_P} in the~SM), and can be bounded by the corrections to the stationarity in Eq.~\eqref{eq:Cmm}, by considering the distance to  $\rho\equiv \rho(t)$  within the metastable regime, 
	\begin{eqnarray}\label{eq:Cp2}
	\Cp&\leq& \inf_{t''\leq t\leq t'}\lVert e^{t\L} -\P\rVert\equiv \Ctp(t'',t')\leq \Cmm(t'',t').\quad
\end{eqnarray}
Furthermore, the corrections to the stationarity in Eq.~\eqref{eq:Cmm} establish a bound not only on Eq.~\eqref{eq:Cp}, but also  on
the corrections in  Eqs.~(\ref{Expansion0}),~(\ref{Expansion2}),~\eqref{ExpansionOB1}, and~\eqref{ApproxCor}.
In fact, beyond the metastable regime, the corrections in Eq.~\eqref{Expansion2} \emph{decay exponentially}, as in the leading order they can be shown to be bounded by $2\Cmm^n(t'',t')$, where $n$ is an integer such that $t/n$ belongs to the metastable regime~\footnote{For an integer $n\geq 1$ such that $t''\leq t_n\equiv t/n \leq t'$, we have $\lVert\rho(t)-e^{t \LMM}\P[\rho(0)] \rVert =\lVert [e^{t_n \L}-\P]^n(\I-\P)[\rho(0)]\rVert \leq \lVert e^{t_n \L}-\P\rVert^n \lVert \I-\P\rVert $. Since we have ${\lVert \rho(t_n)-\P[\rho(0)]\rVert \leq \Cmm(t'',t')}$ [cf.~Eq.~\eqref{eq:Cmm}], while $\lVert \I-\P\rVert\leq\lVert \I\rVert +\lVert\P\rVert = 2+\Cp $ [cf.~Eq.~\eqref{eq:Cp}], we arrive at $\lVert\rho(t)-e^{t \LMM}\P[\rho(0)] \rVert\leq \Cmm^n(t'',t')(2+\Cp)\lesssim 2\Cmm^n(t'',t')$.}.
Similarly, the corrections to observable averages and correlations in Eqs.~\eqref{ExpansionOB1} and~\eqref{ApproxCor} are bounded by $2\Cmm^n(t'',t')\lVert O\rVert_{\max}$ and $2\Cmm^n(t'',t')\lVert O\rVert_{\max}^2$, respectively, where $\lVert O\rVert_{\max}$ denotes the maximum singular value of $O$. 
We thus conclude that the corrections to the stationarity in Eq.~\eqref{eq:Cmm} are a \emph{figure of merit} in the theory of metastability.  This is further confirmed by the role played by them in the errors of classical  approximations for the structure of the metastable states and the long-time dynamics when classical metastability occurs, which we discuss in later sections.

We note that due to the way the condition in Eq.~\eqref{eq:Cmm2} is formulated, a choice of the metastable regime is not unique.  Indeed, the corrections in Eq.~\eqref{eq:Cmm} grow when $t'$ increases or $t''$ decreases. In particular, extending the length $t'-t''$ of the metastable regime $n$ times leads to the corresponding corrections the stationarity bounded in the leading order by $(n+3) \Cmm(t'',t')$~\footnote{For an integer $n$ such that $t''+n(t'-t'')\leq t\leq t'+n(t'-t'')$, we have $\lVert e^{t\L}-\P\rVert\leq \lVert (\I-\P)e^{t\L}\rVert+\lVert \P e^{t\L}-\P\rVert$, where  $ \lVert (\I-\P)e^{t\L}\rVert\leq  \lVert (\I-\P)e^{t''\L}\rVert \leq (2+\Cp)\Cmm$ due to $\lVert e^{(t-t'')\L}\rVert=1$, while $\lVert \P e^{t\L}-\P\rVert \leq \sum_{n'=1}^n \lVert \P e^{[t-(n'-1)(t'-t'')]\L}- \P e^{[t-n'(t'-t'')]\L}\rVert + \lVert \P e^{t-n(t'-t'')\L}-\P\rVert\leq n \lVert \P [e^{t-n(t'-t'')\L}-\P]\rVert \leq (n +1)(1+\Cp)\Cmm$}. In the rest of this work, we consider a given choice of the metastable regime for an open quantum system displaying metastability and denote $\Cmm(t'',t')$ and $\Ctp(t'',t')$ by $\Cmm$ and $\Ctp$, respectively.
As a pronounced metastable regime is a hallmark of metastability phenomenon, however, some of our results rely on it being much longer than the initial relaxation time: the classical hierarchy of metastable phases discussed in Sec.~\ref{sec:disjoint} on $t'-t''\geq t''$, and the correspondence of coarse-grained quantum trajectories to classical stochastic trajectories discussed in Secs.~\ref{sec:QTraj_Cum} and~\ref{sec:QJMC}  on $ t'-t''\geq n t''$ with $1/n\ll 1$. 

Finally, an analogous approach to Eq.~\eqref{eq:Cmm} can be introduced to formally define the timescales $\tau''$, $\tau'$ and $\tau$; see Sec.~\ref{app:tau_defMM} in the~SM. In particular, the relations $-1/\lambda_{m+1}^R\leq \tau''< t''$, $t'\ll \tau'\leq -1/\lambda_{m}^R$, and $-1/\lambda_{2}^R\leq \tau$  follow (cf.~Sec.~\ref{sec:spectral}).

\subsection{Dissipative phase transitions} \label{sec:DPT}

When metastability is a consequence of approaching a first-order dissipative  phase transition, we have by definition   $\lambda_{m}^{R}/\lambda_{m+1}^{R}\rightarrow 0$ and the ratios of the timescales for the final and the initial relaxation diverge ($\tau'/\tau'',\tau/\tau''\rightarrow \infty$). Therefore, the ratios  $t''/\tau''$ and $\tau'/t'$ for the metastable regime can be chosen arbitrarily large leading to all corrections arbitrarily small, $\Cp, \Cmm\rightarrow 0$; cf.~Sec.~\ref{app:tau_defMM} in the~SM.

\begin{figure*}[t]
	\includegraphics[width=0.75\linewidth]{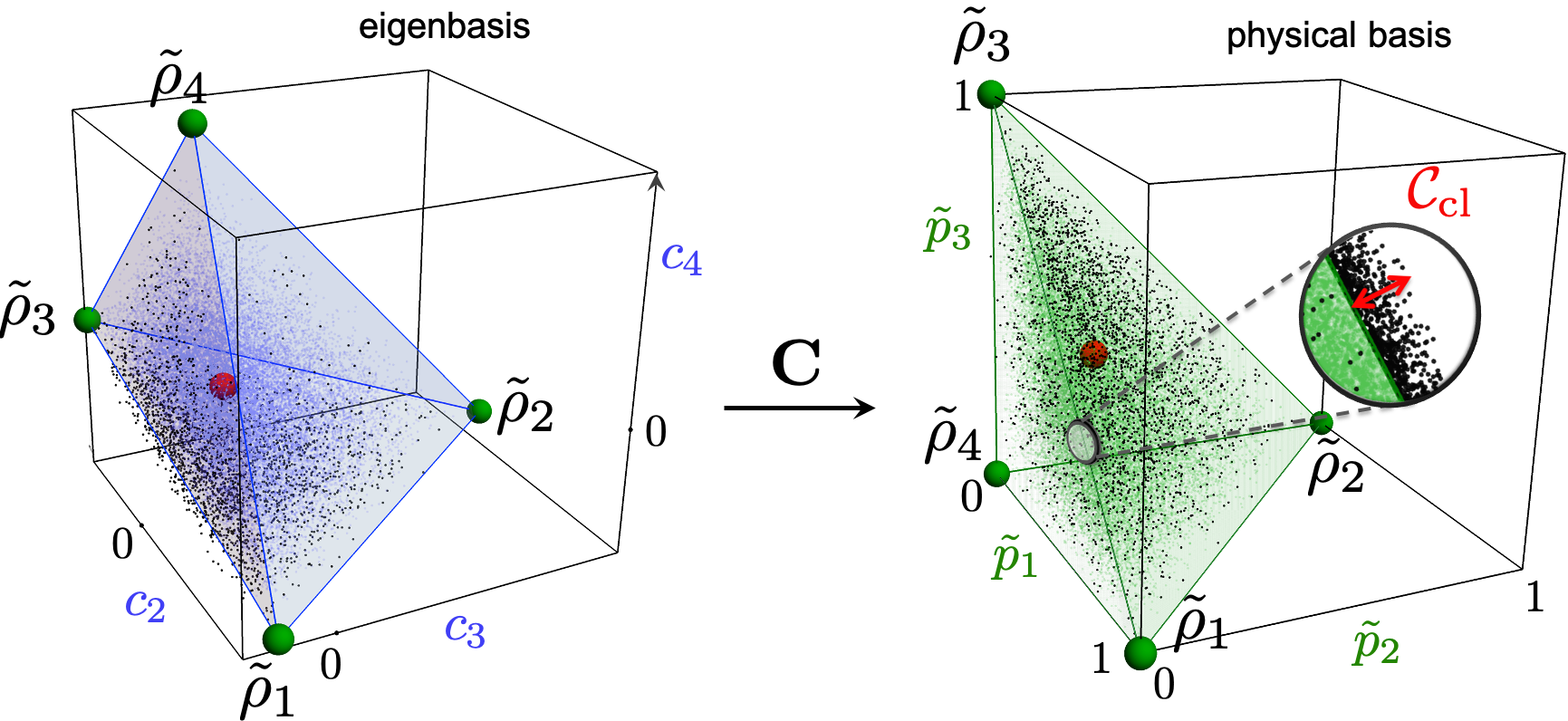}
	\vspace*{-2mm}
	\caption{\label{fig:MM}\textbf{Classical metastability}: \textbf{(left)} In the space of coefficients $(c_2,c_3,c_4)$ [Eq.~\eqref{Expansion}], the MM features the stationary state at $(0,0,0)$ (red sphere) and is approximated by the simplex (blue lines) of $m=4$ metastable phases (green spheres at the vertices). The dots represent metastable states from randomly generated pure states found inside (blue) and outside (black) the simplex. \textbf{(right)} Barycentric coordinates $(\tilde{p}_1,\tilde{p}_2,\tilde{p}_3)$ (with $\tilde{p_4}=1-\sum_{l=1}^3 \tilde{p}_l$) obtained by the transformation $\C$ [Eq.~\eqref{eq:C}] to the physical basis of metastable phases [Eq.~\eqref{eq:rhotilde}] yield probability distributions for states inside the simplex (green), while for states outside the simplex (black) the maximal distance $\Ccl$ becomes the figure of merit for classical metastability [Eq.~\eqref{eq:Ccl}]. 
		\vspace*{-5mm}}
\end{figure*}

\section{Classical metastability in open quantum systems} \label{sec:cMM}

We now introduce the notion of classical metastability, by the virtue of approximation of metastable states by probabilistic mixtures of a finite number of system states.  We show that this definition can be translated into a \emph{geometric criterion} on the decomposition of metastable states in projections of those states.  The corresponding corrections, together with the corrections to the stationarity and the positivity, play the central role in emerging classical approximations to the structure of the metastable manifold, long-time dynamics, and weak symmetries, discussed in Secs.~\ref{sec:disjoint}--\ref{sec:symmetry}. Therefore, our criterion identifies the \emph{parent feature} and the \emph{figure of merit} that govern the phenomenon of classical metastability in open quantum systems.

\subsection{Definition of classical metastability}

We define classical metastability to take place when any state of the system during the metastable regime  ${t''\leq t\leq t'}$ can be approximated as a probabilistic mixture of $m$ states $\rho_l\geq 0$ with $\Tr(\rho_l)=1$, $l=1,...,m$
\begin{equation}\label{eq:classicality}
\rho(t)= \sum_{l=1}^m p_l \rho_l+...,
\end{equation}
where  $p_l\geq 0$ with $\sum_{l=1}^m p_l=1$ represent the probabilities that depend only on an initial system state $\rho(0)$, while $\rho_l$, $l=1,...,m$, are independent from both time and the initial state. That is, the corresponding corrections in the trace norm 
	\begin{equation}\label{eq:classicality2}
	\mathcal{C}(\rho_1,...,\rho_m)\equiv\sup_{\rho(0)} \inf_{p_1,..,p_m}\sup_{t''\leq t \leq t'}\Big\lVert 	\rho(t)- \sum_{l=1}^m p_l \rho_l \Big\rVert, 
\end{equation}
fulfill 
	\begin{eqnarray}\label{eq:classicality3}
	\mathcal{C}(\rho_1,...,\rho_m)\ll 1. 
\end{eqnarray}
Here, the corrections depend on the choice of a metastable regime, but, for simplicity, we do not include it in the notation.
 We refer to $\rho_l$ as metastable phases (although their metastability is not assumed, but it is proven to follow together with their approximate disjointness in Sec.~\ref{sec:disjoint}, where we also discuss their nonuniqueness). The number of phases  in Eq.~\eqref{eq:classicality} is motivated by uniqueness of the decomposition (see also Sec.~\ref{app:defCMM}  of~the~SM) and the structure of first-order phase transitions in classical Markovian dynamics, where $m$ disjoint stationary probability distributions constitute stable phases of the system, and the system is asymptotically found in a probabilistic mixture of those phases, with probabilities depending on the initial system configuration. In later Secs.~\ref{sec:disjoint}--\ref{sec:QTraj} we show that classical properties of metastable phases and long-time dynamics akin to those in proximity to a first-order transition in a classical system follow as well. 

Remarkably, any metastable state in classical Markovian dynamics can be approximated by a probabilistic mixture of approximately disjoint metastable phases~\cite{Gaveau1987,Gaveau1998,Bovier2002,Gaveau2006,Kurchan2016}, whether metastability results from proximity to a first-order phase transition, or from constrained dynamics as in glassy systems. In open quantum dynamics, for the bimodal case $m=2$, it is known that any metastable state is a probabilistic mixture of two approximately disjoint metastable phases~\cite{Macieszczak2016a,Rose2016}. For higher dimensional MMs, however, the general structure is not known. Furthermore, it may be no longer classical~\cite{Macieszczak2016a}, as not only disjoint phases, but also  decoherence free subspaces~\cite{Zanardi1997,Zanardi1997a,Lidar1998} and noiseless subsystems~\cite{Knill2000,Zanardi2000} can be metastable, e.g.,~when perturbed away from a dissipative phase transition at a finite system size~\cite{Baumgartner2008} (see also Supplemental Material in Ref.~\cite{Macieszczak2016a} and  cf.~Refs.~\cite{Zanardi2014,Zanardi2015,Zanardi2016,Popkov2018}). Therefore, it is important to be able to verify whether a MM of an open quantum system  is classical as defined in Eq.~\eqref{eq:classicality}. In this section, we introduce such a systematic approach based on a geometric criterion equivalent to the definition in Eq.~\eqref{eq:classicality}, and refer to it as the  test of classicality.

\subsection{Test of classicality}\label{sec:cMM_test}

 For a given set of $m$ candidate system states, the test of classicality enables one to verify the approximation of  Eq.~\eqref{eq:classicality} and thus the classical metastability. Furthermore, it facilitates a check of whether a given set of $m$ initial states evolve into such metastable phases. Based on this, in Sec.~\ref{sec:algorithm} we introduce an efficient numerical technique delivering candidate states which, with the help of the test of classicality, can be postselected into metastable phases forming classical MMs. 


We first note that the definition of classicality in Eq.~\eqref{eq:classicality} leads to the MM in the space of coefficients being approximated by a simplex   (see Fig.~\ref{fig:MM}). 
When the MM is classical, the coefficients $c_k$ of a general initial state $\rho(0)$  can be approximated from Eq.~\eqref{Expansion}  as $\sum_{l=1}^m p_l c_{k}^{(l)}$, up to $\mathcal{C}+\Ctp$, where $\mathcal{C}$ is the correction in trace norm in Eq.~\eqref{eq:classicality} and $\Ctp$ is given in~Eq.~\eqref{eq:Cp2}~\footnote{We have $|c_{k}^{(l)} - \sum_{l=1}^m p_l c_{k}^{(l)}| =|\Tr( L_k \{\P[\rho(0)]- \sum_{l=1}^m p_l\rho_l\})|\leq \lVert L_k \rVert_{\max} \lVert\P[\rho(0)]- \sum_{l=1}^m p_l\rho_l\rVert \leq \lVert\P[\rho(0)]- \rho(t)\rVert+\lVert\rho(t) - \sum_{l=1}^m p_l\rho_l\rVert$, where we used the left-eigenmatrix normalization $\lVert L_k \rVert_{\max}\leq c_k^{\max}-c_k^{\min}=1$ with $c_k^{\max}$ and $c_k^{\min}$ denoting the maximal and minimal eigenvalue of $L_k$. We then choose time $t$ within the metastable regime for which $\lVert\P[\rho(0)]-\rho(t)\rVert$ is minimal.}. Here, $c_{k}^{(l)} =\Tr(L_k \rho_l)$ represent the metastable phases $\rho_l$  in the coefficient space [cf.~Eq.~\eqref{Expansion}],
\begin{equation}\label{eq:rhotilde}
\trho_l\equiv \P(\rho_l)=\rhoss+\sum_{k=2}^m c_{k}^{(l)} R_k, \quad l=1,...,m.
\end{equation}
Thus, the MM is approximated by a \emph{simplex  in the coefficient space} with vertices given by the metastable phases. For low-dimensional MMs ($m\leq4$), this can be verified visually by projecting a randomly generated set of initial conditions on their metastable states (to sample the MM) and checking that they are found approximately within the chosen simplex  (cf.~Fig.~\ref{fig:MM}). 

%
Motivated by the structure of classical MMs in the coefficient space, we now introduce the \emph{test of classicality}---a geometric way of checking whether degrees of freedom describing metastable states during the metastable regime correspond, approximately, to probability distributions.  
 Degrees of freedom in the MM are described by the coefficients of decomposition into the eigenmodes $R_k$, $k=1,...,m$, so that, with $c_1=1$, their number is $m-1$. Motivated by Eq.~\eqref{eq:classicality}, here we instead consider the decomposition in the new basis given by the projections of metastable phases in Eq.~\eqref{eq:rhotilde}, which is encoded by the transformation
 \begin{equation}\label{eq:C}
 (\C)_{kl}\equiv c_{k}^{(l)}, \quad k,l=1,...,m,
 \end{equation}
 so that $ \trho_l=\sum_{k=1}^m(\C)_{kl} \,R_k$. 
In particular, the volume of the corresponding simplex in the coefficient space is $|\!\det{\C}|/(m\!-\!1)!$~\footnote{We have $\det{\C}=\det {\bar{\C}}$, where $(\bar{\C})_{l-1,k-1}=c_{k}^{(l)}-c_{k}^{(1)}$, $k,l=2,..,m$, encodes the coefficients for the simplex with the vertex of $\trho_1$ shifted to the origin.}.
The decomposition of a metastable state in this new basis [cf.~Eq.~\eqref{Expansion}]
\begin{equation}\label{eq:p_k}
\P[\rho(0)]=\sum_{l=1}^m \tilde{p}_l \trho_l,
\end{equation} 
 is given by the barycentric coordinates $\tilde{p}_l=\sum_{k=1}^m(\C^{-1})_{lk} c_k$ of the simplex in the coefficient space, so that $\tilde{p}_l=  \Tr[\tilde{P}_{l}\rho(0)]$ with the new dual basis
\begin{equation}\label{eq:Ptilde}
\tilde{P}_l\equiv\sum_{k=1}^m(\C^{-1})_{lk} \,L_k,\quad l=1,...,m.
\end{equation}
When $\trho_l$ are linearly independent for $l=1,...,m$, $|\!\det{\C}|> 0$ and $\C$ is invertible so that Eq.~\eqref{eq:Ptilde} is well defined. In this case, $\Tr(\tilde{P}_{k}\trho_l)=\sum_{n=1 }^m{(\C^{-1})}_{kn}{(\C)}_{nl}= (\C^{-1}\C )_{kl} =\delta_{kl}$ and the normalization of the dual basis in Eq.~\eqref{eq:Ptilde} is fixed by the traces of metastable states in Eq.~\eqref{eq:rhotilde} being $1$.

Although for the barycentric coordinates we have $\sum_{l=1}^m\tilde p_l=1$, and thus $\sum_{l=1}^m \tilde{P}_l=\mathds{1}$, they do not in general correspond to probability distributions. Indeed, they are not all positive whenever a metastable state lies outside the simplex in the coefficient space corresponding to $\trho_l$, $l=1,...,m$ (see Fig.~\ref{fig:MM}), as the distance in L1 norm of barycentric coordinates to the simplex is given by $\lVert\pt-\p\rVert_1=\lVert\pt\rVert_1-1$, where $\p$ is the closest probability distribution to barycentric coordinates $\pt$ [here, $(\p)_l=p_l$ and $(\pt)_l=\tilde{p}_l$, $l=1,...,m$]. Nevertheless, when  the \emph{maximum distance} to the simplex (cf.~Fig.~\ref{fig:MM}), 
\begin{equation}\label{eq:Ccl}
\Ccl(\trho_1,...,\trho_m)\equiv \max_{\rho(0)}\,\lVert\pt\rVert_1-1,
\end{equation}
is small, it follows that the metastability is classical, with the corrections in Eq.~\eqref{eq:classicality} bounded as
\begin{eqnarray}\label{eq:Ccl2}
\left\lVert\rho(t)-\sum_{l=1}^m p_l \rho_l\right\rVert
&\lesssim& \Ccl(\trho_1,...,\trho_m)+\Cp+\Cmm,
\end{eqnarray} 
where $t''\leq t\leq t'$ and  $\lesssim$ stands for $\leq$ in the leading order of the corrections  (see Sec.~\ref{app:test} in the~SM),
while $p_l$ is chosen as the closest probability distribution to the barycentric coordinates, $\rho_l$ as the closest state to $\trho_l$, so that $\lVert \trho_l-\rho_l\rVert\leq \Cp$ [cf.~Eq.~\eqref{eq:Cp}], and $\Cmm$ bounds the approximation of $\rho(t)$ by the projection on the low-lying modes [cf.~Eqs.~\eqref{eq:Cmm} and~\eqref{eq:p_k}]. Similarly, the average distance to the simplex can be considered (cf.~Refs.~\cite{Zyczkowski2001,Zyczkowski2003} and see Sec.~\ref{app:test} in the~SM).

Finally, the corrections in~Eq.~\eqref{eq:Ccl}, which we refer to as the \emph{corrections to the classicality},  can be efficiently estimated using the dual basis,
\begin{equation}\label{eq:Ctcl}
\Ccl(\trho_1,...,\trho_m)\leq 2\sum_{l=1}^m(-\tilde p_{l}^{\min})\equiv\Ctcl(\trho_1,...,\trho_m),
\end{equation}
where $\tilde{p}_l^{\min}\leq 0$ is the minimum eigenvalue of $\tilde{P}_l$ in Eq.~\eqref{eq:Ptilde}, so that 
\begin{equation}\label{eq:Ctcl2}
\Ctcl(\trho_1,...,\trho_m) \leq m\,\Ccl(\trho_1,...,\trho_m).
\end{equation}
Apart from being easy to compute, $\Ctcl(\trho_1,...,\trho_m)$ also carries the \emph{operational meaning} of being an upper bound on the distance of the operators $\tilde{P}_l$ to the set of POVMs; cf.~Sec.~\ref{sec:DF} and see Sec.~\ref{app:Ptilde} in the~SM.

\subsection{Figures of merit}


From Eq.~\eqref{eq:Ccl}, we obtain a criterion for verification of whether for a given set of states, the MM can be approximated as a probabilistic mixture of the corresponding metastable states [Eq.~\eqref{eq:rhotilde}]. In particular, 
 whenever
  \begin{equation}\label{eq:Ccl3}
  \Ccl(\trho_1,...,\trho_m)\ll 1,
  \end{equation}
   the metastability is classical.
   Moreover, it can be shown that $\Ccl(\trho_1,...,\trho_m)\lesssim \mathcal{C}(\rho_1,...,\rho_m)+\Ctp$, provided that $m[\mathcal{C}(\rho_1,...,\rho_m)+\Ctp]\ll 1$, which also implies $\Ctcl(\trho_1,...,\trho_m)\ll 1$  (cf.~Eqs.~\eqref{eq:classicality2} and~\eqref{eq:Ctcl}, and see Sec.~\ref{app:test} in the~SM).  Since for the classical metastability we have Eq.~\eqref{eq:classicality3}, assuming these corrections decrease when changing a dynamical parameter or when increasing system size, but $m$ remains constant, Eq.~\eqref{eq:Ccl3} follows. We thus conclude that it is a necessary and sufficient condition for classical metastability. 
   
    Interestingly, the bimodal case of $m=2$ is always classical as the metastable phases $\trho_1$ and $\trho_2$ leading to  $\Ccl(\trho_1,\trho_2)=\Ctcl(\trho_1,\trho_2)=0$ can be constructed explicitly~\cite{Macieszczak2016a,Rose2016}. For higher dimensional MMs, the presence of classical  metastability can be uncovered using the bound in Eq.~\eqref{eq:Ctcl} for candidate states generated by
   the numerical approaches of Sec.~\ref{sec:numerics}, see, e.g., Ref.~\cite{Rose2020}. In particular, approaching an effectively classical first-order dissipative phase transition with a finite $m$ requires the possibility of choosing  $m$ candidate states such that $\Ccl(\trho_1,...,\trho_m)\rightarrow0$  (cf.~Sec.~\ref{sec:DPT}).  
   Importantly,  the condition in Eq.~\eqref{eq:Ccl3} is independent from the presence of weak symmetries (which, nevertheless, can be efficiently incorporated; cf.~Sec.~\ref{sec:symmetry_test}).

 In Sec.~\ref{sec:disjoint},  we show that the metastable phases in Eq.~\eqref{eq:rhotilde} are approximately disjoint, while the operators in Eq.~\eqref{eq:Ptilde} take the role of basins of attractions. Moreover, in Sec.~\ref{sec:Leff}, we explain how the long-time dynamics toward the stationary state corresponds approximately to classical stochastic dynamics between metastable phases. Since corrections in those results depend only on the corrections to the stationarity, the positivity, and the classicality defined in Eqs.~\eqref{eq:Cmm},~\eqref{eq:Cp}, and~\eqref{eq:Ccl}, these quantities can be viewed as a complete set of \emph{figures of merit characterizing classical metastability} in open quantum systems. As in the most of this work, we consider a given choice of $m$ states, we denote  $\mathcal{C}(\rho_1,...,\rho_m)$ by $\mathcal{C}$
 	and $\Ccl(\trho_1,...,\trho_m)$ by $\Ccl$ for simplicity.

\section{Classical metastable phases}\label{sec:disjoint}

In Sec.~\ref{sec:cMM}, we introduced the definition of classical metastability of when MMs of open quantum systems can be approximated as probabilistic mixtures of a set of states. We now show that in this case, those states are necessarily metastable and constitute a physical basis of the MM as distinct phases of the system. To this aim, we demonstrate that the probabilities that represent the degrees of freedom in the MM can be accessed with negligible disturbance, so that the metastable phases can be distinguished with a negligible error. We further argue that their supports and basins of attraction are approximately disjoint,  in analogy to first-order phase transitions and metastability  in classical Markovian systems~\cite{Gaveau2006}. Finally, we also discuss how, in the case of any further separation in the low-lying spectrum, later MMs are necessarily classical as well.

\subsection{Physical representation of metastable manifold}

We begin by noting that phases given in Eq.~\eqref{eq:classicality} are uniquely defined up to the so far considered corrections when the condition in Eq.~\eqref{eq:classicality3} is fulfilled. Indeed, for states $\rho_l$ in Eq.~\eqref{eq:classicality}, the distance to their projections $\tilde{\rho}_l$ in Eq.~\eqref{eq:rhotilde} is bounded by  $2(\mathcal{C}+\Ctp)$ when $\Ctcl\ll 1$ (see Sec.~\ref{app:test} in the~SM), so that they are metastable. It then follows that their distance to the states chosen  in Eq.~\eqref{eq:Ccl2} as closest states to the projections $\trho_l$ in Eq.~\eqref{eq:rhotilde} is bounded by $\lesssim2(\mathcal{C}+\Ctp)+\Cp$. Finally, for two different sets of $m$ metastable phases corresponding to different projections in Eq.~\eqref{eq:rhotilde} and the corrections to the classicality  $\Ccl$ and $\Ccl'$, which fulfill Eq.~\eqref{eq:Ccl3}, the distance in trace norm between the projection of a metastable phase in one set to the closest projection of a metastable phase in the other set is bounded by $\lesssim \Ccl+\Ccl'+\min(\Ccl,\Ccl')$ (see Sec.~\ref{app:phases} in the~SM). 


Furthermore, in contrast to the right eigenmodes of the master operator with $\Tr(R_k)=0$ for $k=2,...,m$ [from $L_1=\mathds{1}$ and $\Tr(L_k R_l)=\delta_{kl}$], the projections of metastable phases in Eq.~\eqref{eq:rhotilde} feature normalized trace, $\Tr(\trho_l)=1$, $l=1,...,m$, are Hermitian, and approximately positive [see Sec.~\ref{sec:master} and cf.~Eq.~\eqref{eq:Cp}]. Moreover, when the condition in Eq.~\eqref{eq:Ccl3} is fulfilled, any metastable state is approximated well by their probabilistic mixture [cf.~Eq.~\eqref{eq:Ccl2} and Fig.~\ref{fig:MM}]. Thus, the projections in Eq.~\eqref{eq:rhotilde} can be considered as physical basis of the MM and \emph{approximate metastable phases}.

While the left low-lying eigenmodes $L_k$, $k=2,...,m$, describe quantities conserved in the system during the initial relaxation and  the metastable regime [cf.~Eq.~\eqref{ExpansionOB1} for $t\leq t'$, where $b_k=\delta_{kl}$ for $O=L_l$] the dual basis operators in Eq.~\eqref{eq:Ptilde} determine the decomposition of a metastable state into the basis in Eq.~\eqref{eq:rhotilde}, and as such, when the condition in Eq.~\eqref{eq:Ccl3} is fulfilled, they represent \emph{approximate basins of attraction} for metastable phases (see also Sec.~\ref{app:Ptilde} in the~SM). Importantly, via barycentric coordinates in Eq.~\eqref{eq:ptilde}, they define \emph{order parameters} that distinguish the metastable phases, $\Tr(\tilde{P}_{k}\trho_l)= \delta_{kl}$, with system observable averages being their linear combinations [cf.~Eq.~\eqref{ExpansionOB1}].  

Finally, the  barycentric coordinates are a physical representation of $m$-1 \emph{degrees of freedom} present in the metastable regime, as they approximate probability distributions.
As a consequence, we next show that they are classical from an operational perspective of measuring the system.

\subsection{Classical degrees of freedom}\label{sec:DF}

We now argue that in the case of classical metastability, the degrees of freedom determining the MM can be accessed with a negligible disturbance of metastable states. It then follows that metastable phases can be distinguished with a negligible error. We also discuss consequences for measurements of system observables.

The MM is determined by the coefficients $c_k$ of decomposition into low-lying eigenmodes $L_k$,  $k=2,...,m$ [cf.~Eq.~\eqref{Expansion}],  or, equivalently,  by the barycentric coordinates $\tilde{p}_l$ of decomposition into the projections $\trho_l$ of metastable phases, $l=1,...,m$ [cf.~Eqs.~\eqref{eq:rhotilde} and~\eqref{eq:p_k}]. Furthermore, system states can be probed by  POVMs ($P_l=P_l^\dagger$, $P_l\geq 0$, $\sum_{l} P_l=\mathds{1}$), including von Neumann measurements ($P_k P_l =\delta_{kl} P_l$) corresponding to measuring system observables.  Considering the POVM
\begin{equation}\label{eq:POVM0}
P_l\equiv \frac{\tilde{P_l}-\tilde{p}_l^{\min}\mathds{1}}{1+\frac{\Ctcl}{2}}, \quad l=1,...,m,
\end{equation}  
the distance of the probability distribution $(\mathbf{p})_l\equiv \Tr[P_l \rho(0)]$, $l=1,...,m$, to the barycentric coordinates is bounded by $\Vert \pt-\mathbf{p}\rVert_1\lesssim\Ctcl$ (cf.~Eq.~\eqref{eq:Ctcl} and see Sec.~\ref{app:Ptilde} in the~SM). Therefore, by measuring a metastable state $\rho(t)$, the barycentric coordinates can be accessed with the error up to $\Ctcl+\Cmm$ [cf.~Eq.~\eqref{eq:Cmm}], while the  state can be reconstructed by
preparing the closest state $\rho_l$ to $\trho_l$ upon obtaining $l$th outcome,   with the resulting \emph{disturbance}~\footnote{We note that often the diamond norm, $\lVert\mathcal{E}-\mathcal{I} \rVert_{\diamond}\equiv \lVert\mathcal{E}\otimes \mathcal{I} -\mathcal{I}\otimes\mathcal{I}\rVert$, is used instead of induced trace norm, $\lVert\mathcal{E}-\mathcal{I}\rVert\leq \lVert\mathcal{E}-\mathcal{I} \rVert_{\diamond}$, to quantify the disturbance caused by a quantum channel $\mathcal{E}$ with respect to the identity channel $\mathcal{I}$.}
\begin{equation}
\lVert \rho(t)-\sum_{l=1}^m \Tr[P_l \rho(t)]\,\rho_l \rVert\lesssim \Ctcl+\Cp+\Cmm,
\end{equation}
where $t''\leq t\leq t'$.
This result should be contrasted with the case of measuring a general system state where no information is available without disturbance (see, e.g., Ref.~\cite{Wolf2012}). 

The minimal average error of distinguishing equally probable two states according to Holevo-Helstrom theorem is determined by  the distance in the trace norm as $1/2 -  \lVert \rho^{(1)} -\rho^{(2)}\rVert/4$ (see, e.g., Ref.~\cite{Nielsen2010}). For metastable states, this error is approximately determined by the distance between their barycentric coordinates as
\begin{subequations}\label{eq:B_trace1}
	\begin{align}
 \lVert \rho^{(1)}\!(t) -\rho^{(2)}\!(t)\rVert&\gtrsim\lVert \pt^{(1)}\! -\pt^{(2)} \rVert_1 \Big(1-\frac{\Ctcl}{2}\Big)-2\Cmm,\\
  \lVert \rho^{(1)}\!(t) -\rho^{(2)}\!(t)\rVert&\lesssim\lVert \pt^{(1)}\! -\pt^{(2)} \rVert_1 \left(1+\Cp\right)+2\Cmm,
  \end{align}
\end{subequations}
where 
 the first bound corresponds to the error when measuring the POVM  in Eq.~\eqref{eq:POVM0}. In particular, a pair of metastable phases can be distinguished with the error $\lesssim[\min(\Ctcl/2,\Ccl)+\Cp]/2$, since 
\begin{eqnarray}\label{eq:B_trace2}
\lVert \rho_k -\rho_l\rVert&\gtrsim&2\left(1-\Ccl -\Cp\right),\quad k\neq l,
\end{eqnarray}
for $\rho_l$ being the closest state to $\trho_l$ in Eq.~\eqref{eq:rhotilde}, which corresponds to a measuring the POVM with two elements: $P\equiv (\tilde{P}_l-\tilde{p}_l^{\min}\mathds{1})/(\tilde{p}_l^{\max}-\tilde{p}_l^{\min})$ and $\mathds{1}-P$. 
For $\rho_l$ that projects on $\trho_l$, the bound in Eq.~\eqref{eq:B_trace2} reduces to $\gtrsim 2(1-\Ccl)$. For derivations, see  Sec.~\ref{app:norm} in the~SM.

Finally, during the metastable regime, $t''\leq t\leq t'$, the probability distribution for any measurement of the system is approximated as a probabilistic mixture of probability distribution for individual metastable phases, as
\begin{equation} 
	\sum_{k} \Big|\Tr [P_k \rho(t)] - \sum_{l=1}^m p_l \Tr [P_k \rho_l] \Big|\leq \Big\lVert \rho(t) -\sum_{l=1}^m p_l \rho_l \Big\rVert,
\end{equation}	
where $\{P_k\}_k$ is a POVM and the right-hand side is bounded by corrections in Eqs.~\eqref{eq:classicality} or~\eqref{eq:Ccl2}, depending on the choice of metastable phases. This conditional structure of the probability distribution, however, is not directly related to the classicality, since it is present for any system state being a probabilistic mixture of not necessarily orthogonal states, while, as we argue next, metastable phases are approximately disjoint.


\subsection{Approximate disjointness of metastable phases} 

Below we show that the metastable phases in Eqs.~\eqref{eq:classicality} and~\eqref{eq:Ccl2} are approximately disjoint, that is, they describe states restricted to distinct regions of the system space. Furthermore, we also find that their basins of attraction are approximately disjoint. 

First,  note that the distance in the trace norm equals $2$ only for disjoint  (mutually orthogonal, states). Therefore, the bound in Eq.~\eqref{eq:B_trace2} implies that the  \emph{metastable phases are approximately disjoint}. This is further corroborated by similar bounds on scalar products of $\sqrt{\rho_l}$ or $\rho_l$; see Sec.~\ref{app:disjoint}  of~the~SM.

Second, to capture approximately disjoint supports of metastable phases, we consider  the subspaces $\mathcal{H}_l$ defined as the space spanned by eigenstates of $\tilde{P}_l$ in Eq.~\eqref{eq:Ptilde} with eigenvalues equal or above $1/2$, $l=1,...,m$. We have (see Sec.~\ref{app:disjoint} in the~SM)
\begin{eqnarray}\label{eq:B_support01}
	\Tr\left( \mathds{1}_{\mathcal{H}_k}\, \rho_l\right)&\lesssim& \Ccl + 2\Cp,\quad k\neq l\\\label{eq:B_support02}
	\Tr\left( \mathds{1}_{\mathcal{H}_l}\, \rho_l\right) &\gtrsim& 1-\Ccl - 2\Cp,
\end{eqnarray}
where $\rho_l$ is the closest state to $\trho_l$, $k,l=1,...,m$. 
Furthermore, we also have [cf.~Eq.~\eqref{eq:Ctcl}] 
\begin{equation}\label{eq:B_support03}
\sum_{\substack{1\leq k\leq m:\\k\neq l}}\Tr\left( \mathds{1}_{\mathcal{H}_k}\, \rho_l\right)\lesssim \Ctcl + 2\Cp.
\end{equation}
The bounds in Eqs.~(\ref{eq:B_support01})-(\ref{eq:B_support03}) support the statement that the metastable phases reside in approximately disjoint areas of the state space. 
The bounds in Eqs.~(\ref{eq:B_support01})-(\ref{eq:B_support03})  also hold well for $\trho_l$  in Eq.~\eqref{eq:rhotilde} as $|\Tr( \mathds{1}_{\mathcal{H}_k}\, \rho_l)-\Tr( \mathds{1}_{\mathcal{H}_k}\, \tilde{\rho_l})|\leq \Cp$, while for the states in Eq.~\eqref{eq:classicality} that project on $\trho_l$, they are further reduced to  $\lesssim \Ccl$, $\gtrsim 1-\Ccl$ and $\lesssim \Ctcl$, respectively. For the bimodal case of open quantum dynamics, $m=2$, approximate disjointness was already argued in Refs.~\cite{Macieszczak2016a,Rose2016}. 

Finally, the subspace $\mathcal{H}_l$ in Eqs.~\eqref{eq:B_support01}--\eqref{eq:B_support03} captures not only majority of $\rho_l$ support, but by its definition also the corresponding basin of attraction, i.e., the initial states which evolve into metastable states close to $\trho_l$, i.e., with $1-\tilde{p}_l\ll 1$. Indeed, in Sec.~\ref{app:disjoint}  of~the~SM, we show that  $\Tr[ \mathds{1}_{\mathcal{H}_l}\, \rho(0)]\gtrsim 1-2|1-\tilde{p}_l|-\Ccl$. Furthermore, we have $\Tr[ \mathds{1}_{\mathcal{H}_k}\, \rho(0)] \leq 2 |1-\tilde{p}_l|+\Ccl$ for $k\neq l$, and  $\sum_{1\leq k\leq m:\,k\neq l}\Tr[ \mathds{1}_{\mathcal{H}_k}\, \rho(0)]\leq 2 |1-\tilde{p}_l|+\Ctcl $.  
Thus, we conclude that the \emph{basins of attractions are approximately disjoint}.
We note, however, that in general subspaces $\mathcal{H}_l$  themselves are not disjoint as they feature states that decay into multiple metastable phases  (it can be shown that subspaces spanned by the  $\tilde{P_l}$ eigenstates with the eigenvalues separated from $0$ and $1$ by distance $\gg \Ccl+\Cp$, $l=1,...,m$, can be neglected in the support of metastable phases; see Sec.~\ref{app:disjoint} in the~SM).

\subsection{Classical hierarchy of metastable phases}\label{sec:hierarchy}

A second metastable regime corresponds to a further separation in the low-lying spectrum of the master operator in Eq.~\eqref{eq:L}, $\lambda_{m_2}^R/\lambda_{m_2+1}^R\ll 1$ with $m_2<m$ (cf.~Ref.~\cite{Gaveau1999}).  In Sec.~\ref{app:hierarchy}  of~the~SM, we show that metastable states during the second metastable regime,  $t''_2 \leq t\leq t'_2$,  also form a classical MM, i.e., are mixtures of $m_2$ metastable phases provided that $t'_2 \geq 2 t''_2$. Those $m_2$ metastable phases are  approximately disjoint mixtures of $m$ metastable phases of the first MM, and their supports as well as their basins of attraction are approximately disjoint [cf.~Eqs.~(\ref{eq:B_trace2})-(\ref{eq:B_support03})]. Therefore, each metastable phase of the first MM evolves approximately into a single metastable phase in the second MM, unless the second MM is not supported on that phase (the phase belongs to the decay subspace). 

These results are a direct consequence of long-time dynamics in a classical MM being well approximated by classical stochastic dynamics, which we discuss in next, as metastable states of classical stochastic dynamics are known to be mixtures of as many metastable phases as the number of low-lying modes~\cite{Gaveau1998,Gaveau2006}.

\section{Classical long-time dynamics}\label{sec:Leff}

The definition of classical metastability in Eq.~\eqref{eq:classicality} determines not only the structure of metastable states.  Remarkably, as a consequence of the long-time relaxation toward the stationary state effectively taking place inside the MM,  the long-time dynamics is  approximately classical as well. We prove that it corresponds to \emph{classical stochastic dynamics} occurring between disjoint metastable phases and can be accessed by measuring averages or time-correlations of system observables whenever  metastable phases differ in the averages. We also discuss the role played by a further separation of timescales in the long-time dynamics, i.e., another metastable regime. Finally, we show that stochastic transitions between metastable phases can be observed directly by means of continuous measurements of quanta emitted during system interaction with the environment, provided that metastable phases differ in the average measurement rates. In that case, the statistics of integrated continuous measurement is generally \emph{multimodal} for times within the metastable regime, which for times after the final relaxation can lead to a high fluctuations rate, reminiscent of the proximity to a dynamical phase transition~\cite{Garrahan2010}.

\subsection{Classical average dynamics of system and observables}\label{sec:EffDynAv}

\subsubsection{Long-time dynamics}

\begin{figure}[t]
	\includegraphics[width=0.95\linewidth]{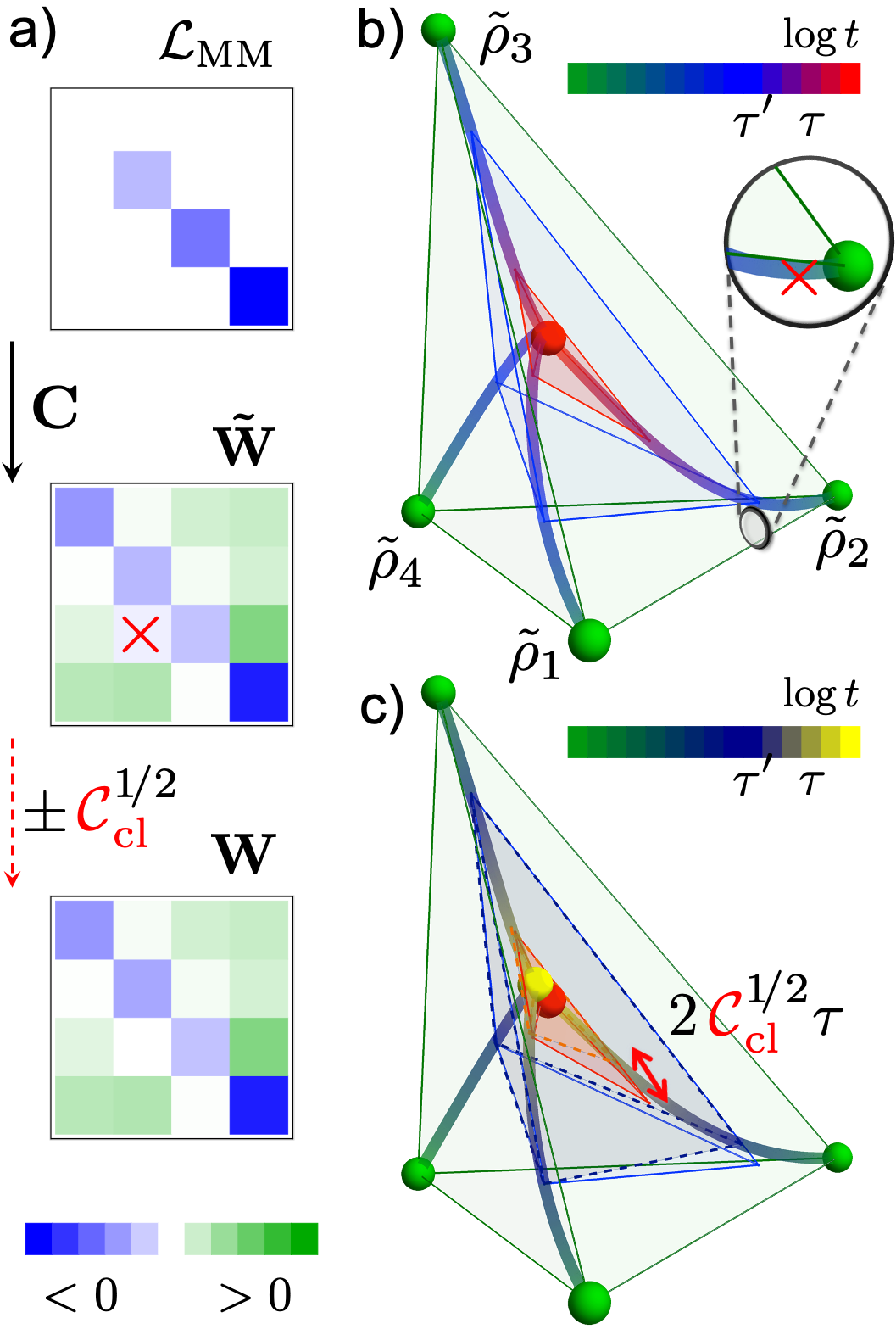}
	\vspace{-2mm}
	\caption{\label{fig:EffDyn}\textbf{Classical long-time dynamics}: \textbf{(a)} The long-time dynamics [Eq.~\eqref{Expansion2}] can be understood as dynamics between metastable phases [Eq.~\eqref{eq:ptilde}], governed by the trace-preserving generator $\Wt$ [Eq.~\eqref{eq:Wtilde}], which can be approximated by a classical stochastic generator $\W$ [Eqs.~\eqref{eq:W} and~\eqref{eq:deltaW}], which is both trace preserving and positive; here a negative transition rate from $\trho_2$ to $\trho_3$  (marked by red cross) is put to $0$. \textbf{(b)} The long-time dynamics in the barycentric coordinates (cf.~Fig.~\ref{fig:MM}): green simplex corresponds to $t\ll\tau'$, blue to $t=\tau'$, and red to $t=\tau$, while the stationary state is marked by red sphere.	Positive dynamics corresponds to the simplex of metastable phases mapped onto itself, which requires all metastable phases to be mapped inside the simplex at all times. Here, $\trho_2$ initially acquires a negative probability $\tilde{p}_3(t)$ at small $t$ [red cross in the inset; cf.~$\Wt$ in panel (a)]. \textbf{(c)} Approximating by $\W$ alters the dynamics, with corrections increasing in time [Eq.~\eqref{eq:etW_CL}]; blue dashed simplex corresponds to $t=\tau'$ and orange dashed to $t=\tau$ [cf.~panel~(b)]. This ultimately leads to a different stationary state (yellow sphere) [cf.~Eq.~\eqref{eq:pss_CL}], which is close to the true stationary state when Eq.~\eqref{eq:cond_CL} is fulfilled.
		\vspace*{-3mm}}
\end{figure}

From Eq.~\eqref{Expansion2} the evolution for times $t\geq t''$ effectively takes place on the MM with the effective generator  $\L_{\text{MM}}$ defined in Eq.~\eqref{eq:Leff}. This generator can be expressed in the basis of the metastable phases [Eqs.~\eqref{eq:rhotilde} and~\eqref{eq:Ptilde}] as 
\begin{equation}\label{eq:Wtilde}
(\Wt)_{kl}\equiv \Tr[\tilde{P}_k \LMM (\trho_l)],
\end{equation}
where $k,l=1,..,m$, and thus $\Wt= \C^{-1}\mathbf{\Lambda}\C$ with $(\mathbf{\Lambda})_{kl}\equiv\lambda_{k}\delta_{kl}$ [cf.~Eq.~\eqref{eq:C} and see Fig.~\ref{fig:EffDyn}\textcolor{blue}{(a)}]. The dynamics of the system state within the MM is then determined by the dynamics of the barycentric coordinates
 \begin{equation}\label{eq:ptilde}
 \pt(t)= e^{t \Wt} \pt,
 \end{equation}
where $(\pt)_{l}\equiv\Tr[\tilde{P}_l \rho(0)]$, so that $\P[\rho(t)]=\sum_{l=1}^m[\pt(t)]_{l} \trho_l$ [cf.~Eq.~\eqref{eq:p_k} and see Fig.~\ref{fig:EffDyn}\textcolor{blue}{(b)}]. 

By definition, the long-time evolution in Eq.~\eqref{eq:Leff} transforms the MM onto itself, see, e.g., Fig.~\ref{fig:Spectrum}\textcolor{blue}{(b)}. This does not guarantee, however, that the simplex of $m$ metastable phases is transformed onto itself,  as the evolution may cause states inside the simplex to evolve toward states outside, and thus an initial probability distribution (positive barycentric coordinates) acquiring some negative values at later times [see the inset in Fig.~\ref{fig:EffDyn}\textcolor{blue}{(b)}]. Therefore, the dynamics generated  by $\Wt$ is in general not positive [cf.~Fig.~\ref{fig:EffDyn}\textcolor{blue}{(a)}]. Nevertheless, as we discuss below, when the simplex of metastable phases is a good approximation for the MM in the sense of the condition in Eq.~\eqref{eq:Ccl3}, $\Wt$ is well approximated by a generator of stochastic classical dynamics between metastable phases. 

\subsubsection{Classical generator}

Dynamics generated by $\Wt$ conserves the probability, as from $\sum_{k=1}^m \tilde{P}_k =\mathds{1}$ we have $\sum_{k=1}^m(\Wt)_{kl}=0$ (cf.~Sec.~\ref{app:classical} in the~SM). Furthermore, it can be shown to be approximately positive, with $\Wt$ approximated by the closest classical stochastic generator $\W$ (cf.~Fig.~\ref{fig:EffDyn}\textcolor{blue}{(a)} and see Secs.~\ref{app:Leff_approx} and~\ref{app:Leff_delta} in the~SM),
\begin{eqnarray}\label{eq:W}
({\W})_{kl}&\equiv&\max [(\Wt)_{kl},0],\quad k\neq l,\\\nonumber
({\W})_{ll}&\equiv&(\Wt)_{ll} + \sum_{k\neq l}\min [(\Wt)_{kl},0],
\end{eqnarray}
$k,l=1,...,m$, as
\begin{equation}\label{eq:deltaW}
\frac{ \lVert\Wt-\W\rVert_1}{ \lVert\Wt\rVert_1}\lesssim  2\sqrt{\Ccl},
\end{equation} 
where the norm $\lVert \mathbf{X}\rVert_1\equiv\max_{1\leq l\leq m}  \sum_{k=1}^m|( \mathbf{X})_{kl}|$~\footnote{$\lVert \mathbf{X}\rVert_1$ is  the matrix norm induced by the L1 norm of vectors. For a vector $\mathbf{v}$ and a matrix $\mathbf{X}$, we have $ \lVert \mathbf{X}\mathbf{v}\rVert_1=\sum_{k=1}^m |\sum_{l=1}^m X_{kl} v_l|\leq \sum_{k,l=1}^m |X_{kl} v_l|\leq  (\max_{1\leq l\leq m}\sum_{k=1}^m |X_{kl}|) \sum_{l=1}^m |v_l| \equiv \lVert \mathbf{X}\rVert_1 \lVert\mathbf{v}\rVert_1 $, and the inequality can be saturated by considering basis vectors.} and $(1-\Cp-\Ccl)  \lVert \LMM\rVert \lesssim \lVert \Wt   \rVert_1  \leq (1+\Ctcl/2) \lVert \LMM\rVert$ (see Sec.~\ref{app:norm} in the~SM). From Eq.~\eqref{eq:deltaW} the normalized distance between the generators is bounded as
\begin{eqnarray}
\Delta_+\equiv\frac{\lVert\Wt-\W\rVert_1}{\lVert\Wt\rVert_1+\lVert\W\rVert_1}&\lesssim& \sqrt{\Ccl}. \label{Delta}
\end{eqnarray}
Note that columns of ${\W}$ sum to $0$, and then negativity of its diagonal terms follows from the positivity of the off-diagonal terms, so that dynamics generated by $\W$ is indeed positive and probability-conserving (cf.~Sec.~\ref{app:classical} in the~SM). For the bimodal case of $m=2$, the MM is always classical with $\Ccl=0$,  and thus $\Wt=\W$ is exactly a generator of stochastic classical dynamics~\cite{Rose2016}.

\subsubsection{Classical system dynamics}

 We now discuss how the dynamics generated by $\Wt$ is approximated by the classical dynamics generated by $\W$. We also discuss conditions for the stationary state to be approximated in terms of stationary distribution of $\W$.
 
  In Sec.~\ref{app:Leff_etW}  of~the~SM, we show it follows from Eq.~\eqref{eq:deltaW} that
\begin{eqnarray}
\lVert e^{t\Wt} -e^{t \W} \rVert_1 &\lesssim& 2\sqrt{\Ccl} \,t\,\lVert \Wt\rVert_1. \label{eq:etW_CL}
\end{eqnarray}
Therefore, for times $ t\lVert \Wt\rVert_1\ll 1/\sqrt{\Ccl}$ the effective dynamics in the MM is well approximated by the classical dynamics, as $\lVert\pt(t)-\p(t) \rVert_1 \leq \lVert e^{t\Wt} -e^{t \W} \rVert_1\lVert\pt\rVert_1\lesssim\lVert e^{t\Wt} -e^{t \W} \rVert_1$ [where $\p(t)= e^{t \W} \pt$, cf.~Eq.~\eqref{eq:ptilde}; $\pt$ can further replaced by the closest probability distribution with additional corrections bounded by $\Ccl$, which are of the higher-order for times after the metastable regime, e.g., $t\geq 1/\lVert \Wt\rVert_1$]; see Fig.~\ref{fig:EffDyn}\textcolor{blue}{(c)}. This also holds true for the corresponding density matrices (see~Sec.~\ref{app:norm}  of~the~SM).  

When the approximation in Eq.~\eqref{eq:etW_CL} holds 
for times after the relaxation in the MM, which requires
\begin{equation}\label{eq:cond_CL}
	\tau\lVert \Wt\rVert_1\ll \frac{1}{\sqrt{\Ccl}},
\end{equation} 
the stationary state $\rhoss$ described within the MM by $(\ptss)_{k}=\Tr(\tilde{P}_k \rhoss)$ is well approximated by the stationary probability $\pss$ of the classical dynamics $\W$~\footnote{We note that an analogous approximation of the stationary state can be also obtained by considering non-Hermitian perturbation theory for $\pss$ with respect to perturbation $\W-\Wt$ of $\Wt$ (see Sec.~\ref{app:Leff_pss} in the~SM).}; cf.~Fig.~\ref{fig:EffDyn}\textcolor{blue}{(c)}. Indeed, 
\begin{equation}\label{eq:pss_CL}	 
\lVert\ptss-\pss \rVert_1 \lesssim \lVert \Ptss -e^{t\Wt} \rVert_1+ 2\sqrt{\Ccl} \,t\,\lVert \Wt\rVert_1,
\end{equation}
where $\Ptss$ denotes the projection on $\ptss$. Therefore, $\lVert\ptss-\pss \rVert_1 \ll 1$  follows provided that  $t\lVert \Wt\rVert_1\ll 1/\sqrt{\Ccl}$ for $t$ such that $\lVert \Ptss -e^{t\Wt} \rVert_1\ll 1$~\footnote{Due to the exponential decay of $\lVert \Ptss -e^{n\tau\Wt} \rVert_1\leq \lVert \Ptss -e^{\tilde{\tau}\Wt} \rVert_1^n$ and $\lVert \Ptss -e^{\tau\Wt} \rVert_1\leq (1+\Ctcl/2) \lVert e^{\tau\L}-\P_\text{ss} \rVert$, this will be typically implied by Eq.~\eqref{eq:cond_CL}; cf.~Sec.~\ref{app:tau_defMM}~c in the~SM.}.  
 As a corollary of Eq.~\eqref{eq:pss_CL},	the stationary probability distribution $\pss$ of classical dynamical generator $\W$ in Eq.~\eqref{eq:W} is \emph{unique}. Thus, the  \emph{classical dynamics is ergodic} with the average time spent in $l$th metastable phase equal $(\pss)_l$, $l=1,...,m$. Furthermore, the approximation also holds true for the distance in the trace norm of the corresponding density matrices  (see~Sec.~\ref{app:norm} in the~SM).

Similarly, not only the stationary state but all eigenmodes of the long-time dynamics in the MM can be approximated by those of the classical stochastic dynamics. In particular, in  Sec.~\ref{app:Leff}  of~the~SM, we discuss approximation of the pseudoinverse of $\Wt$ in Eq.~\eqref{eq:Wtilde} by the pseudoinverse of $\W$ in Eq.~\eqref{eq:W}, a result which plays an important role in the approximation of quantum trajectory statistics that we discuss in Sec.~\ref{sec:QTraj}.

We note that the quality of the classical approximations for the structure of the long-time dynamics depends not only on the corrections $\Ccl$ within the metastability regime [Eq.~\eqref{eq:Ccl}], but also on the timescale of the final relaxation (cf.~Eq.~\eqref{eq:pss_CL} and Secs.~\ref{app:Leff_pss} and \ref{app:Leff_R} in the~SM). This is due to the fact that the approximation in Eq.~\eqref{eq:deltaW} captures the fastest among the low-lying modes, while the final relaxation timescale is governed by the slowest among them. In particular, in the case of another metastable regime~\cite{Gaveau1999}, which corresponds to further separation in the spectrum of the master operator in Eq.~\eqref{eq:L}, the condition in Eq.~\eqref{eq:cond_CL} may generally not be valid. 
For example, when a classical first-order phase transition occurs at finite system size, in its proximity $1/(\tau\lVert \Wt\rVert_1)$ is finite when the degeneracy of $m$ stable phases is lifted in the same order $1/(\tau\lVert \Wt\rVert_1)$, so that time $t$ can be chosen $\tau\lVert \Wt\rVert_1 \ll t\lVert \Wt\rVert_1 \ll 1/\sqrt{\Ccl}$ leading to $\lVert\ptss-\pss \rVert_1\ll 1$. But when the perturbation away from the transition lifts the degeneracy of $m$ phases in several different orders (so that $\Ccl$ is of a lower order in the perturbation than $1/\tau$), Eq.~\eqref{eq:cond_CL} is no longer fulfilled (see Sec.~\ref{app:PT} in the~SM).  
We discuss next  how the approximation of the long-time dynamics by classical stochastic dynamics can be refined to take into account the hierarchy of metastabilities. 

Finally, we note it is also possible to approximate $e^{t\Wt}$ by discrete classical dynamics. This again leads to corrections scaling linearly in time, but proportional to $\Ccl$ rather than $\sqrt{\Ccl}$; see Sec.~\ref{app:Leff_discrete} in the~SM.

\subsubsection{Hierarchy of classical long-time dynamics}

When there exists a second metastable regime in the system dynamics, $t''_2\leq t\leq t'_2$, the corresponding metastable states of the system are simply approximated by the projection on the low-lying modes of the classical stochastic dynamics in Eq.~\eqref{eq:W} provided that $t_2''\ll 1/\sqrt{\Ccl}$ [cf.~Eqs.~\eqref{Expansion} and~\eqref{eq:etW_CL}].  When, $t'_2 \geq 2 t''_2$, after the second metastable regime, $t\geq t'_2$, the system dynamics toward the stationary state is approximated by classical dynamics taking place only between $m_2$ metastable phases of the second MM [cf.~Eqs.~\eqref{Expansion} and~\eqref{eq:etW_CL}, and see Sec.~\ref{sec:hierarchy}]. Moreover, when that approximation holds also after the final relaxation,
the system stationary state $\rhoss$  is well approximated by the stationary distribution of that \emph{classical} dynamics [cf.~Eq.~\eqref{eq:pss_CL}]. For further discussion, see Sec.~\ref{app:hierarchy} in the~SM.

\subsubsection{Classical observable dynamics} \label{sec:ObsD_CL}

We now argue how at times after the initial relaxation, the classical long-time dynamics can be observed in the behavior of expectation values or autocorrelations of general system observables. In particular, it can be directly accessed by measuring  the dual basis in Eq.~\eqref{eq:Ptilde}.

For times $t\geq t''$, the dynamics of the average for an observable $O$ depends only on the evolution of the distribution between the metastable phases,
\begin{eqnarray}\nonumber
	\left\langle{O}(t)\right\rangle&=&\tilde{\textbf{o}}^{T}e^{t\Wt}\,\pt+...=\tilde{\textbf{o}}^{T}\pt(t)+...
	\\
	&=&\tilde{\textbf{o}}^{T}e^{t\W}\,\pt+...=\tilde{\textbf{o}}^{T}\p(t)+...,
	\label{ExpansionOB1_CL}
\end{eqnarray}
where $(\tilde{\textbf{o}})_{l}=\Tr(O\trho_{l})$, $l=1,...,m$, are the averages of the observable $O$ in the metastable phases. The first line corresponds  to~Eq.~\eqref{ExpansionOB1}, while the second line follows from Eq.~\eqref{eq:etW_CL} introducing additional corrections bounded in the leading order by $2t\lVert \Wt\rVert_1\sqrt{\Ccl}\max_{1\leq l\leq m} |(\tilde{\textbf{o}})_{l}|$.

 Similarly, the autocorrelation
\begin{eqnarray}\nonumber
\left\langle{O}(t) O(0)\right\rangle_\text{ss}-\left\langle O\right\rangle_\text{ss}^2&=&\tilde{\textbf{o}}^{T}{e}^{t\Wt}\,{\tilde{\textbf{O}}}\,\ptss-(\tilde{\textbf{o}}^{T}\ptss)^2+...\qquad\,\\
&=&\tilde{\textbf{o}}^{T}{e}^{t\W}\,{\tilde{\textbf{O}}}\,\pss-(\tilde{\textbf{o}}^{T}\pss)^2+...,
\label{ApproxCor_CL}
\end{eqnarray}
where $(\tilde{\textbf{O}})_{kl}=\textrm{Tr}[\tilde{P}_{k}\mathcal{O}({\trho}_{l})]$ (cf.~Ref.~\cite{Rose2016}). The first line corresponds to Eq.~\eqref{ApproxCor}, while the second line follows from Eq.~\eqref{eq:etW_CL} with the additional corrections bounded by $\max_{1\leq l\leq m} |(\tilde{\textbf{o}})_{l}[|\lVert \tilde{\textbf{O}}\rVert_{1}(2t\lVert \Wt\rVert_1\sqrt{\Ccl}+\lVert\ptss-\pss \rVert_1)+ 2 |\langle O\rangle_\text{ss}|\, \lVert\ptss-\pss \rVert_1] $ in the leading order. 

Therefore, when metastable phases differ in observable averages [up to the correction in Eq.~\eqref{ExpansionOB1}], the long-time dynamics can be observed by measuring the observable average or autocorrelation. For example, for an observable chosen as a dual basis operator $O=\tilde{P}_l$ in Eq.~\eqref{eq:Ptilde}, we simply have $\tilde{\textbf{o}}^{T}\p(t)=[\p(t)]_l$, 	$l=1,...,m$. Furthermore, when the approximation in Eq.~\eqref{eq:etW_CL} holds for times after the final relaxation [cf.~Eq.~\eqref{eq:cond_CL}], the dynamics of averages and autocorrelations of all system observables is effectively classical  [cf.~Eq.~\eqref{eq:pss_CL}]. In particular, if the  measurement of an observable $O$ is noninvasive, i.e., does not disrupt basins of attractions of metastable phases,  $(\tilde{\textbf{O}})_{kl}\approx\delta_{kl} (\tilde{\mathbf{o}})_{l}$, $k,l=1,...,m$, the long-time dynamics leads to the decay of the autocorrelations exactly as the decay of the autocorrelation of $\tilde{\mathbf{o}}$ in the classical dynamics:   from the observable $\tilde{\textbf{o}}$ variance in $\pss$ during the metastable regime, toward $0$ achieved after the final relaxation (cf.~Ref.~\cite{Smirne2018}). This is the case for the dual basis operators in Eq.~\eqref{eq:Ptilde}, with the distance between the matrices bounded by $\lesssim3(\Cp+\Ctcl+\Ccl/2)$ (see Sec.~\ref{app:Ptilde_corr} in the~SM). 
Finally, higher-order correlations, also between different observables, can be analogously approximated by correlations in classical dynamics.

\subsection{Classical characteristics of quantum trajectories}\label{sec:QTraj}

\begin{figure*}[t]
	\includegraphics[width=\linewidth]{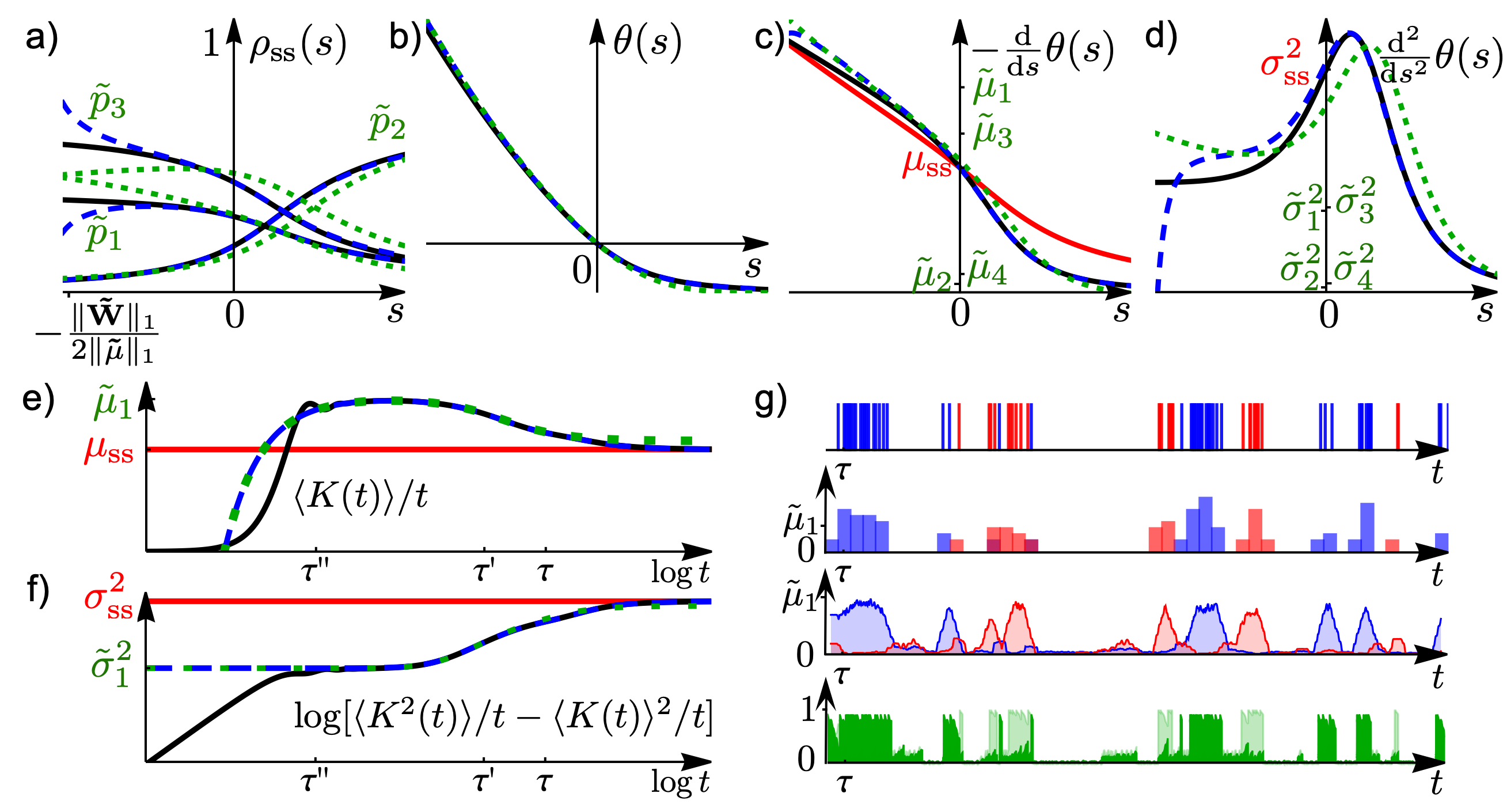}
	\vspace*{-2mm}
	\caption{\label{fig:Trajectories}\textbf{Classical features of quantum trajectories}:  \textbf{(a,b)} Approximating the biased operator of jump activity $\L_s$ [Eq.~\eqref{eq:Ls}] by the biased classical generator $\W_s$ [Eq.~\eqref{eq:Ws}] gives the approximations of: $\rhoss(s)$ (black solid) by the maximal eigenvector of $\W_s$ (green short-dashed) [panel (a); we plot $\Tr[\tilde{P}_l\rhoss(s)]$, $l=1,...,m$; cf.~Eq.~\eqref{eq:Ptilde}], $\theta(s)$ by the cumulant generating function of total activity in classical trajectories [panel (b)], which are valid for $s$ in the perturbative regime of the statistics captured by $m=4$ slow modes (blue dashed) [Eq.~\eqref{eq:Wstilde}], as given by Eqs.~\eqref{eq:rhoss_s} and~\eqref{eq:theta_s_2}. 
		\textbf{(c,d)} Similarly, $-d  \theta(s)/d s=e^{-s}\sum_j\Tr[J_j^\dagger J_j \rhoss(s)]-(e^{-s}-1)\sum_j\Tr[J_j^\dagger J_j\,d\rhoss(s)/d s]$ (black solid), with the first term being the activity of $\rhoss(s)$ (red solid), is captured by the first derivative of the classical cumulant generating function (green short-dashed), up to non-Poissonian contribution to fluctuations in metastable phases [cf.~Eq.~\eqref{eq:Kvar_t_approx_CL}].
		This contribution can be neglected for internal activity dominating classical dynamics, in which case the leading contribution to fluctuations is the result of long timescales of mixing between metastable phases rather than fluctuations within, as demonstrated in panel (d).
	\textbf{(e,f)}	Not only asymptotically, but already for times after the initial relaxation, the average rate and the fluctuation rate of jump number $K(t)$ (black solid) 
	 can be approximated by the constant contribution $\tilde{K}$ from before the metastable regime  and  the contribution from the dynamics within the MM (blue dashed),  with the latter approximated by the corresponding rate for classical total activity $K_\text{cl}(t)$ (green short-dashed) [cf.~Eqs.~\eqref{eq:Kav_t_approx_CL} and~\eqref{eq:Kvar_t_approx_CL}]. 
	 \textbf{(g)} Coarse-graining of jump records (top) in time gives values close to metastable phases activity (upper center) [cf.~panel~(c)], up to fluctuations which decrease with grain size (here $\delta t=0.7\tau'$; cf.~Sec.~\ref{sec:intermittence}). In turn, they capture the average activity of the conditional system state $|\psi(t)\rangle$ (lower center; we plot running average over $\delta t$), and the metastable phase support where $|\psi(t)\rangle$ is found [bottom; we plot $\langle \psi(t)|\tilde{P}_l|\psi(t)\rangle$, $l=1,3$]. 
		\vspace*{-5mm}}
\end{figure*}

In Sec.~\ref{sec:EffDynAv}, we showed that the dynamics of the average system state can be approximated  with the classical dynamics generated by a classical stochastic generator. Now we argue that this relation pertains to individual experimental realizations of system evolution~\cite{Gardiner2004}. Therefore, stochastic transitions between metastable phases can be observed in continuous measurement records or the system state sampled in QJMC simulations~\cite{Molmer92,Dum1992,Molmer93,Plenio1998,Daley2014} so called quantum trajectories (see Fig.~\ref{fig:Trajectories}). First, we show that statistics of quantum trajectories can be directly related to the statistics of classical stochastic trajectories. Second, we argue how coarse-graining in time returns classical trajectories between metastable phases, which for metastable phases differing in activity is the mechanism behind the phenomena of intermittence~\cite{Garrahan2010,Ates2012} and dynamical heterogeneity~\cite{Lesanovsky2013,Olmos2012}, and leads to multimodal distribution of integrated measurement records during the metastable regime. Finally, we explain how system metastability can manifest itself as proximity to a first-order dynamical phase transition in the ensemble of quantum trajectories~\cite{Garrahan2010}.

\subsubsection{Statistics of quantum trajectories}

Quantum trajectories describe the system state conditioned on a continuous measurement record, e.g., counting or homodyne measurement of photons emitted by the system due to action of jump operators in Eq.~\eqref{eq:L}. In particular, the statistics of the total number of jumps in a quantum trajectory (total number of detected photons) is encoded by the biased or ``tilted'' master operator~\cite{Garrahan2010,Ates2012}
\begin{equation}\label{eq:Ls}
\L_s(\rho) =  \L(\rho) +\left({e}^{-s}-1\right)\J(\rho), 
\end{equation}
where $\J(\rho)\equiv\sum_j  J_j \rho J_j^\dagger$. That is, $\Theta(s,t)\equiv\ln(\Tr \{e^{t\L_s}[\rho(0)]\})$ is the cumulant generating function for the number $K(t)$ of jumps that occurred until time $t$ for quantum trajectories initialized in $\rho(0)$. The rates of the asymptotic statistics are determined then by $\theta(s)\equiv\lim_{t\rightarrow\infty} \Theta(s,t)/t$, which is simply the eigenvalue of $\L_s$ with the largest real part. We denote the associated (positive) eigenmatrix as $\rhoss(s)$ and choose the normalization $\Tr[\rhoss(s)]=1$, so that $\theta(s)=\Tr\{\L_s [\rhoss(s)]\}=(e^{-s}-1) \sum_j \Tr[J_j^\dagger J_j \rhoss(s)]$.
Then, $\rhoss(s)$ is the average asymptotic state of the system in trajectories with the probability biased by the factor $e^{-sK(t)}$, while the derivatives of $\theta(s)$ correspond to asymptotic rate of the corresponding cumulants. In particular, for a unique stationary state $\rhoss$ and  the bias $|s|$ small enough with respect to the gap $-\mathrm{Re}(\lambda_2)$,  $\rhoss(s)= \rhoss+...$, while 
\begin{eqnarray}\label{eq:theta_s}
\theta(s)&=& (e^{-s}-1)\, \mu_\text{ss}+...,\\
k(s)&\equiv&-\frac{d }{d s}\theta(s)= {e}^{-s} \,\mu_\text{ss}+...\label{eq:k_s}
\end{eqnarray}
where  $\mu_\text{ss}\equiv\sum_j \Tr[J_j^\dagger J_j \rhoss] $ (cf.~Ref.~\cite{Kato1995}), 
so that the asymptotic jump rate---the asymptotic activity---is determined by the stationary state.

 The nonanalyticities of $\theta(s)$ can be recognized as \emph{dynamical phase transitions}~\cite{Garrahan2010}, in analogy to nonanalyticities of the free energy in equilibrium statistical mechanics.  In particular, a first-order dynamical phase transition occurs at $s_c$ for which the maximal eigenvalue of $\L_{s_c}$ is not unique, so that the asymptotic activity $k(s)=-d  \theta(s)/{d s}$ is no longer continuous, but features a jump at $s_c$~\cite{Garrahan2010,Garrahan2011,Ates2012,Lesanovsky2013}. 
 
 Similarly, statistics for integrated homodyne current and for time-integral of system observables are considered~\cite{Levitov1996,Nazarov2003,Esposito2009,Flindt2009,Hickey2012,Hickey2013} (see also Secs.~\ref{app:homodyne} and~\ref{app:time_int} in the~SM).

\subsubsection{Classical tilted generator} \label{sec:QTraj_Ws}

We now present our first result regarding classicality of quantum trajectories. We argue that the tilted master operator in Eq.~\eqref{eq:Ls} can be approximated by a tilted classical generator encoding the statistics in stochastic trajectories of the classical dynamics in Eq.~\eqref{eq:W}.  This leads to classical approximations for the asymptotic rate of the cumulant generating function and for the asymptotic system state in biased quantum trajectories. 

The statistics of total activity \cite{Lecomte2007,Garrahan2007,Maes2020} in classical dynamics is encoded by a biased or ``tilted'' classical generator [for reviews see Refs.~\cite{Garrahan2018,Jack2019};
cf.~Eqs.~\eqref{eq:W} and~\eqref{eq:Ls}]  
\begin{equation}\label{eq:Ws}
\W_s= \W+  ({e}^{-s}-1 )\,(\Jbf+\bm{\mut}^\text{in}),
\end{equation}
where $(\Jbf)_{kl}\equiv(1-\delta_{kl})(\W)_{kl}$, $k,l=1,...,m$ encodes the transition rates in the classical dynamics, while $(\bm{\mut}^\text{in})_{kl}\equiv\delta_{kl} [\mut_l+(\Wt)_{ll}]$ with $\mut_l\equiv\sum_{j}\Tr (  {J}_{j}^{\dagger}{J}_{j} \,\trho_l)$, $k,l=1,...,m$, encodes the average internal activity in metastable phases (which here is assumed Poissonian distributed; cf.~Sec.~\ref{app:classical} in the~SM).

In Sec.~\ref{app:Ws}  of~the~SM, we show that the tilted classical generator $\W_s$  approximates, in the metastable phase basis, the tiled master operator $\L_s$ when the latter is restricted to the low-lying modes [cf.~Eq.~\eqref{eq:Wtilde}],   
 \begin{equation}\label{eq:Wstilde}
 (\Wt_s)_{kl}\equiv \Tr[\tilde{P}_k \L_s (\trho_l)],
 \end{equation}
 $k,l=1,...,m$,
 with the corrections bounded as~\footnote{$H$ can be further replaced by $H-c\mathds{1}$, with $c$ being a real constant and the norm minimized with respect to $c$ in the corrections; see Sec.~\ref{app:Ws} in the~SM.}
 \begin{eqnarray}\label{eq:Wstilde2}
 &&\lVert \Wt_s- \W_s\rVert \lesssim 2{e}^{-s}\lVert \Wt \rVert_1 \sqrt{\Ccl} \\\nonumber&&\qquad+\left|e^{-s}-1\right|  m\,\Big\lVert H +\frac{i}{2} \sum_{j} {J}_{j}^{\dagger} J_{j}\Big\rVert_{\max}  \sqrt{2\Ccl+4\Cp}.
\end{eqnarray}
 For dynamics of classical systems with metastability, or, more generally, for the basis of metastable phases in Eq.~\eqref{eq:rhotilde} commuting with the dual basis in Eq.~\eqref{eq:Ptilde},  $m\sqrt{2\Ccl+4\Cp}$ in Eq.~\eqref{eq:Wstilde2} can be reduced to $2(\Ctcl+\Cp)$.

Since the biased dynamics $\L_s$ in Eq.~\eqref{eq:Ls} can be considered as the perturbation of the master operator $\L$ in Eq.~\eqref{eq:L} with $({e}^{-s}-1 )\J$, for bias $|s|$ much smaller than  the separation to the fast eigenmodes, $\lambda_m^R -\lambda^R_{m+1}$, the $m$ low-lying eigenmodes and eigenvalues of $\L_s$ in Eq.~\eqref{eq:Ls} are approximated by those of $\P\L_s\P$~\cite{Kato1995} or, in the metastable phase basis, those of $\Wt_s$ in Eq.~\eqref{eq:Wstilde}. From Eq.~\eqref{eq:Wstilde2}, it then follows that the average asymptotic state in biased quantum trajectories is approximated by the asymptotic probability distribution in biased classical trajectories,
\begin{equation}\label{eq:rhoss_s}
\rhoss(s)=\sum_{l=1}^m [\pss(s)]_l  \,\trho_l+...,
\end{equation}
where $\pss(s)$ is the maximal eigenmode of $\W_{s}$ [see Fig.~\ref{fig:Trajectories}\textcolor{blue}{(a)}], while the asymptotic rate of the cumulant generating function is approximated via the asymptotic total activity in biased classical trajectories,
\begin{equation}
\label{eq:theta_s_2}
\theta(s)= (e^{-s}-1)\, \sum_{l=1}^m [\pss(s)]_l\,\mut_l^\text{tot}+...,
\end{equation}
where $\mut^\text{tot}_l\equiv\sum_{k=1}^m(\Jbf)_{kl}+\mut^\text{in}_l$, $l=1,...,m$, are the average total activities in the metastable phases (cf.~Sec.~\ref{app:var_ms} in the~SM), so that the right-hand side of Eq.~\eqref{eq:theta_s_2} is the maximal eigenvalue of $\W_s$  [see Fig.~\ref{fig:Trajectories}\textcolor{blue}{(b)}]. For corrections from non-Hermitian perturbation theory~\cite{Kato1995}, see Sec.~\ref{app:Lsj}~in the~SM. One can consider similarly approximating  $d\theta(s)/ds$ and $d^2\theta(s)/ds^2$, but additional contributions arise from non-Poissonian fluctuations in metastable phases (see Sec.~\ref{sec:QTraj_Cum} and Sec.~\ref{app:Lsj}~in the~SM).  Nevertheless, those can be neglected when the internal activities of metastable phases dominate transition rates of classical dynamics [see~Figs.~\ref{fig:Trajectories}\textcolor{blue}{(c)} and~\ref{fig:Trajectories}\textcolor{blue}{(d)}], making Eq.~\eqref{eq:Wstilde2} the crucial result in the link between the system metastability and its proximity to a first-order dynamical phase transition, which we discuss in Sec.~\ref{sec:metaDPT}.

In Secs.~\ref{app:homodyne} and~\ref{app:time_int}  of~the~SM, we show that in the presence of classical metastability generators of statistics of integrated homodyne current  and time-integral of system observables can be similarly linked to generators of statistics in classical trajectories, but with respect to time-integrals of their average value in metastable phases.

\subsubsection{Classical cumulants}   \label{sec:QTraj_Cum}
We now discuss how the dynamics of the first and the second cumulants of the jump number, directly accessible in experiments via counting measurement, are governed by the classical long-time dynamics for times after the initial relaxation. In particular, we argue how the asymptotic activity and fluctuation rate are approximated by the total activity and total fluctuation rates in the classical dynamics. These results establish a further correspondence between statistics of quantum and classical trajectories for times during and after the metastable regime, and,  even asymptotically, they do not directly follow from Eq.~\eqref{eq:theta_s_2} as cumulants are encoded by derivatives of the rate function [cf.~Figs.~\ref{fig:Trajectories}\textcolor{blue}{(c)} and~\ref{fig:Trajectories}\textcolor{blue}{(d)}].  \\

\emph{Classical dynamics of first cumulant}.
 For times $t\geq t''$ such that $ t \lVert\Wt \rVert_1\ll 1/\sqrt{\Ccl}$, the rate of average jump number is approximated by the time-integral of the total activity in classical trajectories, $K_\text{cl}(t)$, whose statistics in encoded by $\W_s$ of Eq.~\eqref{eq:Ws}, together with the constant contribution $\tilde{K}$ to the jump number accumulated before the metastable regime (cf.~Fig.~\ref{fig:Trajectories}\textcolor{blue}{(e)} and see Sec.~\ref{app:var_t}  of~the~SM),
\begin{eqnarray}\label{eq:Kav_t_approx_CL}
\frac{\langle K(t) \rangle}{t}&=& \frac{\langle K_\text{cl}(t)\rangle}{t}+\frac{\tilde{K}}{t}+...,
\\\nonumber
&\equiv& \frac{1}{t}\!\int_0^t \!\!d  t_1 \!\sum_{l=1}^m\!\big[ \bm{\mut}^\text{tot}  \p(t_1)\big]_l\!-\frac{\Tr\{\J\S\Q[\rho(0)]\}}{t}+\!...,
\end{eqnarray}
where $(\bm{\mut^\text{tot}})_{kl}\equiv\delta_{kl} \mut^\text{tot}_l$, $\p(t)=e^{t\W}\pt$, $\S$ is the pseudoinverse of the master operator $\L$ in Eq.~\eqref{eq:L},  $\Q\equiv\I-\P$ is the projection on the fast-modes of the dynamics [cf.~Eq.~\eqref{Expansion}]. Therefore, similarly as for averages of system observables  [Eq.~\eqref{ExpansionOB1_CL}], the classical dynamics can be observed by measuring $\langle K(t)\rangle$ when the metastable phases differ in the total activity.
 
 When the approximation in Eq.~\eqref{eq:Kav_t_approx_CL} holds for time $t$ after the final relaxation  [cf.~Fig.~\ref{fig:EffDyn}\textcolor{blue}{(c)}], the asymptotic activity in quantum trajectories is approximated by the asymptotic total activity of classical trajectories  [cf.~Fig.~\ref{fig:Trajectories}\textcolor{blue}{(e)}]
 	\begin{eqnarray}\nonumber
 	\mu_\text{ss}&\equiv&\lim_{t\rightarrow\infty}\frac{\langle K\!(t) \rangle}{t}
 	= \sum_j\Tr(J_j^\dagger J_j\rhoss) 
 	\\&=&\lim_{t\rightarrow\infty}\frac{\langle K_\text{cl}(t) \rangle}{t}+...= \sum_{l=1}^m\big(\bm{\mut^\text{tot}}\pss\big)_l +...,
 	\label{eq:Kav}
 	\end{eqnarray}
 	where $\pss$ is the stationary distribution of the classical dynamics $\W$. 
 The corrections are bounded by $\lesssim\max_{1\leq l\leq m}\! |\mut_l|\lVert\ptss-\pss\rVert_1+\lVert\Wt\rVert\sqrt{\Ccl}$ [as $\Tr[\J(\rhoss)] =\sum_{l=1}^m (\bm{\mut}\ptss)_l$; cf.~Eqs.~\eqref{eq:pss_CL} and~\eqref{eq:deltaW}]. \\

\emph{Classical dynamics of second cumulant}. 
For times such that  $\lVert \S\Q\rVert \lVert\J \rVert\ll t \lVert \bm{\mut}\rVert_1$ and $t \lVert\Wt \rVert_1\ll \min(1/\sqrt{\Ccl}, \sqrt{\lVert \Rt\rVert_1 \lVert \Wt\rVert_1 /\sqrt{\Ccl} })$,  where $\Rt$ denotes the pseudoinverse of the long-time-dynamics generator $\Wt$ in Eq.~\eqref{eq:Wtilde}, the rate of fluctuations of jump number is approximated by the rate of fluctuations of total activity in classical trajectories, corrected by non-Poissonian fluctuations in metastable phases and by the contribution to the average from before the metastable regime (see Fig.~\ref{fig:Trajectories}\textcolor{blue}{(f)} and Sec.~\ref{app:var_t} in the~SM)
\begin{eqnarray}\label{eq:Kvar_t_approx_CL}
&&\frac{\langle K^2(t) \rangle\!-\!\langle K(t) \rangle^2}{t}=\frac{\langle K_\text{cl}^2(t) \rangle\!-\!\langle K_\text{cl}(t) \rangle^2}{t}+\frac{	\langle\Delta_\text{cl}(t)\rangle}{t}\qquad\\\nonumber
&&\quad-\sum_{k,l=1}^m\tilde{p}_k\tilde{p}_l \frac{\langle K_\text{cl}^{(k)}(t)\rangle-\langle K_\text{cl}^{(l)}(t)\rangle}{t}\big(\tilde{K}_k-\tilde{K}_l\big)+...,
\end{eqnarray}
where 
\begin{eqnarray} \label{eq:activity2_CL}
&&\langle K^2_\text{cl}(t) \rangle = \!\!\int_0^t \!\!d  t_1 \sum_{l=1}^m\big[\bm{\mut}^\text{tot}  \p(t_1)\big]_l\\\nonumber
&&\qquad+ 2 \!\!\int_0^t \!\!d  t_1\!\!\int_0^{t-t_1}\!\!\!\!\!\! d  t_2 \sum_{l=1}^m[ (\Jbf\!+\!\bm{\mut}^\text{in}) e^{t_2\W}(\Jbf\!+\!\bm{\mut}^\text{in}) \p(t_1)]_l, 
\end{eqnarray}	
and we denoted
\begin{eqnarray}\label{eq:delta_CL}
\langle\Delta_\text{cl}(t)\rangle&\equiv&\!\!\int_0^t \!\!d  t_1 \sum_{l=1}^m\big[ \bm{\delta_{\tilde{\sigma}^2}}  \p(t_1)\big]_l,
\end{eqnarray}
with $(\bm{\delta_{{\tilde{\sigma}}^2}})_{kl}\equiv-\delta_{kl}\, 2\Tr[\J\S\Q \J (\trho_l)]$,  $k,l=1,...,m$, so that $(\bm{\tilde{\sigma}^\text{tot}})^2\equiv \bm{\mut^\text{tot}}+\bm{\delta_{{\tilde{\sigma}}^2}}$ are rates of total fluctuations in metastable phases (see Sec.~\ref{app:var_ms} in the~SM), $\langle K_\text{cl}^{(l)}(t)\rangle$ is the average of $K_\text{cl}(t)$ for $l$th metastable phase in Eq.~\eqref{eq:Kav_t_approx_CL}, i.e., $(\pt)_k=\delta_{kl}$, and $\tilde{K}_l\equiv \Tr\{\tilde{P}_l\J\S\Q[\rho(0)]\}/ \Tr[\tilde{P}_l\rho(0)]$ is the contribution to the jump number from before the metastable regime conditioned on the metastable phase that the system evolves into. Therefore, similarly as for autocorrelations of system observables  [Eq.~\eqref{ApproxCor_CL}], the classical dynamics can be observed even for the stationary state, by measuring fluctuations of $ K(t)$ whenever the metastable phases differ in the total activity.
 
When the approximation in Eq.~\eqref{eq:Kvar_t_approx_CL} is valid for times after the final relaxation, the asymptotic fluctuation rate~\cite{Gammelmark2014,Macieszczak2016a} 
\begin{eqnarray}\label{eq:Kvar}
\sigma_\text{ss}^2&\equiv&\lim_{t\rightarrow\infty}\frac{\langle K^2(t) \rangle\!-\!\langle K(t) \rangle^2}{t}\\\nonumber
&=&\sum_j \Tr(J_j^\dagger J_j\rhoss) - 2\Tr[\J\S\J (\rhoss)]
\end{eqnarray}
is approximated by the asymptotic rate of fluctuations of total activity in classical trajectories, corrected by non-Poissonian contribution to fluctuations in metastable phases (cf.~Fig.~\ref{fig:Trajectories}\textcolor{blue}{(f)} and see Secs.~\ref{app:MM_P} and~\ref{app:var_ss} in the~SM)
\begin{eqnarray}\label{eq:Kvar2}
\sigma_\text{ss}^2
&=&\lim_{t\rightarrow\infty}\Big[\frac{\langle K_\text{cl}^2(t) \rangle\!-\!\langle K_\text{cl}(t) \rangle^2}{t} +\frac{	\langle\Delta_\text{cl}(t)\rangle}{t}\Big]+...\\\nonumber
&=&\!\sum_{l=1}^m \!\big\{\!\big[\bm{\mut^\text{tot}}\! -\!2 \big(\Jbf\!+\!\bm{\mut^\text{in}}\big)\!\R\big(\Jbf\!+\!\bm{\mut^\text{in}}\big) \!+\!\bm{\bm{\delta_{{\tilde{\sigma}}^2}}} \big]\pss\big\}_l\!+\!...
\\\nonumber
&=& \!\sum_{l=1}^m \!\big[\!\big( \bm{\mut^\text{tot}} \! -\!2 \bm{\mut^\text{tot}} \R\bm{\mut^\text{tot}}  \!+\!\bm{\bm{\delta_{{\tilde{\sigma}}^2}}} \big)\pss\big]_l\!+\!...,
\end{eqnarray}
where $\R$ denotes the pseudoinverse of the classical stochastic generator $\W$ in Eq.~\eqref{eq:W} and the last equality follows by noting that $\Jbf+\bm{\mut^\text{in}}=\W+\bm{\mut^\text{tot}}$ and thus $(\Jbf+\bm{\mut^\text{in}})\pss=\bm{\mut^\text{tot}} \pss$. \\

\emph{Other statistics}.
 Similarly to Eqs.~\eqref{eq:Kav_t_approx_CL},~\eqref{eq:Kvar_t_approx_CL},~\eqref{eq:Kav} and~\eqref{eq:Kvar2}, the first and the second cumulants for integrated homodyne current  or for time-integrals of system observables can be related to the statistics in classical dynamics with respect to observables given by the corresponding averages for metastable phases (see Secs.~\ref{app:homodyne} and~\ref{app:time_int} in the~SM). Furthermore, the integrals of average and autocorrelations of system measurements  in Eqs.~\eqref {ExpansionOB1_CL} and~\eqref{ApproxCor_CL} can be approximated analogously.

\subsubsection{Classical dynamics of quantum trajectories}  \label{sec:intermittence}

For systems exhibiting metastability in the system dynamics,  individual evolutions of the system over time typically exhibit \emph{intermittence} (distinct periods of jump activity isolated in time) or \emph{dynamical heterogeneity}  (distinct periods of jump activity isolated both in time and space) in the emission measurement record or time-integral of observables [see Fig.~\ref{fig:Spectrum}\textcolor{blue}{(e)}]. We now explain that these features can be understood in terms of classical dynamics between the metastable phases whose differences in internal (global or local) jump activity dominate transition rates of that dynamics (see also Refs.~\cite{Macieszczak2016,Rose2016}). To this aim, using results of Sec.~\ref{sec:QTraj_Cum}, we establish a \emph{direct} relation between classical trajectories and time-coarse-grained records of continuous measurements. Therefore, we prove that metastability can be observed not only on average [cf.~Eqs.~\eqref{ExpansionOB1_CL} and~\eqref{ApproxCor_CL}], but also in individual realizations of continuous measurement experiments and individual samples of QJMC simulations provided that the metastable regime is long enough. 
	
As a corollary of our results, integrated continuous measurements can be used to distinguish metastable phases  during the metastable regime, as their distribution is \emph{multimodal} with distinct modes corresponding to the metastable phases differing in the rate of the measurement average. Interestingly, during the metastable regime all continuous measurements lead to negligible disturbance of the system as on average they simply correspond to the system dynamics, so that the disturbance is bounded by $2\Cmm$  [cf.~Eq.~\eqref{eq:Cmm} and~Sec.~\ref{sec:DF}].\\

\emph{Time-coarse-grained measurement records as classical trajectories}. We focus here on the measurement of total number of jumps that occur in the system, but analogous arguments hold for the  measurements of local jump activity (see Sec.~\ref{app:Lsj} in the~SM) and of homodyne current (cf.~Sec.~\ref{app:homodyne} in the~SM).

 Consider course-graining in time of a record of jump counting measurement, with the activity in time bins $\delta t$
\begin{equation}
k(n)\equiv \frac{K[(n\!+\!1)\delta t]-K\!(n\delta t)}{\delta t}, 
\end{equation}
for $n=0,1,2,...$. We argue that  time-coarse-grained measurement records can be interpreted as classical trajectories between metastable phases when the internal activity dominates the long time dynamics, $\lVert \bm{\mut}\rVert_1\gg \lVert \Wt\rVert_1$, and $\delta t$ is chosen long enough within the metastable regime, as in this case the activity typically attains only values of the internal activities in metastable phases [see Fig.~\ref{fig:Trajectories}\textcolor{blue}{(g)}].

From Eq.~\eqref{eq:Kav_t_approx_CL}, for $\delta t\leq t'$ such that $\lVert \S\Q\rVert \lVert \J\rVert\ll \delta t \lVert \bm{\mut}\rVert_1$ and $t \lVert\Wt \rVert_1\ll 1/\sqrt{\Ccl}$, the average activity in trajectories originating in $\rho_\text{cond}(t)$ at time $t=n\delta t$ is approximated as
\begin{equation}\label{eq:kav_ms}
\langle k(n)\rangle_{\rho_\text{cond}(t)}=\sum_{l=1}^m\mut_l^\text{in}\tilde{p}_l(n)+...,
\end{equation}
where $\tilde{p}_l(n)\equiv\Tr[\tilde{P}_l \rho_\text{cond}(t)]$ determines the metastable state.
 Similarly, from Eq.~\eqref{eq:Kvar_t_approx_CL}, the variance  
\begin{eqnarray}\label{eq:kvar_ms}
&&\langle k^2(n)\rangle_{\rho_\text{cond}(t)}\!-\!\langle k(n)\rangle_{\rho_\text{cond}(t)}^2=\!\sum_{l=1}^m \tilde{p}_l(n)\,\frac{ (\tilde{\sigma}_l^\text{tot})^2}{\delta t}\qquad\quad\\\nonumber
&&\qquad\qquad\quad\,\,\,+ \!\sum_{k,l=1}^m \!\tilde{p}_k(n)\, \tilde{p}_l(n)\,  \frac{(\mut_k^\text{in}\!-\!\mut_l^\text{in})^2}{2}\\\nonumber
&&\qquad\qquad\quad\,\,\,+ \!\sum_{k,l=1}^m \!\tilde{p}_k(n)\, \tilde{p}_l(n)\, \frac{(\mut_k^\text{in}\!-\!\mut_l^\text{in})(\tilde{K}_k\!-\!\tilde{K}_l)}{\delta t}+...
\end{eqnarray}
(by assuming $(1+\Ctcl/2)\Cmm\ll 1$; see Sec.~\ref{app:var_ms} in the~SM).

 When the conditional state evolves into a single metastable phase $\P[\rho_\text{cond}(t)]=\trho_l$, the average activity is approximated by the activity $\mut_l^\text{in}$ of $l$th metastable phase [cf.~Fig.~\ref{fig:Trajectories}\textcolor{blue}{(e)}] and its variance $(\tilde{\sigma}_l^\text{tot})^2/\delta t$ decays inversely with the increasing time-bin length $\delta t$  [cf.~Fig.~\ref{fig:Trajectories}\textcolor{blue}{(f)}]. Therefore, for long enough metastable regime, $\delta t$ can be chosen so that the fluctuations between measurement records become negligible, and the activity typically takes values approximately equal the average $\mut_l^\text{in}$ [cf.~Fig.~\ref{fig:Trajectories}\textcolor{blue}{(g)} (center)]. 
For $\rho_\text{cond}(t)$ evolving  into a mixture of metastable phases, however, a constant term is present in Eq.~\eqref{eq:kvar_ms} because of a \emph{multimodal distribution} of the activity number in $n$th time bin. Namely, when $\Ctcl\ll 1$,  the distribution can be approximated, up to corrections $2(\Cmm +\Ctcl)+\Ccl$, as a mixture, with probabilities approximating $\tilde{p}_l(n)$, of distributions with averages equal internal activities of metastable phase, $\mut_l^\text{in}$, and variances inversely proportional to $\delta t$, $l=1,...,m$.
This is proved in Sec.~\ref{app:distribution}  of~the~SM, by postselecting trajectories in terms of probability of the final state in the time bin, $\rho_\text{cond}(t+\delta t)$, evolving (on average) into a metastable phase $\trho_l$, which, formally, corresponds to performing at $t+\delta t$ the measurement in Eq.~\eqref{eq:POVM0} that approximates $\tilde{P}_l$ in Eq.~\eqref{eq:Ptilde} [cf.~Fig.~\ref{fig:Trajectories}\textcolor{blue}{(g)} (bottom)].

We conclude that, for long enough $\delta t$, the activity $k(n)$, $n=0,1,2,...$, takes in typical measurement records only $m$ values  $\mut_l^\text{in}$, $l=1,...,m$, corresponding to the internal activities of $m$ metastable phases  (approximately, up to fluctuations decaying inversely in $\delta t$). For the bimodal case $m=2$, see also Ref.~\cite{Rose2016}.  \\

\emph{Dynamics of time-coarse-grained measurement records as classical long-time dynamics}.
We now argue that transitions in coarse-grained measurement records are captured by the generator of the effective long-time dynamics $\W$ [Eq.~\eqref{eq:W}]. 	In particular, the effective lifetime of the $l$th metastable phase in coarse-grained  trajectories is approximated by
$\tau_l\equiv -1/(\W)_{ll}$, $l=1,...,m$. 

From the discussion above, for an initial state $\rho(0)$, the distribution of activity $k(0)$ can be approximated, up to small fluctuations, by a probability distribution over metastable phase activities $\mut_l^\text{in}$, with probabilities approximated by $\tilde{p}_l=\Tr[\tilde{P}_l\rho(0)]$, $l=1,...,m$. Analogously, the distribution of the activity $k(n)$ in a later $n$th time bin is approximated by $\tilde{p}_k(n)=[(e^{\delta t\Wt})^n\pt]_k$, where $t=n\delta t$, which is further approximated by $[(e^{\delta t\W})^n\pt]_k$, $k=1,...,m$ (cf.~Eq.~\eqref{eq:etW_CL}; corrections can be further reduced to $n\Ccl$ by considering discrete stochastic dynamics; see Sec.~\ref{app:Leff_discrete} in the~SM). Therefore, the transition matrix, i.e., the probability of observing $k(n)\approx\mut_k^\text{in}$  conditioned on the observation of the initial activity $k(0)\approx\mut_l^\text{in}$, is approximated by the classical dynamics transition matrix $(e^{\delta t\W})_{kl}$ (or a discrete stochastic dynamics; see Sec.~\ref{app:Leff_discrete} in the~SM).  This relation is further corroborated by Eqs.~\eqref{eq:Kav_t_approx_CL} and~\eqref{eq:Kvar_t_approx_CL}, with the average and variance of the integrated activity, $\sum_{n=0}^{\lfloor t/\delta t\rfloor} k(n)=K(\lfloor t/\delta t\rfloor \delta t)/\delta t$,  approximately governed by the classical long-time dynamics $\W$ [cf.~Figs.~\ref{fig:Trajectories}\textcolor{blue}{(e)} and~\ref{fig:Trajectories}\textcolor{blue}{(f)}].\\

\subsubsection{Classical metastability and dynamical phase transitions}  \label{sec:metaDPT}

Finally, we explain how classical metastability can manifest itself as proximity to a first-order dynamical phase transition in the ensemble of quantum trajectories~\cite{Garrahan2010}, i.e., to a first-order nonanalyticity of $\theta(s)$. \\

\emph{Metastable phases as eigenmodes of tilted generator}. Building on the results of Sec.~\ref{sec:QTraj_Ws}, when the differences in the activity of the metastable phases dominate the transition rates of the classical dynamics  between them, we can approximate $\Wt_s$ in Eq.~\eqref{eq:Wstilde} as
\begin{eqnarray}\label{eq:Wstilde3}
\Wt_s&=& \W-  h_s \bm{\mut}^\text{in} +...\equiv\W_{h_s}+...,
\end{eqnarray}
where $h_s\equiv1-{e}^{-s}$ and $\W_{h_s}$ encodes the statistics of the time-integral of the observable $\bm{\mut}^\text{in}$ in  classical trajectories, rather than their activity (cf.~Sec.~\ref{app:classical} in the~SM). The  corrections in Eq.~\eqref{eq:Wstilde3} additional to Eq.~\eqref{eq:Wstilde2} are $\lesssim h_s\lVert \Wt \rVert_1/2$ [replacing  $ \bm{\mut}^\text{in} $ by $\bm{\mut}$ or $\bm{\mut}^\text{tot}$ doubles them]. 
Furthermore, for the bias large enough, so that $h_s$ is finite,
the contribution from $\W$ in $\W_{h_s}$ of Eq.~\eqref{eq:Wstilde3} can be neglected. In that case, if the bias is still negligible with respect to the gap to the fast eigenmodes, $\lambda_m^R -\lambda^R_{m+1}$, $m$ low-lying eigenmodes of $\L_s$ are simply approximated by the metastable phases and the corresponding eigenvalues and their derivatives approximated analogously to Eqs.~\eqref{eq:theta_s} and~\eqref{eq:k_s}.
In particular, the maximal eigenmode corresponds to the metastable phase with the maximum (for $s<0$) or minimum activity (for $s>0$), and
\begin{eqnarray}\label{eq:theta_s2}
\theta(s)&=&  (e^{-s}-1)\,\mut_{l(s)}+...,
\\\label{eq:k_s2}
k(s)&=&  e^{-s}\,\mut_{l(s)}+...,
\end{eqnarray}
where
\begin{equation}
\mut_{l(s)}\equiv\begin{cases}
\max_{1 \leq l\leq m}\mut_l^\text{in}, & s<0,\\
\min_{1 \leq l\leq m} \mut_l^\text{in}, & s>0.
\end{cases} 
\end{equation}
(cf.~Figs.~\ref{fig:Trajectories}\textcolor{blue}{(a)},~\ref{fig:Trajectories}\textcolor{blue}{(b)}, and~\ref{fig:Trajectories}\textcolor{blue}{(c)}, and  see Sec.~\ref{app:Lsj} in the~SM for corrections from non-Hermitian perturbation theory~\cite{Kato1995}). This is a key observation for the numerical method we introduce in Sec.~\ref{sec:QJMC} to find metastable phases as well as for the relation of metastability to  dynamical phase transitions we explain next. \\

\emph{Metastability as proximity to first-order dynamical phase transitions}.
 For metastable phases differing in activity (or observable averages or homodyne current), Eq.~\eqref{eq:k_s2} implies a sharp change in the derivative of $\theta(s)$, i.e., $-k(s)$, close to $s=0$ [see Fig.~\ref{fig:Trajectories}\textcolor{blue}{(c)}]. This sharp change can be interpreted as the proximity to a first-order dynamical phase transition~\cite{Garrahan2010,Garrahan2011,Ates2012,Lesanovsky2013}. An analogous argument was made for the classical Markovian dynamics in Ref.~\cite{Kurchan2016}.

A sharp change in $k(s)$  around $s=0$, implies in turn a large second derivative of $\theta(s)$ [see Fig.~\ref{fig:Trajectories}\textcolor{blue}{(d)}]. In particular,  $d ^2\theta(s)/d s^2$ at $s=0$  determines the rate of fluctuations in jump number, which can be approximated as [cf.~Eqs.~\eqref{eq:Kvar2} and~\eqref{eq:Wstilde3}]
\begin{equation}
\sigma_\text{ss}^2
=\sum_{l=1}^m \Big[(\bm{\tilde{\sigma}}^\text{in})^2\pss-2 \bm{\mut^\text{in}}\R\bm{\mut^\text{in}}\pss\Big]_l+...,
\label{eq:Kvar3}
\end{equation}
where $ (\bm{\tilde{\sigma}}^\text{in})^2\equiv \bm{\mut^\text{in}}+\bm{\delta_{\tilde{\sigma}^2}}$ and the additional corrections are bounded by $\lVert \Wt\rVert_1 (1 + 2 \max_{1\leq l\leq m}|\mut_l|\lVert \Rt\rVert_1)$. The fluctuation rate is indeed large for the stationary state being a mixture of metastable phases with different activities  [cf.~Figs.~\ref{fig:Trajectories}\textcolor{blue}{(a)} and~\ref{fig:Trajectories}\textcolor{blue}{(d)}]. This is a consequence of long timescales of the effective classical dynamics between metastable phases which govern the intermittence in emission records~\cite{Ates2012,Macieszczak2016}, and are captured by the resolvent in the second term of Eq.~\eqref{eq:Kvar3}. 
In contrast, when the stationary state corresponds to a single metastable phase (so that $\R\bm{\mut}^\text{in}\pss\propto \R\pss =0$), the fluctuation rate is finite as fluctuations originate inside that metastable phase alone [up to corrections of Eq.~\eqref{eq:Kvar3}]. A large second derivative of $\theta(s)$ occurs then away from $s=0$ at intermediate (negative or positive) $s$ values. 

In terms of phase-transition phenomenology, the proximity of a first-order dynamical phase transition manifests itself in a multimodal distribution of a dynamical quantity (i.e., the jump number) in (biased) trajectories for times within the metastability regime, while at longer times in the coexistence, within individual trajectories, of active and inactive regimes that can be considered as dynamical phases [cf.~Fig.~\ref{fig:Trajectories}\textcolor{blue}{(g)}]. These dynamical phases correspond directly to metastable phases (cf.~Sec.~\ref{sec:intermittence}). \\

\emph{Other statistics}.
Similar results to Eqs.~\eqref{eq:theta_s2} and~\eqref{eq:k_s2}, and thus the relation of metastability to dynamical phase transitions, also follow for: individual jump activity (see Sec.~\ref{app:Lsj} in the~SM), integrated homodyne current (see Sec.~\ref{app:homodyne} in the~SM and cf.~Ref.~\cite{Hickey2012}) and time-integrals of system observables (see Sec.~\ref{app:time_int} in the~SM and  cf.~Ref.~\cite{Hickey2013}).

\section{Classical weak symmetries}\label{sec:symmetry}

Here, we discuss how weak symmetries, i.e., symmetries of the master operator in Eq.~\eqref{eq:L}, are inherited by  MMs and long-time dynamics. For classical metastability, we find that nontrivial symmetries are necessarily discrete as they  correspond to classical symmetries, i.e., approximate permutations of metastable phases, which are inherited by the classical long-time dynamics.
Since first-order dissipative phase transitions occurring in thermodynamic limit manifest themselves as metastability for finite system size, our results pave a way for understanding symmetry breaking in open quantum systems  (see also Ref.~\cite{Minganti2018}).

By a
\emph{weak symmetry} we refer to the generator of the system dynamics $\L$ obeying a symmetry on the master operator level,
\begin{equation}\label{eq:weak}
[\L,\U]=0,
\end{equation}
where $\U(\rho)\equiv U\rho \,U^\dagger$ with a unitary operator $U$ of the symmetry (see Refs.~\cite{Baumgartner2008,Buvca2012,Albert2014}). As we consider a unique stationary state, we are interested in the case when the symmetry operator $U$ is not itself conserved by the dynamics, so that in general $\L^\dagger(U)\neq 0$ (as the number of distinct stationary states is the same as the number of linearly independent conserved quantities~\cite{Baumgartner2008,Albert2014}). For example, $\U$ can describe the translation symmetry in homogeneous dissipative systems with periodic boundary conditions. 

From Eq.~\eqref{eq:weak} it follows that $\L$ is block diagonal in the operator basis of eigenmatrices of $\U$. Therefore, the eigenmatrices of $\L$, $R_k$ (and $L_k$ of $\L^\dagger$) can be simultaneously chosen as eigenmatrices of $\U$ (and $\U^\dagger$), in which case $\U(R_k)=e^{i\delta\phi_k}R_k$ [$\U^\dagger(L_k)=e^{i\phi_k} L_k$], where $\phi_k$ equals a difference in arguments of $U$ eigenvalues $(\text{mod}\, 2\pi)$ (cf.~Fig.~\ref{fig:Symmetry} and see Sec.~\ref{app:symmetry1} in the~SM).

\subsection{Symmetry and general metastability}

We first discuss how a weak symmetry in Eq.~\eqref{eq:weak} affects the structure of a general MM and the long-time dynamics within it.

\subsubsection{Symmetry of metastable manifolds}

As the set of all density matrices is invariant under any symmetry, its image under the dynamics featuring a weak symmetry is also symmetric at any time $t$, $\U\{e^{t\L}[\rho(0)]\}=e^{t\L}\{\U[\rho(0)]\}$. In particular, a unique stationary state achieved asymptotically is necessarily symmetric $\U(\rhoss)=\rhoss$ (or,  in the case of degeneracy, the manifold of stationary states is invariant).  Similarly, the set of system states during the metastable regime, i.e., the set of metastable states, is invariant under the symmetry $\U$. This can be seen from the MM being determined by the projection $\P$ on the low-lying modes in Eq.~\eqref{Expansion}, which in the presence of the weak symmetry fulfills 
\begin{equation}\label{eq:weak_P}
[\P,\U]=0
\end{equation}
[cf.~Eq.~\eqref{eq:weak}]. This is a direct consequence of the modes of $\L$ being eigenmatrices of the symmetry, so that the coefficients gain a phase under the symmetry, $\Tr\{L_k\U[\rho(0)]\}=\Tr[\U^\dagger(L_k)\rho(0)]= e^{i\phi_k} \Tr[L_k\rho(0)]=e^{i\phi_k}  c_k$, and thus $\P \{\U[\rho(0)]\}=\sum_{k=1}^m e^{i\phi_k}\, \Tr[L_k\rho(0)]\,R_k=\sum_{k=1}^m \Tr[L_k\rho(0)]\,\U(R_k) =\U\{\P[\rho(0)]\}$; see Fig.~\ref{fig:Symmetry}\textcolor{blue}{(a)}. 

\subsubsection{Symmetry of long-time dynamics} 

The weak symmetry in Eq.~\eqref{eq:weak} together with the symmetry of the MM in Eq.~\eqref{eq:weak_P} yields the symmetry of the long-time dynamics in the MM   in Eq.~\eqref{eq:Leff} as
\begin{equation}\label{eq:weak_LMM}
[\LMM,\U]=0=[\LMM,\U_\text{MM}],
\end{equation}
where $\U_\text{MM}\equiv \P\U\P$; see Fig.~\ref{fig:Symmetry}\textcolor{blue}{(b)}. 
This follows since $[\P\L\P,\U]=\P[\L,\U]\P=0$  and $[\P\L\P,\P\U\P]=\P[\L,\U]\P=0$ from Eq.~\eqref{eq:weak_P} and $[\L,\P]=0$.

\subsection{Symmetry and classical metastability}

We now explain how weak symmetries for classical metastability necessarily correspond to  approximate permutations of metastable phases and thus any nontrivial continuous weak symmetries of low-lying modes preclude classical metastability. We also show the set of metastable phases can be chosen invariant under the symmetry, in which case, the sets of supports and basins of attractions of metastable phases are also invariant. Furthermore, both the long-time dynamics and its classical approximation are then symmetric with respect to the corresponding permutation. This restricts the structure of the low-lying eigenmodes, including the stationary state, and, in turn, simplifies the test of classicality introduced in Sec.~\ref{sec:cMM}.

\subsubsection{Approximate symmetry of metastable phases}

 The symmetry $\U$ in Eq.~\eqref{eq:weak} transforms the projections $\trho_1$, ..., $\trho_m$ in Eq.~\eqref{eq:rhotilde} into $\U(\trho_1)$, ..., $\U(\trho_1)$, which are also projections of system states [e.g., $\U(\rho_1)$, ..., $\U(\rho_m)$ for states in Eq.~\eqref{eq:classicality}]. In the space of coefficients, the symmetry transformation is unitary, and does not change distances. Therefore, as the simplex with vertices corresponding to $\trho_1$, ..., $\trho_m$ approximates well the MM in the space of coefficients, so does the simplex of the transformed new vertices. In fact, it can be shown that the corrections of classicality in Eq.~\eqref{eq:Ccl} are the same for both choices (see Sec.~\ref{app:symmetry2_inv}  of~the~SM). We thus expect that the new vertices to be close to the those of metastable phases.  

Indeed, it can be shown that the set of metastable phases is approximately invariant under the symmetries of the dynamics. In Sec.~\ref{app:symmetry2_inv}  of~the~SM, we prove that the action of the symmetry on the metastable phases
\begin{equation}\label{eq:U}
(\Ubf)_{kl}\equiv \Tr[\tilde{P}_k \,\U (\trho_l)]= \Tr[\tilde{P}_k \,\U_\text{MM} (\trho_l)],
\end{equation}
$k,l=1,...,m$, can be understood as an approximate \emph{permutation of metastable phases}, that is,
\begin{equation}\label{eq:UP}
\lVert\Ubf^n-\Pibf^n\rVert_1\lesssim 3\,\Ccl,
\end{equation}
where $\Pibf$ is a permutation matrix and $n=1,2,...$ are powers of the transformation~\footnote{In Eq.~\eqref{eq:UP} we require $n'\Ccl\ll 1$ for all prime factors $n'$ of $n$ (see Sec.~\ref{app:symmetry2_inv} in the~SM).}. Therefore, from Eq.~\eqref{eq:UP} we obtain that $\trho_l$ is approximately transformed into $\pi^n(l)$ under symmetry applied $n$ times,  $\left\lVert \U^n(\trho_l)-\trho_{\pi^n(l)} \right\rVert\lesssim3\Ccl$,  where $\pi$ is the permutation corresponding to $\Pibf$. Similarly for $\rho_l$ being the closest state to $\trho_l$ we have $\left\lVert \U^n(\rho_l)-\rho_{\pi^n(l)} \right\rVert\lesssim3\Ccl+2\Cp$ [cf.~Eq.~\eqref{eq:Cp}].

\subsubsection{No continuous symmetries} 

We now argue that any continuous weak symmetry acts trivially on the low-lying modes of the master operator when metastability is classical. A continuous weak symmetry is a symmetry $[\U_\phi,\L]=0$ [cf.~Eq.~\eqref{eq:weak}] for all $\phi$, where $\U_\phi\equiv e^{\phi \G}$ with $\G(\rho)\equiv i[G,\rho]$ for a Hermitian operator $G$. For a small enough $\phi$, $\U_\phi$ is approximated by the identity transformation, and therefore for such values of $\phi$ we have $\Pibf=\Ibf$ in Eq.~\eqref{eq:UP} with $n=1$. Since $\Ubf_\phi^n=\Ubf_{n\phi}$, from Eq.~\eqref{eq:UP} the symmetry $\Ubf_{\phi} $ is approximated by $\Ibf$ for any $\phi$. But this is only possible when $\Ubf_\phi=\Ibf$, i.e., the symmetry leaves each metastable phase invariant, otherwise the corrections, as given by the Taylor series,  could accumulate beyond $3\Ccl\ll1$  (see Sec.~\ref{app:symmetry2_C}  in the~SM for a formal proof). Therefore, all slow eigenmodes of the dynamics must be invariant as well. As a corollary, we obtain that \emph{any nontrivial continuous symmetry of slow eigenmodes precludes classical metastability}.

\begin{figure}[t]
	\includegraphics[width=\linewidth]{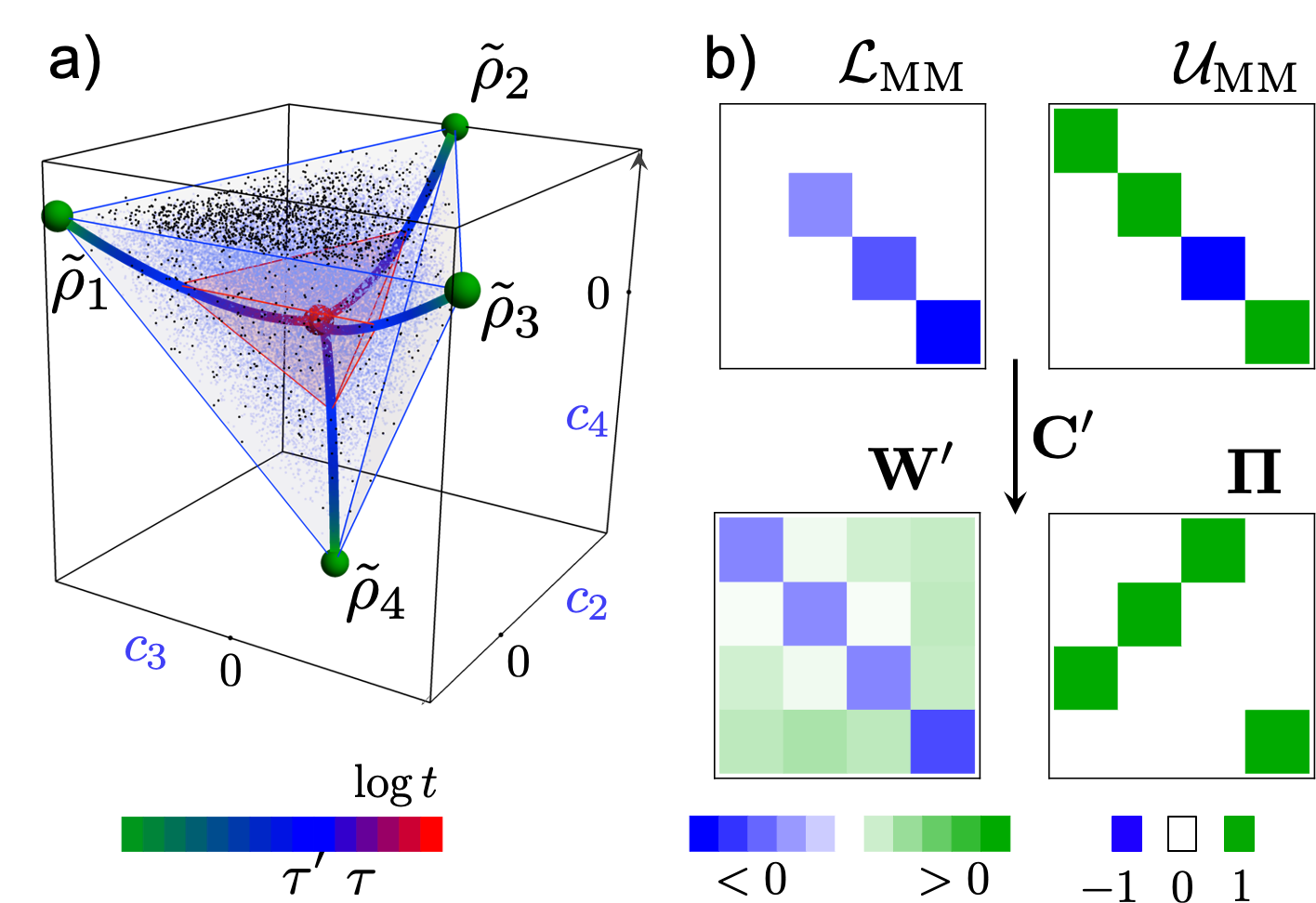}
	\vspace*{-2mm}
	\caption{\label{fig:Symmetry}\textbf{Symmetry of dynamics and classical metastability}: \textbf{(a)} A weak symmetry leads to the MM being symmetric under the corresponding transformation of coefficients, here $c_3\mapsto -c_3$ [cf.~$\U_\text{MM}$ in panel (b)], which is also preserved by the long-time dynamics (blue simplex at $t\ll\tau'$, red simplex at $t=\tau'$).  States invariant under the symmetry, e.g., metastable phases $\trho_2$ and $\trho_4$ and the stationary state $\rhoss$ (red sphere), necessarily feature $c_3=0$. \textbf{(b)} The transformation $\C'$ to the symmetric set of metastable phases in Eqs.~\eqref{eq:rhotilde_U} and~\eqref{eq:rhotilde_U2} yields the classical long-time dynamics $\W'$ symmetric with respect to the permutation $\mathbf{\Pi}$ that corresponds to the action of the symmetry on metastable phases, and in this case swaps $\trho_1$ and $\trho_3$  [cf.~Eqs.~\eqref{eq:weak_LMM} and~\eqref{eq:weak_W2}].
		\vspace*{-5mm}}
\end{figure}

\subsubsection{Symmetric set of metastable phases}

 We now show that the set of metastable phases can be chosen invariant under the action of a weak symmetry. 
 For a discrete symmetry, there exist a smallest nonzero integer $D$  such that $\U^D\P=\P$. We then have $\mathbf{U}^{D}=\Ibf$, and thus from Eq.~\eqref{eq:UP} also $\mathbf{\Pi}^{D}=\Ibf$. Let be $\pi$ be a permutation associated with $\Pibf$, that is, $(\Pibf)_{kl}=\delta_{k \pi(l)}$, $k,l=1,...,m$. For each cycle in the permutation, we choose an element $l$ and define    
\begin{equation}\label{eq:rhotilde_U}
\trho_l'\equiv \frac{d_l}{D} \sum_{n=0}^{\frac{D}{d_l}-1} \U^{n d_l}(\trho_l),
\end{equation}
where $d_l$ is the length of the cycle $\pi^{d_l}(l)=l$ (and thus $D$ is divisible by $d_l$), while for the other elements of that cycle we define
\begin{equation}\label{eq:rhotilde_U2}
\trho_{\pi^n(l)}'\equiv \U^{n}(\trho_l'),\quad n=1,...,d_l-1,
\end{equation}
and denote the transformation from the eigenmodes to this basis as $\C'$ [cf.~Eq.~\eqref{eq:C}].
This gives a \emph{symmetric set of metastable states},
\begin{eqnarray}\label{eq:rhotilde_U3}
\U(\trho_{l}')&\equiv& \trho_{\pi(l)}', \quad l=1,...,m,
\end{eqnarray}
for which the distance to system states is again bounded by $\Cp$ in Eq.~\eqref{eq:Cp}. Furthermore, from Eq.~\eqref{eq:UP} it can be shown that $\left\lVert \trho_l'-\trho_{l} \right\rVert\lesssim 6\Ccl$, $l=1,...,m$, and the corresponding corrections to the classicality, $\Ccl'$,  defined analogously to Eq.~\eqref{eq:Ccl}, can increase at most by $\lesssim 6\Ccl$ (see Sec.~\ref{app:symmetry2_set}  in the~SM for the proofs).  Therefore, without loss of generality, the set of metastable phases can be considered invariant under the symmetry.  In Sec.~\ref{sec:algorithm}, we show how symmetric sets of $m$ candidate sets can be generated efficiently. 
 

In the invariant basis of metastable phases, the action of the symmetry is exactly the permutation [see Fig.~\ref{fig:Symmetry}\textcolor{blue}{(b)}]
\begin{equation}\label{eq:Pi}
\Ubf'\equiv \Pibf,
\end{equation}
where $(\Ubf')_{kl}\equiv \Tr[\tilde{P}_k' \,\U (\trho_l')]= \Tr[\tilde{P}_k' \,\U_\text{MM} (\trho_l')]$,  $k,l=1,...,m$,  and $\tilde{P}_l'$ is the dual basis to $\trho_l'$ in Eqs.~\eqref{eq:rhotilde_U} and~\eqref{eq:rhotilde_U2}, $l=1,...,m$; that is,  $\Tr(\tilde{P}_k' \,\trho_l')=\delta_{kl}$, from which it follows $\U(\tilde{P}'_{l})=\tilde{P}'_{\pi(l)}$ [cf.~Eq.~\eqref{eq:rhotilde_U3}].

Finally, the set of corresponding basins of attraction is symmetric. That is, not only supports of the metastable phases in a cycle are connected by the symmetry operator $\U$, but also their basins of attractions. Indeed,  for $\tilde{p}'_l=\Tr[\tilde{P}'_{l} \rho(0)]$, we have $\tilde{p}'_{l}=\Tr\{\U(\tilde{P}'_{l}) \U[\rho(0)]\}=\Tr\{\tilde{P}'_{\pi(l)} \U[\rho(0)]\}$, $l=1,...,m$. Thus, when $\rho(0)$ belongs to the basis of attraction of $\trho_l$, i.e., $|1-\tilde{p}'_l|\ll 1$, $\U[\rho(0)]$ belongs to the basis of attraction of $\trho_{\pi(l)}$. Similarly,  $\U(\mathds{1}_{\mathcal{H}'_l})=\mathds{1}_{\mathcal{H}'_{\pi(l)}}$  for $\mathcal{H}'_l$ defined for $\tilde{P}'_{l}$  as in Eqs.~\eqref{eq:B_support01}--\eqref{eq:B_support03}.

\subsubsection{Symmetry of classical long-time dynamics}

 \emph{Permutation symmetry}. The weak symmetry of the long-time dynamics in Eq.~\eqref{eq:weak_LMM} in the basis of the metastable phases reads
\begin{equation}\label{eq:weak_Wt}
[\Wt,\Ubf]=0
\end{equation}
 [cf.~Eqs.~\eqref{eq:Wtilde} and~\eqref{eq:U}]. For the set of $m$ metastable phases chosen invariant under the symmetry  [Eqs.~\eqref{eq:rhotilde_U} and~\eqref{eq:rhotilde_U2}], the classical stochastic dynamics between metastable phases $\W'$ that approximates the long-time-dynamics $\Wt'$ [Eq.~\eqref{eq:W}] also features the weak symmetry with respect to the permutation $\Pibf$, 
 \begin{equation}\label{eq:weak_W2}
[\Wt',\Pibf]=0=[\W',\Pibf]
\end{equation}
[cf.~Eq.~\eqref{eq:Pi} and  Fig.~\ref{fig:Symmetry}\textcolor{blue}{(b)}  and see Sec.~\ref{app:symmetry_W1} in the~SM for the proof]. \\

\emph{Structure of low-lying eigenmodes}. 
	We now show that, as a consequence of the symmetry in Eq.~\eqref{eq:weak_W2},  the long-time dynamics may couple only  plane waves over cycles of metastable phases with the same momentum, which results in the low-lying eigenmodes being their linear combinations. In particular, a unique stationary state is composed of uniform mixtures of states in cycles,
\begin{eqnarray}
\label{eq:rhoss_U}
\rhoss=\sum_{l} (\pt_\text{ss}')_l\,\frac{1}{d_l}\sum_{n=0}^{d_l-1} \trho_{\pi^n(l)}',
\end{eqnarray}
where  $l$ runs over cycles representatives with $d_l$ denoting the length of the corresponding cycle  [cf.~Eqs.~\eqref{eq:rhotilde_U}--\eqref{eq:rhotilde_U3}], and $(\pt_\text{ss}')_l=\Tr(\tilde{P}_l'\rhoss)$ corresponds to the stationary distribution of approximately classical dynamics of the symmetric degrees of freedom (see Fig.~\ref{fig:Symmetry}\textcolor{blue}{(a)} and cf.~Secs.~\ref{app:classical_symmetry} and~\ref{app:symmetry_W2} in the~SM).

In the presence of the weak symmetry in Eq.~\eqref{eq:weak_W2}, the long-time dynamics generator $\Wt'$ is block diagonal in an eigenbasis of $\Pibf$, which we can choose as plane waves over the  cycles in the corresponding permutation $\pi$.  Thus, the weak symmetry limits the number of free parameters of $\Wt'$ to the sum of squared degeneracies of the symmetry eigenvalues, i.e., the plane-wave momenta (less $1$ from  the trace-preservation condition), and results in the eigenvectors of $\Wt'$ being linear combination of the plane waves with the same momenta. In particular, $\Wt'$ restricted to the symmetric plane waves, i.e., the uniform mixtures of metastable phases in each cycle, governs the long-time dynamics of symmetric states, which is trace-preserving and approximately positive with the corrections $\lesssim 2\sqrt{\Ccl'}$ (see Sec.~\ref{app:symmetry_W2} in the~SM).

Eigenvectors of $\Wt'$ correspond directly to the low-lying eigenmodes of the master operator $\L$, as they determine the coefficients in the basis of the metastable phases [cf.~Eqs.~\eqref{eq:rhotilde_U}--\eqref{eq:rhotilde_U3}],
\begin{eqnarray}
\label{eq:RkLk}
R_k=\sum_{l=1}^m(\C'^{-1})_{kl}\, \trho'_l,\quad
L_k=\sum_{l=1}^m(\C')_{kl} \,\tilde{P}'_l,
\end{eqnarray}
where $k=1,...,m$.
In particular, the left eigenvector of $\Wt'$ corresponding to the eigenmode $L_k$ is simply the vector of $k$th coefficient for the metastable phases [cf.~Eq.~\eqref{eq:C}]. 
Analogously,
 the plane waves correspond to the eigenmodes of $\U_\text{MM}$ [cf.~Eq.~\eqref{eq:RkLk}], 
\begin{subequations}
\label{eq:RkLk_U}
	\begin{align}
\label{eq:Rk_U}
R_{\pi^{j}(l)}'&\equiv\frac{1}{d_l}\sum_{n=0}^{d_l-1} \left(e^{ -i 2\pi \frac{j}{d_l} }\right)^n\trho_{\pi^n(l)}', 
\\
\label{eq:Lk_U}
L_{\pi^{j}(l)}'&\equiv\sum_{n=0}^{d_l-1} \left(e^{ i 2\pi \frac{j}{d_l} }\right)^n\tilde{P}_{\pi^n(l)}', 
	\end{align}
\end{subequations}
with $j=0,1,...,d_l-1$, $d_l$ being the length of the considered cycle, and  $e^{ i 2\pi j/d_l} $ the corresponding symmetry eigenvalue [cf.~Eqs.~\eqref{eq:rhotilde_U} and~\eqref{eq:rhotilde_U2}]. Therefore, the low-lying modes are their linear combinations,
\begin{eqnarray}
\label{eq:RkLk_C_U}
R_k=\sum_{l=1}^m(\C_\Ubf^{-1})_{kl}\,  R_{l}',\quad
L_k=\sum_{l=1}^m(\C_\Ubf)_{kl} L_{l}',
\end{eqnarray}
where $\C_\Ubf$ is the transformation from the basis of the low-lying eigenmodes to the basis of Eq.~\eqref{eq:RkLk_U},
\begin{eqnarray}\label{eq:C_U}
(\C_\Ubf)_{k\pi^{j}(l)}&=& c_k'^{(l)}  \quad\text{if} \quad e^{i\phi_k}=e^{ i 2\pi \frac{j}{d_l}}\quad\\\nonumber
(\C_\Ubf)_{k\pi^{j}(l)}&=&0\quad\quad\text{otherwise},
\end{eqnarray}
with $k=1,2,...,m$, $j=0,...,d_l-1$, and $c_k'^{(l)}=\Tr(L_k\rho'_l)$ [cf.~Eq.~\eqref{eq:C} and Fig.~\ref{fig:Symmetry}\textcolor{blue}{(a)}]. 
Importantly, $\C_\Ubf$ is \emph{block diagonal} in the eigenspaces of $\Pibf$, so that $R_k$ and $L_k$ are only linear combinations of the eigenmatrices in Eq.~\eqref{eq:RkLk_U} that fulfill $e^{ i 2\pi j/d_l}=e^{i\phi_k}$. In particular, the number of symmetric low-lying modes equals the number of cycles in the permutation, the corresponding block of $\C_\Ubf$ is determined by coefficients for uniform mixtures of metastable phases in each cycle, and the symmetric stationary state is given by Eq.~\eqref{eq:rhoss_U}.
Furthermore, when the symmetry eigenvalue $e^{i\phi_k}$ is unique among low-lying spectrum modes, $R_k$ and $L_k$ are necessarily proportional to Eqs.~\eqref{eq:Rk_U} and~\eqref{eq:Lk_U}, but this is not the case for multiple cycles in general, e.g., for symmetric eigenmodes.
For a single cycle, however, all symmetry eigenvalues are unique, so that $\C_\Ubf$ is diagonal and determined by the coefficients of a single candidate phase. Thus, all low-lying eigenmodes are determined uniquely and the stationary state is an equal mixture of $m$ metastable phases [cf.~Eq.~\eqref{eq:rhoss_U}], as discussed in Ref.~\cite{Minganti2018}.\\

\emph{(No) conservation of symmetry}. Finally, note that the symmetry of the MM can take place without the unitary operator $U$ being conserved during the initial relaxation and the metastable regime $\U$, $\P^\dagger(U)=U$, analogously as is the case for the symmetric manifold of stationary states~\cite{Baumgartner2008,Albert2014,Gough2015}.  Indeed, $U$ itself is a symmetric operator, $\U^\dagger(U)=U$, and thus, if conserved, it is spanned by the symmetric eigenmodes, $U=\sum_l u_l \sum_{n=0}^{d_l-1}\tilde{P}_{\pi^n(l)}$, where $u_l\equiv\Tr(U\trho_l)$ and  $l$ runs over cycles representatives. For a MM with a single cycle, however, the conservation would lead to contradiction, as it would imply a trivial symmetry, $U\propto \mathds{1}$, with $m$, rather than one, cycles.

\subsubsection{Symmetric test of classicality}\label{sec:symmetry_test}

We now use the structure of eigenmodes of the dynamics in the presence of symmetry in Eqs.~\eqref{eq:RkLk_C_U} and~\eqref{eq:C_U} to simplify the test of classicality introduced in Sec.~\ref{sec:cMM_test}. It is important to note beforehand that Eq.~\eqref{eq:RkLk_U} forms a valid basis for any symmetric set of $m$ candidate states that are linearly independent. Thus, to verify whether candidate states indeed correspond to $m$ metastable phases, the test of classicality is necessary even in the case of a single cycle (see Sec.~\ref{app:symmetry_example} in the~SM for an example).

 Exploiting the structure of the eigenmodes, the test of classicality can be simplified as follows. 
 First, only coefficients of cycle representatives are needed to construct $\C_\Ubf$ in Eq.~\eqref{eq:C_U}. Second,  as $\C_\Ubf$ is block diagonal, to find the dual basis to the plane waves [Eq.~\eqref{eq:Lk_U}], only matrices of the size of the permutation eigenspaces need to be inverted~\footnote{When momenta of plane waves over candidate state cycles do not correspond to arguments of symmetry eigenvalues for the low-lying eigenmodes, $\C_\Ubf$ is not invertible and $\!\det\C_\Ubf=0$.}. The dual basis to metastable phases in~Eq.~\eqref{eq:Ptilde} can then be found by the inverse transformation to Eq.~\eqref{eq:Lk_U}, that is, with the coefficients as in Eq.~\eqref{eq:Rk_U}. 
Finally, to estimate the corrections to the classicality as in Eq.~\eqref{eq:Ctcl}, it is enough to consider the elements of the dual basis corresponding to the chosen cycle representatives, $\tilde{P}'_{l}$ [Eq.~\eqref{eq:Lk_U} averaged over $j$], and multiply their contribution by $d_l$~\footnote{This follows from the spectrum of any operator unchanged under the action of the symmetry, and $\U [\tilde{P}'_{\pi^n(l)}]=\tilde{P}'_{\pi^{n+1}(l)}$. 	Alternatively, the minimal eigenvalue can be found from the wave-plane basis by considering averages of $\sum_{n=0}^{d_l-1} (e^{-i 2\pi\frac{k}{d}})^n \U^{\dagger n}(|\psi{\rangle\langle}\psi|)$ maximized over ${|\psi\rangle}$, which can be restricted to $U^{d_l}$ symmetric states [cf.~Eq.~\eqref{eq:Lk_U}].}.


\section{Unfolding classical metastability numerically}\label{sec:numerics}

With the theory of classical metastability now established, we turn to the question of how to efficiently uncover the structure of a MM and long-time dynamics in a given open quantum system governed by a master equation. We introduce two numerical methods to analyze the classical metastability in such systems with or without weak symmetries. The first approach in Sec.~\ref{sec:algorithm} requires diagonalizing the master operator and its low-lying eigenmodes. It verifies the presence of classical metastability, delivers the set of metastable phases, and uncovers the structure of the long-time dynamics. The second approach in Sec.~\ref{sec:QJMC} instead utilizes quantum trajectories with probabilities biased according to their activity, so that the metastable phases with the extreme activity are found.

\subsection{Metastable phases from master operator spectrum}  \label{sec:algorithm}

Efficient algorithms to uncover the structure of the stationary state manifold~\cite{Blume2008,Blume2010} (which utilize Ref.~\cite{Holbrook2004}; see also Ref.~\cite{Choi2006}) rely on the exact diagonalization of the master operator and the von Neumann algebra structure of the stationary left eigenmodes which arises when they are restricted to the maximal support of stationary states. This algebraic structure, however, is not generally present for the low-lying left eigenmodes, as visible, e.g., in first-order corrections to the formerly stationary modes when they are perturbed away from a dissipative phase transition at a finite system size by excitation of decaying modes (see Supplemental Material of Refs.~\cite{Macieszczak2016a}).

Here, we introduce a general approach which delivers the set metastable phases and the structure of the long-time dynamics when the metastability is classical. Similarly to the algorithm in Refs.~\cite{Blume2008,Blume2010}, it is based on the low-lying left eigenmodes of the master operator in Eq.~\eqref{eq:L}, but their connection to basins of attractions is utilized by the observation that extreme values of the corresponding coefficients are achieved by metastable phases (cf.~Fig.~\ref{fig:MM}), which thus correspond to the projections on the MM of the extreme eigenstates of left eigenmodes. To guarantee that all metastable phases are found, random rotations of low-lying modes are employed until the corrections to the classicality are small. This way, the algorithm remains efficient, as it does not simply probe the whole space system space of pure states (cf.,~e.g.,~Ref.~\cite{Zurek1993}).
Furthermore, symmetries of the dynamics can be exploited to simplify the method. 
We also consider how observable averages distinguishing metastable phases (i.e., order parameters) can be utilized. 
Finally, we note that this method has been recently successfully applied in Ref.~\cite{Rose2020} to the open quantum East model \cite{Olmos2012} featuring a hierarchy of metastabilities and translation symmetry.

\begin{figure}[t]
	\includegraphics[width=\linewidth]{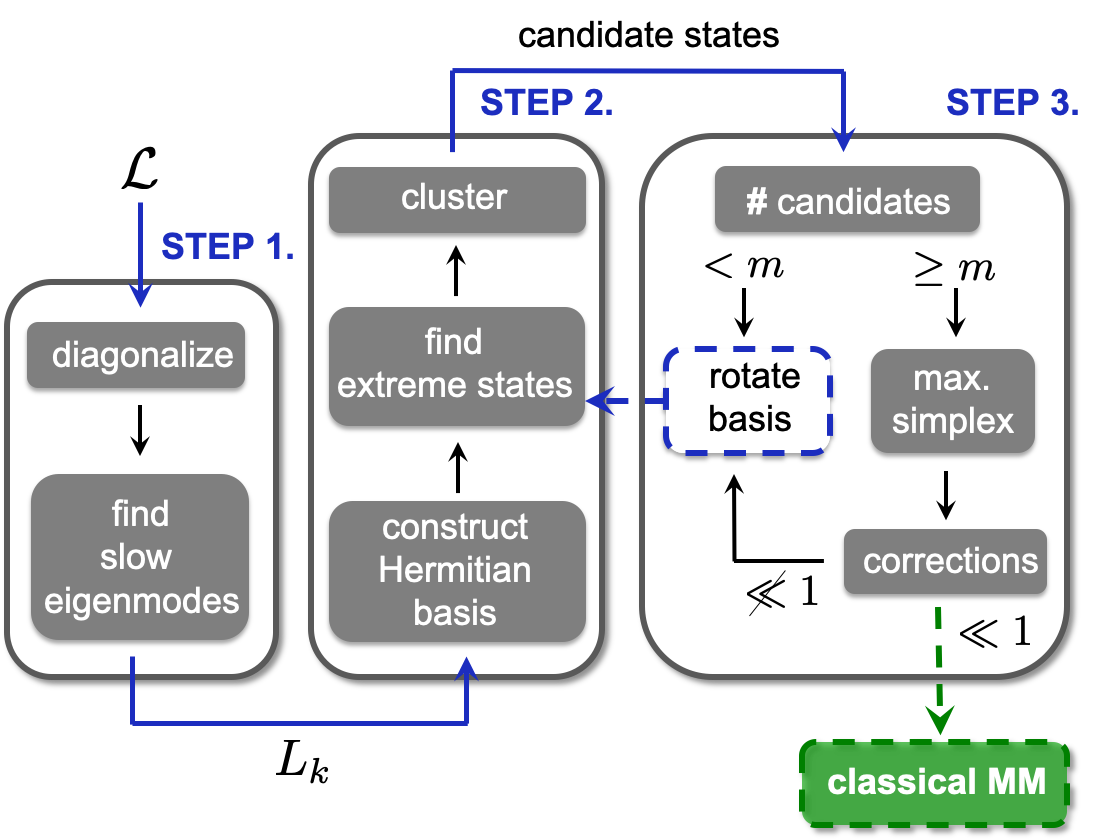}
	\caption{ \label{fig:Algorithm}\textbf{Metastable phases construction}. We sketch the approach to construct a candidate set of $m$ metastable phases within a given MM that provides the best classical approximation in terms of corrections in Eq.~\eqref{eq:Ccl}. The construction can be refined by considering  symmetries of the dynamics.}
	\vspace*{-5mm}
\end{figure}

\subsubsection{Metastable phases construction}

 Our approach consists of the following steps (see also Fig.~\ref{fig:Algorithm}):
\begin{enumerate}
	\item Diagonalize $\L$ to find the left eigenmatrices  $L_k$ below the gap in the spectrum, $k=2,...,m$.
	\item Construct candidate metastable states: 
	\begin{itemize}
	\item[-] diagonalize the (rotated) eigenmatrices $L_k$, 
	\item[-] choose the eigenstates associated to their extreme eigenvalues as initial states for candidate metastable states,
	\item[-] discard repetitions in candidate metastable states---cluster in the coefficients space.
	\end{itemize}
	\item Find best candidate metastable states:
	\begin{enumerate}
		\item If the number of candidate states $\geq m$, then choose the set of $m$ states providing the simplex with the largest volume, i.e., the largest $|\text{det}\C|$ [cf.~Eq.~\eqref{eq:C}] and calculate the corresponding corrections to the classicality in Eq.~\eqref{eq:Ccl}, as can be easily bounded by Eq.~\eqref{eq:Ctcl}. 
		\item If the number of candidate states $< m$, or the corrections to the classicality in Step~3i are not negligible, then enlarge the set of candidates obtained from Step~3 by considering a random rotation of the basis of the left eigenmatrices in Step~2.     
	\end{enumerate}
\end{enumerate}

Step~1 in the above construction provides the low-dimensional description of the MM, and, as explained in Sec.~\ref{sec:cMM}, allows for testing the approximation of the MM  as mixtures of $m$ candidate states. We choose Hermitian $L_k$ replacing conjugate pairs of eigenmodes $L_k$, $L_k^\dagger$~\cite{Note2} by
\begin{eqnarray}\label{eq:LRLI}
L_k^R\equiv\frac{e^{-i\varphi_k}L_k+e^{i\varphi_k}L_k^\dagger}{\sqrt{2}},\quad L_k^I\equiv\frac{e^{-i\varphi_k}L_k-e^{i\varphi_k}L_k^\dagger}{\sqrt{2}i},\qquad
\end{eqnarray}
where $e^{-i\varphi_k}$ is an arbitrary phase.  Step~2 relies on the result in Sec.~\ref{app:algorithm_Lk} of~the~SM, that metastable states arising from extreme eigenstates of the dynamics eigenmodes can be used to approximate metastable phases in classical MMs, as long as only a single metastable phase is close to the extreme value of the corresponding coefficient $c_k$ (which we refer to as the case without degeneracy); cf.~Figs.~\ref{fig:MM} and~\ref{fig:Symmetry}\textcolor{blue}{(a)}. We then discard any repetitions in the set of candidate states (to treat all coefficients on equal footing, we set the normalization $c_k^{\max}-c_k^{\min}=1$, where $c_k^{\max}$ and $c_k^{\min}$ are extreme eigenvalues of $L_k$). Indeed, for a given left basis we obtain $2(m\!-\!1)$ candidate metastable states corresponding to $2(m\!-\!1)$ extreme eigenvalues of the basis elements, which may provide up to $m$ metastable phases. In the case without degeneracy, each candidate corresponds to one of $m$ metastable phases. In the case with degeneracy, some of extreme eigenstates may correspond to mixtures of metastable phases: provided that the set of candidate states features all metastable phases, such a candidate state should be discarded in Step~3i  (this relies on the result from Sec.~\ref{app:algorithm_vol} in the~SM, that the simplex of metastable phase is approximately the largest simplex inside the MM; cf.~Fig.~\ref{fig:MM}).
However, such mixtures may cause less than $m$ candidates to remain after clustering, or result in a set of phases which provides a poor approximation to the true MM; even without degeneracy, it is possible that some metastable phases may reside on the interior of the hypercube defined by the extreme values of the coefficients, and as such will not appear in the set of candidate states taken from extreme eigenvalues of the eigenmodes. Nevertheless,
random rotations in Step~3ii ensure that each metastable phase is eventually \emph{exposed}, i.e., a basis in which that metastable phase achieves an extreme coefficient value without degeneracy is eventually considered (with the probability $1$ achieved exponentially in the number of rotations; cf.~Sec.~\ref{app:algorithm_rot} in the~SM).   When the set of metastable phases leading to small corrections to the classicality is found, the dual basis in Eq.~\eqref{eq:Ptilde} can be constructed, and the long-time dynamics decomposed as in Eq.~\eqref{eq:Wtilde}.

Naturally, instead of considering distances between candidate states in the space of coefficients, as used in Steps~2 and~3i, candidate states can be clustered with respect to the trace distance in the space of density matrices, while the best $m$ candidate states can be chosen to achieve minimal corrections to the classicality instead of the maximal simplex volume. These modifications, however, require working with operators on the system Hilbert space, rather than on the space of coefficients, and thus are in general more expensive numerically [for classical MMs, $m\leq \text{dim}(\mathcal{H})$], while not necessary for MMs with nonnegligible volumes (cf.~Secs.~\ref{app:algorithm_rot} and~\ref{app:algorithm_vol} in the~SM).

Our approach will not deliver a set of metastable phases with negligible corrections to the classicality  whenever the MM is not classical. In particular, quantum MMs~\cite{Macieszczak2016a} featuring decoherence free subspaces~\cite{Zanardi1997,Zanardi1997a,Lidar1998} or noiseless subsystems~\cite{Zanardi2000,Knill2000}: which, e.g., at $m=4$ amount to Bloch-sphere in the coefficient space, rather than a tetrahedron (see also Ref.~\cite{Bengtsson2006}), cannot be approximated as probabilistic mixtures of $m$ metastable phases. Even for classical metastability emerging in many-body open quantum systems, the approach relies on the condition in Eq.~\eqref{eq:Ccl3}, which may be fulfilled only at larger system sizes, when the low-lying part of the master operator spectrum to be sufficiently separated from the fast modes. In this case, our approach may not succeed for smaller system sizes with  less pronounced metastability.


While degeneracies are unlikely to occur in a generic model, they typically appear in the presence a hierarchy of metastabilities or as a consequence of symmetries of the dynamics, which we discuss in detail below. In particular, in the presence of a weak symmetry, not only the degeneracy can be efficiently remedied, but also the search for candidates states made even more efficient.

\subsubsection{Construction for hierarchy of metastable manifolds}


In the presence of hierarchy of metastabilities with a further separation at $m_2<m$ in the spectrum of the master operator,  the degeneracy appears as a consequence of the fact that the simplex of $m$ metastable phases of the first MM, when projected onto the coefficients  $(c_2,...,c_{m_2})$,  is approximated by a simplex with $m_2$ vertices corresponding to $m_2$ metastable phases of the second MM. 
 This requires (at least) $m_2$ metastable phases in the first MM to evolve directly into $m_2$ metastable phases of the second MM.  Each of other $m-m_2$ metastable phases of the first MM either evolves into a single phase of the second MM, or it belongs to the decay subspace in which case it in general evolves into a mixture of $m_2$ metastable phases of the second MM. In the former case, the metastable phases in the first MM that evolve into the same phase in the second MM are degenerate in the coefficients  $(c_2,...,c_{m_2})$. In the latter case, they do cannot take extreme values of those coefficients, even after a rotation of the first $m_2$ modes  (see Secs.~\ref{app:hierarchy_def} and~\ref{app:algorithm_hierarchy} in the~SM). Nevertheless, rotations of \emph{all} $m$ modes allow  both for the degeneracy to be lifted and for every metastable phase to take an extreme value in one of the coefficients.

\subsubsection{Construction for metastable manifolds with symmetries}

In the presence of a weak symmetry $\U$ [Eq.~\eqref{eq:weak}], the master operator $\L$ is block diagonal in the eigenspaces of $\U$, which simplifies its diagonalization~\cite{Baumgartner2008,Buvca2012,Albert2014}. 
As a consequence, the low-lying eigenmodes of the dynamics are chosen as linear combinations of the plane waves over the cycles induced by the symmetry on $m$ metastable phases [cf.~Eq.~\eqref{eq:RkLk_C_U}]. In particular, for an eigenmode $L_k$, $k=2,...,m$ with a symmetry eigenvalue $e^{i\phi_k}$,  $\U^\dagger(L_k)=e^{i\phi_k} L_k$, and the minimal integer $n_k>0$ leading to $e^{in_k\phi_k}=1$,  $L_k$ is  supported on cycles with the length equal $n_k$ or larger but divisible by $n_k$.  The latter case, of subcycles with length $n_k$, leads to degeneracy of the coefficient $c_k$ for the metastable phases connected by $\U^{n_k}$, with its extreme eigenstates generically projecting onto their uniform mixture  [cf.~Eq.~\eqref{eq:Lk_U} and Fig.~\ref{fig:Symmetry}\textcolor{blue}{(a)}, where $\trho_1$ and $\trho_3$ are degenerate in $c_2$ and $c_4$ ($n_2=n_4=1$)]. Nevertheless, coefficient degeneracy can be remedied and the structure of the low-lying eigenmodes can be used to actually enhance the introduced approach, as we now explain (see also Sec.~\ref{app:algorithm_symmetry} in the~SM).


 %
 We note that when $\text{GCF}(n_k,n_l)<n_k$ for all $n_l\neq n_k$, $k,l=2,...,m$, the eigenmode $L_k$ is supported only on cycles of length $n_k$, as there are no longer cycles with the length divisible by $n_k$. The symmetry $\U^{n_k}$ then acts trivially on the corresponding metastable phases and the above discussed degeneracy is absent. Analogously to the general case, any plane wave in $L_k$ can then be exposed by random rotations of the eigenmodes with the same symmetry eigenvalue  $e^{i\phi_k}$, while at least a single metastable phase from each cycle can be obtained by considering extreme eigenstates of both $L_k^R$ and $L_k^I$~\footnote{For $n_k$ not divisible by $4$, otherwise see Sec.~\ref{app:algorithm_symmetry3} in the~SM.}. Other metastable phases in the considered cycles can be recovered by applying the symmetry $n_k-1$ times, so there is no need to consider eigenmodes $L_{l}$ with a different symmetry eigenvalue supported on the same cycles (i.e., $L_l$ with $e^{i\phi_k}\neq e^{i\phi_k}$ but $n_l=n_k$). 

Furthermore, metastable phases in cycles with the length corresponding to subcycles, e.g., invariant metastable phases, can also be found. For $L_l$ such that $n_l$ is divides only $n_k>n_l$ for the above  considered eigenmodes $L_k$, when the degeneracy of $e^{i\phi_l}$ equals the number of already considered cycles with $n_k$ divisible by $n_l$ (i.e., the sum of the corresponding symmetry eigenvalue degeneracy for all such $n_k$ values), $L_l$ is supported on the already considered cycles. Otherwise, $L_l$ features new cycles with the length $n_l$, which can be unfolded, as before, by rotations of all eigenmodes with the same symmetry eigenvalue as $e^{i\phi_l}$ and considering both $L_l^R$ and $L_l^I$~\cite{Note16}. Here, equal mixtures of already considered phases connected by $\U^{n_k/n_l}$ will also be found, but such candidate states will not lead to the maximal volume simplex. Again, by applying symmetry $\U$ the full (sub)cycles can be recovered, and other eigenmodes $L_j$ with $n_j=n_l$, but a different symmetry eigenvalue can be discarded. Analogous results hold for the remaining eigenmodes, but with respect to $L_l$ and $e^{i\phi_l}$ degeneracy.

In summary, the set of eigenmodes considered in Step~2 is significantly reduced, with its size equal to the number of cycles and the subcycles with other cycle's length. Furthermore, only rotations of eigenmodes with the same symmetry eigenvalue are necessary in Step~3ii. Importantly, as the eigenstates of $L_k^R$ and $L_k^I$ and the corresponding metastable states are invariant under $\U^{n_k}$, that is, generate cycles of candidate states with the length dividing $n_k$, following the prescription above, we arrive at an invariant set of candidate states. This invariance can be maintained by clustering whole cycles of candidate states rather than individual states in Step~2. Then, without loss of generality,  in Step 3i, instead of considering subsets of all candidate states, we can choose candidate states as sets of cycles with their lengths summing to $m$. In that case, the volume of the simplex can be efficiently calculated as $ |\!\det\C_\Ubf|/(m-1)!\prod_l\sqrt{d_l}$, where $\C_\Ubf$ is the block-diagonal matrix in Eq.~\eqref{eq:C_U} and the product, which runs over cycle representatives, is the same for all sets of linearly independent candidates~\cite{Note14}, while the corrections to the classicality can be efficiently calculated with the symmetric test of classicality of Sec.~\ref{sec:symmetry_test}.

\subsubsection{Construction utilizing order parameters}

Instead of considering the eigenmodes of the dynamics, we can choose a left eigenbasis formed by a set of $m$ observables $O_l$, $l=1,...,m$ projected onto the low-lying eigenmodes, i.e., $\P^\dagger(O_l)=\sum_{k=1}^m  b_k^{(l)} L_k $, where $b_k^{(l)}\equiv\Tr(O_l R_k)$ [cf.~Eq.~\eqref{Expansion}], provided that those projections are linearly independent.  In this case, the extreme eigenstates of $\P^\dagger(O_k)$ will give metastable states attaining extreme values in the average of the observable $O_k$. Those metastable states will correspond to metastable phases, up to degeneracy of metastable phase averages of $O_k$ (in particular, in the presence of nontrivial weak symmetry of low-lying eigenmodes, it is necessary to consider observables breaking the symmetry). Among others, this can be helpful when the volume of MM in the space of coefficients is negligible.  
 In the next section, we extend this approach by considering continuous measurements instead of system observables.

\subsection{Metastable phases from biased quantum trajectories}  \label{sec:QJMC}

In some systems, the metastability can be a collective effect emerging as the system size increases~\cite{Rose2016}.
If large system sizes are required for prominent metastability, then exact diagonalization may not be feasible. Therefore, we now introduce an alternative numerical approach to finding metastable phases in classical MMs using QJMC simulations~\cite{Molmer92,Dum1992,Molmer93,Plenio1998,Daley2014} and biased sampling in the framework of large deviation theory (see Ref.~\cite{Touchette2009} for a review). In classical stochastic dynamics biased sampling can be efficiently incorporated into the generation of trajectories, with techniques such as transition path sampling~\cite{Hedges2009} and cloning~\cite{Giardina2011}. 

Trajectories of the biased master equation $\L_s$ in Eq.~\eqref{eq:Ls} can be viewed as trajectories of $\L$ with their probability multiplied by $e^{-s K(t)}$, where $K(t)$ is the total number of jumps occurring in a quantum trajectory of length $t$. The maximal eigenmode $\rhoss(s)$ of $\L_s $ corresponds then to the asymptotic system state in quantum trajectories averaged with the biased probability. In Sec.~\ref{sec:metaDPT}, we argued that $\rhoss(s)$  can approximate metastable phases of extreme activity for appropriately chosen $s$ when the activity dominates the transition rates of long-time dynamics 
[cf.~Figs.~\ref{fig:Trajectories}\textcolor{blue}{(a)} and~\ref{fig:Trajectories}\textcolor{blue}{(c)}]. Thus, if the efficient biased sampling could be generalized to QJMC sampling, metastable phases with extreme activity could be accessed via time-average of a biased trajectory over time-length within the metastable regime~\footnote{The timescale $\tau(s)$  of the relaxation toward $\rhoss(s)$  in the biased dynamics  $e^{t\L_s}[\rho(0)]/\Tr \{e^{t\L_s}[\rho(0)]\}$ belongs to the metastable regime when Eqs.~\eqref{eq:theta_s2} and~\eqref{eq:k_s2} hold. Indeed, due to the separation also present in the spectrum of $\L_s$
$\tau(s)$ is approximated by the relaxation time of $\W_{h_s}$ of Eq.~\eqref{eq:Wstilde3} and negligible corrections from fast modes of $\L$ give $\tau(s)\geq t''$. Moreover, as $\W$ can be neglected in Eq.~\eqref{eq:Wstilde3}], $\tau(s)\leq t'$.}.

Similarly as in the case of degeneracy of coefficients, when more than a single metastable phase corresponds to the maximum or the minimum activity, $\rhoss(s)$ corresponds to a mixture of the metastable phases with the extreme value (e.g., when both $\L$ and $\J$ obey a symmetry that is broken in the MM, the mixture is symmetric). Nevertheless, the discussion in Sec.~\ref{sec:metaDPT} is analogous for the activity of individual jumps, and thus a further distinction between metastable phases can be enabled this way, e.g.,  by breaking the translation symmetry of $\L_s$ in the case of identical local jumps  (see Sec.~\ref{app:Lsj} in the~SM). Finally, we note that in the case of general metastability, this approach will return the metastable states corresponding to the extreme values of jump activity.


\section{Conclusions and outlook}\label{sec:conclusion}

In this paper, we formulated a comprehensive theory for the emergence of classical metastability for open quantum systems whose dynamics is governed by a master operator. We showed that classical metastability is characterized by the approximation of metastable states as probabilistic mixtures of $m$ metastable phases, where $m$ is the number of low-lying modes of the master operator. Namely, in terms of the corresponding corrections, metastable phases are approximately disjoint, while the long-time dynamics, both on average and in individual quantum trajectories, is approximately governed by an effective classical stochastic generator. Furthermore, any nontrivial weak symmetries present at long times are necessarily discrete as they correspond to approximate permutations of metastable phases, under which the classical dynamics is invariant. To investigate metastability for a given open quantum system, we introduced the test of classicality---an approach to verify the approximation of the MM by a set of candidate metastable phases. We also developed a complementary numerical approach to deliver sets of candidate metastable phases. Since that approach requires diagonalization of the master operator---a difficult task in systems of larger size---we also discussed an alternative based on the concept of biased trajectory sampling.

The techniques we introduced here allow us to achieve a complete understanding of classical metastability emerging in an open quantum system. A concrete application of the methods described here to a many-body system of an open quantum East model \cite{Olmos2012}  is given in Ref.~\cite{Rose2020}, where despite the stationary state being analytic in dynamics parameters,  the dynamics is found to  feature a hierarchy of classical metastabilities, with metastable phases breaking the translation symmetry of the model and their number increasing with the system size. This structure is then shown to be analogous to metastability in the classical East model, but with an effective temperature and coherent excitations.

While our general approach relies only on the Markovian approximation of system dynamics, non-Markovian effects could be includedvia the chain partial diagonalization~\cite{Burkey1984,Chin2010,Woods2014} (see also Ref.~\cite{Iles2014}). However, an open quantum system and its environment form an isolated system, which raises a question about the relation of classical metastability to  separation of timescales in closed system dynamics, which generically leads to prethermalization of subsystem states~\cite{Gring2012}. Furthermore, as in this work no assumptions were made on the structure of system Hamiltonian or jump operators, it should be explored how metastable phases and their basins of attraction can be further characterized in systems with local interactions and local dissipation (see also Ref.~\cite{Landa2020}).
Finally, it remains an open question what is the structure of general quantum metastability and how it can be efficiently investigated, e.g., to uncover metastable coherences~\cite{Macieszczak2016a}. Extending the methods described here to this general case  would inform, among others, the study of a general structure of first-order dissipative phase transitions in open quantum systems.

\bigskip

\acknowledgments
K.M.\ thanks M.\ Gu\ifmmode \mbox{\c{t}}\else \c{t}\fi{}\ifmmode \u{a}\else \u{a}\fi{} for helpful discussions. K.M.\ gratefully acknowledges support from a Henslow Research Fellowship. We acknowledge financial support from EPSRC Grant No.\ EP/R04421X/1, from the H2020-FETPROACT-2014 Grant No. 640378 (RYSQ), and by University of Nottingham Grant No.\ FiF1/3.
I.L.\ acknowledges support by the ``Wissenschaftler-R\"uckkehrprogramm GSO/CZS'' of the Carl-Zeiss-Stiftung, through the Deutsche Forschungsgemeinsschaft
(DFG, German Research Foundation) under Project No. 435696605 and
through the European Union’s H2020 research and innovation programme
[Grant Agreement No. 800942 (ErBeStA)]. We are grateful for access to the University of Nottingham Augusta HPC service.


%


\onecolumngrid
\clearpage
\begin{center}
	\textbf{\large Theory of classical metastability in open quantum systems\\Supplemental Material}
\end{center}
\smallskip

\addcontentsline{toc}{section}{Supplemental Material}

\makeatletter
\setcounter{section}{0}
\setcounter{equation}{0}
\setcounter{figure}{0}
\setcounter{table}{0}

\makeatother
\makeatletter
\renewcommand{\thesection}{\Alph{section}}
\renewcommand{\thesubsection}{\arabic{subsection}}
\renewcommand{\thesubsubsection}{\alph{subsubsection}}
\renewcommand{\theequation}{\thesection\arabic{equation}}
\renewcommand{\thefigure}{\thesection\arabic{figure}}
\renewcommand{\bibnumfmt}[1]{[S#1]}
\renewcommand{\citenumfont}[1]{S#1}

\begin{center}
	\begin{minipage}{0.79\textwidth}
		\small
		In this Supplemental Material, we provide proofs of the results in "Theory of classical metastability in open quantum systems". The open quantum system which serves as the illustration of our general results in Figs.~\ref{fig:Spectrum}--\ref{fig:Symmetry} of the main text is discussed in Sec.~\ref{app:DPT}, together with classical metastability arising in proximity to dissipative phase transitions at finite size. The derivations for the quantitative approach to metastability are given in Sec.~\ref{app:MM}.
		Then proofs are given of the following results: classical metastability in Sec.~\ref{app:cMM}, classical metastable phases in Sec.~\ref{app:phases}, classical long-time dynamics in Sec.~\ref{app:Leff_general}, classical hierarchy of metastabilities in Sec.~\ref{app:hierarchy}, and classical weak symmetries in Sec.~\ref{app:symmetry}.	Finally, the~correctness of the new numerical approach introduced in the main text is argued in Sec.~\ref{app:algorithm}.  	\bigskip	\bigskip
	\end{minipage}
\end{center}
\twocolumngrid



\section{Classical metastability in proximity to dissipative phase transition at finite size}\label{app:DPT}

Here, we first describe the model we illustrate our general theory of metastability in Markovian open quantum systems with in Figs.~\ref{fig:Spectrum}--\ref{fig:Symmetry} of the main text. This example belongs to a class of systems where classical metastability is a consequence of perturbing dynamical parameters away  from effectively classical dissipative phase transitions occurring at a finite size. Second, we find and discuss the general structure of metastable phases and long-time dynamics for this class.

\subsection{Example in Figs.~\ref{fig:Spectrum}--\ref{fig:Symmetry} of the main text}
\label{app:model}

\subsubsection{Model}
\begin{figure}
	\includegraphics[width=0.8\linewidth]{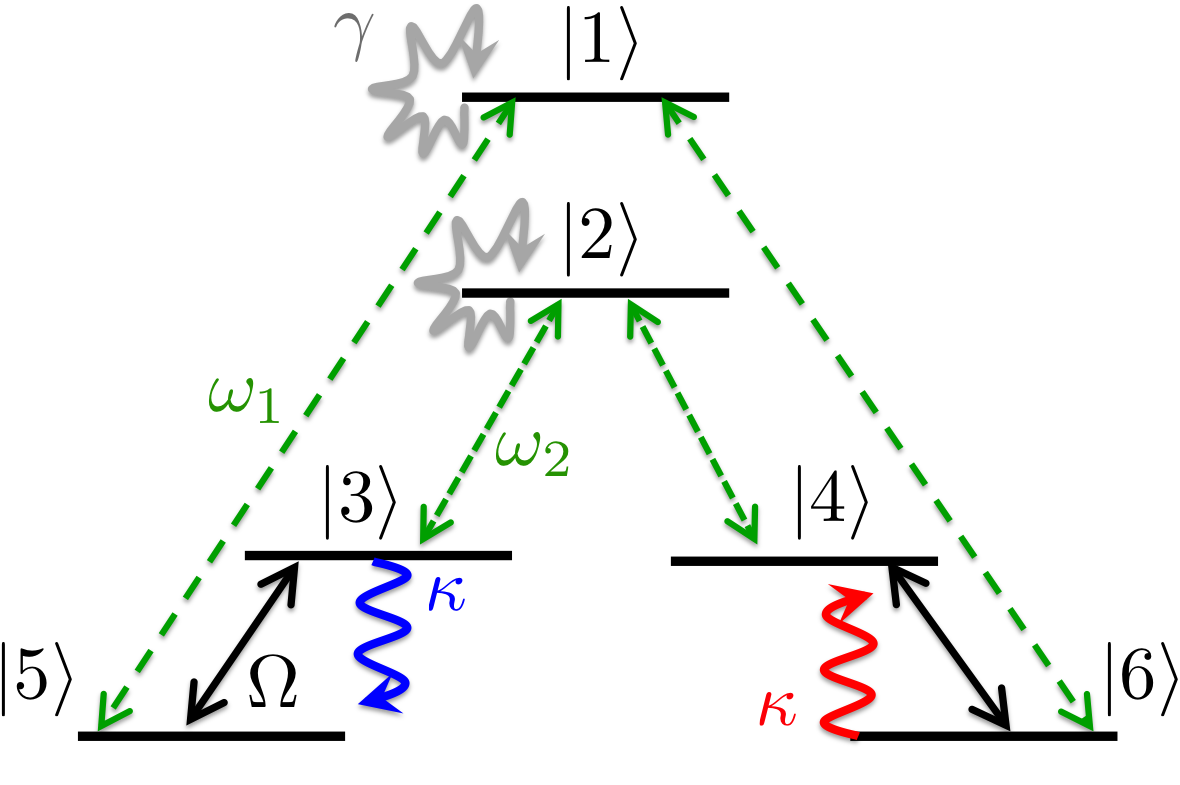}
	\caption{\label{fig:Model}\textbf{Example of open quantum dynamics} defined in Eqs.~\eqref{eq:H_model},~\eqref{eq:J12_model} and~\eqref{eq:J34_model}.
		Four phases with distinct support arise in the perturbative limit of Eq.~\eqref{eq:cond_model} when dynamics associated with $\omega_1$ and $\omega_2$ (green dashed arrows) can be neglected.  When the jump from $|6\rangle$ to $|4\rangle$ (red arrow)  is reverted, the dynamics features a discrete swap symmetry [cf.~$J_4$ in Eq.~\eqref{eq:J34_model}].
		\vspace*{-5mm}}
\end{figure}

In Figs.~\ref{fig:Spectrum}--\ref{fig:Trajectories} of the main text, we choose a system of $6$ levels $|j\rangle$, $j=1,2,...,6$, connected in the Markovian dynamics governed by the master operator  in Eq.~\eqref{eq:L}, with the Hamiltonian
\begin{eqnarray}\label{eq:H_model}
	H=
	\omega_1\! \left( \sigma_{13}^x +\sigma_{14}^x\right)
	+\omega_2\! \left( \sigma_{25}^x +\sigma_{26}^x\right)
	+\Omega \left( \sigma_{35}^x +\sigma_{46}^x\right)\!, \qquad
\end{eqnarray}
where $\sigma_{jk}^x\equiv |j\rangle\!\langle k| +|k\rangle\!\langle j| $,
and the dissipative jumps
\begin{eqnarray}\label{eq:J12_model}
	J_1=\sqrt{\gamma}\,\sigma_{11},&\quad& J_2=\sqrt{\gamma}\,\sigma_{22},\\
	\label{eq:J34_model}
	J_3=\sqrt{\kappa}\,\sigma_{35}^\dagger,&\quad& J_4=\sqrt{\kappa}\,\sigma_{46},
\end{eqnarray}
where $\sigma_{jk}= |j\rangle\!\langle k|$,  as depicted in Fig.~\ref{fig:Model}. We assume $J_3$ and $J_4$ correspond to emission of quanta [cf.~Fig.~\ref{fig:Spectrum}\textcolor{blue}{(e)} in the main text].  

In Fig.~\ref{fig:Symmetry} of the main text, we replace $J_4$ in Eq.~\eqref{eq:J34_model} by $J_4^\dagger$. This introduces a discrete weak symmetry under the simultaneous swap $|3\rangle\leftrightarrow|4\rangle$ and $|5\rangle\leftrightarrow|6\rangle$  (cf.~Fig.~\ref{fig:Model} and see Sec.~\ref{sec:symmetry} in the main text). 

\subsubsection{Dissipative phase transition} \label{app:model_DPT}

In the limit 
\begin{equation}\label{eq:cond_model}
	|\omega_1|,|\omega_2|\ll |\Omega|, \kappa, \gamma,
\end{equation}
the dynamics is in the proximity of a dissipative phase transition with four independent sectors corresponding to: two decoupled pure states $|1\rangle$ and $|2\rangle$, whose coherences between two decoupled states are not stable because of dephasing caused by $J_1$ and $J_2$ in Eq.~\eqref{eq:J34_model}, and  two effective two-level atoms, which are formed by $|3\rangle$ and $|5\rangle$,  and by $|4\rangle$ and $|6\rangle$ (see Fig.~\ref{fig:Model}). Note that there is no decay subspace.
Due to the perturbation by the Hamiltonian $\delta H=\omega_1 ( \sigma_{13}^x +\sigma_{14}^x)
+\omega_2( \sigma_{25}^x +\sigma_{26}^x)$, the long-time dynamics arises in the second order in $\omega_{1}$ and $\omega_{2}$, while metastable phases arise from the stationary states, perturbed already in the first order~(see Sec.~\ref{app:PT1}). 
Therefore, the condition in Eq.~\eqref{eq:cond_model} leads to the classical metastability with $m=4$. This is the limit considered in  Figs.~\ref{fig:Spectrum}--\ref{fig:Symmetry} of the main text (see below for parameter values).

\subsubsection{Plot parameters and numerical results}

In Figs.~\ref{fig:Spectrum}--\ref{fig:Trajectories} of  the main text, we choose $\omega_1=2.5\omega_2$ and $\omega_2=0.2\kappa$, while $\Omega=1.5\kappa$ and $\gamma= 0.75 \kappa$. The low-lying eigenvalues are $\lambda_2=-0.0785\kappa$, $\lambda_3=-0.1564\kappa$ and $\lambda_4=-0.2886\kappa$ with the next eigenvalue $\lambda_5=-0.4297\kappa$. Although the separation is not significant, the choice of such set of parameters is motivated by clearly illustrating approximations in the theory of classical metastability. The corrections to the classicality are bounded by $\Ctcl=0.2369$ in Eq.~\eqref{eq:Ctcl}, which we get for the metastable phases obtained from initial states chosen as extreme eigenstates of $L_2$, $L_3$ and $L_4$ (normalized so that $\langle 1|L_k|1\rangle>0$ and $c_{k}^{\max}-c_{k}^{\min}=1$, $k=2,3,4$) that give the maximal simplex (cf.~Fig.~\ref{fig:MM} in the main text). In particular, in Figs.~\ref{fig:MM}--\ref{fig:Trajectories} of the main text, $\trho_1$ is supported mostly on $|3\rangle$ and $|5\rangle$,  $\trho_2$ on $|1\rangle$, $\trho_3$  on $|4\rangle$ and $|6\rangle$, while $\trho_4$ on $|2\rangle$ (cf.~Fig.~\ref{fig:Model}), and the stationary state decomposes as $\pt_\text{ss}=(0.2763,0.1689,0.4021,0.1528)^T$.
The initial state considered in Figs.~\ref{fig:Spectrum}\textcolor{blue}{(c)},~\ref{fig:Spectrum}\textcolor{blue}{(d)},~\ref{fig:Spectrum}\textcolor{blue}{(e)},~\ref{fig:Trajectories}\textcolor{blue}{(e)},~\ref{fig:Trajectories}\textcolor{blue}{(f)}, and~\ref{fig:Trajectories}\textcolor{blue}{(g)} of the main text, is $\rho(0)=|3\rangle\!\langle{3}|$.
In Figs.~\ref{fig:Spectrum}\textcolor{blue}{(c)} and~\ref{fig:Spectrum}\textcolor{blue}{(d)} of the main text, the chosen observable is the activity of jumps in Eq.~\eqref{eq:J12_model}, $O= J_3^\dagger J_3+J_4^\dagger J_4=\kappa (\sigma_{33}+\sigma_{44})$, which in trajectories in Figs.~\ref{fig:Spectrum}\textcolor{blue}{(e)} and~\ref{fig:Trajectories}\textcolor{blue}{(g)} of the main text are represented by blue ($J_3$) and red ($J_4$). Therefore, $\langle K(t)\rangle$ in Fig.~\ref{fig:Trajectories}\textcolor{blue}{(e)} of the main text corresponds to the time-integral of $\langle O(t)\rangle$. In terms of the asymptotic rates, we have $\mu_\text{ss}=0.3115\kappa$ and $\sigma^2_\text{ss}=1.057\kappa$, while the metastable phase rates are  $\mut_1=0.5129\kappa$,   $\mut_2=0.01832\kappa$, $\mut_3=0.3971\kappa$, $\mut_4=0.0458\kappa$, and $\tilde{\sigma}_1^2=0.3868\kappa$, $\tilde{\sigma}_2^2=0.01876\kappa$, $\tilde{\sigma}_3^2=0.4026\kappa$, $\tilde{\sigma}_4^2=0.03852\kappa$.


In Figs.~\ref{fig:Spectrum}\textcolor{blue}{(a)},~\ref{fig:Spectrum}\textcolor{blue}{(c)},~\ref{fig:Spectrum}\textcolor{blue}{(d)},~\ref{fig:Trajectories}\textcolor{blue}{(e)}, and~\ref{fig:Trajectories}\textcolor{blue}{(f)} of the main text, we choose reduced $\omega_2=0.025\kappa$, to obtain a more pronounced separation of the eigenvalues:  $\lambda_2=-0.0008522\kappa$, $\lambda_3=-0.001114\kappa$, $\lambda_4=-0.004682\kappa$ and $\lambda_5=-0.5000\kappa$. In this case, the activity of the stationary state is characterized by $\mu_\text{ss}=0.3137\kappa$ and $\sigma^2_\text{ss}=69.03\kappa$.
Considering the extreme eigenstates of $(L_2+L_3)/2$, $(L_2-L_3)/2$ (rotation of coefficients by $\pi/4$) and $L_4$ as initial states for candidate metastable phases and the maximal simplex, we obtain $\Ctcl=0.001703$, while $\mut_1=0.4739\kappa$,   $\mut_2=0.00004521\kappa$, $\mut_3=0.4732\kappa$, $\mut_4=0.0002876\kappa$, and $\tilde{\sigma}_1^2=0.4034\kappa$, $\tilde{\sigma}_2^2= 0.00003849\kappa$, $\tilde{\sigma}_3^2=0.4027\kappa$, $\tilde{\sigma}_4^2=0.0002447\kappa$. Here, $\pt_\text{ss}=(0.1833,0.4792,0.1907,0.1467)^T$ and, thus, a significant increase in $\sigma^2_\text{ss}=1.057\kappa$ is due to longer timescales of the dynamics within the metastable manifold (MM) (cf.~Sec.~\ref{sec:metaDPT} in the main text).
In this case $\Wt=\W$ and, to compare the classical approximation to the rate of jump average and fluctuations in Figs.~\ref{fig:Trajectories}\textcolor{blue}{(e)} and~\ref{fig:Trajectories}\textcolor{blue}{(f)} of the main text, we rescale $\Wt$ obtained for $\omega_2=0.025\kappa$ by the square of the frequency $\omega_2$ ratio ($=0.125^2$).


In Fig.~\ref{fig:Symmetry} of the main text, we replace $J_4$ by $J_4^\dagger$ and again choose $\omega_2=0.2\kappa$ leading to $\lambda_2=-0.10099\kappa$, $\lambda_3=-0.14641\kappa$, $\lambda_4=-0.22028\kappa$ and $\lambda_5=-0.458922\kappa$ and $\Ctcl=0.1597$, and the stationary state with $\mu_\text{ss}=0.2499\kappa$ and $\sigma^2_\text{ss}=1.1790\kappa$. Considering again the extreme eigenstates of $L_2$, $L_3$ and $L_4$ and the maximal simplex,  in Fig.~\ref{fig:Symmetry}\textcolor{blue}{(a)} of the main text, we again have that $\trho_1$ is supported mostly on $|3\rangle$ and $|5\rangle$,  $\trho_2$ on $|1\rangle$, $\trho_3$  on $|4\rangle$ and $|6\rangle$, while $\trho_4$ on $|2\rangle$ (cf.~Fig.~\ref{fig:Model}), with $\mut_1=0.4781\kappa$,   $\mut_2= 0.0297\kappa$, $\mut_3=0.4781\kappa$, $\mut_4=-0.01771\kappa$, and $\tilde{\sigma}_1^2=0.4187\kappa$, $\tilde{\sigma}_2^2= 0.0284\kappa$, $\tilde{\sigma}_3^2=0.4802\kappa$, $\tilde{\sigma}_4^2=-0.0181\kappa$. Here, the stationary state decomposes as $\pt_\text{ss}=(0.2605,0.1947,0.2605,0.2843)^T$. 
We observe that the symmetry increases the separation in the master operator spectrum and reduces the corrections to the classicality. We find it also leads to $\Wt=\W$ [cf.~Fig.~\ref{fig:Symmetry}\textcolor{blue}{(b)} in the main text].


Finally, in Figs.~\ref{fig:Spectrum}--\ref{fig:Symmetry}, due to the simplicity of the model, we approximate the timescales as $\tau=-1/\lambda_2$, $\tau'=-1/\lambda_4$, and $\tau''=-1/\lambda_5$; cf.~Sec.~\ref{app:tau_defMM}. Furthermore, in Figs.~\ref{fig:Spectrum}\textcolor{blue}{(b)},~\ref{fig:EffDyn}\textcolor{blue}{(b)}, and~\ref{fig:Symmetry}\textcolor{blue}{(a)}, for $t\ll \tau'$ we choose $t=0$.

\subsection{General case}\label{app:PT}

Here, we consider perturbing dynamics of an open quantum system that features multiple disjoint stationary states.  First, we show that the corrections to the positivity in Eq.~\eqref{eq:Cp} and the corrections to the classicality in Eq.~\eqref{eq:Ccl} are of the second order in the perturbation. Second, we find that the long dynamics arises in the second order, with corrections to its positivity at most of the forth order. 
We also discuss the effectively classical statistics of quantum trajectories and explain how a hierarchy of classical metastabilities arises when degeneracy of zero eigenvalues of the master equation  is partially lifted in the second order.

\subsubsection{Dissipative phase transition and its proximity}\label{app:PT0}

\emph{Unperturbed dynamics}. We consider dynamics with $m$ disjoint stationary states $\rho_1,...,\rho_m$. Let $\mathcal{H}$ denote the system Hilbert space and $\mathcal{H}_l$, $l=1,...,m$, the respective orthogonal supports of the stationary states, so that $\mathcal{H}=\oplus_{l=1}^m\mathcal{H}_l \oplus \mathcal{D}$, where $\mathcal{D}$ is the decay subspace. The dual basis $P_l$, $l=1,...,m$ to the stationary states then fulfills $0\leq P_l$, $\sum_{l=1}^m P_l=\mathds{1}$, and
\begin{equation}\label{eq:PT_Pl}
	P_l=\mathds{1}_{\mathcal{H}_l}+	\mathds{1}_{\mathcal{D}} P_l\mathds{1}_{\mathcal{D}}.
\end{equation}	
It also follows that~\cite{Baumgartner2008,Wolf2012,Albert2016}
\begin{eqnarray}\label{eq:PT_JH}
	J_j^{(0)}&=&\sum_{l=1}^m \Big[\mathds{1}_{\mathcal{H}_l} J_j^{(0)} \mathds{1}_{\mathcal{H}_l}\!+\mathds{1}_{\mathcal{H}_l}J_j^{(0)} \mathds{1}_{\mathcal{D}}\! +\mathds{1}_{\mathcal{D}}J_j^{(0)} \mathds{1}_{\mathcal{D}}\Big]\!,\\\nonumber
	H_\text{eff}&=&\sum_{l=1}^m \Big[\mathds{1}_{\mathcal{H}_l} H_\text{eff}^{(0)} \mathds{1}_{\mathcal{H}_l}\!+\mathds{1}_{\mathcal{H}_l} H_\text{eff}^{(0)} \mathds{1}_{\mathcal{D}}\!+\mathds{1}_{\mathcal{D}}H_\text{eff}^{(0)} \mathds{1}_{\mathcal{D}}\Big]\!,\qquad
\end{eqnarray}
where $H^{(0)}_\text{eff}=H^{(0)}-i \sum_j J_j^{(0)\dagger} J_j^{(0)}/2$ is referred to as the  effective Hamiltonian. We denote the corresponding master operator as $\L^{(0)}$, the projection on the stationary states as $\P^{(0)}$, and the reduced resolvent at $0$ as $\S^{(0)}$ [$\S^{(0)}\L^{(0)}=\I-\P^{(0)}=\L^{(0)}\mathcal{S}^{(0)}$ and $\S^{(0)}\P^{(0)}=0=\P^{(0)}\S^{(0)}$]. 


In the discussion of classical metastability, we will consider states that evolve into $\rho_l$ with probability $0$, $l=1,...,m$. Those states are supported in the eigenspace of $P_l$ corresponding to the eigenvalue $0$ , which consists of $\oplus_{k\ne l} \mathcal{H}_k$  and a subspace in $\mathcal{D}$, which we denote $\mathcal{D}_{l,0}$. It follows that $( \mathds{1}_{\mathcal{H}_l}+ \mathds{1}_{\mathcal{D}_{l,0}^\perp}) J_j^{(0)} \mathds{1}_{\mathcal{D}_{l,0}}=0=( \mathds{1}_{\mathcal{H}_l}+ \mathds{1}_{\mathcal{D}_{l,0}^\perp})  H_\text{eff}^{(0)} \mathds{1}_{\mathcal{D}_{l,0}}$, where $\mathcal{D}_{l,0}^\perp$ denotes the orthogonal complement of $\mathcal{D}_{l,0}$ within $\mathcal{D}$. 
We will also consider states that evolve into $\rho_l$ with probability $1$, $l=1,...,m$. Those states are analogously supported in the eigenspace of $P_l$ corresponding to the eigenvalue  $1$, which consists of $\mathcal{H}_l$  and a subspace in $\mathcal{D}$, which we denote $\mathcal{D}_{l,1}$. We have $( \sum_{k\neq l}\mathds{1}_{\mathcal{H}_k}+ \mathds{1}_{\mathcal{D}_{l,1}^\perp}) J_j^{(0)} \mathds{1}_{\mathcal{D}_{l,1}}=0=(  \sum_{k\neq l}\mathds{1}_{\mathcal{H}_k}+ \mathds{1}_{\mathcal{D}_{l,1}^\perp})  H_\text{eff}^{(0)} \mathds{1}_{\mathcal{D}_{l,1}}$, and $\mathcal{D}_{l,1}$ for $l=1,...,m$ are orthogonal subspaces.
\\

\emph{Perturbation}. We consider perturbing the Hamiltonian and jumps as $H=H^{(0)}+H^{(1)}$ and $J_j=J_j^{(0)}+J_j^{(1)}$, where
\begin{eqnarray}\label{eq:PT_JH1}
	\mathds{1}_{\mathcal{H}_l} J_j^{(1)} \mathds{1}_{\mathcal{H}_l}=0\\
	\mathds{1}_{\mathcal{H}_l} H^{(1)} \mathds{1}_{\mathcal{H}_l}=0
\end{eqnarray}
for $l=1,...,m$, as other perturbations can be included in $H^{(0)}$ and $J_j^{(0)}$ without changing their structure with respect to the subspaces $\mathcal{H}_l$, $l=1,...,m$, and $\mathcal{D}$, as well as $\mathcal{D}_l$, $l=1,...,m$ [here, we could also assume $J_j^{(1)} \mathds{1}_{\cap_{l=1}^m \mathcal{D}_l^\perp} =0= H^{(1)} \mathds{1}_{\cap_{l=1}^m \mathcal{D}_l^\perp}$]. 	

This leads to the first- and second-order perturbation of the master operator $\L=\L^{(0)}+\L^{(1)}+\L^{(2)}$, where [cf.~Eq.~\eqref{eq:L}]
\begin{eqnarray}\label{eq:PT_L1}
	\L^{(1)}&\equiv&-i[ H^{(1)},\rho]+\sum_{j}\bigg[{J}_{j}^{(1)}\rho{J}_{j}^{(0)\dagger}+{J}_{j}^{(0)}\rho{J}_{j}^{(1)\dagger}\qquad\quad\\\nonumber
	\label{eq:PT_L2}
	&&\qquad\qquad\qquad\quad\,\,-\frac{1}{2}\left\{{J}_{j}^{(1)\dagger}{J}_{j}^{(0)}+{J}_{j}^{(0)\dagger}{J}_{j}^{(1)},\rho\right\}\bigg],\\
	\L^{(2)}&\equiv&\sum_{j}\bigg[{J}_{j}^{(1)}\rho\,{J}_{j}^{(1)\dagger}-\frac{1}{2}\left\{{J}_{j}^{(1)\dagger}{J}_{j}^{(1)},\rho\right\}\bigg].
\end{eqnarray}

\subsubsection{Perturbation theory}\label{app:PT1}
To discuss the classical metastability arising due to perturbations in Eqs.~\eqref{eq:PT_L1} and~\eqref{eq:PT_L2}, we consider the non-Hermitian perturbation theory in finitely-dimensional spaces (see Chapter~2 in~\cite{Kato1995}). In particular, we exploit the perturbation series for the projection on the low-lying eigenmodes arising from the stationary states. \\

\emph{Metastable phases}. We choose the candidate states as the stationary states of the unperturbed dynamics leading to (cf.~Eq.~\eqref{eq:rhotilde} and see Eq.~(2.14) in Ref.~\cite{Kato1995})
\begin{eqnarray}\label{eq:PT_rhotilde}
	\trho_l
	&=& \rho_l - \S^{(0)} \L^{(1)}\!(\rho_l)- \S^{(0)} \L^{(2)}\!(\rho_l)
	\\\nonumber
	&&+\S^{(0)}\!\L^{(1)}\S^{(0)}\! \L^{(1)}(\rho_l)-\P^{(0)}\!\L^{(1)}[\S^{(0)}]^2\! \L^{(1)}(\rho_l)+...,
\end{eqnarray}  
where $l=1,...,m$ and ... denotes corrections of the third and higher order. The corresponding (normalised) dual basis is given by  [cf.~Eq.~\eqref{eq:Ptilde}]
\begin{eqnarray}\label{eq:PT_Ptilde}
	&&\tilde{P}_l=P_l - \S^{(0)\dagger} \L^{(1)\dagger}\!(P_l)- \S^{(0)\dagger} \L^{(2)\dagger}\!(P_l)
	\\\nonumber
	&&\qquad+\S^{(0)\dagger}\!\L^{(1)\dagger}\S^{(0)\dagger}\! \L^{(1)\dagger}(P_l)+....
\end{eqnarray} 

From Eq.~\eqref{eq:PT_rhotilde}, we have $\Tr(\trho_l)=1+...$ [actually, this holds to all orders as $\mathds{1}$ is a conserved quantity for $\L$, and thus $\P^\dagger(\mathds{1})=\mathds{1}$]. Furthermore, as $\L^{(1)}$ in Eq.~\eqref{eq:PT_L1} can only create coherences between $\mathcal{H}_l$ and $\oplus_{k\neq l}\mathcal{H}_k\oplus \mathcal{D}$ [cf.~Eqs.~\eqref{eq:PT_JH} and~\eqref{eq:PT_JH1}], while
$\S^{(0)}$ maps such coherences onto themselves and operators on $\mathcal{H}_l$ (the latter only from coherences between $\mathcal{H}_l$ and $\oplus_{k\neq l}\mathcal{H}_k\oplus \mathcal{D}_{l,0}^\perp$), where eigenvalues of $\rho_l$ are strictly larger than $0$, it follows that $\lVert \trho_l\rVert_1-1$ is at most of the second order. Therefore, the corrections to the positivity in Eq.~\eqref{eq:Cp} are at most of the second order [cf.~Eq.~\eqref{eq:PT_Ccl} below].

Furthermore, it also follows that $\Tr [(\mathds{1}_{\mathcal{H}_k}+\mathds{1}_{\mathcal{D}_{k,1}}) \trho_l]-\delta_{kl}$ is at most of the second order, so that the corrections to disjointness of the supports and the basins of attraction of the metastable phases are of the second order. Similarly, since subspaces corresponding to the sum of eigenspaces of $\tilde{P}_k$ with the eigenvalues  $>1/2$, considered in Sec.~\ref{sec:disjoint} of the main text are included in the perturbed sum of eigenspaces of $\tilde{P}_k$ with the eigenvalues $\geq 1/2$, their overlap with $\rho_l$ changes from $\delta_{kl}$ at most in the second order. Finally, if the corrections are of the second order, the bounds in Eqs.~\eqref{eq:B_support01}--\eqref{eq:B_support03} are saturated in terms of the figure of merit,  since we show next that the corrections to the classicality, like the corrections to the positivity, are at most of the second order. 
\\

\emph{Corrections to classicality}. We now argue that the corrections to the classicality arise at most in the second order of the perturbation. We have [cf.~Eq.~\eqref{eq:PT_Ptilde}]
\begin{eqnarray}\label{eq:PT_Ccl}
	\tilde{p}_l&\equiv&\Tr[\tilde{P}_l\rho(0)] 
	\\\nonumber&=&p_l-\Tr \{P_l\L^{(1)}\S^{(0)}\![\rho(0)]\}-\Tr \{P_l\L^{(2)}\S^{(0)}\![\rho(0)]\}\qquad\\\nonumber&&+\Tr\{P_l\L^{(1)}\S^{(0)}\! \L^{(1)}\S^{(0)}\![\rho(0)]\}+...,
\end{eqnarray}
where $p_l\equiv\Tr[P_l\rho(0)]$, so that  the barycentric coordinates for the metastable phases in Eq.~\eqref{eq:PT_rhotilde} in general feature first-order corrections.
The corrections to classicality in Eq.~\eqref{eq:Ccl}, however, are determined by the corrections in Eq.~\eqref{eq:PT_Ccl} for states projected by $\P^{(0)}$ onto the surface of the simplex of probability distributions, that is, featuring $p_l=0$ for some $l$. This requires $\rho(0)$ to be supported inside the $0$-eigenspace of $P_l$, i.e.,  $\oplus_{k\neq l}\mathcal{H}_k\oplus \mathcal{D}_{l,0}$. From the structure of the dynamics discussed in Sec.~\ref{app:PT0}, this leads to $\S^{(0)}[\rho(0)]$ supported within the same subspace. Furthermore, as $\L^{(1)}$ in Eq.~\eqref{eq:PT_L1} can only create coherences from $\oplus_{k\neq l}\mathcal{H}_k\oplus \mathcal{D}_{l,0}$ to $\mathcal{H}_l\oplus \mathcal{D}_{l,0}^\perp$ [cf.~Eqs.~\eqref{eq:PT_JH} and~\eqref{eq:PT_JH1}], 	the first-order corrections are not present in this case $\Tr\{P_l\L^{(1)}\S^{(0)}\![\rho(0)]\}=0$ [cf.~Eq.~\eqref{eq:PT_Ptilde}]. We conclude that $\Ccl>0$ arises, at earliest, in the second order (for example, see~Sec.~\ref{app:model_DPT}). This also holds for $\Ctcl$ in Eq.~\eqref{eq:Ctcl} [cf.~Eq.~\eqref{eq:Ctcl2}].
Finally, we note that the perturbations restricted to $J_j^{(1)}=\sum_{l=1}^m \mathds{1}_{\mathcal{H}_l}J_j^{(1)} \mathds{1}_{\mathcal{D}}$ and $H^{(1)}=\sum_{l=1}^m \mathds{1}_{\mathcal{H}_l}H^{(1)} \mathds{1}_{\mathcal{D}}$, do not lift the degeneracy of $m$ stationary states, but lead to  reduced subspaces $\mathcal{D}_{l,0}$ and thus $\tilde{p}_l\geq0$, so that $\Ccl=0=\Ctcl$. 
\\


\emph{Long-time dynamics}.
For the generator of the long-time dynamics in Eq.~\eqref{eq:Wtilde}, we have [cf.~Eqs.~\eqref{eq:PT_rhotilde} and~\eqref{eq:PT_Ptilde}]
\begin{eqnarray}\label{eq:PT_Wtilde}
	&&(\Wt)_{kl}
	=\Tr [P_k\L^{(2)}(\rho_l)]-\Tr [P_k\L^{(1)}\S^{(0)}\!\L^{(1)}(\rho_l)]\qquad\\\nonumber&&-\Tr [P_k\L^{(1)}\S^{(0)}\!\L^{(2)}(\rho_l)]-\Tr [P_k\L^{(2)}\S^{(0)}\!\L^{(1)}(\rho_l)]\\\nonumber&&+\Tr [P_k\L^{(1)}\S^{(0)}\!\L^{(1)}\S^{(0)}\!\L^{(1)}(\rho_l)]+...,
\end{eqnarray}
where ... denotes corrections of the fourth and higher order. The second-order dynamics in the first line, $\Wt^{(2)}$, is trace-preserving and and can be shown to be positive (see below and cf.~Supplemental Material of Ref.~\cite{Macieszczak2016a}), and thus we denote them as $\W^{(2)}$.	

While the third-order corrections are generally present, breaking the positivity of the dynamics requires $[\Wt^{(3)}]_{kl}\neq 0$ when $[\W^{(2)}]_{kl}=0$. This is not possible, and, thus, the corrections to the positivity of $\Wt$ are at most of forth order. Indeed, let us rescale the perturbations $J_j^{(1)}$ and $H^{(1)}$ by a parameter $\delta$. It can be shown that $\lim_{t\rightarrow \infty}[\partial_\delta |_{\delta=0} \Tr( P_k e^{t\L}\rho_l)]/t=[\Wt^{(1)}]_{kl}$, $\lim_{t\rightarrow \infty}[\partial_\delta^2 |_{\delta=0} \Tr( P_k e^{t\L}\rho_l)] /t=2[\Wt^{(2)}]_{kl}$ and  $\lim_{t\rightarrow \infty}[\partial_\delta^3 |_{\delta=0} \Tr( P_k e^{t\L}\rho_l) ]/t=6[\Wt^{(3)}]_{kl}$. Since $\Tr( P_k e^{t\L}\rho_l)\geq 0$ from $P_k\geq 0$ and $\rho_l\geq 0$, while $\Tr( P_k e^{t\L}\rho_l)|_{\delta=0}=\delta_{kl}$, it follows that $[\Wt^{(1)}]_{kl}=0$ (there are no first-order corrections to the dynamics),  $[\Wt^{(2)}]_{ll}\leq 0$ and $[\Wt^{(2)}]_{kl}\geq 0$ for $k\neq l$ (the second-order dynamics is positive). Analogously, when  $[\Wt^{(2)}]_{kl}=0$, it follows that $[\Wt^{(3)}]_{kl}=0$.
\\

\emph{Statistics of quantum trajectories}. For the tilted generator  in Eq.~\eqref{eq:Wstilde}, which encodes the statistics of jumps in quantum trajectories, we have [cf.~Eqs.~\eqref{eq:PT_rhotilde} and~\eqref{eq:PT_Ptilde} and see Eq.~\eqref{eq:Wstilde2}]
\begin{eqnarray}\label{eq:PT_Wstilde}
	&&(\Wt_s)_{kl}=[\W^{(2)}]_{kl}\!+\!(e^{-s}\!-\!1)[\Jbf^{(2)}\!\!+\! \delta_{kl} \mut_{l}^\text{in} ]\qquad\\\nonumber
	&&-(e^{-s}\!-\!1) \times\\\nonumber
	&&((1\!-\!\delta_{kl})\Tr\{P_k \mathcal{H}_\text{eff}^{(0)}[\trho_l^{(2)}]\}\!+\!\delta_{kl}\Tr\{(P_l\!-\!\mathds{1}) \mathcal{H}_\text{eff}^{(0)}[\trho_l^{(2)}]\}\\\nonumber
	&&+(1\!-\!\delta_{kl})\Tr\{P_k \mathcal{H}_\text{eff}^{(1)}[\trho_l^{(1)}]\}\!+\!\delta_{kl}\Tr\{(P_l\!-\!\mathds{1}) \mathcal{H}_\text{eff}^{(1)}[\trho_l^{(1)}]\}\\\nonumber
	&& +\Tr[\tilde{P}_k^{(2)}\!\mathcal{H}_\text{eff}^{(0)}(\rho_l)]\!+\!\Tr\{\tilde{P}_k^{(1)}  \!\mathcal{H}_\text{eff}^{(0)}[\trho_l^{(1)}]\}\!+\!\Tr[\tilde{P}_k^{(1)}  \!\mathcal{H}_\text{eff}^{(1)}(\trho_l)])\\\nonumber
	&&+...,
\end{eqnarray}
where $\mathcal{H}_\text{eff}^{(0)}(\rho)\equiv- i H_\text{eff}^{(0)}\rho +\rho H_\text{eff}^{(0)\dagger}$, $\mathcal{H}_\text{eff}^{(1)}(\rho)\equiv- i [H^{(1)},\rho] -[{J}_{j}^{(1)\dagger}{J}_{j}^{(0)}+{J}_{j}^{(0)\dagger}{J}_{j}^{(1)}] \rho +\rho[{J}_{j}^{(1)\dagger}{J}_{j}^{(0)}+{J}_{j}^{(0)\dagger}{J}_{j}^{(1)}] \}/2$, 
$\mut_{l}^\text{in}=\mut_{l}+[\W^{(2)}]_{ll}$ and $\mut_{l}=\mu_{l}+\mut_{l}^{(1)}+\mut_{l}^{(2)}+...\equiv \Tr[ {J}_{j}^{(0)\dagger}{J}_{j}^{(0)} \rho_l]+\Tr[ {J}_{j}^{(0)\dagger}{J}_{j}^{(0)} \trho_l^{(1)}]+\{\Tr[ {J}_{j}^{(0)\dagger}{J}_{j}^{(0)} \trho_l^{(2)}]+\Tr[ {J}_{j}^{(0)\dagger}{J}_{j}^{(1)}\trho_l^{(1)}]+\Tr[ {J}_{j}^{(1)\dagger}{J}_{j}^{(0)} \trho_l^{(1)}]+\Tr[ {J}_{j}^{(1)\dagger}{J}_{j}^{(1)} \rho_l]\}+...$ with  $\mu_{l}$ encoding the stationary activity in the unperturbed dynamics inside $\mathcal{H}_l$, $(\Jbf^{(2)})_{kl}\equiv (1-\delta_{kl})[\W^{(2)}]_{kl}$, and ... denotes corrections of the third and higher order.

Similarly, for the statistics of the homodyne measurement we have (see Sec.~\ref{app:homodyne})
\begin{eqnarray}\label{eq:PT_Wrtilde}
	&&	(\Wt_r)_{kl}=[\W^{(2)}]_{kl}-r\,\delta_{kl} \tilde{x}_{l} + \frac{r^2}{8}\\\nonumber
	&&-r ((1\!-\!\delta_{kl})\Tr\{P_k \X^{(0)}[\trho_l^{(2)}]\}\!+\!\delta_{kl}\Tr\{(P_l\!-\!\mathds{1}) \X^{(0)}[\trho_l^{(2)}]\}\\\nonumber
	&&+(1\!-\!\delta_{kl})\Tr\{P_k \X^{(1)}[\trho_l^{(1)}]\}\!+\!\delta_{kl}\Tr\{(P_l\!-\!\mathds{1}) \X^{(1)}[\trho_l^{(1)}]\}\\\nonumber
	&& +\Tr[\tilde{P}_k^{(2)}\! \X^{(0)}(\rho_l)]\!+\!\Tr\{\tilde{P}_k^{(1)}  \!\X^{(0)}[\trho_l^{(1)}]\}\!+\!\Tr[\tilde{P}_k^{(1)}  \!\X^{(1)}(\trho_l)])\\\nonumber
	&&+...,
\end{eqnarray}
where  $\X^{(0)}(\rho)\equiv\sum_j[e^{-i \varphi}J_j^{(0)} \rho+e^{i \varphi}\rho J_j^{(0)\dagger} ]$, $\X^{(1)}(\rho)\equiv \sum_j[e^{-i \varphi}J_j^{(1)} \rho+e^{i \varphi}\rho J_j^{(1)\dagger} ]$, and $\tilde{x}_l=x_l+\tilde{x}_l^{(1)}+\tilde{x}_l^{(2)}+...\equiv\sum_j
\Tr\{[e^{-i \varphi}J_j^{(0)} +e^{i \varphi}J_j^{(0)\dagger} ]\rho_l\}/2+\Tr\{[e^{-i \varphi}J_j^{(0)} +e^{i \varphi}J_j^{(0)\dagger} ]\trho_l^{(1)}\}/2+(\Tr\{[e^{-i \varphi}J_j^{(0)} +e^{i \varphi}J_j^{(0)\dagger} ]\trho_l^{(2)}\}/2+\Tr\{[e^{-i \varphi}J_j^{(1)} +e^{i \varphi}J_j^{(1)\dagger} ]\trho_l^{(1)}\})+...$ with $x_l$ being the asymptotic rate of integrated homodyne current in the unperturbed dynamics inside $\mathcal{H}_l$. 

Finally, for the statistics of  time-integrals of system observables (see Sec.~\ref{app:time_int})
\begin{eqnarray}\label{eq:PT_Whtilde}
	&&(\Wt_h)_{kl}=[\W^{(2)}]_{kl}-h\,\delta_{kl}\tilde{m}_{l}\\\nonumber
	&& -h ( (1\!-\!\delta_{kl})\Tr\{P_k \M[\trho_l^{(1)}]\}\!+\!\delta_{kl} \Tr\{(P_l\!-\!\mathds{1}) \M[\trho_l^{(1)}]\}\\\nonumber
	&&\qquad\!+\!\Tr[\tilde{P}_k^{(1)}\! \M(\rho_l)] ) \\\nonumber
	&&-h ((1\!-\!\delta_{kl})\Tr\{P_k \M[\trho_l^{(2)}]\}\!+\!\delta_{kl}\Tr\{(P_l\!-\!\mathds{1}) \M[\trho_l^{(2)}]\}\\\nonumber
	&&\qquad \!+\!\Tr[\tilde{P}_k^{(2)}\! \M(\rho_l)]\!+\!\Tr\{\tilde{P}_k^{(1)}  \!\M[\trho_l^{(1)}]\})
	+...,
\end{eqnarray}
where $\M(\rho)\equiv M\rho + \rho M$, $\tilde{m}_l=m_l+\tilde{m}_l^{(1)}+\tilde{m}_l^{(2)}+...\equiv\Tr(M\rho_l)+\Tr[M\trho_l^{(1)}]+\Tr[M\trho_l^{(2)}]+...$ and $m_l$ is the asymptotic observable average in the unperturbed dynamics inside $\mathcal{H}_l$.

Note that there are no first-order corrections to the difference to corresponding generator for the classical dynamics for continuous measurements [Eqs.~\eqref{eq:PT_Wstilde} and~\eqref{eq:PT_Wrtilde}; cf. Eqs.~\eqref{eq:Ws} and~\eqref{eq:Wrtilde2}]. Furthermore,	if there is no decay subspace, the first-order corrections are not present in the rates $\mu_l$ and $x_l$.  Finally, if the first-order corrections in second and third line of Eq.~\eqref{eq:PT_Whtilde} are present, the corrections to the classicality are necessarily of the second order and the corresponding \emph{bound derived in Sec.~\ref{app:Ws} is saturated} in terms of the order of the figure of merit [this requires $\mathds{1}_{\mathcal{H}_l} M\mathds{1}_{\mathcal{H}_l}\neq 0$ for some $l$; cf. Eqs.~\eqref{eq:PT_JH1},~\eqref{eq:PT_Wstilde}, and~\eqref{eq:PT_Wrtilde}]. 
\\

\emph{Hierarchy of metastabilities}. When the classical second-order dynamics in Eq.~\eqref{eq:PT_Wtilde} features $m_2$ stationary probability distributions $\p_{l_2}$,  $l_2=1,...,m$, we have 
\begin{eqnarray}\label{eq:PT_W2}
	\sum_{\substack{k\in S_{k_2}\\l\in S_{l_2}}}	[\W^{(2)}]_{kl}&=&0,\quad k_2\neq l_2\\\nonumber
	\sum_{\substack{k\in D\\l\in S_{l_2}}}	[\W^{(2)}]_{kl}&=&0,
\end{eqnarray}
where sets $S_{l_2}$ consist of the labels $l=1,...,m$ for which $(\p_{l_2})_{l}>0$ and $D$ is the complement of $\cup_{l_2=1}^{m_2} S_{l_2}$ [cf.~Eq.~\eqref{eq:PT_JH}]. We denote the corresponding projection as $\Pbf_2^{(0)}$ and the reduced resolvent for $\W^{(2)}$ at $0$ as $\R_2^{(0)}$. Note that the third-order corrections $\Wt^{(3)}$ do not change the structure above or the positivity of the dynamics [cf.~Eq.~\eqref{eq:PT_JH1}]. 

The metastable phases in the second metastable manifold arise from $\rho_{2,l_2}\equiv\sum_{l=1^m}(\p_{l_2})_l \rho_l$, $l_2=1,...,m_2$. The first-order  corrections are as given in Eq.~\eqref{eq:PT_rhotilde} and by $\sum_{l=1^m} [\pt_{l_2}^{(1)}]_l \rho_l$, where $\pt_{l_2}^{(1)}=-\R_2^{(0)}  \Wt^{(3)}\p_{l_2}$. The latter contribution does not affect the positivity of metastable phases since it only redistributes the probability within the supports $S_{l_2}$ of $\p_{l_2}$, $l_2=1,...,m_2$. Therefore, the corrections to positivity are at most of the second order. Similarly, the projections on $\rho_{2,l_2}$ are $P_{2,l_2}\equiv\sum_{l=1^m}(\mathbf{v}_{l_2})_l P_l$, $l_2=1,...,m_2$, where are the dual vectors to  $m_2$ stationary probability distributions, $\mathbf{v}_{k_2}^\text{T}\p_{l_2}=\delta_{k_2 l_2}$, so that  $\mathbf{v}_{l_2}\geq 0$, $\sum_{l_2=1}^{m_2}{\mathbf{v}_{l_2}}=1$, and $(\mathbf{v}_{l_2})_{l}=1$ for $l\in S_{l_2}$ [cf.~Eq.~\eqref{eq:PT_Pl}]. Their first-order perturbation is as given in Eq.~\eqref{eq:PT_Ptilde} and by $\sum_{l=1^m} [\mathbf{\tilde{v}}_{l_2}^{(1)}]_l P_l$, where $\mathbf{\tilde{v}}_{l_2}^{(1)}=-\R_2^{(0)\text{T}}  \Wt^{(3)\text{T}}\mathbf{v}_{l_2}$ does not affect the the positivity of $\mathbf{\tilde{v}}_{l_2}$. We conclude that the corrections to the classicality are at most of the second order.

The dynamics after the second metastable regime  is governed by a generator $\Wt_{2}$ at most of the fourth order [cf. Eq.~\eqref{eq:PT_Wtilde}],
\begin{equation}
	\Wt_{2}=\Pbf_2^{(0)} \Wt^{(4)} \Pbf_2^{(0)}  +...,
\end{equation}
where $\Wt^{(4)}$ is the fourth-order correction to $\Wt$ and ... denotes the corrections of the fifth and higher order. This is due to the fact that $\Pbf_2^{(0)} \Wt^{(3)} \R_2^{(0)}  \Wt^{(3)}\Pbf_2^{(0)} =0$, as the third-order corrections do not facilitate the dynamics between supports of $\rho_{2,l_2}$, $l_2=1,...,m$.

\section{Metastability in open quantum systems}\label{app:MM}

In this section, we first discuss the properties of the projection on the low-lying modes in Eq.~\eqref{Expansion}. We then discuss a quantitative relation of the metastable regime to the spectrum of the master operator in the context of corrections in Eq.~\eqref{eq:Cmm}. Finally, we consider timescales of the initial relaxation and the long-time dynamics defined in analogy to Eq.~\eqref{eq:Cmm}.

\subsection{Projection on low-lying modes} \label{app:MM_P}

\subsubsection{Derivation of  Eq.~\eqref{eq:Cp}  in the main text}\label{app:Cp_proof}
As the projection $\P$ on the MM in~Eq.~\eqref{Expansion} preserves the trace, the closest state (density matrix) diagonal in the eigenbasis of $\P[\rho(0)]$ is that with the spectrum given by a closest probability distribution to the spectrum of $\P[\rho(0)]$ [cf.~the discussion in Sec.~\ref{app:p_distance}, in particular Eqs.(\ref{eq:p_distance_bound})-(\ref {eq:p_distance})], leading to the distance in the trace norm equal $\lVert \P[\rho(0)]\rVert-1$.	As for any density matrix $\rho$ from the triangle inequality we have
\begin{eqnarray}
	\lVert \P[\rho(0)]\rVert-1 &\leq& \lVert \P[\rho(0)]-\rho\rVert+\lVert \rho\rVert-1\\\nonumber&=&\lVert \P[\rho(0)]-\rho\rVert,
\end{eqnarray}
the choice of $\rho$ as diagonal in the eigenbasis of $ \P[\rho(0)]$ with the eigenvalues given as in Eq.~\eqref{eq:p_distance_opt} is optimal.

\subsubsection{Bound on metastability of states closest to Eq.~\eqref{Expansion} in the main text}
We now show that for state $\rho$ chosen as the closest state to $\P[\rho(0)]$ we have
\begin{equation}\label{eq:metastability_projections} 
	\lVert \rho-\P(\rho)\rVert \leq (2+\Cp) \Cp \lesssim 2\Cp.
\end{equation}
Indeed, we have [cf.~Eq.~\eqref{eq:Cp}]	
\begin{equation}
	\lVert \rho-\P[\rho(0)]\rVert \leq \Cp
\end{equation} 
and \begin{eqnarray}
	\lVert \P(\rho)-\P[\rho(0)]\rVert &\leq&  \lVert \P\rVert  \,\lVert\rho-\P[\rho(0)]\rVert 
	\\\nonumber&\leq& (1+\Cp) \Cp,
\end{eqnarray} 
where we used $\P^2=\P$. These lead to Eq.~\eqref{eq:metastability_projections} thanks to the triangle inequality.

\subsection{Metastable regime}	\label{app:defMM} 

We now relate the metastable regime directly to the spectrum of the master operator (cf.~Sec.~\ref{sec:spectral}).
For the metastable regime, we have [cf.~Eq.~\eqref{eq:Cmm} and see below]
\begin{subequations}\label{eq:meta_lambda}
	\begin{align}
		\label{eq:t''_lambda}
		e^{t''\lambda_{m+1}^R}&\leq\lVert e^{t''\L}-\P\rVert\leq \Cmm,\\
		\label{eq:t'_lambda}
		1-e^{t'\lambda_m^R}&\leq \lVert e^{t'\L}-\P\rVert\leq\Cmm,\\
		\label{eq:t''_lambda_I}
		|\sin(t'\lambda_k^I)| &\leq \frac{\lVert e^{t' \L}-\P\rVert}{1- \lVert e^{t' \L}-\P\rVert}\lesssim \Cmm.
	\end{align}
\end{subequations}
Therefore, 
\begin{subequations}\label{eq:meta_lambda2}
	\begin{align}
		\label{eq:t''_lambda2}
		t''&\geq \frac{\ln(\Cmm)}{\lambda_{m+1}^R},\\
		\label{eq:t'_lambda2}
		t'&\lesssim \frac{-\Cmm}{\lambda_m^R},\\
		\label{eq:t'_lambda_I2}
		t'&\lesssim \frac{\Cmm} {|\lambda_k^I|}, \quad k\leq m.
	\end{align}
\end{subequations}
where the last inequality holds for  $t'-t''\geq t''$.
Indeed, note that  $|\sin(t\lambda_k^I)| \lesssim \Cmm$ for $k\leq m$ is valid for all times within the metastable regime, so that there exists an integer $l$ such that $|t\lambda_k- l\pi|\lesssim\Cmm$. When $t'-t''\geq t''$  this implies $l=0$ [this also holds when $t''/n \leq t'-t''\leq t''/(n-1)$ with $n\geq 2$ provided that $n\Cmm\ll 1$].\\

\emph{Derivation}. The results in Eq.~\eqref{eq:meta_lambda} follow directly from the following bounds,
\begin{subequations}\label{eq:t_lambda_all}
	\begin{align}
		\label{eq:t_lambda0}
		e^{t\lambda_k^R}  &\leq \lVert e^{t \L}-\P_\text{ss}\rVert,\\
		\label{eq:t_lambda1}
		e^{t\lambda_k^R}  &\leq \lVert e^{t \L}-\P\rVert,\quad k> m,\\
		\label{eq:t_lambda2}
		1-e^{t\lambda_k^R}  &\leq \lVert e^{t \L}-\P\rVert,\quad k\leq m,\\
		\label{eq:t_lambda3}
		|\sin(t\lambda_k^I)|&\leq \frac{\lVert e^{t \L}-\P\rVert}{1- \lVert e^{t \L}-\P\rVert}, \quad k\leq m.
	\end{align}		
\end{subequations}
Moreover, in Eq.~\eqref{eq:t_lambda1} can be also replaced by $ \lVert (\I-\P)e^{t \L}\rVert$.

To derive Eq.~\eqref{eq:t_lambda_all} we repeatedly use the fact that $|\Tr\{O \mathcal{X} [\rho(0)]\}|\leq \lVert O\rVert_{\max}\lVert  \mathcal{X}\rVert$ for an observable $O$ and a superoperator $\mathcal{X}$. 

First, for a real eigenvalue $\lambda_k=\lambda_k^*$, $L_k$ can be chosen Hermitian ($L_k=L_k^\dagger$) and we consider $O=L_k$ and choose $\rho(0)$ as the pure state corresponding  to the eigenvalue of $L_k$ with the maximal eigenvalue.  By considering then $\mathcal{X}=e^{t \L}-\P_\text{ss}$ and $\mathcal{X}=e^{t \L}-\P$, we obtain  Eq.~\eqref{eq:t_lambda_all}. 

Second, for a complex eigenvalue $\lambda_k$ with a corresponding non-Hermitian eigenmode $L_k$, we consider $\rho(0)$ corresponding to the maximum $|c_k|=|\Tr[L_k\rho(0)]|$. For an observable $L_k(\phi)\equiv(e^{i \phi} L_k+e^{-i \phi} L_k^\dagger)/2$, we have $\lVert L_k^R\rVert_{\max}\leq |c_k|$.
Indeed, otherwise there would exist an initial state $\rho'(0)$ with $c'_k=\Tr[L_k\rho'(0)]$ such that $|\Tr[L_k^R\rho'(0)]|=|\text{Re}(c'_k e^{i\phi})|>|c_k|$, but this requires $|c'_k|>|c_k|$. Let  $\phi_k$ be such that $e^{i \phi_k}c_k= |c_k|$. We then have $\Tr[L_k(\phi_k-t\lambda_k^I )\rho(t)]=|c_k| e^{t\lambda_k^R}$, which leads to Eqs.~\eqref{eq:t_lambda0} and~\eqref{eq:t_lambda1} [as well as	$|\cos(t\lambda_k^I)-e^{t\lambda_k^R} | \leq \lVert e^{t \L}-\P\rVert$ for $k\leq m$]. Furthermore, $\Tr[L_k(\phi_k)\rho(t)]=|c_k| \cos(t\lambda_k^I) e^{t\lambda_k^R}$, which leads to Eq.~\eqref{eq:t_lambda2}  as $1-e^{t\lambda_k^R}\leq|1-\cos(t\lambda_k^I)e^{t\lambda_k^R} |\leq \lVert e^{t \L}-\P\rVert$ for $k\leq m$ [as well as $|\cos(t\lambda_k^I)e^{t\lambda_k^R}  | \leq \lVert e^{t \L}-\P_\text{ss}\rVert$ and	$|\cos(t\lambda_k^I)e^{t\lambda_k^R}  | \leq \lVert e^{t \L}-\P\rVert$ for $k>m$]. Finally, from  $\Tr[L_k(\phi_k-\pi/2)\rho(t)]=|c_k| \sin(t\lambda_k^I)e^{t\lambda_k^R}$ we arrive at	$|\sin(t\lambda_k^I)|	e^{t\lambda_k^R} \leq \min[\lVert e^{t \L}-\P_\text{ss}\rVert,\lVert e^{t \L}-\P\rVert]$. For $k\leq m$ this together with Eq.~\eqref{eq:t_lambda2}  leads to Eq.~\eqref{eq:t_lambda3}.

Third, for an eigenvalue $\lambda_k$ corresponding to a Jordan block, we consider $L_k$ corresponding to its last mode, so that $\Tr \{L_k \L[\rho(0)]\}=c_k \lambda_k$ (i.e., in this case $L_k$ is an eigenmode of $\L$). Then, the above discussion applies directly. 

\subsection{Relaxation times}  \label{app:tau_defMM}

Here, we first discuss the definitions of the initial relaxation time $\tau''$ and the smallest timescale $\tau'$ of the long-time consistent with Eq.~\eqref{eq:Cmm}. We then consider their relation to  the master operator eigenvalues.

\subsubsection{Definitions}

We define the \emph{final relaxation time} $\tau$ in terms of the closeness to the stationary state
\begin{equation}\label{eq:tau1}
	\sup_{\rho(0)}\, \lVert \rho(t) -\rhoss \rVert =\lVert e^{\tau\L} -\P_\text{ss} \rVert = c,
\end{equation}
where $\P_\text{ss}$ is the projection on the stationary state and $c<1$ is a constant typically chosen as $1/e$. Due to the contractivity of the dynamics, this timescale is uniquely defined and for $t\leq \tau $ we also have $\lVert \rho(t)-\rhoss\rVert\leq c$ [we have $\lVert e^{(t-\tau)\L} \rVert \leq 1$].

We can analogously define the \emph{initial relaxation time} $\tau''$ as 
\begin{equation}\label{eq:tau''1}
	\lVert e^{\tau''\L} -\P\rVert =c'',\quad \tau''\leq t'',
\end{equation}
where we require $c''>\Cmm$ for $\tau''$ to exist [e.g., valid when $c''=1/e$; cf.~Eq.~\eqref{eq:Cmm2}].

Similarly, the \emph{smallest timescale of the long-time dynamics} $\tau'$ can be defined as
\begin{equation}\label{eq:tau'1}
	\lVert e^{\tau'\L} -\P\rVert =1-c',\quad t'\leq \tau'\leq \tau,
\end{equation}
where we require $\, \lVert \P_\text{ss}-\P \rVert \geq 1-c'+\lVert e^{t'\L}-\P\rVert$ for $\tau'> t'$  to exist, and $c+c'\leq 1$ to also get $\tau'\leq \tau$. \\

\emph{Other definitions}. We note that when the relaxation time is instead of Eq.~\eqref{eq:tau1} defined as the inverse of the maximal rate of an upper exponential bound on approaching the stationary state
\begin{equation}\label{eq:tau2}
	\lVert e^{t\L} -\P_\text{ss} \rVert \leq e^{-\frac{t}{\tau}},
\end{equation}
where $\P_\text{ss}$ is the projection on the stationary state,  we can define the initial relaxation analogously with respect to times only before the metastable regime as
\begin{equation}\label{eq:tau''2}
	\lVert e^{t\L} -\P \rVert \leq e^{-\frac{t}{\tau''}}, \quad t\leq t''.
\end{equation}
The smallest timescale of the long-time dynamics $\tau'$ can be instead defined as the inverse of the minimal rate of a lower exponential bound on departing from a metastable state, so that
\begin{equation}\label{eq:tau'2}
	\lVert e^{t\L} -\P \rVert \leq 1-e^{-\frac{t}{\tau'}}, \quad t'\leq t.
\end{equation}
We have that for $c=c'=c''=1/e$ the timescales defined in Eqs.~\eqref{eq:tau1}, \eqref{eq:tau''1}, and~\eqref{eq:tau'1} are lower bounds on the definitions in Eqs.~\eqref{eq:tau2}, \eqref{eq:tau''2}, and~\eqref{eq:tau'2}.


We note that, as $e^{t\L}-\P_\text{ss}=(\I-\P_\text{ss})e^{ t \L}$, another definition of  the initial relaxation time  $\tau''$ that correspond to Eq.~\eqref{eq:tau1} is
\begin{equation}\label{eq:tau''3}
	\lVert (\I-\P)e^{\tau''\L}\rVert =c'',\quad \tau''\leq t'',
\end{equation}
where we assume $c''\geq(2+\Cp)\Cmm$ for $\tau''$ to exist. Note that, since $\lVert (\I-\P)e^{t\L}\rVert\leq (2+\Cp)\lVert e^{t\L}-\P\rVert$ and $\lVert e^{t\L}-\P\rVert\leq \lVert (\I-\P)e^{t\L}\rVert + \lVert \P e^{t\L}-\P\rVert\lesssim \lVert (\I-\P)e^{t\L}\rVert +\Cmm$, for the initial relaxation time defined in Eq.~\eqref{eq:tau''1} can be bounded from below and above by the relaxation time in Eq.~\eqref{eq:tau''3} corresponding to $c''/(2+\Cp)$ and $c''-\Cmm$, respectively, rather than $c''$.

Finally, the smallest timescale of the dynamics can be considered as $1/\lVert\L\rVert$, as $\lVert e^{t\L}-\I\rVert\leq e^{t\lVert \L\rVert}-1\lesssim t\lVert \L\rVert$ from the Taylor series expansion. 
Analogously, we can define 
\begin{equation}\label{eq:tau'3}
	\tau'=\frac{1}{\lVert\LMM\rVert}.
\end{equation}

\subsubsection{Relation to master operator spectrum}	

For the definitions of the timescales $\tau$, $\tau'$, and $\tau''$ in terms of the projections on the low-lying modes in Eqs.~\eqref{eq:tau''1} and~\eqref{eq:tau'1}, from Eq.~\eqref{eq:t_lambda_all} we have
\begin{subequations}\label{eq:tau_lambda_all}
	\begin{align}
		\label{eq:tau_lambda}
		\tau&\geq  \frac{\ln(c)}{\lambda_2^R},\\
		\label{eq:tau''_lambda}
		\tau''&\geq  \frac{\ln(c'')}{\lambda_{m+1}^R},\\
		\label{eq:tau'_lambda}
		\tau'&\leq  \frac{\ln(c')}{\lambda_m^R},
		\\
		\label{eq:tau'_lambda_I}
		\tau'&\leq  \frac{-\ln(c')}{|\lambda_k^I|},\quad k\leq m,
	\end{align}
\end{subequations}
where the last inequality holds for $t'-t''\geq t''$  or $t''/n \leq t'-t''\leq t''/(n-1)$ with $n\geq 2$ provided that $n\Cmm\ll 1$; cf.~Eq.~\eqref{eq:t'_lambda_I2}.

\emph{Other definitions}. For $\tau$, $\tau''$ and $\tau'$ defined in Eqs.~\eqref{eq:tau2}, ~\eqref{eq:tau''2}, and~\eqref{eq:tau'2}, respectively $\ln(c)$, $\ln(c'')$ and $\ln(c'')$ are all replaced by $-1$ in Eq.~\eqref{eq:tau_lambda_all}. 
For $\tau''$ defined in Eq.~\eqref{eq:tau''3}, Eq.~\eqref{eq:tau''_lambda} holds directly. Finally, below we show that 
\begin{equation}\label{eq:tau'3_lambda}
	1/\lVert \LMM\rVert\leq 1/\max_{k\leq m}|\lambda_{k}|\leq -1/\lambda_{m}^R,
\end{equation}
which provides an upper bounds for the definition in Eq.~\eqref{eq:tau'3}. 
\\

\emph{Derivation of Eq.~\eqref{eq:tau'3_lambda}}.
The lower bound on the norm of the master operator projected on the low-lying modes, $\max_{2\leq k\leq m}|\lambda_k|\leq\lVert\LMM\rVert$,  can be obtained by considering system states and observables as in the derivation of Eq.~\eqref{eq:t_lambda_all}. Indeed, let $\varphi_k$  such that that  $e^{i\varphi_k}\lambda_k=|\lambda_k|$. This gives $\Tr\{ L_k(\phi_k+\varphi_k) \L[\rho(0)]\}=|c_k||\lambda_k|$, and thus $|c_k||\lambda_k|\leq \lVert L_k(\phi_k+\varphi_k)\rVert_{\max} \lVert \LMM  \rho(0)\rVert \leq |c_k|\lVert\LMM\rVert$.

\subsubsection{Relation to metastable regime}

Here, we discuss the relation between timescales $\tau''$ and $\tau'$  and the metastable regime. For all definitions, we obtain $\tau''<t''$ and $t'\ll \tau'$. \\

First, we consider the speed of dynamics in the trace norm given by $\lVert \L e^{t \L}\rVert$. In particular, we make use of the fact that  $\lVert e^{t_1 \L}-e^{t_2 \L}\rVert \leq \int_{t_2}^{t_1}dt\lVert \L e^{t \L}\rVert$ follows from the triangle inequality. 

At any time, the speed is bounded as $\lVert \L e^{t \L}\rVert \leq \lVert \L\rVert$ since $\lVert  e^{t \L}\rVert=1$. Therefore, for $\tau''$ in Eq.~\eqref{eq:tau''1},
\begin{eqnarray}
	t''-\tau''\geq \frac{c''-\lVert e^{t'' \L}-\P\rVert}{\lVert \L\rVert}\geq \frac{c''-\Cmm}{\lVert \L\rVert}.
\end{eqnarray}
Analogously, by  replacing  $c''$ with $1/e$, we obtain a bound for $\tau''$ in Eq.~\eqref{eq:tau''2}. 	

For times after the metastable regime, the speed is bounded by $\lVert \L e^{t' \L}\rVert\lesssim\lVert \LMM\rVert + 2\Cmm^2 \lVert \L \rVert$ when $t'\geq 2 t''$ ~\footnote{For an integer $n\geq 1$ such that $t''\leq t_n\equiv t/n \leq t'$, we have $\lVert\rho(t)-e^{t \LMM}\P[\rho(0)] \rVert =\lVert [e^{t_n \L}-\P]^n(\I-\P)[\rho(0)]\rVert \leq \lVert e^{t_n \L}-\P\rVert^n \lVert \I-\P\rVert $. Since we have ${\lVert \rho(t_n)-\P[\rho(0)]\rVert \leq \Cmm}$ [cf.~Eq.~\eqref{eq:Cmm}], while $\lVert \I-\P\rVert\leq\lVert \I\rVert +\lVert\P\rVert = 2+\Cp $ [cf.~Eq.~\eqref{eq:Cp}], we arrive at $\lVert\rho(t)-e^{t \LMM}\P[\rho(0)] \rVert\leq \Cmm^n(2+\Cp)\lesssim 2\Cmm^n$.}. Therefore, $\tau'-t'\geq (c'-\lVert e^{t' \L}-\P\rVert)/\lVert \LMM\rVert\geq (c'-\Cmm)/(\lVert \LMM\rVert+ 2\Cmm^2 \lVert \L \rVert)$ for $\tau'$ in Eq.~\eqref{eq:tau'1}. 
Using Eq.~\eqref{eq:t'4} below we then obtain
\begin{equation}
	\frac{\tau'}{t'}\gtrsim \frac{c'+2\Cmm^2 t' \lVert \L \rVert}{\Cmm(1+2\Cmm t' \lVert \L \rVert)},
\end{equation}
where the right-hand side is $\gg 1$ when $\Cmm^2 t' \lVert \L \rVert \ll c'$. The bound also holds for $\tau'$ defined in Eq.~\eqref{eq:tau'2} when $c'$ is replaced by $1/e$. Finally, for the definition in Eq.~\eqref{eq:tau'2}, we simply have $\tau'/t'\gtrsim 1/\Cmm$ as
\begin{equation}\label{eq:t'4}
	t'\lVert \LMM\rVert\lesssim\Cmm,
\end{equation}
which we prove below for the metastable regime such that $t'-t''\geq t''$ [or $t''/n \leq t'-t''\leq t''/(n-1)$, where $n\geq 2$ allows for $n\Cmm\ll 1$].\\

We can also find a quantitative relation between the initial relaxation time in Eq.~\eqref{eq:tau''3} and the metastable regime. By noting that $\lVert (\I-\P) e^{n t\L}\rVert \leq \lVert (\I-\P) e^{ t\L}\rVert^n $. Therefore, for $\tau''$ defined in Eq.~\eqref{eq:tau''3} with $c''<1$ we have
\begin{eqnarray}
	\frac{t''}{\tau''}&\leq& \lceil\log_{\lVert (\I-\P) e^{\tau''\L}\rVert } \lVert (\I-\P) e^{t''\L}\rVert\rceil
	\\\nonumber& \leq& \log_{c''} \lVert (\I-\P) e^{t''\L}\rVert,
\end{eqnarray} 
where we assumed $c''<1$.
For $\tau''$ defined in Eq.~\eqref{eq:tau''1}, the bound holds as well with $c''$ replaced by $c''(2+\Cp)$ (assumed $<1$), while  for $\tau''$ in Eq.~\eqref{eq:tau''2} by $(2+\Cp)/e$.

We note that an analogous result holds for times after the final relaxation, that is, 
\begin{eqnarray}
	\frac{t}{\tau}&\leq& \lceil\log_{\lVert e^{\tau\L}-\P_\text{ss}\rVert } \lVert e^{t\L}-\P_\text{ss}\rVert\rceil
	\\\nonumber&=& \lceil \log_{c} \lVert e^{t\L}-\P_\text{ss}\rVert\rceil
\end{eqnarray} 
for $t\geq \tau$, where $\tau$ is defined in Eq.~\eqref{eq:tau1} and we assume $c<1$. For $\tau$ defined in Eq.~\eqref{eq:tau2}, $c$ above is instead replaced by $1/e$ and the equality by $\leq$.\\ 

\emph{Derivation of Eq.~\eqref{eq:t'4}}.   	We have  
$ \lVert e^{t\LMM}\P-\P\rVert\leq \lVert \P\rVert\lVert e^{t\L}-\P\rVert\leq (1+\Cp)\Cmm$ for $t''\leq t\leq t'$ and a similar bound exists for time $t\leq t'-t''$, as follows. We  have $ \lVert e^{(t+t'')\LMM}\P-e^{t\LMM}\P\rVert\leq \lVert e^{t\L}\rVert\lVert e^{t''\LMM}\P-\P\rVert\leq (1+\Cp) \Cmm$, so that $ \lVert e^{t\LMM}\P-\P\rVert\leq \lVert e^{(t+t'')\LMM}\P-e^{t\LMM}\P\rVert+\lVert e^{t''+t\LMM}\P-\P\rVert\leq 2(1+\Cp)\Cmm$ for $t\leq t'-t''$ (in particular, when $t'-t''\geq t''$ this holds for all $t\leq t''$). Next, let us consider time $t$ such that $(2+c)\Cmm= t\lVert \LMM\rVert$, where $\Cmm\ll c\leq 1$, so that 
$\lVert e^{t\LMM}\P-\P\rVert\approx t\lVert \LMM\rVert=(2+c)\Cmm$. It then follows that $t$ fulfills $t> t'$ or $t'-t''< t< t''$. The former case directly implies Eq.~\eqref{eq:t'4},  while the latter case leads to a contradiction, as when $t'-t''\geq t''$ it is not possible, but for $t''/n \leq t'-t''\leq t''/(n-1)$, we have $ 0< t''-t< (n-1)t$ so that $ (2+c)\Cmm < t''\lVert \LMM\rVert< n (2+c)\Cmm $ which, when $n\Cmm\ll 1$, is in the contradiction with $\lVert e^{t''\LMM}\P-\P\rVert\lesssim \Cmm$.
This ends the proof of Eq.~\eqref{eq:t'4}.

\section{Classical metastability in open quantum systems}\label{app:cMM}

In this section, we first discuss the correctness of the definition of classical metastability in Eq.~\eqref{eq:classicality} in terms of the number of metastable phases. We then consider the test of classicality and prove that  $\Ccl$ in Eq.~\eqref{eq:Ccl} is the maximal distance of the barycentric coordinates of a metastable state from a given simplex of candidate metastable phases when measured by $L_1$-norm. We further prove the bound on the corrections in Eq.~\eqref{eq:Ccl2}. We also derive a similar bound on the average distance. Finally, we consider the optimality of the metastable phase construction in Eq.~\eqref{eq:Ccl2} in the context of corrections Eq.~\eqref{eq:classicality}. 	As a corollary of derivations presented in this section, the second line of Eq.~\eqref{eq:Cp} in the main text follows.

\subsection{Definition of classical metastability}  \label{app:defCMM}

In Eq.~\eqref{eq:classicality} we assumed the number of states to be equal to the number $m$ of low-lying modes in the spectrum of the master operator in Eq.~\eqref{eq:L}.   We now justify this assumption.

A higher number of states than $m$ Eq.~\eqref{eq:classicality} necessarily leads to linearly dependent matrices after the projection onto the low-lying modes as in Eq.~\eqref{eq:rhotilde}. Therefore, the decomposition of $\P[\rho(0)]$ into such states in the space of coefficients in not unique [even with the additional assumption of the (approximate) positivity of decomposition]. Therefore, when the states in Eq.~\eqref{eq:classicality} are well approximated by their projection on the low-lying modes in Eq.~\eqref{eq:rhotilde}, we conclude that there are no more than $m$ states for the decomposition to be uniquely defined [uniqueness here is defined up to corrections in Eq.~\eqref{eq:classicality}]. Furthermore, when some of the states in Eq.~\eqref{eq:classicality} differ significantly from their projection on the low-lying modes, i.e., are not metastable,  not all probabilistic mixtures of the states in Eq.~\eqref{eq:classicality} are metastable, and thus the probabilities, although possibly uniquely defined, do not represent degrees of freedom of the MM. Moreover, although in this case the candidate states can be replaced by the states closest to their projections in Eq.~\eqref{eq:rhotilde} [with an increase of corrections in Eq.~\eqref{eq:classicality} by at most $\Cmm+\Ctp+\Cp$; see Eq.~\eqref{eq:classicality_corr4} below], due  to linear dependency in the space of coefficients, this again will lead to a nonunique decomposition. 

A lower number of states than $m$ in Eq.~\eqref{eq:classicality} indicates degeneracy of the description of the MM in the space of coefficients (see e.g., Ref.~\cite{Gaveau2006}), leading to the effective lower dimension of the metastable state manifold than $m-1$. This case will be discussed elsewhere. 


\subsection{Test of classicality} \label{app:test}

\subsubsection{Distance of barycentric coordinates to probability distributions}   \label{app:p_distance}

We now show that the distance of barycentric coordinates $\pt=(\tilde{p}_1,...,\tilde{p}_m)^T$ to the set of probability distributions is given by $ \lVert \pt\rVert_1-1$. 

For any probability distribution $\p=(p_1,...,p_m)^T$, we have  in the L1-norm 
\begin{eqnarray}\label{eq:p_distance_bound}
	\lVert \pt-\p\rVert_1 &\equiv& \sum_{l=1}^m \left|\tilde{p}_l-p_l\right|\\\nonumber
	&=& \sum_{l:\, \tilde{p}_l< 0} \left(-\tilde{p}_l+p_l\right) +\sum_{l:\, \tilde{p}_l\geq 0} \left|\tilde{p}_l-p_l\right|\\\nonumber
	&\geq & \sum_{l:\, \tilde{p}_l< 0} \left(-\tilde{p}_l+p_l\right) +\Bigg|\sum_{l:\, \tilde{p}_l\geq 0} \left(\tilde{p}_l-p_l\right)\Bigg| 
	\\\nonumber
	&= &  2\sum_{l:\, \tilde{p}_l< 0} \left(-\tilde{p}_l+p_l\right) \\\nonumber
	&\geq & 2\sum_{l=1}^m\max\left(-\tilde p_{l},0\right)\equiv \lVert \pt\rVert_1-1,
\end{eqnarray}
where in the second and last line we used the positivity of $p_l\geq 0 $, $l=1,...,m$, the  third line follows from the triangle inequality, and the fourth line follows from $\sum_{l=1}^m \tilde{p}_l=1=\sum_{l=1}^m p_l$ (note that the final result corresponds to the triangle inequality). We now construct the probability distribution for which the lower bound is saturated. We define $\Delta = \sum_{k=1}^m\max\left(-\tilde p_{k},0\right)$ and
\begin{eqnarray}\label{eq:p_distance_opt}
	p_l&\equiv& 0 \quad\text{if}\quad  \tilde{p}_l\leq 0,\\\nonumber
	p_l&\equiv&\tilde{p}_l- \min\Big[\tilde{p}_l,\Delta -\sum_{\substack{k<l:\\\tilde{p}_k>0}} (\tilde{p}_k-p_k)\Big]\!\quad\text{if}\quad  \tilde{p}_l> 0,
\end{eqnarray}
$l=1,...,m$.
In Eq.~\eqref{eq:p_distance_opt} all the negative coordinates in $(\tilde{p}_1,...,\tilde{p}_m)$ are replaced by $0$ [which saturates the second inequality in Eq.~\eqref{eq:p_distance_bound}]. To obtain the sum of probabilities equal $1$, the remaining nonzero positive coordinates of $\pt$, which sum up to $1+\Delta$, are individually reduced while keeping their positivity [which guarantees saturation of the first inequality in Eq.~\eqref{eq:p_distance_bound}], in Eq.~\eqref{eq:p_distance_opt} we chose to set the remaining $\tilde p_{l}>0$ either to $0$ or reduced by the remaining difference of the probability sum to $1$. This gives 
\begin{equation}\label{eq:p_distance}
	\lVert \pt-\p\rVert_1 = \lVert \pt\rVert_1-1.
\end{equation}
Note that the choice of optimal $\p$ in Eq.~\eqref{eq:p_distance_opt} is generally not unique. \\

We conclude that the \emph{maximal distance} of barycentric coordinates to the simplex of probability distributions over all initial states of the system is given by  Eq.~\eqref{eq:Ccl}. As a barycentric coordinate $\tilde{p}_l$ is bounded from below by the minimal eigenvalue of $\tilde P_l$ in the dual basis, from Eq.~\eqref{eq:p_distance} we arrive at Eq.~\eqref{eq:Ctcl}.  \\

The \emph{average distance} $\Cbcl$ for uniformly distributed pure initial states of the system is bounded by 
\begin{equation}\label{eq:Ccl_av}
	\Cbcl\equiv \max_{\rho(0)}\overline{\lVert\pt\rVert_1}-1\leq 2\sum_{l=1}^m\frac{-\Tr(\tilde P_{l}^{-})}{\text{dim}(\mathcal{H})}
\end{equation}
with $\tilde P_{l}^{-}$ being $\tilde P_{l}$ restricted to its negative eigenvalues and $\text{dim}(\mathcal{H})$ the dimension of the system Hilbert space. By construction we have $\Cbcl\leq\Ccl$, and  $\Ctcl\leq \text{dim}(\mathcal{H})\Cbcl$ also holds [cf.~Eqs.~\eqref{eq:Ctcl} and~\eqref{eq:Ctcl2}]. 	
Indeed, $\tilde{p}_{l}$ can be bounded from below by overlap with $\tilde P_l^-$. As $\tilde p_{l}=\Tr[ \tilde P_l\rho(0)]$, while $P_l^-\leq 0$, we have $\max\left(-\tilde p_{l},0\right)\leq-\Tr( P_l^-\rho(0))$, with the right-hand side being linear in $\rho(0)$. Averaging $\rho(0)$ over initial states (uniformly distributed pure states or mixed states in Hilbert-Schmidt metric~\cite{Zyczkowski2001,Zyczkowski2003}) gives $\bar{\rho}=\mathds{1}/\text{dim}(\mathcal{H})$, and thus we arrive at Eq.~\eqref{eq:Ccl_av}.

Furthermore, the average corrections in Eq.~\eqref{eq:classicality} are bounded as
\begin{eqnarray}\label{eq:classicality_av}
	\overline{\bigg\lVert\rho(t)-\sum_{l=1}^m p_l \rho_l\bigg\rVert}&\lesssim&  \Cbcl+\Cp+\Cmm,
\end{eqnarray} 
[see the derivation of Eq.~\eqref{eq:Ccl2} below].

\subsubsection{Derivation of Eq.~\eqref{eq:Ccl2} in the main text}

From the triangle inequality we have for the classical approximation of metastable states in the trace norm
\begin{eqnarray}
	&&\left\lVert\rho(t) -\sum_{l=1}^m p_l \rho_l\right\rVert
	\leq \left\lVert\rho(t) -\sum_{l=1}^m \tilde{p}_l \trho_l \right\rVert\\\nonumber
	&&+\left\lVert\sum_{l=1}^m \tilde{p}_l \trho_l -\sum_{l=1}^m p_l \trho_l\right\rVert+\left\lVert\sum_{l=1}^m p_l \trho_l -\sum_{l=1}^m p_l \rho_l\right\rVert \\\nonumber
	&&\leq\Cmm+\sum_{l=1}^m |\tilde{p}_l-p_l|  \left\lVert\trho_l\right\rVert +\sum_{l=1}^m p_l  \left\lVert\trho_l-\rho_l\right\rVert \\\nonumber
	&&\leq\Cmm+ \left(1+\Cp\right)\lVert \pt-\p\rVert_1+\Cp,
\end{eqnarray}
where in the second inequality we used Eq.~\eqref{eq:Cmm}, while the third inequality follows from $\lVert\trho_l-\rho_l\rVert\leq \Cp$ for $\rho_l$ being the closest state to $\trho_l$ [cf.~Eq.~\eqref{eq:Cp}] and $\lVert\trho_l\rVert\leq\lVert\trho_l-\rho_l\rVert+\lVert\rho_l\rVert\leq 1+\Cp$. Therefore, using Eq.~\eqref{eq:Ccl} we arrive at Eq.~\eqref{eq:Ccl2}.

\subsubsection{Optimality of test of classicality}

We now consider how the construction of classical approximation in Eq.~\eqref{eq:Ccl2} compares to a given set of states in Eq.~\eqref{eq:classicality}.\\

\emph{Bound on $\Ccl$ in terms of corrections in Eq.~\eqref{eq:classicality}}. We now bound the corrections to the classicality in Eq.~\eqref{eq:Ccl} by the corrections in Eq.~\eqref{eq:classicality}.
First, we have
\begin{eqnarray}\label{eq:classicality_corr2}
	\lVert\pt-\p \rVert_1 &=&\sum_{l=1}^m\Big|\Tr\Big\{  \tilde{P}_l \P[\rho(0)]- \tilde{P}_l \sum_{k=1}^m p_k\rho_k  \Big\} \Big|\\\nonumber
	&=& \Big(1+\frac{\Ctcl}{2}\Big)\! \sum_{l=1}^m \Big|\Tr\Big\{  P_l \P[\rho(0)]-P_l\!\sum_{k=1}^m p_k\rho_k\Big\} \Big|\\\nonumber
	&\leq& \Big(1+\frac{\Ctcl}{2}\Big)\! \sum_{l=1}^m \Tr\Big\{  P_l \Big|\P[\rho(0)]-\!\sum_{k=1}^m p_k\rho_k\Big|\Big\} \\\nonumber
	&= &   \Big(1+\frac{\Ctcl}{2}\Big) \Big\lVert \P[\rho(0)]-\sum_{l=1}^m p_l\rho_l\Big\rVert
\end{eqnarray}
where $[\pt(t)]_l\equiv \Tr[ \tilde{P}_l \rho(t)]$, and $P_l$ is the POVM defined in Eq.~\eqref{eq:POVM0}, $l=1,....,m$ , so that $P_l\geq 0$, which we use in the inequality, while $\sum_{l=1}^m P_l=\mathds{1}$, which leads to the last equality. Furthermore, 
\begin{eqnarray}\nonumber
	\Big\lVert \P[\rho(0)]-\sum_{l=1}^m p_l\rho_l\Big\rVert &\leq& \lVert  \P[\rho(0)]-\rho(t)\rVert\\\nonumber&&+\Big\lVert \rho(t)-\sum_{l=1}^m p_l\rho_l\Big\rVert\\\label{eq:classicality_corr}
	&\leq& \Ctp+\mathcal{C},
\end{eqnarray}
where $\mathcal{C}$ are the maximal corrections in Eq.~\eqref{eq:classicality} over the set of metastable states and time $t$ is chosen as in Eq.~\eqref{eq:Cp2}.
Therefore, 
\begin{eqnarray}
	\Ccl&\leq& \max_{\rho(0)}\lVert\pt-\p \rVert_1\leq  \Big(1+\frac{\Ctcl}{2}\Big)\left(\Ctp+\mathcal{C}\right),
\end{eqnarray}
so that [cf.~Eq.~\eqref{eq:Ccl}]
\begin{eqnarray}
	\frac{m\Ccl}{1+m\Ccl}&\leq& m\left(\Ctp+\mathcal{C}\right),
\end{eqnarray}
and thus 
\begin{eqnarray}
	\Ccl&\lesssim& \Ctp+\mathcal{C},
\end{eqnarray}
when $m(\Ctp+\mathcal{C})\ll1$, in which case $\Ctcl\ll 1$ follows 
[cf.~Eq.~\eqref{eq:Cp2} and see Sec.~\ref{app:norm}].\\

\emph{Bound on the stationarity of states in Eq.~\eqref{eq:classicality}}. We now prove that for states in Eq.~\eqref{eq:classicality}  we have

\begin{equation}\label{eq:classicality_corr3}
	\lVert \rho_l -\P(\rho_l)\rVert \lesssim 2 \left(\Ctp+ \mathcal{C}\right),
\end{equation}
when $\Ctcl\ll 1$ or $m(\Ctp+\mathcal{C})\ll1$. Therefore,  in this case the states in Eq.~\eqref{eq:classicality} are \emph{metastable}.

Consider $\rho_l$ in Eq.~\eqref{eq:classicality} as an initial state of the system, $l=1,...,m$. We denote $\rho_l(t)$ the corresponding state at time $t$. By definition, there exists a  probability distribution $p_k^{(l)}$, $k=1,...m$ such that for all times within the metastability regime 
\begin{equation} \label{eq:C0}
	\Big\lVert \rho_l(t)-\sum_{k=1}^m p_k^{(l)} \rho_k \Big\rVert \leq \mathcal{C},
\end{equation}
and therefore from Eq.~\eqref{eq:classicality_corr}
\begin{eqnarray}
	\Big\lVert\P(\rho_l) -\sum_{k=1}^m p_k^{(l)} \rho_k \Big\rVert &\leq&  \Ctp +\mathcal{C} .
\end{eqnarray}    
We also have [cf.~Eqs.~\eqref{eq:classicality_corr} and~\eqref{eq:classicality_corr2}]
\begin{eqnarray}
	\Big\lVert \rho_l-\sum_{k=1}^m p_k^{(l)} \rho_k  \Big\rVert &\leq& |1-p_l^{(l)}|\lVert \rho_l \rVert+\sum_{k\neq l}p_k^{(l)}  \lVert \rho_k \rVert \quad\\\nonumber&=& 	\lVert\pt^{(l)}-\p^{(l)} \rVert_1 \\\nonumber&\leq&   \Big(1+\frac{\Ctcl}{2}\Big)\left(\Ctp+\mathcal{C}\right),
\end{eqnarray}
and, thus, we arrive at
\begin{eqnarray}\nonumber
	\Big\lVert \rho_l-\P[\rho_l] \Big\rVert &\leq&\Big\lVert \rho_l-\sum_{k=1}^m p_k^{(l)} \rho_k  \Big\rVert+\Big\lVert\P(\rho_l) -\sum_{k=1}^m p_k^{(l)} \rho_k \Big\rVert
	\\&\leq& \Big(2+\frac{\Ctcl}{2}\Big)\left(\Ctp+\mathcal{C}\right).
\end{eqnarray}\\

\emph{Bound on corrections for metastable phases replaced by their projections}. We have
\begin{eqnarray}\nonumber
	\Big\lVert \P[\rho(0)]-\sum_{l=1}^m p_l\P(\rho_l)\Big\rVert &\leq& \lVert\P\rVert \Big\lVert \P[\rho(0)]-\sum_{l=1}^m p_l\rho_l\Big\rVert\\
	&\leq& (1+\Cp) (\Ctp+\mathcal{C}),
\end{eqnarray}
where in the second inequality we used Eq.~\eqref{eq:classicality_corr}. 
Therefore, from the triangle inequality
\begin{eqnarray}\nonumber
	\Big\lVert \rho(t)-\sum_{l} p_l\P(\rho_l)\Big\rVert &\leq&  \big\lVert \rho(t)-\P[\rho(0)]\big\rVert\\\nonumber&&+\Big\lVert \P[\rho(0)]-\sum_{l=1}^m p_l\P(\rho_l)\Big\rVert\\\nonumber
	&\leq& \Cmm+(1+\Cp) (\Ctp+\mathcal{C})\\
	&\lesssim &\Cmm+\Ctp+\mathcal{C}.\label{eq:classicality_corr4}
\end{eqnarray} \\

\emph{Bound on corrections for metastable phases replaced by closest states to their projections}. Let $\rho_l'$  be the closest state to the projection $\P(\rho_l) $ of $\rho_l$ in Eq.~\eqref{eq:classicality}, so that $\lVert\P(\rho_l) - \rho_l'\rVert \leq \Cp$ [cf.~Eq.~\eqref{eq:Cp}]. From the triangle inequality we then have [cf.~Eq.~\eqref{eq:classicality_corr4}]
\begin{eqnarray}\label{eq:classicality_corr5}
	\Big\lVert \rho(t)-\sum_{l} p_l\rho_l'\Big\rVert &\leq & \Big\lVert \P[\rho(0)]-\sum_{l=1}^m p_l\P(\rho_l)\Big\rVert\\\nonumber&&+ \sum_{l=1}^m p_l\left\lVert\P(\rho_l)-\rho_l'\right\rVert\\\nonumber
	&\lesssim& \Cmm+\Ctp+\mathcal{C}+\Cp.
\end{eqnarray} 
Equations~\eqref{eq:classicality_corr4} and~\eqref{eq:classicality_corr5} actually hold for any number of states in Eq.~\eqref{eq:classicality} (see the discussion on the correctness of classicality definition in Sec.~\ref{app:defCMM}). We conclude that corrections in Eq.~\eqref{eq:classicality} can increase in the leading order by at most $ \Cmm+\Ctp+\Cp$ when the states  in Eq.~\eqref{eq:classicality} are replaced by closest states to their projections.\\

\section{Classical metastable phases}\label{app:phases}

Here, we prove  the bounds in Eqs.~\eqref{eq:B_trace1}--\eqref{eq:B_support02} of the main text. We begin by discussing the properties of the dual basis. We then derive relations between distances measured by trace norm for density matrices  and L1 norm for barycentric coordinates leading to Eqs.~\eqref{eq:B_trace1} and~\eqref{eq:B_trace2}. We also prove related bounds on scalar products of metastable phases. Finally, we derive Eqs.~\eqref{eq:B_support01} and~\eqref{eq:B_support02}.

\subsection{Properties of dual basis in classical metastable manifolds}\label{app:Ptilde}

Here, we discuss properties of minimum and maximum eigenvalues of the dual basis in Eq.~\eqref{eq:Ptilde} and find corresponding bounds on its norms in terms of $\Ccl$ in Eq.~\eqref{eq:Ccl}. We also show that $\Ctcl$ in Eq.~\eqref{eq:Ctcl} can be understood a distance to operators in a certain POVM and discuss cross-correlations of the elements of the dual basis in metastable states.

\subsubsection{Properties of dual basis}
Let $\tilde{p}_{l}^\text{max}$ be the maximal eigenvalue of $\tilde{P}_l$, $\rho^{\max(l)}$ be the density matrix of the corresponding  eigenstate. Let $[\pt^{\max(l)}]_k=\tilde{p}_k^{\max(l)}= \Tr[\tilde{P}_k\,\rho^{\max(l)}]$, and $\p^{\max(l)}$ be the closest probability distribution to $\pt^{\max(l)}$ [cf.~Eq.~\eqref{eq:p_distance}]. We then have
\begin{eqnarray}\nonumber
	\tilde{p}_{l
	}^\text{max}&=& 1-\sum_{k\neq l}\tilde{p}_k^{\max(l)}   \\\label{eq:Ptilde_max}
	&\leq& 1 - \sum_{k\neq l}\max(-\tilde{p}_k^{\max(l)},0)\leq 1+\frac{\Ccl}{2}.
\end{eqnarray}
Furthermore, for the state $\rho_l$ from which $\trho_l$ is obtained by the projection on the low-lying modes [cf.~Eq.~\eqref{eq:rhotilde}] we have, by the definition,   $\tilde{p}_{l
}^\text{max}\geq\Tr(\tilde{P}_l\rho_l)=1$, so that from Eq.~\eqref{eq:Ptilde_max}
\begin{eqnarray}\label{eq:Ptilde_max2}
	|1-\tilde{p}_{l
	}^\text{max}|&\leq& \frac{\Ccl}{2}.
\end{eqnarray}
Let $\tilde{p}_{l}^\text{min}$ be the minimal eigenvalue of $\tilde{P}_l$. We obtain that $\tilde{p}_{l}^\text{min}\leq 0$ by considering $k\neq l$,  $\tilde{p}_{l}^\text{min}\leq\Tr(\tilde{P}_l\rho_k)=0$ and thus  [cf.~Eq.~\eqref{eq:Ccl}]
\begin{equation}\label{eq:Ptilde_min}
	-\tilde{p}_{l}^\text{min}\leq \frac{\Ccl}{2}
\end{equation}
as well as [cf.~Eq.~\eqref{eq:Ctcl}]
\begin{equation}
	\Ctcl=\sum_{l=1}^m(-\tilde{p}_{l}^\text{min}).
\end{equation}
As the norm $\big\lVert\tilde{P}_l\big\rVert_{\max}$ is by definition  equal the maximal or minus the minimal eigenvalue of $\tilde{P}_l$, we conclude
\begin{equation}\label{eq:Ptilde_norm}
	\big\lVert\tilde{P}_l\big\rVert_{\max} \leq 1+\frac{\Ccl}{2} .
\end{equation}

\subsubsection{Distance to POVM}
Consider operators [cf.~Eq.~\eqref{eq:POVM0}]
\begin{equation}\label{eq:POVM}
	P_l\equiv \frac{\tilde{P}_l-\tilde{p}_{l}^\text{min}\mathds{1}}{1+\frac{\Ctcl}{2}},\quad l=1,...,m.
\end{equation}
We have that $P_l\geq 0$ and $\sum_{l=1}^m P_l=\mathds{1}$, so that these operators constitute a POVM, i.e., a set of operators for which $p_l\equiv \Tr[P_l\rho(0)]$, $l=1,...,m$, corresponds to a probability distribution for any initial state $\rho(0)$. 

We now estimate the distance of this probability distribution to the barycentric coordinates $\tilde{p}_l\equiv \Tr[\tilde{P}_l\rho(0)]$ [cf.~Eq.~\eqref{eq:Ptilde}]. We have
\begin{eqnarray}\label{eq:POVMdistance}
	\lVert \pt -\p\rVert_1&=&\frac{1}{1+\frac{\Ctcl}{2}}\sum_{l=1}^m \left|-\frac{\Ctcl}{2} \tilde{p}_l-\tilde{p}_{l}^\text{min}\right|\qquad\\\nonumber
	&\leq&\frac{\frac{\Ctcl}{2}}{1+\frac{\Ctcl}{2}}\left(\sum_{l=1}^m |\tilde{p}_l|+1\right)\\\nonumber
	&\leq&\frac{\frac{\Ctcl}{2}}{1+\frac{\Ctcl}{2}}(2+\Ccl)\lesssim \Ctcl,
\end{eqnarray}
where the first inequality corresponds to the triangle inequality.

\subsubsection{Cross-correlations of dual basis}\label{app:Ptilde_corr}

Below we show that the cross-correlation for  the dual basis measurement is diagonal in metastable states,
\begin{eqnarray}\label{eq:POVMcorr}
	&&	\sum_{k=1}^m |\Tr\{\tilde{\P}_k \tilde{\P}_l \P[\rho(0)]\}-\delta_{k,l}\tilde{p}_l|\\\nonumber&&\lesssim|\tilde{p}_l|\Big(3\Cp+\Ctcl+\frac{3\Ccl}{2}\Big)+|1-\tilde{p}_l|\Big( \Cp+3\Ctcl+\frac{\Ccl}{2}\Big),
\end{eqnarray}
where $\tilde{\P}_l$ is the superoperator describing the action of measuring $\tilde{P}_l$ [see Eq.~\eqref{ApproxCor} in the main text] and $\tilde{p_l}=\Tr[\tilde{P}_l\rho(0)]=\Tr\{\tilde{P}_l\P[\rho(0)]\}$.	This further leads to 
\begin{eqnarray}\label{eq:POVMcorr}
	&&\max_{1\leq n\leq m}	\sum_{k=1}^m |\Tr[\tilde{\P}_k \tilde{\P}_l (\trho_n)]-\delta_{k,l}\delta_{n,l}|\\\nonumber&&\lesssim\max\Big(3\Cp+\Ctcl+\frac{3\Ccl}{2}, \Cp+3\Ctcl+\frac{\Ccl}{2}\Big),
\end{eqnarray}
which approximation is responsible for the classical dynamics of autocorrelations discussed in Sec.~\ref{sec:ObsD_CL} of the main text. 
The results here can be seen as a consequence of the approximate disjointness of the supports of metastable phases proven in Sec.~\ref{app:disjoint_supp}. \\

\emph{Derivation of Eq.~\eqref{eq:POVMcorr}}. For the POVM elements defined in Eq.~\eqref{eq:POVM}, and $\rho'_n$ being the closest state to $\trho_n$ [cf.~Eq.~\eqref{eq:Cp}], $n=1,...,m$, 
we have  
\begin{eqnarray}
	&&\sum_{k=1}^m |\Tr[{\P}_k {\P}_l (\rho'_n)]-\delta_{k,l}\delta_{n,l}|\\\nonumber
	&&=\Tr[(\mathds{1}-{P}_l) {\P}_l (\rho'_n)]+|\Tr[{P}_l {\P}_l (\rho'_n)]-\delta_{n,l}]|\\\nonumber
	&&=\Tr[(\mathds{1}-{P}_l) P_l \rho'_n]+|\Tr(P_l ^2\rho'_n)-\delta_{n,l}|
\end{eqnarray}
where we used $\Tr[{\P}_k {\P}_l (\rho'_n)]=\Tr[{P}_k {\P}_l (\rho'_n)]$, the positivity of ${P}_k$ and ${\P}_l (\rho'_n)$, $\sum_{l=1}^m P_l=\mathds{1}$, and the fact that the bases of $P_l$ and $\mathds{1}-P_l$ are the same. Moreover, 
\begin{eqnarray}
	\Tr[(\mathds{1}\!-\!{P}_l) P_l \rho'_n]&\leq& \min\{\Tr(P_l \rho'_n),\!\Tr[(\mathds{1}\!-\!{P}_l) \rho'_n]\},\qquad\,\,\,\,\\
	\Tr[{P}_l ^2 (\rho_n')]&\leq &	\Tr({P}_l \rho'_n),
\end{eqnarray}
where we used $P_l, \mathds{1}-P_l\leq \mathds{1}$, while $|\Tr(P_l ^2\rho'_l)-1|=1-\Tr(P_l ^2\rho'_l)= \Tr[(\mathds{1}-P_l)\rho'_l]+\Tr[(\mathds{1}-P_l)P_l\rho'_l]$. Therefore, 
\begin{eqnarray}
	&&\sum_{k=1}^m |\Tr[{P}_k {\P}_l (\rho_n')]|\leq 2	\Tr({P}_l \rho_n'),\\
	&&\sum_{k=1}^m |\Tr[{P}_k {\P}_l (\rho_l')]|\leq 3\Tr[(\mathds{1}\!-\!{P}_l) \rho_l'].
\end{eqnarray}
Noting that for $n\neq l$, $\Tr(P_l \rho_n')=[\Tr(\tilde{P}_l \rho_n')-\tilde{p}_l^{\min}]/(1+\Ctcl/2)\leq [(1+\Ccl/2)\Cp+\Ccl/2]/(1+\Ctcl/2)$, while $\Tr[(\mathds{1}-P_l) \rho_l']=1-[\Tr(\tilde{P}_l \rho_l')-\tilde{p}_l^{\min}]/(1+\Ctcl/2)\leq 1-[1-\Cp\Ccl/2]/(1+\Ctcl/2)=(\Ctcl+\Cp\Ccl)/(2+\Ctcl)$, we arrive at
\begin{eqnarray}
	&&\sum_{k=1}^m |\Tr[{P}_k {\P}_l (\rho_n')]|\leq 2\Cp+\frac{2\Ccl}{2+\Ctcl} \lesssim2 \Cp+ \Ccl,\qquad\\
	&&\sum_{k=1}^m |\Tr[{P}_k {\P}_l (\rho_l')]|\lesssim \frac{3\Ctcl}{2+\Ctcl}\lesssim\frac{3}{2}\Ctcl.
\end{eqnarray}
where for the second inequalities in both lines we assumed $\Ctcl\ll 1$. Finally, we have
\begin{eqnarray}
	&&\sum_{k=1}^m |\Tr[{\P}_k {\P}_l (\rho_n')]- \Tr[\tilde{\P}_k \tilde{\P}_l (\trho_n)]| \\\nonumber
	&&\leq \sum_{k=1}^m |\Tr
	\{{P}_k {\P}_l [\rho_n'-(1+\Ctcl/2)^2\trho_n]\}| \\\nonumber
	&&\quad+ (-\tilde {p}_l^\text{min})\Big(1+\frac{\Ctcl}{2}\Big) \sum_{k=1}^m |\Tr
	({P}_k \trho_n)|\\\nonumber
	&&\quad+\frac{\Ctcl}{2} \Big(1+\frac{\Ctcl}{2}\Big) |\Tr
	({P}_l \trho_n)|,
\end{eqnarray}
where we used the fact that the bases of $P_l$ and $\tilde{P}_l$ are the same. Noting that $|\Tr
\{{P}_k {\P}_l [\rho_n'-(1+\Ctcl/2)^2\trho_n]\}| \leq \Tr
\{{P}_k |{\P}_l [\rho_n'-(1+\Ctcl/2)^2\trho_n]|\}  $ and similarly $|\Tr
({P}_k \trho_n)|\leq \Tr
({P}_k |\trho_n|)$, we have
\begin{eqnarray}
	&&\sum_{k=1}^m |\Tr[{\P}_k {\P}_l (\rho_n')]- \Tr[\tilde{\P}_k \tilde{\P}_l (\trho_n)]| \\\nonumber
	&&\leq \lVert\rho'_n-\trho_n\rVert+\Ctcl(1-\Ctcl/4)\lVert\trho_n\rVert\\\nonumber
	&&\quad+\Big(-\tilde {p}_l^\text{min}+\frac{\Ctcl}{2}\Big)\Big(1+\frac{\Ctcl}{2}\Big) \lVert\trho_n\rVert\\\nonumber
	&&\leq \Cp +\Ctcl(1+\Cp)\\\nonumber
	&&\quad +\frac{\Ccl+\Ctcl}{2}\Big(1+\frac{\Ctcl}{2}\Big) (1+\Cp)
	\\\nonumber
	&&\lesssim\Cp+\frac{\Ccl+3\Ctcl}{2}.
\end{eqnarray}
where we used $\lVert \P_l[\rho'_n-(1+\Ctcl/2)^2\trho_n]\rVert\leq  \lVert \P_l\rVert\lVert\rho'_n-(1+\Ctcl/2)^2\trho_n\rVert\leq \lVert\rho'_n-\trho_n\rVert+\Ctcl(1-\Ctcl/4)\lVert\trho_n\rVert$ as $\lVert \P_l\rVert\leq 1$ from the complete-positivity of $\P_l$.

\subsection{Trace-norm vs. L1-norm in classical metastable manifolds}\label{app:norm}
Here, we discuss how the distances measured in the trace norm for density matrices and by L1 norm for barycentric coordinates of the MM are related. In particular, we derive Eqs.~\eqref{eq:B_trace1} and~\eqref{eq:B_trace2} of the main text. We also discuss the norms of the long-time generator  and the relaxation times.

\subsubsection{Distance between metastable states}
We consider two initial states $\rho^{(1)}\!(0)$ and $\rho^{(2)}\!(0)$ projected on low-lying eigenmodes as $\trho^{(1)}$ and $\trho^{(2)}$ [see Eq.~\eqref{Expansion}], respectively, that are described by barycentric coordinates $\pt^{(1)}$ and $\pt^{(2)}$. 

For L1 norm, we have [cf.~Eq.~\eqref{eq:classicality_corr2}]
\begin{eqnarray}\label{eq:rho1_CL}
	&&\left\lVert \pt^{(1)} -\pt^{(2)}   \right\rVert_1= \sum_{l=1}^m|[\pt^{(1)}]_l-[\pt^{(2)}]_l|\\\nonumber
	&&\qquad= \sum_{l=1}^m  \left|\Tr[\tilde{P}_l \trho^{(1)}]-\Tr[\tilde{P}_l\trho^{(2)}] \right|\\\nonumber
	&&\qquad= \left(1+\frac{\Ctcl}{2}\right) \sum_{l=1}^m  \left|\Tr(P_l \trho^{(1)}]-\Tr[P_l\trho^{(2)}] \right|\\\nonumber
	&&\qquad\leq  \left(1+\frac{\Ctcl}{2}\right)   \min\left(\lVert \trho^{(1)}-\trho^{(2)}\rVert, \lVert \rho^{(1)}(0)-\rho^{(2)}(0)\rVert\right),
\end{eqnarray}
where $P_l$ is the POVM defined in Eq.~\eqref{eq:POVM}. The first inequality in Eq.~\eqref{eq:B_trace1} follows by observing $\lVert \rho(t)-\P[\rho(0)]\rVert\leq \Cmm$ for time $t$ within the metastable regime [cf.~Eq.~\eqref{eq:Cmm}].

For the trace norm, we have
\begin{eqnarray}
	\left\lVert \trho^{(1)}-\trho^{(2)}\right\rVert&=&  \left\lVert\sum_{l=1}^m \trho_l \left[\Tr(\tilde{P}_l \trho^{(1)})-\Tr(\tilde{P}_l\trho^{(2)})\right] \right\rVert \nonumber\\
	&\leq &(1+\Cp) \sum_{l=1}^m \left|\Tr(\tilde{P}_l \trho^{(1)})-\Tr(\tilde{P}_l\trho^{(2)})\right|  \nonumber\\
	&=& (1+\Cp) \,\lVert \pt^{(1)} -\pt^{(2)}   \rVert_1, 
	\label{eq:rho2_CL}
\end{eqnarray}
where the inequality corresponds to the triangle inequality and $\lVert\trho_k\rVert \leq 1+\Cp $ follows from Eq.~\eqref{eq:Cp}. Therefore, the distance of the states  $\rho'^{(1)}$ and $\rho'^{(2)}$ closest to $\trho^{(1)}$ and $\trho^{(2)}$, respectively, is bounded as [cf.~Eq.~\eqref{eq:Cp}]
\begin{equation} \label{eq:rho3_CL}
	\left\lVert \rho'^{(1)}-\rho'^{(2)}\right\rVert\leq (1+\Cp) \,\lVert \pt^{(1)} -\pt^{(2)}   \rVert_1+2\Cp.
\end{equation}
Similarly, the distance between states that project on $\trho^{(1)}$ and $\trho^{(2)}$  during metastable regime is bounded by  $(1+\Cp) \,\lVert \pt^{(1)} -\pt^{(2)}   \rVert_1+2\Cmm$ [cf.~Eq.~\eqref{eq:Cmm}], which gives the second inequality in Eq.~\eqref{eq:B_trace1}.

Finally, we note that, since the distance in the trace norm is contractive under quantum dynamics~\cite{Nielsen2010}, we have that the following series of inequalities
\begin{eqnarray}
	\lVert \rho^{(1)}\!(0)-\rho^{(2)}\!(0)\rVert& \geq &\lVert \rho^{(1)}(t)-\rho^{(2)}(t)\rVert 
	\\\nonumber&\geq& \lVert \trho^{(1)}-\trho^{(2)}\rVert -2\Cmm 
	\\\nonumber&\geq& \lVert \rho'^{(1)}-\rho'^{(2)}\rVert -2\Cmm-2\Cp.
\end{eqnarray}
In particular, states that project on $\trho^{(1)}$ and $\trho^{(2)}$ do not need to be metastable. Therefore, only the distance between their projections [or metastable states, e.g., as in Eqs.~\eqref{eq:classicality} or~\eqref{eq:Ccl2}] is bounded in Eq.~\eqref{eq:rho3_CL}.

\subsubsection{Distance between metastable phases}
From Eqs.~\eqref{eq:rho1_CL} and~\eqref{eq:rho2_CL}, we obtain that the distance between the projections of the metastable phases in Eq.~\eqref{eq:rhotilde} is bounded from below as

\begin{equation}\label{eq:phases_CL}
	\frac{4}{ 2+\Ctcl} \leq  \min\left(\lVert \trho_l-\trho_k\rVert, \lVert \rho_k-\rho_l\rVert\right) \quad k\neq l,
\end{equation}
where $k,l=1,...,m$ and $\rho_l$ is the state that projects on $\trho_l$  [cf.~Eq.~\eqref{eq:classicality}]. This follows from $\lVert \pt_{l} -\pt_{k} \rVert_1=2$ with $(\pt_{l})_k\equiv \Tr(\tilde{P}_k\trho_l)=\delta_{kl}$, $k,l=1,...,m$. Note that the upper bound from Eq.~\eqref{eq:rho2_CL} is trivial [cf.~Eq.~\eqref{eq:Cp}].
Similarly, for the states $\rho_l'$ closest to the projections in Eq.~\eqref{eq:rhotilde}, we have [cf.~Eq.~\eqref{eq:Cp}]
\begin{equation}\label{eq:phases2_CL}
	\frac{4}{2+\Ctcl} -2\Cp\leq \left\lVert \rho_k'-\rho_l'\right\rVert,  \quad k\neq l.
\end{equation}

In analogy to Eq.~\eqref{eq:rho1_CL}, by considering the POVM of two elements: $P\equiv (\tilde{P}_l-\tilde{p}_l^{\min}\mathds{1})/(\tilde{p}_l^{\max}-\tilde{p}_l^{\min})$ and $\mathds{1}-P=(\tilde{p}_l^{\max}\mathds{1}-\tilde{P}_l)/(\tilde{p}_l^{\max}-\tilde{p}_l^{\min})$, we obtain 
\begin{eqnarray}
	&&\frac{2}{\tilde{p}_l^{\max}-\tilde{p}_l^{\min}}\leq    \min\left(\lVert \trho_l-\trho_k\rVert, \lVert \rho_k-\rho_l\rVert\right),\,\,\, k\neq l, \qquad\
\end{eqnarray}
so that [cf.~Sec.~\ref{app:Ptilde}]
\begin{eqnarray}\label{eq:phases3_CL}
	&& \frac{2}{1+\Ccl} \leq    \min\left(\lVert \trho_l-\trho_k\rVert, \lVert \rho_k-\rho_l\rVert\right),\quad k\neq l, \quad
\end{eqnarray}	
while  for the states $\rho_l'$ closest to the projections in Eq.~\eqref{eq:rhotilde}, we have [cf.~Eq.~\eqref{eq:Cp}]
\begin{equation}\label{eq:phases4_CL}
	\frac{2}{1+\Ccl} -2\Cp\leq \left\lVert \rho_k'-\rho_l'\right\rVert,  \quad k\neq l,
\end{equation}
which corresponds to the bound in Eq.~\eqref{eq:B_trace2} of the main text.

Therefore, we conclude that the metastable phases are approximately disjoint with respect to the trace norm (for the bimodal case $m=2$, see also~Ref.~\cite{Rose2016}).

\subsubsection{Norms of long-time generator}

We now compare the induced trace norm for $\LMM$ in Eq.~\eqref{eq:Leff} and the induced $L1$ norm for $\Wt$ in Eq.~\eqref{eq:Wtilde}.

Since $\mathds{1}$ is $0$ eigenvector of $\LMM$ due its trace preservation, analogously to Eq.~\eqref{eq:rho1_CL}, we have
\begin{eqnarray}\label{eq:Leff1_CL}
	\lVert \Wt   \rVert_1&\equiv& \max_{1\leq l\leq m} \sum_{k=1}^m| \Tr[ \tilde{P}_k \LMM (\trho_l) ]|\\\nonumber
	&=&  \left(1+\frac{\Ctcl}{2}\right) \max_{1\leq l\leq m}  \sum_{k=1}^m| \Tr[ P_k \LMM (\trho_l) ]|\\\nonumber
	&\leq &  \left(1+\frac{\Ctcl}{2}\right)   \max_{1\leq l\leq m} \lVert \LMM (\trho_l)\rVert\\\nonumber
	&\leq &  \left(1+\frac{\Ctcl}{2}\right) \lVert \LMM\rVert.
\end{eqnarray}
Similarly, as  $\LMM$ maps any state onto the low-lying modes, analogously to Eq.~\eqref{eq:rho2_CL}, we also have
\begin{eqnarray}	\label{eq:Leff2_CL}
	\left\lVert \LMM\right\rVert&\equiv & \max_{\rho(0)} \left\lVert\sum_{k,l=1}^m \trho_k\,\Wt_{kl}(\pt)_l\right\rVert\\\nonumber
	&\leq & \max_{1\leq k \leq m} \lVert \rho_k   \rVert\max_{\rho(0)}  \lVert \pt   \rVert_1   \lVert \Wt   \rVert_1 \\ \nonumber
	&\leq& (1+\Cp)(1+\Ccl) \,\lVert \Wt   \rVert_1,
\end{eqnarray}
where $(\pt)_l=\Tr[\tilde{P}_l \rho(0)]$.

\subsubsection{Relaxation time}
We now show when the relaxation time towards the stationary state  $\rhoss$ in the trace norm and  the relaxation of metastable state within the classical MM are approximately the same.

We have
\begin{eqnarray}\label{eq:tau1_CL}
	\left\lVert e^{t\Wt} -\Ptss    \right\rVert_1&=&\max_{1\leq l\leq m}  \sum_{k=1}^m\left|(e^{t\Wt})_{kl}-(\ptss)_l\right|\\\nonumber
	&\leq &   \left(1+\frac{\Ctcl}{2}\right) \max_{1\leq l\leq m} \left\lVert e^{t\L}(\rho_l)-\rhoss\right\rVert  \\\nonumber
	&\leq&   \left(1+\frac{\Ctcl}{2}\right) \left\lVert e^{t\L}  -\P_\text{ss}\right\rVert. 
\end{eqnarray}
In the second line we used  Eq.~\eqref{eq:rho1_CL} by noting that, for a given $l$, the vector $[(e^{t\Wt})_{1l},...,(e^{t\Wt})_{ml}]$ corresponds to the barycentric coordinates of $e^{t\LMM}(\trho_l)$, which is the projection onto the low-lying modes of the state $e^{t\L}(\rho_l)$ [cf.~Eq.~\eqref{eq:rhotilde}]. 	

Similarly, we have
\begin{eqnarray}\label{eq:tau2_CL}
	\left\lVert (e^{t\L}  -\P_\text{ss})\right\rVert&\leq &\left\lVert (e^{t\L}  -\P_\text{ss})\P\right\rVert\\\nonumber&&+\left\lVert (e^{t\L}  -\P_\text{ss})(\I-\P)\right\rVert
\end{eqnarray}
with
\begin{eqnarray}\label{eq:tau3_CL}
	&&\left\lVert (e^{t\L}  -\P_\text{ss})\P\right\rVert \leq (1+\Ccl) \max_{1\leq l\leq m} \lVert e^{t\LMM}(\trho_l)-\rhoss\rVert\qquad\quad\\\nonumber
	&&\qquad\qquad\leq (1+\Ccl)(1+\Cp) \max_{1\leq l\leq m}  \sum_{k=1}^m\left|(e^{t\Wt})_{kl}-(\ptss)_l\right| \\\nonumber
	&&\qquad\qquad= (1+\Ccl)(1+\Cp) \left \lVert e^{t\Wt} -\Ptss   \right \rVert_1
\end{eqnarray}
and 
\begin{eqnarray}\label{eq:tau4_CL}
	\left\lVert (e^{t\L}  -\P_\text{ss})(\I-\P)\right\rVert &=& \left\lVert e^{t\L} (\I-\P)\right\rVert \qquad\quad\\\nonumber
	&=& \left\lVert e^{t\L}- e^{t\LMM}\P\right\rVert\\\nonumber
	&\leq& 2\Cmm^n.
\end{eqnarray}
In Eq.~\eqref{eq:tau3_CL}, in the first inequality we used $\P[\rho(0)]= \sum_{l=1}^m \tilde{p}_l \trho_l$, where $\sum_{l=1}^m|\tilde{p}_l|\leq (1+\Ccl)$ for any initial state $\rho(0)$ [cf.~Eq.~\eqref{eq:Ccl}], and in the second inequality Eq.~\eqref{eq:rho2_CL}. In Eq.~\eqref{eq:tau4_CL}, the inequality holds for times $t\geq t''$ and $n$ such that $t/n$ belongs to the metastable regime,   $t''\leq  t/n \leq t'$.

When $\Ctcl\ll 1$,  from Eqs.~\eqref{eq:tau1_CL}--\eqref{eq:tau4_CL}, the decay to the stationary state after the metastable regime [cf.~Eqs.~\eqref{Expansion0} and~\eqref{Expansion}] is equally well captured by the decay of the probabilities between metastable phases. Therefore, from Eq.~\eqref{eq:tau1_CL}, the relaxation time $\tilde{\tau}$ with respect to L1-norm is approximately equal the relaxation time $\tau$ with respect to the trace norm,
\begin{equation}
	\tau \approx\tilde{\tau}.
\end{equation} 
For example, in the case of a perturbation away from a classical first-order phase transition occurring for a finite system size, both relaxation times are of the same order in the perturbation. When $\Ctcl\ll 1$ does not hold,  the relaxation time for the barycentric coordinates of the MM is generally longer, $\tau \leq\tilde{\tau}$; cf.~Eqs.~\eqref{eq:tau2_CL}--\eqref{eq:tau4_CL}.

\subsection{Orthogonality and disjointness of phases in classical metastable manifolds}\label{app:disjoint}

\subsubsection{Bounds on scalar products of metastable phases}

We first derive an upper bound on the scalar product between square roots of the states $\rho_l$ that project on $\trho_l$ [cf.~Eqs.~\eqref{eq:classicality} and~\eqref{eq:rhotilde}], 
\begin{eqnarray}\label{eq:B_disjoint}
	\Tr \left(\sqrt{\rho_k} \sqrt{\rho_l}\right)&\lesssim&\sqrt{2\Ccl},\quad k\neq l.
\end{eqnarray}
where $k,l=1,...,m$. For the states  $\rho_l'$ closest  $\trho_l$ in Eq.~\eqref{eq:rhotilde} [cf.~Eq.~\eqref{eq:Cp}], we instead have 
\begin{eqnarray}\label{eq:B_disjoint2}
	\Tr \left(\sqrt{\rho'_k} \sqrt{\rho'_l}\right)&\lesssim&\sqrt{2\Ccl +4\Cp},\quad k\neq l.
\end{eqnarray}
Since $\sqrt{\rho'_l}$ and $\sqrt{\rho'_l}$ are positive and normalized in the scalar product, the bounds in Eqs.~\eqref{eq:B_disjoint} and~\eqref{eq:B_disjoint2} imply that the  \emph{metastable phases are approximately disjoint} (see also Sec.~\ref{sec:disjoint} of the main text).

Second, we prove the following bounds on the scalar product of the metastable phases
\begin{eqnarray}\label{eq:B_orth}
	|\Tr (\rho_k \rho_l)|&\lesssim&\! \sqrt{ \frac{\Ccl}{2}}\! \left(\lVert\rho_k \rVert_{\max}\!+\!\lVert\rho_l \rVert_{\max}\right)\!,\,\,\,k\neq l,\qquad
\end{eqnarray}
and
\begin{eqnarray}\label{eq:B_orth2}
	|\Tr (\rho'_k \rho'_l)|&\lesssim&\! \sqrt{ \frac{\Ccl}{2}\! +\!\Cp}\! \left(\lVert\rho'_k \rVert_{\max}\!+\!\lVert\rho'_l \rVert_{\max}\right)\!,\,\,k\neq l,\qquad\,\,\,\,
\end{eqnarray}
where $k,l=1,...,m$. The same inequality as in Eq.~\eqref{eq:B_orth2} also holds for 	$\Tr (\trho_k \trho_l)$.  
Note that the scalar product of metastable phases is affected by mixedness of both phases,  $|\Tr (\rho_k \rho_l)| \leq  \sqrt{ \Tr (\rho_k^2)}\sqrt{\Tr (\rho_l^2)} $, which may enhance their approximate orthogonality.  \\

\emph{Proof of Eqs.~\eqref{eq:B_disjoint} and~\eqref{eq:B_disjoint2}}. For any states $\rho_k$ and $\rho_l$, we have
\begin{eqnarray}\nonumber
	&&(\tilde{p}_{k}^\text{max} -\tilde{p}_{k}^\text{min})\Tr (\sqrt{\rho_k}\,\sqrt{\rho_l})=\Tr [\sqrt{\rho_k}\, (\tilde{P}_k-\tilde{p}_{k}^\text{min}\mathds{1})\sqrt{\rho_l}]\\
	&&\qquad\qquad\qquad\qquad+\Tr [\sqrt{\rho_k}\, (\tilde{p}_{k}^\text{max}\mathds{1} -\tilde{P}_k )\sqrt{\rho_l}], \qquad\label{eq:D_disjoint1}
\end{eqnarray}
where $\tilde{p}_{k}^\text{min}$ and $\tilde{p}_{k}^\text{max}$ are minimal and maximal eigenvalues of $\tilde{P}_k$, which are introduced to obtain products with positive operators. 
From the Cauchy-Schwarz inequality we further obtain, 
\begin{eqnarray}\label{eq:D_disjoint2}
	&&|\Tr [\sqrt{\rho_k}\, (\tilde{P}_k-\tilde{p}_{k}^\text{min}\mathds{1})\sqrt{\rho_l}]| \\\nonumber
	&&\leq \sqrt{\Tr [ (\tilde{P}_k-\tilde{p}_{k}^\text{min}\mathds{1})^2\rho_l\,]}  \\\nonumber
	&&\leq  \sqrt{\tilde{p}_{k}^\text{max}-\tilde{p}_{k}^\text{min}} \sqrt{\Tr [ (\tilde{P}_k-\tilde{p}_{k}^\text{min}\mathds{1})\rho_l\,]}, 
\end{eqnarray}
where in the third line we used $(\tilde{P}_k-\tilde{p}_{k}^\text{min}\mathds{1})^2\leq  (\tilde{p}_{k}^\text{max}-\tilde{p}_{k}^\text{min}) (\tilde{P}_k-\tilde{p}_{k}^\text{min}\mathds{1})$.
Analogously, we also have	
\begin{eqnarray}\label{eq:D_disjoint3}
	&&|\Tr [\sqrt{\rho_k}\, (\tilde{p}_{k}^\text{max}\mathds{1} -\tilde{P}_k )\sqrt{\rho_l}] |
	\\\nonumber
	&&\leq \sqrt{\Tr [\rho_k\, (\tilde{p}_{k}^\text{max}\mathds{1} -\tilde{P}_k )^2]} 
	\\\nonumber
	&&\leq \sqrt{\tilde{p}_{k}^\text{max}-\tilde{p}_{k}^\text{min}}\sqrt{\Tr [\rho_k\, (\tilde{p}_{k}^\text{max}\mathds{1} -\tilde{P}_k )]} .
\end{eqnarray}	
Therefore, 
\begin{eqnarray}\label{eq:D_disjoint4}
	\Tr (\sqrt{\rho_k}\,\sqrt{\rho_l})&\leq&\sqrt{ \frac{\Tr [ (\tilde{P}_k-\tilde{p}_{k}^\text{min}\mathds{1})\rho_l\,]} {1-\frac{\Ccl}{2}}} \\\nonumber
	&&+\sqrt{\frac{\Tr [\rho_k\, (\tilde{p}_{k}^\text{max}\mathds{1} -\tilde{P}_k )]}{1-\frac{\Ccl}{2}}}, \qquad
\end{eqnarray}
where we used $  |\tilde{p}_{k}^\text{max}-\tilde{p}_{k}^\text{min}|\geq |\tilde{p}_{k}^\text{max}|-|\tilde{p}_{k}^\text{min}|\geq 1-\Ccl/2$ [cf.~Eqs.~\eqref{eq:Ptilde_max2} and~\eqref{eq:Ptilde_min}].

For the states $\rho_l$ that project onto $\trho_l$ [cf.~Eqs.~\eqref{eq:classicality} and~\eqref{eq:rhotilde}], $\trho_l=\P(\rho_l)$,  $|\Tr[ (\tilde{P}_k-\tilde{p}_{k}^\text{min}\mathds{1})\,\rho_l]|=|\tilde{p}_{k}^\text{min}|\leq \Ccl/2$  [cf.~Eq.~\eqref{eq:Ptilde_min}], while $|\Tr [\rho_k\, (\tilde{p}_{k}^\text{max}\mathds{1} -\tilde{P}_k )]|=|\tilde{p}_{k}^\text{max}-1|\leq \Ccl/2$ [cf.~Eq.~\eqref{eq:Ptilde_max2}]. Therefore, from Eq.~\eqref{eq:D_disjoint4}, we arrive at
\begin{eqnarray}\label{eq:B_disjoint_full}
	\Tr (\sqrt{\rho_k}\,\sqrt{\rho_l})&\leq &  \sqrt{\frac{2\Ccl}{1-\frac{\Ccl}{2}}},
\end{eqnarray}	 
which in the leading order gives Eq.~\eqref{eq:B_disjoint}. 

For the states closest to the projections $\trho_l$ [cf.~Eqs.~\eqref{eq:classicality}], which we denote $\rho_l'$, $l=1,...,m$, we have that $\Tr [ (\tilde{P}_k-\tilde{p}_{k}^\text{min}\mathds{1})\rho_l'\,]\leq |\tilde{p}_{k}^\text{min}|+ \lVert \tilde{P}_k \rVert_{\max}  \lVert \rho_l'-\trho_l \rVert \leq $ and $\Tr [\rho_k\, (\tilde{p}_{k}^\text{max}\mathds{1} -\tilde{P}_k )]\leq |1-\tilde{p}_{k}^\text{max}|+\lVert \tilde{P}_k \rVert_{\max}  \lVert \rho'_l-\trho_l \rVert$, where $|\tilde{p}_{k}^\text{min}|\leq \Ccl/2$ [cf.~Eq.~\eqref{eq:Ptilde_min}], $|1-\tilde{p}_{k}^\text{max}|\leq\Ccl/2$  [cf.~Eq.~\eqref{eq:Ptilde_max2}], $\lVert \tilde{P}_k \rVert_{\max}  \leq 1+\Ccl/2$ [cf.~Eq.~\eqref{eq:Ptilde_norm}], and $\lVert \rho'_l-\trho_l \rVert\leq\Cp$ [cf.~Eq.~\eqref{eq:Cp}]. Therefore,  from Eq.~\eqref{eq:D_disjoint4}, we arrive at  [cf.~Eq.~\eqref{eq:B_disjoint_full}]
\begin{eqnarray}\label{eq:B_disjoint_full2}
	\Tr (\sqrt{\rho'_k}\,\sqrt{\rho'_l})&\leq &  \sqrt{\frac{2\Ccl+2\left(2+\Ccl\right)\Cp}{1-\frac{\Ccl}{2}}},
\end{eqnarray}
where $\Cp$  is a bound from above on the distance $\lVert \rho_l-\trho_l \rVert $ [cf.~Eq.~\eqref{eq:Cp}]. In the leading order Eq.~\eqref{eq:B_disjoint_full2} gives Eq.~\eqref{eq:B_disjoint2}. \\

\emph{Proof of Eqs.~\eqref{eq:B_orth} and~\eqref{eq:B_orth2}}. 
In analogy to Eq.~\eqref{eq:D_disjoint1}, for any states $\rho_k$ and $\rho_l$,  we have
\begin{eqnarray}
	(\tilde{p}_{k}^\text{max}-\tilde{p}_{k}^\text{min})\Tr [\rho_k\,\rho_l]&=&\Tr [\rho_k\, (\tilde{P}_k-\tilde{p}_{k}^\text{min}\mathds{1})\,\rho_l]\qquad\\\nonumber
	&+&\Tr [\rho_k\, (\tilde{p}_{k}^\text{max}\mathds{1} -\tilde{P}_k )\,\rho_l].
\end{eqnarray}	
From the Cauchy-Schwarz inequality we obtain 
\begin{eqnarray}\label{eq:orth_d1}
	&&|\Tr [\rho_k\, (\tilde{P}_k-\tilde{p}_{k}^\text{min})\mathds{1}\,\rho_l]| \\\nonumber
	&&\leq\sqrt{	\Tr [(\tilde{P}_k-\tilde{p}_{k}^\text{min}\mathds{1})^2\rho_l]} \, \sqrt{\Tr (\rho_k^2\rho_l)}\\\nonumber
	&&\leq  \sqrt{\tilde{p}_{k}^\text{max}-\tilde{p}_{k}^\text{min}}\sqrt{	\Tr [(\tilde{P}_k-\tilde{p}_{k}^\text{min}\mathds{1})\rho_l]} \,\lVert\rho_k \rVert_{\max},
\end{eqnarray} 
and, similarly,
\begin{eqnarray}\label{eq:orth_d2}
	&&|\Tr [\rho_k\, (\tilde{p}_{k}^\text{max}\mathds{1} -\tilde{P}_k )\,\rho_l] | \\\nonumber
	&&\leq	\sqrt{\Tr [\rho_k\, (\tilde{p}_{k}^\text{max}\mathds{1}-\tilde{P}_k)^2]} \, \sqrt{\Tr (\rho_k\, \rho_l^2)}\\\nonumber
	&&\leq \sqrt{\tilde{p}_{k}^\text{max}-\tilde{p}_{k}^\text{min}} 	\sqrt{\Tr [\rho_k\, (\tilde{p}_{k}^\text{max}\mathds{1}-\tilde{P}_k)]} \,\lVert\rho_l \rVert_{\max}
\end{eqnarray}	
[cf.~Eqs.~\eqref{eq:D_disjoint2} and~\eqref{eq:D_disjoint3} and see Eq.~\eqref{eq:vN2} below]. 

For the states $\rho_l$ that project onto $\trho_l$ [cf.~Eqs.~\eqref{eq:classicality} and~\eqref{eq:rhotilde}],  $\trho_l=\P(\rho_l)$, we have [cf.~Eq.~\eqref{eq:B_disjoint_full}]
\begin{eqnarray}\label{eq:orth_d5}
	\Tr (\rho_k\,\rho_l)&\leq &  \sqrt{\frac{\frac{\Ccl}{2}}{1-\frac{\Ccl}{2}}} \left(\lVert\rho_k \rVert_{\max}+\lVert\rho_l \rVert_{\max}\right),
\end{eqnarray}
which in the leading order gives Eq.~\eqref{eq:B_orth}.

For the states closest to the projections $\trho_l$ [cf.~Eqs.~\eqref{eq:classicality}], which we denote $\rho_l'$, $l=1,...,m$, we have [cf.~Eq.~\eqref{eq:B_disjoint_full2}]
\begin{eqnarray}
	&&|\Tr (\rho_k \rho_l)|\!\leq\!	\sqrt{\frac{\frac{\Ccl}{2}\!+\!\left(1\!+\!\frac{\Ccl}{2}\right)\!\Cp}{1-\frac{\Ccl}{2}}} \!\left(\lVert\rho_k \rVert_{\max}\!+\!\lVert\rho_l \rVert_{\max}\right)\!,\qquad\,\,\,\,
	\label{eq:orth_d3}
\end{eqnarray}
which in the leading order gives  Eq.~\eqref{eq:B_orth2}.

Finally, 
\begin{eqnarray}\nonumber
	&&|\Tr (\rho'_k\,\rho'_l)|\leq |\Tr (\rho'_k\,\rho'_l)-\Tr (\trho_k \trho_l)|+|\Tr (\trho_k \trho_l)|,
\end{eqnarray}
where [cf.~Eq.~\eqref{eq:vN2} below]
\begin{eqnarray}\nonumber
	&&|\Tr (\rho'_k\,\rho'_l)-\Tr (\trho_k \trho_l)|=|\Tr [(\rho'_k-\trho_k) \rho'_l]+\\\nonumber
	&&\qquad\quad+\Tr [\rho'_k(\rho'_l-\trho_l) ]+\Tr [(\rho'_k -\trho_k)(\rho'_l -\trho_l) ]|\\\nonumber
	&&\qquad\leq \lVert\rho'_k-\trho_k \rVert \lVert\rho'_l \rVert_{\max}+\lVert\rho'_l-\trho_l \rVert \lVert\rho'_k \rVert_{\max} \\\nonumber&&\qquad\quad+\lVert\rho'_k -\trho_k \rVert\lVert\rho'_l-\trho_l  \rVert
	\\ &&\qquad\leq  \Cp (\lVert\rho_k \rVert_{\max}+\lVert\rho_l \rVert_{\max}+\Cp),\quad
	\label{eq:B_orth0}
\end{eqnarray}
which contributes to $\Tr (\trho_k \trho_l)$ in the higher order than $\Tr (\trho_k \trho_l)$ [cf.~Eq.~\eqref{eq:orth_d3}].\\

\emph{A bound on the trace of product of operators}. In the proofs above, we make use of the following results
\begin{equation}\label{eq:vN2}
	|\Tr(XY)|\leq \lVert XY \rVert \leq \min\left(\lVert X\rVert_{\max}\lVert Y\rVert,\lVert X\rVert\lVert Y\rVert_{\max} \right).
\end{equation}
Here, the first inequality follows from the definition of the trace norm, while the second inequality from the H\"older inequality for Schatten norms. 

\subsubsection{Proof of Eqs.~(\ref{eq:B_support01})-(\ref{eq:B_support03}) in the main text}\label{app:disjoint_supp}

We now prove Eqs.~(\ref{eq:B_support01})-(\ref{eq:B_support03}). To this aim, we first derive bounds on the support of general system states. These bounds can also be used to  allows show approximate disjointness of basins of attractions. We also discuss the role of the decay subspace.  \\

From the discussion in Sec.~\ref{app:Ptilde}, there exist at least one eigenvalue of $\tilde{P}_l$ greater or equal $1$, and one less or equal $0$. Let $\mathcal{H}_l$ be the sum of the eigenspaces of $\tilde{P}_l$  with eigenvalue above or equal $\Delta$, where $0\leq \Delta\leq 1$. We will consider the overlap of system states with $\mathcal{H}_l$.  \\

\emph{Bounds for general states}. 	For an initial system state $\rho(0)$, we have
\begin{eqnarray}\label{eq:B_support1}
	\Tr[ \tilde{P_l} \rho(0)]\!&\leq& \Tr[ \mathds{1}_{\mathcal{H}_l} \rho(0)] p_l^{\max} \!\!+ \!\{1\!-\!\Tr[ \mathds{1}_{\mathcal{H}_l} \rho(0)]\}\Delta \qquad\,\,\,\,\,\,\,\\ \!&\leq&\Tr[ \mathds{1}_{\mathcal{H}_l} \rho(0)] \!\left(\! 1\!+\!\frac{\Ccl}{2}\!\right)\! + \left\{1\!-\!\Tr[ \mathds{1}_{\mathcal{H}_l} \rho(0)]\right\}\Delta
	\nonumber
\end{eqnarray}
where $\tilde{p}_{l}^\text{max}\geq 1+\Ccl/2$ is the maximal eigenvalue of $\tilde{P}_l$, and 
\begin{eqnarray} \nonumber 
	\Tr[ \tilde{P_l}\, \rho(0)]&\geq&  \Tr[ \mathds{1}_{\mathcal{H}_l} \rho(0)]  \Delta \!+\!  \{1\!-\!\Tr[ \mathds{1}_{\mathcal{H}_l} \rho(0)]\}\,
	\tilde{p}_{l}^\text{min}\\&\geq&   \Tr[ \mathds{1}_{\mathcal{H}_l} \rho(0)]  \Delta \!-\! \{1\!-\!\Tr[ \mathds{1}_{\mathcal{H}_l}\rho(0)]\}\frac{\Ccl}{2},\qquad\quad\label{eq:B_support2}
\end{eqnarray}
where $\tilde{p}_{l}^\text{min}\geq -\Ccl/2$ is the minimal eigenvalue of $\tilde{P}_l$.
Therefore, from Eq.~\eqref{eq:B_support1} we obtain
\begin{equation}\label{eq:B_support3}
	\Tr[ \mathds{1}_{\mathcal{H}_l} \rho(0)]\geq \frac{\Tr[\tilde{P_l} \rho(0)]- \Delta}{1-\Delta+\frac{\Ccl}{2}}
\end{equation}
and from Eq.~\eqref{eq:B_support2}
\begin{equation}\label{eq:B_support4}
	\Tr[ \mathds{1}_{\mathcal{H}_l} \rho(0)]\leq \frac{\Tr[ \tilde{P_l} \rho(0)]+\frac{\Ccl}{2} }{\Delta+\frac{\Ccl}{2}}\leq \frac{\Tr[ \tilde{P_l} \rho(0)]+\frac{\Ccl}{2} }{\Delta},
\end{equation}
as well as [cf.~Eq.~\eqref{eq:Ctcl}]
\begin{equation}\label{eq:B_support5}
	\sum_{k\neq l}\Tr[ \mathds{1}_{\mathcal{H}_k} \rho(0)]< \frac{1-\Tr[ \tilde{P_l} \rho(0)]+\frac{\Ctcl}{2} }{\Delta},
\end{equation}
where we used $\sum_{k\neq l}\Tr[ \tilde{P_l} \rho(0)]= 1-\Tr[ \tilde{P_l}\, \rho(0)]$. These bounds are used to argue disjointness of basins of attraction in Sec.~\ref{sec:disjoint} of the main text. \\

\emph{Proof of Eqs.~(\ref{eq:B_support01})-(\ref{eq:B_support03})}.  For the states $\rho_l$ that project onto $\trho_l$ [cf.~Eqs.~\eqref{eq:classicality} and~\eqref{eq:rhotilde}], 
$\trho_l=\P(\rho_l)$, we have
\begin{eqnarray} 
	\Tr (\tilde{P}_l \rho_l)&=&1,\\
	\Tr (\tilde{P}_l \rho_k)&=&0,\quad k\neq l,
\end{eqnarray}
where $k,\l=1,...,m$. Therefore, from Eq.~\eqref{eq:B_support3}, we obtain
\begin{eqnarray}
	\Tr( \mathds{1}_{\mathcal{H}_l}\, \rho_l) &\geq& \frac{1- \Delta}{1-\Delta+\frac{\Ccl}{2}},
\end{eqnarray}
while from Eqs.~\eqref{eq:B_support4} and~\eqref{eq:B_support5}, 
\begin{eqnarray}
	\Tr( \mathds{1}_{\mathcal{H}_k}\, \rho_l)&\leq& \frac{ \Ccl }{2\Delta},\quad k\neq l,
	\\
	\sum_{k\neq l}	\Tr( \mathds{1}_{\mathcal{H}_k}\, \rho_l)&\leq& \frac{ \Ctcl }{2\Delta},
\end{eqnarray}
which for $\Delta=1/2$ in the leading order gives the bounds discussed in Sec.~\ref{sec:disjoint} of the main text.

For states closest to the projections $\trho_l$ in Eq.~\eqref{eq:rhotilde}, which we denote here by $\rho_l'$, we have
\begin{eqnarray}\label{eq:B_support6}
	\Tr (\tilde{P}_l \rho_l')&\geq&1 - \left(1+\frac{\Ccl}{2}\right) \Cp,\\
	\label{eq:B_support7}
	|\Tr (\tilde{P}_k \rho_l')|&\leq& \left(1+\frac{\Ccl}{2}\right) \Cp,\quad k\neq l,
\end{eqnarray}
where $k,l=1,...,m$.
This follows from $\Tr (\tilde{P}_k \rho'_l)= |\Tr [\tilde{P}_k (\rho'_l-\trho_l)] +\Tr (\tilde{P}_k\trho_l)$ and $\Tr (\tilde{P}_k\trho_l)=\delta_{kl}$ and $|\Tr [\tilde{P}_k (\rho_l-\tilde\rho_l)]|\leq \lVert \tilde{P}_k \rVert_{\max} \lVert \rho'_l-\trho_l \Vert\leq (1+\frac{\Ccl}{2}) \Cp$ [cf.~Eqs.~\eqref{eq:Cp} and~\eqref{eq:Ptilde_norm}]. 
Therefore, from Eq.~\eqref{eq:B_support3}, we obtain
\begin{eqnarray}
	\Tr( \mathds{1}_{\mathcal{H}_l}\, \rho'_l) &\geq& \frac{1- \Delta- \left(1+\frac{\Ccl}{2}\right) \Cp}{1-\Delta+\frac{\Ccl}{2}},\qquad
\end{eqnarray}
while Eqs.~\eqref{eq:B_support4} and~\eqref{eq:B_support5}, 
\begin{eqnarray}
	\Tr( \mathds{1}_{\mathcal{H}_k}\, \rho'_l)&\leq& \frac{ \left(1+\frac{\Ccl}{2}\right)\Cp+\frac{\Ccl}{2} }{\Delta}, \quad k\neq l,
	\\
	\sum_{k\neq l}\Tr( \mathds{1}_{\mathcal{H}_k}\, \rho'_l)&\leq& \frac{ \left(1+\frac{\Ccl}{2}\right)\Cp+\frac{\Ctcl}{2} }{\Delta}.
\end{eqnarray}
These bounds, in the leading order, correspond to Eqs.~(\ref{eq:B_support01})-(\ref{eq:B_support03}) when $\Delta=1/2$ is chosen. 

Finally, for the truncated metastable phases we simply have $|\Tr( \mathds{1}_{\mathcal{H}_k}\, \rho_l)-\Tr( \mathds{1}_{\mathcal{H}_k}\, \tilde{\rho_l})|\leq  \lVert\rho_l- \tilde{\rho_l}\rVert\leq \Cp$, which introduces corrections of the same order.\\

\emph{Decay subspace}. 
%
We now show that the subspaces spanned by eigenvectors of $\tilde{P_l}$ with the eigenvalues separated from $0$ and $1$ by $\gg \Ccl$ (or by $\gg \Ccl+\Cp$), $l=1,...,m$, which generally evolve into nontrivial mixture of metastable phases, can be neglected in the support of metastable phases, i.e., they belong to the (approximate) decay subspace.
Therefore, the subspaces $\mathcal{H}_l$ considered in Sec.~\ref{sec:disjoint}, that is with $\Delta=1/2$, feature states that can be neglected in the support of metastable phases. This leads to $\mathcal{H}_l$ being generally not disjoint.

Let $\mathcal{K}_l$ be a subspace of the system space spanned by the eigenvectors of $\tilde{P}_l$ with the eigenvalues between $\Delta_1$ and $\Delta_2$, where $0\leq \Delta_1\leq \Delta_2\leq 1$. Considering $\Delta=\Delta_2$ in Eq.~\eqref{eq:B_support3}, we obtain
\begin{equation}\label{eq:B_decay1}
	\Tr[ \mathds{1}_{\mathcal{K}_l} \rho(0)]\leq 1-	\Tr[ \mathds{1}_{\mathcal{H}_l} \rho(0)]\leq 1-\frac{\Tr[ \tilde{P_l}\rho(0)]- \Delta_2}{1-\Delta_2+\frac{\Ccl}{2}}.
\end{equation}
Considering $\Delta=\Delta_1$ in Eq.~\eqref{eq:B_support4}, we get
\begin{equation}\label{eq:B_decay2}
	\Tr[ \mathds{1}_{\mathcal{K}_l} \rho(0)]\leq\Tr[ \mathds{1}_{\mathcal{H}_l} \rho(0)] \leq \frac{\Tr[ \tilde{P_l} \rho(0)]+\frac{\Ccl}{2} }{\Delta_1+\frac{\Ccl}{2}}
\end{equation}
and the same in Eq.~\eqref{eq:B_support5},
\begin{equation}\label{eq:B_decay3}
	\sum_{k\neq l}\Tr[ \mathds{1}_{\mathcal{K}_k} \rho(0)]\leq\sum_{k\neq l}\Tr[ \mathds{1}_{\mathcal{H}_k} \rho(0)] \leq \frac{1-\Tr[ \tilde{P_l} \rho(0)]+\frac{\Ctcl}{2} }{\Delta_1}.
\end{equation}

Therefore, for the state $\rho_l$ that projects on $\trho_l$ in Eq.~\eqref{eq:rhotilde}, $l=1,...,m$,  Eq.~\eqref{eq:B_decay1} gives
\begin{eqnarray}\label{eq:B_decay4}
	\Tr( \mathds{1}_{\mathcal{K}_l}\, \rho_l)&\leq& 1-\frac{1- \Delta_2}{1-\Delta_2+\frac{\Ccl}{2}}
	\lesssim \frac{\Ccl}{2(1-\Delta_2)},\qquad
\end{eqnarray}
where in the second inequality we assumed $1\!-\!\Delta_2\gg \Ccl$, while from Eq.~\eqref{eq:B_decay2} 
\begin{eqnarray}\label{eq:B_decay5}
	\Tr( \mathds{1}_{\mathcal{K}_k}\, \rho_l)&\leq& \frac{\Ccl }{2\Delta_1+\Ccl}\leq \frac{\Ccl }{2\Delta_1},\quad k\neq l,
\end{eqnarray}
and from Eq.~\eqref{eq:B_decay3}
\begin{equation}\label{eq:B_decay6}
	\sum_{k\neq l}\Tr( \mathds{1}_{\mathcal{K}_k}\, \rho_l) \leq \frac{\Ctcl }{2\Delta_1}.
\end{equation}
Similarly, for the closest state $\rho_l'$ to  $\trho_l$ we have [cf.~Eqs.~\eqref{eq:B_support6} and~\eqref{eq:B_support7}]
\begin{eqnarray}\label{eq:B_decay4'}
	\Tr( \mathds{1}_{\mathcal{K}_l}\, \rho'_l)&\leq& 1-\frac{1- \Delta_2-(1+\frac{\Ccl}{2})\Cp}{1-\Delta_2+\frac{\Ccl}{2}}\\\nonumber&\lesssim& \frac{\Ccl+2\Cp}{2(1-\Delta_2)},\qquad
\end{eqnarray}
where in the second inequality we assumed $1\!-\!\Delta_2\gg \Ccl$,
\begin{eqnarray}\label{eq:B_decay5'}
	\Tr( \mathds{1}_{\mathcal{K}_k}\, \rho'_l)&\leq& \frac{\Ccl +(2+\Ccl)\Cp }{2\Delta_1+\Ccl}\\\nonumber
	&\lesssim& \frac{\Ccl +2\Cp}{2\Delta_1},\quad k\neq l,\qquad
\end{eqnarray}
and
\begin{equation}\label{eq:B_decay6'}
	\sum_{k\neq l}\Tr( \mathds{1}_{\mathcal{K}_k}\, \rho'_l) \leq \frac{ \Ctcl+(2+\Ccl)\Cp }{2\Delta_1}\lesssim  \frac{ \Ctcl +2\Cp}{2\Delta_1}.
\end{equation}
We conclude that for  $\Delta_1\gg\Ctcl$ and $1\!-\!\Delta_2\gg \Ccl$ (or $\Delta_1\gg \Ctcl+2\Cp$ and $1\!-\!\Delta_2\gg \Ccl+2\Cp$), the subspaces $\mathcal{K}_l$, $l=1,...,m$ can be neglected in the support of metastable phases $\rho_1$, ..., $\rho_m$ (or $\rho'_1$, ..., $\rho'_m$).

Importantly, by its definition, $\mathcal{K}_l$ exists outside the basins of attraction, $l=1,...,m$, as its states evolve into nontrivial mixtures of metastable phases. Indeed, for a state $\rho(0)$ supported on $\mathcal{K}_l$, we have $\Delta_1\leq \tilde{p}_l \leq \Delta_2$.  When $\Delta_1\gg\Ctcl$ and $1\!-\!\Delta_2\gg \Ccl$, we have that  $\tilde{p}_l \geq \Delta_1$ is nonnegligible and so is the sum of other barycentric coordinates, as   $ \sum_{k\neq l} \tilde{p}_k=1\!-\!\tilde{p}_l \geq 1\!-\!\Delta_2$. Thus, the  metastable state $\P[\rho(0)]$ corresponding to $\rho(0)$ is a mixture of $\trho_l$ with at least one other metastable phase $\trho_k$, $k\neq l$.

\subsection{Nonuniqueness of phases in classical metastable manifolds}\label{app:nonunique}
Here, we discuss the nonuniqueness of the choice of metastable phases. We prove that considering an alternate set of $m$ metastable phases leading to the corrections to the classicality $\Ccl'$, the distance of the alternative metastable phases to closest phase in the original set is bounded by $\Ccl+\Ccl'+\min(\Ccl,\Ccl')$.\\

We consider two sets of linearly independent  metastable phases, $\trho_1$, ..., $\trho_m$ and  $\trho_1'$, ..., $\trho_m'$ with the corrections to the classicality $\Ccl$ and $\Ccl'$, respectively. We can represent the metastable phases using barycentric coordinates as $\trho_l'=\sum_{k=1}^m\tilde{p}_k^{(l)} \trho_k $ and $\trho_k=\sum_{n=1}^m\tilde{p}_n'^{(k)} \trho_n' $, so that from the linear independence from $\trho_l'=\sum_{k=1}^m\sum_{n=1}^m \tilde{p}_k^{(l)} \tilde{p}_n'^{(k)} \trho_n' $ we have
\begin{eqnarray} \label{eq:MPinv1_2bases}
	\sum_{k=1}^m\tilde{p}_k^{(l)}\tilde{p}_n'^{(k)}	&=& \delta_{nl},
\end{eqnarray}
and, analogously, 
\begin{eqnarray} \label{eq:MPinv1_2bases_a}
	\sum_{l=1}^m\tilde{p}_n^{(l)}\tilde{p}_l'^{(k)}	&=& \delta_{nk}.
\end{eqnarray}
Therefore, from the triangle inequality
\begin{eqnarray} \label{eq:MPinv2_2bases}
	\sum_{k=1}^m\left|\tilde{p}_k^{(l)}\right|\left|\tilde{p}_l'^{(k)}\right|\geq 1
\end{eqnarray}
and
\begin{eqnarray} \label{eq:MPinv2_2bases_a}
	\sum_{l=1}^m\left|\tilde{p}_k^{(l)}\right|\left|\tilde{p}_l'^{(k)}\right|\geq 1.
\end{eqnarray}

From the inequality in Eq.~\eqref{eq:MPinv2_2bases}, by noting that 	$\sum_{k=1}^m|\tilde{p}_k^{(l)}|\leq 1+\Ccl$ [cf.~Eq.~\eqref{eq:Ccl}] and $|\tilde{p}_l'^{(k)}|\leq 1+\Ccl'/2$ [cf.~Eq.~\eqref{eq:Ptilde_max}],  we observe 
\begin{eqnarray} \label{eq:MPinv4_2bases}
	\left|\tilde{p}_k^{(l)}\right|\left|\tilde{p}_l'^{(k)}\right|+\left(1+\Ccl-	|\tilde{p}_k^{(l)}|\right)\left(1+\frac{\Ccl'}{2}\right) \geq 1,\qquad
\end{eqnarray}
and thus
\begin{eqnarray} \label{eq:MPinv5_2bases}
	\left|\tilde{p}_l'^{(k)}\right|\geq 1+\frac{\Ccl'}{2} -\frac{2\Ccl+\Ccl'+\Ccl\Ccl'}{2	|\tilde{p}_k^{(l)}|}.
\end{eqnarray}
Analogously, from Eq.~\eqref{eq:MPinv2_2bases_a} we also have
\begin{eqnarray} \label{eq:MPinv7_2bases}
	|\tilde{p}_k^{(l)}|\geq 1+\frac{\Ccl}{2} -\frac{2\Ccl'+\Ccl+\Ccl\Ccl'}{2	|\tilde{p}_l'^{(k)}|}.
\end{eqnarray}

Since $\sum_{k=1}^m|\tilde{p}_k^{(l)}|\leq 1+\Ccl$, from the lower bound in Eq.~\eqref{eq:MPinv2_2bases}, we also have $\max_{1\leq k\leq m} |\tilde{p}_l'^{(k)}|\geq 1/(1+\Ccl)$ and we can remove the absolute value as $\Ccl\ll 1$ [cf.~Eq.~\eqref{eq:Ptilde_min}]. Choosing $k$ corresponding to the maximum, that is,
\begin{eqnarray} \label{eq:MPinv9a_2bases}
	\tilde{p}_l'^{(k)}\gtrsim 1-\Ccl,\qquad \sum_{n\neq l}\left|\tilde{p}_n'^{(k)}\right|\lesssim \Ccl+\Ccl',
\end{eqnarray}
where the second inequality follows from $\sum_{l=1}^m|\tilde{p}_l'^{(k)}|\leq 1+\Ccl'$, 
we have from Eq.~\eqref{eq:MPinv7_2bases}
\begin{eqnarray} \label{eq:MPinv9_2bases}
	\tilde{p}_k^{(l)}\gtrsim 1-\Ccl', \qquad \sum_{n\neq k}\left|\tilde{p}_k^{(l)}\right|\lesssim \Ccl+\Ccl'.
\end{eqnarray}

In the trace norm, we have 
\begin{eqnarray}
	\left\lVert\trho_l'- \trho_{k}\right\rVert&\leq& \left|1-\tilde{p}_k^{(l)}\right|\left\lVert \trho_k\right\rVert+ \sum_{n\neq k} \left|\tilde{p}_n^{(l)}\right|\left\lVert \trho_n \right\rVert\\\nonumber
	&\leq& (1+\Cp) \left(\left|1-\tilde{p}_{k}^{(l)}\right|+\sum_{n\neq k} \left|\tilde{p}_n^{(l)}\right|\right) 
\end{eqnarray}
where we used the fact that $\lVert \trho_n \rVert\leq 1+\Cp$ from Eq.~\eqref{eq:Cp} [see also Eq.~\eqref{eq:rho2_CL}]. Analogously, by considering the decomposition in the barycentric coordinates of $\rho_1'$, ..., $\rho_m'$, we arrive at the upper bound by	$\leq (1+\Cp) (2|1-\tilde{p}_{l}'^{(k)}|+\Ccl')$. Therefore, using Eqs.~\eqref{eq:MPinv9a_2bases} and~\eqref{eq:MPinv9_2bases}, we find
\begin{eqnarray}
	\left\lVert\trho_l'- \trho_{k}\right\rVert	&\lesssim& \Ccl+\Ccl'+\min(\Ccl,\Ccl'),
	\label{eq:MPinv12_2bases}
\end{eqnarray}
while for the metastable phases chosen as the closest states to the projections $\trho_l$ and $\trho_k$, the corrections can increase by $2\Cp$ [cf.~Eq.~\eqref{eq:Cp}].

Finally, we observe that $k$ is uniquely defined for each $l$, i.e., is a \emph{function} of $l$. Indeed, 
this follows directly from two inequalities Eq.~\eqref{eq:MPinv9_2bases}. Furthermore, this function is \emph{bijective} from the second inequality in  Eq.~\eqref{eq:MPinv9a_2bases}.

\section{Classical long-time dynamics}\label{app:Leff_general}

Here,  first review of properties of classical stochastic dynamics in Sec.~\ref{app:classical}. 
We then prove that the long-time dynamics is effectively classical as discussed in Sec.~\ref{sec:Leff} of the main text. First, in Sec.~\ref{app:Leff}, we consider the dynamics of the average system state within a classical MM, and derive bounds on its approximation by classical dynamics governed by a classical stochastic generator. Second, in Sec.~\ref{app:Ls}, we prove that classical trajectories of that classical stochastic generator capture the statistics of quantum trajectories for times after the initial relaxation.

\subsection{Classical stochastic dynamics} \label{app:classical}
Here, we review properties of classical stochastic dynamics and statistics of its trajectories. We also discuss weak symmetries.

\subsubsection{Positivity and probability conservation}
Let $l=1,...,m$ label $m$ configurations of a classical system and $\p=(p_1,...,p_m)^T$ be a vector of the corresponding probabilities ($0\leq p_l\leq 1$, $l=1,...,m$ and $\sum_{k=1}^m p_k =1$). Time-homogeneous dynamics which preserves the positivity of the probability vector is generated by a matrix $\W$ [cf.~Eq.~\eqref{eq:Leff}]
\begin{equation} \label{eq:pt}
	\frac{d}{dt}\p(t)= \W \p(t)
\end{equation}
such that
\begin{equation}\label{eq:Wpos}
	(\W)_{ll}\leq 0, \quad (\W)_{kl}\geq 0 \,\,\text{for} \,\, k\neq l,
\end{equation}
where $k,l=1,...,m$, while the preservation of the total probability requires 
\begin{equation} \label{eq:Wtrace}
	\sum_{k=1}^m (\W)_{kl}=0.
\end{equation}
for all $l=1,...,m$.

\subsubsection{Stochastic trajectories} 
The matrix $W$ can be considered as a Markovian generator of stochastic trajectories of system configurations. The waiting time for a transition from the $k$th configuration is distributed exponentially with the rate $-(W)_{ll}$, so that the average lifetime is 
\begin{equation}\label{eq:Wtau}
	\tau_l=-\frac{1}{(\W)_{ll}},
\end{equation}	
while the probability that the transition takes place from $l$th to $k$th configuration is proportional to $(\W)_{kl}$ [equals $-(\W)_{kl}/(\W)_{ll}$], $k,l=1,...,m$. 

The cumulants of the statistics of number of transitions in a stochastic trajectory [cf.~Eq.~\eqref{eq:Ls}] is encoded by the maximal eigenvalue of the biased operator
\begin{equation}
	\W_s=\W+ (e^{-s}-1)\Jbf,
\end{equation}
where $\Jbf=\W+\bm{\mu}$ with 
\begin{equation}
	(\bm{\mu})_{kl}= \delta_{kl} \mu_l=- \delta_{kl}(\W)_{ll}
\end{equation}
being a diagonal matrix of activities. The (minus) first derivative of the maximal eigenvalue determines the average activity 
\begin{equation}\label{av_CL}
	\mu_\text{ss}= \sum_{l=1}^m(\Jbf\pss)_l= \sum_{l=1}^m(\pss)_l \mu_l,
\end{equation}
where $\pss$ is the stationary probability of $\W$, i.e., $\W\pss=\bm{0}$, and $(\pss)_l$ corresponds to the average time spent in the $l$th configuration, $l=1,..,m$. The second derivative the rate of fluctuations of the number of transitions in stochastic trajectories
\begin{eqnarray}\label{var_CL}
	\sigma_\text{ss}^2 &=&\sum_{l=1}^m(\Jbf\pss)_l -2 \sum_{l=1}^m(\Jbf\R  \Jbf\pss)_l \\\nonumber
	&=& \sum_{l=1}^m(\bm{\mu}\pss)_l -2 \sum_{l=1}^m(\bm{\mu}\R  \bm{\mu}\pss)_l ,
\end{eqnarray}
where $\R$ is the pseudoinverse of $\W$. 
In the second line in Eq.~\eqref{var_CL}, we used $(\Jbf-\bm{\mu})\pss=\W \pss=\mathbf{0}$ as well as $\sum_{l=1}^m(\Jbf)_{lk}=(\bm{\mu})_k$. Equations~\eqref{av_CL} and~\eqref{var_CL} follow from the perturbation theory for $\W_s$ with respect to $(e^{-s}-1)\Jbf$.

Similarly, the cumulants of the statistics of a time-integral of system observable which takes value $m_k$ in the $k$th configurations is encoded by the maximal eigenvalue of the biased operator  [cf.~Eq.~\eqref{eq:Lh}]
\begin{equation}
	\W_s=\W- h\m,
\end{equation}
where the diagonal matrix of the system observable 
\begin{equation}
	(\m)_{kl}= \delta_{kl} m_l.
\end{equation}
The  rate of the average time-integral 
\begin{equation}\label{av_CL2}
	m_\text{ss}= \sum_{l=1}^m(\m\pss)_l=\sum_{l=1}^m(\pss)_l m_l,
\end{equation}
while the rate of fluctuations of time-integral
\begin{equation}\label{var_CL2}
	\delta_\text{ss}^2 =-2 \sum_{l=1}^m(\m\R  \m\,\pss)_l ,
\end{equation}
Eqs.~\eqref{av_CL2} and~\eqref{var_CL2} follow from the perturbation theory for $\W_h$ with respect to $h\m$. 

\subsubsection{Weak symmetries} \label{app:classical_symmetry}
Stochastic dynamics features weak symmetry $\Pibf$ when
\begin{equation}\label{eq:weak_W}
	[\W,\Pibf]=0,
\end{equation}
where $\Pibf$ is a permutation matrix between $m$ system configurations. 

The symmetry in Eq.~\eqref{eq:weak_W} is equivalent to
\begin{equation}\label{eq:weak_W1}
	(\W)_{kl}= (\W)_{\pi(k) \pi(l)},
\end{equation}
where $k,l=1,..m$ and $\pi$ denotes the permutation corresponding to $\Pibf$, that is $(\Pibf)_{kl}=\delta_{k \pi(l)}$, $k,l=1,...,m$. From Eq.~\eqref{eq:weak_W1}, all configurations that belong to the same cycle, feature the same decay rate, $(\W)_{kk}= (\W)_{\pi^n(k) \pi^n(k)}$, $n=1,2,...$. Furthermore,  for two different cycles of length $d_1$ and $d_2$, with  $d_{1,2}$ being the greatest common factor of $d_1$ and $d_2$, transitions from the first to the second cycle are the same for all elements of a subcycle of length $d_1/d_{1,2}$ and all elements of a subcycle $d_2/d_{1,2}$  (and analogously for transitions from the second to the first cycle). In particular, the transition rates from (or to) an invariant configuration $l$ [$\pi(l)=l$] are the same for all the elements of a cycle, $(\W)_{kl}= (\W)_{\pi^n(k) l}$ $[(\W)_{lk}= (\W)_{l \pi^n(l)}]$, $n=1,2,...$.

From Eq.~\eqref{eq:weak_W}, the generator $\W$ is \emph{block diagonal} in an eigenbasis of $\Pibf$, which can be chosen as plane waves over the  cycles in the permutation $\pi$.  Thus, the number of free parameters of $\W$ is limited to the sum of squared degeneracies of $\Pibf$ eigenvalues, i.e., the plane-wave momenta [less $1$ from  the trace-preservation condition in Eq.~\eqref{eq:Wtrace}]. Furthermore, the eigenvectors of $\W$ can be chosen as eigenvectors of $\Pibf$, in which case they are linear combination of the plane waves with the same eigenvalue.

Finally, the symmetric eigenspace of $\Pibf$ is spanned by  plane waves with $0$ momenta that are uniform mixtures of system configurations in each cycle, and thus its dimension equals number of cycles in the permutation $\pi$. Moreover, the effective dynamics between those mixtures, i.e., dynamics of symmetric probabilities, is also stochastic, and given by [cf.~Eq.~\eqref{eq:pt}]
\begin{equation}\label{eq:pt_0}
	\frac{d}{dt}\p_0(t)= {\W\!}_0\, \p_0(t)
\end{equation}
where $[\p_0(t)]_l=\sum_{n=0}^{d_l-1}[\p(t)]_{\pi^n(l)}$ and the index $l$ runs over representatives of cycles with $d_l$ denoting the corresponding cycle length, while
\begin{eqnarray}\label{eq:W_0}
	({\W\!}_0)_{kl}&=& \frac{1}{d_l} \sum_{n_k=0}^{d_k-1}\sum_{n_l=0}^{d_l-1} (\W)_{\pi^{{n\!}_k}\!(k) \,\pi^{{n\!}_l}\!(l)}.
\end{eqnarray}
Indeed, $\p_0(t)$ is a probability vector, while ${\W\!}_0$ preserves the positivity and the total probability [cf.~Eqs.~\eqref{eq:Wpos} and~\eqref{eq:Wtrace}]. In particular, the lifetime of a uniform mixture of system configurations in a given cycle is generally effectively extended, as microscopic transitions to other elements of the same cycle do not contribute to its effective dynamics, $({\W\!}_0)_{ll}=(\W)_{ll}+\sum_{n=1}^{d_l-1} (\W)_{\pi^n\!(l)\,l}\geq (\W)_{ll} $ [cf.~Eq.~\eqref{eq:Wtau}].
Analogously, when multiple symmetries are present, the dynamics of probabilities invariant under all corresponding permutations is also stochastic.


\subsection{Classical dynamics of average system state} \label{app:Leff}

We now show that the long-time dynamics of the average system state is effectively classical. We first derive the best continuous approximation of the generator of the long-time dynamics in Eq.~\eqref{eq:Wtilde} by classical stochastic dynamics between metastable phases in Eq.~\eqref{eq:W}. We then prove an upper bound Eq.~\eqref{eq:deltaW} on the distance between the two generators. We use this result to show the closeness of the effective dynamics to that generated by a classical stochastic generator in Eq.~\eqref{eq:etW_CL} and of the resulting stationary states in Eq.~\eqref{eq:pss_CL}.	Similarly, all timescales of the dynamics in the MM can be approximated by the classical stochastic dynamics. Here, we consider here an approximation of the pseudoinverse of the long-time dynamics generator. Finally, we also discuss the approximation of the long-time dynamics by discrete, rather than continuous, positive and trace-preserving dynamics, and prove results analogous  to Eqs.~\eqref{eq:deltaW},~\eqref{eq:etW_CL} and~\eqref{eq:pss_CL}.

\subsubsection{Best classical stochastic approximation of long-time dynamics generator} \label{app:Leff_approx}

We now prove that $\W$ defined in Eq.~\eqref{eq:W} is the closest classical stochastic generator to $\Wt$ in Eq.~\eqref{eq:Wtilde} with respect to the matrix norm induced by the L1 vector norm.
For any probability-conserving and positive generator $\W$, we have
\begin{equation}\label{eq:W_distance_bound0}
	\lVert \Wt-\W\rVert_1 \equiv \max_{1\leq l\leq m}\sum_{k=1}^m \left|(\Wt)_{kl}-(\W)_{kl}\right|
\end{equation}
while  [cf.~Eq.~\eqref{eq:p_distance_bound}]
\begin{eqnarray}\nonumber
	\sum_{k=1}^m \left|(\Wt)_{kl}-(\W)_{kl}\right| 
	&=& \sum_{\substack{k:\,  (\Wt)_{kl}< 0,\\k\neq l}} \left[-(\Wt)_{kl}+(\W)_{kl}\right]\\\nonumber
	&&+\sum_{\substack{k:\,  (\Wt)_{kl}\geq 0,\\\text{or}\,k= l}} \left|(\Wt)_{kl}-(\W)_{kl}\right|\\\nonumber
	&\geq&  \sum_{\substack{k:\,  (\Wt)_{kl}< 0,\\k\neq l}} \left[-(\Wt)_{kl}+(\W)_{kl}\right]\\\nonumber &&+\Bigg|\sum_{\substack{k:\,  (\Wt)_{kl}\geq 0,\\\text{or}\,k= l}} \left[(\Wt)_{kl}-(\W)_{kl}\right]\Bigg|
	\\\nonumber
	&=& 2 \sum_{\substack{k:\,  (\Wt)_{kl}< 0,\\k\neq l}} \left[-(\Wt)_{kl}+(\W)_{kl}\right]\\
	&\geq& 2\sum_{k\neq l}  \max [-(\Wt)_{kl},0],\label{eq:W_distance_bound}
\end{eqnarray}
where in the second and last line we used the positivity of $(\W)_{kl}\geq 0 $, for $k\neq l$, the  third line follows from the triangle inequality, and the fourth line follows from the probability conservation in both generators $\sum_{k=1}^m (\Wt)_{kl}=0=\sum_{k=1}^m (\W)_{kl}$. Finally, for $\W$ defined in Eq.~\eqref{eq:W} the bound in Eq.~\eqref{eq:W_distance_bound} is saturated [cf.~Eq.~\eqref{eq:p_distance}]
\begin{equation}\label{eq:W_distance}
	\lVert\Wt-\W\rVert_1\equiv \max_{1\leq l\leq m}2\sum_{k\neq l}  \max [-(\Wt)_{kl},0] .
\end{equation}

\subsubsection{Derivation of Eq.~\eqref{eq:deltaW} in the main text}\label{app:Leff_delta}

The vector $\pt$ of barycentric coordinates between metastable phases evolves as
\begin{equation}\label{eq:ptW}
	\pt(t)= e^{t\Wt} \pt(0)=  \pt(0) + t \Wt\,\pt(0) + \frac{t^2}{2!} \Wt^2\,\pt(0)+...,
\end{equation}
[cf.~Eqs.~\eqref{Expansion1},~\eqref{Expansion} and~\eqref{eq:ptilde}]. As the corresponding state belongs to the MM, at any time $t$ its distance from the simplex of metastable phases is bounded by $\Ccl$ in~Eq.~\eqref{eq:Ccl}
\begin{equation}\label{eq:ptC}
	\lVert \pt(t)\rVert_1-1 =2 \sum_{l=1}^m \max [-\tilde p_{l}(t),0]\leq\Ccl.
\end{equation}	
Let us consider time $t$ such that  
\begin{equation}\label{eq:tW}
	t \lVert  \Wt\rVert =c\, \sqrt{\Ccl}, 
\end{equation}
where $c$ is a constant, so that $c\, \sqrt{\Ccl} \ll 1$. In this case, the dynamics in  Eq.~\eqref{eq:ptW} can be approximated to the linear order with corrections
\begin{equation}\label{eq:tnorm}
	\left\lVert \pt(t)-\pt(0)- t \Wt\,\pt(0)\right\rVert_1 \lesssim \frac{c^2}{2}\,\Ccl. 
\end{equation}
In particular, for the system initially in $\rho_l$ that projects onto $\tilde\rho_l$ in Eq.~\eqref{eq:rhotilde}, i.e., $\tilde{p}_k(0)=\delta_{kl}$, from Eq.~\eqref{eq:tnorm} we have
\begin{equation}\label{eq:tnorm_k}
	\left|\tilde{p}_l(t)-1-t (\Wt)_{ll}\right|+ \sum_{k\neq l} \left|\tilde{p}_k(t)- t (\Wt)_{kl} \right| \lesssim \frac{c^2}{2}\,\Ccl.
\end{equation}
Since from the definition of the absolute value 
\begin{equation}
	-t(\Wt)_{kl}\leq -\tilde{p}_k(t)+ \left|\tilde{p}_k(t)- t (\Wt)_{kl} \right|	
\end{equation}		
and, from the monotonicity of $ \max(x,0)$  in $x$, 
\begin{equation}
	t\max [-(\Wt)_{kl},0]\leq  \max [-\tilde{p}_k(t),0]+\left|\tilde{p}_k(t)- t (\Wt)_{kl} \right|,
\end{equation}	 
we arrive at [cf.~Eqs.~\eqref{eq:ptC} and~\eqref{eq:tnorm_k}]
\begin{eqnarray}\label{CtW}
	t\sum_{k\neq l} \max [-(\Wt)_{kl},0]&\leq& \sum_{k\neq l} \max [-\tilde{p}_k(t),0] \\\nonumber&&+ \sum_{k\neq l} \left|\tilde{p}_k(t)- t (\Wt)_{kl} \right| \\\nonumber&\lesssim& \frac{1+c^2}{2}\,\Ccl. 
\end{eqnarray}
For the classical stochastic generator $\W$ defined in Eq.~\eqref{eq:W}, we have from Eq.~\eqref{eq:tW} that
\begin{equation}
	t \lVert\Wt-\W\rVert_1\equiv t\,\max_{1\leq l\leq m}2\sum_{k\neq l}  \max [-(\Wt)_{kl},0]  \lesssim (1+c^2)\,\Ccl,
\end{equation}
so that $\W$ approximates well $\Wt$ 
\begin{eqnarray}\nonumber
	\frac{\lVert\Wt-\W\rVert_1}{\lVert\Wt\rVert_1} =\frac{t\lVert\Wt-\W\rVert_1}{t\lVert\Wt\rVert_1}  
	& \lesssim& \frac{1+c^2}{c}\,\sqrt{\Ccl} , 
\end{eqnarray}
and for the (optimal) choice $c=1$, we obtain Eq.~\eqref{eq:deltaW}.

\subsubsection{Derivation of Eq.~\eqref{eq:etW_CL}  in the main text}\label{app:Leff_etW}

We have
\begin{eqnarray}
	e^{t\Wt}-e^{t \W}&=& \int_{0}^t d t_1\, \frac{d } {d t_1} e^{(t-t_1)\W} e^{t_1 \Wt}\\\nonumber
	&=&\int_{0}^t d t_1\,e^{(t-t_1)\W} (\Wt-\W)e^{t_1 \Wt},
\end{eqnarray}
which gives
\begin{eqnarray}
	\lVert e^{t\Wt} -e^{t \W} \rVert_1 &\leq& \int_{0}^td t_1\,\lVert e^{(t-t_1)\W}\rVert_1  \lVert \Wt-\W\rVert_1 \lVert e^{t_1 \Wt}\rVert_1\nonumber\\
	&\leq& t   \lVert \Wt-\W\rVert_1 (1+\Ccl) \nonumber\\
	&\lesssim& 2\sqrt{\Ccl} \,t\,\lVert \Wt\rVert_1.
\end{eqnarray}
In the second inequality we used the fact that $\W$ generates positive trace preserving dynamics and thus $\lVert e^{t \W}\rVert_1\nonumber=1$, while $\Wt$ transforms any probability vectors not further away than $\Ccl$ from the simplex [cf.~Eq.~\eqref{eq:Ccl}], and thus $\lVert e^{t \Wt}\rVert_1\nonumber\leq 1+\Ccl$. The last inequality follows from Eq.~\eqref{eq:deltaW}.

\subsubsection{Derivation of Eq.~\eqref{eq:pss_CL}  in the main text}\label{app:Leff_pss}
For the stationary states of the open quantum dynamics described within the MM by $\ptss$  and the stationary state $\pss$ of the classical stochastic dynamics $\W$, we have 
\begin{eqnarray}\label{eq:pss_CL_proof}
	\lVert\ptss-\pss \rVert_1 &\leq& \lVert \ptss- e^{t\Wt} \pss\rVert_1	+\lVert e^{t\Wt} \pss-\pss\rVert_1\nonumber\\
	&\leq& \lVert e^{t\Wt} -\Ptss \rVert_1+\lVert e^{t\Wt} -e^{t \W} \rVert_1 \nonumber\\
	&\lesssim& \lVert  e^{t\Wt} -\Ptss\rVert_1+ 2\sqrt{\Ccl} \,t\,\lVert \Wt\rVert_1.\qquad
\end{eqnarray}
In the first line we used the triangle inequality. In the second line we used that by definition of the projection $\Ptss$  on the stationary probability distribution  $\Ptss\pss= \ptss$, and further exploited $e^{t\W} \pss=\pss$ and the definition of the induced norm. The last inequality follows from Eq.~\eqref{eq:etW_CL}. Note that we did not assume uniqueness of $\pss$. Therefore, when $t$ can be chosen so that $ \lVert e^{t\Wt}-\Ptss  \rVert_1\ll 1$, while $ t\lVert \Wt\rVert_1\ll 1/\sqrt{\Ccl}$, we obtain $\lVert\ptss-\pss \rVert_1\ll 1$, but this requires 
\begin{equation}\label{eq:cond2_CL}
	\tau \lVert \Wt\rVert_1\leq \tilde{\tau} \lVert \Wt\rVert_1 \ll\frac{1}{\sqrt{\Ccl}},
\end{equation} 
where $\tilde{\tau}$ is the relaxation time with respect to $\lVert e^{t\Wt}-\Ptss \rVert_1$ rather than $\lVert e^{t\L}-\P_\text{ss}\rVert$  ($\tilde{\tau}\approx\tau$ when $\Ctcl\ll 1$; cf.~Sec.~\ref{app:norm}).

We note that a related result can be obtained in the non-Hermitian perturbation theory~\cite{Kato1995}, where in the first order
\begin{equation}\label{eq:pss_CL_PT}
	\ptss-\pss=\Rt (\W-\Wt)\ptss+...,
\end{equation}
where $\Rt$ is the pseudoinverse of $\Wt$,  $\Rt \Wt= \Wt\Rt = \Ibf-\Ptss $, with $\Ibf$ being the identity matrix (see also Sec.~\ref{app:classical}). Therefore, the first-order corrections can be bounded as [cf.~Eq.~\eqref{eq:deltaW}]
\begin{eqnarray}
	\lVert\ptss-\pss \rVert_1 &\leq& \lVert\Rt \rVert_1 \lVert\W-\Wt \rVert_1\lVert\ptss \rVert_1 +...\\\nonumber
	&\lesssim & 4 t \lVert\Wt \rVert_1 \sqrt{\Ccl} +....
\end{eqnarray}
For the estimate on $\lVert\Rt \rVert_1\lesssim 2 t$, where $t\geq \tilde{\tau}$ is such that $\lVert e^{t\Wt} -\Ptss \rVert_1\ll 1$,  see Eq.~\eqref{eq:Rnorm} below. Therefore, when the right-hand-side of Eq.~\eqref{eq:pss_CL} is negligible, so are  the first-order corrections in Eq.~\eqref{eq:pss_CL_PT}.

\subsubsection{Approximation of dynamics resolvent}\label{app:Leff_R}

\emph{Resolvent}. We now consider the resolvent $\Rt$ of the dynamics $\Wt$ at $0$, i.e., the pseudoinverse of $\Wt$,  $\Rt \Wt= \Wt\Rt = \Ibf-\Ptss $, where $\Ibf$ is the identity matrix (see also Sec.~\ref{app:classical}). In Eqs.~\eqref{eq:DeltaR} and~\eqref{eq:cond2_CL} below, we show it can be approximated by the resolvent $\R$ of the classical dynamics $\W$ in Eq.~\eqref{eq:W}.

We have
\begin{equation}
	\Rt= -\int_{0}^\infty d t_1\left(e^{t_1\Wt}-\Ptss\right), \label{eq:R_CL}
\end{equation}
and thus 
\begin{eqnarray}
	\left\lVert\Rt\,+ \int_{0}^t d t_1\left(e^{t_1\Wt}-\Ptss\right)\right\rVert_1\!\!&\leq&\left\lVert\int_{t}^\infty d t_1\left(e^{t_1\Wt}-\Ptss\right)\right\rVert_1\,\,\nonumber\\
	&\leq& \lVert e^{t \Wt}-\Ptss\rVert_1\lVert\Rt\rVert_1, \qquad\,\,\,\label{eq:R1_CL}
\end{eqnarray}\\
where in the second inequality we used the fact $e^{t \Wt}\Ptss=\Ptss e^{t \Wt}=\Ptss$ so that $(e^{t \Wt}-\Ptss)\int_{0}^\infty d t_1(e^{t_1\Wt}-\Ptss)=\int_{t}^\infty d t_1(e^{t_1\Wt}-\Ptss)$. Equation~\eqref{eq:R1_CL} holds analogously for the dynamics with $\W$. Furthermore, from Eqs.~\eqref{eq:etW_CL} and~\eqref{eq:pss_CL} we have
\begin{eqnarray}\label{eq:R2_CL}
	&&\left\lVert\int_{0}^t d t_1\left(e^{t_1\Wt}-\Ptss\right)-\int_{0}^t d t_1\left(e^{t_1\W}-\Pss\right)\right\rVert_1\\\nonumber
	&&\qquad\qquad\leq 	
	\int_{0}^t d t_1\lVert e^{t_1\W}- e^{t_1\W}\rVert_1 + t\lVert \Ptss- \Pss\rVert_1 , \\\nonumber
	&&\qquad\qquad\lesssim 3 \sqrt{\Ccl} t^2 \lVert\Wt\rVert_1 + t \lVert \Ptss-e^{t\Wt} \rVert_1.
\end{eqnarray}
The first inequality follows from the triangle inequality. In the last inequality we used the fact $\lVert \Ptss- \Pss\rVert_1 =\lVert \ptss-\pss\rVert_1$ and Eq.~\eqref{eq:pss_CL_proof}. 
Therefore, by applying the triangle inequality  we arrive at
\begin{eqnarray}
	&&\lVert \Rt-\R\rVert_1\\
	&&\leq \left\lVert\int_{0}^t d t_1\left(e^{t_1\Wt}\!\!-\Ptss\right)-\int_{0}^t d t_1\left(e^{t_1\W}\!\!-\Pss\right)\!\right\rVert_1\nonumber\\
	&&+ \left\lVert\Rt\,+\! \int_{0}^t \!\!d t_1\left(e^{t_1\Wt}\!\!-\Ptss\right)\right\rVert_1\!\!+\left\lVert\R\,+ \!\int_{0}^t\!\! d t_1\left(e^{t_1\W}\!\!-\Pss\right)\!\right\rVert_1\nonumber\\
	&&\lesssim
	\sqrt{\Ccl}\! \left(\!10 t^2 +4 t\lVert\Rt\rVert_1 \!\right)\! \lVert\Wt\rVert_1\! + 3(t +\lVert\Rt\rVert_1)\lVert e^{t\Wt}\!\! -\Ptss \rVert_1 \nonumber\\ \nonumber
	&&+  \left(2\lVert e^{t\Wt}\! -\Ptss \rVert_1 +4\sqrt{\Ccl} \,t\lVert \Wt\rVert_1\right) \lVert \Rt-\R\rVert_1. 
\end{eqnarray}
In the last inequality we used Eqs.~\eqref{eq:R1_CL} and~\eqref{eq:R2_CL} together with $ \lVert e^{t\W} -\Pss \rVert_1 \leq  \lVert e^{t\Wt} -\Ptss \rVert_1+\lVert e^{t\Wt} - e^{t\W} \rVert_1+ \lVert \Ptss- \Pss\rVert_1 \lesssim 2 (\lVert e^{t\Wt} -\Ptss \rVert_1 +2\sqrt{\Ccl} \,t\lVert \Wt\rVert_1)$ and $	\lVert \R\rVert_1 \leq	\lVert \Rt\rVert_1+	\lVert \Rt-\R\rVert_1$. Therefore, 
\begin{eqnarray}\label{eq:DeltaR}	
	&&\frac{\lVert \Rt-\R\rVert_1}{\lVert \Rt\rVert_1}\lesssim\\\nonumber
	&&\frac{
		\sqrt{\Ccl} (10 t^2/ \lVert \Rt\rVert_1\! + \!4 t ) \lVert\Wt\rVert_1\!+ 3(t /\lVert \Rt\rVert_1\! +\!1)\lVert e^{t\Wt} \!\!-\Ptss \rVert_1}{1-2 \lVert e^{t\Wt}\!\! -\Ptss \rVert_1 -4\sqrt{\Ccl} \,t\,\lVert \Wt\rVert_1 }.
\end{eqnarray}
The above inequality holds for any time $t$. When time $t$ can be chosen  as in the case of the discussion of Eq.~\eqref{eq:pss_CL}, that $t \lVert \Wt\rVert_1\ll 1/\sqrt{\Ccl}$ holds for $\lVert e^{t\Wt} -\Ptss \rVert_1 \ll 1 $, the leading corrections in the right-hand side of Eq.~\eqref{eq:DeltaR} are given by the numerator. In this case, the closeness of resolvents occurs when $t$ can be further chosen so that  $t\lVert e^{t\Wt} -\Ptss \rVert_1 / \lVert \Rt\rVert_1\ll 1 $ [note that $t^2 \lVert \Wt\rVert_1 /\rVert_1\lVert\Rt \rVert_1\ll 1/\sqrt{\Ccl}$  follows from $t\lVert \Wt\rVert_1 \ll 1/\sqrt{\Ccl}$ for $t\geq\lVert\Rt \rVert_1$, while we have $\lVert\Rt \rVert_1\lesssim 2 t$ for any $t$ such that $\lVert e^{t\Wt} -\Ptss \rVert_1\ll 1$; see Eq.~\eqref{eq:Rnorm}]. This, in turn, requires that the relaxation time
\begin{equation}\label{eq:cond2_CL}
	\tilde{\tau} \lVert \Wt\rVert_1 \ll\min\Bigg[\frac{1}{\sqrt{\Ccl}},  \sqrt{\frac{\lVert\Wt \rVert_1\lVert\Rt \rVert_1}{\sqrt{\Ccl}}}\Bigg].
\end{equation} 
In fact, this condition on the relaxation time will typically imply existence of time such that $t \lVert \Wt\rVert_1\ll 1/\sqrt{\Ccl}$ and  $t\lVert e^{t\Wt} -\Ptss \rVert_1 / \lVert \Rt\rVert_1\ll 1 $, as $n\tau \lVert e^{n\tau\Wt} -\Ptss \rVert_1 \leq  n\tau \lVert e^{\tau\Wt} -\Ptss \rVert_1^n$, which decays to $0$ as $n\rightarrow \infty$ (exponentially at larger $n$).
\\

\emph{Discussion of the results in the proximity to a first-order phase transition}. We note that both Eqs.~\eqref{eq:cond_CL} and~\eqref{eq:cond2_CL} are fulfilled for all perturbations away from a first-order phase transition in a finite-size system, provided that the degeneracy of stationary states is lifted in the same order.  Indeed, in this case $\lVert \Wt\rVert_1$, $1/\lVert \Rt\rVert_1$ and $1/\tilde{\tau}$ are of the same order in the perturbation, while $\Ccl$ is of the same or higher order  (see Sec.~\ref{app:PT1}).
Thus, $\tilde{\tau}\lVert \Wt\rVert_1 $ and $\lVert \Rt\rVert_1\lVert \Wt\rVert_1 $ are of zero-order, that is, they remain finite in the limit of the perturbation decreasing to $0$, while the upper limits in Eqs.~\eqref{eq:cond_CL} and~\eqref{eq:cond2_CL} diverge in this limit. In fact, in this case, time $t$ can be chosen so that  $\tilde{\tau} \lVert \Wt\rVert_1 \ll t  \lVert \Wt\rVert_1\ll \min[1/\sqrt{\Ccl},  \sqrt{\lVert\Wt \rVert_1\lVert\Rt \rVert_1/\sqrt{\Ccl}}]$ thus leading to 	$\lVert\ptss-\pss \rVert_1 \ll 1$ and $\lVert \Rt-\R\rVert_1/\lVert \Rt\rVert_1$ [cf.~Eq.~\eqref{eq:pss_CL} and~\eqref{eq:DeltaR}].\\

\emph{Bounds on the norm of the resolvent}. For the relaxation time $\tau$ defined 
as in Eq.~\eqref{eq:tau1}, we have that the resolvent $\S$ of $\L$ fulfills 
\begin{equation}
	\label{eq:Snorm}
	\lVert S\rVert\leq \frac{2\tau}{1- \lVert e^{\tau\L}-\P_\text{ss}\rVert}
	\lesssim 2t,
\end{equation}
where in the second inequality we consider $t \geq \tau $ for which $\lVert e^{t\L} -\P_\text{ss} \rVert\ll 1$.
This follows from $\S= -\int_0^\infty d t(e^{t\L}-\P_\text{ss})$ [cf.~Eq.~\eqref{eq:R_CL}], and thus $(\I-e^{t\L})\S=-\int_0^t d t_1(e^{t_1\L}-\P_\text{ss})$. Indeed, we have
$(1- \lVert e^{\tau\L}-\P_\text{ss}\rVert)\lVert\S\rVert \leq (1- \lVert e^{t\L}-\P_\text{ss}\rVert)\lVert\S\rVert \leq \lVert\S\rVert-\lVert e^{t\L}\S\rVert \leq \lVert(\I-e^{t\L})\S\rVert=\lVert\int_0^t (e^{t\L}-\P_\text{ss})\rVert\leq 2 t$. Analogously, we have 
\begin{equation}\label{eq:Rnorm}
	\lVert \Rt\rVert\lesssim \frac{2 \tilde{\tau}}{1-2\lVert e^{\tilde{\tau}\Wt} -\Ptss \rVert_1}\lesssim  2t,
\end{equation}
where the second inequality holds for $t\geq\tilde{\tau}$  such that $\lVert e^{t\Wt} -\Ptss \rVert_1\ll 1$ (cf.~Sec.~\ref{app:norm}).

The bound in Eq.~\eqref{eq:Snorm} and~\eqref{eq:Rnorm} also hold when the relaxation times $\tau$ and $\tau''$ are instead defined as the rate of exponential bound on the approach to the stationary state and the MM, i.e., by Eqs.~\eqref{eq:tau2} and~\eqref{eq:tau''2}. But in this case we also have [cf.~Eq.~\eqref{eq:R_CL}]
\begin{eqnarray}\label{eq:Snorm2}
	\lVert S \rVert\leq  \int_{0}^\infty d t_1\ \lVert e^{t_1\L}-\P_\text{ss}\rVert=\tau.\qquad
\end{eqnarray}
Similarly, within the MM [see Eq.~\eqref{eq:tau1_CL}]
\begin{equation}\label{eq:tau2_CL}
	\lVert e^{t\Wt} -\Ptss \rVert_1 \lesssim \Big(1+\frac{\Ctcl}{2}\Big) e^{-\frac{t}{\tau}},
\end{equation}
which leads to the norm of the resolvent bounded as 
\begin{eqnarray}\label{eq:Rnorm2}
	\lVert\Rt \rVert_1\leq  \int_{0}^\infty d t_1\ \lVert e^{t_1\Wt}-\Ptss\rVert_1\lesssim    \Big(1+\frac{\Ctcl}{2}\Big) \tau.\qquad
\end{eqnarray}
Therefore, when $\Ctcl\ll1$, for example, if $m$ does not scale with $\Ccl$ so that $m\Ccl\ll 1$ (e.g., $m$ is independent of the system size), the norm of resolvent is bounded by the relaxation time.
Furthermore, the condition in Eq.~\eqref{eq:cond_CL} is implied by the condition in Eq.~\eqref{eq:cond2_CL}. 

Finally, we have lower bounds in terms of the master operator spectrum [cf.~Eq.~\eqref{eq:tau'3_lambda}]
\begin{equation}
	\label{eq:Snorm3}
	\lVert S\rVert\geq \max_{2\leq k\leq D^2}\frac{1}{|\lambda_k|}
\end{equation}
and, analogously, 
\begin{equation}
	\label{eq:Rnorm3}
	\lVert\Rt \rVert_1\gtrsim \max_{2\leq k\leq m}\frac{1-\Cp-\Ccl}{|\lambda_k|}.
\end{equation}
as   $ \lVert \P\S\rVert \leq (1+\Cp)(1+\Ccl)\lVert\Rt \rVert_1$ [cf.~Eq.~\eqref{eq:Leff2_CL}].

\subsubsection{Classical discrete approximation of long-time dynamics} \label{app:Leff_discrete}

Below, we show how dynamics $e^{t\Wt}$ at any time $t$ can be approximated by classical discrete dynamics $\T_t$ [see Eqs.~\eqref{eq:Tt} and~\eqref{eq:Tt2} for the definition] as  
\begin{equation}\label{eq:T_CL}
	\lVert e^{t\Wt}- \T_t \rVert_1\leq \Ccl
\end{equation} 
[cf.~Eq.~\eqref{eq:deltaW}].
At later times $nt$, where $n$ is an integer, we then have the approximation [cf.~Eq.~\eqref{eq:etW_CL}]
\begin{equation}\label{eq:Tn_CL}
	\lVert e^{n t\Wt}- \T_t^n \rVert_1\lesssim n \Ccl.
\end{equation} 
For $t$ chosen long enough, so $\lVert \Ptss -e^{nt\Wt} \rVert_1\ll 1$ for an integer $n\ll 1/\Ccl$, which requires [cf.~Eq.~\eqref{eq:cond_CL}]
\begin{equation}\label{eq:cond_T_CL}
	\frac{t}{\tau}\geq\frac{t}{\tilde{\tau}}\gg \Ccl,
\end{equation} 
the stationary state $\rhoss$ described within the MM by $(\ptss)_{k}=\Tr(\tilde{P}_k \rhoss)$ is well approximated by the stationary probability $\pss$ of the classical discrete dynamics $\T_t$ [cf.~Eq.~\eqref{eq:pss_CL}] as
\begin{equation}\label{eq:pss_T_CL}	 
	\lVert\ptss-\pss \rVert_1 \lesssim \lVert \Ptss -e^{nt\Wt} \rVert_1+ n\Ccl.
\end{equation}\\

\emph{Derivation} of Eq.~\eqref{eq:T_CL}. This result follows from the fact that the dynamics must map any metastable state to another state within the MM. 

The closest positive trace preserving dynamics to $e^{t\Wt}$ is as follows.  Let $\Delta_l\equiv\sum_{k=1}^m \max[-(e^{t\Wt})_{kl},0]$, $l=1,...,m$. We then define [cf.~Eq.~\eqref{eq:p_distance_opt}]
\begin{equation}\label{eq:Tt}
	({\T_t})_{kl}\equiv0\quad \text{if}\quad (e^{t\Wt})_{kl}\leq 0,
\end{equation} 
and otherwise, when $(e^{t\Wt})_{kl}> 0$,
\begin{eqnarray}\label{eq:Tt2}
	({\T_t})_{kl}&\equiv& (e^{t\Wt})_{kl}\\\nonumber&&- \min\!\Big[(e^{t\Wt})_{kl},\Delta_l -\!\!\!\!\!\!\!\!\!\sum_{\substack{n<k:\\(e^{t\Wt})_{nl}>0}} \!\!\!\!\!\!\!\big[(e^{t\Wt})_{nl} -({\T_t})_{nl}\big]\Big],
\end{eqnarray} 
$k=1,...,m$. We then have [cf.~Eq.~\eqref{eq:p_distance}]
\begin{equation}
	\lVert e^{t\Wt}- \T_t \rVert_1=2\max_{1\leq l\leq m} \Delta_l=\lVert e^{t\Wt}\rVert_1-1\leq \Ccl,
\end{equation}
where in the last inequality we used $\lVert  e^{ t \Wt} \rVert_1\leq 1+\Ccl$, as the long-time dynamics  transforms the MM onto itself. This gives Eq.~\eqref{eq:T_CL}.

Finally, $\T_t$ defined in Eqs.~\eqref{eq:Tt} and~\eqref{eq:Tt2} is optimal, as for any discrete trace preserving dynamics $\T$ we have $\lVert \T\rVert_1=1$ and thus from triangle inequality
\begin{equation}
	\lVert e^{t\Wt}- \T \rVert_1\geq \lVert e^{t\Wt}\rVert_1 -\lVert \T \rVert_1 =\lVert e^{t\Wt}\rVert_1 -1.
\end{equation}
\\

\emph{Derivation} of Eq.~\eqref{eq:Tn_CL}. We have
\begin{eqnarray}\nonumber
	\lVert e^{n t\Wt}- \T_t^n \rVert_1&=&\left \lVert \sum_{k=1}^{n} e^{(k-1) t\Wt} (e^{t\Wt}- \T_t)\T_t^{n-k} \right\rVert_1 \\\nonumber
	&\leq& \lVert e^{t\Wt}- \T_t\rVert_1 \!\sum_{k=1}^{n} \lVert  e^{ (k-1)t \Wt} \rVert_1\lVert\T_t \rVert_1^{n-k}\\
	&\leq&  n \Ccl \left[1+\Ccl\right]\lesssim n \Ccl,
	\label{eq:Tn_CL_proof}
\end{eqnarray}
where in the second inequality we used Eq.~\eqref{eq:T_CL}, the fact that $\lVert\T_t \rVert_1=1$ from the positivity and the trace-preservation, and $\lVert  e^{ kt\Wt} \rVert_1\leq 1+\Ccl$ as $e^{ kt \L}$ transforms the MM onto itself.\\

\emph{Derivation} of Eq.~\eqref{eq:pss_T_CL}. We have
\begin{eqnarray}\label{eq:pss_T_CL_proof}
	\lVert\ptss-\pss \rVert_1 &\leq& \lVert \ptss- e^{nt\Wt} \pss\rVert_1	+\lVert e^{nt\Wt} \pss-\pss\rVert_1\nonumber\\
	&\leq& \lVert \Ptss -e^{nt\Wt} \rVert_1+\lVert e^{nt\Wt} -\T_t^n \rVert_1 \nonumber\\
	&\lesssim& \lVert \Ptss -e^{nt\Wt} \rVert_1+ n\Ccl.\qquad
\end{eqnarray}
In the second line we used that the projection $\Ptss$  on the stationary probability distribution  $\Ptss\pss= \ptss$, and further exploited $\T_t \pss=\pss$ and the definition of the induced norm. The last inequality follows from Eq.~\eqref{eq:Tn_CL}. We did not assume uniqueness of $\pss$.

\subsection{Classical statistics of quantum trajectories}\label{app:Ls}

Here, we first consider generalized statistics of jump number where individual jumps are taken into account (Sec.~\ref{app:Lsj}). Second, we discuss classical approximations of the statistics for a homodyne measurement~\cite{Hickey2012} (Sec.~\ref{app:homodyne}) and for time-integral of system-observables~\cite{Levitov1996,Nazarov2003,Esposito2009,Flindt2009,Hickey2012,Hickey2013} (Sec.~\ref{app:time_int}) in the presence of classical metastability. Third,	we derive corrections to the approximation of generators of  statistics of jump number [Eq.~\eqref{eq:Wstilde2}],  time-integral of homodyne current [Eq.~\eqref{eq:Wrtilde2}] and system observables  [Eq.~\eqref{eq:Whtilde2}] (Sec.~\ref{app:Ws}). We then prove the relation of the statistics of jumps and system-observables in quantum trajectories to statistics of classical dynamics for finite (Secs.~\ref{app:var_t} and~\ref{app:var_ms}) and asymptotic time (Sec.~\ref{app:var_ss}). Finally, we discuss multimodal distributions of continuous measurement results during the metastable regime (Sec.~\ref{app:distribution}). For simplicity, we denote here an initial state of the system as $\rho$ instead of $\rho(0)$ and omit parentheses in the action of superoperators on system states, e.g., $\rho(t)=e^{t\L}\rho$ instead of $e^{t\L}[\rho(0)]$.

\subsubsection{Activity in quantum trajectories}\label{app:Lsj}

\emph{Total activity}. In Sec.~\ref{sec:QTraj}  of the main text, we discussed the approximation in Eq.~\eqref{eq:Wstilde2} of the operator $\L_s$ in Eq.~\eqref{eq:Ls} by the classical operator $\W_s$ in Eq.~\eqref{eq:Ws}. This led to the approximation of the maximal eigenvalue $\theta(s)$ of $\L_s$ eigenmode and the corresponding eigenmode   $\rhoss(s)$ by the maximal eigenvalue and the corresponding eigenmode $\W_s$. Here, we discuss the corrections for those approximations.

We consider $\L\P+(e^{-s}-1)\J$ as a perturbation of $\L\Q$, by means of the degenerate non-Hermitian perturbation theory~\cite{Kato1995}, where $\Q\equiv\I-\P$ is the projection on the fast modes of $\L$  [cf.~Eq.~\eqref{Expansion}], with a further (higher order) approximation of $\L\P+(e^{-s}-1)\P\J\P=\P\L_s\P$ by $\mathcal{W}_s$, which is the superoperator corresponding to $\W_s$. The degeneracy of zero-eigenspace $\P$ of $\Q\L$ is lifted by $\mathcal{W}_s$ with the maximal eigenvalue $\theta^{(1)}\!(s)$ [cf.~Eq.~\eqref{eq:theta_s_2}] corresponding to the eigenmode $\rho_\text{ss}^{(0)}\!(s)\equiv \sum_{l=1}^m [\pss(s)]_l  \,\trho_l$ [cf.~Eq.~\eqref{eq:rhoss_s}]. The higher order corrections are given by $\theta^{(2)}\!(s)\equiv\Tr\{L_1^{(0)}\!(s) [-(e^{-s}-1)^2 \J\S\Q\J+(\L_s-\mathcal{W}_s)][\rho_\text{ss}^{(0)}\!(s)]\}$ and $\rho_\text{ss}^{(1)}\!(s)\equiv-(e^{-s}-1) \S\Q \J [\rho_\text{ss}^{(0)}\!(s)]+(e^{-s}-1)^2[\mathcal{W}_s-\theta^{(1)}\!(s)\P]^{+} \J\S\Q\J [\rho_\text{ss}^{(0)}\!(s)] -[\mathcal{W}_s-\theta^{(1)}\!(s)\P]^{+}(\L_s-\mathcal{W}_s) [\rho_\text{ss}^{(0)}\!(s)]$, where the first and the second terms correspond to the corrections from the fast modes [cf.~Eq.~\eqref{eq:Wstilde}], while the third term originate from the approximation in Eq.~\eqref{eq:Wstilde2}. 
Here, $^+$ denotes the pseudoinverse and $\S=\L^+$, while $L_1^{(0)}\!(s)$ is the dual eigenmatrix to   $\rho_\text{ss}^{(0)}\!(s) $.  The normalization $\Tr[\rhoss(s)]=1$ can be further achieved by additional correction $-\Tr[\rho_\text{ss}^{(1)}\!(s)] \rho_\text{ss}^{(0)}\!(s)$.  Analogously, other $m-1$ low-lying modes of $\L_s$ can be approximated by eigenmodes of $\W_s$.	Finally, we note that the corrections can be adjusted to include nonperturbative change of fast modes, by considering perturbation of $\L\Q+(e^{-s}-1)\Q\J\Q=\Q\L_s\Q$ instead of $\L\Q$, which leads to replacing $\S\Q$ by $[\L\Q +(e^{-s}-1)\Q\J\Q]^+=\S\Q +(e^{-s}-1)\S\J\S+...$ in the above corrections, but this change will only contribute in a higher order than already considered.

Similarly, one can consider approximating the first and the second derivatives of $\theta(s)$, using the fact that $\theta(s)=(e^{-s}-1)\Tr[\J \rhoss(s)]$ and thus $\theta'(s)=-e^{-s}\Tr[\J \rhoss(s)]+(e^{-s}-1)\Tr[\J \rhoss'(s)]$ and $\theta''(s)=e^{-s}\Tr[\J \rhoss(s)]-2e^{-s}(e^{-s}-1)\Tr[\J \rhoss'(s)]+(e^{-s}-1)\Tr[\J \rhoss''(s)]$. In particular, $\theta'(s)=-e^{-s}\Tr[\J \rhoss(s)]+(e^{-s}-1)e^{-s}\Tr[\J\S_s\J \rhoss(s)]-(e^{-s}-1)e^{-s}\Tr[\S_s\J \rhoss(s)]\Tr[\J \rhoss(s)]$, where $\S_s$ is the resolvent of $\L_s$ at $\theta(s)$, while the last term follows from the chosen normalization $\Tr[\rhoss(s)]=1$. We can then approximate $\theta'(s)=-e^{-s} \sum_{l=1}^m[\bm{\mut} \p_\text{ss}(s)]_{l}+(e^{-s}-1)e^{-s} \sum_{l=1}^m[(\Jbf+\bm{\mut^\text{in}})\R_s(\Jbf+\bm{\mut^\text{in}}) \p_\text{ss}]_l-(e^{-s}-1)e^{-s}\sum_{k,l=1}^m [\R_s(\Jbf+\bm{\mut^\text{in}}) \p_\text{ss}(s)]_{k}[\bm{\mut} \p_\text{ss}(s)]_{l}-(e^{-s}-1)e^{-s}\sum_{l=1}^m[\bm{\delta_{\tilde{\sigma}^2}} \p_\text{ss}(s)]_{l}/2+...$, where $\R_s$ is the resolvent  of $\W_s$ at $\theta^{(1)}(s)$. The last term corresponds to the non-Poissonian contribution from fluctuations in the metastable phases [cf.~Eq.~\eqref{eq:Kvar_t_approx_CL}], while the rest is the first-derivative of the classical cumulant generating function for the total activity, i.e., the first derivative of the maximal eigenvalue $\theta^{(1)}(s)$ of $\W_s$. 
\\

\emph{Local activity}. The joint statistics of the numbers of individual jumps is encoded by [cf.~Eq.~\eqref{eq:Ls}]
\begin{equation}\label{eq:Lsj}
	\L_{\mathbf{s}}(\rho) \equiv -i[H,\rho]+\!\sum_{j}\!\left(\!{e}^{-s_j}\!{J}_{j}\rho{J}_{j}^{\dagger}-\frac{1}{2}\{{J}_{j}^{\dagger}{J}_{j},\rho\}\!\right)\!,
\end{equation}
where $(\mathbf{s})_j={s}_{j}$ encodes bias values for individual jumps. 

In the presence of classical metastability and with internal activity dominating the long-time dynamics,  the projection of $\L_{\mathbf{s}}$ on the low-lying modes [cf.~Eq.~\eqref{eq:Wstilde}]
\begin{equation}\label{eq:Wsjtilde}
	(\Wt_{\mathbf{s}})_{kl}\equiv \Tr[\tilde{P}_k \L_{\mathbf{s}} (\trho_l)],
\end{equation}
$k,l=1,...,m$, can be approximated as  [cf.~Eq.~\eqref{eq:Wstilde3}]
\begin{equation}\label{eq:Wsjtilde2}
	(\Wt_{\mathbf{s}})_{kl}=\W-  \sum_j h_{s_j}\bm{\mut_j}+...\equiv \W_\mathbf{h_{s}}+...,
\end{equation}
where $(\mathbf{h_{s}})_j\equiv1-{e}^{-s_j}$ and the individual activities
\begin{equation}
	\left(\bm{\mut_j}\right)_{kl}=\Tr\left(J_j^\dagger J_j\,\trho_l\right),
\end{equation}
$k,l=1,...,m$. The corrections in Eq.~\eqref{eq:Wsjtilde2} [and thus also  Eq.~\eqref{eq:Wstilde3}] are  bounded in the leading order  by $2\lVert \Wt \rVert_1 \sqrt{\Ccl} +\max_{j}|e^{-s_j}-1| [\lVert \Wt \rVert_1+  m\,\lVert H +i \sum_{j} {J}_{j}^{\dagger} J_{j}/2\rVert_{\max}  \sqrt{2\Ccl+4\Cp}]$~\footnote{$H$ can be further replaced by $H-c\mathds{1}$, with $c$ being a real constant and the norm minimized with respect to $c$ in the corrections; see Sec.~\ref{app:Ws}.}. For the derivation, see Eq.~\eqref{eq:deltaW} and Sec.~\ref{app:Ws}. 

Therefore, $\theta(\mathbf{s})$ is approximated by the maximal eigenvalue of $\W_\mathbf{h_{s}}$ in Eq.~\eqref{eq:Wsjtilde}, while the corresponding density matrix   
\begin{equation}\label{eq:rhoss_hsj}
	\rhoss(\mathbf{s})=\sum_{l=1}^m [\pss(\mathbf{h_{s}})]_l  \,\trho_l+...\equiv \rho_\text{ss}^{(0)}(\mathbf{s})+...,
\end{equation}
where $\pss(\mathbf{h_{s}})$ is the maximal eigenmode of $\W_\mathbf{h_{s}}$.
To obtain the relevant corrections, we consider the degenerate non-Hermitian perturbation theory with $\L\P+\sum_j(e^{-s_j}-1)\J_j$ as a perturbation of $\L\Q$, and $\L\P+\sum_j(e^{-s_j}-1)\P\J_j\P$ further approximated by $\mathcal{W}_{\mathbf{h_{s}}}$, where $\J_j(\rho)\equiv J_j\rho J_j^\dagger$, and $\mathcal{W}_{\mathbf{h_{s}}}$   denotes the superoperator corresponding to $\W_\mathbf{h_{s}}$. Thus, the corrections in Eq.~\eqref{eq:rhoss_hsj} are given by $\rho_\text{ss}^{(1)}\!(\mathbf{s})\equiv \S\Q(\sum_j h_{s_j}\J_j) [\rho_\text{ss}^{(0)}\!(\mathbf{h_{s}})]+[\mathcal{W}_{\mathbf{h_{s}}}-\theta^{(1)}\!(\mathbf{h_{s}})\P]^{+}  (\sum_j h_{s_j}\J_j) \S\Q(\sum_j h_{s_j}\J_j)[\rho_\text{ss}^{(0)}\!(\mathbf{h_{s}})] -[\mathcal{W}_{\mathbf{h_{s}}}-\theta^{(1)}\!(\mathbf{h_{s}})\P]^{+}(\L_\mathbf{s}-\mathcal{W}_{\mathbf{h_{s}}}) [\rho_\text{ss}^{(0)}\!(\mathbf{h_{s}})]$ and  the normalization $\Tr[\rhoss(s)]=1$ can be achieved by additional correction $-\Tr[\rho_\text{ss}^{(1)}\!(s)] \rho_\text{ss}^{(0)}\!(s)$. 
Here, $\theta^{(1)}\!(\mathbf{h_{s}})$ denotes the maximal eigenvalue of $\W_\mathbf{h_{s}}$, which approximates $\theta(\mathbf{s})$, with the higher order correction given by $\theta^{(2)}\!(\mathbf{h_{s}})\equiv\Tr\{L_1^{(0)}\!(\mathbf{h_s})[-(\sum_j h_{s_j}\J_j) \S\Q(\sum_j h_{s_j}\J_j)+(\L_\mathbf{s}-\mathcal{W}_\mathbf{h_s})][\rho_\text{ss}^{(0)}\!(\mathbf{h_{s}})] \}$, where $L_1^{(0)}\!(\mathbf{h_{s}})$ is the dual eigenmatrix to  $\rho_\text{ss}^{(0)}\!(\mathbf{h_{s}})$ in Eq.~\eqref{eq:rhoss_hsj}.  Analogously, other $m-1$ low-lying modes of $\L_\mathbf{s}$ can be approximated by eigenmodes of $\W_\mathbf{h_{s}}$.\\

\emph{Metastability and dynamical phase transitions in activity}. For the bias $\lVert\mathbf{s}\rVert_1$ large enough [but small with respect to the gap to fast modes of the dynamics $\lambda_m^R -\lambda^R_{m+1}$], the contribution from $\W$ in $\W_\mathbf{h_s}$ of Eq.~\eqref{eq:Wsjtilde2} can be neglected. In that case, $m$ low-lying eigenmodes and eigenvalues of $\L_s$ are approximated as metastable phases $\trho_l$ in Eq.~\eqref{eq:rhotilde}.

In particular, for a \emph{homogeneous bias}, $s_j=s$,  and approximation in Eq.~\eqref{eq:Wstilde3} we  have the correction $h_s \S\Q \J (\trho_l)-h_s (\tilde{\M}^\text{in}-\mut_l^\text{in} \P)^{+}\J\S\Q \J (\trho_l) +(\tilde{\M}^\text{in}-\mut_l^\text{in} \P)^{+}(\L_s+h_s\tilde{\M}^\text{in})  (\trho_l)/h_s$ for the eigenmode $R_l(s)= \trho_l+...$, the correction $-h_s^2\Tr[ \tilde{P}_l\J\S\Q \J (\trho_l)]  +(\Wt)_{ll}-h_s(\Jt-\bm{\mut}^{\text{in}})_{ll}$ to the corresponding eigenvalue $\theta_l(s)=(e^{-s}-1)\mut_l+...$, and the correction $e^{-s}\{-2h_s \Tr[ \tilde{P}_l\J\S\Q \J (\trho_l)]  -h_s(\Jt-\bm{\mut}^{\text{in}})_{ll}- \sum_k (\Wt)_{lk}(\Wt)_{kl}/(\mut_k^{\text{in}}-\mut_l^{\text{in}}) /h_s^2\}$  to its derivative, $-k_l(s)=-e^{-s}\mut_l+...$.  Here, $\tilde{\M}^\text{in}$ is the superoperator corresponding to $\bm{\mut}^\text{in}$.


\subsubsection{Homodyne current in quantum trajectories}\label{app:homodyne}

We now discuss the relation of the homodyne measurement statistics to the classical dynamics between metastable phases.	\\

\emph{Statistics of homodyne current}. We consider emissions of quanta associated with jump occurrence and denote by $dB_j^\dagger(t)$ the creation operator for the quanta emitted with the action of jump $J_j$ at time $t$. The statistics of the \emph{homodyne current} measured with $dX(t)\equiv\sum_j[e^{i\varphi_j} dB_j^\dagger(t)+e^{-i\varphi_j}dB_j(t)]/2$ corresponds to the biased ``tilted'' master equation~\cite{Hickey2012} 
\begin{equation}\label{eq:Lr}
	\L_r(\rho) \equiv  \L(\rho) - r \X(\rho) +\frac{r^2}{8}, 
\end{equation}
where $\X(\rho)\equiv\sum_j   (e^{-i\varphi_j}{J}_{j}\rho+\,e^{i\varphi_j}\rho{J}_{j}^{\dagger})/2$.
That is, $\Theta(r,t)\equiv\ln\{\Tr \left[e^{t\L_r}(\rho)\right]\}$ is the cumulant generating function for the integrated value of the measured homodyne current until time $t$. The asymptotic statistics is then determined by $\theta(r)\equiv\lim_{t\rightarrow \infty}\Theta(r,t)/t$, which equals the eigenvalue  of $\L_r$ with the largest real part. \\

\emph{Classical tilted generator for homodyne current}.   in the presence of classical metastability, for the bias parameter $r$ smaller than the separation to the fast eigenmodes of $\L$, the maximal eigenmode of $\L_r$ in Eq.~\eqref{eq:Lr} can be approximated as a maximal eigenmode of $\P\L_r\P$, which in the basis of metastable phase corresponds to [cf.~Eq.~\eqref{eq:Wstilde}]
\begin{equation}\label{eq:Wrtilde}
	(\Wt_r)_{kl}\equiv \Tr[\tilde{P}_k \L_r (\trho_l)],
\end{equation}
$k,l=1,...,m$. We have $\Wt_r= \Wt-   r\,	\Xt +\frac{r^2}{8}$, where $(\Xt)_{kl}\equiv\Tr[\tilde{P}_k\X(\trho_l)]$.  $\Xt$ can be approximated by the diagonal matrix of the observable averages in the metastable phases 
\begin{equation}\label{eq:average_hom}
	(\xt)_{kl}\equiv\delta_{kl}\tilde{x}_l,
\end{equation}
where $\tilde{x}_l\equiv\sum_j\Tr[(e^{-i\varphi_j}{J}_{j}+e^{i\varphi_j}{J}_{j}^{\dagger} )\trho_l]/2$, $k,l=1,...,m$. 
Therefore, 
\begin{equation}\label{eq:Wrtilde2}
	\Wt_r= \W-   r\,\xt+ \frac{r^2}{8}+...\equiv \W_r+\frac{r^2}{8}+...
\end{equation}
The corrections in Eq.~\eqref{eq:Wrtilde2} can be  bounded in the leading order  by $ 2\sqrt{\Ccl}\lVert \Wt \rVert_1 +|r| \,m\,\lVert  \sum_j e^{-i\varphi_j} {J}_{j}  \rVert_{\max}  \sqrt{\Ccl/2+\Cp  }$ [cf.~Eq.~\eqref{eq:Wstilde2}]. Here, 	$\W_{r}$ encodes the statistics of time-integral of the system observable $\xt$ of Eq.~\eqref{eq:average_hom} in classical trajectories. From Eq.~\eqref{eq:Wrtilde2}, $\theta(r)$ is approximated by the maximal eigenvalue of $\W_{r}$ in Eq.~\eqref{eq:Wrtilde2} shifted by  $r^2/8$. Furthermore, the maximal eigenmode $\rhoss(r)$ of $\L_r$ is approximated by the corresponding $\Wt_{r}$  eigenmode $\pss(r)$ [cf.~Eq.~\eqref{eq:rhoss_h}] as
\begin{equation}\label{eq:rhoss_r}
	\rhoss(r)=\sum_{l=1}^m [\pss(r)]_l  \,\trho_l+...\equiv\rho_\text{ss}^{(0)}\!(r)+....
\end{equation}
The corrections in Eq.~\eqref{eq:rhoss_r} are given by $\rho_\text{ss}^{(1)}\!(r)\equiv r\S\Q \X [\rho_\text{ss}^{(0)}\!(r)]+r^2[\mathcal{W}_r-\theta^{(0)}\!(r)\P]^{+}\X\S\Q\X [\rho_\text{ss}^{(0)}\!(r)] -[\mathcal{W}_r-\theta^{(0)}\!(r)\P]^{+}(\L-r\X-\mathcal{W}_r) [\rho_\text{ss}^{(0)}\!(r)]$ in the second order of the degenerate non-Hermitian perturbation theory with respect to  $\L\P-r\X$ as a perturbation of $\L\Q+r^2/8$ and with a further approximation of $\L\P-r\P\X\P$ by $\mathcal{W}_r$. Here,  $\mathcal{W}_r$ denotes the superoperator corresponding to $\W_r$, $\Q\equiv\I-\P$ is the projection on the fast modes of $\L$ [cf.~Eq.~\eqref{Expansion}],  $^+$ stands for a pseudoinverse and $\S=\L^+$, while $\theta^{(1)}\!(r)$ denotes the maximal eigenvalue of $\W_r$. The normalization  $\Tr[\rhoss(r)]=1$ can be achieved by considering the additional correction $-\Tr[\rho_\text{ss}^{(1)}\!(r)]\rho_\text{ss}^{(0)}\!(r)$.
The correction $\theta(r)-\theta^{(1)}\!(r)-{r^2}/{8}$, is given by $\theta^{(2)}\!(r)\equiv\Tr\{L_1^{(0)}\!(r) [-r^2\X\S\Q\X+(\L-r\X-\mathcal{W}_r) ][\rho_\text{ss}^{(0)}\!(r)]\}$, where $L_1^{(0)}\!(r)$ is the dual eigenmatrix to   $\rho_\text{ss}^{(0)}\!(r) $ in Eq.~\eqref{eq:rhoss_r}.\\

\emph{Classical cumulants of homodyne current}. For times $t\geq t''$ such that $ t\lVert \Wt\rVert_1\ll\sqrt{\Ccl}$, the rate of the average integrated homodyne current can be approximated as [cf.~Eq.~\eqref{eq:Kav_t_approx_CL}]
\begin{eqnarray}\nonumber
	\frac{\langle X(t) \rangle}{t} &=&\frac{1}{t}\!\int_0^t \!\!d  t_1\! \sum_{l=1}^m\!\big( \xt e^{t_1\W} \pt\big)_l-\frac{\Tr\{\X\S\Q[\rho(0)]\}}{t}+\!...\\\label{eq:Xav_t_approx_CL}
	&\equiv& \frac{\langle X_\text{cl}(t)\rangle}{t}+\frac{\tilde{X}}{t}+....
\end{eqnarray}
The first term in Eq.~\eqref{eq:Xav_t_approx_CL} is the rate of the time-integral $X_\text{cl}(t)$ in classical trajectories of the average homodyne current in metastable phases, whose statistics in encoded by $\W_r$ in Eq.~\eqref{eq:Wrtilde2} (see Sec.~\ref{app:var_ms}). The second term represents the constant contribution to the integrated homodyne current from before the metastable regime. 	When time $t$ in Eq.~\eqref{eq:Xav_t_approx_CL} can be chosen after the final relaxation, the asymptotic rate is approximated as [cf.~Eq.~\eqref{eq:Kav}]
\begin{eqnarray}\label{eq:Xav}
	x_\text{ss}\equiv\lim_{t\rightarrow\infty}\frac{\langle X\!(t) \rangle}{t}&=&\lim_{t\rightarrow\infty}\frac{\langle X_\text{cl}(t) \rangle}{t}+...\\\nonumber &=&\sum_{l=1}^m (\pss)_l\, \tilde{x}_l+...,
\end{eqnarray}
with corrections bounded in the leading order by $\max_{1\leq l\leq m}\! |\tilde{x}_l|\lVert\ptss-\pss\rVert_1$ [note that $x_\text{ss}=\sum_{l=1}^m (\xt\ptss)_l$ and cf.~Eqs.~\eqref{eq:pss_CL}].

For times such that  $\lVert \S\Q\rVert\lVert\X \rVert\ll t\lVert\xt \rVert_1$ and $t \lVert\Wt \rVert_1 \ll\min[1/\sqrt{\Ccl}, \sqrt{ \lVert \Rt\rVert_1 \lVert \Wt\rVert_1 /\sqrt{\Ccl} }]$, the rate of fluctuations of integrated homodyne current is approximated as [cf.~Eq.~\eqref{eq:Kvar2}]
\begin{eqnarray}\nonumber
	&&\frac{\langle X^2(t) \rangle\!-\!\langle X(t) \rangle^2}{t}=\frac{\langle X_\text{cl}^2(t) \rangle\!-\!\langle X_\text{cl}(t) \rangle^2}{t}+ \langle\chi^2_\text{cl}(t)\rangle\qquad\quad\\
	&&\quad-\sum_{k,l=1}^m\tilde{p}_k\tilde{p}_l \frac{\langle X_\text{cl}^{(k)}\!(t)\rangle-\langle X_\text{cl}^{(l)}\!(t)\rangle}{t}\big(\tilde{X}_k-\tilde{X}_l\big)+...,
	\label{eq:Xvar_t_approx_CL}
\end{eqnarray}
where the fluctuations of the average homodyne current in classical trajectories 
\begin{equation} \label{eq:X2_CL}
	\langle X^2_\text{cl}(t) \rangle = -2 \!\!\int_0^t \!\!d  t_1\!\!\int_0^{t-t_1}\!\!\!\!\!\! d  t_2 \sum_{l=1}^m (\xt e^{t_2\W}\xt e^{t_1\W} \pt)_l, 
\end{equation}	
we denoted by
\begin{eqnarray}\label{eq:Chi_CL}
	\langle\chi^2_\text{cl}(t)\rangle&\equiv&\frac{\int_0^t \!\!d  t_1 \sum_{l=1}^m\big( \bm{\tilde{\chi}^2} e^{t_1\W} \pt\big)_l}{t}
\end{eqnarray}
the rate of the time-integral of fluctuation rates in metastable phases $(\bm{\tilde{\chi}}^2)_{kl}\equiv\delta_{kl}\{1/2- 2\Tr[\X\S\Q \X (\trho_l)]\}$ (see Sec.~\ref{app:var_ms}), $\langle X_\text{cl}^{(l)}\rangle$ is the average of time-integral of the homodyne current for $l$th metastable phase in Eq.~\eqref{eq:Xav_t_approx_CL}, i.e., $(\pt)_k=\delta_{kl}$, $k,l=1,...,m$, and $\tilde{X}_l\equiv \Tr\{\tilde{P}_l\X\S\Q[\rho(0)]\}/ \Tr[\tilde{P}_l\rho(0)]$ is the average contribution to the integrated homodyne current from before the metastable regime conditioned on the metastable phase that the system evolves into. Therefore, fluctuations of the homodyne current stem from classical transitions between metastable phases with differing average homodyne current and fluctuations inside the metastable phases, corrected by the average from before the metastable regime. When time $t$ in Eq.~\eqref{eq:Xvar_t_approx_CL} can be chosen after the final relaxation, the asymptotic rate of homodyne current fluctuations 
\begin{equation}\label{eq:Xvar}
	\chi_\text{ss}^2\equiv\lim_{t\rightarrow\infty}\frac{\langle X^2(t) \rangle\!-\!\langle X(t) \rangle^2}{t}= \frac{1}{2}- 2\Tr[\X\S\X (\rhoss)],
\end{equation}
is approximated as [cf.~Eq.~\eqref{eq:Wrtilde2}] 
\begin{eqnarray}\label{eq:Xvar2}
	\chi_\text{ss}^2&=&\lim_{t\rightarrow\infty}\Big[\frac{\langle X_\text{cl}^2(t) \rangle\!-\!\langle X_\text{cl}(t) \rangle^2}{t}+\langle\chi^2_\text{cl}(t)\rangle\Big]\qquad\\\nonumber	
	&=&-2 \sum_{l=1}^m\big[ \big( \bm{\xt}\R\bm{\xt}+ \bm{\tilde{\chi}^2}\big)\pss \Big]_l+...
\end{eqnarray}
(cf.~Sec.~\ref{app:classical}). The corrections to Eqs.~\eqref{eq:Xav_t_approx_CL} and~\eqref{eq:Xvar_t_approx_CL} are given in Sec.~\ref{app:var_t}, and to Eq.~\eqref{eq:Xvar} in Sec.~\ref{app:var_ss}.\\

\emph{Classical metastability and dynamical phase transitions in homodyne current}.  When metastable phases differ in average homodyne current, for the bias $|r|$ large enough $\W$ can be neglected in Eq.~\eqref{eq:Wrtilde2}, and Eq.~\eqref{eq:rhoss_r} approximated as [cf.~Sec.~\ref{sec:metaDPT} in the main text]
\begin{equation}\label{eq:rhoss_r2}
	\rhoss(r)=\trho_l+...,
\end{equation}
where $l$ is such that $\tilde{x}_l$ is maximal (for negative $r$) or minimal (for positive $r$)  among metastable phases. The corrections given in the lowest order of the non-Hermitian perturbation theory are $r\Q\S\X(\trho_l)- r(\tilde{\X}-\tilde{x}_l\P)^+ \X\S\Q\X(\trho_l)+ (\tilde{\X}-\tilde{x}_l\P)^+[\L-r(\X-\tilde{\X})](\trho_l)/r$, where $\tilde{\X}$ is the superoperator corresponding to $\xt$ in Eq.~\eqref{eq:average_hom}. Furthermore, 
\begin{eqnarray}\label{eq:theta_r2}
	\theta(r)&=&  -r \tilde{x}_l+\frac{r^2}{8}+...,\\\label{eq:k_r2}
	k(r)&=&  -\tilde{x}_l+\frac{r}{4}+...,
\end{eqnarray}
with the corrections given in the lowest order of the non-Hermitian perturbation theory by $-r^2 \Tr[\tilde{P}_l \X\S\Q\X (\trho_l)]+(\Wt)_{ll}- r (\Xt-\xt)_{ll}$ and by $-2 r \Tr[\tilde{P}_l \X\Q\S\X (\trho_l)] -\sum_{k\neq l}(\Wt)_{lk}(\Wt)_{kl}/(\tilde{x}_k-\tilde{x}_l)/r^2-(\Xt-\xt)_{ll}$, respectively. Therefore, the derivative of $\theta(r)$ undergoes a sharp change around $r=0$ when metastable phases differ in average homodyne current. This can be interpreted as a proximity to a dynamical phase transition~\cite{Hickey2012}. Furthermore, if the stationary state $\rhoss$ is different from the metastable phase with the maximal or the minimal homodyne current, i.e., features contributions from metastable phases with a different average of the homodyne current, the current fluctuations in Eq.~\eqref{eq:Xvar} are large [cf.~Eq.~\eqref{eq:Xvar2}]. 

\subsubsection{Time-integrals of system observables in quantum trajectories}\label{app:time_int}

We now discuss the relation of the statistics of time-integrated system observables to the classical dynamics between metastable phases.	\\

\emph{Statistics of time-integrated system observables}. The statistics of a system observable $M$ time-integrated over quantum trajectory corresponds to the biased ``tilted'' master operator~\cite{Levitov1996,Nazarov2003,Esposito2009,Flindt2009,Hickey2013}  
\begin{equation}\label{eq:Lh}
	\L_h(\rho)\equiv  \L-h\mathcal{M}(\rho),
\end{equation}
where $ \mathcal{M}(\rho)\equiv(M\rho+\rho M)/2$.
Here, $\Theta(h,t)\equiv\ln\{\Tr \left[e^{t\L_h}(\rho)\right]\}$ is the generating function for time-ordered cumulants of the integral of $M$ until time $t$. The asymptotic statistics is then determined by $\theta(h)\equiv\lim_{t\rightarrow\infty}\Theta(h,t)/t$, which equals the eigenvalue of $\L_h$ with the largest real part.\\

\emph{Classical tilted generator for time-integrated system observables}.  In the presence of classical metastability, for the bias parameter $h$ smaller than the separation of the slow eigenmodes of $\L$ to the fast ones, the maximal eigenmode can be approximated as a maximal eigenmode of $\P\L_h\P$, which in the basis of metastable phase corresponds to [cf.~Eq.~\eqref{eq:Wstilde}]
\begin{equation}\label{eq:Whtilde}
	(\Wt_h)_{kl}= \Tr[\tilde{P}_k \L_h (\trho_l)],
\end{equation}
$k,l=1,...,m$. We have $\Wt_h= \Wt-   h\,	\Mt$, where $(\Mt)_{kl}\equiv\Tr[\tilde{P}_k  \M( \trho_l)]$ and $\Mt$ can be approximated by the diagonal matrix of the observable averages in the metastable phases 
\begin{equation}\label{eq:average_M}
	(\mt)_{kl}=\delta_{kl}\tilde{m}_l,
\end{equation}
$k,l=1,...,m$, where $\tilde{m}_l\equiv\Tr (  M \trho_l)$, 
and thus
\begin{equation}\label{eq:Whtilde2}
	\Wt_h= \W-   h\,\mt+...\equiv \W_h+....
\end{equation}
The corrections in Eq.~\eqref{eq:Whtilde2} can be  bounded in the leading order  by $2 \sqrt{\Ccl}\lVert \Wt \rVert_1 +|h| \,m\,\lVert M\rVert_{\max} \sqrt{\Ccl/2+\Cp  }$ [see Eq.~\eqref{eq:deltaW} and Sec.~\ref{app:Ws}]. For dynamics of classical systems, $m\sqrt{\Ccl/2+\Cp}$ can be replaced by $\Ctcl+m\,\Cp$ in the corrections to Eqs.~\eqref{eq:Whtilde2}. 	

$\W_{h}$ in Eq.~\eqref{eq:Whtilde2} encodes the statistics of time-integral of the system observable $\mt$  in Eq.~\eqref{eq:average_M} in classical trajectories (cf.~Sec.~\ref{app:classical}). Therefore, for pronounced enough classical metastability ($\Ccl,\Cp\ll1$), $\theta (h)$ is approximated by the maximal eigenvalue of $\W_{h}$ in Eq.~\eqref{eq:Whtilde2}. Furthermore, the maximal eigenmode $\rhoss(h)$ of $\L_h$ in Eq.~\eqref{eq:Lh} is approximated by the maximal $\Wt_h$ eigenmode $\pss(h)$ [cf.~Eq.~\eqref{eq:rhoss_s}]
\begin{equation}\label{eq:rhoss_h}
	\rhoss(h)=\sum_{l=1}^m [\pss(h)]_l  \,\trho_l+...\equiv\rho_\text{ss}^{(0)}\!(h)+....
\end{equation}
The corrections in Eq.~\eqref{eq:rhoss_h} are given by $\rho_\text{ss}^{(1)}\!(h)\equiv h\S\Q \M [\rho_\text{ss}^{(0)}\!(h)] +h^2[\mathcal{W}_h-\theta^{(1)}\!(0)\P]^{+} \M\S\Q\M[\rho_\text{ss}^{(0)}\!(h)] -[\mathcal{W}_h-\theta^{(1)}\!(h)\P]^{+}(\L_{h}-\mathcal{W}_{h}) [\rho_\text{ss}^{(0)}\!(h)]$ in the degenerate non-Hermitian perturbation theory with respect to $\L\P -h\M$ as a perturbation of $\L\Q$, with a further approximation of $\L\P-h\P\M\P$ by $\mathcal{W}_h$. Here, $\mathcal{W}_h$ denotes the  superoperator corresponding to $\W_h$, $^+$ stands for the pseudoinverse, $\S=\L^+$ and $\Q\equiv\I-\P$ is the projection on the fast modes of $\L$ [cf.~Eq.~\eqref{Expansion}]. The maximal eigenvalues of $\W_h$ denoted as $\theta^{(1)}\!(h)$ approximates $\theta(h)$, with the higher order correction $\theta^{(2)}\!(h)\equiv\Tr\{L_1^{(0)}\!(h)[-h^2\M \S\Q\M+(\L_h-W_h)] [\rho_\text{ss}^{(0)}\!(h)]\}$, where $L_1^{(0)}\!(h)$ is the dual eigenmatrix to   $\rho_\text{ss}^{(0)}\!(h) $. Finally, the normalization  $\Tr[\rhoss(h)]=1$ can be achieved by considering the additional correction $-\Tr[\rho_\text{ss}^{(1)}\!(h)]\rho_\text{ss}^{(0)}\!(h)$.\\

\emph{Classical cumulants of time-integrated system observables}. For times $t\geq t''$ such that $t\lVert \Wt\rVert_1\ll\sqrt{\Ccl}$, the rate of the average integrated system observable can be approximated as [cf.~Eq.~\eqref{eq:Kav_t_approx_CL}]
\begin{eqnarray}\nonumber
	\frac{\langle M(t) \rangle}{t} &=&	\frac{1}{t}\!\int_0^t \!\!d  t_1\! \sum_{l=1}^m\!\big( \mt e^{t_1\W} \pt\big)_l-\frac{\Tr\{\M\S\Q[\rho(0)]\}}{t}+\!...\\\label{eq:Mav_t_approx_CL}
	&\equiv& \frac{\langle M_\text{cl}(t)\rangle}{t}+\frac{\tilde{M}}{t}+....
\end{eqnarray}
The first term in Eq.~\eqref{eq:Mav_t_approx_CL} is the rate of the time-integral $M_\text{cl}(t)$ in classical trajectories of the average observable in metastable phases, whose statistics in encoded by $\W_h$ in Eq.~\eqref{eq:Whtilde2} (see Sec.~\ref{app:var_ms}). The second term represents the constant contribution to the time-integral from before the metastable regime. When time $t$ in Eq.~\eqref{eq:Mav_t_approx_CL} can be chosen after the final relaxation, the asymptotic rate is approximated as [cf.~Eq.~\eqref{eq:Kav}]
\begin{eqnarray}\label{eq:Mav}
	m_\text{ss}\equiv\lim_{t\rightarrow\infty}\frac{\langle M\!(t) \rangle}{t}&=&\lim_{t\rightarrow\infty}\frac{\langle M_\text{cl}(t) \rangle}{t}+...\\\nonumber &=&\sum_{l=1}^m (\pss)_l\, \tilde{m}_l+...,
\end{eqnarray}
with corrections bounded in the leading order by $\max_{1\leq l\leq m}\! |\tilde{m}_l|\lVert\ptss-\pss\rVert_1$ [note that $m_\text{ss}=\sum_{l=1}^m (\mt\ptss)_l$ and cf.~Eqs.~\eqref{eq:pss_CL}].

For times such that   $\lVert \S\Q\rVert\lVert\M \rVert\ll t\lVert\mt \rVert_1$ and $ t \lVert\Wt \rVert_1\ll \min[\sqrt{\Ccl}, \sqrt{ \lVert \Rt\rVert_1 \lVert \Wt\rVert_1 /\sqrt{\Ccl} }]$ the rate of time-ordered fluctuations of the integrated system observable is approximated as [cf.~Eq.~\eqref{eq:Kvar2}]
\begin{eqnarray}\nonumber
	&&\frac{\overline{\langle M^2(t) \rangle}\!-\!\langle M(t) \rangle^2}{t}=\frac{\langle M_\text{cl}^2(t) \rangle\!-\!\langle M_\text{cl}(t) \rangle^2}{t}+ \langle\delta^2_\text{cl}(t)\rangle\quad\\
	&&-\!\sum_{k,l=1}^m\!\tilde{p}_k\tilde{p}_l \frac{\langle M_\text{cl}^{(k)}\!(t)\rangle-\langle M_\text{cl}^{(l)}\!(t)\rangle}{t}\big(\tilde{M}_k-\tilde{M}_l\big)+...,\qquad\label{eq:Mvar_t_approx_CL}
\end{eqnarray}
where the fluctuations of the average system observable in classical trajectories 
\begin{equation} \label{eq:M2_CL}
	\langle M^2_\text{cl}(t) \rangle = -2 \!\!\int_0^t \!\!d  t_1\!\!\int_0^{t-t_1}\!\!\!\!\!\! d  t_2 \sum_{l=1}^m (\mt e^{t_2\W}\mt e^{t_1\W} \pt)_l, 
\end{equation}	
we denoted by
\begin{eqnarray}\label{eq:Mdelta_CL}
	\langle\delta^2_\text{cl}(t)\rangle&\equiv&\frac{\int_0^t \!\!d  t_1 \sum_{l=1}^m\big( \bm{\tilde{\delta}^2} e^{t_1\W} \pt\big)_l}{t}
\end{eqnarray}
the rate of the time-integral of time-ordered fluctuation rates in metastable phases $(\bm{\tilde{\delta}}^2)_{kl}\equiv-2\delta_{kl} \Tr[\M\S\Q \M (\trho_l)]$ (see Sec.~\ref{app:var_ms}), $\langle M_\text{cl}^{(l)}\rangle$ is the average of time-integral of system observable in classical trajectories for $l$th metastable phase in Eq.~\eqref{eq:Mav_t_approx_CL}, i.e., $(\pt)_k=\delta_{kl}$, $k,l=1,...,m$, and $\tilde{M}_l\equiv \Tr\{\tilde{P}_l\M\S\Q[\rho(0)]\}/ \Tr[\tilde{P}_l\rho(0)]$ is the average contribution to the integrated system observable from before the metastable regime conditioned on the metastable phase that the system evolves into. When time $t$ in Eq.~\eqref{eq:Mvar_t_approx_CL} can be chosen after the final relaxation, the asymptotic rate of time-ordered fluctuations
\begin{equation}\label{eq:Mvar}
	\chi_\text{ss}^2\equiv\lim_{t\rightarrow\infty}\frac{\overline{\langle M^2(t) \rangle}\!-\!\langle M(t) \rangle^2}{t}=- 2\Tr[\M\S\M (\rhoss)],
\end{equation}
is approximated as [cf.~Eq.~\eqref{eq:Wrtilde2}] 
\begin{eqnarray}\label{eq:Mvar2}
	\chi_\text{ss}^2&=&\lim_{t\rightarrow\infty}\Big[\frac{\langle M_\text{cl}^2(t) \rangle\!-\!\langle M_\text{cl}(t) \rangle^2}{t}+\langle\delta^2_\text{cl}(t)\rangle\Big]\qquad\\\nonumber	
	&=&-2 \sum_{l=1}^m\big[ \big( \bm{\mt}\R\bm{\mt}+ \bm{\tilde{\delta}^2}\big)\pss \Big]_l+...
\end{eqnarray}
(cf.~Sec.~\ref{app:classical}). The corrections to Eqs.~\eqref{eq:Mav_t_approx_CL} and~\eqref{eq:Mvar_t_approx_CL} are given in Sec.~\ref{app:var_t}, and to Eq.~\eqref{eq:Mvar} in Sec.~\ref{app:var_ss}.\\

\emph{Classical metastability and dynamical phase transitions in time-integrated system observables}.  When metastable phases differ in observable averages, for the bias $|h|$ large enough $\W$ can be neglected in Eq.~\eqref{eq:Whtilde2}, and Eq.~\eqref{eq:rhoss_h} approximated as [cf.~Sec.~\ref{sec:metaDPT} in the main text]
\begin{equation}\label{eq:rhoss_h2}
	\rhoss(h)=\trho_l+...,
\end{equation}
where $l$ is such that $\tilde{m}_l$ is maximal (for negative $h$) or minimal (for positive $h$) among metastable phases. The corrections given in the lowest order of the non-Hermitian perturbation theory are $h\S\Q\M(\trho_l) -h(\tilde{\M}-\tilde{m}_l\P)^+ \M\S\Q\M(\trho_l)+ (\tilde{\M}-\tilde{m}_l\P)^+ [\L-h(\M-\tilde{\M})](\trho_l)/h$, where $\tilde{\M}$ is the superoperator corresponding to $\mt$ in Eq.~\eqref{eq:average_hom}. Furthermore, the maximal eigenvalue of $\L_h$ and its derivative can be approximated by 
\begin{eqnarray}\label{eq:theta_h2}
	\theta(h)&=&  -h \tilde{m}_l+...,\\\label{eq:k_h2}
	k(h)&=&  -\tilde{m}_l+...,
\end{eqnarray}
with the corrections given in the lowest order of the non-Hermitian perturbation theory by $-h^2 \Tr[\tilde{P}_l \M\S\Q\M (\trho_l)]+(\Wt)_{ll}- h(\Mt-\mt)_{ll}$ and by $-2 h \Tr[\tilde{P}_l \M\S\Q\M (\trho_l)] - \sum_{k\neq l} (\Wt)_{lk}(\Wt)_{kl}/(\tilde{m}_k-\tilde{m}_l)/h^2 -(\Mt-\mt)_{ll}$, respectively. Therefore, the derivative of $\theta(h)$ undergoes a sharp change around $h=0$ when metastable phases differ in observable averages, which can be interpreted as a proximity to a dynamical phase transition~\cite{Hickey2013}. When the stationary state features contributions from metastable phases with different observable averages, the fluctuations in Eq.~\eqref{eq:Mvar} are large [cf.~Eq.~\eqref{eq:Mvar2}].

\subsubsection{Corrections in approximations of $\Wt_s$, $\Wt_h$, and $\Wt_r$} \label{app:Ws}

In this section, we give and prove corrections in approximations of $\Wt_s$, $\Wt_h$, and $\Wt_r$. We also discuss how the order of approximation is changed for the metastability in classical stochastic dynamics.\\

\emph{Corrections in classical approximation of $\Wt_s$}.
For $\Wt_s$ in Eq.~\eqref{eq:Wstilde}, we have
\begin{equation}\label{eq:Wstilde0}
	\Wt_s=\Wt+ (e^{-s}-1)\Jt,
\end{equation}
where 
\begin{equation}\label{eq:Jt}
	(\Jt)_{kl}\equiv\sum_{j}\Tr(\tilde{P}_k  {J}_{j}\trho_l\,{J}_{j}^{\dagger}).
\end{equation}
can be related to the jump activity 
\begin{equation}\label{eq:activity}
	(\bm{\mut})_{kl}\equiv\delta_{kl}\,\mut_l,
\end{equation} as
\begin{eqnarray}\label{eq:bound_D}
	&& | ( \Jt  )_{kl}-( \Wt  )_{kl} -\delta_{kl}\,\mut_k | \\\nonumber
	&&\leq\Big\lVert H-c\mathds{1} +\frac{i}{2} \sum_{j} {J}_{j}^{\dagger} J_{j}\Big\rVert_{\max} \sqrt{2\Ccl +4  \Cp}, 
\end{eqnarray}
where $c$ is a real constant chosen to minimize the operator norm and $k,l=1,...,m$. Equation~\eqref{eq:bound_D} illustrates that jump operators considered in the activity can lead to transitions between metastable phases, but not at rates higher than the rates in the effective dynamics $\Wt$  (up to corrections to the positivity of metastable phases and their projections).  Therefore, together with Eq.~\eqref{eq:deltaW} [and analogously $\lVert \Jbf-(\Wt+\bm{\mut}-\bm{\mut}^\text{in})\rVert\lesssim \lVert \Wt \rVert_1\sqrt{\Ccl}$],  we obtain in Eq.~\eqref{eq:Wstilde2}
\begin{eqnarray}\label{eq:bound_Dfull}
	&&\lVert\Wt_s- \W_{s}\rVert_1\lesssim   ({e}^{-s}+1)\sqrt{\Ccl}\lVert \Wt \rVert_1\\\nonumber
	&&+|{e}^{-s}-1 |\,  m\,\lVert H-c\mathds{1} +\frac{i}{2} \sum_{j} {J}_{j}^{\dagger} J_{j}\rVert_{\max}  \sqrt{2\Ccl   +4\Cp}.
\end{eqnarray}   

We also show for individual jumps that [cf.~Eq.~\eqref{eq:bound_D}]
\begin{eqnarray}\label{eq:bound_D2}
	&&\Bigg| \sum_j| \Tr(\tilde{P}_k  {J}_{j}\trho_l\,{J}_{j}^{\dagger}) -\delta_{kl}  \Tr({J}_{j}^{\dagger}{J}_{j}\trho_l)|- (   \Wt  )_{kl} \Bigg|\qquad\qquad\\\nonumber
	&&\leq   \Big\lVert H-c\mathds{1} +\frac{i}{2} \sum_{j} {J}_{j}^{\dagger} J_{j}\Big\rVert_{\max}  \sqrt{2\Ccl+4\Cp   },
\end{eqnarray}
which bounds the corrections in Eq.~\eqref{eq:Wsjtilde2}.
\\

\emph{Corrections in classical approximation of $\Wt_r$}.
For $\Wt_r$ in Eq.~\eqref{eq:Wrtilde2}, we show that
\begin{equation}
	(\Xt)_{kl}\equiv\sum_j\Tr[\tilde{P}_k(e^{-i\varphi_j}{J}_{j}\trho_k+e^{i\varphi_j}\trho_l{J}_{j}^{\dagger})]/2
\end{equation}
is approximated by the diagonal matrix $\xt$ of the observable averages in metastable phases in Eq.~\eqref{eq:average_hom} as   
\begin{equation}\label{eq:bound_D3}
	| ( \Xt  )_{kl} -(\xt)_{kl} |\leq   \lVert \sum_j e^{-i\varphi_j} {J}_{j}\rVert_{\max}  \sqrt{\Ccl/2+\Cp   },
\end{equation} 
which together with Eq.~\eqref{eq:deltaW} gives
\begin{eqnarray}\label{eq:bound_D3full}
	&&\lVert\Wt_r- \W+   r\,\xt-\frac{r^2}{8}\rVert_1\lesssim  2\sqrt{\Ccl}\lVert \Wt \rVert_1\\\nonumber
	&&\qquad\qquad+|r| m\,\lVert \sum_j e^{-i\varphi_j} {J}_{j}\rVert_{\max} \sqrt{\Ccl/2+\Cp  }.
\end{eqnarray}   \\

\emph{Corrections in classical approximation of $\Wt_h$}.
Similarly, for $\Wt_h$ in Eq.~\eqref{eq:Whtilde} we show below that 
\begin{equation}
	(\Mt)_{kl}\equiv\Tr(\tilde{P}_k  \{M, \trho_l\})/2
\end{equation}
is approximated by the diagonal matrix $\mt$ of the observable averages in metastable phases  in Eq.~\eqref{eq:average_M} as    
\begin{equation}\label{eq:bound_D4}
	| ( \Mt  )_{kl} -(\mt)_{kl} |\leq   \lVert M\rVert_{\max}  \sqrt{\Ccl/2+\Cp}.
\end{equation}  
Therefore, together with Eq.~\eqref{eq:deltaW},
\begin{eqnarray}\label{eq:bound_D4full}
	&&\lVert\Wt_h- \W+   h\,\mt\rVert_1\lesssim 2\sqrt{\Ccl}\lVert \Wt \rVert_1 \\\nonumber
	&&\qquad\qquad+|h| m\,\lVert M\rVert_{\max}  \sqrt{\Ccl/2+\Cp},
\end{eqnarray}   
which leads to Eq.~\eqref{eq:Whtilde2}.\\

\emph{Derivation} of Eq.~\eqref{eq:bound_D}. We want to compare $\Jt$ in Eq.~\eqref{eq:Jt} with the effective dynamics in the MM [cf.~Eq.~\eqref{eq:Wtilde}]
\begin{eqnarray}\label{eq:bound_der1}
	&& (\Wt )_{kl} = \sum_{j} \Tr (  \tilde{P_k}\, {J}_{j}\trho_l {J}_{j}^{\dagger} ) \\\nonumber
	&&\qquad -i\,\Tr \Big[\tilde{P_k} \Big(H-c\mathds{1}-\frac{i}{2}\sum_{j} {J}_{j}^{\dagger} {J}_{j} \Big)\,\trho_l\Big ]+\text{h.c.},
\end{eqnarray}
where $c$ is an arbitrary real constant. For $k\neq l$, the difference of Eqs.~\eqref{eq:Jt} and~\eqref{eq:bound_der1} corresponds to the second line of Eq.~\eqref{eq:bound_der1}, which size we now estimate. 

For a matrix $X$, we have
\begin{eqnarray}\label{eq:bound_der2}
	\Tr (  \tilde{P}_k \,  X\,\trho_l  )&=& \Tr [  (\tilde{P}_k-\tilde{p}_{k}^\text{min}\mathds{1} )\,  X \,\rho_l   ] \\\nonumber
	& +&\tilde{p}_{k}^\text{min} \Tr (X \,\rho_l ) + \Tr [\tilde{P}_k \,X\,(\rho_l -\trho_l )] ,
\end{eqnarray}
where $\tilde{p}_{k}^\text{min} $ is the most negative eigenvalue  of $\tilde{P}_k$, and $\rho_l$ is the closest state to $\trho_l$ [cf.~Eq.~\eqref{eq:Cp}]. The terms in the second line of~\eqref{eq:bound_der2} can be bounded as [cf.~Eq.~\eqref{eq:vN2}]
\begin{eqnarray}\label{eq:bound_der3a}
	|\tilde{p}_{k}^\text{min}\Tr (X\,\rho_l   )|&\leq& |\tilde{p}_{k}^\text{min}|  \lVert X \rVert_{\max},\\\label{eq:bound_der3b}
	\Tr [ \tilde{P}_k \,X \, (\rho_l -\trho_l )    ]&\leq&    \lVert \tilde{P}_k \rVert_{\max}  \lVert X \rVert_{\max} \lVert \rho_l-\trho_l  \rVert.\qquad
\end{eqnarray}
The first line in~\eqref{eq:bound_der2} can be bounded as follows. 
From the Cauchy-Schwarz inequality with respect to the operators $\sqrt{\rho_l}(\tilde{P}_k-\tilde{p}_{k}^\text{min}\mathds{1})$ and $ X\,\sqrt{\rho_l}$ we have
\begin{eqnarray}\label{eq:bound_der4}
	&& \lvert\Tr [  (\tilde{P}_k-\tilde{p}_{k}^\text{min}\mathds{1} )\,X\,\rho_l    ] \rvert\\\nonumber
	&&  \leq  \sqrt{\Tr [  (\tilde{P}_k-\tilde{p}_{k}^\text{min}\mathds{1} )^2\,\rho_l ]} \sqrt{\Tr (X^\dagger X \, \rho_l    ) } 
	\\\nonumber
	&& \leq  \sqrt{\tilde{p}_{k}^\text{max}-\tilde{p}_{k}^\text{min}}  \sqrt{ |\tilde{p}_{k}^\text{min} |+\Tr ( \tilde{P}_k\,\rho_l ) }    \lVert X \rVert_{\max},
\end{eqnarray}
where in the second line we used $(\tilde{P}_k-\tilde{p}_{k}^\text{min}\mathds{1})^2\leq  (\tilde{p}_{k}^\text{max}-\tilde{p}_{k}^\text{min}) (\tilde{P}_k-\tilde{p}_{k}^\text{min}\mathds{1})$ and $|\Tr[ (\tilde{P}_k-\tilde{p}_{k}^\text{min}\mathds{1})\,\rho_l]|\leq |\tilde{p}_{k}^\text{min}| + | \Tr(\tilde{P}_k \rho_l) |$. 
%
%

For off-diagonal terms ($k\neq l$), we further obtain  
\begin{equation}
	| \Tr(\tilde{P}_k \rho_l) |=| \Tr\{\tilde{P}_k (\rho_l-\trho_l )|\leq \lVert \tilde{P}_k \rVert_{\max}   \lVert \rho_l-\trho_l  \rVert.
\end{equation}
Therefore, from~\eqref{eq:bound_der2} using Eqs.~(\ref{eq:bound_der3a})-(\ref{eq:bound_der4}) we obtain
\begin{eqnarray}
	&& 	|  ( \Jt   )_{kl} - ( \Wt   )_{kl} |\leq  2 \Big\lVert H-c\mathds{1} - \frac{i}{2}\sum_j {J}_{j}^{\dagger} {J}_{j}\Big\rVert_{\max} \\\nonumber
	&&\quad\times\sqrt{|\tilde{p}_{k}^\text{min}|+  \lVert \tilde{P}_k \rVert_{\max}   \lVert \rho_l-\trho_l  \rVert } \\\nonumber
	&& \quad\times \Big(   \sqrt{|\tilde{p}_{k}^\text{min}|+\lVert \tilde{P}_k \rVert_{\max}  \lVert \rho_l-\trho_l  \rVert }+\sqrt{\tilde{p}_{k}^\text{max}-\tilde{p}_{k}^\text{min} }  \Big ), \\\nonumber
	&& \quad \lesssim \Big\lVert H-c\mathds{1} - \frac{i}{2}\sum_j {J}_{j}^{\dagger} {J}_{j}\Big\rVert_{\max} \,\sqrt{2\Ccl+4\Cp},
\end{eqnarray}
where in the last line we used $|\tilde{p}_{k}^\text{min}|\leq \Ccl/2$ [cf.~Eq.~\eqref{eq:Ptilde_min}],   $  |\tilde{p}_{k}^\text{max}-\tilde{p}_{k}^\text{min}|\geq |\tilde{p}_{k}^\text{max}|-|\tilde{p}_{k}^\text{min}|\geq 1-\Ccl/2$ [cf.~Eq.~\eqref{eq:Ptilde_max2}] and $\lVert \tilde{P}_k \rVert_{\max}  \leq 1+\Ccl/2$ [cf.~Eq.~\eqref{eq:Ptilde_norm}], [cf.~Eq.~\eqref{eq:B_disjoint}].

Similarly, for diagonal terms [cf.~Eq.~\eqref{eq:activity}]
\begin{eqnarray}\label{eq:activity2}
	&&  ( \Jt   )_{ll}  - \mut_l = - \sum_{j} \Tr [ ( \mathds{1}-  \tilde{P_l} ){J}_{j}\trho_l {J}_{j}^{\dagger}  ],
\end{eqnarray}
while 	from $\sum_{l=1}^m \tilde{P_l}=\mathds{1}$ we also have
\begin{eqnarray}
	&& -(\Wt )_{ll} = \sum_{k\neq l} (\Wt )_{kl}=\sum_{j} \Tr [ ( \mathds{1}-  \tilde{P_l} )\, {J}_{j}\trho_l {J}_{j}^{\dagger} ] \\\nonumber
	&&\qquad -i\,\Tr \Big[( \mathds{1}-  \tilde{P_l} ) \Big(H-c\mathds{1}-\frac{i}{2}\sum_{j} {J}_{j}^{\dagger} {J}_{j} \Big)\,\trho_l\Big ]+\text{h.c.}.
\end{eqnarray}
Therefore, by replacing $\tilde{P}_l$ by $\mathds{1}-  \tilde{P}_l$ in Eq.~\eqref{eq:bound_der4} and noting that 
\begin{eqnarray}
	&&| \Tr[(\mathds{1}-\tilde{P}_l) \rho_l] |=| \Tr[(\mathds{1}-\tilde{P}_l) (\rho_l-\trho_l )|\\\nonumber
	&&\leq \lVert \mathds{1}-\tilde{P}_l \rVert_{\max}   \lVert \rho_l-\trho_l  \rVert\leq \left(1+\frac{\Ccl}{2}\right) \lVert \rho_l-\trho_l  \rVert
\end{eqnarray}
from $|\tilde{p}_{l}^\text{min}|\leq \Ccl/2$ [cf.~Eq.~\eqref{eq:Ptilde_min}], we obtain 
\begin{eqnarray}\label{eq:bound_D_final}
	&& 	|  ( \Jt   )_{ll} - \mut_l -( \Wt   )_{ll} | \\\nonumber
	&&\quad \lesssim \Big\lVert H-c\mathds{1} - \frac{i}{2}\sum_j {J}_{j}^{\dagger} {J}_{j}\Big\rVert_{\max} \,\sqrt{2\Ccl+4\Cp}.
\end{eqnarray}
Eq.~\eqref{eq:bound_D} follows by the triangle inequality. Note that $\lVert X\rVert_{\max}$ in the bound in Eq.~\eqref{eq:bound_D} can be replaced by $\sqrt{\max_{1\leq l \leq m} \Tr(X^\dagger X\trho_l)+\lVert X\rVert_{\max}^2\Cp}$ [cf.~Eq.~\eqref{eq:bound_der4}].	\\

\emph{Derivation} of Eq.~\eqref{eq:bound_D2}.
For activities related to individual jumps, we note that
\begin{eqnarray}
	&&\Tr [ (  \tilde{P_k}-\tilde{p}_{k}^\text{min} ){J}_{j}\rho_l\, {J}_{j}^{\dagger} ] =| \Tr [ (  \tilde{P_k}-\tilde{p}_{k}^\text{min} ) {J}_{j}\rho_l\, {J}_{j}^{\dagger} ] |,\qquad\quad\\\nonumber
	&& \Tr [ (  \tilde{p}_{l}^\text{max}-\tilde{P_l} ){J}_{j}\rho_l\, {J}_{j}^{\dagger} =|\Tr [ (  \tilde{p}_{l}^\text{max}-\tilde{P_l} ){J}_{j}\rho_l\, {J}_{j}^{\dagger}|,
\end{eqnarray}
which follows from the positivity of $\tilde{P_k}-\tilde{p}_{k}^\text{min}$ and $\tilde{p}_{l}^\text{max}-\tilde{P_l}$ and $\rho_l$. Therefore, 
\begin{eqnarray}
	&&\Big|\sum_j| \Tr (  \tilde{P_k} \,{J}_{j}\trho_l\, {J}_{j}^{\dagger} ) - (\bm{\mut})_{kl} | -[(\Jt)_{kl}-(\bm{\mut}_j)_{kl}] \Big|\qquad\qquad\\\nonumber
	&& \lesssim \!\!\left[|\tilde{p}_{k}^\text{min} |(1\!-\!\delta_{kl})+|\tilde{p}_{l}^\text{max}\!\!-\!1|\delta_{kl} +(\tilde{p}_{k}^\text{max}\!\!-\tilde{p}_{k}^\text{min}) \lVert\rho_l-\trho_l \rVert \right]\\\nonumber
	&&\qquad\times\Big[\sum_{j} |\mut_l^{(j)}| +|\mut_l|\Big],
\end{eqnarray}
where $(\bm{\mut}_j)_{kl}\equiv\mut_l^{(j)}\delta_{kl}\equiv \Tr (  {J}_{j}^{\dagger} {J}_{j}\trho_l)\delta_{kl}  $.
As $|\tilde{p}_{k}^\text{min}|\leq \Ccl/2$, $|\tilde{p}_{k}^\text{max}-1|\leq \Ccl/2$, $(\tilde{p}_{k}^\text{max}-\tilde{p}_{k}^\text{min} )\leq (1+2\Ccl)$, and $\lVert\rho_l-\trho_l \rVert\leq \Cp$,  the corrections are of a higher order than the corrections in Eq.~\eqref{eq:bound_D} and thus can be neglected, giving rise to Eq.~\eqref{eq:bound_D2}.  \\

\emph{Derivation} of Eq.~\eqref{eq:bound_D3}. By considering Eqs.~(\ref{eq:bound_der2})-(\ref{eq:bound_der4}) and Eq.~\eqref{eq:activity2} for $X=\sum_j e^{-i\varphi_j} {J}_{j}/2  $ and $X^\dagger$, and noting that $\lVert X\rVert_{\max}=\lVert X^\dagger\rVert_{\max}$  we arrive at the result. Furthermore, $\lVert X\rVert_{\max}$ in the bound can be replaced by $\sqrt{\max_{1\leq l \leq m} \Tr(X^\dagger X\trho_l)+\lVert X\rVert_{\max}^2\Cp}$ [cf.~Eq.~\eqref{eq:bound_der4}].\\

\emph{Derivation} of Eq.~\eqref{eq:bound_D4}. By considering Eqs.~(\ref{eq:bound_der2})-(\ref{eq:bound_der4}) and Eq.~\eqref{eq:activity2} for $X=M/2$ we arrive at the result. Note that $\lVert M\rVert_{\max}$ in the bound in Eq.~\eqref{eq:bound_D3} can be replaced by $\sqrt{\max_{1\leq l \leq m} \Tr(M^\dagger M\trho_l)+\lVert M\rVert_{\max}^2\Cp}$ [cf.~Eq.~\eqref{eq:bound_der4}].\\

\emph{Corrections for classical stochastic dynamics}. We now show that metastability in dynamics of probability distributions rather than density matrices (see Sec.~\ref{app:classical}) leads to linear order of corrections in Eqs.~\eqref{eq:bound_D},~\eqref{eq:bound_D2} and~\eqref{eq:bound_D3}.  

To bound terms $\Tr [  \tilde{P}_k\,X\,\trho_l   ]=\Tr [  X\,\trho_l\,\tilde{P}_k] $, we can again make use of the inequality in Eq.~\eqref{eq:vN2} with the choice $Y=\trho_l \tilde{P}_k$
\begin{equation}
	\lvert\Tr [  \tilde{P}_k\,X\,\trho_l   ] \rvert\leq \lVert X\rVert_{\max}  \lVert \tilde{P}_k   \trho_l \rVert .
\end{equation}
Below we show that for commuting $\trho_l$ and $\tilde{P}_k$ [e.g., classical stochastic dynamics] we have
\begin{equation}
	\sum_{k\neq l} \lVert  \tilde{P}_k \trho_l  \rVert\lesssim \Ctcl+\Cp.
\end{equation}
Therefore, we arrive at [cf.~Eq.~\eqref{eq:bound_D_final}]
\begin{eqnarray}
	&&\lVert \Jt- \W- \bm{\mut}\rVert_1\\\nonumber
	&&\lesssim 2 \,\lVert H -c\mathds{1}+\frac{i}{2} \sum_{j} {J}_{j}^{\dagger} J_{j}\rVert_{\max}   (\Ctcl+\Cp)
\end{eqnarray} 
and [cf.~Eq.~\eqref{eq:bound_D3full}]
\begin{eqnarray}
	&&\lVert \Mt- \mt\rVert_1\lesssim \lVert M\rVert_{\max}  (\Ctcl+\,\Cp).
\end{eqnarray}

Indeed, we have
\begin{eqnarray}
	&&\sum_{k\neq l} \lVert   \tilde{P}_k\trho_l  \rVert \equiv \sum_{k\neq l} \sum_{n} |\tilde{p}_k(n) | |\trho_l(n)|\\\nonumber
	&&=\sum_{k\neq l} \sum_{n}|\trho_l(n)| [|\tilde{p}_k(n) |-\tilde{p}_k(n)]\\\nonumber
	&&\quad+\sum_n [|\trho_l(n)|-\trho_l(n)] \sum_{k\neq l} \tilde{p}_k(n) \\\nonumber
	&&\leq  (1+\Cp)\,  \sum_{k\neq l} (-2\tilde{p}_k^{\min}) + \Cp \max_n |1-\tilde{p}_l(n)|\\\nonumber
	&&\le1(1+\Cp)\Ctcl +(1+\Ccl/2)\Cp\lesssim  \Ctcl +\Cp,
\end{eqnarray}
where $\trho_l(n)$  and $\tilde{p}_k(n) $ are the diagonal entries of $\trho_l$ and $\tilde{P}_k$, respectively, in their eigenbasis, $n=1, ..., \text{dim}(\mathcal{H})$. In the second line we used $\sum_{n}\trho_l(n)\tilde{p}_k(n)=\Tr(\tilde{P}_k\trho_l)=\delta_{kl}$. In the first inequality  we used  $\sum_{n}|\trho_l(n)|\leq 1+\Cp$ [cf.~Eq.~\eqref{eq:Cp}], so that also $\sum_n |\trho_l(n) |-\trho_l(n)= \sum_n |\trho_l(n) |-1 \leq \Cp$,  as well as $|\tilde{p}_k(n) |-\tilde{p}_k(n)= 2\max[-\tilde{p}_k(n),0] \leq -2\tilde{p}_k^{\min} $ [cf.~Eq.~\eqref{eq:Ccl}] and $\sum_{k\neq l} \tilde{p}_k(n)=1-\tilde{p}_l(n)$ (see Sec.~\ref{app:Ptilde}).

\subsubsection{Rates of average and fluctuations in quantum trajectories after initial relaxation} \label{app:var_t}

Here, we give corrections to the classical approximations of the average and fluctuations rates of jump number after the metastable regime given in Eqs.~\eqref{eq:Kav_t_approx_CL} and~\eqref{eq:Kvar_t_approx_CL}. We also discuss the cases of homodyne measurement and time-integral of system observables introduced in Secs.~\ref{app:homodyne} and~\ref{app:time_int}.	Derivations rely on the master operator in Eq.~\eqref{eq:L} being diagonalizable and can be found at the end of this section. \\

\emph{Corrections to average rate of jump number}. For time $t\geq t''$, the rate of average jump number is approximated as in Eq.~\eqref{eq:Kav_t_approx_CL} with the following corrections
\begin{eqnarray}\label{eq:Kav_t_approx_CL_corr}
	&&\Big|\frac{\langle K(t) \rangle}{t} - \frac{\langle K_\text{cl}(t)\rangle+\langle\tilde{K}\rangle}{t}\Big|\\\nonumber
	&&\lesssim 2\Cmm^n\frac{\lVert\S\Q\rVert}{t} \lVert\J\rVert+\lVert\bm{\mut}-\bm{\mut^\text{tot}}\rVert_1\\\nonumber
	&&\quad+ \frac{\lVert\bm{\mut}\rVert_1 }{t}\int_{0}^t d t_1 \lVert e^{t_1\Wt}-e^{t_1\W}\rVert
	\\\nonumber
	&&\lesssim 2\Cmm^n\frac{\lVert\S\Q\rVert}{t} \lVert\J\rVert+\sqrt{\Ccl}\lVert\Wt \rVert_1\big(1+t\lVert\bm{\mut}\rVert_1),
\end{eqnarray}
where an integer $n\geq 1$ is chosen so that $t/n$ belongs to the metastable regime,  $\lVert\J\rVert=\lVert\sum_j J_j^\dagger J_j\rVert_{\max}$, $\lVert\bm{\mut}\rVert_1=\max_{1\leq l\leq m}|\mut_l|$, and $\Q\equiv\I-\P$ is the projection on the fast modes of $\L$, so that $\lVert\S\Q\rVert\lesssim 2t''$ [see~Eq.~\eqref{eq:SQnorm}].  The first correction arises from the approximation of the dynamics of fast modes by their complete decay, while the second correction stems from replacing the long-time dynamics in Eq.~\eqref{eq:Wtilde} by the classical stochastic dynamics in Eq.~\eqref{eq:W} [in the second inequality we used Eq.~\eqref{eq:deltaW} and~\eqref{eq:etW_CL}], so that we require
\begin{equation}\label{eq:cond_av_t}
	t \lVert\Wt \rVert_1 \ll 1/\sqrt{\Ccl}.
\end{equation}   
When
\begin{equation}\label{eq:cond_av_t_1}
	t \lVert \bm{\mut}\rVert_1\gg \lVert\S\Q\rVert \lVert\J\rVert,
\end{equation}
the contribution $\tilde{K}/t$ from before the metastable regime can be neglected and the approximation in Eq.~\eqref{eq:Kav_t_approx_CL} holds well [provided that  $\langle K_\text{cl}(t)\rangle/t$ is of the order of $\lVert \bm{\mut}\rVert_1$, as $\lVert\Wt \rVert_1\lesssim \lVert\bm{\mut}\rVert_1$]. In particular, when time $t$  can be chosen after the final relaxation, we obtain the classical approximation of the asymptotic average rate in Eq.~\eqref{eq:Kav}. \\


\emph{Corrections to fluctuation rate of jump number}. Similarly, for $t\geq t''$, the rate of fluctuations in jump number is approximated as in Eq.~\eqref{eq:Kvar_t_approx_CL} with the corrections 
\begin{eqnarray}\label{eq:Kvar_t_approx_CL_corr} 
	&&\bigg|\frac{\langle K^2(t) \rangle\!-\!\langle K(t) \rangle^2}{t}-\frac{\langle K_\text{cl}^2(t) \rangle\!-\!\langle K_\text{cl}(t) \rangle^2+\langle\Delta_\text{cl}(t)\rangle}{t}\\\nonumber
	&&\quad-\sum_{k,l=1}^m\tilde{p}_k\tilde{p}_l \frac{\langle K_\text{cl}^{(k)}(t)\rangle-\langle K_\text{cl}^{(l)}(t)\rangle}{t}\big(\tilde{K}_k-\tilde{K}_l\big)
	\bigg|
	\\\nonumber
	&&\lesssim   \frac{\lVert\S\Q\rVert}{t}\lVert \J\rVert\Big[1+  \lVert\S\Q\rVert\big(4\lVert \J\P\rVert
	+3\lVert \J\rVert\big)\big]
	\\\nonumber&&\quad+4 \Cmm^n  \bigg(\lVert \bm{\mut}\rVert_1+ \frac{\lVert\S\Q\rVert}{t}\lVert \J\rVert\bigg) \lVert \J\rVert \lVert\S\Q\rVert \\\nonumber
	&&\quad + t \lVert\bm{\mut}-\bm{\mut^\text{tot}}\rVert_1 \min(\lVert\Jt\rVert_1,\lVert\J\P\rVert)\\\nonumber
	&&\quad+  t \lVert\bm{\mut}\rVert_1\big(\lVert\Jt-\Wt-\bm{\mut}\rVert_1+   \frac{1}{2}\lVert\W-\Wt\rVert_1\big)\\\nonumber
	&&\quad + \frac{4}{t}\!\int_{0}^t\!\! d t_1 \int_{0}^{t-t_1} \!\!\!\!\!\!d t_2 \lVert e^{t_2\Wt}\!\!-e^{t_2\W}\rVert_1 \min (\lVert\bm{\mut}\rVert_1\lVert\Jt\rVert_1,\lVert\J\P\rVert^2)\\\nonumber
	&&\quad+ \bigg[t \lVert\bm{\mut}-\bm{\mut^\text{tot}}\rVert_1  +\int_{0}^t \!\!d t_1 \lVert e^{t_1\Wt}\!\!-e^{t_1\W}\rVert_1 \lVert\bm{\mut}\rVert_1\bigg]2\lVert\bm{\mut}\rVert_1
	\\\nonumber
	&&\quad +\lVert\bm{\mut}-\bm{\mut^\text{tot}}\rVert_1+\frac{1}{t}\int_{0}^t \!\!d t_1 \lVert e^{t_1\Wt}\!\!-e^{t_1\W}\rVert_1 \lVert\bm{\tilde{\sigma}}^2\rVert_1 \\\nonumber
	&&\quad +\bigg[\lVert\bm{\mut}-\bm{\mut^\text{tot}}\rVert_1+\frac{1}{t}\int_{0}^t \!\!d t_1 \lVert e^{t_1\Wt}\!\!-e^{t_1\W}\rVert_1\lVert\bm{\mut}\rVert_1\bigg]\\\nonumber
	&&\qquad \qquad\times 4\min(\lVert \bm{\tilde{K}}\pt\rVert_1,\lVert\J\rVert\lVert\S\Q\rVert)\\\nonumber
	&&\lesssim   \frac{\lVert\S\Q\rVert}{t}\lVert \J\rVert\Big[1+  \lVert\S\Q\rVert\big(4\lVert \J\P\rVert
	+3\lVert \J\rVert\big)\big]
	\\\nonumber&&\quad+4 \Cmm^n  \bigg(\lVert \bm{\mut}\rVert_1+ \frac{\lVert\S\Q\rVert}{t}\lVert \J\rVert\bigg) \lVert \J\rVert \lVert\S\Q\rVert \\\nonumber
	&&\quad +\sqrt{\Ccl} \lVert \Wt\rVert_1  t \big[\min(\lVert\Jt\rVert_1,\lVert\J\P\rVert)+\lVert\bm{\mut}\rVert_1\big]\\\nonumber
	&&\quad +\sqrt{2\Ccl\! +\!4  \Cp} m \Big\lVert H \!+\!\frac{i}{2}\! \sum_{j} {J}_{j}^{\dagger} J_{j}\Big\rVert_{\max}  \,t \lVert\bm{\mut}\rVert_1\\\nonumber
	&&\quad +\frac{2}{3} \sqrt{\Ccl} \lVert \Wt\rVert_1 t^2 \min (\lVert\bm{\mut}\rVert_1\lVert\Jt\rVert_1,\lVert\J\P\rVert^2)\\\nonumber
	&&\quad+\sqrt{\Ccl} \lVert \Wt\rVert_1 \big( 2 + t\lVert \bm{\mut}\rVert_1\big)t\lVert \bm{\mut}\rVert_1  \\\nonumber
	&&\quad +\sqrt{\Ccl} \lVert \Wt\rVert_1 (1+ t\lVert\bm{\tilde{\sigma}}^2\rVert_1) \\
	&&\quad + 2\sqrt{\Ccl} \lVert \Wt\rVert_1  \big(2+t\lVert \bm{\mut}\rVert_1\big) \min(\lVert \bm{\tilde{K}}\pt\rVert_1,\lVert\J\rVert\lVert\S\Q\rVert),\nonumber
\end{eqnarray}
where
\begin{equation}\label{eq:sigmat}
	\big(\bm{\tilde{\sigma}}^2\big)_{kl} \equiv \delta_{kl}[\mut_l-2\Tr (\J \S\Q\J \trho_l)],
\end{equation}
$k,l=1,...,m$, which can be interpreted as fluctuation rate of metastable phases (see Sec.~\ref{app:var_ms}),
and 
\begin{equation}\label{eq:Ktilde2}
	(\bm{\tilde{K}})_{kl}\equiv\delta_{kl} \tilde{K}_l=\delta_{kl}\frac{\Tr(\tilde{P}_l\J\S\Q\rho)}{\Tr(\tilde{P}_l\rho)}
\end{equation}
encodes the contribution from before the metastable regime. We further have $\lVert \J\P\rVert\lesssim \lVert \bm{\mut}\rVert_1+2\lVert \J\rVert \Cp$~\footnote{For $\lVert \J\P\rVert\leq\lVert (1+\Ccl) \bm{\mut}\rVert_1+\Cp\lVert \J\rVert$, we note that $\lVert\J \P(\rho)\rVert\leq \Tr \J(\rho') +\lVert\J \rVert \lVert \rho'-\P\rho' \rVert \leq | \sum_{l=1}^m \mut_l \tilde{p}_l |+2\lVert\J \rVert \lVert \rho'-\P\rho' \rVert$, and choose $\rho'$ as the closest state to $\P(\rho)$ [cf.~Eq.~\eqref{eq:Cp}]}, as well as $\lVert \Jt\rVert_1\leq (1+\Ctcl)\lVert \J\P\rVert$ and $\lVert \bm{\tilde{K}}\pt\rVert_1\leq (1+\Ctcl)\lVert \J\rVert \lVert\S\Q\rVert$ [cf.~Eq.~\eqref{eq:rho1_CL}].

In the second inequality of Eq.~\eqref{eq:Kvar_t_approx_CL_corr}  we used Eqs.~\eqref{eq:etW_CL} and~\eqref{eq:bound_D}.  The first and second lines of corrections in the first inequality arise from the  approximation of the dynamics of fast modes by the full decay and neglecting any constant contribution to the fluctuations, while the rest of corrections originate from replacing the long-time dynamics in Eq.~\eqref{eq:Wtilde} by the classical stochastic approximation in Eq.~\eqref{eq:W} (and considering the space of metastable phases with L1 norm, or the space of density matrices with trace norm; cf.~Sec.~\ref{app:norm}). In particular, we can identify that the third to fifth lines correspond to the corrections to the non-Poissonian classical fluctuations of the activity, $[\langle K_\text{cl}^2(t) \rangle\!-\!\langle K_\text{cl}(t) \rangle^2-\langle K_\text{cl}(t) \rangle]/t$, the sixth line is the correction to the time-integral of fluctuation rate inside metastable phases, $[\langle K_\text{cl}(t) \rangle+\langle\Delta_\text{cl}(t)\rangle]/t$,  while the seventh and eighth lines describe corrections to the contribution from before the metastable regime. 


When the corrections in Eq.~\eqref{eq:Kvar_t_approx_CL_corr}  are negligible in comparison with the corresponding (positive) contributions to the fluctuation rate, we obtain \emph{sufficient conditions} on the classical approximation in Eq.~\eqref{eq:Kvar_t_approx_CL}. As we discuss below, the contributions to the fluctuation rate can be bounded in terms of operator norms leading to generalized conditions, which are sufficient for the classical approximation to hold whenever the bounds saturate (for the opposite case, see e.g., Sec.~\ref{app:var_ms}).  In particular, if those conditions are fulfilled at time $t$ after the final relaxation, the asymptotic fluctuation rate can be approximated in terms of the classical dynamics [see Eq.~\eqref{eq:Kvar2} and cf.~Sec.~\ref{app:var_ss}]. Finally, if there exists a leading contribution, e.g., non-Poissonian classical fluctuations characteristic of a first-order dynamical phase transition (see Sec.~\ref{sec:metaDPT} in the main text), the conditions can be relaxed.

The contribution to the time-integral of fluctuation rate inside metastable phases  $[\langle K_\text{cl}(t) \rangle+\langle\Delta_\text{cl}(t)\rangle]/t$ is bounded by $\lVert\bm{\tilde{\sigma}}^2\rVert_1$, while the contribution to the fluctuation rate from before the metastable regime [the second line of Eq.~\eqref{eq:Kvar_t_approx_CL_corr}] is bounded by $4 \lVert\bm{\mut} \rVert_1 \min(\lVert \bm{\tilde{K}}\pt\rVert_1,\lVert\J\rVert\lVert\S\Q\rVert)$ (see derivation below). Therefore, the first and the second lines of corrections are negligible in comparison with those bounds,  when
\begin{equation}\label{eq:cond_var_t_1}
	t \min(\lVert\bm{\mut}\rVert_1,\lVert\bm{\tilde{\sigma}}^2\rVert_1) \gg \lVert\S\Q\rVert \lVert \J\rVert
\end{equation}
[cf.~Eq.~\eqref{eq:cond_av_t_1}].
Furthermore, the corrections to  the time-integral of fluctuation rate inside metastable phases  and contribution from before the metastable regime in the last two lines are negligible in comparison with the corresponding bounds when
\begin{equation}\label{eq:cond_var_t_2}
	t\lVert\Wt \rVert_1\ll \frac{1}{\sqrt{\Ccl}},
\end{equation}
as well as
\begin{equation}\label{eq:cond_var_t_3}
	\frac{\lVert\Wt \rVert_1}{\lVert\bm{\tilde{\sigma}}^2\rVert_1}\ll\frac{1}{\sqrt{\Ccl}},
\end{equation}
[which is the condition for replacing $\bm{\tilde{\sigma}}^2$ by $(\bm{\tilde{\sigma}}^\text{tot})^2$, while $\bm{\tilde{\mu}}$ can be replaced by $\bm{\tilde{\mu}}^\text{tot}$ in the contribution from before the metastable regime as $\lVert\Wt \rVert_1/\lVert\bm{\mut}\rVert_1\lesssim 2$]. 
Note that Eq.~\eqref{eq:cond_var_t_3} can be viewed as condition on the maximal anti-bunching of fluctuations inside metastable phases.  Finally, the corrections in the third to fifth lines of Eq.~\eqref{eq:Kvar_t_approx_CL}  are the corrections to $[\langle K_\text{cl}^2(t) \rangle\!-\langle K_\text{cl}(t) \rangle^2-\langle K_\text{cl}(t) \rangle]/t\leq t  \lVert \bm{\mut}\rVert_1 [\min(\lVert\Jt\rVert_1,\lVert\J\P\rVert)+\lVert \bm{\mut}\rVert_1]$. They are negligible in comparison with this bound when Eq.~\eqref{eq:cond_var_t_2} is fulfilled together with
\begin{equation}\label{eq:cond_var_t_4}
	\frac{ m \lVert H +\frac{i}{2} \sum_{j} {J}_{j}^{\dagger} J_{j}\rVert_{\max}}{ \min(\lVert\Jt\rVert_1,\lVert\J\P\rVert)}\ll \frac{1}{\sqrt{2\Ccl +4  \Cp}}.
\end{equation}
It can also be shown that 
$[\langle K_\text{cl}^2(t) \rangle\!-\langle K_\text{cl}(t) \rangle^2-\langle K_\text{cl}(t) \rangle]/t\leq    2\lVert\Rt \rVert_1 \lVert \bm{\mut}\rVert_1  [\min(\lVert\Jt\rVert_1,\lVert\J\P\rVert)+ \lVert \bm{\mut}\rVert_1] (1 + 4 \lVert\Rt \rVert_1/t)$ 
[cf.~derivation of Eq.~\eqref{eq:Kvar_t_approx_CL} below], which requires 
\begin{equation}\label{eq:cond_var_t_5}
	t\lVert\Wt \rVert_1 \ll\sqrt{\frac{\lVert\Rt \rVert_1\lVert\Wt \rVert_1} {\sqrt{\Ccl}}},
\end{equation} 
which is a stricter condition than  Eq.~\eqref{eq:cond_var_t_2} for $t\geq 4\lVert\Rt \rVert_1$. For those times, no additional condition beyond Eq.~\eqref{eq:cond_var_t_4} is required, as the corresponding corrections can be shown to be bounded by $2\lVert\Rt \rVert_1 (1 + 4 \lVert\Rt \rVert_1/t)\lVert\bm{\mut}\rVert_1\lVert\Jt-\Wt-\bm{\mut}\rVert_1$. \\

\emph{Corrections to average rate of time-integrated homodyne current}. Similarly to Eq.~\eqref{eq:Kav_t_approx_CL_corr}, for the integrated homodyne current we obtain that the corrections in Eq.~\eqref{eq:Xav_t_approx_CL} are bounded by 
\begin{eqnarray}\label{eq:Xav_t_approx_CL_corr}
	&&\Big|\frac{\langle X(t) \rangle}{t} - \frac{\langle X_\text{cl}(t)\rangle+\langle\tilde{X}\rangle}{t}\Big|\\\nonumber
	&&\lesssim 2\Cmm^n\frac{\lVert\S\Q\rVert}{t} \lVert\X\rVert+ \frac{\lVert\xt\rVert_1 }{t}\int_{0}^t d t_1 \lVert e^{t_1\Wt}-e^{t_1\W}\rVert
	\\\nonumber
	&&\lesssim 2\Cmm^n\frac{\lVert\S\Q\rVert}{t} \lVert\X\rVert+t\sqrt{\Ccl}\lVert\Wt \rVert_1\lVert\xt\rVert_1,
\end{eqnarray}
where an integer $n\geq 1$ is chosen so that $t/n$ belongs to the metastable regime, $\lVert\X\rVert\leq \lVert  \sum_{j} e^{-i\varphi_j}J_{j}\rVert_{\max}$ and $\lVert\xt\rVert_1=\max_{1\leq l\leq m}|\tilde{x}_l|$.\\

\emph{Corrections to fluctuation rate of time-integrated homodyne current}. Similarly to  Eq.~\eqref{eq:Kvar_t_approx_CL_corr}, for the integrated homodyne current we obtain that the corrections in Eq.~\eqref{eq:Xvar_t_approx_CL} are bounded by
\begin{eqnarray}\label{eq:Xvar_t_approx_CL_corr}
	&&\bigg|\frac{\langle X^2(t) \rangle\!-\!\langle X(t) \rangle^2}{t}-\frac{\langle X_\text{cl}^2(t) \rangle\!-\!\langle X_\text{cl}(t) \rangle^2}{t}\qquad\\\nonumber
	&&\quad-\langle\chi^2_\text{cl}(t)\rangle\-\!\sum_{k,l=1}^m\!\tilde{p}_k\tilde{p}_l \frac{\langle X_\text{cl}^{(k)}(t)\rangle-\langle X_\text{cl}^{(l)}(t)\rangle}{t}\big(\tilde{X}_k-\tilde{X}_l\big)
	\bigg|
	\\\nonumber&&	
	\lesssim  \frac{\lVert\S\Q\rVert^2}{t}\lVert \X\rVert\big(4\lVert \X\P\rVert+3\lVert \X\rVert\big)
	\\\nonumber&&\quad+4 \Cmm^n  \bigg(\lVert \xt\rVert_1+ \frac{\lVert\S\Q\rVert}{t}\lVert \X\rVert\bigg) \lVert \X\rVert \lVert\S\Q\rVert\\\nonumber
	&&\quad+ t \lVert\xt\rVert_1\lVert\Xt-\xt\rVert_1\\\nonumber
	&&\quad + \frac{4}{t}\!\int_{0}^t \!\!d t_1 \!\int_{0}^{t-t_1} \!\!\!\!\!d t_2 \lVert e^{t_2\Wt}\!-e^{t_2\W}\rVert_1 \\\nonumber
	&&\qquad \qquad\times \min(\lVert\xt\rVert_1\lVert\Xt\rVert_1,\lVert\X\P\rVert^2)\\\nonumber
	&&\quad +2\int_{0}^t \!\!d t_1 \lVert e^{t_1\Wt}\!\!-e^{t_1\W}\rVert_1 \lVert\xt\rVert_1^2 
	\\\nonumber
	&&\quad + \frac{1}{t} \int_{0}^t\! d t_1 \lVert e^{t_1\Wt}\!-e^{t_1\W}\rVert_1 \lVert\bm{\tilde{\chi}^2}\rVert_1\\\nonumber
	&&\quad +\frac{4}{t}\int_{0}^t \!\!d t_1 \lVert e^{t_1\Wt}\!\!-e^{t_1\W}\rVert_1\lVert\xt\rVert_1 \lVert\X\rVert\lVert\S\Q\rVert
	\\\nonumber&&
	\lesssim  \frac{\lVert\S\Q\rVert^2}{t}\lVert \X\rVert\big(4\lVert \X\P\rVert+3\lVert \X\rVert\big)
	\\\nonumber&&\quad+4 \Cmm^n  \bigg(\lVert \xt\rVert_1+ \frac{\lVert\S\Q\rVert}{t}\lVert \X\rVert\bigg) \lVert \X\rVert \lVert\S\Q\rVert\\\nonumber
	&&\quad + \sqrt{\frac{\Ccl}{2} + \Cp} m\Big\lVert  \sum_{j} e^{-i\varphi_j}J_{j}\Big\rVert_{\max}\,t \lVert\bm{\xt}\rVert_1\\\nonumber
	&&\quad +\frac{2}{3} \sqrt{\Ccl} \lVert \Wt\rVert_1 t^2  \min(\lVert\xt\rVert_1\lVert\Xt\rVert_1,\lVert\X\P\rVert^2)\\\nonumber
	&&\quad+\sqrt{\Ccl} \lVert \Wt\rVert_1 t^2\lVert \xt\rVert_1^2  \\\nonumber
	&&\quad +\sqrt{\Ccl} \lVert \Wt\rVert_1 t\lVert\bm{\tilde{\chi}}^2\rVert_1 +2 \sqrt{\Ccl} \lVert \Wt\rVert_1  t\lVert \xt\rVert_1 \lVert\X\rVert\lVert\S\Q\rVert,
\end{eqnarray}
where 
\begin{equation}\label{eq:chit}
	\big(\bm{\tilde{\chi}}^2\big)_{kl} \equiv \delta_{kl}\Big[\frac{1}{2}- 2 \Tr (\X \S\Q \X \trho_l)\Big],
\end{equation}
and $\lVert \xt\rVert_1\leq \lVert \Xt\rVert_1\leq (1+\Ctcl)\lVert \X\P\rVert$  [cf.~Eq.~\eqref{eq:rho1_CL}].
In the first inequality, the first line of corrections in Eq.~\eqref{eq:Xvar_t_approx_CL_corr} arises from neglecting the dynamics of fast modes, while the rest of corrections originate from replacing the long-time dynamics by the classical stochastic dynamics. 
In the second inequality we used Eqs.~\eqref{eq:etW_CL} and~\eqref{eq:bound_D3}. \\

\emph{Corrections to average rate of time-integrated system observables}.  Analogously to Eq.~\eqref{eq:Xav_t_approx_CL_corr}, for the time-integral of the system observable the corrections in Eq.~\eqref{eq:Mav_t_approx_CL} are bounded by 
\begin{eqnarray}\label{eq:Mav_t_approx_CL_corr}
	&&\Big|\frac{\langle M(t) \rangle}{t} - \frac{\langle M_\text{cl}(t)\rangle+\langle\tilde{M}\rangle}{t}\Big|\\\nonumber
	&&\lesssim 2\Cmm^n\frac{\lVert\S\Q\rVert}{t} \lVert\M\rVert+t\sqrt{\Ccl}\lVert\Wt \rVert_1\lVert\mt\rVert_1,
\end{eqnarray}
where an integer $n\geq 1$ is chosen so that $t/n$ belongs to the metastable regime, $\lVert\M\rVert\leq \lVert  M\rVert_{\max}$ and $\lVert\mt\rVert_1=\max_{1\leq l\leq m}|\tilde{m}_l|$.
\\

\emph{Corrections to fluctuation rate of time-integrated system observables}. Analogously to Eq.~\eqref{eq:Xvar_t_approx_CL_corr} for the time-integral of a system observable we obtain that the corrections to the classical approximation of time-ordered correlations in Eq.~\eqref{eq:Mvar_t_approx_CL} are bounded by
\begin{eqnarray}\label{eq:Mvar_t_approx_CL_corr}
	&&\bigg|\frac{\overline{\langle M^2(t) \rangle}\!-\!\langle M(t) \rangle^2}{t}-\frac{\langle M_\text{cl}^2(t) \rangle\!-\!\langle M_\text{cl}(t) \rangle^2}{t}\qquad\quad\\\nonumber
	&&\quad-	\langle\delta^2_\text{cl}(t)\rangle-\!\sum_{k,l=1}^m\!\tilde{p}_k\tilde{p}_l \frac{\langle M_\text{cl}^{(k)}(t)\rangle-\langle M_\text{cl}^{(l)}(t)\rangle}{t}\big(\tilde{M}_k-\tilde{M}_l\big)
	\bigg|
	\\\nonumber&&	
	\lesssim  \frac{\lVert\S\Q\rVert^2}{t}\lVert \M\rVert\big(4\lVert \M\P\rVert+3\lVert \M\rVert\big)
	\\\nonumber&&\quad+4 \Cmm^n  \bigg(\lVert \mt\rVert_1+ \frac{\lVert\S\Q\rVert}{t}\lVert \M\rVert\bigg) \lVert \M\rVert \lVert\S\Q\rVert\\\nonumber
	&&\quad +  \sqrt{\frac{\Ccl}{2} + \Cp}m\lVert  M\rVert_{\max}  \,t\lVert\bm{\mt}\rVert_1\\\nonumber
	&&\quad +\sqrt{\Ccl} \lVert \Wt\rVert_1\Big\{ t^2 \Big[\frac{2}{3}\min(\lVert\xt\rVert_1\lVert\Xt\rVert_1,\lVert\X\P\rVert^2)+\lVert \mt\rVert_1  ^2\Big]\\\nonumber
	&&\qquad\qquad\qquad+t\lVert\bm{\tilde{\delta}^2}\rVert_1 +2 t\lVert\mt\rVert_1\lVert\M\rVert\lVert\S\Q\rVert\Big\}
	,
\end{eqnarray}
where 
\begin{equation}\label{eq:deltat}
	\big(\bm{\tilde{\delta}}^2\big)_{kl} \equiv -2 \delta_{kl}\Tr (\M \S\Q \M \trho_l).
\end{equation}
and  $\lVert \mt\rVert_1\leq \lVert \Mt\rVert_1\leq (1+\Ctcl)\lVert \M\P\rVert$  [cf.~Eq.~\eqref{eq:rho1_CL}]. In the  inequality we used Eqs.~\eqref{eq:etW_CL} and~\eqref{eq:bound_D4}. 
\\

Below we drive Eqs.~\eqref{eq:Kav_t_approx_CL_corr}, \eqref{eq:Kvar_t_approx_CL_corr}, \eqref{eq:Xav_t_approx_CL_corr}, \eqref{eq:Xvar_t_approx_CL_corr}, \eqref{eq:Mav_t_approx_CL_corr}, and~\eqref{eq:Mvar_t_approx_CL_corr}. We start by introducing several useful facts.
\\

\emph{Exact formulas for cumulants of finite-time distribution}.	By noting that 
\begin{equation} \label{eq:Theta_s_t}
	e^{\Theta(s,t)}=\Tr[ e^{t\L_s}(\rho)]	
\end{equation}
is the moment generating function for the number of jumps $K(t)$ detected up to time $t$ [cf.~Eq.~\eqref{eq:Ls}], we have 
\begin{eqnarray}\label{eq:Kav_t}
	\langle K(t) \rangle&=& \int_0^t d  t_1  \Tr(\J e^{t_1 \L} \rho) \qquad\quad\!
	\\\nonumber&=& t \Tr(\J\rhoss)+\Tr [\J \S (e^{t \L}-\I)  \rho],
	\\
	\langle K^2(t) \rangle&=&\int_0^t d  t_1   \Tr(\J e^{t_1 \L} \rho) \label{eq:Kvar_t} \\\nonumber 
	&+&2 \int_0^t d  t_1\int_0^{t-t_1} d  t_2  \Tr (\J e^{t_2\L}\J e^{t_1\L} \rho)
	\\\nonumber &=& t \Tr(\J\rhoss)+\Tr [\J \S (e^{t \L}-\I)  \rho] \\\nonumber 
	&+& 2  \Tr (\J \rhoss) \int_0^t d  t_1 (t-t_1)\Tr (\J e^{t_1\L} \rho)\\\nonumber 
	&+&2 \int_0^t d  t_1\Tr \{\J \S[e^{(t-t_1)\L}-\I]\J e^{t_1\L} \rho\},
\end{eqnarray}
as $\partial_{s}e^{t \L_s}=\int_0^t d  t_1  e^{(t-t_1) \L_s}\partial_{s}\L_s e^{t_1 \L_s}=-e^{-s}\int_0^t d  t_1  e^{(t-t_1) \L_s}\J e^{t_1 \L_s}$ [cf.~Eq.~\eqref{eq:Ls}].

Similarly, for the integrated homodyne current $X(t)$ we have that 
\begin{equation} \label{eq:Theta_r_t}
	e^{\Theta(r,t)}=\Tr[ e^{t\L_r}(\rho)]	
\end{equation}	
is a moment generating function [cf.~Eq.~\eqref{eq:Lr}] and thus [cf.~Eqs.~\eqref{eq:Kav_t} and~\eqref{eq:Kvar_t}]
\begin{eqnarray}\label{eq:Xav_t}
	\langle X(t)\rangle & =& t \Tr(\X\rhoss) + \Tr [\X \S (e^{t \L}\!-\I) \rho], \\
	\label{eq:Xvar_t}
	\langle X^2(t)\rangle&=&\frac{t}{2}+2  \Tr (\X \rhoss)\!\! \int_0^t\!\! d  t_1 (t-t_1)\Tr (\X e^{t_1\L} \rho)\\\nonumber
	&&+2\! \int_0^t \!\!\! d t_1  \Tr \{\X \S[e^{(t-t_1)\L}\!-\I]\X e^{t_1 \L} \rho\}.\qquad\qquad
\end{eqnarray}	 
Analogous formulas hold for time-ordered moments of time-integral of system observables with $\X$ replaced by $\M$ and $1/2$ removed in Eq.~\eqref{eq:Xvar_t}  [cf.~Eq.~\eqref{eq:Lh}]. 
\\

\emph{Useful facts about norms}. We now prove bounds on norms of operators appearing in Eq.~\eqref{eq:Kav_t} and Eq.~\eqref{eq:Kvar_t}. During the metastable regime, $t''\leq t\leq t'$, from the Taylor series we have  [see Eq.~\eqref{eq:t'4}]
\begin{equation}\label{eq:approx_1}
	t\lVert\LMM\rVert \approx \lVert (e^{t \L}-\I)\P\rVert\leq(1+\Cp)\Cmm\lesssim\Cmm.
\end{equation}
Note that it follows that Eq.~\eqref{eq:approx_1} holds also before the metastable regime, that is, for $t\leq t''$.

For times $ t\geq t''$, let an integer $n\geq 1$ be such that $t/n$ belongs to the metastable regime. We have~\cite{Note1}
\begin{eqnarray}\label{eq:approx_2}
	\lVert  e^{t\L} \Q \rVert&=& \lVert  (e^{\frac{t}{n}\L}\!-\!\P)^n \Q \rVert\leq \lVert  (e^{\frac{t}{n}\L}\!-\!\P)\rVert^n \lVert\Q\rVert\qquad\qquad\\\nonumber&\leq& \Cmm^n (2+\Cp)\lesssim 2\Cmm^n \leq 2\Cmm.
\end{eqnarray}

Finally, we have that for any time within metastable regime, $t''\leq t\leq t'$,
\begin{equation}
	\label{eq:SQnorm}
	\lVert S\Q\rVert\lesssim 2t.
\end{equation}
Eq.~\eqref{eq:SQnorm} follows from $\S=\int_0^\infty d t (e^{t\L}-\P_\text{ss})$ [cf.~Eq.~\eqref{eq:R_CL}], and thus $(\I-e^{t\L})\S=\int_0^t d t_1(e^{t_1\L}-\P_\text{ss})$ so that $ (1-2\Cmm)\lVert\S\Q\rVert\leq (1- \lVert e^{t\L}\Q\rVert)\lVert\S\Q\rVert \leq \lVert\S\Q\rVert-\lVert e^{t\L}\S\Q\rVert \leq \lVert(\I-e^{t\L})\S\Q\rVert=\lVert\int_0^t e^{t\L}\Q\rVert\leq \lVert\Q\rVert t\leq 2 t$ [cf.~Eq.~\eqref{eq:Snorm}].  \\

\emph{Derivation} of Eq.~\eqref{eq:Kav_t_approx_CL_corr}. For times $t\geq t''$, we have [cf.~Eq.~\eqref{eq:Kav_t}]
\begin{eqnarray}\nonumber
	&&\left| \langle K(t) \rangle-t\mu_\text{ss} - \,\Tr[\J \S (e^{t\L}-\I)\P\rho]+\Tr(\J \S \Q\rho)\right| \qquad\\&&
	=\big|\Tr(\J \S e^{t\L}\Q\rho)\big|\lesssim 2 \Cmm^n \lVert \J\rVert \lVert\S\Q\rVert \lVert\Q\rho\rVert,	\label{eq:Kvar_approx1_CL}
\end{eqnarray}
where we used Eq.~\eqref{eq:approx_2} with $n$ such that $t''\leq t/n\leq t'$. Furthermore, we recognize
\begin{eqnarray}\label{eq:activity_CL0}
	&&t\mu_\text{ss} +\Tr[\J \S (e^{t\L}-\I)\P\rho]\\\nonumber
	&&=\sum_{l=1}^m \big\{\bm{\mut}\big[t\ptss+\Rt(e^{t\Wt}-\Ibf)\pt\big]\big\}_l\\\nonumber
	&&=\int_{0}^t\!\! d t_1 \sum_{l=1}^m \big(\bm{\mut}e^{t_1\Wt}\pt\big)_l,
\end{eqnarray}
which from Eqs.~\eqref{eq:etW_CL} leads to Eq.~\eqref{eq:Kav_t_approx_CL_corr} [note that in Eq.~\eqref{eq:Kvar_approx1_CL} $\lVert\Q\rho\rVert$ can be simply replaced by $1$].\\

\emph{Derivation} of Eq.~\eqref{eq:Kvar_t_approx_CL_corr}. We have [cf.~Eq.~\eqref{eq:Kvar_t}]
\begin{eqnarray}
	&&\int_0^t d  t_1 \int_0^{t-t_1} d  t_2  \Tr (\J e^{t_2\L}\P\J e^{t_1 \L} \Q\rho)\\\nonumber
	&&=\int_0^t d  t_2 \int_{0}^{t-t_2} d  t_1  \Tr (\J e^{t_2\L}\P\J e^{t_1 \L} \Q\rho)
	\\\nonumber
	&&=\int_0^t d  t_2  \Tr \{\J e^{t_2\L}\P\J \S [e^{(t-t_2) \L}-\I]) \Q\rho\}
\end{eqnarray}
and	
\begin{eqnarray}\label{eq:Kvar_approx2_CL}
	&& \Bigg| \int_0^t d  t_2  \Tr \{\J e^{t_2\L}\P\J \S [e^{(t-t_2) \L}-\I]) \Q\rho\}\\\nonumber&&\quad+ t\mu_\text{ss}\Tr(\J \S\Q\rho) - \Tr[\J \S(e^{t \L}-\I) \P\J \S\Q\rho] \\\nonumber&&\quad+\Tr (\J e^{t\L}\P\J \S^2   \Q\rho)\Bigg|\\\nonumber&&\approx\big|
	\Tr(\J e^{t \L}\L \P\J \S^3\Q\rho)-\Tr(\J \P\J \S^2 e^{t\L}\Q\rho)\big|
	\\\nonumber&&\lesssim (2\Cmm^n+\Cmm) \lVert \S\Q\rVert^2\lVert\J\P\rVert\lVert\J\rVert\lVert\Q\rho\rVert,
\end{eqnarray}
where in the approximation we used $\int_{0}^t d t_2\,e^{t_2\lambda_k} e^{(t-t_2)\lambda_l}=(e^{t\lambda_k}-e^{t\lambda_l})/(\lambda_k-\lambda_l)\approx -e^{t\lambda_k}/\lambda_l-\lambda_k e^{t\lambda_k}/\lambda_l^2+e^{t\lambda_l}/\lambda_l$ for $|\lambda_k|\ll |\lambda_l|$ [we have $|\lambda_k|\ll -\lambda_l^R$ for $k\leq m<l$; cf.~Sec.~\ref{app:defMM}]. In the inequality we used  $\lVert e^{t\L}\P\rVert\leq1+\Cp$ together with Eq.~\eqref{eq:approx_1} for time $ \lVert\S \Q\rVert$.	
Analogously to Eq.~\eqref{eq:Kvar_approx2_CL} we have 
\begin{eqnarray}\label{eq:Kvar_approx4_CL}
	&& \Bigg| \int_0^t d  t_1 \Tr \{\J \S[e^{(t-t_1)\L}-\I]\Q\J e^{t_1 \L} \P\rho\}\\\nonumber&&
	\quad+ t\Tr (\J \S\Q\J \rhoss) + \Tr[\J \S\Q\J \S(e^{t \L}-\I) \P\rho] 	\\\nonumber&&
	\quad- 	 \Tr(\J \S^2\Q\J e^{t\L}\P\rho) \Bigg| \\\nonumber&&\lesssim\big|\Tr (\J \S^2\Q\J e^{t\L}\L\P\rho) -\Tr (\J \S^3 e^{t\L}\Q\J \P\rho) \bigg|\\\nonumber&&
	\lesssim(2\Cmm^n+\Cmm)\lVert\J\rVert\lVert\J\P\rVert  \lVert\S \Q\rVert^2.
\end{eqnarray} 
Finally, from Eq.~\eqref{eq:approx_2}, we have
\begin{eqnarray}\label{eq:Kvar_approx6_CL}
	&& \Bigg| \int_0^t d  t_1 \Tr \{\J \S[e^{(t-t_1)\L}-\I]\Q\J e^{t_1 \L} \Q\rho\}\\\nonumber&&
	\quad+ \Tr (\J \S\Q\J \S\Q\rho) \Bigg|\\\nonumber
	&& =\Big| \!\! \int_{0}^t\!\!\!  \Tr [\J\S e^{(t-t_1)\L}\Q\J e^{t_1\L}\Q\rho] \!+\! \Tr (\J \S\Q \J\S e^{t\L}\Q\rho)  \Big| \\\nonumber&&
	\lesssim \big(6\Cmm^n+4\sqrt{\Cmm^n}\big)\lVert \J\rVert^2\lVert\S\Q\rVert^2 \lVert\Q\rho\rVert,
\end{eqnarray}
where in the last inequality we used $\int_{0}^t d t_1  \Tr [\J\S e^{(t-t_1)\L}\Q\J e^{t_1\L}\Q\rho] =(\int_{0}^{\frac{t}{2}} +\int_{\frac{t}{2}}^{t}) d t_1  \Tr [\J\S e^{(t-t_1)\L}\Q\J e^{t_1\L}\Q\rho]$, while $\Tr [\J\S e^{(t-t_1)\L}\Q\J e^{t_1\L}\Q\rho]$ can be bounded by $\lVert\J\rVert^2\lVert\S\Q\rVert\lVert e^{(t-t_1)\L}\Q\rVert \lVert\Q\rho\rVert$ or $\lVert\J\rVert^2\lVert\S\Q\rVert \lVert e^{t_1\L}Q\rVert \lVert\Q\rho\rVert$ (used in the first and the second integrals, respectively). We also assumed that $n\geq 2$, so that $t/2\geq t''$ [cf.~Eq.~\eqref{eq:approx_2}].

From Eqs.~(\ref{eq:Kvar_approx1_CL})-(\ref{eq:Kvar_approx6_CL}) we arrive at [cf.~Eqs.~\eqref{eq:Kav_t} and~\eqref{eq:Kvar_t}]
\begin{eqnarray}\label{eq:Kvar_t_approx_CL0}
	&&\bigg|\langle K^2(t) \rangle\!-\!\langle K(t) \rangle^2  \qquad \\\nonumber
	&&\quad-t\mu_\text{ss} - \,\Tr[\J \S (e^{t\L}-\I)\P\rho]+\Tr(\J \S \Q\rho)\\\nonumber
	&&\quad -2 \int_0^t d  t_1\int_0^{t-t_1} d  t_2  \Tr (\J e^{t_2\L}\P\J e^{t_1\L} \P\rho)\\\nonumber
	&&\quad+\big\{t\mu_\text{ss} + \,\Tr[\J \S (e^{t\L}-\I)\P\rho]-\Tr(\J \S \Q\rho)\big\}^2 
	\\\nonumber
	&&\quad+2 t\mu_\text{ss}\Tr(\J \S\Q\rho) + 2\Tr[\J \S(e^{t \L}-\I) \P\J \S\Q\rho]\\\nonumber
	&&\quad+2\Tr (\J e^{t\L}\P\J \S^2   \Q\rho)	\\\nonumber
	&&\quad+2t \Tr (\J \S\Q\J \rhoss) + 2\Tr[\J \S\Q\J \S(e^{t \L}-\I) \P\rho]\\\nonumber
	&&\quad+ 2\Tr (\J \S^2\Q\J e^{t\L}\P\rho)+2 \Tr (\J \S\Q\J \S\Q\rho) \bigg|
	\\\nonumber&&	
	\lesssim  (4\Cmm^n+2\Cmm)\lVert \J\P\rVert\lVert \J\rVert  \lVert\S\Q\rVert^2\big(1+\lVert\Q\rho\rVert\big)\\\nonumber
	&&\quad+\big(12\Cmm^n+8\sqrt{\Cmm^n}\big)\lVert \J\rVert^2\lVert\S\Q\rVert^2 \lVert\Q\rho\rVert
	\\\nonumber
	&&\quad+2 \Cmm^n \lVert \J\rVert  \lVert\S\Q\rVert\lVert\Q\rho\rVert
	\\\nonumber
	&&\quad+4 \Cmm^n \lVert \J\rVert \langle K(t)\rangle \lVert\S\Q\rVert\lVert\Q\rho\rVert,
\end{eqnarray}
where $n$ is such that $t''\leq t/n\leq t'$.
For $t\gg\lVert\S\Q\rVert$ [cf.~Eq.~\eqref{eq:cond_var_t_1}], we can further neglect the constant contribution, which gives
\begin{eqnarray}\label{eq:Kvar_t_approx_CL1}
	&&\bigg|\frac{\langle K^2(t) \rangle\!-\!\langle K(t) \rangle^2}{t}-\mu_\text{ss} - \frac{\Tr[\J \S (e^{t\L}-\I)\P\rho]}{t}\qquad\\\nonumber
	&&\quad -\frac{2}{t} \int_0^t d  t_1\int_0^{t-t_1} d  t_2  \Tr (\J e^{t_2\L}\P\J e^{t_1\L} \P\rho)\\\nonumber
	&&\quad+\frac{\big\{t\mu_\text{ss} +\Tr[\J \S (e^{t\L}-\I)\P\rho]\big\}^2}{t} \\\nonumber
	&&\quad-2  \Tr[\J \S (e^{t\L}-\I)\P\rho]\frac{\Tr(\J \S \Q\rho)}{t} \\\nonumber
	&&\quad +2 \frac{\Tr[\J \S(e^{t \L}-\I) \P\J \S\Q\rho]}{t}\\\nonumber
	&&\quad+2 \Tr (\J \S\Q\J \rhoss) + 2\frac{\Tr[\J \S\Q\J \S(e^{t \L}-\I) \P\rho]}{t} \bigg|
	\\\nonumber&&	
	\lesssim   \frac{\lVert\S\Q\rVert}{t}\lVert \J\rVert\lVert\Q\rho\rVert+2\frac{\lVert\S\Q\rVert^2}{t}\lVert \J\rVert\lVert \J\P\rVert(1+\lVert\Q\rho\rVert)
	\\\nonumber&&\quad+\frac{\lVert\S\Q\rVert^2}{t}\lVert \J\rVert^2\lVert\Q\rho\rVert(2+\lVert\Q\rho\rVert)
	\\\nonumber&&\quad+4 \Cmm^n  \bigg(\lVert \bm{\mut}\rVert_1+ \frac{\lVert\S\Q\rVert}{t}\lVert \J\rVert\bigg) \lVert \J\rVert \lVert\S\Q\rVert \lVert\Q\rho\rVert
\end{eqnarray}
We note that in Eqs.~(\ref{eq:Kvar_approx1_CL})-(\ref{eq:Kvar_approx6_CL}) $\lVert\Q\rho\rVert$ can be simply replaced by $1$.

We recognize
\begin{eqnarray}\nonumber
	&&t\mu_\text{ss} + \Tr[\J \S (e^{t\L}-\I)\P\rho]=\int_0^t d  t_1  \sum_{l=1}^m \big(\Jt e^{t_1\Wt}\pt\big)_l\qquad\\
	&&\qquad= \int_0^t d  t_1  \sum_{l=1}^m \big(\bm{\mut} e^{t_1\Wt}\pt\big)_l =\langle \tilde{K}_\text{cl}(t)\rangle, \label{eq:Kvar_t_approx_CL2}
\end{eqnarray}
where $\tilde{K}_\text{cl}(t)$ is the time-integral of the variable with the cumulant generating function encoded by $\Wt_s$ in Eq.~\eqref{eq:Wstilde}, and, similarly,
\begin{eqnarray}\label{eq:Kvar_t_approx_CL3}
	&&2\int_0^t d  t_1\int_0^{t-t_1} d  t_2  \Tr (\J e^{t_2\L}\P\J e^{t_1\L} \P\rho)\qquad\qquad\\\nonumber
	&&\quad=2\int_0^t d  t_1\int_0^{t-t_1} d  t_2   \sum_{l=1}^m \big(\Jt e^{t_2\Wt}\Jt e^{t_1\Wt}\pt\big)_l\\\nonumber
	&&\quad=2\int_0^t d  t_1\int_0^{t-t_1} d  t_2   \sum_{l=1}^m \big(\bm{\mut} e^{t_2\Wt}\Jt e^{t_1\Wt}\pt\big)_l\\\nonumber
	&&\quad=\langle \tilde{K}_\text{cl}^2(t)\rangle-\langle \tilde{K}_\text{cl}(t)\rangle. 
\end{eqnarray}
Furthermore,
\begin{eqnarray}\label{eq:Kvar_t_approx_CL4}
	&&t\mu_\text{ss} + \,\Tr[\J \S (e^{t\L}-\I)\P\rho]\qquad\qquad\\\nonumber
	&&-2 t\Tr (\J \S\Q\J \rhoss)-2  \Tr[\J \S\Q\J \S(e^{t \L}-\I) \P\rho]\qquad\qquad\\\nonumber
	&&\quad=\sum_{l=1}^m \!\big\{\bm{\tilde{\sigma}}^2\big[t\ptss\!+\!\Rt\big(e^{t\Wt}\!\!-\Ibf\big)\pt\big]\!\big\}_l\\\nonumber
	&&\quad=\int_0^t d  t_1 \sum_{l=1}^m \big(\bm{\tilde{\sigma}}^2 e^{t_1\Wt}\pt\big)_l =\langle \tilde{K}_\text{cl}(t)\rangle+\langle \Delta_\text{cl}(t)\rangle
\end{eqnarray}
[cf.~Eq.~\eqref{eq:sigmat}] and
\begin{eqnarray}\label{eq:Kvar_t_approx_CL5}
	&&2  \Tr[\J \S (e^{t\L}-\I)\P\rho]\Tr(\J \S \Q\rho) \\\nonumber
	&&\quad -2 \Tr[\J \S(e^{t \L}-\I) \P\J \S\Q\rho]\\\nonumber
	&&=-2\int_0^t d  t_1 \sum_{l=1}^m \big(\bm{\mut} e^{t_1\Wt}\pt\big)_l \sum_{k=1}^m \big(\bm{\tilde{K}}\pt\big)_k\\\nonumber
	&&\quad  + 2\int_0^t d  t_1 \sum_{l=1}^m \big(\bm{\mut} e^{t_1\Wt}\bm{\tilde{K}}\pt\big)_l \\\nonumber
	&&=\sum_{k,l=1}^m\tilde{p}_k\tilde{p}_l \big[\langle \tilde{K}_\text{cl}^{(k)}(t)\rangle-\langle \tilde{K}_\text{cl}^{(l)}(t)\rangle\big]\big(\tilde{K}_k-\tilde{K}_l\big),
\end{eqnarray}
where $\bm{\tilde{K}}$ is defined in Eq.~\eqref{eq:Ktilde2} and $\langle \tilde{K}_\text{cl}^{(l)}(t)\rangle$ is $\langle \tilde{K}_\text{cl}(t)\rangle$ for the system in the $l$th metastable phase during metastable regime, $\P(\rho)=\rho_l$, $k,l=1,...,m$.
This, together with Eq.~\eqref{eq:activity_CL0}, leads to Eq.~\eqref{eq:Kvar_t_approx_CL_corr} by replacing the long time-dynamics generator $\Wt$ by the stochastic dynamics generator $\W$ [cf.~Eqs.~\eqref{eq:deltaW},~\eqref{eq:etW_CL} and~\eqref{eq:bound_D}].\\

\emph{Derivation} of Eqs.~\eqref{eq:Xav_t_approx_CL_corr} and~\eqref{eq:Mav_t_approx_CL_corr}.    	We replace $\J$ by $\X$ and $\M$, in Eq.~\eqref{eq:Kvar_approx1_CL}, and $\bm{\mut}$ by $\xt$ and $\mt$ in Eq.~\eqref{eq:activity_CL0}, respectively.\\

\emph{Derivation} of Eqs.~\eqref{eq:Xvar_t_approx_CL_corr} and~\eqref{eq:Mvar_t_approx_CL_corr}.    	From Eqs.~\eqref{eq:Xav_t} and~\eqref{eq:Xvar_t}, using Eqs.~(\ref{eq:Kvar_approx1_CL})-(\ref{eq:Kvar_approx6_CL}), we arrive at [cf.~Eq.~\eqref{eq:Kvar_t_approx_CL0}]
\begin{eqnarray}\label{eq:Xvar_t_approx_CL0}
	&&\bigg|\langle X^2(t) \rangle\!-\!\langle X(t) \rangle^2 -\frac{t}{2} \qquad \\\nonumber
	&&\quad -2 \int_0^t d  t_1\int_0^{t-t_1} d  t_2  \Tr (\X e^{t_2\L}\P\X e^{t_1\L} \P\rho)\\\nonumber
	&&\quad+\big\{tx_\text{ss} + \,\Tr[\X \S (e^{t\L}-\I)\P\rho]-\Tr(\X \S \Q\rho)\big\}^2 
	\\\nonumber
	&&\quad+ 2 tx_\text{ss} \Tr(\X \S\Q\rho) +2 \Tr[\X \S(e^{t \L}-\I) \P\X \S\Q\rho]\\\nonumber
	&&\quad+2\Tr (\X e^{t\L}\P\X \S^2   \Q\rho)	\\\nonumber
	&&\quad+2t \Tr (\X \S\Q\X \rhoss) + 2\Tr[\X \S\Q\X \S(e^{t \L}-\I) \P\rho]\\\nonumber
	&&\quad+2 \Tr (\X \S^2\Q\X e^{t\L}\P\rho)+ 2\Tr (\X \S\Q\X \S\Q\rho) \bigg|
	\\\nonumber&&	
	\lesssim  (4\Cmm^n+2\Cmm)\lVert \X\P\rVert\lVert \X\rVert  \lVert\S\Q\rVert^2\big(1+\lVert\Q\rho\rVert\big)\\\nonumber
	&&\quad+\big(12\Cmm^n+8\sqrt{\Cmm^n}\big)\lVert \X\rVert^2\lVert\S\Q\rVert^2 \lVert\Q\rho\rVert
	\\\nonumber
	&&\quad+4 \Cmm^n \lVert \X\rVert  \langle X(t)\rangle\lVert\S\Q\rVert\lVert\Q\rho\rVert,
\end{eqnarray}
where $n$ is such that $t''\leq t/n\leq t'$. Thus,
[cf.~Eq.~\eqref{eq:Kvar_t_approx_CL1}]
\begin{eqnarray}\label{eq:Xvar_t_approx_CL1}
	&&\bigg|\frac{\langle X^2(t) \rangle-\langle X(t) \rangle^2}{t}\qquad\\\nonumber
	&&\quad -\frac{2}{t} \int_0^t d  t_1\int_0^{t-t_1} d  t_2  \Tr (\X e^{t_2\L}\P\X e^{t_1\L} \P\rho)\\\nonumber
	&&\quad+\frac{\big\{t x_\text{ss} +\Tr[\X \S (e^{t\L}-\I)\P\rho]\big\}^2}{t} \\\nonumber
	&&\quad-2  \Tr[\X \S (e^{t\L}-\I)\P\rho]\frac{\Tr(\X \S \Q\rho)}{t} \\\nonumber
	&&\quad +2 \frac{\Tr[\X \S(e^{t \L}-\I) \P\X \S\Q\rho]}{t}\\\nonumber
	&&\quad+2 \Tr (\X \S\Q\X \rhoss) + 2\frac{\Tr[\X \S\Q\X \S(e^{t \L}-\I) \P\rho]}{t} \bigg|
	\\\nonumber&&	
	\lesssim  \frac{\lVert\S\Q\rVert^2}{t}\lVert \X\rVert\big[ 2\lVert \X\P\rVert(1+\lVert\Q\rho\rVert)+\lVert \X\rVert\lVert\Q\rho\rVert(2+\lVert\Q\rho\rVert)\big]
	\\\nonumber&&\quad+4 \Cmm^n  \bigg(\lVert \xt\rVert_1+ \frac{\lVert\S\Q\rVert}{t}\lVert \X\rVert\bigg) \lVert \X\rVert \lVert\S\Q\rVert \lVert\Q\rho\rVert
\end{eqnarray}

Eq.~\eqref{eq:Mvar_t_approx_CL_corr} follows analogously, by replacing $\X$ by $\M$ an dropping $t/2$ and $1/2$ in Eqs.~\eqref{eq:Xvar_t_approx_CL0}  and~\eqref{eq:Xvar_t_approx_CL1}, respectively.

\subsubsection{Rates of average and fluctuations in quantum trajectories during metastable regime} \label{app:var_ms}

We now discuss rates of average and fluctuations within metastable regime for jump number [Eq.~\eqref{eq:Ls}], integrated homodyne current [Eq.~\eqref{eq:Lr}] and time-integral of a system observable  [Eq.~\eqref{eq:Lh}]. We give corrections to the classical approximation of the rate of average and fluctuations as given in Eqs.~\eqref{eq:Kav},~\eqref{eq:Kvar2} for jump number, Eqs.~\eqref{eq:Xav} and~\eqref{eq:Xvar2} for integrated homodyne current, and Eqs.~\eqref{eq:Mav} and~\eqref{eq:Mvar2} for time-integral of system-observables (see also Ref.~\cite{Macieszczak2016}). Derivations rely on the master operator in Eq.~\eqref{eq:L} being diagonalizable and can be found at the end of this section.\\

\emph{Rates of average and fluctuations of jump number during metastable regime}. For a general initial state $\rho$, the rate of average number of jumps within the metastable regime, $t''\leq t\leq t'$, is approximated as [cf.~Eq.~\eqref{eq:Kav_t_approx_CL_corr}] 
\begin{eqnarray}\label{eq:Kav_t_approx}
	&&\bigg|\frac{\langle K(t) \rangle}{t} - \sum_{l=1}^m \tilde{p}_l\mut_l +\frac{\tilde{K}}{t}\bigg|\\\nonumber&&\lesssim \Cmm\Big\{\frac{1}{2}  \min\Big[\lVert\J\P\rVert,\Big(1+\frac{\Ctcl}{2}\Big) \lVert\bm{\mut}\rVert_1 \Big]  +   \lVert \J\rVert \frac{\lVert\S\Q\rVert}{t}\Big\},
\end{eqnarray}
where $\tilde{p}_l=\Tr(\tilde{P}_l \rho)$ [cf.~Eq.~\eqref{eq:Ptilde}],
\begin{equation}\label{eq:activity_ms}
	\mut_l\equiv\Tr\big(\sum_j\!J_j^\dagger J_j \trho_l\big),
\end{equation}
and $ \tilde{K} \equiv - \Tr ( \J  \S  \Q\rho)$ is the contribution to the number of jumps from before the metastable regime. We have $\lVert\bm{\mut}\rVert_1=\max_{1\leq l \leq m}|\mut_l|$, $\lVert\J\rVert=\lVert\sum_j\!J_j^\dagger J_j\rVert_{\max}$, and $\lVert \J\P\rVert\leq \lVert \J\rVert$ can be replaced by $\lVert \bm{\mut}\rVert_1+2\Cp\lVert \J\rVert$~\cite{Note3}. We can further replace  $\mut_{l}$  by $\mut_{l}^\text{tot}$ introducing additional corrections bounded by $\sqrt{\Ccl}\lVert \Wt\rVert_1$. If 
\begin{equation}\label{eq:av_cond}
	t'\lVert\bm{\mut}\rVert_1\geq t\lVert\bm{\mut}\rVert_1  \gg \lVert\S\Q\rVert\lVert \J\rVert,
\end{equation}
the contribution $ \tilde{K}/t$ from before the metastable regime can be neglected [cf.~Eq.~\eqref{eq:cond_av_t_1}].

For time $t$ in Eq.~\eqref{eq:av_cond}, the rate of fluctuations of jump number is approximated as [cf.~Eq.~\eqref{eq:Kvar_t_approx_CL_corr}] 
\begin{eqnarray}\label{eq:Kvar_t_approx}
	&&\bigg|\frac{\langle K^2(t) \rangle-\langle K(t) \rangle^2}{t} \\\nonumber
	&&\quad- t\!\sum_{k,l=1}^m \!\! \Big[\frac{1}{2} \tilde{p}_k \tilde{p}_l(\mut_{k}-\mut_{l})^2  - \mut_{k}(\Wt)_{kl}\,\tilde{p}_l\Big]\\\nonumber
	&&\quad- \!\sum_{l=1}^m  \tilde{p}_l \tilde{\sigma}^2_{l}- \!\sum_{k,l=1}^m \!\! \tilde{p}_k \tilde{p}_l(\mut_{k}-\mut_{l})(\tilde{K}_{k}-\tilde{K}_{l}) \bigg|
	\\\nonumber
	&&\lesssim   \frac{\lVert\S\Q\rVert}{t}\lVert \J\rVert\Big[1+  \lVert\S\Q\rVert\big(4\lVert \J\P\rVert
	+3\lVert \J\rVert\big)\big]
	\\\nonumber&&\quad+4 \Cmm  \bigg(\lVert \bm{\mut}\rVert_1+ \frac{\lVert\S\Q\rVert}{t}\lVert \J\rVert\bigg) \lVert \J\rVert \lVert\S\Q\rVert 
	\\\nonumber&&\quad+ t^2\min\big(\lVert \LMM\rVert \lVert\J\P\rVert^2,\lVert \Wt\rVert_1 \lVert\bm{\mut}\rVert_1\lVert\Jt\rVert_1\big) 
	\\\nonumber&&\quad+ t^2\min\big(\lVert \LMM\rVert \lVert\J\P\rVert,\lVert \Wt\rVert_1 \lVert\bm{\mut}\rVert_1\big) \lVert\bm{\mut}\rVert_1\\\nonumber
	&&\quad+ t\lVert\bm{\mut}\rVert_1  m\,\Big\lVert H \!+\!\frac{i}{2}\! \sum_{j}\!\! {J}_{j}^{\dagger} J_{j}\Big\rVert_{\max} \!\! \sqrt{2\Ccl +4  \Cp}
	\\\nonumber&&\quad+\frac{1}{2}\Cmm\min\Big[\lVert\J\P\rVert\big(1+2\lVert \J\rVert\lVert\S\Q\rVert\big), \Big( 1+\frac{\Ctcl}{2}\Big) \lVert\bm{\tilde{\sigma}^2}\rVert_1 \Big]\\\nonumber&&\quad+2 \Cmm \min\Big[\lVert\J\P\rVert\lVert \J\rVert\lVert\S\Q\rVert,\Big( 1+\frac{\Ctcl}{2}\Big)  \lVert\bm{\mut}\rVert_1\lVert\bm{\tilde{K}}\pt\rVert_1\Big] ,
\end{eqnarray}
where 
\begin{equation}\label{eq:sigma_ms}
	\tilde{\sigma}_l^2\equiv \mut_l - 2 \Tr (\J \S\Q \J \trho_l),
\end{equation}
[cf.~Eq.~\eqref{eq:sigmat}] and $\tilde{K}_l \equiv- \Tr ( \tilde{P}_l\J  \S  \Q\rho)/\Tr (\tilde{P}_l\rho)$,  $l=1,...,m$, is the contribution to the number of jumps from before the metastable regime, conditioned on which the metastable phase the state evolves into.  We have $(1-\Cp-\Ccl)\lVert \LMM\rVert\lesssim\lVert \Wt\rVert_1\leq (1+\Ctcl/2)\lVert \LMM\rVert$ (cf.~Sec.~\ref{app:norm}).

The corrections in Eq.~\eqref{eq:Kvar_t_approx} to be negligible require conditions~\eqref{eq:cond_var_t_1},~\eqref{eq:cond_var_t_2} and Eq.~\eqref{eq:cond_var_t_4}, together with a new condition $(1+\Ctcl/2)\Cmm \ll 1$. Indeed, the first and the second lines of corrections require  Eqs.~\eqref{eq:cond_var_t_1} and~\eqref{eq:cond_var_t_2} [cf.~Eq.~\eqref{eq:Kvar_t_approx_CL_corr}]. The quadratic-in-time corrections in Eq.~\eqref{eq:Kvar_t_approx} originate from the linear term in the fluctuation rate bounded by $t\min(\lVert\J\P\rVert^2,\lVert\bm{\mut}\rVert_1\lVert\Jt\rVert_1)+t\lVert\bm{\mut}\rVert_1^2$. As $t\lVert \LMM\rVert\lesssim \Cmm$ [Eq.~\eqref{eq:approx_1}] and $\lVert \Wt\rVert_1\leq (1+\Ctcl/2)\lVert \LMM\rVert$, the third and fourth lines of corrections require $\min(1+\Ctcl/2,\lVert\J\P\rVert/\lVert\bm{\mut}\rVert_1)\,\Cmm\lesssim \min(1+\Ctcl/2,1+2\Cp\lVert \J\rVert/\lVert\bm{\mut}\rVert_1)\,\Cmm\ll 1$ [cf.~Eq.~\eqref{eq:approx_1} and see~\cite{Note3}] to be negligible in comparison. This condition also guarantees that the seventh line [corrections to the fluctuation rate from before the metastable regime bounded by $4  \min(\lVert\J\P\rVert\lVert\J\rVert\lVert\S\Q\rVert,\lVert\bm{\mut}\rVert_1\lVert\bm{\tilde{K}}\pt\rVert_1)$] is negligible. The fifth line of corrections requires Eq.~\eqref{eq:cond_var_t_4} [cf.~Eq.~\eqref{eq:Kvar_t_approx_CL_corr}]. Finally, the corrections in the sixth line are negligible  when $\min[1+\Ctcl/2,\lVert\J\P\rVert\max (1,\lVert\S\Q\rVert\lVert \J\rVert)/\lVert\bm{\tilde{\sigma}}^2\rVert_1]\,\Cmm\ll 1$.

We can further replace    $\mut_{l}$  by $\mut_{l}^\text{tot}$, $\Wt$ by $\W$, and $\tilde{\sigma}_{l}^2$ by $(\tilde{\sigma}_{l}^\text{tot})^2$ [cf.~Eq.~\eqref{eq:deltaW}] in Eqs.~\eqref{eq:Kvar_t_approx}, introducing additional corrections bounded by $\sqrt{\Ccl}\lVert \Wt\rVert_1[ t(4\lVert\bm{\mut}\rVert_1+\lVert\Wt\rVert_1)+1 +4\min(\lVert\J\rVert\lVert\S\Q\rVert, \lVert\bm{\tilde{K}}\pt\rVert_1)]$, which to be negligible require Eq.~\eqref{eq:cond_var_t_3}. 
Furthermore,  when  the internal activity dominates transition rates in the long-time dynamics,
\begin{equation}\label{eq:tot_2_in_approx}
	\lVert \Wt\rVert_1\ll \lVert \bm{\mut}\rVert_1,
\end{equation}
we can neglect the contribution from the long-time dynamics (non-Poissonian fluctuations contribution to the classical fluctuation rate); we then obtain Eq.~\eqref{eq:kvar_ms}.
Alternatively, this contribution can be neglected when 
\begin{equation}\label{eq:var_ms_cond}
	\frac{\min \big[\lVert \J\P\rVert,\big(1+\frac{\Ctcl}{2}\big)\lVert\bm{\mut}\rVert_1 \big]}{\lVert\bm{\tilde{\sigma}}^2\rVert_1} \ll \frac{1}{\Cmm},
\end{equation}
which condition can be viewed as the lower bound on the anti-bunching of fluctuations inside metastable phases.\\

\emph{Rates of average and fluctuations of jump number in metastable phases}. For the initial state $\rho$ chosen as the closest state to the projection $\trho_l$ of $l$th metastable phase  [Eq.~\eqref{eq:rhotilde}], we have 
\begin{eqnarray} \label{eq:Kav_ms}
	&&\bigg|\!\frac{\langle K(t) \rangle}{t} \!- \mut_{l}\bigg|\lesssim 2\Cp \lVert\J\rVert\frac{\lVert\S\Q\rVert}{t}\\\nonumber
	&&\quad+\Big(\frac{1}{2}\Cmm+ 2\Cp\Big) \min\Big[\lVert\J\P\rVert,\!\Big(1\!+\!\frac{\Ctcl}{2}\Big) \lVert\bm{\mut}\rVert_1 \Big]
\end{eqnarray}
Therefore, if $ \min(1+\Ctcl/2,\lVert\bm{\mut}\rVert_1/\lVert \J\P\rVert) (\Cmm+\Cp)\ll 1$, and there exists time $t\lVert\bm{\mut}\rVert_1\gg\Cp\lVert\S\Q\rVert   \lVert\J\rVert$ [cf.~Eq.~\eqref{eq:av_cond}], the corrections are negligible, and $\mut_l$ in Eq.~\eqref{eq:activity_ms} is the activity of $l$th metastable phase  (see also Sec.~\ref{app:distribution}).

For the initial state chosen as the closest state to the projection $\trho_l$ of $l$th metastable phase, we also have 
\begin{eqnarray}\label{eq:Kvar_ms}
	&&\bigg|\frac{\langle K^2(t) \rangle\!-\!\langle K(t) \rangle^2}{t}-\tilde{\sigma}_l^2 - t \sum_{k=1}^m \mut_k (\Wt)_{kl} \bigg|
	\\\nonumber&&	
	\lesssim   2\frac{\lVert\S\Q\rVert^2}{t}\lVert \J\rVert\lVert \J\P\rVert
	\\\nonumber&&\quad+ t^2\min\big(\lVert \LMM\rVert \lVert\J\P\rVert^2,\lVert \Wt\rVert_1 \lVert\bm{\mut}\rVert_1\lVert\Jt\rVert_1\big) 
	\\\nonumber&&\quad+ t^2\min\big(\lVert \LMM\rVert \lVert\J\P\rVert,\lVert \Wt\rVert_1 \lVert\bm{\mut}\rVert_1\big) \lVert\bm{\mut}\rVert_1
	\\\nonumber
	&&\quad+ t\lVert\bm{\mut}\rVert_1  m\,\Big\lVert H \!+\!\frac{i}{2}\! \sum_{j}\!\! {J}_{j}^{\dagger} J_{j}\Big\rVert_{\max} \!\! \sqrt{2\Ccl +4  \Cp}
	\\\nonumber&&\quad+ t \Cp\min\Big[3 \lVert\J\P\rVert^2,\Big( 1+\frac{\Ctcl}{2}\Big)  \lVert\bm{\mut}\rVert_1\big(\lVert\Jt\rVert_1+2\lVert\bm{\mut}\rVert_1\big)\Big] 
	\\\nonumber&&\quad+\big(\frac{1}{2}\Cmm+\Cp\Big)\\\nonumber&&
	\quad\qquad\times\min\Big[\lVert\J\P\rVert\big(1+2\lVert \J\rVert\lVert\S\Q\rVert), \Big( 1+\frac{\Ctcl}{2}\Big)  \lVert\bm{\tilde{\sigma}^2}\rVert_1 \Big]\\\nonumber
	&&\quad+4\Cp \lVert\J\P\rVert\lVert\J\rVert\lVert\S\Q\rVert.
\end{eqnarray}
The contribution independent of long-time dynamics, $(\delta_{\bm{\tilde{\sigma}}^2})_l\equiv\tilde{\sigma}_l^2-(\mut)_l$ [cf.~Eq.~\eqref{eq:sigma_ms}], can thus be viewed as the non-Poissonian contribution to fluctuations of $l$th metastable phase (see also Sec.~\ref{app:distribution}).   We note that $\tilde{\sigma}_l^2$ can be further replaced by $(\tilde{\sigma}_l^\text{tot})^2$ introducing additional corrections bounded by $\sqrt{\Ccl}\lVert \Wt\rVert_1$ [cf.~Eq.~\eqref{eq:deltaW}], which requires Eq.~\eqref{eq:cond_var_t_5}.

The corrections in Eq.~\eqref{eq:Kvar_ms} to be negligible in comparison with $\lVert\bm{\tilde{\sigma}}^2\rVert_1$, imply a lower and upper bounds on time $t$, the condition $(1+\Ctcl/2)\Cmm\ll 1$, as well as the following condition on the anti-bunching of fluctuations inside metastable phases,
\begin{equation}\label{eq:var_ms_cond0}
	\frac{\lVert \J\P\rVert\lVert \J\rVert\lVert\S\Q\rVert}{\lVert\bm{\tilde{\sigma}}^2\rVert_1} \ll \frac{1}{\Cp},
\end{equation}
A similar condition in Eq.~\eqref{eq:var_ms_cond} guarantees that we can neglect the linear-in-time contribution (non-Poissonian fluctuations in the classical dynamics). 

We note that, alternatively, for the approximations in Eqs.~\eqref{eq:Kav_ms} and~\eqref{eq:Kvar_ms} we can consider an initial state $\rho$ such that $\P(\rho)=\trho_l$, in which case the corrections are bounded as in Eqs.~\eqref{eq:Kav_t_approx_CL_corr} and~\eqref{eq:Kvar_t_approx_CL_corr}, respectively. \\

\emph{Rates of average and fluctuations of time-integrated homodyne current during metastable regime}. During the metastable regime, we have  for the homodyne current [cf.~Eq.~\eqref{eq:Xav_t_approx_CL_corr}] 
\begin{eqnarray}\label{eq:Xav_t_approx}
	&&\bigg|\frac{\langle X(t) \rangle}{t} - \sum_{l=1}^m \tilde{p}_l\tilde{x}_l +\frac{\tilde{X}}{t}\bigg|\\\nonumber&&\lesssim\Cmm\bigg\{\frac{1}{2} \min\Big[\lVert\X\P\rVert,\!\Big(1\!+\!\frac{\Ctcl}{2}\Big)\lVert\xt\rVert_1 \Big] +2\frac{ \lVert\S\Q\rVert }{t}  \lVert\X\rVert\!\bigg\},
\end{eqnarray}
where 
\begin{equation}\label{eq:x_ms}
	\tilde{x}_{l}\equiv\Tr\big[\sum_j (e^{-i\varphi_j} J_j+e^{i\varphi_j} J_j^\dagger) \trho_l\big],
\end{equation}
with $\lVert\xt\rVert_1=\max_{1\leq l \leq m}|\tilde{x}_{l}|\leq (1+\Ctcl)\lVert\X\P\rVert$ [cf.~Eq.~\eqref{eq:rho1_CL}] and $\lVert\X\rVert\leq \lVert\sum_j e^{-i\varphi_j} J_j\rVert_{\max}$, while
[cf.~Eq.~\eqref{eq:Xvar_t_approx_CL_corr}] 
\begin{eqnarray}\label{eq:Xvar_t_approx}
	&&\bigg|\frac{\langle X^2(t) \rangle-\langle X(t) \rangle^2}{t}- \frac{t}{2} \!\sum_{k,l=1}^m \!\! \tilde{p}_k \tilde{p}_l(\tilde{x}_{k}-\tilde{x}_l)^2   \\\nonumber
	&&\quad- \!\sum_{l=1}^m  \tilde{p}_l \tilde{\chi}^2_{l}- \!\sum_{k,l=1}^m \!\! \tilde{p}_k \tilde{p}_l(\tilde{x}_{k}-\tilde{x}_{l})(\tilde{X}_{k}-\tilde{X}_{l}) \bigg|
	\\\nonumber	
	&&\lesssim   \frac{\lVert\S\Q\rVert^2}{t}\lVert \X\rVert \big(4\lVert \X\P\rVert +3\lVert \X\rVert \big)
	\\\nonumber	
	&&\quad+4 \Cmm \lVert\xt\rVert_1\lVert \X\rVert \lVert\S\Q\rVert
	\\\nonumber&&\quad+ t^2\min\big(\lVert \LMM\rVert \lVert\X\P\rVert^2,\lVert \Wt\rVert_1 \lVert\xt\rVert_1\lVert\Xt\rVert_1\big) 
	\\\nonumber&&\quad+ t^2\min\big(\lVert \LMM\rVert \lVert\X\P\rVert,\lVert \Wt\rVert_1 \lVert\xt\rVert_1\big) \lVert\xt\rVert_1\\\nonumber
	&&\quad+ t\lVert\xt\rVert_1  m\,\Big\lVert \sum_{j} e^{-\varphi_j}{J}_{j}\Big\rVert_{\max} \sqrt{\frac{\Ccl}{2} + \Cp}
	\\\nonumber&&\quad+2 \Cmm \lVert\X\P\rVert\lVert \X\rVert\lVert\S\Q\rVert
	\\\nonumber&&\quad+\frac{1}{2}\Cmm\min\Big[2\lVert\X\P\rVert\lVert \X\rVert\lVert\S\Q\rVert, \Big( 1+\frac{\Ctcl}{2}\Big) \lVert\bm{\tilde{\chi}^2}\rVert_1 \Big],
\end{eqnarray}
where [cf.~Eq.~\eqref{eq:chit}]
\begin{equation}\label{eq:chi_ms}
	\tilde{\chi}_l^2\equiv \frac{1}{2}- 2 \Tr (\X \S\Q \X \trho_l)
\end{equation}
and $\tilde{X}_l \equiv- \Tr ( \tilde{P}_l\X  \S  \Q\rho)/\Tr (\tilde{P}_l\rho)$,  $l=1,...,m$.

For  the initial state chosen as the closest state to the projection $\trho_l$ of $l$th metastable phase, we further have [cf.~Eq.~\eqref{eq:Kav_ms}] 
\begin{eqnarray}\label{eq:Xav_ms}
	&&\bigg|\!\frac{\langle X(t) \rangle}{t} \!- \tilde{x}_{l}\bigg|\lesssim 2\Cp \lVert\X\rVert\frac{\lVert\S\Q\rVert}{t}\! \\\nonumber
	&&\quad+\Big(\frac{1}{2}\Cmm+ 2\Cp\Big)\min\Big[\lVert\X\P\rVert,\!\Big(1\!+\!\frac{\Ctcl}{2}\Big)\lVert\xt\rVert_1 \Big]\qquad 
\end{eqnarray}
as well as [cf.~Eq.~\eqref{eq:Kvar_ms}] 
\begin{eqnarray}\label{eq:Xvar_ms}
	&&\bigg|\frac{\langle X^2(t) \rangle\!-\!\langle X(t) \rangle^2}{t}-\tilde{\sigma}_l^2 \bigg|
	\\\nonumber&&
	\lesssim   2\frac{\lVert\S\Q\rVert^2}{t}\lVert \X\rVert\lVert \X\P\rVert 
	\\\nonumber&&\quad+ t^2\min\big(\lVert \LMM\rVert \lVert\X\P\rVert^2,\lVert \Wt\rVert_1 \lVert\xt\rVert_1\lVert\Xt\rVert_1\big) 
	\\\nonumber&&\quad+ t^2\min\big(\lVert \LMM\rVert \lVert\X\P\rVert,\lVert \Wt\rVert_1 \lVert\xt\rVert_1\big) \lVert\xt\rVert_1
	\\\nonumber
	&&\quad+ t\lVert\xt\rVert_1  m\,\Big\lVert \sum_{j} e^{-\varphi_j}{J}_{j}\Big\rVert_{\max}  \sqrt{\frac{\Ccl}{2} + \Cp}
	\\\nonumber&&\quad+ t \Cp\min\Big[3 \lVert\X\P\rVert^2,\Big( 1+\frac{\Ctcl}{2}\Big) \lVert\xt\rVert_1\big(\lVert\Xt\rVert_1+2\lVert\xt\rVert_1\big)\Big] 
	\\\nonumber&&\quad+\big(\frac{1}{2}\Cmm+\Cp\Big)
	\\\nonumber&&\quad\qquad\times\min\Big[2\lVert\X\P\rVert\lVert \X\rVert\lVert\S\Q\rVert, \Big( 1+\frac{\Ctcl}{2}\Big) \lVert\bm{\tilde{\chi}^2}\rVert_1 \Big]\\\nonumber
	&&\quad+4\Cp \lVert\X\P\rVert \lVert \X\rVert\lVert\S\Q\rVert.
\end{eqnarray}
Therefore, provided that  corrections in Eq.~\eqref{eq:Xav_ms}  and~\eqref{eq:Xvar_ms} are negligible, we can conclude that $\tilde{x}_l$ in Eq.~\eqref{eq:x_ms} and $\tilde{\sigma}_l^2$ in Eq.~\eqref{eq:chi_ms} are the rates of average and fluctuations, respectively, of an integrated homodyne current in $l$th metastable phase. \\

\emph{Rates of average and fluctuations of time-integrated system observable during metastable regime}.  Analogously, for the time-ordered integral of a system observable $M$ we have [cf.~Eq.~\eqref{eq:Mav_t_approx_CL_corr}] 
\begin{eqnarray}\label{eq:Mav_t_approx}
	&&\bigg|\frac{\langle M(t) \rangle}{t} - \sum_{l=1}^m \tilde{p}_l\tilde{m}_l +\frac{\tilde{M}}{t}\bigg|\\\nonumber&&\lesssim\Cmm\bigg\{\frac{1}{2} \min\Big[\lVert\M\P\rVert,\Big(1+\frac{\Ctcl}{2}\Big) \lVert\mt\rVert_1 \Big] +2\frac{ \lVert\S\Q\rVert }{t}  \lVert\M\rVert\!\bigg\},
\end{eqnarray}
where 
\begin{equation}\label{eq:m_ms}
	\tilde{m}_{l}\equiv\Tr[M \trho_l],
\end{equation}
$\lVert\mt\rVert_1=\max_{1\leq l \leq m}|\tilde{m}_{l}|\leq (1+\Ctcl)\lVert\M\P\rVert$ [cf.~Eq.~\eqref{eq:rho1_CL}] and $\lVert\M\rVert\leq \lVert M\rVert_{\max}$, while [cf.~Eq.~\eqref{eq:Mvar_t_approx_CL_corr}] 
\begin{eqnarray}\label{eq:Mvar_t_approx}
	&&\bigg|\frac{\overline{\langle M^2(t) \rangle}-\langle M(t) \rangle^2}{t}- \frac{t}{2} \!\sum_{k,l=1}^m \!\! \tilde{p}_k \tilde{p}_l(\tilde{m}_{k}-\tilde{m}_l)^2 \qquad  \\\nonumber
	&&\quad- \!\sum_{l=1}^m  \tilde{p}_l \tilde{\delta}^2_{l}- \!\sum_{k,l=1}^m \!\! \tilde{p}_k \tilde{p}_l(\tilde{m}_{k}-\tilde{m}_{l})(\tilde{M}_{k}-\tilde{M}_{l}) \bigg|
	\\\nonumber	
	&&\lesssim   \frac{\lVert\S\Q\rVert^2}{t}\lVert \M\rVert \big( 4\lVert \M\P\rVert +3\lVert \M\rVert \big)
	\\\nonumber	
	&&\quad+4 \Cmm \lVert\mt\rVert_1\lVert \M\rVert \lVert\S\Q\rVert
	\\\nonumber&&\quad+ t^2\min\big(\lVert \LMM\rVert \lVert\M\P\rVert^2,\lVert \Wt\rVert_1 \lVert\mt\rVert_1\lVert\Mt\rVert_1\big) 
	\\\nonumber&&\quad+ t^2\min\big(\lVert \LMM\rVert \lVert\M\P\rVert,\lVert \Wt\rVert_1 \lVert\mt\rVert_1\big) \lVert\mt\rVert_1\\\nonumber
	&&\quad+ t\lVert\mt\rVert_1  m\,\lVert M\rVert_{\max}  \sqrt{\frac{\Ccl}{2} + \Cp}
	\\\nonumber&&\quad+2 \Cmm\lVert\M\P\rVert \lVert \M\rVert\lVert\S\Q\rVert
	\\\nonumber&&\quad+\frac{1}{2}\Cmm\min\Big[2\lVert\M\P\rVert\lVert \M\rVert\lVert\S\Q\rVert, \Big( 1+\frac{\Ctcl}{2}\Big) \lVert\bm{\tilde{\delta}^2}\rVert_1 \Big],
\end{eqnarray}   
where [cf.~Eq.~\eqref{eq:deltat}]
\begin{equation}\label{eq:delta_ms}
	\tilde{\delta}_l^2\equiv - 2 \Tr (\M \S\Q \M \trho_l).
\end{equation}
and $\tilde{M}_l \equiv- \Tr ( \tilde{P}_l\M \S  \Q\rho)/\Tr (\tilde{P}_l\rho)$,  $l=1,...,m$. 

For  the initial state chosen as the closest state to the projection $\trho_l$ of $l$th metastable phase, we further have [cf.~Eq.~\eqref{eq:Xav_ms}] 
\begin{eqnarray}\label{eq:Mav_ms}
	&&\bigg|\!\frac{\langle M(t) \rangle}{t} \!- \tilde{m}_{l}\bigg|\lesssim2\Cp \lVert\M\rVert\frac{\lVert\S\Q\rVert}{t}\!  \\\nonumber&&\quad+\Big(\frac{1}{2}\Cmm+ 2\Cp\Big)\min\Big[\lVert\M\P\rVert,\Big(1+\frac{\Ctcl}{2}\Big) \lVert\mt\rVert_1 \Big]  
\end{eqnarray}
and [cf.~Eq.~\eqref{eq:Xvar_ms}] 
\begin{eqnarray}\label{eq:Mvar_ms}
	&&\bigg|\frac{\overline{\langle M^2(t) \rangle}\!-\!\langle M(t) \rangle^2}{t}-\tilde{\delta}_l^2 \bigg|
	\\\nonumber&&
	\lesssim   2\frac{\lVert\S\Q\rVert^2}{t}\lVert \M\rVert\lVert \M\P\rVert
	\\\nonumber&&\quad+ t^2\min\big(\lVert \LMM\rVert \lVert\M\P\rVert^2,\lVert \Wt\rVert_1 \lVert\mt\rVert_1\lVert\Mt\rVert_1\big) 
	\\\nonumber&&\quad+ t^2\min\big(\lVert \LMM\rVert \lVert\M\P\rVert,\lVert \Wt\rVert_1 \lVert\mt\rVert_1\big) \lVert\mt\rVert_1
	\\\nonumber
	&&\quad+ t\lVert\mt\rVert_1  m\,\lVert M\rVert_{\max}  \sqrt{\frac{\Ccl}{2} + \Cp}
	\\\nonumber&&\quad+ t \Cp\min\Big[3 \lVert\M\P\rVert^2,\Big( 1+\frac{\Ctcl}{2}\Big) \lVert\mt\rVert_1\big(\lVert\Mt\rVert_1+2\lVert\mt\rVert_1\big)\Big] 
	\\\nonumber&&\quad+\big(\frac{1}{2}\Cmm+\Cp\Big)\\\nonumber&&\quad\qquad\times\min\Big[2\lVert\M\P\rVert\lVert \M\rVert\lVert\S\Q\rVert, \Big( 1+\frac{\Ctcl}{2}\Big) \lVert\bm{\tilde{\delta}^2}\rVert_1 \Big]\\\nonumber
	&&\quad+4\Cp \lVert\M\P\rVert\lVert \M\rVert\lVert\S\Q\rVert.
\end{eqnarray}
Therefore, provided that  corrections in Eq.~\eqref{eq:Mav_ms}  and~\eqref{eq:Mvar_ms} are negligible, we can conclude that $\tilde{m}_l$ in Eq.~\eqref{eq:m_ms} and $\tilde{\delta}_l^2$ in Eq.~\eqref{eq:delta_ms} are the rates of average and fluctuations, respectively, of a time-integrated system-observable in $l$th metastable phase. \\

\emph{Derivation} of Eqs.~\eqref{eq:Kav_t_approx} and~\eqref{eq:Kvar_t_approx}.
We consider time within the metastable regime, $t'' \leq t\leq t'$, to investigate the cumulants in Eqs.~\eqref{eq:Kvar_approx1_CL} and~\eqref{eq:Kvar_t_approx_CL1} with respect to the metastable state of the system.  Therefore, within the MM, we can expand the long-time dynamics in the Taylor series, while the contribution from the outside is bounded as in Eq.~\eqref{eq:approx_2}.

We have
\begin{eqnarray}\label{eq:Kvar_approx1}
	&&\left| \frac{\langle K(t) \rangle}{t} - \Tr(\J \P\rho)+\frac{\Tr(\J \S \Q\rho)}{t}\right| \qquad\\&&
	\lesssim \Cmm\Big\{\frac{1}{2}   \min\Big[\lVert\J\P\rVert,\Big(1+\frac{\Ctcl}{2}\Big) \lVert\bm{\mut}\rVert_1 \Big]  +   \lVert \J\rVert \frac{\lVert\S\Q\rVert}{t} \lVert\Q\rho\rVert\Big\},	 \nonumber
\end{eqnarray}
from Eq.~\eqref{eq:Kvar_t_approx_CL1} and
\begin{eqnarray}\label{eq:Kvar_approx4}
	&&\Bigg|\int_0^t\!\!d  t_1  \Tr (\J e^{t_1\L} \P\rho)-t\int_0^t \Tr (\J \P\rho) \Bigg|
	\\\nonumber&&\lesssim\Bigg|\int_0^t\!\! d  t_1 \,t_1 \Tr (\J\L \P\rho) \big]\Bigg|
	\\\nonumber&&\lesssim \frac{t^2}{2}\min\big(\lVert \LMM\rVert \lVert\J\P\rVert, \lVert \Wt\rVert_1 \lVert\bm{\mut}\rVert_1\big).
\end{eqnarray}
where we used the Taylor series $e^{t\L}\P=\P+t\L\P+...$  [note that $\lVert\Q\rho\rVert$ can be replaced by $1$ in derivation of Eq.~\eqref{eq:Kvar_approx1_CL}]. We also used Eq.~\eqref{eq:approx_1} so that  $ t \lVert\LMM\rVert\lVert\J\P\rVert\lesssim  \Cmm\lVert\J\P\rVert $ or  $t\lVert\bm{\mut}\rVert_1 \lVert \Wt\rVert_1 \leq  \Cmm (1+\Ctcl/2) \lVert\bm{\mut}\rVert_1$ via Eq.~\eqref{eq:Leff1_CL}. If the metastable regime is long enough, for time in Eq.~\eqref{eq:av_cond}  [cf.~Eq.~\eqref{eq:SQnorm}], we can further neglect the contribution from before the metastable regime, 
\begin{eqnarray}\label{eq:Kvar_approx1b}
	&&\left| \frac{\langle K(t) \rangle}{t} - \Tr(\J \P\rho)\right| \qquad\\&&
	\lesssim  \frac{1}{2}\Cmm  \min\Big[\lVert\J\P\rVert,\Big(1+\frac{\Ctcl}{2}\Big) \lVert\bm{\mut}\rVert_1 \Big] +   \lVert \J\rVert \frac{\lVert\S\Q\rVert}{t} \lVert\Q\rho\rVert.	\nonumber 
\end{eqnarray}

Similarly, if time $t\leq t'$ can be chosen as Eq.~\eqref{eq:cond_var_t_1}, we have
\begin{eqnarray}\label{eq:Kvar_approx2}
	&&\bigg|\frac{\langle K^2(t) \rangle\!-\!\langle K(t) \rangle^2}{t}-t \Tr (\J \P\J  \P\rho)+t\Tr(\J \P\rho)^2\qquad \\\nonumber
	&&\quad-\Tr(\J \P\rho)+2\Tr(\J \S\Q\J\P\rho)\\\nonumber
	&&\quad +2 \Tr(\J \P\J \S\Q\rho)-2  \Tr(\J \P\rho)\Tr(\J \S \Q\rho) \bigg|
	\\\nonumber&&	
	\lesssim   \frac{\lVert\S\Q\rVert}{t}\lVert \J\rVert\big[\lVert\Q\rho\rVert+\lVert \J\rVert\lVert\S\Q\rVert(2+4\lVert\Q\rho\rVert+\lVert\Q\rho\rVert^2)\big]
	\\\nonumber&&\quad+4 \Cmm \lVert\bm{\mut}\rVert_1\lVert \J\rVert \lVert\S\Q\rVert \lVert\Q\rho\rVert
	\\\nonumber&&\quad+ t^2\min\big(\lVert \LMM\rVert \lVert\J\P\rVert^2,\lVert \Wt\rVert_1 \lVert\bm{\mut}\rVert_1\lVert\Jt\rVert_1\big) 
	\\\nonumber&&\quad+ t^2\min\big(\lVert \LMM\rVert \lVert\J\P\rVert,\lVert \Wt\rVert_1 \lVert\bm{\mut}\rVert_1\big) \lVert\bm{\mut}\rVert_1
	\\\nonumber&&\quad+\frac{t}{2}\min\big[\lVert \LMM\rVert \lVert\J\P\rVert\big(1+2\lVert \J\rVert\lVert\S\Q\rVert), \lVert \Wt\rVert_1 \lVert\bm{\tilde{\sigma}^2}\rVert_1 \big]\\\nonumber&&\quad+2 t \big(\lVert \LMM\rVert \lVert\J\P\rVert,\lVert \Wt\rVert_1 \lVert\bm{\mut}\rVert_1\big) \lVert \J\rVert\lVert\S\Q\rVert\lVert\Q\rho\rVert. 
\end{eqnarray}
In Eq.~\eqref{eq:Kvar_approx2} the first two lines of corrections correspond to neglecting the dynamics of the fast modes as in Eq.~\eqref{eq:Kvar_t_approx_CL1}. The third line corresponds to    [cf.~Eq.~\eqref{eq:Kvar_t_approx_CL3}]
\begin{eqnarray}\nonumber
	&&\Bigg|2\!\int_0^t\!\!d  t_1 \!\!\int_0^{t-t_1}\!\!\!\! \!\!d  t_2 \Tr (\J e^{t_2\L}\P\J e^{t_1\L} \P\rho)-t^2\Tr (\J \P\J  \P\rho) \Bigg|
	\\\nonumber&&\lesssim\Bigg|2\int_0^t\!\! d  t_1 \!\!\int_0^{t-t_1}\!\!\!\!\!\! d t_2\, \big[t_2 \Tr (\J \L\P\J \P\rho)+t_1 \Tr (\J \P\J\L \P\rho) \big]\Bigg|
	\\&& \lesssim t^3\min\big(\lVert \LMM\rVert \lVert\J\P\rVert^2,\lVert \Wt\rVert_1 \lVert\bm{\mut}\rVert_1\lVert \Jt\rVert_1\big).
\end{eqnarray}
The fourth line is the correction from the square of the time-integral of the activity of metastable phases in Eq.~\eqref{eq:Kvar_approx4}  [cf.~Eq.~\eqref{eq:Kvar_t_approx_CL2}].
The fifth line are corrections  from the time-integral of the fluctuation rate of metastable phases [cf.~Eq.~\eqref{eq:Kvar_t_approx_CL4}] and and the sixth line corresponds to corrections to the contribution from before metastable regime  [cf.~Eq.~\eqref{eq:Kvar_t_approx_CL5}]
\begin{eqnarray}\nonumber
	&&2 \Bigg|\int_0^t\!\!d  t_1 \big[ \Tr(\J e^{t_1\L}\P\rho) \Tr(\J \S \Q\rho) -\Tr(\J e^{t_1\L}\P\J \S\Q\rho) \big]\\\nonumber
	&&	\quad- 2t \big[ \Tr(\J \P\rho) \Tr(\J \S \Q\rho) -\Tr(\J\P\J \S\Q\rho) \big] \Bigg|
	\\\nonumber&& \lesssim 2 t^2 \min\big(\lVert \LMM\rVert \lVert\J\P\rVert\lVert \J\rVert\lVert\S\Q\rVert\lVert\Q\rho\rVert,\\&& \qquad\qquad\qquad\qquad\qquad\qquad\lVert \Wt\rVert_1 \lVert\bm{\mut}\rVert_1\lVert\bm{\tilde{K}}\pt\rVert_1\big) .
\end{eqnarray}
We can further estimate the linear-in-time contribution to the rate of fluctuations as
\begin{eqnarray}\label{eq:Kvar_approx2b}
	&&t \big[\Tr (\J \P \J \P\rho)\!-\!\Tr(\J \P\rho)^2\big] \\\nonumber
	&&\quad=t\!\sum_{k,l=1}^m \!\! \Big[\frac{1}{2} \tilde{p}_k \tilde{p}_l(\mut_{k}-\mut_{l})^2  - \mut_{k}(\Wt)_{kl}\,\tilde{p}_l\Big]+...,
\end{eqnarray}
where the corrections are given by
\begin{eqnarray}
	&&t\Big|\sum_{l=1}^m \big[\bm{\mut}(\Jt-\bm{\mut}-\Wt)\pt\big]_l\Big|\\\nonumber
	&&\lesssim t\lVert\bm{\mut}\rVert_1 \,m \Big\lVert H \!+\!\frac{i}{2}\! \sum_{j}\!\! {J}_{j}^{\dagger} J_{j}\Big\rVert_{\max} \!\!\!\! \sqrt{2\Ccl +4  \Cp},
\end{eqnarray}
and for the inequality we used Eq.~\eqref{eq:bound_D}. Equation~\eqref{eq:approx_1} further leads to Eq.~\eqref{eq:Kvar_t_approx}.\\

\emph{Derivation} of Eqs.~\eqref{eq:Xav_t_approx},~\eqref{eq:Xvar_t_approx},~\eqref{eq:Mav_t_approx}, and~\eqref{eq:Mvar_t_approx}. To obtain Eqs.~\eqref{eq:Xav_t_approx} and~\eqref{eq:Xvar_t_approx}, we use Eqs.~\eqref{eq:Xav_t_approx_CL_corr} and~\eqref{eq:Xvar_t_approx_CL1} and follow the steps in derivation of Eqs.~\eqref{eq:Kav_t_approx} and~\eqref{eq:Kvar_t_approx}. The derivations of Eqs.~\eqref{eq:Mav_t_approx} and~\eqref{eq:Mvar_t_approx} are analogous.\\

\emph{Derivation} of Eqs.~\eqref{eq:Kav_ms} and~\eqref{eq:Kvar_ms}. To consider the rate of average and fluctuations of the number of jumps inside a metastable state, we choose the initial state $\rho$ as the closest state to the projection of the metastable phase $\trho_l$ in Eq.~\eqref{eq:rhotilde} and time $t$ within metastable regime. In this case, we have that $\lVert Q\rho \rVert=\lVert Q(\rho-\trho_l)\rVert\leq \lVert Q\rVert \Cp\lesssim 2\Cp $ and $\lVert \P\rho-\trho_l \rVert\lesssim \Cp$, and thus Eq.~\eqref{eq:Kav_ms} follows from Eq.~\eqref{eq:Kvar_approx1}.  Similarly, from~\eqref{eq:Kvar_approx2} we obtain Eq.~\eqref{eq:Kvar_ms}. Derivations of Eqs.~(\ref{eq:Xav_ms})-(\ref{eq:Mvar_ms}) are analogous.

\subsubsection{Asymptotic rates of fluctuations in quantum trajectories} \label{app:var_ss}

Here, we give bounds on the corrections in the classical approximations of the asymptotic fluctuation rates in Eqs.~\eqref{eq:Kvar2},~\eqref{eq:Kvar3},~\eqref{eq:Mvar2} and~\eqref{eq:Xvar2}. \\

\emph{Classical approximation of asymptotic rate of jump number fluctuations}. 
The corrections in Eq.~\eqref{eq:Kvar2} originate from replacing $\Jt$ by $\Jbf+\bm{\mut^\text{in}}$  and $\Rt$ by $\R$ in Eq.~\eqref{eq:Kvar}. Therefore, the corrections can be bounded as
\begin{eqnarray}\label{eq:Kvar2_approx}
	&&\Bigg|\sigma_\text{ss}^2-
	\!\sum_{l=1}^m (\bm{\mut^\text{tot}}\pss-2 \bm{\mut^\text{tot}}\R\bm{\mut^\text{tot}}\pss +\bm{\bm{\delta_{{\tilde{\sigma}}^2}}} \pss)_l\Bigg|
	\\\nonumber
	&&=\Bigg|
	\!\sum_{l=1}^m \big[\bm{\mut}\ptss-2 \bm{\mut}\Rt\Jt\ptss -\big(\bm{\mut^\text{tot}}\pss   -2 \bm{\mut^\text{tot}}\R\bm{\mut^\text{tot}}\pss \big)\big]_l\Bigg|
	\\\nonumber
	&&\lesssim\lVert \bm{\mut}\rVert_1\lVert\ptss-\pss\rVert_1\! + \sqrt{\Ccl}\lVert \Wt\rVert_1\\\nonumber
	&&\quad\!+ 2\lVert \bm{\mut}\rVert_1\lVert \Rt\rVert_1 \Big[\lVert \bm{\mut}\rVert_1\lVert\ptss-\pss\rVert_1\! + \sqrt{\Ccl}\lVert \Wt\rVert_1\\\nonumber
	&&\qquad\qquad\qquad\qquad+
	m\Big\lVert H +\frac{i}{2} \sum_{j} {J}_{j}^{\dagger} J_{j}\Big\rVert_{\max}\!\!\!\!  \sqrt{2\Ccl +4  \Cp}\\\nonumber
	&&\qquad\qquad\qquad\qquad+  \lVert \bm{\mut}\rVert_1 \frac{\lVert \Rt-\R\rVert_1}{\lVert \Rt\rVert_1} \Big], 
\end{eqnarray}
where we replaced $\Jt=\ptss$ by $\bm{\mut}\ptss$ in the second line introducing corrections $\lVert(\Jt-\bm{\mut})\ptss\rVert=\lVert(\Jt-\Wt-\bm{\mut})\ptss\rVert\leq \lVert\Jt-\Wt-\bm{\mut}\rVert (1+\Ccl)\lesssim m\lVert H +i \sum_{j} {J}_{j}^{\dagger} J_{j}/2\rVert_{\max} \sqrt{2\Ccl +4  \Cp}$ [cf.~Eq.~\eqref{eq:bound_D}] and further used $\lVert\bm{\mut}^\text{tot}-\bm{\mut}\rVert_1 \leq \sqrt{\Ccl}\lVert \Wt\rVert_1$ [cf.~Eqs.~\eqref{eq:W} and~\eqref{eq:deltaW}].

The first line of corrections in Eq.~\eqref{eq:Kvar2_approx} is negligible in comparison with  $\lVert\bm{\tilde{\sigma}}^2\rVert_1$, which bounds the asymptotic rate of time-integral of fluctuations inside the metastable phases, when Eq.~\eqref{eq:cond_var_t_3} is fulfilled together with  $\lVert \bm{\mut}\rVert_1/\lVert\bm{\tilde{\sigma}}^2\rVert_1\ll 1/\lVert\ptss-\pss\rVert_1$ [cf.~Eq.~\eqref{eq:cond_CL}]. The rest of corrections are negligible in comparison with $2\lVert\Rt\rVert_1 \lVert\bm{\mut}\rVert_1\lVert\Jt\rVert_1$, which bounds the non-Poissonian classical fluctuations of activity between metastable phases, when Eqs.~\eqref{eq:cond_var_t_4},~\eqref{eq:cond_CL} and~\eqref{eq:cond2_CL} hold true. Thus, if the bounds are of the same order as the contributions to the asymptotic fluctuations rate,  Eqs.~\eqref{eq:cond_CL},~\eqref{eq:cond2_CL}~\eqref{eq:cond_var_t_3} and~\eqref{eq:cond_var_t_4} are \emph{sufficient conditions} for the classical approximation in Eq.~\eqref{eq:Kvar2}.

When the internal activity dominates transition rate of the-long time dynamics, as assumed in Secs.~\ref{sec:intermittence} and~\ref{sec:metaDPT} of the main text, from Eq.~\eqref{eq:Kvar2} we obtain Eq.~\eqref{eq:Kvar3}. The additional corrections beyond those in Eq.~\eqref{eq:Mvar2_approx} are 
\begin{eqnarray}\nonumber
	&&\Bigg|
	\!\sum_{l=1}^m \big[\big(\bm{\mut^\text{tot}}\!\!-\!\bm{\mut^\text{in}}\big)\pss\!-2 \big(\bm{\mut^\text{tot}}\R\bm{\mut^\text{tot}}\!\!-\!\bm{\mut^\text{in}}\R\bm{\mut^\text{in}}\big)\pss \big]_l\Bigg|
	\\
	&&\lesssim \lVert \Wt\rVert_1 \big(1 + 2\lVert \bm{\mut}\rVert_1\lVert \Rt\rVert_1 \big),\label{eq:Kvar3_approx} 
\end{eqnarray}
where we used $\lVert\bm{\mut}^\text{tot}-\bm{\mut}^\text{in}\rVert_1 \leq \lVert \W\rVert_1/2\lesssim \lVert \Wt\rVert_1/2$ [cf.~Eq.~\eqref{eq:deltaW}] and $\lVert \R\rVert_1\lesssim \lVert \Rt\rVert_1$ .\\

\emph{Classical approximation of asymptotic rate of integrated homodyne current fluctuations}.  Similarly to Eq.~\eqref{eq:Mvar2_approx}, the corrections in Eq.~\eqref{eq:Xvar2} originate from replacing $\Xt$ by $\xt$ [cf.~Eq.~\eqref{eq:Wrtilde2}] and $\Rt$ by $\R$ in Eq.~\eqref{eq:Xvar}, so that
\begin{eqnarray}\label{eq:Xvar2_approx}
	&&\Bigg|\chi_\text{ss}^2-\sum_{l=1}^m (\pss)_l\,\tilde{\chi}_l^2-2 \sum_{l=1}^m ( \bm{\xt}\R\bm{\xt}\,\pss )_l\Bigg|
	\\\nonumber
	&&=\Bigg|
	2\!\sum_{l=1}^m \big( \xt\Rt\Xt\,\ptss - \xt\R\xt\,\pss \big)_l\Bigg|
	\\\nonumber
	&&\lesssim\lVert\bm{\xt}\rVert_1\lVert \Rt\rVert_1 \Big[m\,\Big\lVert \sum_j e^{-\varphi_j} J_j\Big\rVert_{\max}  \sqrt{2\Ccl+4\Cp } \\\nonumber
	&& \qquad\qquad\quad\quad+2\lVert\bm{\xt}\rVert_1\Big(\lVert\ptss-\pss\rVert_1\!+\frac{\lVert \Rt-\R\rVert_1}{\lVert \Rt\rVert_1}\Big)\Big], 
\end{eqnarray}
where we used   Eq.~\eqref{eq:bound_D3}. \\

\emph{Classical approximation of asymptotic rate of integrated system observable fluctuations}.
Analogously to Eq.~\eqref{eq:Kvar2_approx}, the corrections in Eq.~\eqref{eq:Mvar2} originate from replacing $\Mt$ by $\mt$ [cf.~Eq.~\eqref{eq:Whtilde2}] and $\Rt$ by $\R$ in Eq.~\eqref{eq:Mvar}, so that
\begin{eqnarray}\label{eq:Mvar2_approx}
	&&\Bigg|\delta_\text{ss}^2-\sum_{l=1}^m (\pss)_l\,\tilde{\delta}_l^2\,-2 \sum_{l=1}^m ( \mt\R\mt\,\pss )_l\Bigg|
	\\\nonumber
	&&=\Bigg|
	2\!\sum_{l=1}^m \big( \mt\Rt\Mt\,\ptss - \mt\R\mt\,\pss \big)_l\Bigg|
	\\\nonumber
	&&\lesssim\lVert\bm{\mt}\rVert_1\lVert \Rt\rVert_1 \Big[m\,\lVert M\rVert_{\max}  \sqrt{2\Ccl+4\Cp} \\\nonumber
	&& \qquad\qquad\quad\quad+2\lVert\bm{\mt}\rVert_1\Big(\lVert\ptss-\pss\rVert_1\!+\frac{\lVert \Rt-\R\rVert_1}{\lVert \Rt\rVert_1}\Big)\Big], 
\end{eqnarray}
where we used Eq.~\eqref{eq:bound_D4}.

\subsubsection{Multimodal distribution of quantum trajectories} \label{app:distribution}

Here, we show that distributions of continuous measurement results are generally multimodal during the metastable regime. This is a consequence of distinct continuous measurement statistics in metastable phases of the system. The modes then correspond to measurement distributions for metastable phases and the probabilities are determined by the decomposition of the metastable state between metastable phases. See also Sec.~\ref{sec:intermittence} in the main text. Proofs here rely on the master operator in Eq.~\eqref{eq:L} being diagonalizable. \\

\emph{Conditional distribution of continuous measurement}. We consider the statistics of jump number conditioned on the final system state at time $t$ in a quantum trajectory evolving into the metastable phase $\trho_l$. This condition can be described as obtaining outcome $l$ the POVM in Eq.~\eqref{eq:POVM} performed on the system at time $t$, which takes place with the probability approximated by $[\p(t)]_l\equiv(e^{t\W} \pt)_l$ up to corrections $\lesssim\sum_{k=1}^m (\bm{v}_l)_k |[(e^{t\Wt}-  e^{t\W}\big)\pt]_k| +\Ctcl |[\p(t)]_l |/2 \big| -\tilde{p}_{l}^\text{min}
\lesssim  2 \sqrt{\Ccl} t\lVert \Wt\rVert_1\lVert  +\Ctcl |[\p(t)]_l |/2 -\tilde{p}_{l}^\text{min}$ [where $(\bm{v}_l)_k\equiv(\delta_{kl}-\tilde{p}_{l}^\text{min})/(1+\Ctcl/2)$, $k=1,...,m$, represents $P_l$ in the metastable phase basis]. The conditional average [see Eq.~\eqref{eq:Kav_t_l}] is approximated by  [cf.~Eq.~\eqref{eq:Kav_t_approx_CL}]
\begin{eqnarray}\label{eq:Kav_t_l_approx}
	&&\langle K(t) \rangle_l=\langle K_\text{cl}(t) \rangle_l+\sum_{k=1}^m  \frac{(e^{t\W})_{lk} \tilde{p}_k}{[\p(t)]_l}\tilde{K}_k+...,\qquad
\end{eqnarray}
where 
\begin{equation}\label{eq:Kcl_l_av}
	\langle K_\text{cl}(t)\rangle_l\equiv\frac{\int_{0}^t d t_1 [  e^{(t-t_1)\W} (\Jbf+\bm{\mut}^\text{in}) e^{t_1\W}\pt ]_l}{[\p(t)]_l}
\end{equation}
is the integral of the total activity in the classical trajectories, conditioned on the system found at time $t$ in $l$th metastable phase, while the second term encodes the contribution from before the metastable regime [cf.~Eq.~\eqref{eq:Kvar_t_approx_CL}] reweighed with the probability of observing outcome $l$ at time $t$. Similarly, the conditional variance [see Eq.~\eqref{eq:Kvar_t_l}] is approximated as [cf.~Eq.~\eqref{eq:Kvar_t_approx_CL}]
\begin{eqnarray}\nonumber
	&&\frac{\langle K^2(t) \rangle_l\!-\!\langle K(t) \rangle_l^2}{t}=\frac{\langle K_\text{cl}^2(t)\rangle_l-\langle K_\text{cl}(t)\rangle_l^2}{t} +\frac{\langle\Delta_\text{cl}(t)\rangle_l}{t}
	\\\nonumber&&+\frac{1}{t}\sum_{k,n=1}^m \frac{(e^{t\W})_{lk} \tilde{p}_k}{[\p(t)]_l} \frac{(e^{t\W})_{ln} \tilde{p}_n}{[\p(t)]_l} 
	\\&&\qquad\quad\times[\langle K_\text{cl}^{(k)}(t)\rangle_l-\langle K_\text{cl}^{(n)}(t)\rangle_l] (\tilde{K}_{k} -\tilde{K}_{n})  	+...,\qquad\qquad
	\label{eq:Kvar_t_l_approx}
\end{eqnarray}
where 
\begin{eqnarray}\label{eq:Kcl_l_sq}
	&&\langle K_\text{cl}^2(t)\rangle_l\equiv\langle K_\text{cl}(t)\rangle_l+\\\nonumber
	&&\frac{2\int_{0}^t\!\!d t_1\!\int_{0}^{t-t_1}\!\!\!\!d t_2 \big[ e^{(t-t_1-t_2)\W} \! (\Jbf+\bm{\mut}^\text{in})  e^{t_2\W} \!(\Jbf+\bm{\mut}^\text{in}) e^{t_1\W}\pt\big]_l}{[\p(t)]_l}
\end{eqnarray}
is the square of integral of activity in the classical trajectories, conditioned on the system found at time $t$ in $l$th metastable phase [cf.~Eq.~\eqref{eq:Kcl_l_av}], 
\begin{equation}\label{eq:Delta_cl_l}
	\langle \Delta_\text{cl}(t)\rangle_l\equiv\frac{\int_{0}^td t_1 \big[ e^{(t-t_1)\W}  \bm{\tilde{\Delta}}_{\bm{\tilde{\sigma}}^2}  e^{t_1\W}\pt\big]_l}{[\p(t)]_{l}},
\end{equation}
with 
\begin{equation}\label{eq:Delta_l}
	(\bm{\tilde{\Delta}}_{\bm{\tilde{\sigma}}^2})_{kl}\equiv-2\Tr(\tilde{P}_k \J\S\Q\J\trho_l)
\end{equation} describing the conditional integral of non-Poissonian fluctuations in metastable phases [cf.~Eq.~\eqref{eq:delta_CL}],
and
\begin{equation}\label{eq:Kcl_kl_av}
	\langle K_\text{cl}^{(k)}(t)\rangle_l\equiv\frac{\int_{0}^td t_1 \big[ e^{(t-t_1)\W}  (\Jbf+\bm{\mut}^\text{in}) e^{t_1\W}\big]_{lk}}{(e^{t\W})_{lk}}
\end{equation}
being the conditional classical average for the system initially in $k$th metastable phase [cf.~Eq.~\eqref{eq:Kcl_l_av}].
For the corrections to Eqs.~\eqref{eq:Kav_t_l_approx} and~\eqref{eq:Kvar_t_l_approx}, see the derivations below.\\

\emph{Multimodal distribution during metastable regime}. During the metastable regime, the outcome $l$ is observed with the probability approximated by $\tilde{p}_l$ up to corrections  $\lesssim\sum_{k=1}^m (\bm{v}_l)_k |(\Wt\pt)_k| +\Ctcl |\tilde{p}_{l} |/2 \big| -\tilde{p}_{l}^\text{min}
\lesssim  \Cmm  +\Ctcl |\tilde{p}_{l} |/2 -\tilde{p}_{l}^\text{min}$.
The rate of the conditional average in Eq.~\eqref{eq:Kav_t_l_approx} is approximated by  [cf.~Eq.~\eqref{eq:Kav_t_approx}]
\begin{eqnarray}\label{eq:Kav_t_l_approx_ms}
	\frac{\langle K(t) \rangle_l}{t}&=&\frac{[(\Wt+\bm{\mut})\pt ]_l}{\tilde{p}_l}+\frac{\tilde{K}_l}{t}+...\qquad\\\nonumber
	&=&\frac{[(\Jbf+\bm{\mut}^\text{in})\pt ]_l}{\tilde{p}_l}+\frac{\tilde{K}_l}{t}+....
\end{eqnarray}
If time $t\leq t'$ can be chosen so that the contribution from before metastable regime can be further neglected, the fluctuation rate is approximated by [cf.~Eq.~\eqref{eq:Kvar_t_approx}]
\begin{eqnarray}\label{eq:Kvar_t_l_approx_ms}
	&&\frac{\langle K^2(t) \rangle_l\!-\!\langle K(t) \rangle_l^2}{t}\\\nonumber
	&&=t\,\frac{\tilde{p}_l[(\Wt+\bm{\mut})^2\pt]_l-[(\Wt+\bm{\mut})\pt ]_l^2}{\tilde{p}_l^2}\quad\,\,\\\nonumber&&\,\,\,+ \frac{[(\Wt+\bm{\mut}\!+\!\bm{\tilde{\Delta}}_{\bm{\tilde{\sigma}}^2} )\pt ]_l}{\tilde{p}_l}+ 2\sum_{k=1}^n \frac{\tilde{p}_k}{\tilde{p}_l} (\Wt)_{lk} (\tilde{K}_{k} -\tilde{K}_{l}) \!+\!...\quad\\\nonumber
	&&=t\,\frac{\tilde{p}_l[(\Jbf+\bm{\mut}^\text{in})^2\pt]_l-[(\Jbf+\bm{\mut}^\text{in})\pt ]_l^2}{\tilde{p}_l^2}\quad\,\,\\\nonumber&&\,\,\,+ \frac{[(\Jbf+\bm{\mut}^\text{in}\!+\!\bm{\tilde{\Delta}}_{\bm{\tilde{\sigma}}^2} )\pt ]_l}{\tilde{p}_l}+ 2\sum_{k=1}^n \frac{\tilde{p}_k}{\tilde{p}_l} (\Jbf)_{lk} (\tilde{K}_{k} -\tilde{K}_{l}) \!+\!...\quad
\end{eqnarray}
For corrections, see the derivation below.

When internal activity dominates transition rates of the long-time dynamics 
\begin{equation}\label{eq:K_l_approx_cond}
	\lVert \Wt\rVert\ll \lVert \bm{\mut}\rVert,
\end{equation}
we obtain from Eqs.~\eqref{eq:Kav_t_l_approx_ms} and~\eqref{eq:Kvar_t_l_approx_ms} \emph{constant rates} of the conditional average and fluctuation rate (see the derivation below)
\begin{eqnarray}\label{eq:Kav_t_l_approx_ms2}
	\frac{\langle K(t) \rangle_l}{t}&=&
	\bm{\mut}_l^\text{in}+...,\qquad\\\label{eq:Kvar_t_l_approx_ms2}
	\frac{\langle K(t)^2 \rangle_l\!-\!\langle K(t) \rangle_l^2}{t}
	&=&\bm{\mut}_l^\text{in} +\frac{(\bm{\tilde{\Delta}}_{\bm{\tilde{\sigma}}^2} \pt )_l}{\tilde{p}_l}+....\qquad
\end{eqnarray}

The result in Eqs.~\eqref{eq:Kav_t_l_approx_ms2} and~\eqref{eq:Kvar_t_l_approx_ms2} shows that the statistics of jump number $K(t)$ with $t$ in the metastable regime, which was derived in Sec.~\ref{app:var_ms}, can be understood as follows.  When 
\begin{equation}\label{eq:p_l_cond}
	\tilde{p}_l\gg \sum_{k=1}^m (\bm{v}_l)_k |(\Wt\pt)_k|+\Ctcl \big|\tilde{p}_l\big|/2 +\big|\tilde{p}_{l}^\text{min}\big|,
\end{equation}
where $(\bm{v}_l)_k\equiv(\delta_{kl}-\tilde{p}_{l}^\text{min})/(1+\Ctcl/2)$, $k=1,...,m$, represents $P_l$ in the metastable phase basis, the conditional distribution features the average approximated by the average jump number from the $l$th metastable phase,  $\mut_l t$, while the fluctuation rate is constant. Therefore, with the probability approximated by $\tilde{p}_l$, and a long enough metastable regime, $t$ can be chosen so that the activity  $k(t)\equiv K(t)/t$ typically takes values close to $\mut_l$ (as  the fluctuations decay inversely in $t$), and thus can be interpreted as a single mode in the jump number distribution. When $\Cmm +\Ctcl\ll 1$, there exists at least one such probability $\tilde{p}_l$. Therefore, up to corrections of the order $ 2(\Cmm +\Ctcl)+\Ccl$ from replacing by $0$ all $\tilde{p}_l$ that do not fulfill Eq.~\eqref{eq:p_l_cond} [cf.~Eq.~\eqref{eq:p_distance_bound}], we can understand the probability distribution of jump number as multimodal, with the overall fluctuation rate, Eq.~\eqref{eq:Kvar_t_approx}, featuring linear in time contribution due to the different averages of the modes. Note that corrections to multimodal distribution are always present, even for ideal classical MMs ($\Ccl,\Ctcl=0$), since the long-time dynamics connects the metastable phases, and we measure $p_l(t)$ rather than $p_l$ ($\Cmm>0$).  

In Secs.~\ref{sec:intermittence} and~\ref{sec:metaDPT} of  the main text, we discuss how measurement of activity of system can be used to identify the metastable phases a stochastic system state corresponds to, provided that metastable phases differ in their internal activity which dominates transition rates of the long-time dynamics. We note that results analogous to the presented above and in Secs.~\ref{sec:intermittence} and~\ref{sec:metaDPT} of  the main text  hold for integrated homodyne current (cf.~Secs.~\ref{app:homodyne} and~\ref{app:var_ms}). Moreover, for the constant rates of the conditional average and fluctuation rate, we do no longer require the condition in Eq.~\eqref{eq:K_l_approx_cond}. \\


\emph{Cumulants of conditional finite-time statistics}.  Before deriving corrections to Eqs.~\eqref{eq:Kav_t_l_approx}, \eqref{eq:Kvar_t_l_approx}, \eqref{eq:Kav_t_l_approx_ms}, \eqref{eq:Kvar_t_l_approx_ms},~\eqref{eq:Kav_t_l_approx_ms2}, and~\eqref{eq:Kav_t_l_approx_ms2}, we introduce cumulant generating functions for conditional statistics.

Let $\{P_l\}_l$ denote a POVM, that is, $P_l$ is Hermitian, $P_l\geq 0$ and $\sum_l P_l=\mathds{1}$ so that $p_l=\Tr(P_l\rho)$ is the probability of obtaining an outcome $l$ in the corresponding generalized measurement on the state $\rho$. The statistics of jump number conditioned on outcome $l$ at time $t$ is encoded by the cumulant generating function [cf.~Eq.~\eqref{eq:Theta_s_t}]
\begin{equation} \label{eq:Theta_s_t_l}
	\Theta_l(s,t)\equiv\ln[\Tr(P_l e^{t\L_s}\rho)],
\end{equation}    
so that the conditional average of the jump number and its square [cf.~Eqs.~\eqref{eq:Kav_t} and~\eqref{eq:Kav_t}]
\begin{eqnarray}\label{eq:Kav_t_l}
	&&\langle K(t) \rangle_l= \frac{\int_0^t d  t_1  \Tr[P_l e^{(t-t_1) \L} \J e^{t_1 \L} \rho] }{\Tr(P_l e^{t\L}\rho)},
	\\
	&&\langle K^2(t) \rangle_l=\frac{\int_0^t d  t_1   \Tr[P_l e^{(t-t_1) \L} \J e^{t_1 \L} \rho]}{\Tr(P_l e^{t\L}\rho)} \label{eq:Kvar_t_l} \\\nonumber 
	&&\qquad+2\frac{ \int_0^t d  t_1\int_0^{t-t_1} d  t_2  \Tr [P_l e^{(t-t_1-t_2) \L} \J e^{t_2\L}\J e^{t_1\L} \rho]}{\Tr(P_l e^{t\L}\rho)}.
\end{eqnarray}\
We are now ready to approximate the conditional statistics in terms of classical dynamics. \\

\emph{Derivation} of Eq.~\eqref{eq:Kav_t_l_approx}.  For a POVM operator $P_l$ being a combination of low-lying modes, $\P^\dagger(P_l)=P_l$, we have that the probability of observing outcome $l$
\begin{eqnarray}
	\big|\Tr( P_l e^{t\L} \rho)\!-\! \bm{v}_l^\text{T} \!e^{t\W} \pt\big| \!&\leq& \!\sum_{k=1}^m \!(\bm{v}_l)_k  \big|\big[\big(e^{t\Wt}\!\!\!-\!  e^{t\W}\big)\pt\big]_k\big|,\qquad\quad
	\label{eq:p_t_l_approx1}
\end{eqnarray}
where $(\bm{v}_l)_k=\Tr(P_l \trho_k)$. 
In particular, choosing $P_l$ as POVM in Eq.~\eqref{eq:POVM}, we have that it corresponds to a classical measurement, $0\leq \bm{v}_l\leq 1$, and further
\begin{eqnarray}
	&&\big|\bm{v}_l^\text{T} e^{t\W} \pt-[\p(t)]_l\big|
	\lesssim \frac{\Ctcl}{2} \big|[\p(t)]_l \big| -\tilde{p}_{l}^\text{min},\qquad\quad
	\label{eq:p_t_l_approx3}
\end{eqnarray}
where $[\p(t)]_l\equiv(e^{t\W} \pt)_l$.
Therefore, the corrections [cf.~Eq.~\eqref{eq:etW_CL}]
\begin{eqnarray}
	\label{eq:p_t_l_approx}
	&&\big|\Tr( P_l e^{t\L} \rho)- [\p(t)]_l\big| \\\nonumber
	&&\lesssim  2 \sqrt{\Ccl} t\lVert \Wt\rVert_1\lVert \bm{v}_l\rVert_{\max} +\frac{\Ctcl}{2} \big|[\p(t)]_l \big| -\tilde{p}_{l}^\text{min},
\end{eqnarray}
where we introduced the max norm $\lVert \bm{v}_l\rVert_{\max}\equiv\max_{1\leq k \leq m} |(\bm{v}_l)_k|\leq 1$, as well as [cf.~Eq.~\eqref{eq:Ctcl}]
\begin{equation}
	\label{eq:p_t_l_approx4}
	\sum_{l=1}^m\big|\Tr( P_l e^{t\L} \rho)- [\p(t)]_l\big| \lesssim 2  \sqrt{\Ccl} t\lVert \Wt\rVert_1 +\Ctcl.
\end{equation}

For the average jump number up to time $t$ in Eq.~\eqref{eq:Kav_t_l} conditioned on obtaining $l$th outcome of a POVM being a linear combination of low-lying modes at time $t\geq t''$ we have [cf.~Eq.~\eqref{eq:Kvar_approx2_CL}]
\begin{eqnarray}\nonumber
	&&\bigg|\int_{0}^t d t_1\Tr[P_l e^{(t-t_1)\L} \J e^{t_1\L}\rho]\\\nonumber
	&&\quad- \int_{0}^t d t_1 \bm{v}_l^\text{T}  e^{(t-t_1)\Wt} \Jt e^{t_1\Wt}\pt + \Tr(P_l e^{t \L}\J\S\Q\rho)\bigg|
	\\\nonumber
	&&\approx \big|\Tr(P_l e^{t \L}\L\J\S^2\Q\rho)-\Tr(P_l\J\S e^{t\L}\Q\rho)\big|\\
	&&\lesssim 2(\Cmm+\Cmm^n) \lVert \J\rVert\lVert \S\Q\rVert \lVert P_l\rVert_{\max} ,
	\label{eq:Kav_t_l_approx1},
\end{eqnarray}
where we used Eq.~\eqref{eq:approx_1} and Eq.~\eqref{eq:approx_2} together with~\eqref{eq:SQnorm} ($n\geq 1$ is an integer such that $t/n$ belongs to the metastable regime). Furthermore, [cf.~Eq.~\eqref{eq:Kvar_t_approx}]
\begin{eqnarray}\label{eq:Kav_t_l_approx2}
	&&\bigg|\int_{0}^t d t_1 \bm{v}_l^\text{T}  e^{(t-t_1)\Wt} \Jt e^{t_1\Wt}\pt \\\nonumber
	&&\quad -\int_{0}^t d t_1 \bm{v}_l^\text{T}  e^{(t-t_1)\W} (\Jbf+\bm{\mut}^\text{in}) e^{t_1\W}\pt\bigg|
	\\\nonumber
	&&\lesssim \sqrt{\Ccl} \lVert \Wt\rVert_1 \big(2 t^2\lVert \Jt\rVert_1 +t\big) \lVert \bm{v}_l\rVert_{\max}\\\nonumber&&\quad+ t  m \Big\lVert H \!+\!\frac{i}{2}\! \sum_{j}\!\! {J}_{j}^{\dagger} J_{j}\Big\rVert_{\max} \!\!\!\! \sqrt{2\Ccl +4  \Cp} \lVert \bm{v}_l\rVert_{\max},
\end{eqnarray}
while $\Tr(P_l e^{t \L}\J\S\Q\rho)=\bm{v}_l^\text{T} e^{t\Wt}\bm{\tilde{K}}  \pt$ [cf.~Eq.~\eqref{eq:Ktilde2}] and
\begin{equation}\label{eq:Kav_t_l_approx2b}
	\big|\bm{v}_l^\text{T} e^{t\Wt}\bm{\tilde{K}}  \pt-\bm{v}_l^\text{T} e^{t\W}\bm{\tilde{K}} \pt\big|
	\lesssim \sqrt{\Ccl} t \lVert \Wt\rVert_1 \lVert \bm{\tilde{K}}\pt\rVert \lVert \bm{v}_l\rVert_{\max}.
\end{equation}
Choosing $P_l$ as POVM in Eq.~\eqref{eq:POVM}, we arrive at [cf.~Eq.~\eqref{eq:p_t_l_approx3}]
\begin{eqnarray}\nonumber
	&&\bigg|\int_{0}^t d t_1 \bm{v}_l^\text{T}  e^{(t-t_1)\W} (\Jbf+\bm{\mut}^\text{in}) e^{t_1\W}\pt\ -[\p(t)]_l\langle K_\text{cl}(t)\rangle_l  \bigg|\\
	&&\lesssim \frac{\Ctcl}{2} \big|[\p(t)]_l\langle K_\text{cl}(t)\rangle_l  \big|+\big|\tilde{p}_{l}^\text{min}\big| \big|\langle K_\text{cl}(t)\rangle \big|,
	\label{eq:Kav_t_l_approx3}
\end{eqnarray}
where $\langle K_\text{cl}(t)\rangle_l$ is given in Eq.~\eqref{eq:Kcl_l_av},
while 
\begin{eqnarray}\label{eq:Kav_t_l_approx3b}
	&&\big|\bm{v}_l^\text{T} e^{t\W}\bm{\tilde{K}}-(e^{t\W}\bm{\tilde{K}} \pt)_l\big|\\\nonumber
	&&\lesssim \frac{\Ctcl}{2} \big|(e^{t\W}\bm{\tilde{K}} \pt)_l  \big|+\big|\tilde{p}_{l}^\text{min}\big| \big|\tilde{K} \big|.
\end{eqnarray}
The \emph{relative} corrections to Eq.~\eqref{eq:Kav_t_l_approx}, that is, the corrections  in the leading order are bounded by the sum of corrections in Eqs.~(\ref{eq:Kav_t_l_approx1})-(\ref{eq:Kav_t_l_approx3b}) divided by $\Tr( P_l e^{t\L} \rho) \langle K(t)\rangle_l=\int_{0}^t d t_1\Tr[P_l e^{(t-t_1)\L} \J e^{t_1\L}\rho]$, plus the correction to the probability given in Eq.~\eqref{eq:p_t_l_approx} divided by $\Tr( P_l e^{t\L} \rho)$. Therefore, if the corrections in Eqs.~(\ref{eq:p_t_l_approx}--\ref{eq:Kav_t_l_approx3b}) are negligible we obtain the classical approximation of  Eq.~\eqref{eq:Kav_t_l_approx}.\\

\emph{Derivation} of Eq.~\eqref{eq:Kvar_t_l_approx}.  For the square of jump number up to time $t$  in Eq.~\eqref{eq:Kvar_t_l} conditioned on obtaining $l$th outcome of a POVM being a linear combination of low-lying modes, at time $t\geq t''$, we show below that
\begin{eqnarray}\nonumber
	&&\bigg|\int_{0}^t d t_1\Tr[P_l e^{(t-t_1)\L} \J e^{t_1\L}\rho]\\\nonumber
	&&\quad+2\int_{0}^t d t_1\int_{0}^{t-t_1} d t_2\Tr[P_l e^{(t-t_1-t_2)\L} \J e^{t_2\L} \J e^{t_1\L}\rho]\\\nonumber
	&&\quad- 2\int_{0}^t d t_1\int_{0}^{t-t_1} d t_2 \bm{v}_l^\text{T}  e^{(t-t_1-t_2)\Wt} \Jt  e^{t_2\Wt}\Jt e^{t_1\Wt}\pt \\\nonumber
	&&\quad - 2\int_{0}^t\!\!d t_1 \bm{v}_l^\text{T}  e^{(t-t_1)\Wt} \Jt e^{t_1\Wt} \bm{\tilde{K}}\pt
	\\\nonumber
	&&\quad -\int_{0}^t\!\!d t_1 \bm{v}_l^\text{T}  e^{(t-t_1)\Wt}  \bm{\tilde{\Sigma}}^2 e^{t_1\Wt} \pt \bigg|
	\\\nonumber
	&&\lesssim 2(t\Cmm+\lVert \S\Q\rVert)(1+ \lVert\Q\rho\rVert) \lVert \S\Q\rVert\lVert\J\P\rVert \lVert\J\rVert \lVert P_l\rVert_{\max} 
	\\
	&&\qquad(1+2 \lVert \S\Q\rVert \lVert\J\rVert)  \lVert \S\Q\rVert \lVert\J\rVert\lVert\Q\rho\rVert \lVert P_l\rVert_{\max},
	\label{eq:Kvar_t_l_approx1}
\end{eqnarray}
where $\bm{\tilde{\Sigma}}^2\equiv \Jt+\bm{\tilde{\Delta}}_{\bm{\tilde{\sigma}}^2}$ [cf.~Eqs.~\eqref{eq:sigmat} and~\eqref{eq:Delta_l}]. 
Indeed, together with Eq.~\eqref{eq:Kav_t_l_approx1}, we have [cf.~Eq.~\eqref{eq:Kvar_approx2_CL}] 
\begin{eqnarray}\nonumber
	&&\bigg|\int_{0}^t d t_1\int_{0}^{t_1} d t_2\Tr[P_l e^{(t-t_1)\L} \J \P e^{(t_1-t_2)\L} \J \Q e^{t_1\L}\rho]\\\nonumber
	&&\quad+  \int_{0}^t\!\!d t_1 \Tr[P_l e^{(t-t_1)\L} \J \P e^{t_1\L} \J \S\Q \rho]\bigg|
	\\\nonumber
	&&\approx \bigg|\int_{0}^t\!\!d t_1 \Tr[P_l e^{(t-t_1)\L} \J \P e^{t_1\L}\L \J \S^2 \Q\rho] \\\nonumber
	&&\quad- \int_{0}^t\!\!d t_1 \Tr[P_l e^{(t-t_1)\L} \J \P  \J \S e^{t_1\L}\Q\rho] \bigg|
	\\\nonumber
	&&\lesssim (t\Cmm+\lVert \S\Q\rVert)\lVert \S\Q\rVert\lVert\J\P\rVert \lVert\J\rVert \lVert\Q\rho\rVert \lVert P_l\rVert_{\max},
\end{eqnarray}
where we used Eq.~\eqref{eq:approx_1} with respect to time $ \lVert\S \Q\rVert$ and $-\Tr[P_l e^{(t-t_1)\L} \J \P e^{t_1\L} \J \S\Q \rho]=\bm{v}_l^\text{T}  e^{(t-t_1)\Wt} \Jt e^{t_1\Wt} \bm{\tilde{K}}\pt$ [see Eq.~\eqref{eq:Ktilde2}], 
and, similarly, 
\begin{eqnarray}\nonumber
	&&\bigg|\int_{0}^t d t_1\int_{0}^{t-t_1} d t_2\Tr[P_l e^{(t-t_1-t_2)\L} \P \J \Q e^{t_2\L} \J \P e^{t_1\L}\rho]\\\nonumber
	&&\quad+  \int_{0}^t\!\!d t_1 \Tr[P_l e^{(t-t_1)\L} \P\J \S\Q \J \P e^{t_1\L}\rho] \bigg|
	\\\nonumber
	&&\approx \bigg|\int_{0}^t\!\!d t_1 \Tr[P_l e^{(t-t_1)\L} \L\P\J \S^2\Q \J \P e^{t_1\L}\rho] \\\nonumber
	&&\quad- \int_{0}^t\!\!d t_1 \Tr[P_l \J \S e^{(t-t_1)\L}\Q \J \P e^{t_1\L}\rho]  \bigg|
	\\\nonumber
	&&\lesssim  (t\Cmm+\lVert \S\Q\rVert)\lVert \S\Q\rVert\lVert\J\rVert \lVert\J\P\rVert \lVert P_l\rVert_{\max},
\end{eqnarray}
with $-2\Tr[P_l e^{(t-t_1)\L} \P\J \S\Q \J \P e^{t_1\L}\rho] =\bm{v}_l^\text{T}  e^{(t-t_1)\Wt}  \bm{\tilde{\Delta}}_{\bm{\tilde{\sigma}}^2} e^{t_1\Wt} \pt$ [see Eq.~\eqref{eq:Delta_l}],
while [cf.~Eqs.~\eqref{eq:Kvar_approx2_CL} and~\eqref{eq:Kvar_approx6_CL}]
\begin{eqnarray}\nonumber
	&&\bigg|\int_{0}^t d t_1\int_{0}^{t-t_1} d t_2\Tr[P_l e^{(t-t_1-t_2)\L} \J \Q e^{t_2\L} \J \Q e^{t_1\L}\rho]\bigg|
	\\\nonumber
	&&\approx \bigg|\int_{0}^t d t_1 \Tr[P_l e^{(t-t_1)\L} \J \S\Q  \J \Q e^{t_1\L}\rho]\bigg|
	\\\nonumber
	&&\approx \big| \Tr(P_l e^{t\L} \J \S\Q  \J \S\Q\rho)\big|
	\\\nonumber
	&&\lesssim  \lVert \S\Q\rVert^2 \lVert\J\rVert^2\lVert\Q\rho\rVert \lVert P_l\rVert_{\max}.
\end{eqnarray}

Furthermore, we observe that  [cf.~Eq.~\eqref{eq:Kav_t_l_approx2}]
\begin{eqnarray}\label{eq:Kvar_t_l_approx2}
	&&\bigg|2\int_{0}^t d t_1\int_{0}^{t-t_1} d t_2 \bm{v}_l^\text{T}  e^{(t-t_1-t_2)\Wt} \Jt  e^{t_2\Wt}\Jt e^{t_1\Wt}\pt  \\\nonumber
	&&-2\!\int_{0}^t\!\!d t_1\!\!\int_{0}^{t-t_1}\!\!\!\!\!\! \!\!d t_2 \bm{v}_l^\text{T}  e^{(t-t_1-t_2)\W} \! (\Jbf+\bm{\mut}^\text{in})  e^{t_2\W} \!(\Jbf+\bm{\mut}^\text{in}) e^{t_1\W}\pt\bigg|
	\\\nonumber
	&&\lesssim   2\sqrt{\Ccl} \lVert \Wt\rVert_1 \lVert \Jt\rVert_1\Big(\frac{4}{3} t^3\lVert \Jt\rVert_1 +t^2\Big) \lVert \bm{v}_l\rVert_{\max}\\\nonumber&&\quad+2 t^2  m \Big\lVert H \!+\!\frac{i}{2}\! \sum_{j}\!\! {J}_{j}^{\dagger} J_{j}\Big\rVert_{\max} \!\!\!\! \sqrt{2\Ccl +4  \Cp} \Big]\lVert \Jt\rVert_{1} \lVert \bm{v}_l\rVert_{\max}.
\end{eqnarray}
Similarly, 
\begin{eqnarray}\label{eq:Kvar_t_l_approx2b}
	&&\bigg|2\int_{0}^t\!\!d t_1 \bm{v}_l^\text{T}  e^{(t-t_1)\Wt} \Jt e^{t_1\Wt} \bm{\tilde{K}}\pt  \\\nonumber
	&&-2\!\int_{0}^t\!\!d t_1 \bm{v}_l^\text{T}  e^{(t-t_1)\W} \! (\Jbf+\bm{\mut}^\text{in})  e^{t_1\W} \bm{\tilde{K}}\pt\bigg|
	\\\nonumber
	&&\lesssim  2 \sqrt{\Ccl} \lVert \Wt\rVert_1 \big(2t^2\lVert \Jt\rVert_1+t\big)\lVert \bm{\tilde{K}}\pt \rVert_1 \lVert \bm{v}_l\rVert_{\max}\\\nonumber&&\quad+2 t  m \Big\lVert H \!+\!\frac{i}{2}\! \sum_{j}\!\! {J}_{j}^{\dagger} J_{j}\Big\rVert_{\max} \!\!\!\! \sqrt{2\Ccl +4  \Cp} \Big]\lVert \bm{\tilde{K}}\pt \rVert_1 \lVert \bm{v}_l\rVert_{\max}.
\end{eqnarray}
and
\begin{eqnarray}\label{eq:Kvar_t_l_approx2c}
	&&\bigg|\int_{0}^t\!\!d t_1 \bm{v}_l^\text{T}  e^{(t-t_1)\Wt}  \bm{\tilde{\Sigma}}^2 e^{t_1\Wt} \pt   \\\nonumber
	&&-\!\int_{0}^t\!\!d t_1 \bm{v}_l^\text{T}  e^{(t-t_1)\W} \! \big(\Jbf+\bm{\mut}^\text{in}+\bm{\tilde{\Delta}}_{\bm{\tilde{\sigma}}^2}\big)  e^{t_1\W}\pt\bigg|
	\\\nonumber
	&&\lesssim  \sqrt{\Ccl} \lVert \Wt\rVert_1 \big(2t^2\lVert \bm{\tilde{\Sigma}}^2\rVert_1+t\big)\lVert \bm{v}_l\rVert_{\max}\\\nonumber
	&&\quad+ t  m \Big\lVert H \!+\!\frac{i}{2}\! \sum_{j}\!\! {J}_{j}^{\dagger} J_{j}\Big\rVert_{\max} \!\!\!\! \sqrt{2\Ccl +4  \Cp} \lVert \bm{v}_l\rVert_{\max}.
\end{eqnarray}

Choosing $P_l$ as POVM in Eq.~\eqref{eq:POVM}, we arrive at [cf.~Eq.~\eqref{eq:Kav_t_l_approx3}]
\begin{eqnarray}\nonumber
	&&\bigg|2\int_{0}^t\!\!d t_1\!\!\int_{0}^{t-t_1}\!\!\!\!\!\! \!\!\!\!d t_2 \bm{v}_l^\text{T}  e^{(t-t_1-t_2)\W} \! (\Jbf+\bm{\mut}^\text{in})  e^{t_2\W} \!(\Jbf+\bm{\mut}^\text{in}) e^{t_1\W}\pt\\\nonumber
	&& \quad-[\p(t)]_l\big[\langle K_\text{cl}^2(t)\rangle_l-\langle K_\text{cl}(t)\rangle_l \big] \bigg|\\\nonumber
	&&\lesssim \frac{\Ctcl}{2} \big|[\p(t)]_l\big[\langle K_\text{cl}^2(t)\rangle_l-\langle K_\text{cl}(t)\rangle_l  \big]\big|\\
	&&\quad+\big|\tilde{p}_{l}^\text{min}\big| \big|\langle K_\text{cl}^2(t)\rangle-\langle K_\text{cl}(t)\rangle  \big|,
	\label{eq:Kvar_t_l_approx3}
\end{eqnarray}
where $\langle K_\text{cl}^2(t)\rangle_l$ is given in Eq.~\eqref{eq:Kcl_l_sq}.
Similarly,
\begin{eqnarray}\nonumber
	&&\bigg|\int_{0}^t\!\!d t_1 \bm{v}_l^\text{T}  e^{(t-t_1)\W} \! (\Jbf+\bm{\mut}^\text{in})  e^{t_1\W} \bm{\tilde{K}}\pt\\\nonumber
	&& \quad-\sum_{k=1}^m \langle K_\text{cl}^{(k)}(t)\rangle_l  (e^{t\W})_{lk} \tilde{K}_{k} \tilde{p}_k\bigg|\\\nonumber
	&&\lesssim \frac{\Ctcl}{2} \bigg|\sum_{k=1}^m \langle K_\text{cl}^{(k)}(t)\rangle_l  (e^{t\W})_{lk} \tilde{K}_{k} \tilde{p}_k\bigg|\\
	&&\quad+\big|\tilde{p}_{l}^\text{min}\big| \bigg|\sum_{k=1}^m \langle K_\text{cl}^{(k)}(t)\rangle \tilde{K}_{k} \tilde{p}_k \bigg|
	\label{eq:Kvar_t_l_approx3b},
\end{eqnarray}
where $\langle K_\text{cl}^{(k)}(t)\rangle_l$ is given in Eq.~\eqref{eq:Kcl_kl_av},
as well as
\begin{eqnarray}\nonumber
	&&\bigg|\int_{0}^t\!\!d t_1 \bm{v}_l^\text{T}  e^{(t-t_1)\W} \! \big(\Jbf+\bm{\mut}^\text{in}+\bm{\tilde{\Delta}}_{\bm{\tilde{\sigma}}^2}\big)  e^{t_1\W}\pt\\\nonumber
	&& \quad-[\p(t)]_{l} \big[\langle K_\text{cl}(t)\rangle_l+\langle\Delta_\text{cl}(t)\rangle_l\big]\bigg|\\\nonumber
	&&\lesssim \frac{\Ctcl}{2} \big|[\p(t)]_{l}\big[\langle K_\text{cl}(t)\rangle_l+\langle\Delta_\text{cl}(t)\rangle_l\big]\big|\\
	&&\quad+\big|\tilde{p}_{l}^\text{min}\big| \big|\langle K_\text{cl}(t)\rangle+\langle \Delta_\text{cl}(t)\rangle \bigg|
	\label{eq:Kvar_t_l_approx3c},
\end{eqnarray}
where $\langle \Delta_\text{cl}(t)\rangle_l$ is defined in Eq.~\eqref{eq:Delta_cl_l}.   Eqs.~(\ref{eq:Kvar_t_l_approx1})-(\ref{eq:Kvar_t_l_approx3c})

Eqs.~(\ref{eq:Kvar_t_l_approx1})-(\ref{eq:Kvar_t_l_approx3c}) describe all corrections to $\Tr( P_l e^{t\L} \rho) \langle K^2(t)\rangle_l$. Therefore, we conclude that the \emph{relative} corrections to Eq.~\eqref{eq:Kvar_t_l_approx} are given by the sum of those corrections plus the corrections to the conditional average in Eq.~\eqref{eq:Kav_t_l_approx} multiplied by  $2\Tr( P_l e^{t\L} \rho) \langle K(t)\rangle_l$, all divided by $\Tr( P_l e^{t\L} \rho) [\langle K^2(t)\rangle_l-\langle K(t)\rangle_l^2]$,  plus the correction to the probability in Eq.~\eqref{eq:p_t_l_approx} divided by $\Tr( P_l e^{t\L} \rho)$. When those corrections are negligible, we arrive the classical approximation in  Eq.~\eqref{eq:Kvar_t_l_approx}.\\

\emph{Derivation} of Eqs.~\eqref{eq:Kav_t_l_approx_ms} and~\eqref{eq:Kvar_t_l_approx_ms}. We now approximate the conditional statistics in terms of classical dynamics during the metastable regime. 

From Eq.~\eqref{eq:p_t_l_approx1} for times within metastable regime, $t''\leq t\leq t'$, we have by the Taylor series expansion [cf.~Eqs.~\eqref{eq:Leff1_CL} and~\eqref{eq:approx_1}]
\begin{eqnarray}\label{eq:p_t_l_approx1_ms}
	&&\big|\Tr( P_l e^{t\L} \rho)- \bm{v}_l^\text{T} \pt\big|\lesssim t \!\sum_{k=1}^m \!(\bm{v}_l)_k  \big|\big(\Wt\pt\big)_k\big|,\qquad\quad
	\\\nonumber&&\quad\lesssim \Cmm \min \Big[\Big(1+\frac{\Ctcl}{2}\Big) \lVert \bm{v}_l\rVert_{\max}, \lVert P_l\rVert_{\max}\Big],
\end{eqnarray}
and, choosing $P_l$ as POVM in Eq.~\eqref{eq:POVM}  [cf.~Eq.~\eqref{eq:p_t_l_approx3}],
\begin{eqnarray}
	&&\big|\bm{v}_l^\text{T} \pt-\tilde{p}_l\big| \lesssim \frac{\Ctcl}{2} \big|\tilde{p}_l \big| -\tilde{p}_{l}^\text{min},\qquad\quad
	\label{eq:p_t_l_approx3_ms}
\end{eqnarray}
so that
\begin{eqnarray}
	\label{eq:p_t_l_approx_ms}
	&&\big|\Tr( P_l e^{t\L} \rho)-\tilde{p}_l\big| \\\nonumber
	&& \lesssim \Cmm \min \Big[\Big(1+\frac{\Ctcl}{2}\Big) \lVert \bm{v}_l\rVert_{\max}, \lVert P_l\rVert_{\max}\Big]+\frac{\Ctcl}{2} \big|\tilde{p}_{l}\big| -\tilde{p}_{l}^\text{min},
\end{eqnarray}
and
\begin{equation}
	\label{eq:p_t_l_approx4_ms}
	\sum_{l=1}^m\big|\Tr( P_l e^{t\L} \rho)-\tilde{p}_l\big| \lesssim \Cmm +\Ctcl.
\end{equation}

Similarly, from Eq.~\eqref{eq:Kav_t_l_approx1} for time within the metastable regime, $t''\leq t\leq t'$, we have [cf.~Eq.~\eqref{eq:Kvar_approx1} and~\eqref{eq:Kvar_approx4}]
\begin{eqnarray}\nonumber
	&&\bigg|\int_{0}^t d t_1\Tr[P_l e^{(t-t_1)\L} \J e^{t_1\L}\rho]- t \bm{v}_l^\text{T}   \Jt \pt + \Tr(P_l \J\S\Q\rho)\bigg|\quad
	\\\nonumber
	&&\lesssim 5\Cmm \lVert \J\rVert\lVert \S\Q\rVert \lVert P_l\rVert_{\max} \\
	&&+ t \Cmm \min\!\Big[\lVert \J\P\rVert\lVert P_l\rVert_{\max},\! \Big(\!1\!+\!\frac{\Ctcl}{2}\!\Big) \lVert \Jt\rVert_1 \lVert \bm{v}_l\rVert_{\max}\Big]\!. 
	\label{eq:Kav_t_l_approx1_ms}
\end{eqnarray}
For $P_l$ in Eq.~\eqref{eq:POVM}, we further have [cf.~Eqs.~\eqref{eq:Kav_t_l_approx3} and~\eqref{eq:Kav_t_l_approx3b}]
\begin{eqnarray}\label{eq:Kav_t_l_approx3_ms}
	&&\bigg|t \bm{v}_l^\text{T}   \Jt \pt - \Tr(P_l \J\S\Q\rho) -t(\Jt\pt )_l -\tilde{K}_l\tilde{p}_l\bigg|\\\nonumber
	&&\lesssim \frac{\Ctcl}{2} \big|t(\Jt\pt )_l +\tilde{K}_l \tilde{p}_l \big|+\big|\tilde{p}_{l}^\text{min}\big| \big|t{\mut} + \tilde{K}\big|,
\end{eqnarray}
where $\mut= \sum_{l=1}^m \mut_l  \tilde{p}_l$. Noting that [cf.~Eq.~\eqref{eq:bound_D}]
\begin{eqnarray}\label{eq:Kav_t_l_approx2_ms}
	&&\bigg|t(\Jt\pt )_l -t[(\Wt+\bm{\mut})\pt ]_l\bigg|\\\nonumber
	&&\lesssim tm \Big\lVert H \!+\!\frac{i}{2}\! \sum_{j}\!\! {J}_{j}^{\dagger} J_{j}\Big\rVert_{\max} \!\!\!\! \sqrt{2\Ccl +4  \Cp} 
\end{eqnarray}
we arrive at the first line of  Eq.~\eqref{eq:Kav_t_l_approx_ms},
and further that $\lVert\Wt+\bm{\mut}-(\Jbf+\bm{\mut}^\text{in})\rVert_1\leq \sqrt{\Ccl} \lVert \Wt\rVert_1 $ [cf.~Eq.~\eqref{eq:deltaW}], at the second line of  Eq.~\eqref{eq:Kav_t_l_approx_ms}. We also note that the contribution in Eq.~\eqref{eq:Kav_t_l_approx_ms} from before the metastable regime can be neglected if $t\leq t'$ is long enough, as the resulting relative correction is bounded by $ \Tr(\tilde{P}_l \J\S\Q \rho)/[\mut_l t+(1+\Ctcl/2)\Cmm]$[cf.~Eqs.~\eqref{eq:Leff1_CL} and~\eqref{eq:approx_1}].

Finally, from~\eqref{eq:Kvar_t_l_approx1} for time within the metastable regime we have [cf.~Eq.~\eqref{eq:rho1_CL}]
\begin{eqnarray}\label{eq:Kvar_t_l_approx1_ms}
	&&\bigg|\int_{0}^t d t_1\Tr[P_l e^{(t-t_1)\L} \J e^{t_1\L}\rho]\\\nonumber
	&&\quad+2\int_{0}^t d t_1\int_{0}^{t-t_1} d t_2\Tr[P_l e^{(t-t_1-t_2)\L} \J e^{t_2\L} \J e^{t_1\L}\rho]\\\nonumber
	&&\quad- t^2 \bm{v}_l^\text{T}   \Jt ^2 \pt - 2t \bm{v}_l^\text{T}   \Jt  \bm{\tilde{K}}\pt -t\bm{v}_l^\text{T}  \bm{\tilde{\Sigma}}^2\pt \bigg|
	\\\nonumber
	&&\lesssim 2(t\Cmm+\lVert \S\Q\rVert)(1+ \lVert\Q\rho\rVert) \lVert \S\Q\rVert\lVert\J\P\rVert \lVert\J\rVert \lVert P_l\rVert_{\max} 
	\\\nonumber
	&&\quad+2 \lVert \S\Q\rVert^2 \lVert\J\rVert^2\lVert\Q\rho\rVert \lVert P_l\rVert_{\max}\\\nonumber
	&&\quad+t^2 \Cmm \min\Big[\lVert \J\P\rVert^2 \lVert P_l\rVert_{\max} ,\Big(1+\frac{\Ctcl}{2}\Big)\lVert\Jt\rVert_1^2\lVert \bm{v}_l\rVert_{\max}\Big]\\\nonumber
	&&\quad+2t \Cmm \min\Big[\lVert \J\P\rVert\lVert \J \rVert \lvert \S\Q\rVert \lVert \Q\rho\rVert \lVert P_l\rVert_{\max},\\\nonumber
	&&\qquad\qquad\qquad\qquad \Big(1+\frac{\Ctcl}{2}\Big) \lVert \Jt\rVert_1 \lVert\bm{\tilde{K}}\pt\rVert_1\lVert \bm{v}_l\rVert_{\max}\Big]\\\nonumber
	&&\quad+t \Cmm \min\Big[2\lVert \J\P\rVert\lVert \J \rVert \lvert \S\Q\rVert \lVert P_l\rVert_{\max} ,\\\nonumber
	&&\qquad\qquad\qquad\qquad \Big(1+\frac{\Ctcl}{2}\Big)\lVert\bm{\tilde{\Sigma}}^2\rVert_1\lVert \bm{v}_l\rVert_{\max}\Big].
\end{eqnarray}
For $P_l$ in Eq.~\eqref{eq:POVM}, we further have
\begin{eqnarray}\label{eq:Kvar_t_l_approx3_ms}
	&&\bigg|t^2 \bm{v}_l^\text{T}   \Jt ^2 \pt + 2t \bm{v}_l^\text{T}   \Jt  \bm{\tilde{K}}\pt +t\bm{v}_l^\text{T}  \bm{\tilde{\Sigma}}^2\pt \\\nonumber
	&&\quad-t^2( \Jt ^2 \pt)_l - 2t ( \Jt  \bm{\tilde{K}}\pt)_l -t(\bm{\tilde{\Sigma}}^2\pt)_l\bigg|\\\nonumber
	&&\lesssim \frac{\Ctcl}{2} t \big|t( \Jt ^2 \pt)_l + 2 ( \Jt  \bm{\tilde{K}}\pt)_l +(\bm{\tilde{\Sigma}}^2\pt)_l \big|
	\\\nonumber
	&&\quad+\big|\tilde{p}_{l}^\text{min}\big| \,t\bigg|\sum_{k=1}^m \big(t\mut_l (\Jt)_{lk} + \mut_k\tilde{K}_k +\tilde{\sigma}_k \big)\tilde{p}_k\bigg|.
\end{eqnarray} 
Note that [cf.~Eq.~\eqref{eq:bound_D}]
\begin{eqnarray}\label{eq:Kvar_t_l_approx2_ms}
	&&t\bigg|t( \Jt ^2 \pt)_l + 2 ( \Jt  \bm{\tilde{K}}\pt)_l +(\bm{\tilde{\Sigma}}^2\pt)_l - t[ (\Wt+\bm{\mut}) ^2 \pt]_l\qquad\\\nonumber
	&&\quad   - 2 [ (\Wt+\bm{\mut}) \bm{\tilde{K}}\pt]_l -[(\Wt+\bm{\mut}+\bm{\tilde{\Delta}}_{\bm{\tilde{\sigma}}^2})\pt]_l\bigg|\\\nonumber
	&&\lesssim tm \Big\lVert H \!+\!\frac{i}{2}\! \sum_{j}\!\! {J}_{j}^{\dagger} J_{j}\Big\rVert_{\max} \!\!\!\! \sqrt{2\Ccl +4  \Cp} \\\nonumber
	&&\qquad\qquad\qquad\qquad\times(2t\lVert \Jt\rVert_1+ \lVert\bm{\tilde{K}}\pt\rVert_1+1).
\end{eqnarray}
Replacing $\Wt+\bm{\mut}$ by $\Jbf+\bm{\mut}^\text{in}$ further introduces corrections bounded by $\sqrt{\Ccl} \lVert \Wt\rVert_1 (2t\lVert \Jt\rVert_1+ \lVert\bm{\tilde{K}}\pt\rVert_1+1)$ [cf.~Eq.~\eqref{eq:deltaW}] and
we obtain the classical approximation of the conditional square of jump number. Using then already derived  Eq.~\eqref{eq:Kav_t_l_approx_ms} we arrive at the approximation of the variance, and thus also the fluctuation rate as given in Eq.~\eqref{eq:Kvar_t_l_approx_ms}.\\

\emph{Derivation} of Eqs.~\eqref{eq:Kav_t_l_approx_ms2} and~\eqref{eq:Kvar_t_l_approx_ms2}. When the contribution from before the metastable regime in Eq.~\eqref{eq:Kav_t_l_approx_ms} is negligible, the relative corrections from neglecting the long-time dynamics are bounded by $\lVert \Wt\rVert/\mut_l$.

Similarly, in the first line of Eq.~\eqref{eq:Kvar_t_l_approx_ms} we have
\begin{eqnarray*}
	&&\tilde{p}_l[(\Wt+\bm{\mut})^2\pt]_l-t[(\Wt+\bm{\mut})\pt ]_l^2 \\
	&&=p_l[(\Wt\bm{\mut}-\bm{\mut}\Wt)\pt]_l+t\tilde{p}_l(\Wt^2\pt)_l-t(\Wt\pt )_l^2,
\end{eqnarray*}
and thus neglecting the long-time dynamics removes any time-dependence from the fluctuation rate, with the neglected terms bounded in the leading order by $2(1+\Ctcl/2)\Cmm \lVert \bm{\mut}\rVert_1/\tilde{p}_l$ [cf.~Eqs.~\eqref{eq:Leff1_CL} and~\eqref{eq:approx_1}]. 
Furthermore, the constant fluctuation rate can be further simplified by neglecting the long-time dynamics in the second term, which leads to corrections bounded by $\lVert \bm{\Wt}\rVert_1/\tilde{p}_l$, and in the third term, with corrections bounded by $\lVert \bm{\Wt}\rVert_1\lVert \bm{\tilde{K}}\pt\rVert_1/\tilde{p}_l$. Therefore, we obtain the approximation in Eq.~\eqref{eq:Kvar_t_l_approx_ms2} whenever the relative corrections $[2(1+\Ctcl/2)\Cmm \lVert \bm{\mut}\rVert_1+\lVert \bm{\Wt}\rVert_1(1+\lVert\bm{\tilde{K}}\pt\rVert_1 )]/ (\bm{\Sigma}^2\pt)_l$ are small.

\section{Classical hierarchy of metastabilities}\label{app:hierarchy}

Here, we discuss the existence of the second metastable regime in the long-time dynamics of the system, which corresponds to a further separation in the real part of the $m$ low-lying eigenvalues of the master operator~\cite{Gaveau1999}. We prove that for the classical metastability of $m$ low-lying modes, the second metastability is classical as well. This is a direct consequence of long-time dynamics in a classical MM being well approximated by classical stochastic dynamics (cf.~Sec.~\ref{sec:Leff} of the main text), as metastable states of classical stochastic dynamics are known to be mixtures of as many metastable phases as the number of low-lying modes~\cite{Gaveau1998,Gaveau2006}. As a result, we show that a hierarchy of metastable phases arises: $m_2$ metastable phases of the second MM are approximately disjoint mixtures of $m$ metastable phases of the first MM, while any metastable phase of the first MM that during the second metastable regime evolves into a mixture, rather than a single phase, necessarily belongs to the decay subspace. Furthermore, we show how the approximation of the long-time dynamics by classical stochastic dynamics can be refined to take into account the hierarchy of metastabilities and hold well  for times up to and beyond the relaxation time, e.g., to  obtain a classical approximation of the stationary state. For a discussion of hierarchy of metastabilities arising in the proximity to a dissipative phase transition at a finite-size, see Sec.~\ref{app:PT}.

\subsection{Hierarchy of metastabilities}\label{app:hierarchy_def}

Apart from the metastable regime $ t''\leq t\leq t' $ leading to the spectrum separation with $\lambda_{m}^R/\lambda_{m+1}^R\ll 1$ as considered in Sec.~\ref{sec:MM} of the main text, we assume here that there occurs a later metastable regime $t_2''\leq t\leq t_2'$  that corresponds to $\lambda_{m_2}^R/\lambda_{m_2+1}^R\ll 1$ where $m_2<m$ (see also Ref.~\cite{Gaveau1999}). 
For times $t\geq t''_2$, the system state can be approximated as 
\begin{equation}\label{Expansion1_MM2}
	\rho(t)=\sum_{k=1}^{m_2} e^{\lambda_m t} c_k \,R_k+...
\end{equation}
[cf.~Eq.~\eqref{Expansion1} and note that $t'\ll t_2''$ as $t'\ll -1/\lambda_{m}^R$ and $t''_2\geq -1/\lambda_{m_2+1}^R$; see~Sec.~\ref{app:tau_defMM}].
Furthermore, during the second metastable regime $t''_2\leq t\leq t'_2$, the decay of $m_2$ slower modes is negligible  and the system states appear stationary (cf.~Eq.~\eqref{Expansion} and Ref.~\cite{Gaveau1999})
\begin{equation}\label{Expansion_MM2}
	\rho(t)=\sum_{k=1}^{m_2} c_k\,R_k+... =\P_2[\rho(0)]+...,
\end{equation}
with $\P_2$ denoting the projection on the second MM. For the corrections to the stationarity in Eq.~\eqref{Expansion_MM2} denoted as $\mathcal{C}_{2,\text{MM}}$ [cf.~Eq.~\eqref{eq:Cmm}] and the corrections to the positivity of projection $\P_2$ as $\mathcal{C}_{2,+}^{(2)}$ [cf.~Eq.~\eqref{eq:Cp}], the errors in Eq.~\eqref{Expansion1_MM2} are bounded by $2\mathcal{C}_{2,\text{MM}}^n$, where $n$ is such that $t''_2\leq  t/n\leq t'_2$ (cf.~Sec.~\ref{sec:MM_Cmm} in the main text).

\subsection{Hierarchy of classical metastable manifolds}

We now show that when the first metastability is classical (see Sec.~\ref{sec:cMM} in the main text), it follows that the second metastability is classical as well, i.e.,  metastable states during the second metastable regime are mixtures of $m_2$ metastable phases. For a related discussion, see Ref.~\cite{Gaveau1999}.\\

As discussed in Sec.~\ref{app:Leff}, the dynamics $e^{t\Wt}$ in the first MM can be approximated at any time $t$ by the discrete positive and trace-preserving dynamics $\T_t$, with the  corrections bounded by $\Ccl$ [see Eqs.~\eqref{eq:T_CL} and~\eqref{eq:Tn_CL}]. For $t\geq t''_2$ and an integer $n$ such that $nt\leq t'_2$, that is, both $t$ and $nt$ belonging to the second metastable regime, we have
\begin{eqnarray}\label{eq:Tn2_CL}
	\lVert \Pt_2- \T_{t}^n \rVert_1 &\leq&  \lVert \Pt_2-e^{n t\Wt}\rVert_1+\lVert e^{n t\Wt}- \T_t^n \rVert_1\qquad\qquad\\\nonumber
	&\lesssim& \tilde{\mathcal{C}}_{2,\text{MM}}+n\Ccl,
\end{eqnarray}
where $\Pt_2$ denotes the projection $\P_2$ on the second MM in the basis of the metastable phases of the first MM and $\tilde{\mathcal{C}}_{2,\text{MM}}$ are corrections to the stationarity of barycentric coordinates during the second metastable regime in L1 norm. The first inequality follows from the triangle inequality, the second from Eqs.~\eqref{eq:rho1_CL} and~\eqref{eq:Tn_CL}. We also have $\tilde{\mathcal{C}}_{2,\text{MM}}\ll (1+\Ctcl/2)\mathcal{C}_{2,\text{MM}}$  (see Sec.~\ref{app:norm}), so that for $\Ctcl\ll 1$ we can further replace $\tilde{\mathcal{C}}_{2,\text{MM}}$ by $\mathcal{C}_{2,\text{MM}}$ [otherwise, if needed, a shorter metastable regime $\tilde{t}''_2\leq t\leq \tilde{t}'_2$ could be considered to decrease the corrections with respect to $L1$ norm].

From Eq.~\eqref{eq:Tn2_CL} we obtain that the discrete classical evolution with the transfer matrix $\T_t$ can be approximated by the same operator for all $n\leq t'_2/t$ such that $n\ll 1/\Ccl$, in which case, the corresponding probability distributions are approximately stationary. If this holds for $n= 2$ (e.g., for $t=t''_2$ when $t'_2\geq 2t''_2$), it follows that there exist a separation in the spectrum of $\T_t$ (see the end of this section) and the probability distributions can be approximated as mixtures of $m_2$ approximately disjoint probability distributions (from Ref.~\cite{Gaveau2006}; note that there are additional assumptions on $\T_t$). Let $\Pbf_t$ be a projection on the MM of $\T_t$ and $\Cmm'$ be the corresponding corrections to the stationarity (in the classical space of $m$ configurations corresponding to metastable phases) [cf.~Eq.~\eqref{eq:Cmm}]. From~Eq.~\eqref{Expansion_MM2} we then have 
\begin{eqnarray}\label{eq:PP_MM2}
	\lVert \Pt_2- \Pbf_t\rVert_1&\leq & \lVert \Pt_2- \T_t \rVert_1+ \lVert\T_t-\Pbf_t \rVert_1\qquad\quad\\\nonumber
	&\lesssim&\tilde{\mathcal{C}}_{2,\text{MM}}+\Ccl+\Cmm'.
\end{eqnarray}

We now consider $m_2$ candidate metastable phases for the second MM. Let $\p_1$, ... $\p_{m_2}$ be the classical metastable phases for $\T_t$ [cf.~Eq.~\eqref{eq:classicality}], which under the projection $\Pbf_t$ on the slow modes of $\T_t$ we denote $\p_{l_2}'\equiv\Pbf_t\p_{l_2} $, $l_2=1,...,m_2$ [cf.~Eq.~\eqref{eq:rhotilde}]. Let $\mathbf{v}'_{k_2}$ be the dual basis to $\p_{l_2}'$, i.e., $\mathbf{v}_{k_2}'^{\text{T}}\p_{l_2}' =\delta_{k_2l_2}$, $k_2,l_2=1,...,m_2$. We introduce the corresponding corrections to the classicality [cf.~Eqs.~\eqref{eq:Ptilde} and~\eqref{eq:Ccl}]
\begin{equation}\label{eq:Ccl'_MM2}
	\Ccl'\equiv 2\max_{1\leq l\leq m}\sum_{k_2=1}^{m_2} \max\left[-(\mathbf{v}_{k_2}')_l,0\right],
\end{equation}
and its upper bound [cf.~Eqs.~\eqref{eq:Ctcl} and~\eqref{eq:Ctcl2}]
\begin{equation}\label{eq:Ccl2'_MM2}
	\Ccl'\leq \Ctcl'\equiv 2\sum_{k_2=1}^{m_2} \left[-\min_{1\leq l\leq m}(\mathbf{v}_{k_2}')_l\right]\leq m_2\Ccl'.
\end{equation}
We choose $m_2$ candidate metastable phases for the second MM as the projections of metastable phases in classical dynamics $\T_t$ [cf.~Eq.~\eqref{eq:rhotilde} and see Sec.~\ref{app:test}]
\begin{equation}\label{eq:rhotilde2}
	\trho_{2,l_2}\equiv \P_2\Big[\sum_{k=1}^m (\p_{l_2})_k \,\rho_k\Big] =\sum_{k=1}^m \big(\Pt_2\p_{l_2}\big)_k \,\trho_k, 
\end{equation}
$l_2=1,...m_2$. 

We now estimate the corresponding corrections to the classicality. Let $\pt_2$ denote the barycentric coordinates of $\P_2[\rho(0)]$ in Eq.~\eqref{Expansion_MM2} in the basis of Eq.~\eqref{eq:rhotilde2} [cf.~Eq.~\eqref{eq:Ptilde}]. We can relate $\pt_2$ to the barycentric coordinates  $\pt'$ in the basis of metastable phases $\p_{l_2}'$ of $\T_t$, as   $\pt_2=\C'^{-1}\pt'$, where 
\begin{equation}\label{eq:C'_MM2}
	(\C')_{kl}\equiv \mathbf{v}_{k_2}'^{\text{T}}\Pt_2 \p_{l_2},
\end{equation}
$k_2,l_2=1,...,m_2$. This transformation is close to identity 
\begin{eqnarray}\label{eq:C2'_MM2}
	\lVert \C'-\Ibf\rVert_1&\equiv& \max_{1\leq l_2 \leq m_2}\sum_{k_2=1}^{m_2} |\mathbf{v}_{k_2}'^{\text{T}}(\Pt_2-\Pbf_t) \p_{l_2}|\\\nonumber
	&=& \Big(1+\frac{\Ctcl'}{2}\Big) \max_{1\leq l_2 \leq m_2}\sum_{k_2=1}^{m_2} |\mathbf{v}_{k_2}^{\text{T}}(\Pt_2-\Pbf_t) \p_{l_2}|
	\\\nonumber
	&\leq &  \Big(1+\frac{\Ctcl'}{2}\Big)\lVert \Pt_2- \Pbf_t\rVert_1
	\\\nonumber
	&\leq & \Big(1+\frac{\Ctcl'}{2}\Big) \big(\tilde{\mathcal{C}}_{2,\text{MM}} +\Ccl+\Cmm'\big),
\end{eqnarray}
where we introduced $\mathbf{v}_{k_2}\equiv[\mathbf{v}'_k-\min_{1\leq l\leq m}(\mathbf{v}_{k_2}')_l\mathbf{i}]/[1+\Ctcl'/2]$, where $(\mathbf{i})_{l}=1$, $l=1,....,m$ [cf.~Eqs.~\eqref{eq:POVM0} and~\eqref{eq:classicality_corr2}]. 
Therefore,  
\begin{eqnarray}
	\left\lVert\pt_2\right\rVert_{1} -1&\leq& \left\lVert\C'^{-1}\right\rVert_{1}\left\lVert\pt'\right\rVert_{1}-1 \\\nonumber
	&\lesssim& (1+\left\lVert\C'-\Ibf\right\rVert_{1})\left\lVert\pt'\right\rVert_{1}-1
	\\\nonumber
	&\leq&   \Big(1+\frac{\Ctcl'}{2}\Big) \big(\tilde{\mathcal{C}}_{2,\text{MM}}+2\Ccl+\Cmm'\big) +\Ccl',
\end{eqnarray}
where we used $\lVert\pt'\rVert_{1}\lesssim 1+\Ccl'+(1+\Ctcl'/2)\Ccl$ [it is $1+\Ccl'$ for barycentric coordinates of projected probability distributions over $m$ metastable phases, cf.~Eq.~\eqref{eq:Ccl'_MM2}, but we project $\pt$ with $\lVert \pt\rVert_1\leq 1+\Ccl$; cf.~Eq.~\eqref{eq:C2'_MM2}]. Thus, we arrive at 
\begin{eqnarray}\label{eq:Ccl_MM2}
	\mathcal{C}_{2,\text{cl}} &\lesssim& \Big(1+\frac{\Ctcl'}{2}\Big)\big(\tilde{\mathcal{C}}_{2,\text{MM}} +2\Ccl+\Cmm'\big)+\Ccl'\qquad\\\nonumber
	&\lesssim& \mathcal{C}_{2,\text{MM}}+2\Ccl+\Cmm'+ \Ccl',
\end{eqnarray}
where the last inequality holds for $\Ctcl\ll 1$ and $\Ctcl'\ll 1$; cf.~Eqs.~\eqref{eq:Ctcl} and~\eqref{eq:Ccl2'_MM2}.
We conclude that when the first MM of an open quantum system is classical, \emph{the second metastable manifold is classical} as well.\\

Finally, we prove that a separation in the spectrum of $\T_t$ follows from $\lVert \T_t-\T_t^2\rVert_1 \ll 1$ [which occurs, e.g., when $t$ can be chosen so that $t''_2\leq t\leq 2t\leq t'_2$; cf.~Eq.~\eqref{eq:Tn2_CL}]. First, for $\mathbf{l}'_k$, being a left eigenvector of $\T_t$ corresponding to an eigenvalue $e^{t\lambda'_k}$, $k=1,...,m$, we have $\mathbf{l}_k'^{\text{T}}(\T_t-\T_t^2) \p =e^{t\lambda'_k}(1-e^{t\lambda'_k}) \mathbf{l}_k'^{\text{T}} \p$, while $|\mathbf{l}_k'^{\text{T}}(\T_t-\T_t^2) \p |\leq \mathbf{l}'_k\rVert_{\max}\lVert  \lVert \T_t-\T_t^2\rVert_1$. Therefore, choosing $(\p)_k=\delta_{kl}$, where $l$ is such that $|(\mathbf{l}'_k)_l|=\lVert\mathbf{l}'_k\rVert_{\max}$, we obtain $e^{t\lambda_k'^R}(1-e^{t\lambda_k'^R})\leq |e^{t\lambda'_k}(1-e^{t\lambda'_k})|\leq \lVert \T_t-\T_t^2\rVert_1$. Second, when $\lVert \T_t-\T_t^2\rVert_1\ll 1$, it follows that $e^{t\lambda_k'^R}\lesssim \lVert \T_t-\T_t^2\rVert_1$ or $1-e^{t\lambda_k'^R}\lesssim \lVert \T_t-\T_t^2\rVert_1$, as the function $x(1-x)$ is close to $0$ only at $x=0,1$. For eigenvalues ordered with a decreasing real part, we thus obtain $e^{t\lambda_k'^R}\ll 1$ for $k> m'_2$ and $1-e^{t\lambda_k'^R}\ll 1$ for $k\leq m_2'$. Finally, for $\T_t$ approximating $e^{t\Wt}$ where $t_2''\leq t\leq t_1''$ , we have that $m_2'=m_2$, as $e^{t\Wt}-\T_t$ can be  considered as a perturbation of  $\T_t$, leading to the first-order correction in the eigenvalue given by  $\mathbf{l}_k'^{\text{T}} (e^{t\Wt}-\T_t) \mathbf{r}_k'$, where $\mathbf{r}_k'$ is a right eigenvector of $\T_t$ and $|\mathbf{l}_k'^{\text{T}} (e^{t\Wt}-\T_t) \mathbf{r}_k'|\leq \lVert\mathbf{l}'_k\rVert_{\max}\lVert \T_t-e^{t\Wt}\rVert_1\lVert\mathbf{r}'_k\rVert_1$. 
Indeed, from Eq.~\eqref{eq:PP_MM2} mixtures of approximately disjoint $m_2'$ probability distributions are described by $m_2$ coefficients, which requires $m_2\geq m_2'$, but $m_2>m_2'$ would correspond to degeneracy of degrees of freedom, we assume does not occur (cf.~Sec.~\ref{app:defCMM}).

\subsection{Hierarchy of classical metastable phases}

From Sec.~\ref{sec:disjoint} of the main text, it follows that supports and basins of attractions of metastable phases in the second MM are approximately disjoint. Here, we discuss how $m_2$ metastable phases of the second MM originate from $m$ metastable phases of the first MM (cf.~Ref.~\cite{Gaveau1999}).
We show that for each metastable phase in the second MM, there exist at least one metastable phases of the first MM that during the second metastable regime evolves directly into it.  
Furthermore, metastable phase in the second MM are approximately disjoint mixtures of $m$ metastable phases of the first MM.
Finally, we argue that any metastable phase in the first MM during the second metastable regime that evolves into a nontrivial mixture of metastable phases in the second MM, the second MM is not supported on that phase, i.e., the phase belongs to the decay subspace.


\subsubsection{Supports and basins of attraction}

Let us consider an initial system state that evolves into $l_2$th metastable phase $\trho_{2,l_2}$ of the second MM, and let  $ \tilde{p}_l^{(l_2)}$, $l=1,...,m$, denote their barycentric coordinates with respect to $m$ metastable phases of the first MM. Furthermore, let  $\tilde{p}_{2,l_2}^{(l)}$, $l_2=1,...,m$, denote the barycentric coordinates of $m$ metastable phases of the first MM in the basis of $m_2$ metastable phases of the second MM. We have
\begin{equation}
	\trho_{2,l_2}=\sum_{l=1}^m \tilde{p}_l^{(l_2)} \sum_{k_2=1}^{m_2} \tilde{p}_{2,k_2}^{(l)} \trho_{2,k_2},
\end{equation}
and thus
\begin{equation}\label{eq:pk_simplex_MM2_0}
	\delta_{k_2,l_2}=\sum_{l=1}^m \tilde{p}_l^{(l_2)}  \tilde{p}_{2,k_2}^{(l)}. 
\end{equation} \\

We first show that for each metastable phase in the second MM, there exist at least one metastable phases of the first MM that during the second metastable regime evolves directly into it.  From Eq.~\eqref{eq:pk_simplex_MM2_0} and the triangle inequality 
\begin{equation} \label{eq:pk_simplex_MM2_9}
	\sum_{l=1}^m | \tilde{p}_l^{(l_2)}| | \tilde{p}_{2,l_2}^{(l)} |	\geq 1.
\end{equation}
Since $\sum_{l=1}^m | \tilde{p}_l^{(l_2)}|\leq 1+\Ccl$ [cf.~Eq.~\eqref{eq:Ccl}] and $\max_{1\leq l\leq m} | \tilde{p}_{2,l_2}^{(l)} | \sum_{l=1}^m | \tilde{p}_l^{(l_2)}| \geq \sum_{l=1}^m | \tilde{p}_l^{(l_2)}| | \tilde{p}_{2,l_2}^{(l)} |$ we have 
\begin{equation} \label{eq:pk_simplex_MM2_12}
	\max_{1\leq l\leq m}  \tilde{p}_{2,l_2}^{(l)} 	\geq \frac{1}{1+\Ccl}\gtrsim 1-\Ccl,
\end{equation}
where we removed absolute value since $\Ccl,\mathcal{C}_{2,\text{cl}}\ll 1$ [cf.~Eq.~\eqref{eq:Ptilde_min}].\\

Second, to discuss supports and basins of attraction,  for each $k_2=1,...,m_2$, we consider a subset $S_{k_2}$ consisting of the labels $l$ for which $\tilde{p}_{2,k_2}^{(l)}\geq \Delta$, where $\mathcal{C}_{2,\text{cl}}/2\leq \Delta\leq 1$.	Noting that $-\mathcal{C}_{2,\text{cl}}/2\leq \tilde{p}_{2,k_2}^{(l)}\leq 1+\mathcal{C}_{2,\text{cl}}/2$, we have
\begin{eqnarray}\label{eq:pk_simplex_MM2_3}
	\sum_{l=1}^m | \tilde{p}_l^{(l_2)}|  |\tilde{p}_{2,k_2}^{(l)}|	&\leq &\Delta \sum_{l\notin S_{k_2}}  | \tilde{p}_l^{(l_2)}| \\\nonumber&&+ \max_{1\leq l\leq m}  \tilde{p}_{2,k_2}^{(l)}	 \sum_{l\in S_{k_2}}   | \tilde{p}_l^{(l_2)}| ,
\end{eqnarray}
Using $\sum_{l\notin S_{k_2}}   | \tilde{p}_l^{(l_2)}|\leq 1+\Ccl -\sum_{l\in S_{k_2}}   | \tilde{p}_l^{(l_2)}|$ and $\max_{1\leq l\leq m}  \tilde{p}_{2,k_2}^{(l)}\leq 1+\mathcal{C}_{2,\text{cl}}/2$, we then obtain 
\begin{eqnarray} \nonumber
	\sum_{l\in S_{l_2}} | \tilde{p}_l^{(l_2)}|	&\geq & \frac{1-\Delta(1+\Ccl) }{1+\frac{\mathcal{C}_{2,\text{cl}}}{2}-\Delta}\qquad\\
	&\gtrsim & 1-\frac{2\Ccl+\mathcal{C}_{2,\text{cl}}}{2(1-\Delta)},\label{eq:pk_simplex_MM2_5}
\end{eqnarray}
where in the second inequality we assumed $1-\Delta\gg \mathcal{C}_{2,\text{cl}}$.

Furthermore, $\sum_{l=1}^m | \tilde{p}_l^{(l_2)}| \tilde{p}_{2,k_2}^{(l)}=\delta_{k_2,l_2}-\sum_{l=1}^m [ \tilde{p}_l^{(l_2)} - | \tilde{p}_l^{(l_2)}| ]\tilde{p}_{2,k_2}^{(l)}$. As  $-\mathcal{C}_{2,\text{cl}}/2\leq \tilde{p}_{2,k_2}^{(l)}\leq 1+\mathcal{C}_{2,\text{cl}}/2$ from Eq.~\eqref{eq:Ccl}, we have $|\sum_{l=1}^m [ \tilde{p}_l^{(l_2)} - | \tilde{p}_l^{(l_2)}| ] \tilde{p}_{2,k_2}^{(l)}|\leq \Ccl (1+\mathcal{C}_{2,\text{cl}}/2)$. Therefore 
\begin{eqnarray} \label{eq:pk_simplex_MM2_2}
	\sum_{l=1}^m | \tilde{p}_l^{(l_2)}|  \tilde{p}_{2,k_2}^{(l)} 	\leq \Ccl+\frac{\Ccl\mathcal{C}_{2,\text{cl}}}{2}, \quad k_2\neq l_2.\qquad\quad
\end{eqnarray}
We also have
\begin{eqnarray} \label{eq:pk_simplex_MM2_4}
	\sum_{l=1}^m | \tilde{p}_l^{(l_2)}|  \tilde{p}_{2,k_2}^{(l)}	&\geq &\Delta \sum_{l\in S_{k_2}}  | \tilde{p}_l^{(l_2)}| \\\nonumber
	&&+\!\min_{1\leq l\leq m} \!\min\!\left[0,\tilde{p}_{2,k_2}^{(l)}\right] \sum_{l\notin S_{k_2}}  | \tilde{p}_l^{(l_2)}|,
\end{eqnarray}
so that from $-\mathcal{C}_{2,\text{cl}}/2\leq \tilde{p}_{2,k_2}^{(l)}\leq 1+\mathcal{C}_{2,\text{cl}}/2$ we have
\begin{eqnarray}\label{eq:pk_simplex_MM2_6}
	\sum_{l\in S_{k_2}} | \tilde{p}_l^{(l_2)}|	&\leq & \frac{2\Ccl+\mathcal{C}_{2,\text{cl}}+\Ccl\mathcal{C}_{2,\text{cl}}}{2\Delta+\mathcal{C}_{2,\text{cl}}}\\\nonumber
	&\lesssim & \frac{2\Ccl+\mathcal{C}_{2,\text{cl}}}{2\Delta}, \quad k_2\neq l_2
\end{eqnarray}
where we assumed $\Delta\gg \mathcal{C}_{2,\text{cl}}/2$.

Eqs.~\eqref{eq:pk_simplex_MM2_5} and~\eqref{eq:pk_simplex_MM2_6} are analogous to Eqs.~\eqref{eq:B_support01} and\eqref{eq:B_support02} in the main text. In particular, Eq.~\eqref{eq:pk_simplex_MM2_5} shows that $S_{l_2}$ captures the support of $l_2$ metastable phases in the second metastable manifold as well as basin of attraction. Furthermore,  due to the classicality of the considered manifold, the subsets $S_{k_2}$, $k_2=1,...,m_2$,  can be chosen disjoint (not only approximately disjoint), e.g., for the choice $\Delta>(1+\mathcal{C}_{2,\text{cl}})/2$, since  $\sum_{k_2=1}^{m_2}|\tilde{p}_{2,k_2}^{(l)}|\leq 1+\mathcal{C}_{2,\text{cl}}$. In that case, we have $\sum_{l\in \cup_{k_2\neq l_2} S_{k_2}} | \tilde{p}_l^{(l_2)}|	=\sum_{k_2\neq l_2 } \sum_{l\in S_{k_2}} | \tilde{p}_l^{(l_2)}|	$ and 
\begin{eqnarray}\label{eq:pk_simplex_MM2_7}
	\sum_{k_2\neq l_2 } \sum_{l\in S_{k_2}} | \tilde{p}_l^{(l_2)}|	&\leq &1+\Ccl-\sum_{l\in S_{l_2}} | \tilde{p}_l^{(l_2)}| \qquad\\\nonumber
	&\lesssim & \Ccl+\frac{2\Ccl+\mathcal{C}_{2,\text{cl}}}{2(1-\Delta)}
\end{eqnarray}
from Eq.~\eqref{eq:pk_simplex_MM2_5}.
\\

Finally,  we show that, when $m(\Ccl+\mathcal{C}_{2,\text{cl}})\ll 1$, a metastable phase of the first MM on which a metastable phase of the second MM is supported evolves directly into it, or equivalently, a metastable phase in the second MM cannot be supported on phases in the first MM that during the second metastable regime evolve into a different phase. 
From Eq.~\eqref{eq:pk_simplex_MM2_2} we have  $| \tilde{p}_l^{(l_2)}|\lesssim (\Ccl+\mathcal{C}_{2,\text{cl}}) /| \tilde{p}_{2,k_2}^{(l)} |$. Thus, a metastable phase in the second MM cannot be supported on phases in the first MM that during the second metastable regime evolve into a different phase, i.e. $\trho_l$ with $l$ such that  $\tilde{p}_{2,k_2}^{(l)} \gg \Ccl+\mathcal{C}_{2,\text{cl}}$, where $k_2\neq l_2$ [for example, cf.~Eq.~\eqref{eq:pk_simplex_MM2_12}]. Similarly,  from Eq.~\eqref{eq:pk_simplex_MM2_9}
by noting that 	$\sum_{l=1}^m | \tilde{p}_l^{(l_2)}|\leq 1+\Ccl$ and $|\tilde{p}_{2,l_2}^{(l)}|	\leq 1+\mathcal{C}_{2,\text{cl}}/2$,  we obtain
\begin{eqnarray}\label{eq:pk_simplex_MM2_10}
	\left| \tilde{p}_l^{(l_2)}\right|\left|\tilde{p}_{2,l_2}^{(l)}\right|+\!\left(1\!+\!\Ccl\!-\!	| \tilde{p}_l^{(l_2)}|\right)\!\Big(1\!+\!\frac{\mathcal{C}_{2,\text{cl}}}{2}\Big)\!\geq\! 1,\qquad\quad
\end{eqnarray}
so that
\begin{eqnarray} \label{eq:pk_simplex_MM2_11}
	\left|\tilde{p}_{2,l_2}^{(l)}\right|\geq 1+\frac{\mathcal{C}_{2,\text{cl}}}{2} -\frac{2\Ccl+\mathcal{C}_{2,\text{cl}}+\Ccl\mathcal{C}_{2,\text{cl}}}{2	| \tilde{p}_l^{(l_2)}|}.\qquad
\end{eqnarray}
Thus, for $l$ such that $| \tilde{p}_l^{(l_2)}|\gg \Ccl+\mathcal{C}_{2,\text{cl}}$, we have that $\tilde{p}_{2,l_2}^{(l)}$ is approximated by $1$, that is, $l$th phase in the first MM evolves directly into $l_2$th phase in the second MM. 
We note, however, that for a highly-dimensional first MM, such probabilities do not need to exist, when $m\Ccl$ or $m\mathcal{C}_{2,\text{cl}}$ is of order $1$ [in the opposite case, $\max_{1\leq l\leq m}|\tilde{p}_{l}^{(l_2)}|\geq 1/m\gg\Ccl, \mathcal{C}_{2,\text{cl}}$].

\subsubsection{Decay subspace}

Below, we prove that any metastable phase in the first MM that evolves into a nontrivial mixture of metastable phases in the second MM, can be neglected in the supports of metastable phases in the second MM. Therefore, the supports of those metastable phases  in the first MM belong to the approximate decay subspace of the system space (cf.~Sec.~\ref{app:disjoint}).\\

For each $k_2=1,....,m_2$, let us consider a subset $D_{k_2}$ consisting of the labels $l$ for which $\Delta_1\leq \tilde{p}_{2,k_2}^{(l)}\leq \Delta_2$, where $0\leq \Delta_1\leq \Delta_2\leq 1$.
Noting that $\sum_{l=1}^m | \tilde{p}_l^{(l_2)}|\leq 1+\Ccl$ for any $\Delta$, and considering $\Delta=\Delta_2$ in Eq.~\eqref{eq:pk_simplex_MM2_5} we obtain
\begin{eqnarray}
	\sum_{l\in D_{l_2}} | \tilde{p}_l^{(l_2)}|&\leq& \sum_{l=1}^m | \tilde{p}_l^{(l_2)}| - \sum_{l\in S_{l_2}} | \tilde{p}_l^{(l_2)}| \nonumber\\
	\label{eq:pk_decay_MM2_1}
	&\lesssim& \Ccl+\frac{2\Ccl+\mathcal{C}_{2,\text{cl}}}{2(1-\Delta_2)},\qquad
\end{eqnarray}
where we assumed $1\!-\!\Delta_2\gg \mathcal{C}_{2,\text{cl}}$.
Considering $\Delta=\Delta_1$ in Eq.~\eqref{eq:pk_simplex_MM2_6}, we get
\begin{equation}\label{eq:pk_decay_MM2_2}
	\sum_{l\in D_{k_2}} | \tilde{p}_l^{(l_2)}|\lesssim\frac{2\Ccl+\mathcal{C}_{2,\text{cl}}}{2\Delta_1}, \quad k_2\neq l_2.
\end{equation}

Furthermore, considering $\Delta_1+\Delta_2\geq 1+\mathcal{C}_{2,\text{cl}}$, we have from the classicality of the second MM that $l$ belonging to the decay subspace $D_{k_2}$ for any $k_2=1,...,m$ does not belong to the support of any metastable phases of the second MM captured by $S_{l_2}$, $l_2=1,,...,m$ provided that $\Delta>\Delta_2$ is considered. 
Therefore, when we choose $1\!-\!\Delta_2\gg \Ccl+\mathcal{C}_{2,\text{cl}}/2$ [which also implies $\Delta_1\gg \Ccl+\mathcal{C}_{2,\text{cl}}/2$ in Eq.~\eqref{eq:pk_decay_MM2_2}], from Eq.~\eqref{eq:pk_simplex_MM2_5} metastable phases $\trho_l$ of the first MM with $l\in\cup_{k_2=1}^{m_2} D_{k_2}$ contribute negligibly to the support of the metastable phases in the second MM. Note that those are the metastable phases that evolve into nontrivial mixtures of $m_2$ metastable phases, as for $l\in D_{l_2}$ we have  $\Delta_1\leq \tilde{p}_{2,l_2}^{(l)} \leq \Delta_2$, so that $\tilde{p}_{2,l_2}^{(l)}\geq \Delta_1$ is nonnegligible and so is the sum of other barycentric coordinates, as   $ \sum_{k_2\neq l_2} \tilde{p}_{2,k_2}^{(l)}=1\!-\!\tilde{p}_{2,l_2}^{(l)} \geq 1\!-\!\Delta_2$.

\subsection{Hierarchy of classical long-time dynamics}

Here, we discuss how the approximation of the long-time dynamics by classical stochastic dynamics can be refined to take into account the hierarchy of metastabilities. We consider both continuous and discrete approximations of the long-time dynamics (cf.~Sec.~\ref{app:Leff}).

\subsubsection{Hierarchy of continuous approximations of classical long-time dynamics}

In general the norm $\lVert \Wt\rVert_1$ is dominated by the fastest transitions between metastable phases, while the final relaxation time $\tilde{\tau}$ describes the longest timescale of the dynamics. As a consequence, in the presence of the second metastable regime, the classical dynamics $\W$ defined in Eq.~\eqref{eq:W} may not approximate the long time-dynamics up to and beyond the relaxation time. Furthermore,  the conditions in Eqs.~\eqref{eq:cond_CL} and~\eqref{eq:cond2_CL} are in general no longer fulfilled, and the stationary state and the resolvent of $\Wt$ in Eq.~\eqref{eq:Wtilde} may no longer approximated well by the stationary state and the resolvent of the classical stochastic dynamics $\W$. 
Nevertheless, we show below how the approximations can be modified to take into account the hierarchy of metastable regimes in the system dynamics. \\

\emph{Approximating the second MM}. Instead of considering the approximation of the long-time dynamics by the discrete dynamics $\T_t$ we can consider a generally weaker approximation by the classical stochastic generator $\W$ in Eq.~\eqref{eq:W}. This gives instead of Eq.~\eqref{eq:Tn2_CL}
\begin{eqnarray}
	\lVert\Pt_2- e^{t\W} \rVert_1 &\leq& \lVert\Pt_2- e^{t\Wt} \rVert_1+\lVert e^{t\W}- e^{t\Wt} \rVert_1\qquad \label{eq:etW_MM2}
\end{eqnarray}
[cf.~Eqs.~\eqref{eq:etW_CL} and~\eqref{eq:pss_CL}]. Therefore, this approximation holds well if for time $t$ after the relaxation towards the second metastable regime , $\lVert\Pt_2- e^{t\Wt} \rVert_1\ll \tilde{\mathcal{C}}_{2,\text{MM}}$, we also have $\lVert e^{t\W}- e^{t\Wt} \rVert_1\ll 1$, as guaranteed, e.g., by 
\begin{equation}\label{eq:cond_CL_MM2}
	t_2''\lVert \Wt\rVert_1\ll  \frac{1}{\sqrt{\Ccl}}.
\end{equation} 
Furthermore, the second metastability corresponds to the metastability in the classical dynamics generated by $\W$, with $\P_2$ approximated by the projection $\Pbf$ on the low-lying eigenmodes of $\W$ since [Eq.~\eqref{eq:PP_MM2}]
\begin{eqnarray}\label{eq:PP_MM2_2}
	\lVert \Pt_2- \Pbf\rVert_1&\leq & \lVert \Pt_2-  e^{t\W}\rVert_1+ \lVert e^{t\W}-\Pbf \rVert_1
\end{eqnarray}
and  $\lVert e^{t\W}-\Pbf \rVert_1\leq \Cmm'$ with $\Cmm'$ denoting the corrections to the stationarity of $e^{t\W}$. 
\\

\emph{Approximating long-time dynamics for $t\geq t'_2$}.  After the second metastable regime, the system dynamics in Eq.~\eqref{Expansion_MM2} is generated by $\P_2\L\P_2$ [cf.~Eq.~\eqref{eq:Leff}], which we denote in the basis of $m_2$ metastable phases by  $\Wt_{2}$ [cf.~Eq.~\eqref{eq:Wtilde}]. $\Wt_{2}$ is approximated by classical stochastic dynamics $\W_{2}$ [cf.~Eqs.~\eqref{eq:W},~\eqref{eq:deltaW} and~\eqref{eq:etW_CL}], so that
\begin{equation}\label{eq:etW_CL_MM2}
	\lVert e^{t\Wt_{2}} -e^{t \W_{2}} \rVert_1 \lesssim 2 \sqrt{\mathcal{C}_{2,\text{cl}}} \,t\lVert \Wt_{2}\rVert_1,
\end{equation}
where $ \mathcal{C}_{2,\text{cl}}$  is in the approximation of the second MM by the simplex of $m_2$ metastable phases and we consider the L1-norm in the basis of $m_2$ metastable phases. Instead, in the basis of $m$ metastable phases of the first MM, we denote the action of $\W_{2}$ by $\Wt_{2,1}$ and we have 
\begin{eqnarray}\nonumber
	\lVert (e^{t\Wt}-e^{t \Wt_{2,1}})\Pt_2  \rVert_1 &\lesssim& 2\lVert\Pt_2  \rVert_1 \sqrt{\mathcal{C}_{2,\text{cl}}} \,t \, \lVert \Wt_{2}\rVert_1.\\\label{eq:etW_CL_MM2_2}
	&\lesssim& 2 \sqrt{\mathcal{C}_{2,\text{cl}}} \,t \, \lVert \Wt_{2}\rVert_1.
\end{eqnarray}
where in the second inequality we neglected $\lVert\Pt_2  \rVert_1-1\leq \tilde{\mathcal{C}}_{2,\text{MM}}$, i.e., the corrections to the positivity of   the projection $\P_2$ on the second MM in L1 norm. We have that $\lVert\Pt_2  \rVert_1-1\leq(1+\Ctcl/2)\mathcal{C}_{2,+}$, so that for $\Ctcl\ll 1$ we can further replace $\lVert\Pt_2  \rVert_1-1$ by the corrections $\mathcal{C}_{2,+}$ in the trace norm [cf.~Eq.~\eqref{eq:Cp} and Sec.~\ref{app:norm}].
Note that $\Wt_{2,1}$ in Eq.~\eqref{eq:etW_CL_MM2_2} is defined on the image of $\Pt_2$ and in general is \emph{not} a classical stochastic generator, as it is probability conserving but only approximately positive. \\

\emph{Approximating the stationary state}.  When there exists time $t$ such that $t \, \lVert \Wt_{2}\rVert_1 \ll 1/\sqrt{\mathcal{C}_{2,\text{cl}}}$ and $\lVert e^{t\Wt_{2}} -\mathbf{\tilde{P}}_\text{\!2,ss} \rVert_1 \ll 1$ [here $\mathbf{\tilde{P}}_\text{\!2,ss}$ denotes the projection $\P_\text{ss}$ on the stationary state $\rhoss$ in the basis of $m_2$ metastable phases], which requires 
\begin{equation}\label{eq:cond_CL_MM2_2}
	\tau\lVert \Wt_{2}\rVert_1\leq \tilde{\tau}_{2}\lVert \Wt_{2}\rVert_1\ll \frac{1}{\sqrt{\mathcal{C}_{2,\text{cl}}}},
\end{equation} 
[$\tilde{\tau}_{2}$ is the relaxation time with respect to $\lVert e^{t\Wt_{2}} -\mathbf{\tilde{P}}_\text{\!2,ss} \rVert_1 $; we have $\tilde{\tau}_{2}\approx\tau$ when $\tilde{\mathcal{C}}_\text{2,cl}\ll 1$] the stationary state is captured by the stationary distribution $\mathbf{p}_\text{2,ss}$ of $\W_{2}$ as 
\begin{equation}\label{eq:pss_CL_MM2}
	\lVert\mathbf{\tilde{p}}_\text{2,ss}-\mathbf{p}_\text{2,ss} \rVert_1 	\lesssim \lVert \mathbf{\tilde{P}}_\text{\!2,ss}-e^{t\Wt_{2}} \rVert_1 + 2 \sqrt{\mathcal{C}_{2,\text{cl}}} \,t\lVert \Wt_{2}\rVert_1
\end{equation}
[cf.~Eqs.~\eqref{eq:pss_CL} and~\eqref{eq:pss_CL_proof}].
In the basis of $m$ metastable phases of the first MM (or in the trace norm), the distance between the corresponding vectors (or matrices) is in the leading order bounded by $\lVert\mathbf{\tilde{p}}_\text{2,ss}-\mathbf{p}_\text{2,ss} \rVert_1$ and thus by Eq.~\eqref{eq:pss_CL_MM2}  (cf.~Sec.~\ref{app:norm}).
\\

\emph{Approximating the dynamics resolvent}.  We now show that the resolvent of the dynamics can be approximated in two steps: for the faster modes by the resolvent of $\W$ and for the slower modes by the resolvent of $\W_2$; see Eq.~\eqref{eq:DeltaR_MM2} below and cf.~Sec.~\ref{app:Leff}.

We have [cf.~Eqs.~\eqref{eq:R_CL} and~\eqref{eq:R1_CL}]
\begin{eqnarray}
	&&\left\lVert\Rt\,+ \int_{0}^{t_2} d t_1\left(e^{t_1\Wt}-\Ptss\right)\Pt_2 \,+ \int_{0}^{t_2} d t_1 e^{t_1\Wt}\left(\Ibf-\Pt_2\right) \right\rVert_1\nonumber\\
	&&\leq \lVert e^{t_2 \Wt}-\Ptss\rVert_1\lVert\Rt\rVert_1. \qquad\,\,\label{eq:R1_CL_MM2}
\end{eqnarray}
The second term in Eq.~\eqref{eq:R1_CL_MM2}  can be approximated by the classical dynamics after the second metastable regime as [cf.~Eqs.~\eqref{eq:R2_CL},~\eqref{eq:etW_CL_MM2} and~\eqref{eq:pss_CL_MM2}]
\begin{eqnarray}
	&&\left\lVert \int_{0}^{t_2} d t_1\left(e^{t_1\Wt}-\Ptss\right)\Pt_2 -  \int_{0}^{t_2} d t_1\left(e^{t_1\W_2}-\mathbf{P}_\text{\!2,ss} \right)\Pt_2  \right\rVert_1\nonumber\\
	&&\lesssim \left\lVert \int_{0}^{t_2} d t_1\left(e^{t_1\Wt_{2}}-\mathbf{\tilde{P}}_\text{\!2,ss}\right) -  \int_{0}^{t_2} d t_1\left(e^{t_1\W_{2}}-\mathbf{P}_\text{\!2,ss} \right) \right\rVert_1\nonumber\\
	&&\lesssim 3 \sqrt{\mathcal{C}_{2,\text{cl}}} t_2^2 \lVert\Wt_{2}\rVert_1 + t_2 \lVert \mathbf{\tilde{P}}_\text{\!2,ss}-e^{t_2\Wt_{2}} \rVert_1,\,\,\label{eq:R2_CL_MM2}
\end{eqnarray}
where in the first line we introduced the projection $\mathbf{P}_\text{\!2,ss}$ on the stationary state of $\W_2$, and the second and third line refer to operators in the basis of $m_2$ metastable phases of the second MM.
For the resolvent $\R_2$ for the dynamics $\W_2$, we have [cf.~Eq.~\eqref{eq:R1_CL_MM2} and Sec.~\ref{app:norm}]
\begin{eqnarray}\label{eq:R3_CL_MM2}
	&&\left\lVert\R_2\,+  \int_{0}^{t_2} d t_1\left(e^{t_1\W_2}-\mathbf{P}_\text{\!2,ss} \right)\Pt_2  \right\rVert_1\\
	&&=\left\lVert\int_{t_2}^\infty d t_1 \left(e^{t_1\W_2}-\mathbf{P}_\text{\!2,ss} \right)\Pt_2  \right\rVert_1\nonumber\\
	&&\lesssim \lVert e^{t_2 \W_2}-\mathbf{P}_\text{\!2,ss}\rVert_1\lVert\R_2\rVert_1\nonumber\\
	&&\lesssim\lVert e^{t_2 \W_{2}}-\mathbf{P}_\text{\!2,ss}\rVert_1\lVert\R_2\rVert_1. \qquad\,\,\nonumber
\end{eqnarray}
The third term in Eq.~\eqref{eq:R1_CL_MM2} can be approximated by the classical dynamics $\W$ as [cf.~Eqs.~\eqref{eq:R2_CL}]
\begin{eqnarray}
	&&\left\lVert \int_{0}^{t} d t_1e^{t_1\Wt}(\Ibf-\Pt_2) - \int_{0}^{t}\!\! d t_1e^{t_1\W}(\Ibf-\Pt_2) \right\rVert_1\nonumber\\
	&&\leq \int_{0}^{t} d t_1  \lVert e^{t_1\Wt}-e^{t_1\W} \rVert_1 \lVert \Ibf-\Pt_2 \rVert_1 \nonumber\\
	&&\lesssim 2 t^2 \sqrt{\Ccl} \lVert\Wt\rVert_1, \qquad\,\,\label{eq:R4_CL_MM2}
\end{eqnarray}
where we used $\lVert \Ibf-\Pt_2 \rVert_1\leq\lVert \Ibf \rVert_1+\lVert \Pt_2-e^{t \Wt} \rVert_1+\lVert e^{t \Wt} \rVert_1\leq 1+ \tilde{\mathcal{C}}_{2,\text{MM}}+1+\Ccl\lesssim 2$ (for $t$ chosen within the second metastable regime), while [cf.~Eq.~\eqref{eq:R1_CL}]
\begin{eqnarray}\label{eq:R5_CL_MM2}
	&&\left\lVert \int_{t}^{t_2} d t_1e^{t_1\Wt}(\Ibf-\Pt_2)\right\rVert_1\\
	&&\leq \lVert e^{t\Wt} (\Ibf-\Pt_2)\rVert_1 \left\lVert \int_{0}^{t_2-t} \!\!\!\!d t_1e^{t_1\Wt}(\Ibf-\Pt_2)\right\rVert_1\nonumber\\\nonumber
	&&\leq \lVert e^{t\Wt} (\Ibf-\Pt_2)\rVert_1 \Big[1\!+\!\lVert e^{(t_2-t)\Wt}\! (\Ibf-\Pt_2)\rVert_1 \Big]\lVert\Rt\rVert_1.\qquad\,\,
\end{eqnarray}
For the resolvent $\R$ of the classical dynamics  $\W$, we further have
\begin{eqnarray}\nonumber
	&&\left\lVert\R (\Ibf-\mathbf{P}) \,+ \!\!\int_{0}^{t} \!\!\!d t_1(e^{t_1\W}-\Pss) [(\Ibf-\mathbf{P})\!+\!(\mathbf{P}-\Pt_2)] \right\rVert_1\\
	&&\leq\lVert e^{t\W} (\Ibf-\mathbf{P})\rVert_1 \lVert \R (\Ibf-\mathbf{P})\rVert_1 +t \lVert\mathbf{P}-\Pt_2\rVert_1. \qquad\,\,\label{eq:R6_CL_MM2}
\end{eqnarray}

From Eqs.~(\ref{eq:R1_CL_MM2})-(\ref{eq:R6_CL_MM2}) we can approximate
\begin{eqnarray}
	&&\lVert \Rt-\R(\Ibf-\mathbf{P})-\R_2\rVert_1\\\nonumber
	&&\lesssim \lVert e^{t_2 \Wt}-\Ptss\rVert_1\lVert\Rt\rVert_1+3 \sqrt{\mathcal{C}_{2,\text{cl}}} t_2^2 \lVert\Wt_{2}\rVert_1\\\nonumber
	&&+ t_2 \lVert \mathbf{\tilde{P}}_\text{\!2,ss}\!-e^{t_2\Wt_{2}}\! \rVert_1\!+\!\lVert e^{t \W_2}\!\!\!\!-\mathbf{P}_\text{\!2,ss}\rVert_1\lVert\R_2\rVert_1
	\!+\!2 t^2 \sqrt{\Ccl} \lVert\Wt\rVert_1\\\nonumber
	&&+\lVert e^{t\Wt} (\Ibf-\Pt_2)\rVert_1[1+\lVert e^{(t_2-t)\Wt} (\Ibf-\Pt_2)\rVert_1 ]\lVert\Rt\rVert_1\\\nonumber
	&&+\lVert e^{t\W} (\Ibf-\mathbf{P})\rVert_1 \lVert \R (\Ibf-\mathbf{P})\rVert_1 +t \lVert\mathbf{P}-\Pt_2\rVert_1.
\end{eqnarray}
We have $\lVert e^{t_2 \W_2}\!\!\!\!-\mathbf{P}_\text{\!2,ss}\!\rVert_1\leq \lVert e^{t_2 \Wt_{2}}\!\!\!\!-\Pt_\text{ss}^{(2)}\!\rVert_1+\lVert e^{t_2 \Wt_{2}}\!\!\!\!-e^{t_2 \W_2}\!\!\rVert_1+\lVert\Pt_\text{ss}^{(2)}\!- \mathbf{P}_\text{\!2,ss}\!\rVert_1\lesssim 2\lVert \mathbf{\tilde{P}}_\text{\!2,ss}-e^{t_2\Wt_{2}} \rVert_1 + 4\sqrt{\mathcal{C}_{2,\text{cl}}} \,t_2\lVert \Wt_{2}\rVert_1$ and, similarly, $\lVert e^{t\W} (\Ibf-\mathbf{P})\rVert_1\leq \lVert (e^{t\W} -e^{t\Wt}) (\Ibf-\mathbf{P})\rVert_1 +\lVert e^{t\Wt}(\Ibf-\Pt_2)\rVert_1+\lVert e^{t\Wt}(\mathbf{P}-\Pt_2)\rVert_1\lesssim 4 t \sqrt{\Ccl} \lVert\Wt\rVert_1+ \lVert e^{t\Wt}(\Ibf-\Pt_2)\rVert_1+\lVert \mathbf{P}-\Pt_2\rVert_1$ [cf.~Eq.~\eqref{eq:etW_MM2}]. Furthermore, $\R_2=\R_2\Pt_2$ and thus $\lVert\R_2\Pt_2\rVert_1\leq\lVert(\Rt-\R_2)\Pt_2\rVert_1+\lVert\Rt\Pt_2\rVert_1\lesssim 2\lVert \Rt-\R(\Ibf-\mathbf{P})-\R_2\rVert_1+\lVert\R(\Ibf-\mathbf{P})\rVert_1\lVert\Pt_2-\mathbf{P}\rVert_1 +2\lVert\Rt\rVert_1 $ [as $\lVert\Pt_2\rVert_1\lesssim 2$ and $\R(\Ibf-\mathbf{P})\Pt_2=\R(\Ibf-\mathbf{P})(\Pt_2-\mathbf{P})$ from $(\Ibf-\mathbf{P})^2=(\Ibf-\mathbf{P})$]. Analogously,  $\lVert\R(\Ibf-\mathbf{P})\rVert_1 \lesssim 2\lVert \Rt-\R(\Ibf-\mathbf{P})-\R_2\rVert_1+ \lVert\R_2\rVert_1\lVert\Pt_2-\mathbf{P}\rVert_1+ 2 \lVert\Rt\rVert_1$ [as $\lVert\Ibf-\mathbf{P}\rVert_1\leq 1+\lVert\mathbf{P}\rVert_1\lesssim 2$ and $\lVert\R_2(\Ibf-\mathbf{P})\rVert_1=\lVert\R_2(\Pt_2-\mathbf{P})\rVert_1\leq  \lVert\R_2\rVert_1\lVert\Pt_2-\mathbf{P}\rVert_1$]. Combining these two results we have $\lVert\R(\Ibf-\mathbf{P})\rVert_1 \lesssim 2(1+\lVert\Pt_2-\mathbf{P}\rVert_1) (\lVert \Rt-\R(\Ibf-\mathbf{P})-\R_2\rVert_1+ \lVert\Rt\rVert_1)/ (1-\lVert\Pt_2-\mathbf{P}\rVert_1)\lesssim 2(\lVert \Rt-\R(\Ibf-\mathbf{P})-\R_2\rVert_1+ \lVert\Rt\rVert_1)$ and, analogously, $\lVert\R_2\rVert_1 \lesssim 2(\lVert \Rt-\R(\Ibf-\mathbf{P})-\R_2\rVert_1+ \lVert\Rt\rVert_1)$ [cf.~Eq.~\eqref{eq:PP_MM2_2}]. Therefore, we arrive at [cf.~Eq.~\eqref{eq:DeltaR}]
\begin{eqnarray}\label{eq:DeltaR_MM2}
	&&\frac{\lVert \Rt-\R(\Ibf-\mathbf{P})-\R_2\rVert_1}{\lVert\Rt\rVert_1}\\\nonumber
	&&\lesssim \bigg\{\lVert e^{t _2\Wt}-\Ptss\rVert_1+t_2 \Big(3 \frac{ t_2}{\lVert\Rt\rVert_1}+8 \Big)\sqrt{\mathcal{C}_{2,\text{cl}}}  \lVert\Wt_{2}\rVert_1\\\nonumber
	&&\quad+ \Big(\frac{t_2}{\lVert\Rt\rVert_1}+4\Big) \lVert \mathbf{\tilde{P}}_\text{\!2,ss}-e^{t_2\Wt_{2}} \rVert_1\\\nonumber
	&&\quad
	+2 t\Big( \frac{t}{\lVert\Rt\rVert_1}+4\Big) \sqrt{\Ccl} \lVert\Wt\rVert_1\\\nonumber
	&&\quad+\lVert e^{t\Wt} (\Ibf-\Pt_2)\rVert_1[3+\lVert e^{(t_2-t)\Wt} (\Ibf-\Pt_2)\rVert_1 ]\\\nonumber
	&&\quad +\Big(\frac{t}{\lVert\Rt\rVert_1}+2\Big) \lVert\mathbf{P}-\Pt_2\rVert_1\bigg\}/
	\\\nonumber
	&&\quad\Big[1-4\lVert \mathbf{\tilde{P}}_\text{\!2,ss}-e^{t_2\Wt_{2}}\! \rVert_1 + 8\sqrt{\mathcal{C}_{2,\text{cl}}} \,t_2\lVert \Wt_{2}\rVert_1\\\nonumber
	&&\quad\quad-8 t\sqrt{\Ccl} \lVert\Wt\rVert_1-2 \lVert e^{t\Wt}(\Ibf-\Pt_2)\rVert_1-2\lVert \mathbf{P}-\Pt_2\rVert_1
	\Big].
\end{eqnarray}
The above inequality holds for any choice of times $t$ and $t_2$ provided that $t\leq t_2$. When $t_2 \, \lVert \Wt_{2}\rVert_1 \ll 1/\sqrt{\mathcal{C}_{2,\text{cl}}}$ and $\lVert e^{t_2\Wt_{2}} -\mathbf{\tilde{P}}_\text{\!2,ss} \rVert_1 \ll 1$ [cf.~Eq.~\eqref{eq:pss_CL_MM2}], as well as, $t \, \lVert \Wt\rVert_1 \ll 1/\sqrt{\Ccl}$ and $\lVert e^{t\Wt}- \Pt_2 \rVert_1 \ll 1$ [$t\geq \tilde{t}''_2$, cf.~Eq.~\eqref{eq:cond_CL_MM2}], the leading corrections in the right-hand side of Eq.~\eqref{eq:DeltaR_MM2} are given by the numerator as $\lVert e^{t\Wt}(\Ibf-\Pt_2)\rVert_1\lesssim 2 \tilde{\mathcal{C}}_{2,\text{MM}}$~\cite{Note1}. In this case, we can obtain $\lVert \Rt-\R(\Ibf-\mathbf{P})-\R_2\rVert_1/\lVert\Rt\rVert_1\ll 1$ by further choosing $t\ll  \lVert \Rt\rVert_1$ [possible as $t$ can be chosen within the second metastable regime so that $t\ll -1/\lambda_k^R$ for $k\geq m_2$ while $(1-\Cp-\Ccl)/|\lambda_k^R|\lesssim \lVert \Rt\rVert_1$ from Eq.~\eqref{eq:Rnorm3}], and assuming that we can further choose $t_2$ so that $t_2 \lVert \mathbf{\tilde{P}}_\text{\!2,ss}-e^{t_2\Wt_{2}} \rVert_1/\lVert \Rt\rVert_1 \ll 1 $ [which implies $t_2^2 \sqrt{\mathcal{C}_{2,\text{cl}}}\lVert\Wt_{2}\rVert_1/\lVert\Rt\rVert_1\ll 1 $]. Note that the last assumption requires [cf.~Eq.~\eqref{eq:cond2_CL}]
\begin{eqnarray}\label{eq:cond2_CL_MM2}
	\tau\lVert \Wt_{2}\rVert_1  &\leq&\tilde{\tau}_{2} \lVert \Wt_{2}\rVert_1  \\\nonumber
	&\ll&\min\Bigg(\frac{1}{\sqrt{\mathcal{C}_{2,\text{cl}}}}, \sqrt{\frac{ \lVert \Rt\rVert_1 \lVert \Wt_{2}\rVert_1}{\sqrt{\mathcal{C}_{2,\text{cl}}}} }\Bigg).\qquad\qquad
\end{eqnarray}

\subsubsection{Hierarchy of discrete approximations of classical long-time dynamics}

We now consider approximation of the long time-dynamics by the classical discrete dynamics  $\T_t$ defined in Eqs.~\eqref{eq:Tt} and~\eqref{eq:Tt2}.	The approximation corrections $\lVert e^{t\Wt}- \T_t\rVert_1$ for time $t$ chosen before the final relaxation are dominated by the fastest transitions between $m$ metastable phases. As a consequence, in the presence of the second metastable regime, for $\T_t$ with time before the relaxation to the second MM, $t< t_2''$, $\T_t^{n}$ in general may not approximate the long time-dynamics for $nt\geq t_2'$ [cf.~Eqs.~\eqref{eq:T_CL} and~\eqref{eq:Tn_CL}].  We now discuss how the approximations can be modified to include the hierarchy of timescales.\\

\emph{Approximating the second MM}. When time $t\leq t_2'$, possibly shorter than the second metastable regime, fulfills [cf.~Eq.~\eqref{eq:cond_T_CL}]
\begin{equation}\label{eq:cond_T_CL_MM2}
	\frac{t}{t''_2}\gg \Ccl,
\end{equation} 
the long-time dynamics $e^{nt\L}$ is well approximated by $\T_t^n$ after the relaxation into the second metastable regime [cf.~Eq.~\eqref{eq:Tn_CL}]. It also follows that the projection of the second MM is approximated by the projection $\Pbf_t$ on the eigenmodes of $\T_t$ with absolute value of eigenvalues close to $1$ as [cf.~Eq.~\eqref{eq:PP_MM2}]
\begin{eqnarray}\label{eq:PP_MM2_3}
	\lVert \Pt_2- \Pbf_t\rVert_1&\leq & \lVert \Pt_2-  \T_t^n\rVert_1+ \lVert \T_t^n-\Pbf_t \rVert_1\qquad\quad\\\nonumber
	&\lesssim& \tilde{\mathcal{C}}_{2,\text{MM}}+n \Ccl+\Cmm'\ll 1,
\end{eqnarray}
while for $n\leq t''_1/t$ we have $\lVert \Pt_2-  \T_t^n\rVert_1\lesssim \tilde{\mathcal{C}}_{2,\text{MM}}+n \Ccl$  and	$\lVert \T_t^n-\Pbf_t \rVert_1\leq \Cmm'$, where $\Cmm'$ denotes the corrections to the stationarity of $\T_t^n$. \\

\emph{Approximating long-time dynamics for $t\geq t'_2$}.  After the second metastable regime, when system states are effectively restricted to the smaller second MM, the system dynamics $e^{t\Wt_{2}}$ in the basis of $m_2$ metastable phases [cf.~Eq.~\eqref{eq:Wtilde}] is approximated by classical discrete dynamics $\T_{2,t}$ [cf.~Eqs.~\eqref{eq:T_CL} and~\eqref{eq:Tn_CL}],
\begin{equation}\label{eq:Tn_CL_MM2}
	\lVert e^{nt\Wt_{2}} -\T_{2,t}^n \rVert_1 \lesssim n\mathcal{C}_{2,\text{cl}},
\end{equation}
where $ \mathcal{C}_{2,\text{cl}}$  is in the approximation of the second MM by the simplex of $m_2$ metastable phases and we consider the L1-norm in the basis of $m_2$ metastable phases. Similarly, in the basis of $m$ metastable phases of the first MM, where we denote the action of $\T_{2,t} $ by $\Tt_{t,2}$ [cf.~Eq.~\eqref{eq:etW_CL_MM2_2}]
\begin{equation}\label{eq:Tn_CL_MM2_2}
	\lVert (e^{nt\Wt}-\Tt_{t,2}^n)\Pt_2  \rVert_1 \lesssim n\mathcal{C}_{2,\text{cl}}.
\end{equation}
Note that $\Tt_{t,2}$ is defined on the image of $\Pt_2$ only. Furthermore, although it is trace preserving, it is not positive.  We note, however, that since $\T_t$ can approximate the long times dynamics at times $nt$, where $n\ll1/\Ccl$ [cf.~Eq.~\eqref{eq:Tn_CL}], for $t$ chosen for long enough the dynamics inside the second MM can be also captured by $\T_t^n$.\\

\emph{Approximating the stationary state}.  For long enough $t$ such that $\lVert e^{n t\Wt_{2}} -\mathbf{\tilde{P}}_\text{\!2,ss}\rVert\ll 1$ is for an integer $n \ll 1/\mathcal{C}_{2,\text{cl}}$, where $\mathbf{\tilde{P}}_\text{\!2,ss}$ denotes the projection $\P_\text{ss}$ on the stationary state $\rhoss$ in the basis of $m_2$ metastable phases, the stationary state is captured by the stationary distribution $\mathbf{p}_\text{2,ss}$ of $\T_{2,t}$ as [cf.~Eqs.~\eqref{eq:pss_T_CL} and~\eqref{eq:pss_T_CL_proof}]
\begin{equation}\label{eq:pss_T_CL_MM2}
	\lVert\mathbf{\tilde{p}}_\text{2,ss}-\mathbf{p}_\text{2,ss} \rVert_1 	\lesssim \lVert \mathbf{\tilde{P}}_\text{\!2,ss}-e^{nt\Wt_{2}} \rVert_1 + n\mathcal{C}_{2,\text{cl}}.
\end{equation}
Furthermore, in the basis of $m$ metastable phases of the first MM (or in the trace norm), the distance between the corresponding vectors (or matrices) is in the leading order bounded by $\lVert\mathbf{\tilde{p}}_\text{2,ss}-\mathbf{p}_\text{2,ss} \rVert_1$ and thus by Eq.~\eqref{eq:pss_T_CL_MM2};  cf.~Sec.~\ref{app:norm}. We note that this requires  $t$ such that [see Eq.~\eqref{eq:cond_T_CL}]
\begin{equation}\label{eq:cond_T_CL_MM2_2}
	\frac{t}{\tau}\geq\frac{t}{\tilde{\tau}_{2}}\gg \mathcal{C}_{2,\text{cl}}.
\end{equation} 
Here, $\tilde{\tau}_{2}$ is the final relaxation time with respect to $\lVert e^{t\Wt_{2}} -\mathbf{\tilde{P}}_\text{\!2,ss} \rVert_1 $; we have $\tilde{\tau}_{2}\approx\tau$ when $\tilde{\mathcal{C}}_\text{2,cl}\ll 1$.

\section{Classical weak symmetries} \label{app:symmetry}

In this section, we discuss the role of weak symmetries in classical metastability. We prove the resulting symmetry properties of classical MMs and the classical long-time dynamics, which are discussed in Sec.~\ref{sec:symmetry} of the main text. We also provide an example of the classicality test in the presence of a discrete weak symmetry. 

\subsection{Symmetries of low-lying eigenmodes} \label{app:symmetry1}
We begin by discussing the symmetry of eigenmodes of the dynamics which follows from a symmetry of the master operator [see~ Eq.~\eqref{eq:weak}]. These results hold for the case of general metastability.

Thanks to the weak symmetry in Eq.~\eqref{eq:weak}, the eigenmatrices of $\L$ can be chosen as eigenmatrices of $U$. Let $\V_k$ be the eigenspace corresponding to an eigenvalue $e^{i\varphi_k}$  of $U$, and let $|\psi_{j_k}^{(k)}\rangle$, $j_k=1,...,\text{dim}(\V_k)$ be the orthonormal basis of the eigenspace. The eigenmatrices of $\U$ with the eigenvalue $1$ correspond to matrices block diagonal in $\oplus_k \V_k$. In contrast, coherences between $\V_k$ and $\V_l$, i.e.,  $|\psi_{j_k}^{(k)}\rangle\!\langle\psi^{(l)}_{j_l}|$ with $j_k=1,...,\text{dim}(\V_k)$ and $j_l=1,...,\text{dim}(\V_l)$, correspond to eigenmatrices of $\U$ with an eigenvalue $e^{i  \phi}=e^{i(\varphi_k-\varphi_l)}$. Therefore, eigenmatrices of $\U$ with an eigenvalue $e^{i \phi}$ are in general composed of coherences between all pairs of eigenspaces that differ  in the arguments of their eigenvalues $\text{mod}\, 2\pi$ by $\phi$. Note that for complex eigenvalues $e^{i\phi}\neq(e^{i \phi})^*$, the eigenmatrices are non-Hermitian.

\subsection{Symmetries of classical metastable manifolds} \label{app:symmetry2}

Here, we discuss symmetry properties of metastable phases in classical MMs in the presence of a weak symmetry. We prove that under a weak symmetry the set of metastable phase undergoes an approximate permutation. In particular, we show that  under a continuous weak symmetry individual metastable phases are necessarily invariant. We also discuss symmetrization of the set of metastable phases and estimate the resulting change in the corrections to the classicality  in Eq.~\eqref{eq:Ccl}.

\subsubsection{Discrete symmetries of classical metastable manifolds} \label{app:symmetry2_inv}

Here, we prove the approximate invariance of the set of metastable phases under any weak symmetry. It follows that the action of any weak symmetry corresponds to an approximate permutation of metastable phases. \\

First, we argue that the set of metastable phases projections $\trho_1$, ..., $\trho_m$ in Eq.~\eqref{eq:rhotilde} is transformed under a weak symmetry approximately onto itself, by the results of Sec.~\ref{app:nonunique}.

Considering  $\trho_1$, ..., $\trho_m$ or $\U(\trho_1)$, ..., $\U(\trho_m)$ as $m$ metastable phases leads to the same corrections in  Eq.~\eqref{eq:Ccl}. Let $\rho(0)$ be an initial state corresponding to $\lVert\pt\rVert_1=1+\Ccl$ for the choice of $\trho_1$, ..., $\trho_m$. Then the barycentric coordinates of $\U[\rho(0)]$ in the basis of $\U(\trho_1)$, ..., $\U(\trho_m)$ are given by $(\pt)_l$, $l=1,...,m$ and thus the corresponding corrections to the classicality are at least $\Ccl$. Conversely, considering the transformation under $\U^\dagger$ of an initial state corresponding to the corrections to the classicality for $\U(\trho_1)$, ..., $\U(\trho_m)$, we obtain a lower bound for $\Ccl$. Thus, the corrections must be the same for both choices of the basis. 

From Sec.~\ref{app:nonunique}, there exist a bijective function $\pi$ of the set $l=1,...,m$, i.e., a \emph{permutation}, such that [cf.~Eq.~\eqref{eq:MPinv12_2bases}]
\begin{eqnarray}\nonumber
	\left\lVert\U(\trho_l)- \trho_{\pi(l)}\right\rVert
	&\lesssim& 3\Ccl,
	\label{eq:MPinv12}
\end{eqnarray}
as a consequence of  the barycentric coordinates of $\U(\trho_l)=\sum_{k=1}^m\tilde{p}_k^{(l)} \trho_k $, where $\tilde{p}_k^{(l)}=\Tr[\tilde P_k\U(\trho_l)]\equiv(\Ubf)_{kl}$  [Eq.~\eqref{eq:U}] obeying [cf.~Eq.~\eqref{eq:MPinv9_2bases}]
\begin{eqnarray} \label{eq:MPinv9}
	\tilde{p}_{\pi(l)}^{(l)}\gtrsim 1-\Ccl,\qquad
	\sum_{k\neq\pi(l)} \left|\tilde{p}_{k}^{(l)}\right|\lesssim 2\Ccl.
\end{eqnarray}	
Therefore, the action of the symmetry on the set of metastable phases is approximated as
\begin{eqnarray}\nonumber
	\left\lVert\Ubf- \Pibf\right\rVert_1&=&\max_{1\leq l\leq m}\left[ \left|1-\tilde{p}_{\pi(l)}^{(l)}\right|+ \sum_{k\neq {\pi(l)}} \left|\tilde{p}_k^{(l)}\right|\right]\\
	&\lesssim& 3\Ccl,
	\label{eq:MPinv10}
\end{eqnarray} 
where we defined
\begin{equation}\label{eq:MPinv11}
	(\Pibf)_{kl}\equiv \delta_{k \pi(l)}.
\end{equation}
From Sec.~\ref{app:nonunique} it also follows that
\begin{eqnarray}\nonumber
	\left\lVert\Ubf^{-1}- \Pibf^{-1}\right\rVert_1
	&\lesssim& 3\Ccl,
	\label{eq:MPinv10}
\end{eqnarray} 
as $\U^\dagger(\trho_k)=\sum_{l=1}^m\tilde{p}_l'^{(k)}\trho_l$, where $\tilde{p}_l'^{(k)}= (\Ubf^{-1})_{lk}$  (from $\U^\dagger =\U^{-1}$) are the barycentric coordinates of $\trho_k$ in the basis of $\U(\trho_n)$, $n=1,...,m$, while $(\Pibf^{-1})_{kl}=(\Pibf)_{lk}.$\\



Second, we argue that a repeated action of the weak symmetry, corresponds to iterations of the same permutation of the metastable phases. Let $n$ be a nonzero integer. We have 
\begin{eqnarray}\label{eq:MPinv10dd}
	\left\lVert \Ubf^{n}-\Pibf^{n} \right\rVert_1 &=&   \left\lVert \sum_{j=1}^n \Ubf^{j-1}(\Ubf-\Pibf) \Pibf^{n-j} \right\rVert_1  
	\\\nonumber&\leq&  \sum_{j=1}^n \lVert \Ubf^j \rVert_1 \lVert \Ubf-\Pibf\rVert_1 \lVert \Pibf \rVert^{n-j} 
	\\\nonumber&\leq&   n \lVert \Ubf-\Pibf\rVert_1  (1+\Ccl) 
	\\\nonumber&\lesssim&  3n \Ccl,
\end{eqnarray}
where we used $\lVert \Ubf^j \rVert_1\leq 1+\Ccl$ (as $\Ubf^j$ transforms the MM onto itself), $\lVert \Pibf \rVert_1=1$, and Eq.~\eqref{eq:MPinv10}. Since $\U^n$ is also a weak symmetry [cf.~Eq.~\eqref{eq:weak}], from Eq.~\eqref{eq:MPinv10} there exist a permutation with the matrix $\Pibf^{(n)}$ such that 
\begin{equation}\label{eq:MPinv10d}
	\left\lVert \Ubf^{n}-\Pibf^{(n)} \right\rVert_1\lesssim 3\Ccl,
\end{equation} 
and thus  from the triangle inequality 
\begin{eqnarray}
	\left\lVert \Pibf^{n}-\Pibf^{(n)} \right\rVert_1&\leq& \left\lVert \Ubf^{n}-\Pibf^{n} \right\rVert_1+\left\lVert \Ubf^{n}-\Pibf^{(n)} \right\rVert_1\qquad\\\nonumber
	&\lesssim& 3(n+1)\Ccl.
\end{eqnarray}
Therefore, for $n\,\Ccl\ll 1$ we can identify
\begin{equation}\label{eq:MPinv11d}
	\Pibf^{(n)} =\Pibf^n,
\end{equation}
so that [cf.~Eq.~\eqref{eq:MPinv10d}]
\begin{equation}\label{eq:MPinv10d2}
	\left\lVert \Ubf^{n}-\Pibf^{n} \right\rVert_1\lesssim 3\Ccl.
\end{equation}   
Furthermore, for $N$ divisible by $n$ using Eq.~\eqref{eq:MPinv10d2} we can retrace its proof with  $\U$ replaced by $\U^n$ and $n$ replaced by $N/n$, to arrive at 
\begin{equation}\label{eq:MPinv10nd}
	\left\lVert \Ubf^{N}-\Pibf^{N} \right\rVert_1\lesssim 3\Ccl
\end{equation}  
[provided that $(N/n)\Ccl\ll 1$]. We conclude that in Eq.~\eqref{eq:MPinv11d} we only require $n'\Ccl\ll 1$ for all prime factors $n'$ of $n$ (and not necessarily $n\,\Ccl\ll 1$).

\subsubsection{No nontrivial continuous symmetries of classical metastable manifolds}  \label{app:symmetry2_C}
As any weak symmetry corresponds to an approximate permutation of metastable phases, we argue that continuous weak symmetries restricted to low-lying modes are necessarily trivial. \\

Let us now consider a continuous symmetry, i.e., $\U_\phi=e^{i\phi\G}$, where
\begin{equation}\label{eq:G}
	\G(\rho)\equiv [G,\rho],
\end{equation}
with $G$ being a Hermitian operator on the system space. In this case, a weak symmetry [cf.~Eq.~\eqref{eq:weak} and see Refs.~\cite{Buvca2012,Albert2014}] takes place when
\begin{equation}\label{eq:weak_G}
	[\G,\L]=0.
\end{equation}

In the basis of metastable phases [cf.~Eq.~\eqref{eq:U}],
\begin{equation}\label{eq:Gbf}
	(\Gbf)_{kl}\equiv\Tr[\tilde{P}_k \G(\trho_l)].
\end{equation}
Let us consider $\phi$ such that 
\begin{equation}\label{eq:Gbf_cond}
	\phi\left\lVert\Gbf\right\rVert_1=\sqrt{\Ccl}.
\end{equation}
From the  Taylor expansion we then have
\begin{equation}\label{eq:Taylor2}
	\left\lVert\Ubf_{\phi}-\Ibf-i\phi\Gbf\right\rVert_1\lesssim \frac{\phi^2}{2}\left\lVert\Gbf\right\rVert_1^2 =\frac{\Ccl}{2},
\end{equation}
where $(\Ibf)_{kl}=\delta_{kl}$ is the identity matrix. From Eq.~\eqref{eq:MPinv10} we also have
\begin{equation}
	\left\lVert\Ubf_{\phi}-\Pibf\right\rVert_1\lesssim 3\Ccl,
\end{equation}
where $\Pibf$ is a permutation matrix. If $\Pibf\neq\Ibf$, we from Eq.~\eqref{eq:Taylor2} and the triangle inequality we arrive at
\begin{eqnarray}\label{eq:Taylor3}
	2-\sqrt{\Ccl}&=& \left\lVert\Pibf-\Ibf\right\rVert_1  -\phi\left\lVert\Gbf\right\rVert_1\\\nonumber
	&\leq& \left\lVert\Pibf-\Ibf-i\phi\Gbf\right\rVert_1\\\nonumber
	&\leq& \left\lVert\Ubf_{\phi}-\Ibf-i\phi\Gbf\right\rVert_1+ \left\lVert\Ubf_{\phi}-\Pibf\right\rVert_1\\\nonumber
	&\lesssim& \Ccl\frac{7}{2} \quad\text{(contradiction!)}
\end{eqnarray}
for $\Ccl\ll 1$.  The choice $\Pibf=\Ibf$, however, similarly leads to 
\begin{eqnarray}\label{eq:Taylor4}
	\sqrt{\Ccl}&=& \phi\left\lVert\Gbf\right\rVert_1\\
	&\leq& \left\lVert\Ubf_{\phi}-\Ibf-i\phi\Gbf\right\rVert_1+ \left\lVert\Ubf_{\phi}-\Ibf\right\rVert_1\\\nonumber
	&\lesssim& \Ccl\frac{7}{2} \quad\text{(contradiction!)}.
\end{eqnarray}
This can only be remedied when Eq.~\eqref{eq:Gbf_cond} is not possible, which implies $\lVert\Gbf\rVert_1=0$ and thus the trivial continuous symmetry of the MM
\begin{equation}
	\Gbf=0, \quad\text{and}\quad \Ubf_\phi=\Ibf. 
\end{equation}
so that all metastable states are invariant under the symmetry $\U_\phi\P=\I$.

\subsubsection{Symmetric set of metastable phases} \label{app:symmetry2_set}

In Sec.~\ref{app:symmetry2_inv}, we showed that the set of metastable phases is approximately invariant under the weak symmetry [Eq.~\eqref{eq:weak}]. We now assume a nontrivial symmetry of the MM and consider replacing $\trho_1$, ..., $\trho_n$ in Eq.~\eqref{eq:rhotilde} by the invariant set of metastable phases. We also provide corrections to the classicality  in Eq.~\eqref{eq:Ccl} for the invariant set of metastable phases.\\

\emph{Symmetric set of metastable phases}. Let $D$ be the smallest nonzero integer such that $\U^D\P=\P$, and thus $\mathbf{U}^{D}=\Ibf$. 
We assume $d\,\Ccl\ll 1$ for all the prime factors  $d$ of $D$ (which follows from $m\,\Ccl\ll 1$), in which case  $\lVert\Pibf^D-\Ibf\rVert_1\leq \lVert \Ubf^{D}-\Pibf^{D}\rVert_1\lesssim 3\Ccl$ [cf.~Eq.~\eqref{eq:MPinv10d2}], and thus $\Pibf^D=\Ibf$.

Let $\trho_l$ belong to an approximate cycle of length $d_l\leq m$, i.e., 
\begin{equation}\label{eq:Pid}
	(\Pibf^{d_l})_{kl}=\delta_{kl}, 
\end{equation} 
$k=1,...,m$, where $\Pibf$ is the permutation matrix that approximates $\Ubf$ in Eqs.~\eqref{eq:MPinv10} and~\eqref{eq:MPinv11}.  In this case, $D$ is divisible by $d_l$ (as $\Pibf^D=\Ibf$) and we can define $\trho_l'$ as in Eq.~\eqref{eq:rhotilde_U}.
As  $\U^D\P=\P$,  we have that $\trho_l'$  is invariant under $\U^{d_l}$,  $\U^{d_l}(\trho_l')=\trho_l'$. We further have that
\begin{equation}\label{eq:rhotilde_U_distance0}
	\left\lVert\trho_l'-\trho_l \right\rVert\leq \frac{d_l}{D} \sum_{n=0}^{\frac{D}{d_l}-1}  \left\lVert\U^{n d_l}(\trho_l)-\trho_l\right\rVert,
\end{equation} 
which we now estimate from above. From Eq.~\eqref{eq:Pid} and Eq.~\eqref{eq:MPinv10nd} (for $N=nd_l$) we have
\begin{eqnarray}
	\sum_{k=1}^m|(\Ubf^{nd_l})_{kl}-(\Ibf)_{kl}|&=& \sum_{k=1}^m  \left|(\Ubf^{nd_l}-\Pibf^{nd_l})_{kl} \right| \\\nonumber&\leq& \left\lVert \Ubf^{nd_l}-\Pibf^{nd_l} \right\rVert_1 \lesssim 3\Ccl,
\end{eqnarray}
and thus [cf.~Eq.~\eqref{eq:MPinv12}]
\begin{eqnarray}\label{eq:MPinv12d}
	\left\lVert\U^{nd_l}(\trho_l)-\trho_l\right\rVert &\leq&  \sum_{k=1}^m|(\Ubf^{nd_l})_{kl}-(\Ibf)_{kl}| \lVert \trho_k\rVert\\\nonumber &\leq& (1+\Ccl)\left\lVert \Ubf^{nd_l}-\Pibf^{nd_l} \right\rVert_1
\end{eqnarray}
leads to [cf.~Eq.~\eqref{eq:rhotilde_U_distance0}]
\begin{equation}\label{eq:rhotilde_U_distance}
	\left\lVert \trho_l'-\trho_l\right\rVert\lesssim \Ccl \,\frac{3(D-d_l)}{D} < 3\Ccl,
\end{equation}
where in the last inequality  we used $d_l\geq 1$.

From Eq.~\eqref{eq:rhotilde_U_distance} we can replace the approximate cycle $\trho_{\pi^n(l)}$,  where $\pi$ is the permutation associated with $\Pibf$ in Eqs.~\eqref{eq:MPinv10} and~\eqref{eq:MPinv11}, by $\U^n(\trho_l')$ $n=1,...,d_l-1$ [cf.~Eq.~\eqref{eq:rhotilde_U2}]. 
Indeed, the distance between the new and old basis
\begin{eqnarray}\nonumber
	\left\lVert\trho'_{\pi^n(l)}-\trho_{\pi^n(l)}\right\rVert &\leq&\left\lVert\trho_l'-\trho_l \right\rVert + \left\lVert\U^n(\trho_l)-\trho_{\pi^n(l)}\right\rVert
	\\
	&\lesssim& 6\,\Ccl,
	\label{eq:rhotilde_U_distance2}
\end{eqnarray}
where in the first line we used $\lVert\U^n(\trho_l) -\trho'_{\pi^n(l)} \rVert=\lVert\U^n(\trho_l) -\U^n(\trho_l') \rVert=\lVert\trho_l -\trho'_{l}\rVert$ as $\U$ is a unitary transformation, while the second line follows from Eq.~\eqref{eq:MPinv10d2} [cf.~Eq.~\eqref{eq:rhotilde_U_distance}]. We can then repeat the construction in Eqs.~\eqref{eq:rhotilde_U} and~\eqref{eq:rhotilde_U2} for each of the approximate cycle between metastable phases under $\U$.

The metastable phase in Eq.~\eqref{eq:rhotilde_U} corresponds to  projection only on the low-lying eigenmodes invariant under $\U^d$, where $d$ is the length of the approximate cycle $\trho_l$ belongs to,
\begin{equation}\label{eq:rhotilde_U4}
	\trho_l'= \sum_{\substack {1\leq k\leq m:\\ (e^{i \phi_k})^{d_l}=1}} c_k^{(l)} R_k 
\end{equation}
[cf.~Eq.~\eqref{eq:rhotilde}]. We note, however, that by construction in Eq.~\eqref{eq:rhotilde_U}, the closest state to $\trho_l'$ in Eq.~\eqref{eq:rhotilde_U} is not further away than $\Cp$ [cf.~Eq.~\eqref{eq:Cp}], since it is bounded by the distance to the state
\begin{equation}\label{eq:rho_U}
	\rho_l'\equiv\frac{d_l}{D} \sum_{n=0}^{\frac{D}{d_l}-1} \U^{n d_l}(\rho_l),
\end{equation}
where $\rho_l$ is the closest state to $\trho_l$, and  
\begin{eqnarray}
	\left\lVert \trho_l' -\rho_l' \right\rVert&\leq&  \frac{d_l}{D} \sum_{n=0}^{\frac{D}{d_l}-1} \left\lVert \U^{n d_l}(\trho_l) - \U^{n d_l}(\rho_l)\right\rVert\\\nonumber
	&=&\left\lVert \trho_l - \rho_l\right\rVert\leq \Cp.
\end{eqnarray}
Note that the state $\rho_l'$ in Eq.~\eqref{eq:rho_U}, in analogy to $\trho_l'$ in Eq.~\eqref{eq:rhotilde_U}, is invariant under $\U^{d_l}$. Furthermore, the elements of the cycle $\U^n(\rho_l')$, $n=0,...,d-1$ are approximated by $\U^{d_l}(\rho_l)$ with the same distance  $\lVert \trho_l - \rho_l\rVert$. 

Similarly, if $\rho_l$ is a state that projected on the low-lying modes gives $\trho_l$, that is, $\P(\rho_l)=\trho_l$ [cf.~Eq.~\eqref{eq:rhotilde}], from Eq.~\eqref{eq:weak_P} we have that $\P(\rho_l')=\rho_l'$, for $\rho_l'$ defined in Eq.~\eqref{eq:rho_U}. Furthermore, $\rho_l'$ is invariant under $\U^{d_l}$ and $\P[\U^n(\rho_l')]=\U^n(\rho_l')$ holds for all elements of the cycle $n=0,1,...,d_l-1$. \\



\emph{Classicality corrections for symmetric set of metastable phases}.  We now consider how the choice of a symmetric set of metastable phases in Eqs.~\eqref{eq:rhotilde_U} and~\eqref{eq:rhotilde_U2} affects the corrections to the classicality in Eq.~\eqref{eq:Ccl}. We are interested in bounding 
\begin{equation} \label{eq:Ccl_U}
	\mathcal{C}'_\text{cl}\equiv \max_{\rho(0)}\left\lVert\pt'\right\rVert_{1}-1,
\end{equation}
where $(\pt')_l=\tilde{p}_l'\equiv \Tr[\tilde{P}'_l\rho(0)]$ with an operator $\tilde{P}'_l$ being an element of the invariant dual basis, i.e., $\Tr(\tilde{P}'_k\trho_l')=\delta_{kl}$, $k,l=1,...,m$.  The transformation from the basis in Eq.~\eqref{eq:rhotilde} to the invariant basis defined in Eqs.~\eqref{eq:rhotilde_U} and~\eqref{eq:rhotilde_U2} is given by [cf.~Eq.~\eqref{eq:C}]
\begin{equation}\label{eq:C'}
	(\C'')_{kl}\equiv \Tr(\tilde{P}_k \trho_l')=(\C'^{-1} \C)_{kl},
\end{equation}
and the dual basis [cf.~Eq.~\eqref{eq:Ptilde}]
\begin{equation}\label{eq:Ptilde_U}
	\tilde{P}_l'\equiv \sum_{k=1}^m\left(\C''^{-1}\right)_{lk}\tilde{P}_k. 
\end{equation}
Note that $(\C'')_{kl}=(d_l/D)\sum_{n=1}^{D/d_l} (\Ubf^{nd_l})_{kl}$ for $l$ as in Eq.~\eqref{eq:rhotilde_U}, while $(\C'')_{k\pi^n(l)}=(\Ubf^{n} \C'')_{kl}$, $n=1,...,d_l-1$ from  Eq.~\eqref{eq:rhotilde_U2}. Therefore, from Eq.~\eqref{eq:Pid} and Eq.~\eqref{eq:MPinv10nd} we obtain $\sum_{k=1}^m|(\C'')_{kl}-\delta_{kl}|\lesssim 3\Ccl$ [for $N=nd_l$], while from Eq.~\eqref{eq:MPinv10d2} we have  $\sum_{k=1}^m|(\C'')_{k\pi^n(l)}-\delta_{k\pi^n(l)}|\leq \sum_{k=1}^m|[\Ubf^n (\C''-\Ibf)]_{kl}|+\sum_{k=1}^m|(\Ubf^n)_{k l}-(\Pibf^n)_{k l}|\lesssim 3\Ccl (\lVert \Ubf^n\rVert_1+1 )\leq 6\Ccl$, where we used $\lVert \Ubf^n\rVert_1\leq \lVert \Ubf^n-\Pibf^n\rVert_1+\lVert \Pibf\rVert_1^n\lesssim 1+3\Ccl$ from Eq.~\eqref{eq:MPinv10d2} and $\lVert \Pibf\rVert_1=1$. We conclude
\begin{equation}
	\lVert \C''- \Ibf\rVert_1 \lesssim  6 \,\Ccl, 
\end{equation}
and thus we can approximate
\begin{equation}\label{eq:C'inv}
	\C''^{-1}= \Ibf-(\C''-\Ibf) +..., 
\end{equation}
with corrections of the order $\Ccl^2$ in the L1 norm. From Eq.~\eqref{eq:C'inv}, 
\begin{eqnarray}\label{eq:Ccl_U_proof}
	\left\lVert\pt'\right\rVert_{1}=\left\lVert\C''^{-1}\pt\right\rVert_{1} &\leq& \left\lVert\C''^{-1}\right\rVert_{1}\left\lVert\pt\right\rVert_{1} \\\nonumber
	&\lesssim& (1+\left\lVert\C''-\Ibf\right\rVert_{1})\left\lVert\pt\right\rVert_{1}
	\\\nonumber
	&\lesssim& \left(1+6\,\Ccl  \right)(1+\Ccl)
\end{eqnarray}
and thus [cf.~Eq.~\eqref{eq:Ccl_U}]
\begin{equation}\label{eq:Ccl_U2}
	\mathcal{C}'_\text{cl}\lesssim 7\,\Ccl,
\end{equation}
as well as [cf.~Eq.~\eqref{eq:Ccl_av}]
\begin{equation} \label{eq:Ctcl_U2}
	\overline{\Ccl}\equiv \overline{\left\lVert\pt'\right\rVert_{1}}-1\lesssim \overline{\Ccl}+6\,\Ccl.
\end{equation}

\subsection{Symmetries  of classical long-time dynamics}\label{app:symmetry_W}

Here, we show that the approximation of long-time dynamics by classical stochastic dynamics in Eq.~\eqref{eq:W} features a permutation symmetry for the set of metastable phases chosen invariant under a weak symmetry. This corresponds to Eq.~\eqref{eq:weak_W2} in the main text. Furthermore,  the long-time dynamics restricted to the symmetric mixtures of metastable phases is also approximated by classical stochastic dynamics.

\subsubsection{Derivation of Eq.~\eqref{eq:weak_W2} in the main text}\label{app:symmetry_W1}

For the set of metastable phases is chosen invariant under a weak symmetry [cf.~Eq.~\eqref{eq:weak}], the corresponding stochastic classical dynamics $\W'$ in Eq.~\eqref{eq:W} that approximates $\Wt'$ [cf.~Eq.~\eqref{eq:deltaW}] fulfills
[cf.~Eq.~\eqref{eq:Pi} and~\eqref{eq:weak_Wt}]
\begin{equation}\label{eq:weak_Wt2}
	[\Wt',\Pibf]=0,
\end{equation}
which is equivalent to [cf.~Eq.~\eqref{eq:weak_W1}]
\begin{equation}\label{eq:weak_Wt3}
	(\Wt')_{kl}=(\Wt')_{\pi(k)\pi(l)},
\end{equation}
where $\pi$ is the permutation corresponding to $\Pibf$ and $k,l=1,...,m$. We now prove that this condition holds for the closest classical stochastic generator $\W'$ as well, as given by  Eq.~\eqref{eq:weak_W2}. 

First, we have
\begin{eqnarray}
	({\W'})_{kl}&\equiv&\max [(\Wt')_{kl},0] \\\nonumber&=&\max [(\Wt')_{\pi(k)\pi(l)},0]\equiv ({\W'})_{\pi(k)\pi(l)}.
\end{eqnarray}
Second,
\begin{eqnarray}
	({\W'})_{ll}&\equiv&(\Wt')_{ll} + \sum_{\substack{1\leq k\leq m:\\k\neq l}}\min [(\Wt')_{kl},0]\\\nonumber
	&=&(\Wt')_{\pi(l)\pi(l)} + \sum_{\substack{1\leq k\leq m:\\k'\neq l}}\min [(\Wt')_{\pi(k)\pi(l)},0]\\\nonumber
	&=&(\Wt')_{\pi(l)\pi(l)} + \sum_{\substack{1\leq k'\leq m:\\k'\neq \pi(l)}}\min [(\Wt')_{k'\pi(l)},0]\\\nonumber
	& \equiv&({\W'})_{\pi(l)\pi(l)}, 
\end{eqnarray}
where we introduced $k'\equiv \pi(k)$. This ends the proof [cf.~Eq.~\eqref{eq:weak_W1}].

\subsubsection{Classical dynamics of symmetric degrees of freedom}\label{app:symmetry_W2}

Here, we consider the long-time dynamics $\Wt'$ in the invariant set of metastable phases [Eq.~\eqref{eq:Wtilde}] restricted to the symmetric subspace of the symmetry $\Pibf$, that is uniform mixtures of metastable phases in each cycle of the permutation $\pi$ corresponding to $\Pibf$ (cf.~Sec.~\ref{sec:symmetry} in the main text). We show that that dynamics is trace-preserving and approximately positive with corrections bounded by $2\sqrt{\Ccl'}$, where $\Ccl'$ denote the corrections to the classicality in Eq.~\eqref{eq:Ccl} for the invariant set of metastable phases [cf.~Eq.~\eqref{eq:deltaW}]. \\

\emph{Classical symmetric degrees of freedom}. Symmetric mixtures of metastable phases, or the symmetric part of states within the MM,  can be described by (cf.~Sec.~\ref{app:classical_symmetry})
\begin{equation}\label{eq:p0tilde}
	(\pt'_0)_l=\sum_{n_l=0}^{d_l-1}(\pt')_{\pi^n(l)}=\Tr\Big[\sum_{n_l=0}^{d_l-1}\!\tilde{P}_{\pi^n(l)}\,\rho(0)\Big],
\end{equation}
where the index $l$ runs over representatives of cycles with $d_l$ denoting the corresponding cycle length, and $\pt$ are barycentric coordinates of the system state [cf.~Eqs.~\eqref{eq:p_k} and~\eqref{eq:Lk_U}], so that the number of parameters equals the number of cycles in the permutation $\pi$, minus $1$ for $\sum_l(\pt_0)_l=1$. Since Eq.~\eqref{eq:p0tilde} corresponds to an invertible linear transformation of the coefficients $c_k=\Tr[L_k\rho(0)]$ for the symmetric low-lying eigenmodes $\U(\L_k)=L_k$, $k=2,...,m$ [cf.~Eq.~\eqref{eq:C_U}],  it is an equivalent representation of the symmetric degrees of freedom.  

Operationally, $\pt_0$ describes the barycentric coordinates of symmetric states in the MM decomposed between uniform mixtures of metastable phases in each cycle, and, from the triangle inequality, we have 
\begin{equation}\label{eq:p0tilde}
	\lVert \pt_0'\rVert_1\equiv\sum_{l}\Big|\sum_{n_l=0}^{d_l-1}(\pt')_{\pi^n(l)}\Big|\leq \lVert \pt'\rVert_1\leq 1+\Ccl'
\end{equation}
[cf.~Eq.~\eqref{eq:Ccl_U}]. Therefore, the corrections to the classicality for symmetric states are not larger than  $\Ccl'$ (in particular, in the case of only two cycles, those corrections will be $0$, as in the bimodal case with $m=2$; cf.~Refs.~\cite{Macieszczak2016a,Rose2016}). Among others,  it then follows that the uniform mixtures of metastable phases in each cycle and their basins of attraction are approximately disjoint (cf.~Sec.~\ref{sec:disjoint} in the main text).\\

\emph{Trace-preserving and approximately positive dynamics}. The dynamics of the symmetric degrees of freedom,
\begin{equation}\label{eq:pt_0}
	\frac{d}{dt}\pt_0(t)= \mathbf{\tilde{W}}'_0 \, \pt_0(t),
\end{equation}
is governed by the generator 
\begin{eqnarray}\label{eq:W_0}
	(\mathbf{\tilde{W}}'_0 )_{kl}&=& \frac{1}{d_l} \sum_{n_k=0}^{d_k-1}\sum_{n_l=0}^{d_l-1} (\Wt')_{\pi^{{n\!}_k}\!(k) \,\pi^{{n\!}_l}\!(l)},
\end{eqnarray}
which is trace preserving as a consequence of the trace preservation of $\Wt$. 

We now prove that the generator in Eq.~\eqref{eq:W_0} is approximately positive up to corrections bounded by $2\sqrt{\Ccl'}$.
As a consequence of the symmetry of the classical stochastic generator  $\W'$ closest to $\Wt'$ [cf.~Eq.~\eqref{eq:weak_W2}], when restricted to the symmetric subspace of $\Pibf$, which we denote $\W'_0$, it remains trace-preserving and completely-positive (cf.~Sec.~\ref{app:classical_symmetry}). From the triangle inequality, we then have
\begin{eqnarray}
	&&\lVert\mathbf{\tilde{W}}'_0 -{\W'\!}_0\rVert_1 \\\nonumber
	&&\equiv \max_{l}\frac{1}{d_l}\sum_{k} \Big| \sum_{n_k=0}^{d_k-1}\sum_{n_l=0}^{d_l-1} (\Wt')_{\pi^{{n\!}_k}\!(k) \,\pi^{{n\!}_l}\!(l)}-(\Wt)_{\pi^{{n\!}_k}\!(k) \,\pi^{{n\!}_l}\!(l)}\Big|
	\\\nonumber
	&&\leq \max_{l}\frac{1}{d_l} \sum_{n_l=0}^{d_l-1} \sum_{k=1}^{m} \Big| (\Wt')_{k \pi^{{n\!}_l}\!(l)}-(\Wt)_{k\pi^{{n\!}_l}\!(l)}\Big|\\\nonumber
	&&\leq \max_{1\leq l\leq m}\sum_{k=1}^{m} \Big| (\Wt')_{k l}-(\Wt)_{kl}\Big|\equiv\lVert{\Wt'}-{\W'}\rVert_1 \lesssim 2\sqrt{\Ccl'}.
\end{eqnarray}
[cf.~Eq.~\eqref{eq:deltaW}]. We note, however, that $\W'_0$ is generally not the closest classical stochastic generator to $\mathbf{\tilde{W}}'_0$ [cf.~Eq.~\eqref{eq:W}]. The distance to the closest classical stochastic generator is instead bounded by $2\sqrt{\max_{\rho(0)}\lVert \pt_0'\rVert_1-1}\leq 2\sqrt{\Ccl'}$ [cf.~Eq.~\eqref{eq:p0tilde}], where $\max_{\rho(0)}\lVert \pt_0'\rVert_1-1$ bounds the distance of  $\pt_0$ to the simplex formed by the uniform mixtures of metastable phases for each cycle  (in particular, for only two cycles, $\mathbf{\tilde{W}}'_0$ is a classical stochastic generator; cf.~Refs.~\cite{Macieszczak2016a,Rose2016}).

\subsection{Example of classicality test with weak symmetry} \label{app:symmetry_example}
Here, we provide a simple example, why, even in the case of the symmetry fully determining the eigenmodes of the dynamics [cf.~Eq.~\eqref{eq:RkLk_U}], the classicality test is needed. \\

We consider a finite system with $m$ disjoint stationary states $\rho_1,...\rho_m$ and the corresponding projections $P_1$,...,$P_m$ [$\Tr(P_{k}\rho_l)=\delta_{kl}$, $\sum_{l=1}^m P_l=\mathds{1}$]. We assume that the stationary states are connected by a weak symmetry as $\U(\rho_l)=\rho_{l+1}$, $l=1,...,m$ (with periodic boundary conditions on the label, $m+1\equiv1$). We have that eigenmatrices of the symmetry fulfill $e^{i\varphi_k} R_{k}= (1/m)\sum_{l=1}^m (e^{-i 2\pi \frac{j_k}{m}})^l \rho_l$ and $e^{-i\varphi_k} L_{k}= \sum_{l=1}^m (e^{i 2\pi \frac{j_k}{m}})^l P_l$ [cf.~Eq.~\eqref{eq:RkLk_U} and the normalization $\lVert L_k\rVert_{\max}=1$], where $e^{i\varphi_k}$ is an arbitrary global phase, $k=1,...,m$, and $e^{i 2\pi \frac{j_k}{m}}$ is the corresponding symmetry $\U$ eigenvalue,   $j_k\in\{0,...,m-1\}$.
As the invariant set of candidate phases, let us choose 
\begin{equation}\label{eq:rho_example}
	\trho_l= p\,\rho_l+\frac{1-p}{m}\sum_{k=1}^m  \rho_k,
\end{equation}
$l=1,...,m$, where $0\leq p\leq 1$. We then have [cf.~Eq.~\eqref{eq:C_U}]
\begin{equation}
	\tilde{P}_l= \frac{1}{p} P_l - \frac{1}{m}\frac{1-p}{p} \mathds{1}
\end{equation}
so that [cf.~Eq.~\eqref{eq:Ccl}]
\begin{equation}
	\Ccl= 2\frac{m-1}{m} \frac{1-p}{p}
\end{equation} 
(achieved for any of $\rho_l$, $l=1,...,m$), which diverges to $\infty$ as $p\rightarrow 0$ and the basis in Eq.~\eqref{eq:rho_example} stops being linearly independent. We note, however, that when expressed in the basis of $\trho_l$, $R_k$ is still proportional (with the factor $\sqrt{m}/p$ for $k\geq 2$) to Eq.~\eqref{eq:Rk_U}, as so is $L_k$ to Eq.~\eqref{eq:Lk_U} when expressed in the basis of $\tilde{P}_l$ (with the factor $p/\sqrt{m}$ for $k\geq 2$). 

In this example, the proportionality factor between $R_k$ and  Eq.~\eqref{eq:Rk_U} [or $L_k$ and  Eq.~\eqref{eq:Lk_U}] indicates that the choice of the basis in Eq.~\eqref{eq:rho_example} is not optimal. In the presence of classical metastability, however, rather than for a classical phase transition as considered in this example, the proportionality factor for correctly identified phases does not equal $1$, as $\trho_l$ in Eq.~\eqref{eq:rhotilde} are only approximately disjoint (see Sec.~\ref{sec:disjoint} in the main text), as well as $\tilde{P}_l$ are not bounded by $1$ or orthogonal (see Sec.~\ref{app:Ptilde}). We would expect, however, that the proportionality factor can be related to the corrections in Eq.~\eqref{eq:Ccl}, and should be close to $1$ for classical metastable phases.

\section{Quantitative analysis of algorithm in Sec.~\ref{sec:algorithm} of the main text}\label{app:algorithm}

Here, we give quantitative analysis of the effectiveness of the numerical approach introduced in Sec.~\ref{sec:algorithm} of the main text, in terms of corrections to the classicality $\Ccl$ in Eq.~\eqref{eq:Ccl} and the assumption of a nonnegligible volume of the MM in the space of coefficients. 

\subsection{Extreme eigenstates of dynamics eigenmodes for metastable phases}\label{app:algorithm_Lk}

In the coefficient space, the MM is well approximated by the simplex $S$ with vertices given by metastable phases coefficients (see Fig.~\ref{fig:MM} in  the main text)
\begin{equation}\label{eq:ck_simplex}
	\min_{\p}\Big|c_k-\sum_{l=1}^m p_l c_k^{(l)}\Big| \leq\lVert L_k\rVert_{\max} \min_{\p}\lVert\pt-\p\rVert_1 \leq \Ccl,
\end{equation} 
where $c_k=\Tr[L_k\rho(0)]$  [cf.~Eq.~\eqref{Expansion}] and $c_k^{(l)}=\Tr(L_k\trho_l)$, while $(\pt)_l=\Tr[\tilde{P}_l\rho(0)]$ [cf.~Eq.~\eqref{eq:Ptilde}], $k,l=1,..,m$. Here, we assume Hermitian $L_k$, by replacing non-Hermitian conjugate pairs of eigenmatrices $L_k$, $L_k^\dagger$ by $L_k^R$ and $L_k^I$ in Eq.~\eqref{eq:LRLI}, and choose the normalization $c_k^{\max}-c_k^{\min}=1$, where $c_k^{\max}$ and $c_k^{\min}$ are extreme eigenvalues of $L_k$ (thus, $\lVert L_k\rVert_{\max}\leq 1$). 

From Eq.~\eqref{eq:ck_simplex}, there exists at least one metastable phase $\trho_{l}$ with $l=l_k^{\max}$ chosen so that the $k$th coefficient is closer than $\Ccl$, i.e., $\delta_k^{(l_k^{\max})}\leq \Ccl$, where $\delta_k^{(l)}\equiv c_k^{\max}-c_k^{(l)}$. Furthermore, the metastable state $\P(\rho_k^{\max})$ for the initial state $\rho_k^{\max}$ chosen as the maximal $L_k$ eigenstate, approximates the closest metastable  $\trho_{l_k^{\max}}$ as
\begin{equation}\label{eq:rhotilde_simplex}
	\big\lVert \trho_{l_k^{\max}}- \P\big(\rho_k^{\max}\big)\big\rVert \lesssim \frac{2\Ccl}{\Delta_k^{(l_k^{\max})}}+\Ccl,
\end{equation}
where $\Delta_k^{\max}\equiv\min_{l\neq l_k^{\max}}\delta_k^{(l)}$ is the distance in $k$th coefficient to the next closest phase (see the derivation below). For $\Delta_k^{\max}\gg \Ccl$, the corrections in Eq.~\eqref{eq:rhotilde_simplex} are negligible. Otherwise $\P(\rho_k^{\max})$ is a mixture of metastable phases with $\delta_k^{(l)}$ of the order of $\Ccl$. The discussion is analogous for the minimal $L_k$ eigenvalue $c_k^{\min}$ with $\delta_k^{(l)}\equiv c_k^{(l)}-c_k^{\min}$.

The corrections in Eq.~\eqref{eq:rhotilde_simplex} also hold for the L1 distance in the barycentric coordinates of the metastable phases. Thus, the corrections to the classicality in the basis chosen from the extreme eigenstate of dynamics modes are bounded by $2\Ccl/(\min_{1\leq l\leq m} \Delta_l)+\Ccl$, where $\Delta_l$ is the minimal distance to the next metastable phase in the coefficients for which $\trho_l$ was extreme, $\Delta_l^{\max}=\min(\min_{k:\,l_k^{\max}=l} \Delta_k^{\max},\min_{k:\,l_k^{\min}=l}\Delta_k^{\min})$ (also including rotated left eigenbases) (cf.~the corrections to the classicality for the symmetric set of metastable phases in Sec.~\ref{app:symmetry2_set}). \\

\emph{Derivation of Eq.~\eqref{eq:rhotilde_simplex}}. The metastable phase $\trho_{l_k^{\max}}$ exists, as $\delta_k^{(l)}\geq 0$  and, thus, from Eq.~\eqref{eq:ck_simplex} we have $\sum_{l=1}^m p_l \delta_k^{(l)}\leq \Ccl$.
Furthermore, the metastable state $\P(\rho_k^{\max})$ is approximated as a mixture of metastable phases with $\delta_k^{(l)}$ of the order of $\Ccl$, since $\sum_{l: \delta_k^{(l)}<\Delta} p_l= 1-\sum_{l: \delta_k^{(l)}\geq\Delta}p_l\geq 1-\sum_{l} p_l\delta_k^{(l)}/\Delta\geq 1-\Ccl/\Delta$ is close to $1$ for $\Delta\gg \Ccl$. Therefore, when there exists only a single metastable phase $\trho_{ l_k^{\max}}$ with $c_k^{( l_k^{\max})}$ in the proximity to $c_k^{\max}$ , we have  $\tilde{p}_{ l_k^{\max}}\geq 1-\Ccl/\Delta$ and $\P(\rho_k^{\max})$ can be replaced by that metastable state as given in Eq.~\eqref{eq:rhotilde_simplex}, as $\lVert \trho_{l}- \P[\rho(0)]\rVert\leq (|1-\tilde{p_l}|+\sum_{l'\neq l}|\tilde{p}_{l'}|)(1+\Cp)$ and $\sum_{l'\neq l}|\tilde{p}_{l'}|\leq 1+\Ccl-|\tilde{p}_l|\leq |1-\tilde{p}_l|+\Ccl$ [cf.~Eq.~\eqref{eq:Ccl}], where $\tilde{p}_l=\Tr[\tilde{P}_l\rho(0)]$.

\subsection{Rotations of eigenmodes to expose metastable phases}\label{app:algorithm_rot}

We now explain how rotations of the basis of eigenmodes allow to solve problem of degeneracy in the coefficients, as well as they ensure that a given metastable phases corresponds to extreme value of one of the coefficients when the volume of the simplex of metastable phases in the coefficient space is nonnegligible.

The required lack of degeneracy (up to order $\Ccl$), between metastable phases with the maximal (minimal) $k$th coefficient in Eq.~\eqref{eq:ck_simplex}, corresponds to $k$th axis in the space of coefficients being normal to the \emph{supporting hyperplane} at $l$th vertex of the simplex in the coefficient space where  $l=l_k^{\max}$ ($l=l_k^{\min}$).  In convex geometry it is known that for any vertex in a simplex, there exists a supporting hyperplane, that is, there exist a \emph{rotation} such that the rotated $k$th axis is perpendicular to a supporting hyperplane. In our case, the lack of degeneracy up to order $\Ccl$, additionally requires  the distance from that hyperplane of other vertices $\delta_k^{(l)}\gg\Ccl$  for $l\neq l_k^{\max}$ ($l\neq l_k^{\min}$). As we argue below, this condition can be translated into a nonnegligible volume of the metastable phases simplex $S$  in the space of coefficients, which is guaranteed by
\begin{equation}\label{eq:cond_simplex}
	\text{Vol}(S)=|\!\det(\C)|/m!\gg \Ccl \frac{s_{m-1}}{m-1}, 
\end{equation}
where $s_{m-1}$ is the maximal volume of a simplex with $m-1$ vertices inside the $(m-1)$-dimensional unit hypercube.

The condition in Eq.~\eqref{eq:cond_simplex} also guarantees that  the separation between any two metastable phases in the space of coefficients equipped with $L2$ norm  is  $\gg\Ccl$, and thus the metastable phases are distinguishable in the space of coefficients  [cf.~Eq.~\eqref{eq:ck_simplex}]. \\

\emph{Derivation} of Eq.~\eqref{eq:cond_simplex}. Let $\bm{\delta \bar{c}}_{l}$ be the component of the shifted coefficient vector $(\bar{\mathbf{c}}_{l})_k\equiv c_k^{(l)}-c_k^{(1)}$ ($l\neq 1$) orthogonal to the $(m\!-\!2)$ dimensional subspace spanned by other vertices, $\bar{\mathbf{c}}_{l'}$  with $l'\neq 1,l$. The distance of other vertices to the supporting hyperplane at $\bar{\mathbf{c}}_l$  normal to $\bm{\delta \bar{c}}_{l}$ is equal to its $L2$ norm $\lVert \bm{\delta \bar{c}}_{l}\rVert_2$. Furthermore, the simplex volume $\text{Vol}(S)=\lVert \bm{\delta \bar{c}}_{l}\rVert_2 \text{Vol}(S_{l})/(m-1)$, where $S_{l}$ is the simplex of all vertices but $l$th one. Indeed, we have that $\text{Vol}(S)=\det\sqrt{\C^T\C}/(m-1)!$, where $(\bar{\C})_{k-1,l-1}=(\mathbf{\bar{c}}_{l})_k$ with  $k,l=2,...,m$~\footnote{We have $\det{\C}=\det {\bar{\C}}$, where $(\bar{\C})_{l-1,k-1}=c_{k}^{(l)}-c_{k}^{(1)}$, $k,l=2,..,m$, encodes the coefficients for the simplex with the vertex of $\trho_1$ shifted to the origin.}, and $\text{Vol}(S_{l})=\det\sqrt{ (\bar{\C}_{l})^T\bar{\C}_{l}}/(m-2)!$, where $\bar{\C}_{l}$ is obtained from $\bar{\C}$ by removing $l$th column (we assumed $l\neq 1$). Therefore, up to rotations, in order for metastable phases to correspond to extreme eigenstates of dynamics eigenmodes, we require a nonnegligible volume of the MM in the coefficient space. Finally, we note that due to the chosen normalization of $L_k$, the MM is enclosed by a $(m-1)$-dimensional unit [cf.~Eq.~\eqref{eq:cond_simplex}]. 

We also note that the distance between two vertices, $k$th and $l$th, is bounded from below by $\bm{\delta \bar{c}}_{l}$ in the coordinate system shifted to the $k$th vertex (earlier $k=1$). Therefore, it the distance between any two vertices with respect to $L2$ norm in the space of coefficients is $\gg \Ccl$ when Eq.~\eqref{eq:cond_simplex} is fulfilled.

\subsection{Maximal simplex in coefficient space as simplex of metastable phases}\label{app:algorithm_vol}

We show below that the volume of a simplex $S'$ with $m$ vertices inside the MM is bounded by  [cf.~Eq.~\eqref{eq:cond_simplex}]
\begin{equation}\label{eq:volume_simplex}
	\text{Vol}(S')\lesssim\text{Vol}(S)+ \Ccl\,\sqrt{m} s_{m-1},
\end{equation}
where $S$ is the simplex of $m$ metastable phases and $\Ccl$ denotes the corresponding corrections to metastability in Eq.~\eqref{eq:Ccl}. Therefore, when $\text{Vol}(S)\gg \Ccl\sqrt{m-1}s_{m-1}$, the simplex of metastable phases is approximately the maximal simplex within the MM. We note that this is a stronger condition than in Eq.~\eqref{eq:cond_simplex}.\\

\emph{Derivation} of  Eq.~\eqref{eq:volume_simplex}. First, we note that the distance in the space of coefficients of any point within the MM to the simplex $S$ of metastable phases is bounded by $\sqrt{m-1} \Ccl$ in $L2$ norm. Second, consider  $l$th vertex in a simplex $S'$ with $m$ vertices. From the derivation of  Eq.~\eqref{eq:cond_simplex}, we have that $\text{Vol}(S')=\lVert \bm{\delta \bar{c}'}_{l}\rVert_2 \text{Vol}(S'_l)/(m-1)$, where $S_{l}'$ is the simplex obtained from $S'$ after removing $l$the vertex and$ \lVert \bm{\delta \bar{c}'}_l\rVert_2$ is the length of the component of $l$th vertex orthogonal to that simplex. Since by replacing $l$the vertex by the closest point in the simplex $S$ of metastable phases with respect to $L2$ norm, the orthogonal component can decrease at most by $\sqrt{m-1} \Ccl$, we obtain that the volume can decrease at most by  $\Ccl s_{m-1}/\sqrt{m-1} $. Repeating the procedure with respect to remaining $m-1$ vertices, we arrive at a simplex of $m$ vertices inside $S$ at the cost of the volume decrease at most by  $\lesssim \Ccl s_{m-1}m/\sqrt{m-1}\leq \sqrt{m}\Ccl s_{m-1}$. Noting that volume of any simplex inside $S$ is less than $\text{Vol}(S)$, we arrive at Eq.~\eqref{eq:volume_simplex}.

\subsection{Hierarchy of metastable manifolds}\label{app:algorithm_hierarchy}

In the presence of hierarchy of metastabilities with a further separation at $m_2<m$ in the spectrum of the master operator, any metastable state during the second metastable regime is approximated as a mixture of $m_2$ metastable phases up to the corrections to the classicality $\mathcal{C}_{2,\text{cl}}$ in the second MM (see Sec.~\ref{app:hierarchy}). In the space of the coefficients $(c_2,...,c_{m_2})$, we have an analogous bound to Eq.~\eqref{eq:ck_simplex},
\begin{eqnarray}
	\min_{\p_2}\Big|c_{k}-\sum_{l_2=1}^{m_2} p_{2,l_2} c_{k}^{(2,l_2)}\Big| &\leq&\lVert L_k\rVert_{\max} \min_{\p_2}\lVert\pt_2-\p_2\rVert_1 \nonumber\\&\leq& \mathcal{C}_{2,\text{cl}},
	\label{eq:ck_simplex_MM2}
\end{eqnarray} 
where $k\leq m_2$,  $c_k^{(2,l_2)}$ denotes the coefficients for $m_2$ metastable phases of the second MM, $\pt_2$ is the vector of barycentric coordinates in their basis, and $(\p_2)_{l_2}=p_{2,l_2} $ is a vector of a probability distribution.



\subsection{Weak symmetries}\label{app:algorithm_symmetry}

Here, we discuss degeneracy of coefficients in classical MMs in the presence of a weak symmetry $\U$ (see Sec.~\ref{sec:symmetry} in the main text) and argue how this degeneracy can be addressed by the metastable construction in Sec.~\ref{sec:algorithm} of the main text. Furthermore, we explain how the construction can be further refined, to exploit the structure of the MM arising due to the weak symmetry.

\subsubsection{Symmetric eigenstates of dynamics eigenmodes}\label{app:algorithm_symmetry1}

For a weak symmetry $\U$ in Eq.~\eqref{eq:weak}, let an eigenmode $L_k$ be chosen such that $\U^{\dagger}(L_k)=e^{i\phi_k}L_k$. Let $n_k> 0$ then denote the minimal integer such that $e^{in_k\phi_k}=1$, i.e.,  $\U^{\dagger n_k}(L_k)=L_k$ (we consider a discrete symmetry, without loss of generality, as relevant for the classical MM). We have that the Hermitian and anti-Hermitian part of $e^{i\varphi_k}L_k$ in Eq.~\eqref{eq:LRLI} commute with the  symmetry applied $n_k$ times, $[L_k^R,U^{n_k}]=0=[L_k^I,U^{n_k}]$. Therefore, eigenstates of $L_k^R$ and $L_k^I$ can be chosen as eigenstates of $U^{n_k}$. Such states and their projections on the MM,  are symmetric under  $\U^{n_k}$, so that under $\U$ they form cycles with length that divides $n_k$.


\subsubsection{Degeneracy of coefficients}\label{app:algorithm_symmetry2}

We now discuss the degeneracy of coefficients in classical MMs for a weak symmetry $\U$. Let an eigenmode $L_k$ be chosen as a symmetry eigenmatrix, $\U^{\dagger}(L_k)=e^{i\phi_k}L_k$ and $n_k>0$ be the minimal integer  such that $e^{in_k\phi_k}=1$. For a state $\rho(0)$, the coefficient $c_k\equiv\Tr[L_k \rho(0)]=\Tr\{L_k \U^{n_k}[\rho(0)]\}$ is the same for all states generated from $\rho(0)$ under $\U^{n_k}$. In particular, as a weak symmetry acts on an invariant set of metastable phases in a classical MM as their permutation $\pi$ all  metastable phases connected under $\pi^{n_k}$ have the same coefficient $c_k$.

Let us now focus on the extreme values of coefficients. From Eqs.~\eqref{eq:RkLk_C_U} and~\eqref{eq:C_U}, an eigenmode $L_k$ is a linear combination of plane waves over cycles with the length divisible by $n_k$,
\begin{equation}\label{eq:Lk_waves}
	L_k=\sum_{l:\, n_{k}|d_l} c_k'^{(l)} L'_{\pi^{j_{kl}}(l)},
\end{equation}
where $l$ indexes cycles, $d_l$ is their  length, and $ j_{kl}\equiv\phi_k d_l/(2\pi) \mod d_l$ [cf.~Eq.~\eqref{eq:Lk_U}].
Since the maximal coefficient of $L_k^R$ can be approximated up to corrections $\Ccl$ by considering states inside the simplex of metastable phases [cf.~Eqs.~\eqref{eq:ck_simplex}], its value corresponds to the maximum value among plane waves  weighted by the  coefficients $c_k'^{(l)}$,
\begin{eqnarray}\label{eq:ck_simplex_U}
	&& c_k^{R(\max)}=\\\nonumber
	&& \max_l \max _{0\leq n\leq n_k-1} \sqrt{2}\,\big|c_k'^{(l)}\big| \cos\big[ n\phi_k+\varphi_k+\arg c_k'^{(l)}\big]+...,
\end{eqnarray}
where $\varphi_k$ is the arbitrary phase chosen to obtain Hermitian $L_k^R$ in Eq.~\eqref{eq:LRLI}, and we used that $e^{in_k\phi_k}=1$. Similarly, the minimum coefficient $c_k^{R(\min)}$ can be approximated by considering minimal values of the cosine in Eq.~\eqref{eq:ck_simplex_U}, while the extreme values of the coefficient $c_k^{I}$ for $L_k^I$ can be approximated by considering the sine instead of the cosine. Therefore, degeneracy in the extreme coefficients $c_k^R$ or $c_k^I$ can arise in two ways:
\begin{enumerate}
	\item[A.] the coefficients $c_k'^{(l)}$ together with choice of $\varphi_k$ lead to degeneracy of extreme values attained by different plane waves,
	\item[B.] some of the cycles contributing to $L_k$ are of length longer than $n_k$, so that they have $d_l/n_k$-fold degeneracy in the amplitude of the corresponding plane wave [cf.~Eq.~\eqref{eq:Lk_U}]. 
\end{enumerate} 
Case B degeneracy is a direct consequence of from the presence of the weak symmetry, as discussed in the first paragraph, and is always present, e.g., in symmetric eigenmodes. Nevertheless, both degeneracy cases can be remedied; see below. 

\subsubsection{Refined metastable phase construction}\label{app:algorithm_symmetry3}

We now explain how to remedy degeneracy of coefficients arising in the presence of a weak symmetry in the approach of Sec.~\ref{sec:algorithm} in the main text. Although any degeneracy can be generally resolved by random rotations  as argued in Sec.~\ref{app:algorithm_rot}, here, we explain how the structure of the MM  imposed by the weak symmetry in the coefficient space can be exploited to simplify the approach; see also the discussion in Sec.~\ref{sec:algorithm} of the main text.\\

\emph{Case A degeneracy}. In general, cycles of various lengths contribute to $L_k$ in~Eq.~\eqref{eq:Lk_waves}. If we consider $L_k$ such that there does not exists $L_l$ with $n_l$ divisible by $n_k$, however, $L_k$ is supported on only cycles with the length $n_k$  (as there are no longer cycles with the length divisible by $n_k$). In particular, the number of cycles of length $n_k$ equals the degeneracy of the symmetry eigenvalue  $e^{i n\phi_k}$ among low-lying eigenmodes. Therefore, the only coefficient  degeneracy that can be present belongs to Case A discussed in Sec.~\ref{app:algorithm_symmetry2}. We now explain how it can be remedied by an appropriate choice of the phase $\varphi_k$ in $L_k^R$ of Eq.~\eqref{eq:LRLI}.

First, let $n=1,2,...,n_k-1$ be such that it is the closest to $1$ among $e^{in\phi_k}$ and with the positive imaginary part. For any $\varphi_k$ in Eq.~\eqref{eq:LRLI}, the difference of the maximal and the next in value coefficient for the metastable phases in $l$th cycle is less than $2\sqrt{2}|c_k'^{(l)}| \sin(\delta\phi_k)^2$, where $\delta\phi_k=(n\phi_k \mod 2\pi)/2$, but it also can be $0$ in the worst case scenario.  By considering both $\varphi_k^{(1)}=0$ and $\varphi_k^{(2)}=\delta\phi_k$ in Eq.~\eqref{eq:LRLI}, however, the bigger among the differences is no less than $2\sqrt{2}|c_k'^{(l)}| \sin(\delta\phi_k/2)\sin(\delta\phi_k)$; this choice can be effectively facilitated by considering both $L_k^R$ and $L_k^I$, whenever $d_l$ is not divisible by $4$.

Second, let $|c_k'^{(l)}|=\max_{l'}|c_{k}'^{(l')}| $ be the maximal weight among the plane waves in Eq.~\eqref{eq:Lk_waves}. 
The minimal difference between maximal coefficients for metastable phases in different cycles is at least $\max[ |c_k'^{(l)}| \cos(\delta\phi_k/2)- \max_{l'\neq l}|c_{k}'^{(l')}|, 0] $ either for $\varphi_k^{(1)}$ or $\varphi_k^{(2)}$. If needed,  this bound can be improved by considering additionally  $\varphi_k^{(n)}=\delta\phi_k/2^n$, $n=1,..,N$  in Eq.~\eqref{eq:LRLI}, in which case $\cos(\delta\phi_k/2)$ is replaced by $\cos(\delta\phi_k/2^{N+1})$.   

Therefore, there is only a single metastable phase corresponding to the maximal coefficient of $L_k^R$ when the maximal weight $|c_k'^{(l)}|$ is nondegenerate up to $\Ccl$ [also when  multiplied by $\cos(\delta\phi_k/2)$] and when $\sin(\delta\phi_k/2)\sin(\delta\phi_k) \gg \Ccl$. The former condition can be achieved for any plane-wave [any $l$ in Eq.~\eqref{eq:Lk_waves}] by a rotation of all $L_k$ with the same symmetry eigenvalue provided that the simplex of the coefficients $|c_k'^{(l)}|$ for those eigenmodes has a nonnegligible volume  (cf.~Sec.~\ref{app:algorithm_rot}). 

Finally, for each cycle, the metastable phase chosen from an extreme eigenstate of (rotated) $L_k^R$ can be used  to recover other elements of the cycle by applying the symmetry $\U$ $n_k$ times. Therefore, there is no need to consider $L_k^I$ or other modes supported on those cycles, i.e., $L_{k'}$ with a different symmetry eigenvalue $e^{i\phi_{k'}}\neq e^{i\phi_k}$  but with $n_{k'}=n_k$. Furthermore,  this choice corresponds to the symmetric set of metastable phases in Eq.~\eqref{eq:rhotilde_U} (cf.~Sec.~\ref{app:algorithm_symmetry1}). 
\\

\emph{Case B degeneracy}. After finding the extreme states of $L_k$ as described in the above paragraph, we still need to consider $L_l$ with the symmetry eigenvalue $e^{i \phi_l}$ and $n_l$ such that there exists $L_k$ with $n_k$ divisible by $n_l$. In that case $L_l$ features Case B degeneracy discussed in Sec.~\ref{app:algorithm_symmetry2} with respect to metastable phases connected by $\pi^{n_k/n_l}$ in the already found cycles [cf.~Eq.~\eqref{eq:Lk_waves}].

We are interested in whether the metastable states on which $L_l$ is supported via plane waves have been already found by considering $L_k$. This is the case when the degeneracy of $e^{i \phi_l}$ among the low-lying eigenmodes equals the number of already considered cycles with their length divisible by $n_l$ (which is equal the sum of degeneracies of $e^{i \phi_k}$ for already considered $L_k$ such that $n_k/n_l$ is an integer; without repetitions, i.e.,  without $e^{i\phi_{k'}}\neq e^{i\phi_k}$  with $n_{k'}=n_k$). 
Otherwise, $L_l$ is supported on cycles that have not been considered yet. In that case, by choosing $L_l$ with $n_l$ such that the only $L_k$ with $n_k$ divisible by $n_l$ have already been considered, we obtain that all new cycles on which $L_l$ is also supported are exactly of the length $n_l$ (and their number is the difference between the degeneracy of $e^{i \phi_l}$ and the sum of degeneracies of relevant $e^{i\phi_k}$). By considering rotations of all eigenmodes with the symmetry eigenvalue $e^{i\phi_l}$, 
a metastable phase in each cycle with the length $n_l$ can be found from extreme eigenstates of rotated eigenmodes, as discussed in Case A above. We note that equal mixtures of already found phases in cycles of length $n_k$ connected by $\pi^{n_k/n_l}$ will also be found here (due to Case B degeneracy; cf.~Sec.~\ref{app:algorithm_symmetry1}), but those candidate states will be discarded by choosing the candidates yielding the maximal volume simplex. Finally, other elements of the cycles with the length $n_l$ can be recovered by applying the symmetry $n_l$ times, while other modes supported on those cycles, i.e., $L_{l'}$ with $e^{i\phi_{l'}}\neq e^{i\phi_{l}}$ but with $n_{l'}=n_l$, do not need to be considered.

If not all choices of the eigenmodes are exhausted in the way discussed above, i.e., there exist $L_{j}$ with $n_{j}<n_l$, such that $n_{j}$ divides  $n_l$, we again repeat the procedure described in the above paragraph, but with respect to all $L_l$ and with $n_l$ that are divisible by $n_{j}$. In particular, $L_j$ may be discarded if is supported on already considered cycles, i.e., degeneracy of $e^{i\phi_j}$ is equal to the sum of degeneracies of all relevant $e^{i\phi_l}$. \\

\emph{Example}. In Fig.~\ref{fig:Symmetry} of the main text, we have one cycle of length~$2$ corresponding to the metastable phases $\trho_1$ and $\trho_3$ and two cycles of length $1$ corresponding to invariant metastable phases $\trho_2$ and $\trho_4$. Here, we would first consider the eigenmode $L_3$ corresponding to the symmetry eigenvalue $-1$ ($n_k=2$), which would give  as candidates approximately $\trho_1$ and $\trho_3$ (twice; due to cycles of the symmetry). We would then be left with two symmetric eigenmodes $L_2$ and $L_4$ (we do not need to consider trivial $L_1=\mathds{1}$), which would give pairs of symmetric candidates approximately as $\trho_2$, $(\trho_1+\trho_3)/2$, and $\trho_2$, $\trho_4$. By clustering those candidates would be reduced to: $\trho_1$, $\trho_2$, $\trho_3$, $(\trho_1+\trho_3)/2$, and $\trho_4$. Finally, by considering the maximal volume simplex, we would obtain approximately the four metastable phases $\trho_1$, $\trho_2$, $\trho_3$ and $\trho_4$.


\begin{thebibliography}{128}%
\makeatletter
\providecommand \@ifxundefined [1]{%
 \@ifx{#1\undefined}
}%
\providecommand \@ifnum [1]{%
 \ifnum #1\expandafter \@firstoftwo
 \else \expandafter \@secondoftwo
 \fi
}%
\providecommand \@ifx [1]{%
 \ifx #1\expandafter \@firstoftwo
 \else \expandafter \@secondoftwo
 \fi
}%
\providecommand \natexlab [1]{#1}%
\providecommand \enquote  [1]{``#1''}%
\providecommand \bibnamefont  [1]{#1}%
\providecommand \bibfnamefont [1]{#1}%
\providecommand \citenamefont [1]{#1}%
\providecommand \href@noop [0]{\@secondoftwo}%
\providecommand \href [0]{\begingroup \@sanitize@url \@href}%
\providecommand \@href[1]{\@@startlink{#1}\@@href}%
\providecommand \@@href[1]{\endgroup#1\@@endlink}%
\providecommand \@sanitize@url [0]{\catcode `\\12\catcode `\$12\catcode
  `\&12\catcode `\#12\catcode `\^12\catcode `\_12\catcode `\%12\relax}%
\providecommand \@@startlink[1]{}%
\providecommand \@@endlink[0]{}%
\providecommand \url  [0]{\begingroup\@sanitize@url \@url }%
\providecommand \@url [1]{\endgroup\@href {#1}{\urlprefix }}%
\providecommand \urlprefix  [0]{URL }%
\providecommand \Eprint [0]{\href }%
\providecommand \doibase [0]{http://dx.doi.org/}%
\providecommand \selectlanguage [0]{\@gobble}%
\providecommand \bibinfo  [0]{\@secondoftwo}%
\providecommand \bibfield  [0]{\@secondoftwo}%
\providecommand \translation [1]{[#1]}%
\providecommand \BibitemOpen [0]{}%
\providecommand \bibitemStop [0]{}%
\providecommand \bibitemNoStop [0]{.\EOS\space}%
\providecommand \EOS [0]{\spacefactor3000\relax}%
\providecommand \BibitemShut  [1]{\csname bibitem#1\endcsname}%
\let\auto@bib@innerbib\@empty
\bibitem [{\citenamefont {Pritchard}\ \emph {et~al.}(2010)\citenamefont
  {Pritchard}, \citenamefont {Maxwell}, \citenamefont {Gauguet}, \citenamefont
  {Weatherill}, \citenamefont {Jones},\ and\ \citenamefont
  {Adams}}]{Pritchard2010}%
  \BibitemOpen
  \bibfield  {author} {\bibinfo {author} {\bibfnamefont {J.~D.}\ \bibnamefont
  {Pritchard}}, \bibinfo {author} {\bibfnamefont {D.}~\bibnamefont {Maxwell}},
  \bibinfo {author} {\bibfnamefont {A.}~\bibnamefont {Gauguet}}, \bibinfo
  {author} {\bibfnamefont {K.~J.}\ \bibnamefont {Weatherill}}, \bibinfo
  {author} {\bibfnamefont {M.~P.~A.}\ \bibnamefont {Jones}}, \ and\ \bibinfo
  {author} {\bibfnamefont {C.~S.}\ \bibnamefont {Adams}},\ }\bibfield  {title}
  {\enquote {\bibinfo {title} {{Cooperative Atom-Light Interaction in a
  Blockaded Rydberg Ensemble}},}\ }\href {\doibase
  10.1103/PhysRevLett.105.193603} {\bibfield  {journal} {\bibinfo  {journal}
  {Phys. Rev. Lett.}\ }\textbf {\bibinfo {volume} {105}},\ \bibinfo {pages}
  {193603} (\bibinfo {year} {2010})}\BibitemShut {NoStop}%
\bibitem [{\citenamefont {Blatt}\ and\ \citenamefont {Roos}(2012)}]{Blatt2012}%
  \BibitemOpen
  \bibfield  {author} {\bibinfo {author} {\bibfnamefont {R.}~\bibnamefont
  {Blatt}}\ and\ \bibinfo {author} {\bibfnamefont {C.~F.}\ \bibnamefont
  {Roos}},\ }\bibfield  {title} {\enquote {\bibinfo {title} {{Quantum
  simulations with trapped ions}},}\ }\href
  {http://dx.doi.org/10.1038/nphys2252} {\bibfield  {journal} {\bibinfo
  {journal} {Nat. Phys.}\ }\textbf {\bibinfo {volume} {8}},\ \bibinfo {pages}
  {277--284} (\bibinfo {year} {2012})}\BibitemShut {NoStop}%
\bibitem [{\citenamefont {Britton}\ \emph {et~al.}(2012)\citenamefont
  {Britton}, \citenamefont {Sawyer}, \citenamefont {Keith}, \citenamefont
  {Wang}, \citenamefont {Freericks}, \citenamefont {Uys}, \citenamefont
  {Biercuk},\ and\ \citenamefont {Bollinger}}]{Britton2012}%
  \BibitemOpen
  \bibfield  {author} {\bibinfo {author} {\bibfnamefont {J.~W.}\ \bibnamefont
  {Britton}}, \bibinfo {author} {\bibfnamefont {B.~C.}\ \bibnamefont {Sawyer}},
  \bibinfo {author} {\bibfnamefont {A.~C.}\ \bibnamefont {Keith}}, \bibinfo
  {author} {\bibfnamefont {C.~C.~J.}\ \bibnamefont {Wang}}, \bibinfo {author}
  {\bibfnamefont {J.~K.}\ \bibnamefont {Freericks}}, \bibinfo {author}
  {\bibfnamefont {H.}~\bibnamefont {Uys}}, \bibinfo {author} {\bibfnamefont
  {M.~J.}\ \bibnamefont {Biercuk}}, \ and\ \bibinfo {author} {\bibfnamefont
  {J.~J.}\ \bibnamefont {Bollinger}},\ }\bibfield  {title} {\enquote {\bibinfo
  {title} {{Engineered two-dimensional Ising interactions in a trapped-ion
  quantum simulator with hundreds of spins}},}\ }\href
  {http://dx.doi.org/10.1038/nature10981} {\bibfield  {journal} {\bibinfo
  {journal} {Nature}\ }\textbf {\bibinfo {volume} {484}},\ \bibinfo {pages}
  {489--492} (\bibinfo {year} {2012})}\BibitemShut {NoStop}%
\bibitem [{\citenamefont {Dudin}\ and\ \citenamefont
  {Kuzmich}(2012)}]{Dudin2012}%
  \BibitemOpen
  \bibfield  {author} {\bibinfo {author} {\bibfnamefont {Y.~O.}\ \bibnamefont
  {Dudin}}\ and\ \bibinfo {author} {\bibfnamefont {A.}~\bibnamefont
  {Kuzmich}},\ }\bibfield  {title} {\enquote {\bibinfo {title} {{Strongly
  Interacting Rydberg Excitations of a Cold Atomic Gas}},}\ }\href {\doibase
  http://dx.doi.org/10.1126/science.1217901} {\bibfield  {journal} {\bibinfo
  {journal} {Science}\ }\textbf {\bibinfo {volume} {336}},\ \bibinfo {pages}
  {887--889} (\bibinfo {year} {2012})}\BibitemShut {NoStop}%
\bibitem [{\citenamefont {Peyronel}\ \emph {et~al.}(2012)\citenamefont
  {Peyronel}, \citenamefont {Firstenberg}, \citenamefont {Liang}, \citenamefont
  {Hofferberth}, \citenamefont {Gorshkov}, \citenamefont {Pohl}, \citenamefont
  {Lukin},\ and\ \citenamefont {Vuletic}}]{Peyronel2012}%
  \BibitemOpen
  \bibfield  {author} {\bibinfo {author} {\bibfnamefont {T.}~\bibnamefont
  {Peyronel}}, \bibinfo {author} {\bibfnamefont {O.}~\bibnamefont
  {Firstenberg}}, \bibinfo {author} {\bibfnamefont {Q.}~\bibnamefont {Liang}},
  \bibinfo {author} {\bibfnamefont {S.}~\bibnamefont {Hofferberth}}, \bibinfo
  {author} {\bibfnamefont {A.~V.}\ \bibnamefont {Gorshkov}}, \bibinfo {author}
  {\bibfnamefont {T.}~\bibnamefont {Pohl}}, \bibinfo {author} {\bibfnamefont
  {M.~D.}\ \bibnamefont {Lukin}}, \ and\ \bibinfo {author} {\bibfnamefont
  {V.}~\bibnamefont {Vuletic}},\ }\bibfield  {title} {\enquote {\bibinfo
  {title} {{Quantum nonlinear optics with single photons enabled by strongly
  interacting atoms}},}\ }\href {http://dx.doi.org/10.1038/nature11361}
  {\bibfield  {journal} {\bibinfo  {journal} {Nature}\ }\textbf {\bibinfo
  {volume} {488}},\ \bibinfo {pages} {57--60} (\bibinfo {year}
  {2012})}\BibitemShut {NoStop}%
\bibitem [{\citenamefont {G{\"u}nter}\ \emph {et~al.}(2013)\citenamefont
  {G{\"u}nter}, \citenamefont {Schempp}, \citenamefont {Robert-de
  Saint-Vincent}, \citenamefont {Gavryusev}, \citenamefont {Helmrich},
  \citenamefont {Hofmann}, \citenamefont {Whitlock},\ and\ \citenamefont
  {Weidem{\"u}ller}}]{Guenter2013}%
  \BibitemOpen
  \bibfield  {author} {\bibinfo {author} {\bibfnamefont {G.}~\bibnamefont
  {G{\"u}nter}}, \bibinfo {author} {\bibfnamefont {H.}~\bibnamefont {Schempp}},
  \bibinfo {author} {\bibfnamefont {M.}~\bibnamefont {Robert-de
  Saint-Vincent}}, \bibinfo {author} {\bibfnamefont {V.}~\bibnamefont
  {Gavryusev}}, \bibinfo {author} {\bibfnamefont {S.}~\bibnamefont {Helmrich}},
  \bibinfo {author} {\bibfnamefont {C.~S.}\ \bibnamefont {Hofmann}}, \bibinfo
  {author} {\bibfnamefont {S.}~\bibnamefont {Whitlock}}, \ and\ \bibinfo
  {author} {\bibfnamefont {M.}~\bibnamefont {Weidem{\"u}ller}},\ }\bibfield
  {title} {\enquote {\bibinfo {title} {{Observing the Dynamics of
  Dipole-Mediated Energy Transport by Interaction-Enhanced Imaging}},}\ }\href
  {http://science.sciencemag.org/content/342/6161/954.abstract} {\bibfield
  {journal} {\bibinfo  {journal} {Science}\ }\textbf {\bibinfo {volume}
  {342}},\ \bibinfo {pages} {954--956} (\bibinfo {year} {2013})}\BibitemShut
  {NoStop}%
\bibitem [{\citenamefont {Schmidt}\ and\ \citenamefont
  {Koch}(2013)}]{Schmidt2013}%
  \BibitemOpen
  \bibfield  {author} {\bibinfo {author} {\bibfnamefont {S.}~\bibnamefont
  {Schmidt}}\ and\ \bibinfo {author} {\bibfnamefont {J.}~\bibnamefont {Koch}},\
  }\bibfield  {title} {\enquote {\bibinfo {title} {{Circuit QED lattices:
  Towards quantum simulation with superconducting circuits}},}\ }\href
  {\doibase 10.1002/andp.201200261} {\bibfield  {journal} {\bibinfo  {journal}
  {Ann. Phys.}\ }\textbf {\bibinfo {volume} {525}},\ \bibinfo {pages}
  {395--412} (\bibinfo {year} {2013})}\BibitemShut {NoStop}%
\bibitem [{\citenamefont {Dalibard}\ \emph {et~al.}(1992)\citenamefont
  {Dalibard}, \citenamefont {Castin},\ and\ \citenamefont
  {M\o{}lmer}}]{Molmer92}%
  \BibitemOpen
  \bibfield  {author} {\bibinfo {author} {\bibfnamefont {J.}~\bibnamefont
  {Dalibard}}, \bibinfo {author} {\bibfnamefont {Y.}~\bibnamefont {Castin}}, \
  and\ \bibinfo {author} {\bibfnamefont {K.}~\bibnamefont {M\o{}lmer}},\
  }\bibfield  {title} {\enquote {\bibinfo {title} {{Wave-function approach to
  dissipative processes in quantum optics}},}\ }\href {\doibase
  10.1103/PhysRevLett.68.580} {\bibfield  {journal} {\bibinfo  {journal} {Phys.
  Rev. Lett.}\ }\textbf {\bibinfo {volume} {68}},\ \bibinfo {pages} {580--583}
  (\bibinfo {year} {1992})}\BibitemShut {NoStop}%
\bibitem [{\citenamefont {Dum}\ \emph {et~al.}(1992)\citenamefont {Dum},
  \citenamefont {Parkins}, \citenamefont {Zoller},\ and\ \citenamefont
  {Gardiner}}]{Dum1992}%
  \BibitemOpen
  \bibfield  {author} {\bibinfo {author} {\bibfnamefont {R.}~\bibnamefont
  {Dum}}, \bibinfo {author} {\bibfnamefont {A.~S.}\ \bibnamefont {Parkins}},
  \bibinfo {author} {\bibfnamefont {P.}~\bibnamefont {Zoller}}, \ and\ \bibinfo
  {author} {\bibfnamefont {C.~W.}\ \bibnamefont {Gardiner}},\ }\bibfield
  {title} {\enquote {\bibinfo {title} {{Monte Carlo simulation of master
  equations in quantum optics for vacuum, thermal, and squeezed reservoirs}},}\
  }\href {\doibase 10.1103/PhysRevA.46.4382} {\bibfield  {journal} {\bibinfo
  {journal} {Phys. Rev. A}\ }\textbf {\bibinfo {volume} {46}},\ \bibinfo
  {pages} {4382--4396} (\bibinfo {year} {1992})}\BibitemShut {NoStop}%
\bibitem [{\citenamefont {M{\o}lmer}\ \emph {et~al.}(1993)\citenamefont
  {M{\o}lmer}, \citenamefont {Castin},\ and\ \citenamefont
  {Dalibard}}]{Molmer93}%
  \BibitemOpen
  \bibfield  {author} {\bibinfo {author} {\bibfnamefont {K.}~\bibnamefont
  {M{\o}lmer}}, \bibinfo {author} {\bibfnamefont {Y.}~\bibnamefont {Castin}}, \
  and\ \bibinfo {author} {\bibfnamefont {J.}~\bibnamefont {Dalibard}},\
  }\bibfield  {title} {\enquote {\bibinfo {title} {{Monte Carlo wave-function
  method in quantum optics}},}\ }\href {\doibase 10.1364/JOSAB.10.000524}
  {\bibfield  {journal} {\bibinfo  {journal} {J. Opt. Soc. Am. B}\ }\textbf
  {\bibinfo {volume} {10}},\ \bibinfo {pages} {524--538} (\bibinfo {year}
  {1993})}\BibitemShut {NoStop}%
\bibitem [{\citenamefont {Plenio}\ and\ \citenamefont
  {Knight}(1998)}]{Plenio1998}%
  \BibitemOpen
  \bibfield  {author} {\bibinfo {author} {\bibfnamefont {M.~B.}\ \bibnamefont
  {Plenio}}\ and\ \bibinfo {author} {\bibfnamefont {P.~L.}\ \bibnamefont
  {Knight}},\ }\bibfield  {title} {\enquote {\bibinfo {title} {{The
  quantum-jump approach to dissipative dynamics in quantum optics}},}\ }\href
  {\doibase 10.1103/RevModPhys.70.101} {\bibfield  {journal} {\bibinfo
  {journal} {Rev. Mod. Phys.}\ }\textbf {\bibinfo {volume} {70}},\ \bibinfo
  {pages} {101--144} (\bibinfo {year} {1998})}\BibitemShut {NoStop}%
\bibitem [{\citenamefont {Daley}(2014)}]{Daley2014}%
  \BibitemOpen
  \bibfield  {author} {\bibinfo {author} {\bibfnamefont {A.~J.}\ \bibnamefont
  {Daley}},\ }\bibfield  {title} {\enquote {\bibinfo {title} {{Quantum
  trajectories and open many-body quantum systems}},}\ }\href {\doibase
  10.1080/00018732.2014.933502} {\bibfield  {journal} {\bibinfo  {journal}
  {Adv. Phys.}\ }\textbf {\bibinfo {volume} {63}},\ \bibinfo {pages} {77--149}
  (\bibinfo {year} {2014})}\BibitemShut {NoStop}%
\bibitem [{\citenamefont {Gangat}\ \emph {et~al.}(2017)\citenamefont {Gangat},
  \citenamefont {I},\ and\ \citenamefont {Kao}}]{Gangat2017}%
  \BibitemOpen
  \bibfield  {author} {\bibinfo {author} {\bibfnamefont {A.~A.}\ \bibnamefont
  {Gangat}}, \bibinfo {author} {\bibfnamefont {T.}~\bibnamefont {I}}, \ and\
  \bibinfo {author} {\bibfnamefont {Y.-J.}\ \bibnamefont {Kao}},\ }\bibfield
  {title} {\enquote {\bibinfo {title} {{Steady States of Infinite-Size
  Dissipative Quantum Chains via Imaginary Time Evolution}},}\ }\href {\doibase
  10.1103/PhysRevLett.119.010501} {\bibfield  {journal} {\bibinfo  {journal}
  {Phys. Rev. Lett.}\ }\textbf {\bibinfo {volume} {119}},\ \bibinfo {pages}
  {010501} (\bibinfo {year} {2017})}\BibitemShut {NoStop}%
\bibitem [{\citenamefont {Torre}\ \emph {et~al.}(2013)\citenamefont {Torre},
  \citenamefont {Diehl}, \citenamefont {Lukin}, \citenamefont {Sachdev},\ and\
  \citenamefont {Strack}}]{Torre2013}%
  \BibitemOpen
  \bibfield  {author} {\bibinfo {author} {\bibfnamefont {E.~G.~D.}\
  \bibnamefont {Torre}}, \bibinfo {author} {\bibfnamefont {S.}~\bibnamefont
  {Diehl}}, \bibinfo {author} {\bibfnamefont {M.~D.}\ \bibnamefont {Lukin}},
  \bibinfo {author} {\bibfnamefont {S.}~\bibnamefont {Sachdev}}, \ and\
  \bibinfo {author} {\bibfnamefont {P.}~\bibnamefont {Strack}},\ }\bibfield
  {title} {\enquote {\bibinfo {title} {{Keldysh approach for nonequilibrium
  phase transitions in quantum optics: Beyond the Dicke model in optical
  cavities}},}\ }\href {\doibase 10.1103/PhysRevA.87.023831} {\bibfield
  {journal} {\bibinfo  {journal} {Phys. Rev. A}\ }\textbf {\bibinfo {volume}
  {87}},\ \bibinfo {pages} {023831} (\bibinfo {year} {2013})}\BibitemShut
  {NoStop}%
\bibitem [{\citenamefont {Sieberer}\ \emph {et~al.}(2016)\citenamefont
  {Sieberer}, \citenamefont {Buchhold},\ and\ \citenamefont
  {Diehl}}]{Sieberer2016}%
  \BibitemOpen
  \bibfield  {author} {\bibinfo {author} {\bibfnamefont {L.~M.}\ \bibnamefont
  {Sieberer}}, \bibinfo {author} {\bibfnamefont {M.}~\bibnamefont {Buchhold}},
  \ and\ \bibinfo {author} {\bibfnamefont {S.}~\bibnamefont {Diehl}},\
  }\bibfield  {title} {\enquote {\bibinfo {title} {{Keldysh field theory for
  driven open quantum systems}},}\ }\href
  {http://stacks.iop.org/0034-4885/79/i=9/a=096001} {\bibfield  {journal}
  {\bibinfo  {journal} {Rep. Prog. Phys.}\ }\textbf {\bibinfo {volume} {79}},\
  \bibinfo {pages} {096001} (\bibinfo {year} {2016})}\BibitemShut {NoStop}%
\bibitem [{\citenamefont {Maghrebi}\ and\ \citenamefont
  {Gorshkov}(2016)}]{Maghrebi2016}%
  \BibitemOpen
  \bibfield  {author} {\bibinfo {author} {\bibfnamefont {M.~F.}\ \bibnamefont
  {Maghrebi}}\ and\ \bibinfo {author} {\bibfnamefont {A.~V.}\ \bibnamefont
  {Gorshkov}},\ }\bibfield  {title} {\enquote {\bibinfo {title}
  {{Nonequilibrium many-body steady states via Keldysh formalism}},}\ }\href
  {\doibase 10.1103/PhysRevB.93.014307} {\bibfield  {journal} {\bibinfo
  {journal} {Phys. Rev. B}\ }\textbf {\bibinfo {volume} {93}},\ \bibinfo
  {pages} {014307} (\bibinfo {year} {2016})}\BibitemShut {NoStop}%
\bibitem [{\citenamefont {Mendoza-Arenas}\ \emph {et~al.}(2016)\citenamefont
  {Mendoza-Arenas}, \citenamefont {Clark}, \citenamefont {Felicetti},
  \citenamefont {Romero}, \citenamefont {Solano}, \citenamefont {Angelakis},\
  and\ \citenamefont {Jaksch}}]{Mendoza-Arenas2016}%
  \BibitemOpen
  \bibfield  {author} {\bibinfo {author} {\bibfnamefont {J.~J.}\ \bibnamefont
  {Mendoza-Arenas}}, \bibinfo {author} {\bibfnamefont {S.~R.}\ \bibnamefont
  {Clark}}, \bibinfo {author} {\bibfnamefont {S.}~\bibnamefont {Felicetti}},
  \bibinfo {author} {\bibfnamefont {G.}~\bibnamefont {Romero}}, \bibinfo
  {author} {\bibfnamefont {E.}~\bibnamefont {Solano}}, \bibinfo {author}
  {\bibfnamefont {D.~G.}\ \bibnamefont {Angelakis}}, \ and\ \bibinfo {author}
  {\bibfnamefont {D.}~\bibnamefont {Jaksch}},\ }\bibfield  {title} {\enquote
  {\bibinfo {title} {{Beyond mean-field bistability in driven-dissipative
  lattices: Bunching-antibunching transition and quantum simulation}},}\ }\href
  {\doibase 10.1103/PhysRevA.93.023821} {\bibfield  {journal} {\bibinfo
  {journal} {Phys. Rev. A}\ }\textbf {\bibinfo {volume} {93}},\ \bibinfo
  {pages} {023821} (\bibinfo {year} {2016})}\BibitemShut {NoStop}%
\bibitem [{\citenamefont {Foss-Feig}\ \emph {et~al.}(2017)\citenamefont
  {Foss-Feig}, \citenamefont {Niroula}, \citenamefont {Young}, \citenamefont
  {Hafezi}, \citenamefont {Gorshkov}, \citenamefont {Wilson},\ and\
  \citenamefont {Maghrebi}}]{Foss-Feig2017}%
  \BibitemOpen
  \bibfield  {author} {\bibinfo {author} {\bibfnamefont {M.}~\bibnamefont
  {Foss-Feig}}, \bibinfo {author} {\bibfnamefont {P.}~\bibnamefont {Niroula}},
  \bibinfo {author} {\bibfnamefont {J.~T.}\ \bibnamefont {Young}}, \bibinfo
  {author} {\bibfnamefont {M.}~\bibnamefont {Hafezi}}, \bibinfo {author}
  {\bibfnamefont {A.~V.}\ \bibnamefont {Gorshkov}}, \bibinfo {author}
  {\bibfnamefont {R.~M.}\ \bibnamefont {Wilson}}, \ and\ \bibinfo {author}
  {\bibfnamefont {M.~F.}\ \bibnamefont {Maghrebi}},\ }\bibfield  {title}
  {\enquote {\bibinfo {title} {{Emergent equilibrium in many-body optical
  bistability}},}\ }\href {\doibase 10.1103/PhysRevA.95.043826} {\bibfield
  {journal} {\bibinfo  {journal} {Phys. Rev. A}\ }\textbf {\bibinfo {volume}
  {95}},\ \bibinfo {pages} {043826} (\bibinfo {year} {2017})}\BibitemShut
  {NoStop}%
\bibitem [{\citenamefont {Casteels}\ and\ \citenamefont
  {Wouters}(2017)}]{Casteels2017}%
  \BibitemOpen
  \bibfield  {author} {\bibinfo {author} {\bibfnamefont {W.}~\bibnamefont
  {Casteels}}\ and\ \bibinfo {author} {\bibfnamefont {M.}~\bibnamefont
  {Wouters}},\ }\bibfield  {title} {\enquote {\bibinfo {title} {{Optically
  bistable driven-dissipative Bose-Hubbard dimer: Gutzwiller approaches and
  entanglement}},}\ }\href {\doibase 10.1103/PhysRevA.95.043833} {\bibfield
  {journal} {\bibinfo  {journal} {Phys. Rev. A}\ }\textbf {\bibinfo {volume}
  {95}},\ \bibinfo {pages} {043833} (\bibinfo {year} {2017})}\BibitemShut
  {NoStop}%
\bibitem [{\citenamefont {Letscher}\ \emph {et~al.}(2017)\citenamefont
  {Letscher}, \citenamefont {Thomas}, \citenamefont {Niederpr\"um},
  \citenamefont {Fleischhauer},\ and\ \citenamefont {Ott}}]{Letscher2017}%
  \BibitemOpen
  \bibfield  {author} {\bibinfo {author} {\bibfnamefont {F.}~\bibnamefont
  {Letscher}}, \bibinfo {author} {\bibfnamefont {O.}~\bibnamefont {Thomas}},
  \bibinfo {author} {\bibfnamefont {T.}~\bibnamefont {Niederpr\"um}}, \bibinfo
  {author} {\bibfnamefont {M.}~\bibnamefont {Fleischhauer}}, \ and\ \bibinfo
  {author} {\bibfnamefont {H.}~\bibnamefont {Ott}},\ }\bibfield  {title}
  {\enquote {\bibinfo {title} {{Bistability Versus Metastability in Driven
  Dissipative Rydberg Gases}},}\ }\href {\doibase 10.1103/PhysRevX.7.021020}
  {\bibfield  {journal} {\bibinfo  {journal} {Phys. Rev. X}\ }\textbf {\bibinfo
  {volume} {7}},\ \bibinfo {pages} {021020} (\bibinfo {year}
  {2017})}\BibitemShut {NoStop}%
\bibitem [{\citenamefont {Jin}\ \emph {et~al.}(2018)\citenamefont {Jin},
  \citenamefont {Biella}, \citenamefont {Viyuela}, \citenamefont {Ciuti},
  \citenamefont {Fazio},\ and\ \citenamefont {Rossini}}]{Jin2018}%
  \BibitemOpen
  \bibfield  {author} {\bibinfo {author} {\bibfnamefont {J.}~\bibnamefont
  {Jin}}, \bibinfo {author} {\bibfnamefont {A.}~\bibnamefont {Biella}},
  \bibinfo {author} {\bibfnamefont {O.}~\bibnamefont {Viyuela}}, \bibinfo
  {author} {\bibfnamefont {C.}~\bibnamefont {Ciuti}}, \bibinfo {author}
  {\bibfnamefont {R.}~\bibnamefont {Fazio}}, \ and\ \bibinfo {author}
  {\bibfnamefont {D.}~\bibnamefont {Rossini}},\ }\bibfield  {title} {\enquote
  {\bibinfo {title} {{Phase diagram of the dissipative quantum Ising model on a
  square lattice}},}\ }\href {\doibase 10.1103/PhysRevB.98.241108} {\bibfield
  {journal} {\bibinfo  {journal} {Phys. Rev. B}\ }\textbf {\bibinfo {volume}
  {98}},\ \bibinfo {pages} {241108} (\bibinfo {year} {2018})}\BibitemShut
  {NoStop}%
\bibitem [{\citenamefont {de~Melo}\ \emph {et~al.}(2016)\citenamefont
  {de~Melo}, \citenamefont {Wade}, \citenamefont {\ifmmode \check{S}\else
  \v{S}\fi{}ibali\ifmmode~\acute{c}\else \'{c}\fi{}}, \citenamefont {Kondo},
  \citenamefont {Adams},\ and\ \citenamefont {Weatherill}}]{Melo2016}%
  \BibitemOpen
  \bibfield  {author} {\bibinfo {author} {\bibfnamefont {N.~R.}\ \bibnamefont
  {de~Melo}}, \bibinfo {author} {\bibfnamefont {C.~G.}\ \bibnamefont {Wade}},
  \bibinfo {author} {\bibfnamefont {N.}~\bibnamefont {\ifmmode \check{S}\else
  \v{S}\fi{}ibali\ifmmode~\acute{c}\else \'{c}\fi{}}}, \bibinfo {author}
  {\bibfnamefont {J.~M.}\ \bibnamefont {Kondo}}, \bibinfo {author}
  {\bibfnamefont {C.~S.}\ \bibnamefont {Adams}}, \ and\ \bibinfo {author}
  {\bibfnamefont {K.~J.}\ \bibnamefont {Weatherill}},\ }\bibfield  {title}
  {\enquote {\bibinfo {title} {{Intrinsic optical bistability in a strongly
  driven Rydberg ensemble}},}\ }\href {\doibase 10.1103/PhysRevA.93.063863}
  {\bibfield  {journal} {\bibinfo  {journal} {Phys. Rev. A}\ }\textbf {\bibinfo
  {volume} {93}},\ \bibinfo {pages} {063863} (\bibinfo {year}
  {2016})}\BibitemShut {NoStop}%
\bibitem [{\citenamefont {Rodriguez}\ \emph {et~al.}(2017)\citenamefont
  {Rodriguez}, \citenamefont {Casteels}, \citenamefont {Storme}, \citenamefont
  {Carlon~Zambon}, \citenamefont {Sagnes}, \citenamefont {Le~Gratiet},
  \citenamefont {Galopin}, \citenamefont {Lema\^{\i}tre}, \citenamefont {Amo},
  \citenamefont {Ciuti},\ and\ \citenamefont {Bloch}}]{Rodriguez2017}%
  \BibitemOpen
  \bibfield  {author} {\bibinfo {author} {\bibfnamefont {S.~R.~K.}\
  \bibnamefont {Rodriguez}}, \bibinfo {author} {\bibfnamefont {W.}~\bibnamefont
  {Casteels}}, \bibinfo {author} {\bibfnamefont {F.}~\bibnamefont {Storme}},
  \bibinfo {author} {\bibfnamefont {N.}~\bibnamefont {Carlon~Zambon}}, \bibinfo
  {author} {\bibfnamefont {I.}~\bibnamefont {Sagnes}}, \bibinfo {author}
  {\bibfnamefont {L.}~\bibnamefont {Le~Gratiet}}, \bibinfo {author}
  {\bibfnamefont {E.}~\bibnamefont {Galopin}}, \bibinfo {author} {\bibfnamefont
  {A.}~\bibnamefont {Lema\^{\i}tre}}, \bibinfo {author} {\bibfnamefont
  {A.}~\bibnamefont {Amo}}, \bibinfo {author} {\bibfnamefont {C.}~\bibnamefont
  {Ciuti}}, \ and\ \bibinfo {author} {\bibfnamefont {J.}~\bibnamefont
  {Bloch}},\ }\bibfield  {title} {\enquote {\bibinfo {title} {{Probing a
  Dissipative Phase Transition via Dynamical Optical Hysteresis}},}\ }\href
  {\doibase 10.1103/PhysRevLett.118.247402} {\bibfield  {journal} {\bibinfo
  {journal} {Phys. Rev. Lett.}\ }\textbf {\bibinfo {volume} {118}},\ \bibinfo
  {pages} {247402} (\bibinfo {year} {2017})}\BibitemShut {NoStop}%
\bibitem [{\citenamefont {Ates}\ \emph {et~al.}(2012)\citenamefont {Ates},
  \citenamefont {Olmos}, \citenamefont {Garrahan},\ and\ \citenamefont
  {Lesanovsky}}]{Ates2012}%
  \BibitemOpen
  \bibfield  {author} {\bibinfo {author} {\bibfnamefont {C.}~\bibnamefont
  {Ates}}, \bibinfo {author} {\bibfnamefont {B.}~\bibnamefont {Olmos}},
  \bibinfo {author} {\bibfnamefont {J.~P.}\ \bibnamefont {Garrahan}}, \ and\
  \bibinfo {author} {\bibfnamefont {I.}~\bibnamefont {Lesanovsky}},\ }\bibfield
   {title} {\enquote {\bibinfo {title} {{Dynamical phases and intermittency of
  the dissipative quantum Ising model}},}\ }\href {\doibase
  10.1103/PhysRevA.85.043620} {\bibfield  {journal} {\bibinfo  {journal} {Phys.
  Rev. A}\ }\textbf {\bibinfo {volume} {85}},\ \bibinfo {pages} {043620}
  (\bibinfo {year} {2012})}\BibitemShut {NoStop}%
\bibitem [{\citenamefont {Landa}\ \emph {et~al.}(2020)\citenamefont {Landa},
  \citenamefont {Schir\'o},\ and\ \citenamefont {Misguich}}]{Landa2020}%
  \BibitemOpen
  \bibfield  {author} {\bibinfo {author} {\bibfnamefont {H.}~\bibnamefont
  {Landa}}, \bibinfo {author} {\bibfnamefont {M.}~\bibnamefont {Schir\'o}}, \
  and\ \bibinfo {author} {\bibfnamefont {G.}~\bibnamefont {Misguich}},\
  }\bibfield  {title} {\enquote {\bibinfo {title} {{Multistability of
  Driven-Dissipative Quantum Spins}},}\ }\href {\doibase
  10.1103/PhysRevLett.124.043601} {\bibfield  {journal} {\bibinfo  {journal}
  {Phys. Rev. Lett.}\ }\textbf {\bibinfo {volume} {124}},\ \bibinfo {pages}
  {043601} (\bibinfo {year} {2020})}\BibitemShut {NoStop}%
\bibitem [{\citenamefont {Weimer}(2015{\natexlab{a}})}]{Weimer2015}%
  \BibitemOpen
  \bibfield  {author} {\bibinfo {author} {\bibfnamefont {H.}~\bibnamefont
  {Weimer}},\ }\bibfield  {title} {\enquote {\bibinfo {title} {{Variational
  Principle for Steady States of Dissipative Quantum Many-Body Systems}},}\
  }\href {\doibase 10.1103/PhysRevLett.114.040402} {\bibfield  {journal}
  {\bibinfo  {journal} {Phys. Rev. Lett.}\ }\textbf {\bibinfo {volume} {114}},\
  \bibinfo {pages} {040402} (\bibinfo {year} {2015}{\natexlab{a}})}\BibitemShut
  {NoStop}%
\bibitem [{\citenamefont {Weimer}(2015{\natexlab{b}})}]{Weimer2015a}%
  \BibitemOpen
  \bibfield  {author} {\bibinfo {author} {\bibfnamefont {H.}~\bibnamefont
  {Weimer}},\ }\bibfield  {title} {\enquote {\bibinfo {title} {{Variational
  analysis of driven-dissipative Rydberg gases}},}\ }\href {\doibase
  10.1103/PhysRevA.91.063401} {\bibfield  {journal} {\bibinfo  {journal} {Phys.
  Rev. A}\ }\textbf {\bibinfo {volume} {91}},\ \bibinfo {pages} {063401}
  (\bibinfo {year} {2015}{\natexlab{b}})}\BibitemShut {NoStop}%
\bibitem [{\citenamefont {Overbeck}\ \emph {et~al.}(2017)\citenamefont
  {Overbeck}, \citenamefont {Maghrebi}, \citenamefont {Gorshkov},\ and\
  \citenamefont {Weimer}}]{Overbeck2017}%
  \BibitemOpen
  \bibfield  {author} {\bibinfo {author} {\bibfnamefont {V.~R.}\ \bibnamefont
  {Overbeck}}, \bibinfo {author} {\bibfnamefont {M.~F.}\ \bibnamefont
  {Maghrebi}}, \bibinfo {author} {\bibfnamefont {A.~V.}\ \bibnamefont
  {Gorshkov}}, \ and\ \bibinfo {author} {\bibfnamefont {H.}~\bibnamefont
  {Weimer}},\ }\bibfield  {title} {\enquote {\bibinfo {title} {{Multicritical
  behavior in dissipative Ising models}},}\ }\href {\doibase
  10.1103/PhysRevA.95.042133} {\bibfield  {journal} {\bibinfo  {journal} {Phys.
  Rev. A}\ }\textbf {\bibinfo {volume} {95}},\ \bibinfo {pages} {042133}
  (\bibinfo {year} {2017})}\BibitemShut {NoStop}%
\bibitem [{\citenamefont {Biondi}\ \emph {et~al.}(2017)\citenamefont {Biondi},
  \citenamefont {Lienhard}, \citenamefont {Blatter}, \citenamefont {Türeci},\
  and\ \citenamefont {Schmidt}}]{Biondi2017}%
  \BibitemOpen
  \bibfield  {author} {\bibinfo {author} {\bibfnamefont {M.}~\bibnamefont
  {Biondi}}, \bibinfo {author} {\bibfnamefont {S.}~\bibnamefont {Lienhard}},
  \bibinfo {author} {\bibfnamefont {G.}~\bibnamefont {Blatter}}, \bibinfo
  {author} {\bibfnamefont {H.~E.}\ \bibnamefont {Türeci}}, \ and\ \bibinfo
  {author} {\bibfnamefont {S.}~\bibnamefont {Schmidt}},\ }\bibfield  {title}
  {\enquote {\bibinfo {title} {{Spatial correlations in driven-dissipative
  photonic lattices}},}\ }\href {\doibase 10.1088/1367-2630/aa99b2} {\bibfield
  {journal} {\bibinfo  {journal} {New J. Phys.}\ }\textbf {\bibinfo {volume}
  {19}},\ \bibinfo {pages} {125016} (\bibinfo {year} {2017})}\BibitemShut
  {NoStop}%
\bibitem [{\citenamefont {Spohn}(1977)}]{Spohn1977}%
  \BibitemOpen
  \bibfield  {author} {\bibinfo {author} {\bibfnamefont {H.}~\bibnamefont
  {Spohn}},\ }\bibfield  {title} {\enquote {\bibinfo {title} {{An algebraic
  condition for the approach to equilibrium of an open N-level system}},}\
  }\href {\doibase 10.1007/BF00420668} {\bibfield  {journal} {\bibinfo
  {journal} {Lett. Math. Phys.}\ }\textbf {\bibinfo {volume} {2}},\ \bibinfo
  {pages} {33--38} (\bibinfo {year} {1977})}\BibitemShut {NoStop}%
\bibitem [{\citenamefont {Evans}(1977)}]{Evans1977}%
  \BibitemOpen
  \bibfield  {author} {\bibinfo {author} {\bibfnamefont {D.~E.}\ \bibnamefont
  {Evans}},\ }\bibfield  {title} {\enquote {\bibinfo {title} {{Irreducible
  quantum dynamical semigroups}},}\ }\href {\doibase 10.1007/BF01614091}
  {\bibfield  {journal} {\bibinfo  {journal} {Comm. Math. Phys.}\ }\textbf
  {\bibinfo {volume} {54}},\ \bibinfo {pages} {293--297} (\bibinfo {year}
  {1977})}\BibitemShut {NoStop}%
\bibitem [{\citenamefont {Schirmer}\ and\ \citenamefont
  {Wang}(2010)}]{Schirmer2010}%
  \BibitemOpen
  \bibfield  {author} {\bibinfo {author} {\bibfnamefont {S.~G.}\ \bibnamefont
  {Schirmer}}\ and\ \bibinfo {author} {\bibfnamefont {X.}~\bibnamefont
  {Wang}},\ }\bibfield  {title} {\enquote {\bibinfo {title} {{Stabilizing open
  quantum systems by Markovian reservoir engineering}},}\ }\href {\doibase
  10.1103/PhysRevA.81.062306} {\bibfield  {journal} {\bibinfo  {journal} {Phys.
  Rev. A}\ }\textbf {\bibinfo {volume} {81}},\ \bibinfo {pages} {062306}
  (\bibinfo {year} {2010})}\BibitemShut {NoStop}%
\bibitem [{\citenamefont {Nigro}(2019)}]{Nigro2019}%
  \BibitemOpen
  \bibfield  {author} {\bibinfo {author} {\bibfnamefont {D.}~\bibnamefont
  {Nigro}},\ }\bibfield  {title} {\enquote {\bibinfo {title} {{On the
  uniqueness of the steady-state solution of the
  Lindblad{\textendash}Gorini{\textendash}Kossakowski{\textendash}Sudarshan
  equation}},}\ }\href {\doibase 10.1088/1742-5468/ab0c1c} {\bibfield
  {journal} {\bibinfo  {journal} {J. Stat. Mech.}\ }\textbf {\bibinfo {volume}
  {2019}},\ \bibinfo {pages} {043202} (\bibinfo {year} {2019})}\BibitemShut
  {NoStop}%
\bibitem [{\citenamefont {Chaikin}\ and\ \citenamefont
  {Lubensky}(2000)}]{Chaikin2000}%
  \BibitemOpen
  \bibfield  {author} {\bibinfo {author} {\bibfnamefont {P.~M.}\ \bibnamefont
  {Chaikin}}\ and\ \bibinfo {author} {\bibfnamefont {T.~C.}\ \bibnamefont
  {Lubensky}},\ }\href@noop {} {\emph {\bibinfo {title} {{Principles of
  condensed matter physics}}}},\ Vol.~\bibinfo {volume} {1}\ (\bibinfo
  {publisher} {Cambridge University Press},\ \bibinfo {year}
  {2000})\BibitemShut {NoStop}%
\bibitem [{\citenamefont {Gaveau}\ and\ \citenamefont
  {Schulman}(1987)}]{Gaveau1987}%
  \BibitemOpen
  \bibfield  {author} {\bibinfo {author} {\bibfnamefont {B.}~\bibnamefont
  {Gaveau}}\ and\ \bibinfo {author} {\bibfnamefont {L.~S.}\ \bibnamefont
  {Schulman}},\ }\bibfield  {title} {\enquote {\bibinfo {title} {{Dynamical
  metastability}},}\ }\href {http://stacks.iop.org/0305-4470/20/i=10/a=031}
  {\bibfield  {journal} {\bibinfo  {journal} {J. Phys. A}\ }\textbf {\bibinfo
  {volume} {20}},\ \bibinfo {pages} {2865} (\bibinfo {year}
  {1987})}\BibitemShut {NoStop}%
\bibitem [{\citenamefont {Gaveau}\ and\ \citenamefont
  {Schulman}(1998)}]{Gaveau1998}%
  \BibitemOpen
  \bibfield  {author} {\bibinfo {author} {\bibfnamefont {B.}~\bibnamefont
  {Gaveau}}\ and\ \bibinfo {author} {\bibfnamefont {L.~S.}\ \bibnamefont
  {Schulman}},\ }\bibfield  {title} {\enquote {\bibinfo {title} {{Theory of
  nonequilibrium first-order phase transitions for stochastic dynamics}},}\
  }\href {\doibase http://dx.doi.org/10.1063/1.532394} {\bibfield  {journal}
  {\bibinfo  {journal} {J. Mat. Phys.}\ }\textbf {\bibinfo {volume} {39}},\
  \bibinfo {pages} {1517} (\bibinfo {year} {1998})}\BibitemShut {NoStop}%
\bibitem [{\citenamefont {Bovier}\ \emph {et~al.}(2002)\citenamefont {Bovier},
  \citenamefont {Eckhoff}, \citenamefont {Gayrard},\ and\ \citenamefont
  {Klein}}]{Bovier2002}%
  \BibitemOpen
  \bibfield  {author} {\bibinfo {author} {\bibfnamefont {A.}~\bibnamefont
  {Bovier}}, \bibinfo {author} {\bibfnamefont {M.}~\bibnamefont {Eckhoff}},
  \bibinfo {author} {\bibfnamefont {V.}~\bibnamefont {Gayrard}}, \ and\
  \bibinfo {author} {\bibfnamefont {M.}~\bibnamefont {Klein}},\ }\bibfield
  {title} {\enquote {\bibinfo {title} {{Metastability and Low Lying Spectra in
  Reversible Markov Chains}},}\ }\href {\doibase 10.1007/s00220-014-2072-3}
  {\bibfield  {journal} {\bibinfo  {journal} {Comm. Math. Phys}\ }\textbf
  {\bibinfo {volume} {228}},\ \bibinfo {pages} {219} (\bibinfo {year}
  {2002})}\BibitemShut {NoStop}%
\bibitem [{\citenamefont {Gaveau}\ and\ \citenamefont
  {Schulman}(2006)}]{Gaveau2006}%
  \BibitemOpen
  \bibfield  {author} {\bibinfo {author} {\bibfnamefont {B.}~\bibnamefont
  {Gaveau}}\ and\ \bibinfo {author} {\bibfnamefont {L.~S.}\ \bibnamefont
  {Schulman}},\ }\bibfield  {title} {\enquote {\bibinfo {title} {{Multiple
  phases in stochastic dynamics: Geometry and probabilities}},}\ }\href
  {\doibase 10.1103/PhysRevE.73.036124} {\bibfield  {journal} {\bibinfo
  {journal} {Phys. Rev. E}\ }\textbf {\bibinfo {volume} {73}},\ \bibinfo
  {pages} {036124} (\bibinfo {year} {2006})}\BibitemShut {NoStop}%
\bibitem [{\citenamefont {Kurchan}(2016)}]{Kurchan2016}%
  \BibitemOpen
  \bibfield  {author} {\bibinfo {author} {\bibfnamefont {J.}~\bibnamefont
  {Kurchan}},\ }\href {https://arxiv.org/abs/0901.1271} {\enquote {\bibinfo
  {title} {{Six out of equilibrium lectures}},}\ } (\bibinfo {year} {2016}),\
  \Eprint {http://arxiv.org/abs/0901.1271} {arXiv:0901.1271} \BibitemShut
  {NoStop}%
\bibitem [{\citenamefont {Macieszczak}\ \emph
  {et~al.}(2016{\natexlab{a}})\citenamefont {Macieszczak}, \citenamefont
  {Guta}, \citenamefont {Lesanovsky},\ and\ \citenamefont
  {Garrahan}}]{Macieszczak2016a}%
  \BibitemOpen
  \bibfield  {author} {\bibinfo {author} {\bibfnamefont {K.}~\bibnamefont
  {Macieszczak}}, \bibinfo {author} {\bibfnamefont {M.}~\bibnamefont {Guta}},
  \bibinfo {author} {\bibfnamefont {I.}~\bibnamefont {Lesanovsky}}, \ and\
  \bibinfo {author} {\bibfnamefont {J.~P.}\ \bibnamefont {Garrahan}},\
  }\bibfield  {title} {\enquote {\bibinfo {title} {{Towards a Theory of
  Metastability in Open Quantum Dynamics}},}\ }\href {\doibase
  10.1103/PhysRevLett.116.240404} {\bibfield  {journal} {\bibinfo  {journal}
  {Phys. Rev. Lett.}\ }\textbf {\bibinfo {volume} {116}},\ \bibinfo {pages}
  {240404} (\bibinfo {year} {2016}{\natexlab{a}})}\BibitemShut {NoStop}%
\bibitem [{\citenamefont {Rose}\ \emph {et~al.}(2016)\citenamefont {Rose},
  \citenamefont {Macieszczak}, \citenamefont {Lesanovsky},\ and\ \citenamefont
  {Garrahan}}]{Rose2016}%
  \BibitemOpen
  \bibfield  {author} {\bibinfo {author} {\bibfnamefont {D.~C.}\ \bibnamefont
  {Rose}}, \bibinfo {author} {\bibfnamefont {K.}~\bibnamefont {Macieszczak}},
  \bibinfo {author} {\bibfnamefont {I.}~\bibnamefont {Lesanovsky}}, \ and\
  \bibinfo {author} {\bibfnamefont {J.~P.}\ \bibnamefont {Garrahan}},\
  }\bibfield  {title} {\enquote {\bibinfo {title} {{Metastability in an open
  quantum Ising model}},}\ }\href {\doibase 10.1103/PhysRevE.94.052132}
  {\bibfield  {journal} {\bibinfo  {journal} {Phys. Rev. E}\ }\textbf {\bibinfo
  {volume} {94}},\ \bibinfo {pages} {052132} (\bibinfo {year}
  {2016})}\BibitemShut {NoStop}%
\bibitem [{\citenamefont {Minganti}\ \emph {et~al.}(2018)\citenamefont
  {Minganti}, \citenamefont {Biella}, \citenamefont {Bartolo},\ and\
  \citenamefont {Ciuti}}]{Minganti2018}%
  \BibitemOpen
  \bibfield  {author} {\bibinfo {author} {\bibfnamefont {F.}~\bibnamefont
  {Minganti}}, \bibinfo {author} {\bibfnamefont {A.}~\bibnamefont {Biella}},
  \bibinfo {author} {\bibfnamefont {N.}~\bibnamefont {Bartolo}}, \ and\
  \bibinfo {author} {\bibfnamefont {C.}~\bibnamefont {Ciuti}},\ }\bibfield
  {title} {\enquote {\bibinfo {title} {{Spectral theory of Liouvillians for
  dissipative phase transitions}},}\ }\href {\doibase
  10.1103/PhysRevA.98.042118} {\bibfield  {journal} {\bibinfo  {journal} {Phys.
  Rev. A}\ }\textbf {\bibinfo {volume} {98}},\ \bibinfo {pages} {042118}
  (\bibinfo {year} {2018})}\BibitemShut {NoStop}%
\bibitem [{\citenamefont {J{\"a}ckle}\ and\ \citenamefont
  {Eisinger}(1991)}]{Jaeckle1991}%
  \BibitemOpen
  \bibfield  {author} {\bibinfo {author} {\bibfnamefont {J.}~\bibnamefont
  {J{\"a}ckle}}\ and\ \bibinfo {author} {\bibfnamefont {S.}~\bibnamefont
  {Eisinger}},\ }\bibfield  {title} {\enquote {\bibinfo {title} {{A
  hierarchically constrained kinetic Ising model}},}\ }\href {\doibase
  10.1007/BF01453764} {\bibfield  {journal} {\bibinfo  {journal} {Z. Phys. B}\
  }\textbf {\bibinfo {volume} {84}},\ \bibinfo {pages} {115--124} (\bibinfo
  {year} {1991})}\BibitemShut {NoStop}%
\bibitem [{\citenamefont {Sollich}\ and\ \citenamefont
  {Evans}(1999)}]{Sollich1999}%
  \BibitemOpen
  \bibfield  {author} {\bibinfo {author} {\bibfnamefont {P.}~\bibnamefont
  {Sollich}}\ and\ \bibinfo {author} {\bibfnamefont {M.~R.}\ \bibnamefont
  {Evans}},\ }\bibfield  {title} {\enquote {\bibinfo {title} {{Glassy
  Time-Scale Divergence and Anomalous Coarsening in a Kinetically Constrained
  Spin Chain}},}\ }\href {\doibase 10.1103/PhysRevLett.83.3238} {\bibfield
  {journal} {\bibinfo  {journal} {Phys. Rev. Lett.}\ }\textbf {\bibinfo
  {volume} {83}},\ \bibinfo {pages} {3238--3241} (\bibinfo {year}
  {1999})}\BibitemShut {NoStop}%
\bibitem [{\citenamefont {Garrahan}\ and\ \citenamefont
  {Chandler}(2002)}]{Garrahan2002}%
  \BibitemOpen
  \bibfield  {author} {\bibinfo {author} {\bibfnamefont {J.~P.}\ \bibnamefont
  {Garrahan}}\ and\ \bibinfo {author} {\bibfnamefont {D.}~\bibnamefont
  {Chandler}},\ }\bibfield  {title} {\enquote {\bibinfo {title} {{Geometrical
  Explanation and Scaling of Dynamical Heterogeneities in Glass Forming
  Systems}},}\ }\href {\doibase 10.1103/PhysRevLett.89.035704} {\bibfield
  {journal} {\bibinfo  {journal} {Phys. Rev. Lett.}\ }\textbf {\bibinfo
  {volume} {89}},\ \bibinfo {pages} {035704} (\bibinfo {year}
  {2002})}\BibitemShut {NoStop}%
\bibitem [{\citenamefont {Sollich}\ and\ \citenamefont
  {Evans}(2003)}]{Sollich2003}%
  \BibitemOpen
  \bibfield  {author} {\bibinfo {author} {\bibfnamefont {P.}~\bibnamefont
  {Sollich}}\ and\ \bibinfo {author} {\bibfnamefont {M.~R.}\ \bibnamefont
  {Evans}},\ }\bibfield  {title} {\enquote {\bibinfo {title} {{Glassy dynamics
  in the asymmetrically constrained kinetic Ising chain}},}\ }\href {\doibase
  10.1103/PhysRevE.68.031504} {\bibfield  {journal} {\bibinfo  {journal} {Phys.
  Rev. E}\ }\textbf {\bibinfo {volume} {68}},\ \bibinfo {pages} {031504}
  (\bibinfo {year} {2003})}\BibitemShut {NoStop}%
\bibitem [{\citenamefont {Binder}\ and\ \citenamefont
  {Kob}(2011)}]{Binder2011}%
  \BibitemOpen
  \bibfield  {author} {\bibinfo {author} {\bibfnamefont {K.}~\bibnamefont
  {Binder}}\ and\ \bibinfo {author} {\bibfnamefont {W.}~\bibnamefont {Kob}},\
  }\href@noop {} {\emph {\bibinfo {title} {{Glassy Materials and Disordered
  Solids: An Introduction to Their Statistical Mechanics (Revised Edition)}}}}\
  (\bibinfo  {publisher} {World Scientific},\ \bibinfo {year}
  {2011})\BibitemShut {NoStop}%
\bibitem [{\citenamefont {Biroli}\ and\ \citenamefont
  {Garrahan}(2013)}]{Biroli2013}%
  \BibitemOpen
  \bibfield  {author} {\bibinfo {author} {\bibfnamefont {G.}~\bibnamefont
  {Biroli}}\ and\ \bibinfo {author} {\bibfnamefont {J.~P.}\ \bibnamefont
  {Garrahan}},\ }\bibfield  {title} {\enquote {\bibinfo {title} {{Perspective:
  The glass transition}},}\ }\href {\doibase 10.1063/1.4795539} {\bibfield
  {journal} {\bibinfo  {journal} {J. Chem. Phys.}\ }\textbf {\bibinfo {volume}
  {138}},\ \bibinfo {eid} {12A301} (\bibinfo {year} {2013})}\BibitemShut
  {NoStop}%
\bibitem [{\citenamefont {Binder}\ and\ \citenamefont
  {Young}(1986)}]{Binder1986}%
  \BibitemOpen
  \bibfield  {author} {\bibinfo {author} {\bibfnamefont {K.}~\bibnamefont
  {Binder}}\ and\ \bibinfo {author} {\bibfnamefont {A.~P.}\ \bibnamefont
  {Young}},\ }\bibfield  {title} {\enquote {\bibinfo {title} {{Spin glasses:
  Experimental facts, theoretical concepts, and open questions}},}\ }\href
  {\doibase 10.1103/RevModPhys.58.801} {\bibfield  {journal} {\bibinfo
  {journal} {Rev. Mod. Phys.}\ }\textbf {\bibinfo {volume} {58}},\ \bibinfo
  {pages} {801--976} (\bibinfo {year} {1986})}\BibitemShut {NoStop}%
\bibitem [{\citenamefont {Olmos}\ \emph {et~al.}(2012)\citenamefont {Olmos},
  \citenamefont {Lesanovsky},\ and\ \citenamefont {Garrahan}}]{Olmos2012}%
  \BibitemOpen
  \bibfield  {author} {\bibinfo {author} {\bibfnamefont {B.}~\bibnamefont
  {Olmos}}, \bibinfo {author} {\bibfnamefont {I.}~\bibnamefont {Lesanovsky}}, \
  and\ \bibinfo {author} {\bibfnamefont {J.~P.}\ \bibnamefont {Garrahan}},\
  }\bibfield  {title} {\enquote {\bibinfo {title} {{Facilitated Spin Models of
  Dissipative Quantum Glasses}},}\ }\href {\doibase
  10.1103/PhysRevLett.109.020403} {\bibfield  {journal} {\bibinfo  {journal}
  {Phys. Rev. Lett.}\ }\textbf {\bibinfo {volume} {109}},\ \bibinfo {pages}
  {020403} (\bibinfo {year} {2012})}\BibitemShut {NoStop}%
\bibitem [{\citenamefont {Lesanovsky}\ \emph {et~al.}(2013)\citenamefont
  {Lesanovsky}, \citenamefont {van Horssen}, \citenamefont {Gu\ifmmode
  \mbox{\c{t}}\else \c{t}\fi{}\ifmmode~\u{a}\else \u{a}\fi{}},\ and\
  \citenamefont {Garrahan}}]{Lesanovsky2013}%
  \BibitemOpen
  \bibfield  {author} {\bibinfo {author} {\bibfnamefont {I.}~\bibnamefont
  {Lesanovsky}}, \bibinfo {author} {\bibfnamefont {M.}~\bibnamefont {van
  Horssen}}, \bibinfo {author} {\bibfnamefont {M.}~\bibnamefont {Gu\ifmmode
  \mbox{\c{t}}\else \c{t}\fi{}\ifmmode~\u{a}\else \u{a}\fi{}}}, \ and\ \bibinfo
  {author} {\bibfnamefont {J.~P.}\ \bibnamefont {Garrahan}},\ }\bibfield
  {title} {\enquote {\bibinfo {title} {{Characterization of Dynamical Phase
  Transitions in Quantum Jump Trajectories Beyond the Properties of the
  Stationary State}},}\ }\href {\doibase 10.1103/PhysRevLett.110.150401}
  {\bibfield  {journal} {\bibinfo  {journal} {Phys. Rev. Lett.}\ }\textbf
  {\bibinfo {volume} {110}},\ \bibinfo {pages} {150401} (\bibinfo {year}
  {2013})}\BibitemShut {NoStop}%
\bibitem [{\citenamefont {Olmos}\ \emph {et~al.}(2014)\citenamefont {Olmos},
  \citenamefont {Lesanovsky},\ and\ \citenamefont {Garrahan}}]{Olmos2014}%
  \BibitemOpen
  \bibfield  {author} {\bibinfo {author} {\bibfnamefont {B.}~\bibnamefont
  {Olmos}}, \bibinfo {author} {\bibfnamefont {I.}~\bibnamefont {Lesanovsky}}, \
  and\ \bibinfo {author} {\bibfnamefont {J.~P.}\ \bibnamefont {Garrahan}},\
  }\bibfield  {title} {\enquote {\bibinfo {title} {{Out-of-equilibrium
  evolution of kinetically constrained many-body quantum systems under purely
  dissipative dynamics}},}\ }\href {\doibase 10.1103/PhysRevE.90.042147}
  {\bibfield  {journal} {\bibinfo  {journal} {Phys. Rev. E}\ }\textbf {\bibinfo
  {volume} {90}},\ \bibinfo {pages} {042147} (\bibinfo {year}
  {2014})}\BibitemShut {NoStop}%
\bibitem [{\citenamefont {Cugliandolo}\ and\ \citenamefont
  {Lozano}(1999)}]{Cugliandolo1999}%
  \BibitemOpen
  \bibfield  {author} {\bibinfo {author} {\bibfnamefont {L.~F.}\ \bibnamefont
  {Cugliandolo}}\ and\ \bibinfo {author} {\bibfnamefont {G.}~\bibnamefont
  {Lozano}},\ }\bibfield  {title} {\enquote {\bibinfo {title} {Real-time
  nonequilibrium dynamics of quantum glassy systems},}\ }\href {\doibase
  10.1103/PhysRevB.59.915} {\bibfield  {journal} {\bibinfo  {journal} {Phys.
  Rev. B}\ }\textbf {\bibinfo {volume} {59}},\ \bibinfo {pages} {915--942}
  (\bibinfo {year} {1999})}\BibitemShut {NoStop}%
\bibitem [{\citenamefont {Rose}\ \emph {et~al.}(2020)\citenamefont {Rose},
  \citenamefont {Macieszczak}, \citenamefont {Lesanovsky},\ and\ \citenamefont
  {Garrahan}}]{Rose2020}%
  \BibitemOpen
  \bibfield  {author} {\bibinfo {author} {\bibfnamefont {D.~C.}\ \bibnamefont
  {Rose}}, \bibinfo {author} {\bibfnamefont {K.}~\bibnamefont {Macieszczak}},
  \bibinfo {author} {\bibfnamefont {I.}~\bibnamefont {Lesanovsky}}, \ and\
  \bibinfo {author} {\bibfnamefont {J.~P.}\ \bibnamefont {Garrahan}},\
  }\href@noop {} {\enquote {\bibinfo {title} {{Hierarchical classical
  metastability in an open quantum East model}},}\ } (\bibinfo {year} {2020}),\
  \Eprint {http://arxiv.org/abs/2010.15304} {arXiv:2010.15304} \BibitemShut
  {NoStop}%
\bibitem [{\citenamefont {Thingna}\ \emph {et~al.}(2016)\citenamefont
  {Thingna}, \citenamefont {Manzano},\ and\ \citenamefont {Cao}}]{Thingna2016}%
  \BibitemOpen
  \bibfield  {author} {\bibinfo {author} {\bibfnamefont {J.}~\bibnamefont
  {Thingna}}, \bibinfo {author} {\bibfnamefont {D.}~\bibnamefont {Manzano}}, \
  and\ \bibinfo {author} {\bibfnamefont {J.}~\bibnamefont {Cao}},\ }\bibfield
  {title} {\enquote {\bibinfo {title} {{Dynamical signatures of molecular
  symmetries in nonequilibrium quantum transport}},}\ }\href {\doibase
  http://dx.doi.org/10.1038/srep28027} {\bibfield  {journal} {\bibinfo
  {journal} {Sci. Rep.}\ }\textbf {\bibinfo {volume} {6}},\ \bibinfo {pages}
  {28027} (\bibinfo {year} {2016})}\BibitemShut {NoStop}%
\bibitem [{\citenamefont {{Lindblad}}(1976)}]{Lindblad1976}%
  \BibitemOpen
  \bibfield  {author} {\bibinfo {author} {\bibfnamefont {G.}~\bibnamefont
  {{Lindblad}}},\ }\bibfield  {title} {\enquote {\bibinfo {title} {{On the
  generators of quantum dynamical semigroups}},}\ }\href {\doibase
  10.1007/BF01608499} {\bibfield  {journal} {\bibinfo  {journal} {Commun. Math.
  Phys.}\ }\textbf {\bibinfo {volume} {48}},\ \bibinfo {pages} {119--130}
  (\bibinfo {year} {1976})}\BibitemShut {NoStop}%
\bibitem [{\citenamefont {Gorini}\ \emph {et~al.}(1976)\citenamefont {Gorini},
  \citenamefont {Kossakowski},\ and\ \citenamefont {Sudarshan}}]{Gorini1976}%
  \BibitemOpen
  \bibfield  {author} {\bibinfo {author} {\bibfnamefont {V.}~\bibnamefont
  {Gorini}}, \bibinfo {author} {\bibfnamefont {A.}~\bibnamefont {Kossakowski}},
  \ and\ \bibinfo {author} {\bibfnamefont {E.~C.~G.}\ \bibnamefont
  {Sudarshan}},\ }\bibfield  {title} {\enquote {\bibinfo {title} {{Completely
  positive dynamical semigroups of N‐level systems}},}\ }\href {\doibase
  10.1063/1.522979} {\bibfield  {journal} {\bibinfo  {journal} {J. Math.
  Phys.}\ }\textbf {\bibinfo {volume} {17}},\ \bibinfo {pages} {821--825}
  (\bibinfo {year} {1976})}\BibitemShut {NoStop}%
\bibitem [{\citenamefont {Gardiner}\ and\ \citenamefont
  {Zoller}(2004)}]{Gardiner2004}%
  \BibitemOpen
  \bibfield  {author} {\bibinfo {author} {\bibfnamefont {C.~W.}\ \bibnamefont
  {Gardiner}}\ and\ \bibinfo {author} {\bibfnamefont {P.}~\bibnamefont
  {Zoller}},\ }\href@noop {} {\emph {\bibinfo {title} {{Quantum Noise}}}},\
  \bibinfo {edition} {3rd}\ ed.,\ Complexity\ (\bibinfo  {publisher}
  {Springer},\ \bibinfo {year} {2004})\BibitemShut {NoStop}%
\bibitem [{\citenamefont {Garrahan}\ and\ \citenamefont
  {Lesanovsky}(2010)}]{Garrahan2010}%
  \BibitemOpen
  \bibfield  {author} {\bibinfo {author} {\bibfnamefont {J.~P.}\ \bibnamefont
  {Garrahan}}\ and\ \bibinfo {author} {\bibfnamefont {I.}~\bibnamefont
  {Lesanovsky}},\ }\bibfield  {title} {\enquote {\bibinfo {title}
  {{Thermodynamics of Quantum Jump Trajectories}},}\ }\href {\doibase
  10.1103/PhysRevLett.104.160601} {\bibfield  {journal} {\bibinfo  {journal}
  {Phys. Rev. Lett.}\ }\textbf {\bibinfo {volume} {104}},\ \bibinfo {pages}
  {160601} (\bibinfo {year} {2010})}\BibitemShut {NoStop}%
\bibitem [{\citenamefont {Baumgartner}\ and\ \citenamefont
  {Narnhofer}(2008)}]{Baumgartner2008}%
  \BibitemOpen
  \bibfield  {author} {\bibinfo {author} {\bibfnamefont {B.}~\bibnamefont
  {Baumgartner}}\ and\ \bibinfo {author} {\bibfnamefont {H.}~\bibnamefont
  {Narnhofer}},\ }\bibfield  {title} {\enquote {\bibinfo {title} {{Analysis of
  quantum semigroups with GKS--Lindblad generators: II. General}},}\ }\href
  {http://stacks.iop.org/1751-8121/41/i=39/a=395303} {\bibfield  {journal}
  {\bibinfo  {journal} {J. Phys. A}\ }\textbf {\bibinfo {volume} {41}},\
  \bibinfo {pages} {395303} (\bibinfo {year} {2008})}\BibitemShut {NoStop}%
\bibitem [{\citenamefont {Bu{\v c}a}\ and\ \citenamefont
  {Prosen}(2012)}]{Buvca2012}%
  \BibitemOpen
  \bibfield  {author} {\bibinfo {author} {\bibfnamefont {B.}~\bibnamefont
  {Bu{\v c}a}}\ and\ \bibinfo {author} {\bibfnamefont {T.}~\bibnamefont
  {Prosen}},\ }\bibfield  {title} {\enquote {\bibinfo {title} {{A note on
  symmetry reductions of the Lindblad equation: transport in constrained open
  spin chains}},}\ }\href {http://stacks.iop.org/1367-2630/14/i=7/a=073007}
  {\bibfield  {journal} {\bibinfo  {journal} {New J. Phys.}\ }\textbf {\bibinfo
  {volume} {14}},\ \bibinfo {pages} {073007} (\bibinfo {year}
  {2012})}\BibitemShut {NoStop}%
\bibitem [{\citenamefont {Albert}\ and\ \citenamefont
  {Jiang}(2014)}]{Albert2014}%
  \BibitemOpen
  \bibfield  {author} {\bibinfo {author} {\bibfnamefont {V.~V.}\ \bibnamefont
  {Albert}}\ and\ \bibinfo {author} {\bibfnamefont {L.}~\bibnamefont {Jiang}},\
  }\bibfield  {title} {\enquote {\bibinfo {title} {{Symmetries and conserved
  quantities in Lindblad master equations}},}\ }\href {\doibase
  10.1103/PhysRevA.89.022118} {\bibfield  {journal} {\bibinfo  {journal} {Phys.
  Rev. A}\ }\textbf {\bibinfo {volume} {89}},\ \bibinfo {pages} {022118}
  (\bibinfo {year} {2014})}\BibitemShut {NoStop}%
\bibitem [{\citenamefont {Garrahan}\ \emph {et~al.}(2011)\citenamefont
  {Garrahan}, \citenamefont {Armour},\ and\ \citenamefont
  {Lesanovsky}}]{Garrahan2011}%
  \BibitemOpen
  \bibfield  {author} {\bibinfo {author} {\bibfnamefont {J.~P.}\ \bibnamefont
  {Garrahan}}, \bibinfo {author} {\bibfnamefont {A.~D.}\ \bibnamefont
  {Armour}}, \ and\ \bibinfo {author} {\bibfnamefont {I.}~\bibnamefont
  {Lesanovsky}},\ }\bibfield  {title} {\enquote {\bibinfo {title} {{Quantum
  trajectory phase transitions in the micromaser}},}\ }\href {\doibase
  10.1103/PhysRevE.84.021115} {\bibfield  {journal} {\bibinfo  {journal} {Phys.
  Rev. E}\ }\textbf {\bibinfo {volume} {84}},\ \bibinfo {pages} {021115}
  (\bibinfo {year} {2011})}\BibitemShut {NoStop}%
\bibitem [{\citenamefont {Carollo}\ \emph {et~al.}(2018)\citenamefont
  {Carollo}, \citenamefont {Garrahan}, \citenamefont {Lesanovsky},\ and\
  \citenamefont {P\'erez-Espigares}}]{Carollo2018}%
  \BibitemOpen
  \bibfield  {author} {\bibinfo {author} {\bibfnamefont {F.}~\bibnamefont
  {Carollo}}, \bibinfo {author} {\bibfnamefont {J.~P.}\ \bibnamefont
  {Garrahan}}, \bibinfo {author} {\bibfnamefont {I.}~\bibnamefont
  {Lesanovsky}}, \ and\ \bibinfo {author} {\bibfnamefont {C.}~\bibnamefont
  {P\'erez-Espigares}},\ }\bibfield  {title} {\enquote {\bibinfo {title}
  {{Making rare events typical in Markovian open quantum systems}},}\ }\href
  {\doibase 10.1103/PhysRevA.98.010103} {\bibfield  {journal} {\bibinfo
  {journal} {Phys. Rev. A}\ }\textbf {\bibinfo {volume} {98}},\ \bibinfo
  {pages} {010103} (\bibinfo {year} {2018})}\BibitemShut {NoStop}%
\bibitem [{Note1()}]{Note1}%
  \BibitemOpen
  \bibinfo {note} {It is possible that the master operator may not be
  completely diagonalizable, in which this case its simplest form is the Jordan
  normal form. Under exponentiation, this gives rise to a polynomial dependence
  on the time, however this is accompanied by the usual exponential evolution.
  Our results in Secs.~\ref {sec:cMM},~\ref {sec:disjoint},~\ref
  {sec:EffDynAv},~\ref {sec:symmetry}, and~\ref {sec:numerics} carry directly
  to that case by assuming that after the initial relaxation, any polynomial
  evolution of fast modes is dominated by the decaying exponential and can be
  neglected [cf.~Eq.~\protect \textup {\hbox {\mathsurround \z@ \protect
  \normalfont (\ignorespaces \ref {Expansion1}\unskip \@@italiccorr )}}], while
  during the metastable regime dynamics of the $m$ slow modes can be replaced
  by no evolution [cf.~Eq.~\protect \textup {\hbox {\mathsurround \z@ \protect
  \normalfont (\ignorespaces \ref {Expansion}\unskip \@@italiccorr )}}]. In
  Sec.~\ref {sec:QTraj}, the corrections to Eqs.~\protect \textup {\hbox
  {\mathsurround \z@ \protect \normalfont (\ignorespaces \ref
  {eq:Kvar_t_approx_CL}\unskip \@@italiccorr )}} and~\protect \textup {\hbox
  {\mathsurround \z@ \protect \normalfont (\ignorespaces \ref
  {eq:kvar_ms}\unskip \@@italiccorr )}} need to be adjusted, together with
  corresponding Secs.~\ref {app:var_t}, \ref {app:var_ms}, and~\ref
  {app:distribution} in the~SM.}\BibitemShut {Stop}%
\bibitem [{\citenamefont {Albert}\ \emph {et~al.}(2016)\citenamefont {Albert},
  \citenamefont {Bradlyn}, \citenamefont {Fraas},\ and\ \citenamefont
  {Jiang}}]{Albert2016}%
  \BibitemOpen
  \bibfield  {author} {\bibinfo {author} {\bibfnamefont {V.~V.}\ \bibnamefont
  {Albert}}, \bibinfo {author} {\bibfnamefont {B.}~\bibnamefont {Bradlyn}},
  \bibinfo {author} {\bibfnamefont {M.}~\bibnamefont {Fraas}}, \ and\ \bibinfo
  {author} {\bibfnamefont {L.}~\bibnamefont {Jiang}},\ }\bibfield  {title}
  {\enquote {\bibinfo {title} {{Geometry and Response of Lindbladians}},}\
  }\href {\doibase 10.1103/PhysRevX.6.041031} {\bibfield  {journal} {\bibinfo
  {journal} {Phys. Rev. X}\ }\textbf {\bibinfo {volume} {6}},\ \bibinfo {pages}
  {041031} (\bibinfo {year} {2016})}\BibitemShut {NoStop}%
\bibitem [{\citenamefont {Bellomo}\ \emph {et~al.}(2017)\citenamefont
  {Bellomo}, \citenamefont {Giorgi}, \citenamefont {Palma},\ and\ \citenamefont
  {Zambrini}}]{Bellomo2017}%
  \BibitemOpen
  \bibfield  {author} {\bibinfo {author} {\bibfnamefont {B.}~\bibnamefont
  {Bellomo}}, \bibinfo {author} {\bibfnamefont {G.~L.}\ \bibnamefont {Giorgi}},
  \bibinfo {author} {\bibfnamefont {G.~M.}\ \bibnamefont {Palma}}, \ and\
  \bibinfo {author} {\bibfnamefont {R.}~\bibnamefont {Zambrini}},\ }\bibfield
  {title} {\enquote {\bibinfo {title} {{Quantum synchronization as a local
  signature of super- and subradiance}},}\ }\href {\doibase
  10.1103/PhysRevA.95.043807} {\bibfield  {journal} {\bibinfo  {journal} {Phys.
  Rev. A}\ }\textbf {\bibinfo {volume} {95}},\ \bibinfo {pages} {043807}
  (\bibinfo {year} {2017})}\BibitemShut {NoStop}%
\bibitem [{Note2()}]{Note2}%
  \BibitemOpen
  \bibinfo {note} {Since ${\protect \mathcal {L}}$ is Hermiticity preserving,
  if $R_k$ ($L_k$) is a right (left) eigenmode of the dynamics with an
  eigenvalue $\lambda _k$, so is $R_k^\dagger $ ($L_k^\dagger $) with the
  eigenvalue $\lambda _k^*$. Thus the set of low-lying modes is invariant under
  the Hermitian conjugation.}\BibitemShut {Stop}%
\bibitem [{\citenamefont {Sciolla}\ \emph {et~al.}(2015)\citenamefont
  {Sciolla}, \citenamefont {Poletti},\ and\ \citenamefont
  {Kollath}}]{Sciolla2015}%
  \BibitemOpen
  \bibfield  {author} {\bibinfo {author} {\bibfnamefont {B.}~\bibnamefont
  {Sciolla}}, \bibinfo {author} {\bibfnamefont {D.}~\bibnamefont {Poletti}}, \
  and\ \bibinfo {author} {\bibfnamefont {C.}~\bibnamefont {Kollath}},\
  }\bibfield  {title} {\enquote {\bibinfo {title} {{Two-Time Correlations
  Probing the Dynamics of Dissipative Many-Body Quantum Systems: Aging and Fast
  Relaxation}},}\ }\href {\doibase 10.1103/PhysRevLett.114.170401} {\bibfield
  {journal} {\bibinfo  {journal} {Phys. Rev. Lett.}\ }\textbf {\bibinfo
  {volume} {114}},\ \bibinfo {pages} {170401} (\bibinfo {year}
  {2015})}\BibitemShut {NoStop}%
\bibitem [{Note3()}]{Note3}%
  \BibitemOpen
  \bibinfo {note} {We restrict the discussion to the real space of Hermitian
  operators. In this case, the induced norm of a superoperator can be shown to
  be achieved for a pure state (i.e., a rank one operator). Let $X$ be a
  Hermitian operator with eigenvalues $x_n$ and projections on the
  corresponding eigenstates denoted as $\rho _n$. For a superoperator $\protect
  \mathcal {Y}$, we have $\delimiter 69645069 \protect \mathcal
  {Y}[X]\delimiter 86422285 /\delimiter 69645069 X\delimiter 86422285 \leq
  (\DOTSB \sum@ \slimits@ _n |x_n| \delimiter 69645069 \protect \mathcal
  {Y}[\rho _n]\delimiter 86422285 ) /(\DOTSB \sum@ \slimits@ _n |x_n|)\leq
  \protect \qopname \relax m{max}_n \delimiter 69645069 \protect \mathcal
  {Y}[\rho _n]\delimiter 86422285 $.}\BibitemShut {Stop}%
\bibitem [{Note4()}]{Note4}%
  \BibitemOpen
  \bibinfo {note} {For an integer $n\geq 1$ such that $t''\leq t_n\equiv t/n
  \leq t'$, we have $\delimiter 69645069 \rho (t)-e^{t {\protect \mathcal
  {L}_\protect \text {MM}}}{\protect \mathcal {P}}[\rho (0)] \delimiter
  86422285 =\delimiter 69645069 [e^{t_n {\protect \mathcal {L}}}-{\protect
  \mathcal {P}}]^n({\protect \mathcal {I}}-{\protect \mathcal {P}})[\rho
  (0)]\delimiter 86422285 \leq \delimiter 69645069 e^{t_n {\protect \mathcal
  {L}}}-{\protect \mathcal {P}}\delimiter 86422285 ^n \delimiter 69645069
  {\protect \mathcal {I}}-{\protect \mathcal {P}}\delimiter 86422285 $. Since
  we have ${\delimiter 69645069 \rho (t_n)-{\protect \mathcal {P}}[\rho
  (0)]\delimiter 86422285 \leq \protect \mathcal {C}_\protect \text
  {MM}(t'',t')}$ [cf.~Eq.~\protect \textup {\hbox {\mathsurround \z@ \protect
  \normalfont (\ignorespaces \ref {eq:Cmm}\unskip \@@italiccorr )}}], while
  $\delimiter 69645069 {\protect \mathcal {I}}-{\protect \mathcal
  {P}}\delimiter 86422285 \leq \delimiter 69645069 {\protect \mathcal
  {I}}\delimiter 86422285 +\delimiter 69645069 {\protect \mathcal
  {P}}\delimiter 86422285 = 2+\protect \mathcal {C}_{+}$ [cf.~Eq.~\protect
  \textup {\hbox {\mathsurround \z@ \protect \normalfont (\ignorespaces \ref
  {eq:Cp}\unskip \@@italiccorr )}}], we arrive at $\delimiter 69645069 \rho
  (t)-e^{t {\protect \mathcal {L}_\protect \text {MM}}}{\protect \mathcal
  {P}}[\rho (0)] \delimiter 86422285 \leq \protect \mathcal {C}_\protect \text
  {MM}^n(t'',t')(2+\protect \mathcal {C}_{+})\lesssim 2\protect \mathcal
  {C}_\protect \text {MM}^n(t'',t')$.}\BibitemShut {Stop}%
\bibitem [{Note5()}]{Note5}%
  \BibitemOpen
  \bibinfo {note} {For an integer $n$ such that $t''+n(t'-t'')\leq t\leq
  t'+n(t'-t'')$, we have $\delimiter 69645069 e^{t{\protect \mathcal
  {L}}}-{\protect \mathcal {P}}\delimiter 86422285 \leq \delimiter 69645069
  ({\protect \mathcal {I}}-{\protect \mathcal {P}})e^{t{\protect \mathcal
  {L}}}\delimiter 86422285 +\delimiter 69645069 {\protect \mathcal
  {P}}e^{t{\protect \mathcal {L}}}-{\protect \mathcal {P}}\delimiter 86422285
  $, where $ \delimiter 69645069 ({\protect \mathcal {I}}-{\protect \mathcal
  {P}})e^{t{\protect \mathcal {L}}}\delimiter 86422285 \leq \delimiter 69645069
  ({\protect \mathcal {I}}-{\protect \mathcal {P}})e^{t''{\protect \mathcal
  {L}}}\delimiter 86422285 \leq (2+\protect \mathcal {C}_{+})\protect \mathcal
  {C}_\protect \text {MM}$ due to $\delimiter 69645069 e^{(t-t''){\protect
  \mathcal {L}}}\delimiter 86422285 =1$, while $\delimiter 69645069 {\protect
  \mathcal {P}}e^{t{\protect \mathcal {L}}}-{\protect \mathcal {P}}\delimiter
  86422285 \leq \DOTSB \sum@ \slimits@ _{n'=1}^n \delimiter 69645069 {\protect
  \mathcal {P}}e^{[t-(n'-1)(t'-t'')]{\protect \mathcal {L}}}- {\protect
  \mathcal {P}}e^{[t-n'(t'-t'')]{\protect \mathcal {L}}}\delimiter 86422285 +
  \delimiter 69645069 {\protect \mathcal {P}}e^{t-n(t'-t''){\protect \mathcal
  {L}}}-{\protect \mathcal {P}}\delimiter 86422285 \leq n \delimiter 69645069
  {\protect \mathcal {P}}[e^{t-n(t'-t''){\protect \mathcal {L}}}-{\protect
  \mathcal {P}}]\delimiter 86422285 \leq (n +1)(1+\protect \mathcal
  {C}_{+})\protect \mathcal {C}_\protect \text {MM}$}\BibitemShut {NoStop}%
\bibitem [{\citenamefont {Zanardi}(1997)}]{Zanardi1997}%
  \BibitemOpen
  \bibfield  {author} {\bibinfo {author} {\bibfnamefont {P.}~\bibnamefont
  {Zanardi}},\ }\bibfield  {title} {\enquote {\bibinfo {title} {{Dissipative
  dynamics in a quantum register}},}\ }\href {\doibase
  10.1103/PhysRevA.56.4445} {\bibfield  {journal} {\bibinfo  {journal} {Phys.
  Rev. A}\ }\textbf {\bibinfo {volume} {56}},\ \bibinfo {pages} {4445--4451}
  (\bibinfo {year} {1997})}\BibitemShut {NoStop}%
\bibitem [{\citenamefont {Zanardi}\ and\ \citenamefont
  {Rasetti}(1997)}]{Zanardi1997a}%
  \BibitemOpen
  \bibfield  {author} {\bibinfo {author} {\bibfnamefont {P.}~\bibnamefont
  {Zanardi}}\ and\ \bibinfo {author} {\bibfnamefont {M.}~\bibnamefont
  {Rasetti}},\ }\bibfield  {title} {\enquote {\bibinfo {title} {{Noiseless
  Quantum Codes}},}\ }\href {\doibase 10.1103/PhysRevLett.79.3306} {\bibfield
  {journal} {\bibinfo  {journal} {Phys. Rev. Lett.}\ }\textbf {\bibinfo
  {volume} {79}},\ \bibinfo {pages} {3306--3309} (\bibinfo {year}
  {1997})}\BibitemShut {NoStop}%
\bibitem [{\citenamefont {Lidar}\ \emph {et~al.}(1998)\citenamefont {Lidar},
  \citenamefont {Chuang},\ and\ \citenamefont {Whaley}}]{Lidar1998}%
  \BibitemOpen
  \bibfield  {author} {\bibinfo {author} {\bibfnamefont {D.~A.}\ \bibnamefont
  {Lidar}}, \bibinfo {author} {\bibfnamefont {I.~L.}\ \bibnamefont {Chuang}}, \
  and\ \bibinfo {author} {\bibfnamefont {K.~B.}\ \bibnamefont {Whaley}},\
  }\bibfield  {title} {\enquote {\bibinfo {title} {{Decoherence-Free Subspaces
  for Quantum Computation}},}\ }\href@noop {} {\bibfield  {journal} {\bibinfo
  {journal} {Phys. Rev. Lett.}\ }\textbf {\bibinfo {volume} {81}},\ \bibinfo
  {pages} {2594--2597} (\bibinfo {year} {1998})}\BibitemShut {NoStop}%
\bibitem [{\citenamefont {Knill}\ \emph {et~al.}(2000)\citenamefont {Knill},
  \citenamefont {Laflamme},\ and\ \citenamefont {Viola}}]{Knill2000}%
  \BibitemOpen
  \bibfield  {author} {\bibinfo {author} {\bibfnamefont {E.}~\bibnamefont
  {Knill}}, \bibinfo {author} {\bibfnamefont {R.}~\bibnamefont {Laflamme}}, \
  and\ \bibinfo {author} {\bibfnamefont {L.}~\bibnamefont {Viola}},\ }\bibfield
   {title} {\enquote {\bibinfo {title} {{Theory of Quantum Error Correction for
  General Noise}},}\ }\href@noop {} {\bibfield  {journal} {\bibinfo  {journal}
  {Phys. Rev. Lett.}\ }\textbf {\bibinfo {volume} {84}},\ \bibinfo {pages}
  {2525--2528} (\bibinfo {year} {2000})}\BibitemShut {NoStop}%
\bibitem [{\citenamefont {Zanardi}(2000)}]{Zanardi2000}%
  \BibitemOpen
  \bibfield  {author} {\bibinfo {author} {\bibfnamefont {P.}~\bibnamefont
  {Zanardi}},\ }\bibfield  {title} {\enquote {\bibinfo {title} {{Stabilizing
  quantum information}},}\ }\href {\doibase 10.1103/PhysRevA.63.012301}
  {\bibfield  {journal} {\bibinfo  {journal} {Phys. Rev. A}\ }\textbf {\bibinfo
  {volume} {63}},\ \bibinfo {pages} {012301} (\bibinfo {year}
  {2000})}\BibitemShut {NoStop}%
\bibitem [{\citenamefont {Zanardi}\ and\ \citenamefont
  {Campos~Venuti}(2014)}]{Zanardi2014}%
  \BibitemOpen
  \bibfield  {author} {\bibinfo {author} {\bibfnamefont {P.}~\bibnamefont
  {Zanardi}}\ and\ \bibinfo {author} {\bibfnamefont {L.}~\bibnamefont
  {Campos~Venuti}},\ }\bibfield  {title} {\enquote {\bibinfo {title} {{Coherent
  Quantum Dynamics in Steady-State Manifolds of Strongly Dissipative
  Systems}},}\ }\href {\doibase 10.1103/PhysRevLett.113.240406} {\bibfield
  {journal} {\bibinfo  {journal} {Phys. Rev. Lett.}\ }\textbf {\bibinfo
  {volume} {113}},\ \bibinfo {pages} {240406} (\bibinfo {year}
  {2014})}\BibitemShut {NoStop}%
\bibitem [{\citenamefont {Zanardi}\ and\ \citenamefont
  {Campos~Venuti}(2015)}]{Zanardi2015}%
  \BibitemOpen
  \bibfield  {author} {\bibinfo {author} {\bibfnamefont {P.}~\bibnamefont
  {Zanardi}}\ and\ \bibinfo {author} {\bibfnamefont {L.}~\bibnamefont
  {Campos~Venuti}},\ }\bibfield  {title} {\enquote {\bibinfo {title}
  {{Geometry, robustness, and emerging unitarity in dissipation-projected
  dynamics}},}\ }\href {\doibase 10.1103/PhysRevA.91.052324} {\bibfield
  {journal} {\bibinfo  {journal} {Phys. Rev. A}\ }\textbf {\bibinfo {volume}
  {91}},\ \bibinfo {pages} {052324} (\bibinfo {year} {2015})}\BibitemShut
  {NoStop}%
\bibitem [{\citenamefont {Zanardi}\ \emph {et~al.}(2016)\citenamefont
  {Zanardi}, \citenamefont {Marshall},\ and\ \citenamefont
  {Campos~Venuti}}]{Zanardi2016}%
  \BibitemOpen
  \bibfield  {author} {\bibinfo {author} {\bibfnamefont {P.}~\bibnamefont
  {Zanardi}}, \bibinfo {author} {\bibfnamefont {J.}~\bibnamefont {Marshall}}, \
  and\ \bibinfo {author} {\bibfnamefont {L.}~\bibnamefont {Campos~Venuti}},\
  }\bibfield  {title} {\enquote {\bibinfo {title} {{Dissipative universal
  Lindbladian simulation}},}\ }\href {\doibase 10.1103/PhysRevA.93.022312}
  {\bibfield  {journal} {\bibinfo  {journal} {Phys. Rev. A}\ }\textbf {\bibinfo
  {volume} {93}},\ \bibinfo {pages} {022312} (\bibinfo {year}
  {2016})}\BibitemShut {NoStop}%
\bibitem [{\citenamefont {Popkov}\ \emph {et~al.}(2018)\citenamefont {Popkov},
  \citenamefont {Essink}, \citenamefont {Presilla},\ and\ \citenamefont
  {Sch\"utz}}]{Popkov2018}%
  \BibitemOpen
  \bibfield  {author} {\bibinfo {author} {\bibfnamefont {V.}~\bibnamefont
  {Popkov}}, \bibinfo {author} {\bibfnamefont {S.}~\bibnamefont {Essink}},
  \bibinfo {author} {\bibfnamefont {C.}~\bibnamefont {Presilla}}, \ and\
  \bibinfo {author} {\bibfnamefont {G.}~\bibnamefont {Sch\"utz}},\ }\bibfield
  {title} {\enquote {\bibinfo {title} {Effective quantum zeno dynamics in
  dissipative quantum systems},}\ }\href {\doibase 10.1103/PhysRevA.98.052110}
  {\bibfield  {journal} {\bibinfo  {journal} {Phys. Rev. A}\ }\textbf {\bibinfo
  {volume} {98}},\ \bibinfo {pages} {052110} (\bibinfo {year}
  {2018})}\BibitemShut {NoStop}%
\bibitem [{Note6()}]{Note6}%
  \BibitemOpen
  \bibinfo {note} {We have $|c_{k}^{(l)} - \DOTSB \sum@ \slimits@ _{l=1}^m p_l
  c_{k}^{(l)}| =|{\protect \mathrm {Tr}}( L_k \protect \{{\protect \mathcal
  {P}}[\rho (0)]- \DOTSB \sum@ \slimits@ _{l=1}^m p_l\rho _l\protect \})|\leq
  \delimiter 69645069 L_k \delimiter 86422285 _{\protect \qopname \relax
  m{max}} \delimiter 69645069 {\protect \mathcal {P}}[\rho (0)]- \DOTSB \sum@
  \slimits@ _{l=1}^m p_l\rho _l\delimiter 86422285 \leq \delimiter 69645069
  {\protect \mathcal {P}}[\rho (0)]- \rho (t)\delimiter 86422285 +\delimiter
  69645069 \rho (t) - \DOTSB \sum@ \slimits@ _{l=1}^m p_l\rho _l\delimiter
  86422285 $, where we used the left-eigenmatrix normalization $\delimiter
  69645069 L_k \delimiter 86422285 _{\protect \qopname \relax m{max}}\leq
  c_k^{\protect \qopname \relax m{max}}-c_k^{\protect \qopname \relax
  m{min}}=1$ with $c_k^{\protect \qopname \relax m{max}}$ and $c_k^{\protect
  \qopname \relax m{min}}$ denoting the maximal and minimal eigenvalue of
  $L_k$. We then choose time $t$ within the metastable regime for which
  $\delimiter 69645069 {\protect \mathcal {P}}[\rho (0)]-\rho (t)\delimiter
  86422285 $ is minimal.}\BibitemShut {Stop}%
\bibitem [{Note7()}]{Note7}%
  \BibitemOpen
  \bibinfo {note} {We have $\protect \qopname \relax m{det}{{\protect \mathbf
  {C}}}=\protect \qopname \relax m{det}{\protect \mathaccentV
  {bar}016{{\protect \mathbf {C}}}}$, where $(\protect \mathaccentV
  {bar}016{{\protect \mathbf {C}}})_{l-1,k-1}=c_{k}^{(l)}-c_{k}^{(1)}$,
  $k,l=2,..,m$, encodes the coefficients for the simplex with the vertex of
  ${\protect \mathaccentV {tilde}07E{\rho }}_1$ shifted to the
  origin.}\BibitemShut {Stop}%
\bibitem [{\citenamefont {\.Zyczkowski}\ and\ \citenamefont
  {Sommers}(2001)}]{Zyczkowski2001}%
  \BibitemOpen
  \bibfield  {author} {\bibinfo {author} {\bibfnamefont {K.}~\bibnamefont
  {\.Zyczkowski}}\ and\ \bibinfo {author} {\bibfnamefont {H.-J.}\ \bibnamefont
  {Sommers}},\ }\bibfield  {title} {\enquote {\bibinfo {title} {{Induced
  measures in the space of mixed quantum states}},}\ }\href
  {http://stacks.iop.org/0305-4470/34/i=35/a=335} {\bibfield  {journal}
  {\bibinfo  {journal} {J. Phys. A}\ }\textbf {\bibinfo {volume} {34}},\
  \bibinfo {pages} {7111} (\bibinfo {year} {2001})}\BibitemShut {NoStop}%
\bibitem [{\citenamefont {\.Zyczkowski}\ and\ \citenamefont
  {Sommers}(2003)}]{Zyczkowski2003}%
  \BibitemOpen
  \bibfield  {author} {\bibinfo {author} {\bibfnamefont {K.}~\bibnamefont
  {\.Zyczkowski}}\ and\ \bibinfo {author} {\bibfnamefont {H.-J.}\ \bibnamefont
  {Sommers}},\ }\bibfield  {title} {\enquote {\bibinfo {title}
  {{Hilbert--Schmidt volume of the set of mixed quantum states}},}\ }\href
  {http://stacks.iop.org/0305-4470/36/i=39/a=310} {\bibfield  {journal}
  {\bibinfo  {journal} {J. Phys. A}\ }\textbf {\bibinfo {volume} {36}},\
  \bibinfo {pages} {10115} (\bibinfo {year} {2003})}\BibitemShut {NoStop}%
\bibitem [{Note8()}]{Note8}%
  \BibitemOpen
  \bibinfo {note} {We note that often the diamond norm, $\delimiter 69645069
  \protect \mathcal {E}-\protect \mathcal {I} \delimiter 86422285 _{\diamond
  }\equiv \delimiter 69645069 \protect \mathcal {E}\otimes \protect \mathcal
  {I} -\protect \mathcal {I}\otimes \protect \mathcal {I}\delimiter 86422285 $,
  is used instead of induced trace norm, $\delimiter 69645069 \protect \mathcal
  {E}-\protect \mathcal {I}\delimiter 86422285 \leq \delimiter 69645069
  \protect \mathcal {E}-\protect \mathcal {I} \delimiter 86422285 _{\diamond
  }$, to quantify the disturbance caused by a quantum channel $\protect
  \mathcal {E}$ with respect to the identity channel $\protect \mathcal
  {I}$.}\BibitemShut {Stop}%
\bibitem [{\citenamefont {Wolf}(2012)}]{Wolf2012}%
  \BibitemOpen
  \bibfield  {author} {\bibinfo {author} {\bibfnamefont {M.~M.}\ \bibnamefont
  {Wolf}},\ }\href
  {https://www-m5.ma.tum.de/foswiki/pub/M5/Allgemeines/MichaelWolf/QChannelLecture.pdf}
  {\emph {\bibinfo {title} {{Quantum channels and Operations, Guided tour}}}}\
  (\bibinfo {year} {2012})\BibitemShut {NoStop}%
\bibitem [{\citenamefont {Nielsen}\ and\ \citenamefont
  {Chuang}(2010)}]{Nielsen2010}%
  \BibitemOpen
  \bibfield  {author} {\bibinfo {author} {\bibfnamefont {M.~A.}\ \bibnamefont
  {Nielsen}}\ and\ \bibinfo {author} {\bibfnamefont {I.~L.}\ \bibnamefont
  {Chuang}},\ }\href@noop {} {\emph {\bibinfo {title} {{Quantum Computation and
  Quantum Information}}}}\ (\bibinfo  {publisher} {Cambridge University Press,
  10th Anniversary ed. edition},\ \bibinfo {year} {2010})\BibitemShut {NoStop}%
\bibitem [{\citenamefont {Gaveau}\ \emph {et~al.}(1999)\citenamefont {Gaveau},
  \citenamefont {Lesne},\ and\ \citenamefont {Schulman}}]{Gaveau1999}%
  \BibitemOpen
  \bibfield  {author} {\bibinfo {author} {\bibfnamefont {B.}~\bibnamefont
  {Gaveau}}, \bibinfo {author} {\bibfnamefont {A.}~\bibnamefont {Lesne}}, \
  and\ \bibinfo {author} {\bibfnamefont {L.~S.}\ \bibnamefont {Schulman}},\
  }\bibfield  {title} {\enquote {\bibinfo {title} {{Spectral signatures of
  hierarchical relaxation}},}\ }\href {\doibase
  http://dx.doi.org/10.1016/S0375-9601(99)00369-2} {\bibfield  {journal}
  {\bibinfo  {journal} {Phys. Lett. A}\ }\textbf {\bibinfo {volume} {258}},\
  \bibinfo {pages} {222 -- 228} (\bibinfo {year} {1999})}\BibitemShut {NoStop}%
\bibitem [{Note9()}]{Note9}%
  \BibitemOpen
  \bibinfo {note} {$\delimiter 69645069 \protect \mathbf {X}\delimiter 86422285
  _1$ is the matrix norm induced by the L1 norm of vectors. For a vector
  $\protect \mathbf {v}$ and a matrix $\protect \mathbf {X}$, we have $
  \delimiter 69645069 \protect \mathbf {X}\protect \mathbf {v}\delimiter
  86422285 _1=\DOTSB \sum@ \slimits@ _{k=1}^m |\DOTSB \sum@ \slimits@ _{l=1}^m
  X_{kl} v_l|\leq \DOTSB \sum@ \slimits@ _{k,l=1}^m |X_{kl} v_l|\leq (\protect
  \qopname \relax m{max}_{1\leq l\leq m}\DOTSB \sum@ \slimits@ _{k=1}^m
  |X_{kl}|) \DOTSB \sum@ \slimits@ _{l=1}^m |v_l| \equiv \delimiter 69645069
  \protect \mathbf {X}\delimiter 86422285 _1 \delimiter 69645069 \protect
  \mathbf {v}\delimiter 86422285 _1 $, and the inequality can be saturated by
  considering basis vectors.}\BibitemShut {Stop}%
\bibitem [{Note10()}]{Note10}%
  \BibitemOpen
  \bibinfo {note} {We note that an analogous approximation of the stationary
  state can be also obtained by considering non-Hermitian perturbation theory
  for $\protect \mathbf {p}_\protect \text {ss}$ with respect to perturbation
  ${\protect \mathbf {W}}-{\protect \mathbf {\protect \mathaccentV
  {tilde}07E{W}}}$ of ${\protect \mathbf {\protect \mathaccentV
  {tilde}07E{W}}}$ (see Sec.~\ref {app:Leff_pss} in the~SM).}\BibitemShut
  {Stop}%
\bibitem [{Note11()}]{Note11}%
  \BibitemOpen
  \bibinfo {note} {Due to the exponential decay of $\delimiter 69645069
  {\protect \mathbf {\protect \mathaccentV {tilde}07E{P}}_\protect \text
  {\protect \tmspace -\thinmuskip {.1667em}\protect \tmspace -\thinmuskip
  {.1667em}ss}}-e^{n\tau {\protect \mathbf {\protect \mathaccentV
  {tilde}07E{W}}}} \delimiter 86422285 _1\leq \delimiter 69645069 {\protect
  \mathbf {\protect \mathaccentV {tilde}07E{P}}_\protect \text {\protect
  \tmspace -\thinmuskip {.1667em}\protect \tmspace -\thinmuskip
  {.1667em}ss}}-e^{\protect \mathaccentV {tilde}07E{\tau }{\protect \mathbf
  {\protect \mathaccentV {tilde}07E{W}}}} \delimiter 86422285 _1^n$ and
  $\delimiter 69645069 {\protect \mathbf {\protect \mathaccentV
  {tilde}07E{P}}_\protect \text {\protect \tmspace -\thinmuskip
  {.1667em}\protect \tmspace -\thinmuskip {.1667em}ss}}-e^{\tau {\protect
  \mathbf {\protect \mathaccentV {tilde}07E{W}}}} \delimiter 86422285 _1\leq
  (1+{\protect \mathaccentV {tilde}07E{\protect \mathcal {C}}_{\protect \text
  {cl}}}/2) \delimiter 69645069 e^{\tau {\protect \mathcal {L}}}-{\protect
  \mathcal {P}}_\protect \text {ss} \delimiter 86422285 $, this will be
  typically implied by Eq.~\protect \textup {\hbox {\mathsurround \z@ \protect
  \normalfont (\ignorespaces \ref {eq:cond_CL}\unskip \@@italiccorr )}};
  cf.~Sec.~\ref {app:tau_defMM}~c in the~SM.}\BibitemShut {Stop}%
\bibitem [{\citenamefont {Smirne}\ \emph {et~al.}(2018)\citenamefont {Smirne},
  \citenamefont {Egloff}, \citenamefont {D{\'{\i}}az}, \citenamefont {Plenio},\
  and\ \citenamefont {Huelga}}]{Smirne2018}%
  \BibitemOpen
  \bibfield  {author} {\bibinfo {author} {\bibfnamefont {A.}~\bibnamefont
  {Smirne}}, \bibinfo {author} {\bibfnamefont {D.}~\bibnamefont {Egloff}},
  \bibinfo {author} {\bibfnamefont {M.~G.}\ \bibnamefont {D{\'{\i}}az}},
  \bibinfo {author} {\bibfnamefont {M.~B.}\ \bibnamefont {Plenio}}, \ and\
  \bibinfo {author} {\bibfnamefont {S.~F.}\ \bibnamefont {Huelga}},\ }\bibfield
   {title} {\enquote {\bibinfo {title} {{Coherence and non-classicality of
  quantum Markov processes}},}\ }\href {\doibase 10.1088/2058-9565/aaebd5}
  {\bibfield  {journal} {\bibinfo  {journal} {Quantum Science and Technology}\
  }\textbf {\bibinfo {volume} {4}},\ \bibinfo {pages} {01LT01} (\bibinfo {year}
  {2018})}\BibitemShut {NoStop}%
\bibitem [{\citenamefont {Kato}(1995)}]{Kato1995}%
  \BibitemOpen
  \bibfield  {author} {\bibinfo {author} {\bibfnamefont {T.}~\bibnamefont
  {Kato}},\ }\href@noop {} {\emph {\bibinfo {title} {{Perturbation Theory for
  Linear Operators}}}}\ (\bibinfo  {publisher} {Springer},\ \bibinfo {year}
  {1995})\BibitemShut {NoStop}%
\bibitem [{\citenamefont {Levitov}\ \emph {et~al.}(1996)\citenamefont
  {Levitov}, \citenamefont {Lee},\ and\ \citenamefont {Lesovik}}]{Levitov1996}%
  \BibitemOpen
  \bibfield  {author} {\bibinfo {author} {\bibfnamefont {L.~S.}\ \bibnamefont
  {Levitov}}, \bibinfo {author} {\bibfnamefont {H.}~\bibnamefont {Lee}}, \ and\
  \bibinfo {author} {\bibfnamefont {G.~B.}\ \bibnamefont {Lesovik}},\
  }\bibfield  {title} {\enquote {\bibinfo {title} {{Electron counting
  statistics and coherent states of electric current}},}\ }\href {\doibase
  10.1063/1.531672} {\bibfield  {journal} {\bibinfo  {journal} {J. Math.
  Phys.}\ }\textbf {\bibinfo {volume} {37}},\ \bibinfo {pages} {4845--4866}
  (\bibinfo {year} {1996})}\BibitemShut {NoStop}%
\bibitem [{\citenamefont {Nazarov}\ and\ \citenamefont
  {Division}(2003)}]{Nazarov2003}%
  \BibitemOpen
  \bibfield  {author} {\bibinfo {author} {\bibfnamefont {Y.~V.}\ \bibnamefont
  {Nazarov}}\ and\ \bibinfo {author} {\bibfnamefont {N.~A. T. O. S.~A.}\
  \bibnamefont {Division}},\ }\href@noop {} {\emph {\bibinfo {title} {{Quantum
  Noise in Mesoscopic Physics, NATO Science Series: Mathematics, Physics and
  Chemistry}}}}\ (\bibinfo  {publisher} {Springer, New York},\ \bibinfo {year}
  {2003})\BibitemShut {NoStop}%
\bibitem [{\citenamefont {Esposito}\ \emph {et~al.}(2009)\citenamefont
  {Esposito}, \citenamefont {Harbola},\ and\ \citenamefont
  {Mukamel}}]{Esposito2009}%
  \BibitemOpen
  \bibfield  {author} {\bibinfo {author} {\bibfnamefont {M.}~\bibnamefont
  {Esposito}}, \bibinfo {author} {\bibfnamefont {U.}~\bibnamefont {Harbola}}, \
  and\ \bibinfo {author} {\bibfnamefont {S.}~\bibnamefont {Mukamel}},\
  }\bibfield  {title} {\enquote {\bibinfo {title} {{Nonequilibrium
  fluctuations, fluctuation theorems, and counting statistics in quantum
  systems}},}\ }\href {\doibase 10.1103/RevModPhys.81.1665} {\bibfield
  {journal} {\bibinfo  {journal} {Rev. Mod. Phys.}\ }\textbf {\bibinfo {volume}
  {81}},\ \bibinfo {pages} {1665--1702} (\bibinfo {year} {2009})}\BibitemShut
  {NoStop}%
\bibitem [{\citenamefont {Flindt}\ \emph {et~al.}(2009)\citenamefont {Flindt},
  \citenamefont {Fricke}, \citenamefont {Hohls}, \citenamefont {Novotn{\'y}},
  \citenamefont {Neto{\v c}n{\'y}}, \citenamefont {Brandes},\ and\
  \citenamefont {Haug}}]{Flindt2009}%
  \BibitemOpen
  \bibfield  {author} {\bibinfo {author} {\bibfnamefont {C.}~\bibnamefont
  {Flindt}}, \bibinfo {author} {\bibfnamefont {C.}~\bibnamefont {Fricke}},
  \bibinfo {author} {\bibfnamefont {F.}~\bibnamefont {Hohls}}, \bibinfo
  {author} {\bibfnamefont {T.}~\bibnamefont {Novotn{\'y}}}, \bibinfo {author}
  {\bibfnamefont {K.}~\bibnamefont {Neto{\v c}n{\'y}}}, \bibinfo {author}
  {\bibfnamefont {T.}~\bibnamefont {Brandes}}, \ and\ \bibinfo {author}
  {\bibfnamefont {R.~J.}\ \bibnamefont {Haug}},\ }\bibfield  {title} {\enquote
  {\bibinfo {title} {{Universal oscillations in counting statistics}},}\ }\href
  {\doibase 10.1073/pnas.0901002106} {\bibfield  {journal} {\bibinfo  {journal}
  {Proc. Natl. Acad. Sci.}\ }\textbf {\bibinfo {volume} {106}},\ \bibinfo
  {pages} {10116--10119} (\bibinfo {year} {2009})}\BibitemShut {NoStop}%
\bibitem [{\citenamefont {Hickey}\ \emph {et~al.}(2012)\citenamefont {Hickey},
  \citenamefont {Genway}, \citenamefont {Lesanovsky},\ and\ \citenamefont
  {Garrahan}}]{Hickey2012}%
  \BibitemOpen
  \bibfield  {author} {\bibinfo {author} {\bibfnamefont {J.~M.}\ \bibnamefont
  {Hickey}}, \bibinfo {author} {\bibfnamefont {S.}~\bibnamefont {Genway}},
  \bibinfo {author} {\bibfnamefont {I.}~\bibnamefont {Lesanovsky}}, \ and\
  \bibinfo {author} {\bibfnamefont {J.~P.}\ \bibnamefont {Garrahan}},\
  }\bibfield  {title} {\enquote {\bibinfo {title} {{Thermodynamics of
  quadrature trajectories in open quantum systems}},}\ }\href {\doibase
  10.1103/PhysRevA.86.063824} {\bibfield  {journal} {\bibinfo  {journal} {Phys.
  Rev. A}\ }\textbf {\bibinfo {volume} {86}},\ \bibinfo {pages} {063824}
  (\bibinfo {year} {2012})}\BibitemShut {NoStop}%
\bibitem [{\citenamefont {Hickey}\ \emph {et~al.}(2013)\citenamefont {Hickey},
  \citenamefont {Genway}, \citenamefont {Lesanovsky},\ and\ \citenamefont
  {Garrahan}}]{Hickey2013}%
  \BibitemOpen
  \bibfield  {author} {\bibinfo {author} {\bibfnamefont {J.~M.}\ \bibnamefont
  {Hickey}}, \bibinfo {author} {\bibfnamefont {S.}~\bibnamefont {Genway}},
  \bibinfo {author} {\bibfnamefont {I.}~\bibnamefont {Lesanovsky}}, \ and\
  \bibinfo {author} {\bibfnamefont {J.~P.}\ \bibnamefont {Garrahan}},\
  }\bibfield  {title} {\enquote {\bibinfo {title} {{Time-integrated observables
  as order parameters for full counting statistics transitions in closed
  quantum systems}},}\ }\href {\doibase 10.1103/PhysRevB.87.184303} {\bibfield
  {journal} {\bibinfo  {journal} {Phys. Rev. B}\ }\textbf {\bibinfo {volume}
  {87}},\ \bibinfo {pages} {184303} (\bibinfo {year} {2013})}\BibitemShut
  {NoStop}%
\bibitem [{\citenamefont {Lecomte}\ \emph {et~al.}(2007)\citenamefont
  {Lecomte}, \citenamefont {Appert-Rolland},\ and\ \citenamefont {van
  Wijland}}]{Lecomte2007}%
  \BibitemOpen
  \bibfield  {author} {\bibinfo {author} {\bibfnamefont {V.}~\bibnamefont
  {Lecomte}}, \bibinfo {author} {\bibfnamefont {C.}~\bibnamefont
  {Appert-Rolland}}, \ and\ \bibinfo {author} {\bibfnamefont {F.}~\bibnamefont
  {van Wijland}},\ }\bibfield  {title} {\enquote {\bibinfo {title}
  {{Thermodynamic formalism for systems with Markov dynamics}},}\ }\href@noop
  {} {\bibfield  {journal} {\bibinfo  {journal} {J. Stat. Phys.}\ }\textbf
  {\bibinfo {volume} {127}},\ \bibinfo {pages} {51} (\bibinfo {year}
  {2007})}\BibitemShut {NoStop}%
\bibitem [{\citenamefont {Garrahan}\ \emph {et~al.}(2007)\citenamefont
  {Garrahan}, \citenamefont {Jack}, \citenamefont {Lecomte}, \citenamefont
  {Pitard}, \citenamefont {van Duijvendijk},\ and\ \citenamefont {van
  Wijland}}]{Garrahan2007}%
  \BibitemOpen
  \bibfield  {author} {\bibinfo {author} {\bibfnamefont {J.~P.}\ \bibnamefont
  {Garrahan}}, \bibinfo {author} {\bibfnamefont {R.~L.}\ \bibnamefont {Jack}},
  \bibinfo {author} {\bibfnamefont {V.}~\bibnamefont {Lecomte}}, \bibinfo
  {author} {\bibfnamefont {E.}~\bibnamefont {Pitard}}, \bibinfo {author}
  {\bibfnamefont {K.}~\bibnamefont {van Duijvendijk}}, \ and\ \bibinfo {author}
  {\bibfnamefont {F.}~\bibnamefont {van Wijland}},\ }\bibfield  {title}
  {\enquote {\bibinfo {title} {{Dynamical First-Order Phase Transition in
  Kinetically Constrained Models of Glasses}},}\ }\href@noop {} {\bibfield
  {journal} {\bibinfo  {journal} {Phys. Rev. Lett.}\ }\textbf {\bibinfo
  {volume} {98}},\ \bibinfo {pages} {195702} (\bibinfo {year}
  {2007})}\BibitemShut {NoStop}%
\bibitem [{\citenamefont {Maes}(2020)}]{Maes2020}%
  \BibitemOpen
  \bibfield  {author} {\bibinfo {author} {\bibfnamefont {C.}~\bibnamefont
  {Maes}},\ }\bibfield  {title} {\enquote {\bibinfo {title} {{Frenesy:
  Time-symmetric dynamical activity in nonequilibria}},}\ }\href {\doibase
  https://doi.org/10.1016/j.physrep.2020.01.002} {\bibfield  {journal}
  {\bibinfo  {journal} {Phys. Rep.}\ }\textbf {\bibinfo {volume} {850}},\
  \bibinfo {pages} {1 -- 33} (\bibinfo {year} {2020})}\BibitemShut {NoStop}%
\bibitem [{\citenamefont {Garrahan}(2018)}]{Garrahan2018}%
  \BibitemOpen
  \bibfield  {author} {\bibinfo {author} {\bibfnamefont {J.~P.}\ \bibnamefont
  {Garrahan}},\ }\bibfield  {title} {\enquote {\bibinfo {title} {{Aspects of
  non-equilibrium in classical and quantum systems: Slow relaxation and
  glasses, dynamical large deviations, quantum non-ergodicity, and open quantum
  dynamics}},}\ }\href {\doibase 10.1016/j.physa.2017.12.149} {\bibfield
  {journal} {\bibinfo  {journal} {Physica A}\ }\textbf {\bibinfo {volume}
  {504}},\ \bibinfo {pages} {130--154} (\bibinfo {year} {2018})}\BibitemShut
  {NoStop}%
\bibitem [{\citenamefont {Jack}(2020)}]{Jack2019}%
  \BibitemOpen
  \bibfield  {author} {\bibinfo {author} {\bibfnamefont {R.~L.}\ \bibnamefont
  {Jack}},\ }\bibfield  {title} {\enquote {\bibinfo {title} {{Ergodicity and
  large deviations in physical systems with stochastic dynamics}},}\ }\href
  {\doibase 10.1140/epjb/e2020-100605-3} {\bibfield  {journal} {\bibinfo
  {journal} {Eur. Phys. J. B}\ }\textbf {\bibinfo {volume} {93}},\ \bibinfo
  {pages} {74} (\bibinfo {year} {2020})}\BibitemShut {NoStop}%
\bibitem [{Note12()}]{Note12}%
  \BibitemOpen
  \bibinfo {note} {$H$ can be further replaced by $H-c\protect \mathds {1}$,
  with $c$ being a real constant and the norm minimized with respect to $c$ in
  the corrections; see Sec.~\ref {app:Ws} in the~SM.}\BibitemShut {Stop}%
\bibitem [{\citenamefont {Gammelmark}\ and\ \citenamefont
  {M\o{}lmer}(2014)}]{Gammelmark2014}%
  \BibitemOpen
  \bibfield  {author} {\bibinfo {author} {\bibfnamefont {S.}~\bibnamefont
  {Gammelmark}}\ and\ \bibinfo {author} {\bibfnamefont {K.}~\bibnamefont
  {M\o{}lmer}},\ }\bibfield  {title} {\enquote {\bibinfo {title} {{Fisher
  Information and the Quantum Cram\'er-Rao Sensitivity Limit of Continuous
  Measurements}},}\ }\href {\doibase 10.1103/PhysRevLett.112.170401} {\bibfield
   {journal} {\bibinfo  {journal} {Phys. Rev. Lett.}\ }\textbf {\bibinfo
  {volume} {112}},\ \bibinfo {pages} {170401} (\bibinfo {year}
  {2014})}\BibitemShut {NoStop}%
\bibitem [{\citenamefont {Macieszczak}\ \emph
  {et~al.}(2016{\natexlab{b}})\citenamefont {Macieszczak}, \citenamefont
  {Gu\ifmmode \mbox{\c{t}}\else \c{t}\fi{}\ifmmode~\u{a}\else \u{a}\fi{}},
  \citenamefont {Lesanovsky},\ and\ \citenamefont
  {Garrahan}}]{Macieszczak2016}%
  \BibitemOpen
  \bibfield  {author} {\bibinfo {author} {\bibfnamefont {K.}~\bibnamefont
  {Macieszczak}}, \bibinfo {author} {\bibfnamefont {M.}~\bibnamefont
  {Gu\ifmmode \mbox{\c{t}}\else \c{t}\fi{}\ifmmode~\u{a}\else \u{a}\fi{}}},
  \bibinfo {author} {\bibfnamefont {I.}~\bibnamefont {Lesanovsky}}, \ and\
  \bibinfo {author} {\bibfnamefont {J.~P.}\ \bibnamefont {Garrahan}},\
  }\bibfield  {title} {\enquote {\bibinfo {title} {{Dynamical phase transitions
  as a resource for quantum enhanced metrology}},}\ }\href {\doibase
  10.1103/PhysRevA.93.022103} {\bibfield  {journal} {\bibinfo  {journal} {Phys.
  Rev. A}\ }\textbf {\bibinfo {volume} {93}},\ \bibinfo {pages} {022103}
  (\bibinfo {year} {2016}{\natexlab{b}})}\BibitemShut {NoStop}%
\bibitem [{Note13()}]{Note13}%
  \BibitemOpen
  \bibinfo {note} {In Eq.~\protect \textup {\hbox {\mathsurround \z@ \protect
  \normalfont (\ignorespaces \ref {eq:UP}\unskip \@@italiccorr )}} we require
  $n'\protect \mathcal {C}_{\protect \text {cl}}\ll 1$ for all prime factors
  $n'$ of $n$ (see Sec.~\ref {app:symmetry2_inv} in the~SM).}\BibitemShut
  {Stop}%
\bibitem [{\citenamefont {Gough}\ \emph {et~al.}(2015)\citenamefont {Gough},
  \citenamefont {Ratiu},\ and\ \citenamefont {Smolyanov}}]{Gough2015}%
  \BibitemOpen
  \bibfield  {author} {\bibinfo {author} {\bibfnamefont {J.~E.}\ \bibnamefont
  {Gough}}, \bibinfo {author} {\bibfnamefont {T.~S.}\ \bibnamefont {Ratiu}}, \
  and\ \bibinfo {author} {\bibfnamefont {O.~G.}\ \bibnamefont {Smolyanov}},\
  }\bibfield  {title} {\enquote {\bibinfo {title} {{Noether’s theorem for
  dissipative quantum dynamical semi-groups}},}\ }\href {\doibase
  10.1063/1.4907985} {\bibfield  {journal} {\bibinfo  {journal} {J. Math.
  Phys.}\ }\textbf {\bibinfo {volume} {56}},\ \bibinfo {pages} {022108}
  (\bibinfo {year} {2015})}\BibitemShut {NoStop}%
\bibitem [{Note14()}]{Note14}%
  \BibitemOpen
  \bibinfo {note} {When momenta of plane waves over candidate state cycles do
  not correspond to arguments of symmetry eigenvalues for the low-lying
  eigenmodes, ${\protect \mathbf {C}}_{\protect \mathbf {U}}$ is not invertible
  and $\protect \tmspace -\thinmuskip {.1667em}\protect \qopname \relax
  m{det}{\protect \mathbf {C}}_{\protect \mathbf {U}}=0$.}\BibitemShut {Stop}%
\bibitem [{Note15()}]{Note15}%
  \BibitemOpen
  \bibinfo {note} {This follows from the spectrum of any operator unchanged
  under the action of the symmetry, and ${\protect \mathcal {U}}[\protect
  \mathaccentV {tilde}07E{P}'_{\pi ^n(l)}]=\protect \mathaccentV
  {tilde}07E{P}'_{\pi ^{n+1}(l)}$. Alternatively, the minimal eigenvalue can be
  found from the wave-plane basis by considering averages of $\DOTSB \sum@
  \slimits@ _{n=0}^{d_l-1} (e^{-i 2\pi \protect \frac {k}{d}})^n {\protect
  \mathcal {U}}^{\dagger n}(|\psi {\delimiter "526930B \delimiter "426830A
  }\psi |)$ maximized over ${|\psi \delimiter "526930B }$, which can be
  restricted to $U^{d_l}$ symmetric states [cf.~Eq.~\protect \textup {\hbox
  {\mathsurround \z@ \protect \normalfont (\ignorespaces \ref {eq:Lk_U}\unskip
  \@@italiccorr )}}].}\BibitemShut {Stop}%
\bibitem [{\citenamefont {Blume-Kohout}\ \emph {et~al.}(2008)\citenamefont
  {Blume-Kohout}, \citenamefont {Ng}, \citenamefont {Poulin},\ and\
  \citenamefont {Viola}}]{Blume2008}%
  \BibitemOpen
  \bibfield  {author} {\bibinfo {author} {\bibfnamefont {R.}~\bibnamefont
  {Blume-Kohout}}, \bibinfo {author} {\bibfnamefont {H.~K.}\ \bibnamefont
  {Ng}}, \bibinfo {author} {\bibfnamefont {D.}~\bibnamefont {Poulin}}, \ and\
  \bibinfo {author} {\bibfnamefont {L.}~\bibnamefont {Viola}},\ }\bibfield
  {title} {\enquote {\bibinfo {title} {{Characterizing the Structure of
  Preserved Information in Quantum Processes}},}\ }\href {\doibase
  10.1103/PhysRevLett.100.030501} {\bibfield  {journal} {\bibinfo  {journal}
  {Phys. Rev. Lett.}\ }\textbf {\bibinfo {volume} {100}},\ \bibinfo {pages}
  {030501} (\bibinfo {year} {2008})}\BibitemShut {NoStop}%
\bibitem [{\citenamefont {Blume-Kohout}\ \emph {et~al.}(2010)\citenamefont
  {Blume-Kohout}, \citenamefont {Ng}, \citenamefont {Poulin},\ and\
  \citenamefont {Viola}}]{Blume2010}%
  \BibitemOpen
  \bibfield  {author} {\bibinfo {author} {\bibfnamefont {R.}~\bibnamefont
  {Blume-Kohout}}, \bibinfo {author} {\bibfnamefont {H.~K.}\ \bibnamefont
  {Ng}}, \bibinfo {author} {\bibfnamefont {D.}~\bibnamefont {Poulin}}, \ and\
  \bibinfo {author} {\bibfnamefont {L.}~\bibnamefont {Viola}},\ }\bibfield
  {title} {\enquote {\bibinfo {title} {{Information-preserving structures: A
  general framework for quantum zero-error information}},}\ }\href {\doibase
  10.1103/PhysRevA.82.062306} {\bibfield  {journal} {\bibinfo  {journal} {Phys.
  Rev. A}\ }\textbf {\bibinfo {volume} {82}},\ \bibinfo {pages} {062306}
  (\bibinfo {year} {2010})}\BibitemShut {NoStop}%
\bibitem [{\citenamefont {Holbrook}\ \emph {et~al.}(2004)\citenamefont
  {Holbrook}, \citenamefont {Kribs},\ and\ \citenamefont
  {Laflamme}}]{Holbrook2004}%
  \BibitemOpen
  \bibfield  {author} {\bibinfo {author} {\bibfnamefont {J.~A.}\ \bibnamefont
  {Holbrook}}, \bibinfo {author} {\bibfnamefont {D.~W.}\ \bibnamefont {Kribs}},
  \ and\ \bibinfo {author} {\bibfnamefont {R.}~\bibnamefont {Laflamme}},\
  }\bibfield  {title} {\enquote {\bibinfo {title} {{Noiseless subsystems and
  the structure of the commutant in quantum error correction}},}\ }\href@noop
  {} {\bibfield  {journal} {\bibinfo  {journal} {Quant. Inf. Proc.}\ }\textbf
  {\bibinfo {volume} {2}},\ \bibinfo {pages} {381--419} (\bibinfo {year}
  {2004})}\BibitemShut {NoStop}%
\bibitem [{\citenamefont {Choi}\ and\ \citenamefont {Kribs}(2006)}]{Choi2006}%
  \BibitemOpen
  \bibfield  {author} {\bibinfo {author} {\bibfnamefont {M.-D.}\ \bibnamefont
  {Choi}}\ and\ \bibinfo {author} {\bibfnamefont {D.~W.}\ \bibnamefont
  {Kribs}},\ }\bibfield  {title} {\enquote {\bibinfo {title} {{Method to Find
  Quantum Noiseless Subsystems}},}\ }\href {\doibase
  10.1103/PhysRevLett.96.050501} {\bibfield  {journal} {\bibinfo  {journal}
  {Phys. Rev. Lett.}\ }\textbf {\bibinfo {volume} {96}},\ \bibinfo {pages}
  {050501} (\bibinfo {year} {2006})}\BibitemShut {NoStop}%
\bibitem [{\citenamefont {Żurek}(1993)}]{Zurek1993}%
  \BibitemOpen
  \bibfield  {author} {\bibinfo {author} {\bibfnamefont {W.~H.}\ \bibnamefont
  {Żurek}},\ }\bibfield  {title} {\enquote {\bibinfo {title} {{Preferred
  States, Predictability, Classicality and the Environment-Induced
  Decoherence}},}\ }\href@noop {} {\bibfield  {journal} {\bibinfo  {journal}
  {Progress of Theoretical Physics}\ }\textbf {\bibinfo {volume} {89}},\
  \bibinfo {pages} {281--312} (\bibinfo {year} {1993})}\BibitemShut {NoStop}%
\bibitem [{\citenamefont {Bengtsson}\ and\ \citenamefont
  {\.Zyczkowski}(2006)}]{Bengtsson2006}%
  \BibitemOpen
  \bibfield  {author} {\bibinfo {author} {\bibfnamefont {C.~I.}\ \bibnamefont
  {Bengtsson}}\ and\ \bibinfo {author} {\bibfnamefont {K.}~\bibnamefont
  {\.Zyczkowski}},\ }\href@noop {} {\emph {\bibinfo {title} {{Geometry of
  Quantum States: An Introduction to Quantum Entanglement}}}}\ (\bibinfo
  {publisher} {Cambridge University Press},\ \bibinfo {year}
  {2006})\BibitemShut {NoStop}%
\bibitem [{Note16()}]{Note16}%
  \BibitemOpen
  \bibinfo {note} {For $n_k$ not divisible by $4$, otherwise see Sec.~\ref
  {app:algorithm_symmetry3} in the~SM.}\BibitemShut {Stop}%
\bibitem [{\citenamefont {Touchette}(2009)}]{Touchette2009}%
  \BibitemOpen
  \bibfield  {author} {\bibinfo {author} {\bibfnamefont {H.}~\bibnamefont
  {Touchette}},\ }\bibfield  {title} {\enquote {\bibinfo {title} {{The large
  deviation approach to statistical mechanics}},}\ }\href {\doibase
  http://dx.doi.org/10.1016/j.physrep.2009.05.002} {\bibfield  {journal}
  {\bibinfo  {journal} {Phys. Rep.}\ }\textbf {\bibinfo {volume} {478}},\
  \bibinfo {pages} {1 -- 69} (\bibinfo {year} {2009})}\BibitemShut {NoStop}%
\bibitem [{\citenamefont {Hedges}\ \emph {et~al.}(2009)\citenamefont {Hedges},
  \citenamefont {Jack}, \citenamefont {Garrahan},\ and\ \citenamefont
  {Chandler}}]{Hedges2009}%
  \BibitemOpen
  \bibfield  {author} {\bibinfo {author} {\bibfnamefont {L.~O.}\ \bibnamefont
  {Hedges}}, \bibinfo {author} {\bibfnamefont {R.~L.}\ \bibnamefont {Jack}},
  \bibinfo {author} {\bibfnamefont {J.~P.}\ \bibnamefont {Garrahan}}, \ and\
  \bibinfo {author} {\bibfnamefont {D.}~\bibnamefont {Chandler}},\ }\bibfield
  {title} {\enquote {\bibinfo {title} {{Dynamic Order-Disorder in Atomistic
  Models of Structural Glass Formers}},}\ }\href {\doibase
  10.1126/science.1166665} {\bibfield  {journal} {\bibinfo  {journal}
  {Science}\ }\textbf {\bibinfo {volume} {323}},\ \bibinfo {pages} {1309--1313}
  (\bibinfo {year} {2009})}\BibitemShut {NoStop}%
\bibitem [{\citenamefont {Giardina}\ \emph {et~al.}(2011)\citenamefont
  {Giardina}, \citenamefont {Kurchan}, \citenamefont {Lecomte},\ and\
  \citenamefont {Tailleur}}]{Giardina2011}%
  \BibitemOpen
  \bibfield  {author} {\bibinfo {author} {\bibfnamefont {C.}~\bibnamefont
  {Giardina}}, \bibinfo {author} {\bibfnamefont {J.}~\bibnamefont {Kurchan}},
  \bibinfo {author} {\bibfnamefont {V.}~\bibnamefont {Lecomte}}, \ and\
  \bibinfo {author} {\bibfnamefont {J.}~\bibnamefont {Tailleur}},\ }\bibfield
  {title} {\enquote {\bibinfo {title} {{Simulating Rare Events in Dynamical
  Processes}},}\ }\href {\doibase 10.1007/s10955-011-0350-4} {\bibfield
  {journal} {\bibinfo  {journal} {J. Stat. Phys.}\ }\textbf {\bibinfo {volume}
  {145}},\ \bibinfo {pages} {787--811} (\bibinfo {year} {2011})}\BibitemShut
  {NoStop}%
\bibitem [{Note17()}]{Note17}%
  \BibitemOpen
  \bibinfo {note} {The timescale $\tau (s)$ of the relaxation toward ${\rho
  _\protect \text {ss}}(s)$ in the biased dynamics $e^{t{\protect \mathcal
  {L}}_s}[\rho (0)]/{\protect \mathrm {Tr}}\protect \{e^{t{\protect \mathcal
  {L}}_s}[\rho (0)]\protect \}$ belongs to the metastable regime when
  Eqs.~\protect \textup {\hbox {\mathsurround \z@ \protect \normalfont
  (\ignorespaces \ref {eq:theta_s2}\unskip \@@italiccorr )}} and~\protect
  \textup {\hbox {\mathsurround \z@ \protect \normalfont (\ignorespaces \ref
  {eq:k_s2}\unskip \@@italiccorr )}} hold. Indeed, due to the separation also
  present in the spectrum of ${\protect \mathcal {L}}_s$ $\tau (s)$ is
  approximated by the relaxation time of ${\protect \mathbf {W}}_{h_s}$ of
  Eq.~\protect \textup {\hbox {\mathsurround \z@ \protect \normalfont
  (\ignorespaces \ref {eq:Wstilde3}\unskip \@@italiccorr )}} and negligible
  corrections from fast modes of ${\protect \mathcal {L}}$ give $\tau (s)\geq
  t''$. Moreover, as ${\protect \mathbf {W}}$ can be neglected in Eq.~\protect
  \textup {\hbox {\mathsurround \z@ \protect \normalfont (\ignorespaces \ref
  {eq:Wstilde3}\unskip \@@italiccorr )}}], $\tau (s)\leq t'$.}\BibitemShut
  {Stop}%
\bibitem [{\citenamefont {Burkey}\ and\ \citenamefont
  {Cantrell}(1984)}]{Burkey1984}%
  \BibitemOpen
  \bibfield  {author} {\bibinfo {author} {\bibfnamefont {R.~S.}\ \bibnamefont
  {Burkey}}\ and\ \bibinfo {author} {\bibfnamefont {C.~D.}\ \bibnamefont
  {Cantrell}},\ }\bibfield  {title} {\enquote {\bibinfo {title} {Discretization
  in the quasi-continuum},}\ }\href {\doibase 10.1364/JOSAB.1.000169}
  {\bibfield  {journal} {\bibinfo  {journal} {J. Opt. Soc. Am. B}\ }\textbf
  {\bibinfo {volume} {1}},\ \bibinfo {pages} {169--175} (\bibinfo {year}
  {1984})}\BibitemShut {NoStop}%
\bibitem [{\citenamefont {Chin}\ \emph {et~al.}(2010)\citenamefont {Chin},
  \citenamefont {Rivas}, \citenamefont {Huelga},\ and\ \citenamefont
  {Plenio}}]{Chin2010}%
  \BibitemOpen
  \bibfield  {author} {\bibinfo {author} {\bibfnamefont {A.~W.}\ \bibnamefont
  {Chin}}, \bibinfo {author} {\bibfnamefont {A.}~\bibnamefont {Rivas}},
  \bibinfo {author} {\bibfnamefont {S.~F.}\ \bibnamefont {Huelga}}, \ and\
  \bibinfo {author} {\bibfnamefont {M.~B.}\ \bibnamefont {Plenio}},\ }\bibfield
   {title} {\enquote {\bibinfo {title} {Exact mapping between system-reservoir
  quantum models and semi-infinite discrete chains using orthogonal
  polynomials},}\ }\href {\doibase 10.1063/1.3490188} {\bibfield  {journal}
  {\bibinfo  {journal} {J. Math. Phys.}\ }\textbf {\bibinfo {volume} {51}},\
  \bibinfo {pages} {092109} (\bibinfo {year} {2010})}\BibitemShut {NoStop}%
\bibitem [{\citenamefont {Woods}\ \emph {et~al.}(2014)\citenamefont {Woods},
  \citenamefont {Groux}, \citenamefont {Chin}, \citenamefont {Huelga},\ and\
  \citenamefont {Plenio}}]{Woods2014}%
  \BibitemOpen
  \bibfield  {author} {\bibinfo {author} {\bibfnamefont {M.~P.}\ \bibnamefont
  {Woods}}, \bibinfo {author} {\bibfnamefont {R.}~\bibnamefont {Groux}},
  \bibinfo {author} {\bibfnamefont {A.~W.}\ \bibnamefont {Chin}}, \bibinfo
  {author} {\bibfnamefont {S.~F.}\ \bibnamefont {Huelga}}, \ and\ \bibinfo
  {author} {\bibfnamefont {M.~B.}\ \bibnamefont {Plenio}},\ }\bibfield  {title}
  {\enquote {\bibinfo {title} {{Mappings of open quantum systems onto chain
  representations and Markovian embeddings}},}\ }\href {\doibase
  10.1063/1.4866769} {\bibfield  {journal} {\bibinfo  {journal} {J. Math.
  Phys.}\ }\textbf {\bibinfo {volume} {55}},\ \bibinfo {pages} {032101}
  (\bibinfo {year} {2014})}\BibitemShut {NoStop}%
\bibitem [{\citenamefont {Iles-Smith}\ \emph {et~al.}(2014)\citenamefont
  {Iles-Smith}, \citenamefont {Lambert},\ and\ \citenamefont
  {Nazir}}]{Iles2014}%
  \BibitemOpen
  \bibfield  {author} {\bibinfo {author} {\bibfnamefont {J.}~\bibnamefont
  {Iles-Smith}}, \bibinfo {author} {\bibfnamefont {N.}~\bibnamefont {Lambert}},
  \ and\ \bibinfo {author} {\bibfnamefont {A.}~\bibnamefont {Nazir}},\
  }\bibfield  {title} {\enquote {\bibinfo {title} {{Environmental dynamics,
  correlations, and the emergence of noncanonical equilibrium states in open
  quantum systems}},}\ }\href {\doibase 10.1103/PhysRevA.90.032114} {\bibfield
  {journal} {\bibinfo  {journal} {Phys. Rev. A}\ }\textbf {\bibinfo {volume}
  {90}},\ \bibinfo {pages} {032114} (\bibinfo {year} {2014})}\BibitemShut
  {NoStop}%
\bibitem [{\citenamefont {Gring}\ \emph {et~al.}(2012)\citenamefont {Gring},
  \citenamefont {Kuhnert}, \citenamefont {Langen}, \citenamefont {Kitagawa},
  \citenamefont {Rauer}, \citenamefont {Schreitl}, \citenamefont {Mazets},
  \citenamefont {Smith}, \citenamefont {Demler},\ and\ \citenamefont
  {Schmiedmayer}}]{Gring2012}%
  \BibitemOpen
  \bibfield  {author} {\bibinfo {author} {\bibfnamefont {M.}~\bibnamefont
  {Gring}}, \bibinfo {author} {\bibfnamefont {M.}~\bibnamefont {Kuhnert}},
  \bibinfo {author} {\bibfnamefont {T.}~\bibnamefont {Langen}}, \bibinfo
  {author} {\bibfnamefont {T.}~\bibnamefont {Kitagawa}}, \bibinfo {author}
  {\bibfnamefont {B.}~\bibnamefont {Rauer}}, \bibinfo {author} {\bibfnamefont
  {M.}~\bibnamefont {Schreitl}}, \bibinfo {author} {\bibfnamefont
  {I.}~\bibnamefont {Mazets}}, \bibinfo {author} {\bibfnamefont {D.~A.}\
  \bibnamefont {Smith}}, \bibinfo {author} {\bibfnamefont {E.}~\bibnamefont
  {Demler}}, \ and\ \bibinfo {author} {\bibfnamefont {J.}~\bibnamefont
  {Schmiedmayer}},\ }\bibfield  {title} {\enquote {\bibinfo {title}
  {{Relaxation and Prethermalization in an Isolated Quantum System}},}\ }\href
  {\doibase 10.1126/science.1224953} {\bibfield  {journal} {\bibinfo  {journal}
  {Science}\ }\textbf {\bibinfo {volume} {337}},\ \bibinfo {pages} {1318--1322}
  (\bibinfo {year} {2012})}\BibitemShut {NoStop}%
\end{thebibliography}

\begin{thebibliography}{25}%
\makeatletter
\providecommand \@ifxundefined [1]{%
 \@ifx{#1\undefined}
}%
\providecommand \@ifnum [1]{%
 \ifnum #1\expandafter \@firstoftwo
 \else \expandafter \@secondoftwo
 \fi
}%
\providecommand \@ifx [1]{%
 \ifx #1\expandafter \@firstoftwo
 \else \expandafter \@secondoftwo
 \fi
}%
\providecommand \natexlab [1]{#1}%
\providecommand \enquote  [1]{``#1''}%
\providecommand \bibnamefont  [1]{#1}%
\providecommand \bibfnamefont [1]{#1}%
\providecommand \citenamefont [1]{#1}%
\providecommand \href@noop [0]{\@secondoftwo}%
\providecommand \href [0]{\begingroup \@sanitize@url \@href}%
\providecommand \@href[1]{\@@startlink{#1}\@@href}%
\providecommand \@@href[1]{\endgroup#1\@@endlink}%
\providecommand \@sanitize@url [0]{\catcode `\\12\catcode `\$12\catcode
  `\&12\catcode `\#12\catcode `\^12\catcode `\_12\catcode `\%12\relax}%
\providecommand \@@startlink[1]{}%
\providecommand \@@endlink[0]{}%
\providecommand \url  [0]{\begingroup\@sanitize@url \@url }%
\providecommand \@url [1]{\endgroup\@href {#1}{\urlprefix }}%
\providecommand \urlprefix  [0]{URL }%
\providecommand \Eprint [0]{\href }%
\providecommand \doibase [0]{http://dx.doi.org/}%
\providecommand \selectlanguage [0]{\@gobble}%
\providecommand \bibinfo  [0]{\@secondoftwo}%
\providecommand \bibfield  [0]{\@secondoftwo}%
\providecommand \translation [1]{[#1]}%
\providecommand \BibitemOpen [0]{}%
\providecommand \bibitemStop [0]{}%
\providecommand \bibitemNoStop [0]{.\EOS\space}%
\providecommand \EOS [0]{\spacefactor3000\relax}%
\providecommand \BibitemShut  [1]{\csname bibitem#1\endcsname}%
\let\auto@bib@innerbib\@empty
\bibitem [{\citenamefont {Baumgartner}\ and\ \citenamefont
  {Narnhofer}(2008)}]{Baumgartner2008}%
  \BibitemOpen
  \bibfield  {author} {\bibinfo {author} {\bibfnamefont {B.}~\bibnamefont
  {Baumgartner}}\ and\ \bibinfo {author} {\bibfnamefont {H.}~\bibnamefont
  {Narnhofer}},\ }\bibfield  {title} {\enquote {\bibinfo {title} {{Analysis of
  quantum semigroups with GKS--Lindblad generators: II. General}},}\ }\href
  {http://stacks.iop.org/1751-8121/41/i=39/a=395303} {\bibfield  {journal}
  {\bibinfo  {journal} {J. Phys. A}\ }\textbf {\bibinfo {volume} {41}},\
  \bibinfo {pages} {395303} (\bibinfo {year} {2008})}\BibitemShut {NoStop}%
\bibitem [{\citenamefont {Wolf}(2012)}]{Wolf2012}%
  \BibitemOpen
  \bibfield  {author} {\bibinfo {author} {\bibfnamefont {M.~M.}\ \bibnamefont
  {Wolf}},\ }\href
  {https://www-m5.ma.tum.de/foswiki/pub/M5/Allgemeines/MichaelWolf/QChannelLecture.pdf}
  {\emph {\bibinfo {title} {{Quantum channels and Operations, Guided tour}}}}\
  (\bibinfo {year} {2012})\BibitemShut {NoStop}%
\bibitem [{\citenamefont {Albert}\ \emph {et~al.}(2016)\citenamefont {Albert},
  \citenamefont {Bradlyn}, \citenamefont {Fraas},\ and\ \citenamefont
  {Jiang}}]{Albert2016}%
  \BibitemOpen
  \bibfield  {author} {\bibinfo {author} {\bibfnamefont {V.~V.}\ \bibnamefont
  {Albert}}, \bibinfo {author} {\bibfnamefont {B.}~\bibnamefont {Bradlyn}},
  \bibinfo {author} {\bibfnamefont {M.}~\bibnamefont {Fraas}}, \ and\ \bibinfo
  {author} {\bibfnamefont {L.}~\bibnamefont {Jiang}},\ }\bibfield  {title}
  {\enquote {\bibinfo {title} {{Geometry and Response of Lindbladians}},}\
  }\href {\doibase 10.1103/PhysRevX.6.041031} {\bibfield  {journal} {\bibinfo
  {journal} {Phys. Rev. X}\ }\textbf {\bibinfo {volume} {6}},\ \bibinfo {pages}
  {041031} (\bibinfo {year} {2016})}\BibitemShut {NoStop}%
\bibitem [{\citenamefont {Kato}(1995)}]{Kato1995}%
  \BibitemOpen
  \bibfield  {author} {\bibinfo {author} {\bibfnamefont {T.}~\bibnamefont
  {Kato}},\ }\href@noop {} {\emph {\bibinfo {title} {{Perturbation Theory for
  Linear Operators}}}}\ (\bibinfo  {publisher} {Springer},\ \bibinfo {year}
  {1995})\BibitemShut {NoStop}%
\bibitem [{\citenamefont {Macieszczak}\ \emph
  {et~al.}(2016{\natexlab{a}})\citenamefont {Macieszczak}, \citenamefont
  {Guta}, \citenamefont {Lesanovsky},\ and\ \citenamefont
  {Garrahan}}]{Macieszczak2016a}%
  \BibitemOpen
  \bibfield  {author} {\bibinfo {author} {\bibfnamefont {K.}~\bibnamefont
  {Macieszczak}}, \bibinfo {author} {\bibfnamefont {M.}~\bibnamefont {Guta}},
  \bibinfo {author} {\bibfnamefont {I.}~\bibnamefont {Lesanovsky}}, \ and\
  \bibinfo {author} {\bibfnamefont {J.~P.}\ \bibnamefont {Garrahan}},\
  }\bibfield  {title} {\enquote {\bibinfo {title} {{Towards a Theory of
  Metastability in Open Quantum Dynamics}},}\ }\href {\doibase
  10.1103/PhysRevLett.116.240404} {\bibfield  {journal} {\bibinfo  {journal}
  {Phys. Rev. Lett.}\ }\textbf {\bibinfo {volume} {116}},\ \bibinfo {pages}
  {240404} (\bibinfo {year} {2016}{\natexlab{a}})}\BibitemShut {NoStop}%
\bibitem [{Note1()}]{Note1}%
  \BibitemOpen
  \bibinfo {note} {{\protect \color {black}{For an integer $n\geq 1$ such that
  $t''\leq t_n\equiv t/n \leq t'$, we have $\delimiter 69645069 \rho (t)-e^{t
  {\protect \mathcal {L}_\protect \text {MM}}}\protect \mathcal {P}[\rho (0)]
  \delimiter 86422285 =\delimiter 69645069 [e^{t_n {\protect \mathcal
  {L}}}-\protect \mathcal {P}]^n({\protect \mathcal {I}}-\protect \mathcal
  {P})[\rho (0)]\delimiter 86422285 \leq \delimiter 69645069 e^{t_n {\protect
  \mathcal {L}}}-\protect \mathcal {P}\delimiter 86422285 ^n \delimiter
  69645069 {\protect \mathcal {I}}-\protect \mathcal {P}\delimiter 86422285 $.
  Since we have ${\delimiter 69645069 \rho (t_n)-\protect \mathcal {P}[\rho
  (0)]\delimiter 86422285 \leq \protect \mathcal {C}_\protect \text {MM}}$
  [cf.~Eq.~\protect \textup {\hbox {\mathsurround \z@ \protect \normalfont
  (\ignorespaces \ref {eq:Cmm}\unskip \@@italiccorr )}}], while $\delimiter
  69645069 {\protect \mathcal {I}}-\protect \mathcal {P}\delimiter 86422285
  \leq \delimiter 69645069 {\protect \mathcal {I}}\delimiter 86422285
  +\delimiter 69645069 \protect \mathcal {P}\delimiter 86422285 = 2+\protect
  \mathcal {C}_{+}$ [cf.~Eq.~\protect \textup {\hbox {\mathsurround \z@
  \protect \normalfont (\ignorespaces \ref {eq:Cp}\unskip \@@italiccorr )}}],
  we arrive at $\delimiter 69645069 \rho (t)-e^{t {\protect \mathcal
  {L}_\protect \text {MM}}}\protect \mathcal {P}[\rho (0)] \delimiter 86422285
  \leq \protect \mathcal {C}_\protect \text {MM}^n(2+\protect \mathcal
  {C}_{+})\lesssim 2\protect \mathcal {C}_\protect \text
  {MM}^n$.}}}\BibitemShut {Stop}%
\bibitem [{\citenamefont {Gaveau}\ and\ \citenamefont
  {Schulman}(2006)}]{Gaveau2006}%
  \BibitemOpen
  \bibfield  {author} {\bibinfo {author} {\bibfnamefont {B.}~\bibnamefont
  {Gaveau}}\ and\ \bibinfo {author} {\bibfnamefont {L.~S.}\ \bibnamefont
  {Schulman}},\ }\bibfield  {title} {\enquote {\bibinfo {title} {{Multiple
  phases in stochastic dynamics: Geometry and probabilities}},}\ }\href
  {\doibase 10.1103/PhysRevE.73.036124} {\bibfield  {journal} {\bibinfo
  {journal} {Phys. Rev. E}\ }\textbf {\bibinfo {volume} {73}},\ \bibinfo
  {pages} {036124} (\bibinfo {year} {2006})}\BibitemShut {NoStop}%
\bibitem [{\citenamefont {\.Zyczkowski}\ and\ \citenamefont
  {Sommers}(2001)}]{Zyczkowski2001}%
  \BibitemOpen
  \bibfield  {author} {\bibinfo {author} {\bibfnamefont {K.}~\bibnamefont
  {\.Zyczkowski}}\ and\ \bibinfo {author} {\bibfnamefont {H.-J.}\ \bibnamefont
  {Sommers}},\ }\bibfield  {title} {\enquote {\bibinfo {title} {{Induced
  measures in the space of mixed quantum states}},}\ }\href
  {http://stacks.iop.org/0305-4470/34/i=35/a=335} {\bibfield  {journal}
  {\bibinfo  {journal} {J. Phys. A}\ }\textbf {\bibinfo {volume} {34}},\
  \bibinfo {pages} {7111} (\bibinfo {year} {2001})}\BibitemShut {NoStop}%
\bibitem [{\citenamefont {\.Zyczkowski}\ and\ \citenamefont
  {Sommers}(2003)}]{Zyczkowski2003}%
  \BibitemOpen
  \bibfield  {author} {\bibinfo {author} {\bibfnamefont {K.}~\bibnamefont
  {\.Zyczkowski}}\ and\ \bibinfo {author} {\bibfnamefont {H.-J.}\ \bibnamefont
  {Sommers}},\ }\bibfield  {title} {\enquote {\bibinfo {title}
  {{Hilbert--Schmidt volume of the set of mixed quantum states}},}\ }\href
  {http://stacks.iop.org/0305-4470/36/i=39/a=310} {\bibfield  {journal}
  {\bibinfo  {journal} {J. Phys. A}\ }\textbf {\bibinfo {volume} {36}},\
  \bibinfo {pages} {10115} (\bibinfo {year} {2003})}\BibitemShut {NoStop}%
\bibitem [{\citenamefont {Nielsen}\ and\ \citenamefont
  {Chuang}(2010)}]{Nielsen2010}%
  \BibitemOpen
  \bibfield  {author} {\bibinfo {author} {\bibfnamefont {M.~A.}\ \bibnamefont
  {Nielsen}}\ and\ \bibinfo {author} {\bibfnamefont {I.~L.}\ \bibnamefont
  {Chuang}},\ }\href@noop {} {\emph {\bibinfo {title} {{Quantum Computation and
  Quantum Information}}}}\ (\bibinfo  {publisher} {Cambridge University Press,
  10th Anniversary ed. edition},\ \bibinfo {year} {2010})\BibitemShut {NoStop}%
\bibitem [{\citenamefont {Rose}\ \emph {et~al.}(2016)\citenamefont {Rose},
  \citenamefont {Macieszczak}, \citenamefont {Lesanovsky},\ and\ \citenamefont
  {Garrahan}}]{Rose2016}%
  \BibitemOpen
  \bibfield  {author} {\bibinfo {author} {\bibfnamefont {D.~C.}\ \bibnamefont
  {Rose}}, \bibinfo {author} {\bibfnamefont {K.}~\bibnamefont {Macieszczak}},
  \bibinfo {author} {\bibfnamefont {I.}~\bibnamefont {Lesanovsky}}, \ and\
  \bibinfo {author} {\bibfnamefont {J.~P.}\ \bibnamefont {Garrahan}},\
  }\bibfield  {title} {\enquote {\bibinfo {title} {{Metastability in an open
  quantum Ising model}},}\ }\href {\doibase 10.1103/PhysRevE.94.052132}
  {\bibfield  {journal} {\bibinfo  {journal} {Phys. Rev. E}\ }\textbf {\bibinfo
  {volume} {94}},\ \bibinfo {pages} {052132} (\bibinfo {year}
  {2016})}\BibitemShut {NoStop}%
\bibitem [{\citenamefont {Hickey}\ \emph {et~al.}(2012)\citenamefont {Hickey},
  \citenamefont {Genway}, \citenamefont {Lesanovsky},\ and\ \citenamefont
  {Garrahan}}]{Hickey2012}%
  \BibitemOpen
  \bibfield  {author} {\bibinfo {author} {\bibfnamefont {J.~M.}\ \bibnamefont
  {Hickey}}, \bibinfo {author} {\bibfnamefont {S.}~\bibnamefont {Genway}},
  \bibinfo {author} {\bibfnamefont {I.}~\bibnamefont {Lesanovsky}}, \ and\
  \bibinfo {author} {\bibfnamefont {J.~P.}\ \bibnamefont {Garrahan}},\
  }\bibfield  {title} {\enquote {\bibinfo {title} {{Thermodynamics of
  quadrature trajectories in open quantum systems}},}\ }\href {\doibase
  10.1103/PhysRevA.86.063824} {\bibfield  {journal} {\bibinfo  {journal} {Phys.
  Rev. A}\ }\textbf {\bibinfo {volume} {86}},\ \bibinfo {pages} {063824}
  (\bibinfo {year} {2012})}\BibitemShut {NoStop}%
\bibitem [{\citenamefont {Levitov}\ \emph {et~al.}(1996)\citenamefont
  {Levitov}, \citenamefont {Lee},\ and\ \citenamefont {Lesovik}}]{Levitov1996}%
  \BibitemOpen
  \bibfield  {author} {\bibinfo {author} {\bibfnamefont {L.~S.}\ \bibnamefont
  {Levitov}}, \bibinfo {author} {\bibfnamefont {H.}~\bibnamefont {Lee}}, \ and\
  \bibinfo {author} {\bibfnamefont {G.~B.}\ \bibnamefont {Lesovik}},\
  }\bibfield  {title} {\enquote {\bibinfo {title} {{Electron counting
  statistics and coherent states of electric current}},}\ }\href {\doibase
  10.1063/1.531672} {\bibfield  {journal} {\bibinfo  {journal} {J. Math.
  Phys.}\ }\textbf {\bibinfo {volume} {37}},\ \bibinfo {pages} {4845--4866}
  (\bibinfo {year} {1996})}\BibitemShut {NoStop}%
\bibitem [{\citenamefont {Nazarov}\ and\ \citenamefont
  {Division}(2003)}]{Nazarov2003}%
  \BibitemOpen
  \bibfield  {author} {\bibinfo {author} {\bibfnamefont {Y.~V.}\ \bibnamefont
  {Nazarov}}\ and\ \bibinfo {author} {\bibfnamefont {N.~A. T. O. S.~A.}\
  \bibnamefont {Division}},\ }\href@noop {} {\emph {\bibinfo {title} {{Quantum
  Noise in Mesoscopic Physics, NATO Science Series: Mathematics, Physics and
  Chemistry}}}}\ (\bibinfo  {publisher} {Springer, New York},\ \bibinfo {year}
  {2003})\BibitemShut {NoStop}%
\bibitem [{\citenamefont {Esposito}\ \emph {et~al.}(2009)\citenamefont
  {Esposito}, \citenamefont {Harbola},\ and\ \citenamefont
  {Mukamel}}]{Esposito2009}%
  \BibitemOpen
  \bibfield  {author} {\bibinfo {author} {\bibfnamefont {M.}~\bibnamefont
  {Esposito}}, \bibinfo {author} {\bibfnamefont {U.}~\bibnamefont {Harbola}}, \
  and\ \bibinfo {author} {\bibfnamefont {S.}~\bibnamefont {Mukamel}},\
  }\bibfield  {title} {\enquote {\bibinfo {title} {{Nonequilibrium
  fluctuations, fluctuation theorems, and counting statistics in quantum
  systems}},}\ }\href {\doibase 10.1103/RevModPhys.81.1665} {\bibfield
  {journal} {\bibinfo  {journal} {Rev. Mod. Phys.}\ }\textbf {\bibinfo {volume}
  {81}},\ \bibinfo {pages} {1665--1702} (\bibinfo {year} {2009})}\BibitemShut
  {NoStop}%
\bibitem [{\citenamefont {Flindt}\ \emph {et~al.}(2009)\citenamefont {Flindt},
  \citenamefont {Fricke}, \citenamefont {Hohls}, \citenamefont {Novotn{\'y}},
  \citenamefont {Neto{\v c}n{\'y}}, \citenamefont {Brandes},\ and\
  \citenamefont {Haug}}]{Flindt2009}%
  \BibitemOpen
  \bibfield  {author} {\bibinfo {author} {\bibfnamefont {C.}~\bibnamefont
  {Flindt}}, \bibinfo {author} {\bibfnamefont {C.}~\bibnamefont {Fricke}},
  \bibinfo {author} {\bibfnamefont {F.}~\bibnamefont {Hohls}}, \bibinfo
  {author} {\bibfnamefont {T.}~\bibnamefont {Novotn{\'y}}}, \bibinfo {author}
  {\bibfnamefont {K.}~\bibnamefont {Neto{\v c}n{\'y}}}, \bibinfo {author}
  {\bibfnamefont {T.}~\bibnamefont {Brandes}}, \ and\ \bibinfo {author}
  {\bibfnamefont {R.~J.}\ \bibnamefont {Haug}},\ }\bibfield  {title} {\enquote
  {\bibinfo {title} {{Universal oscillations in counting statistics}},}\ }\href
  {\doibase 10.1073/pnas.0901002106} {\bibfield  {journal} {\bibinfo  {journal}
  {Proc. Natl. Acad. Sci.}\ }\textbf {\bibinfo {volume} {106}},\ \bibinfo
  {pages} {10116--10119} (\bibinfo {year} {2009})}\BibitemShut {NoStop}%
\bibitem [{\citenamefont {Hickey}\ \emph {et~al.}(2013)\citenamefont {Hickey},
  \citenamefont {Genway}, \citenamefont {Lesanovsky},\ and\ \citenamefont
  {Garrahan}}]{Hickey2013}%
  \BibitemOpen
  \bibfield  {author} {\bibinfo {author} {\bibfnamefont {J.~M.}\ \bibnamefont
  {Hickey}}, \bibinfo {author} {\bibfnamefont {S.}~\bibnamefont {Genway}},
  \bibinfo {author} {\bibfnamefont {I.}~\bibnamefont {Lesanovsky}}, \ and\
  \bibinfo {author} {\bibfnamefont {J.~P.}\ \bibnamefont {Garrahan}},\
  }\bibfield  {title} {\enquote {\bibinfo {title} {{Time-integrated observables
  as order parameters for full counting statistics transitions in closed
  quantum systems}},}\ }\href {\doibase 10.1103/PhysRevB.87.184303} {\bibfield
  {journal} {\bibinfo  {journal} {Phys. Rev. B}\ }\textbf {\bibinfo {volume}
  {87}},\ \bibinfo {pages} {184303} (\bibinfo {year} {2013})}\BibitemShut
  {NoStop}%
\bibitem [{Note2()}]{Note2}%
  \BibitemOpen
  \bibinfo {note} {$H$ can be further replaced by $H-c\protect \mathds {1}$,
  with $c$ being a real constant and the norm minimized with respect to $c$ in
  the corrections; see Sec.~\ref {app:Ws}.}\BibitemShut {Stop}%
\bibitem [{Note3()}]{Note3}%
  \BibitemOpen
  \bibinfo {note} {For $\delimiter 69645069 {\protect \mathcal {J}}\protect
  \mathcal {P}\delimiter 86422285 \leq \delimiter 69645069 (1+\protect \mathcal
  {C}_{\protect \text {cl}}) \protect \bm {{\protect \mathaccentV
  {tilde}07E{\mu }}}\delimiter 86422285 _1+\protect \mathcal {C}_{+}\delimiter
  69645069 {\protect \mathcal {J}}\delimiter 86422285 $, we note that
  $\delimiter 69645069 {\protect \mathcal {J}}\protect \mathcal {P}(\rho
  )\delimiter 86422285 \leq {\protect \mathrm {Tr}}{\protect \mathcal {J}}(\rho
  ') +\delimiter 69645069 {\protect \mathcal {J}}\delimiter 86422285 \delimiter
  69645069 \rho '-\protect \mathcal {P}\rho ' \delimiter 86422285 \leq | \DOTSB
  \sum@ \slimits@ _{l=1}^m {\protect \mathaccentV {tilde}07E{\mu }}_l \protect
  \mathaccentV {tilde}07E{p}_l |+2\delimiter 69645069 {\protect \mathcal
  {J}}\delimiter 86422285 \delimiter 69645069 \rho '-\protect \mathcal {P}\rho
  ' \delimiter 86422285 $, and choose $\rho '$ as the closest state to
  $\protect \mathcal {P}(\rho )$ [cf.~Eq.~\protect \textup {\hbox
  {\mathsurround \z@ \protect \normalfont (\ignorespaces \ref {eq:Cp}\unskip
  \@@italiccorr )}}]}\BibitemShut {NoStop}%
\bibitem [{\citenamefont {Macieszczak}\ \emph
  {et~al.}(2016{\natexlab{b}})\citenamefont {Macieszczak}, \citenamefont
  {Gu\ifmmode \mbox{\c{t}}\else \c{t}\fi{}\ifmmode~\u{a}\else \u{a}\fi{}},
  \citenamefont {Lesanovsky},\ and\ \citenamefont
  {Garrahan}}]{Macieszczak2016}%
  \BibitemOpen
  \bibfield  {author} {\bibinfo {author} {\bibfnamefont {K.}~\bibnamefont
  {Macieszczak}}, \bibinfo {author} {\bibfnamefont {M.}~\bibnamefont
  {Gu\ifmmode \mbox{\c{t}}\else \c{t}\fi{}\ifmmode~\u{a}\else \u{a}\fi{}}},
  \bibinfo {author} {\bibfnamefont {I.}~\bibnamefont {Lesanovsky}}, \ and\
  \bibinfo {author} {\bibfnamefont {J.~P.}\ \bibnamefont {Garrahan}},\
  }\bibfield  {title} {\enquote {\bibinfo {title} {{Dynamical phase transitions
  as a resource for quantum enhanced metrology}},}\ }\href {\doibase
  10.1103/PhysRevA.93.022103} {\bibfield  {journal} {\bibinfo  {journal} {Phys.
  Rev. A}\ }\textbf {\bibinfo {volume} {93}},\ \bibinfo {pages} {022103}
  (\bibinfo {year} {2016}{\natexlab{b}})}\BibitemShut {NoStop}%
\bibitem [{\citenamefont {Gaveau}\ \emph {et~al.}(1999)\citenamefont {Gaveau},
  \citenamefont {Lesne},\ and\ \citenamefont {Schulman}}]{Gaveau1999}%
  \BibitemOpen
  \bibfield  {author} {\bibinfo {author} {\bibfnamefont {B.}~\bibnamefont
  {Gaveau}}, \bibinfo {author} {\bibfnamefont {A.}~\bibnamefont {Lesne}}, \
  and\ \bibinfo {author} {\bibfnamefont {L.~S.}\ \bibnamefont {Schulman}},\
  }\bibfield  {title} {\enquote {\bibinfo {title} {{Spectral signatures of
  hierarchical relaxation}},}\ }\href {\doibase
  http://dx.doi.org/10.1016/S0375-9601(99)00369-2} {\bibfield  {journal}
  {\bibinfo  {journal} {Phys. Lett. A}\ }\textbf {\bibinfo {volume} {258}},\
  \bibinfo {pages} {222 -- 228} (\bibinfo {year} {1999})}\BibitemShut {NoStop}%
\bibitem [{\citenamefont {Gaveau}\ and\ \citenamefont
  {Schulman}(1998)}]{Gaveau1998}%
  \BibitemOpen
  \bibfield  {author} {\bibinfo {author} {\bibfnamefont {B.}~\bibnamefont
  {Gaveau}}\ and\ \bibinfo {author} {\bibfnamefont {L.~S.}\ \bibnamefont
  {Schulman}},\ }\bibfield  {title} {\enquote {\bibinfo {title} {{Theory of
  nonequilibrium first-order phase transitions for stochastic dynamics}},}\
  }\href {\doibase http://dx.doi.org/10.1063/1.532394} {\bibfield  {journal}
  {\bibinfo  {journal} {J. Mat. Phys.}\ }\textbf {\bibinfo {volume} {39}},\
  \bibinfo {pages} {1517} (\bibinfo {year} {1998})}\BibitemShut {NoStop}%
\bibitem [{\citenamefont {Bu{\v c}a}\ and\ \citenamefont
  {Prosen}(2012)}]{Buvca2012}%
  \BibitemOpen
  \bibfield  {author} {\bibinfo {author} {\bibfnamefont {B.}~\bibnamefont
  {Bu{\v c}a}}\ and\ \bibinfo {author} {\bibfnamefont {T.}~\bibnamefont
  {Prosen}},\ }\bibfield  {title} {\enquote {\bibinfo {title} {{A note on
  symmetry reductions of the Lindblad equation: transport in constrained open
  spin chains}},}\ }\href {http://stacks.iop.org/1367-2630/14/i=7/a=073007}
  {\bibfield  {journal} {\bibinfo  {journal} {New J. Phys.}\ }\textbf {\bibinfo
  {volume} {14}},\ \bibinfo {pages} {073007} (\bibinfo {year}
  {2012})}\BibitemShut {NoStop}%
\bibitem [{\citenamefont {Albert}\ and\ \citenamefont
  {Jiang}(2014)}]{Albert2014}%
  \BibitemOpen
  \bibfield  {author} {\bibinfo {author} {\bibfnamefont {V.~V.}\ \bibnamefont
  {Albert}}\ and\ \bibinfo {author} {\bibfnamefont {L.}~\bibnamefont {Jiang}},\
  }\bibfield  {title} {\enquote {\bibinfo {title} {{Symmetries and conserved
  quantities in Lindblad master equations}},}\ }\href {\doibase
  10.1103/PhysRevA.89.022118} {\bibfield  {journal} {\bibinfo  {journal} {Phys.
  Rev. A}\ }\textbf {\bibinfo {volume} {89}},\ \bibinfo {pages} {022118}
  (\bibinfo {year} {2014})}\BibitemShut {NoStop}%
\bibitem [{Note4()}]{Note4}%
  \BibitemOpen
  \bibinfo {note} {We have $\protect \qopname \relax m{det}{{\protect \mathbf
  {C}}}=\protect \qopname \relax m{det}{\protect \mathaccentV
  {bar}016{{\protect \mathbf {C}}}}$, where $(\protect \mathaccentV
  {bar}016{{\protect \mathbf {C}}})_{l-1,k-1}=c_{k}^{(l)}-c_{k}^{(1)}$,
  $k,l=2,..,m$, encodes the coefficients for the simplex with the vertex of
  ${\protect \mathaccentV {tilde}07E{\rho }}_1$ shifted to the
  origin.}\BibitemShut {Stop}%
\end{thebibliography}

%

\end{document}